\documentclass[11pt, twoside]{report}
\usepackage[top=30mm, bottom=30mm, left=30mm, right=30mm]{geometry}
\usepackage[utf8]{inputenc}
\usepackage{url}
\usepackage{epsfig,pict2e}
\usepackage[titletoc,page]{appendix}
\usepackage{tikz-cd}
\usepackage{float}

\usepackage{fancyhdr}
\newcommand{\hypergeometric}[6][\bigg]{\,{}_{#2} F_{#3} #1
( \begin{matrix} #4 \\ #5 \end{matrix}\, #1\vert\, #6 #1)}

\newcommand{\meijer}[6][\bigg]{G^{#2}_{#3} #1
( \begin{matrix} #4 \\ #5 \end{matrix}\, #1\vert\, #6 #1)}

\let\origdoublepage\cleardoublepage
\newcommand{\clearemptydoublepage}{%
  \clearpage
  {\pagestyle{empty}\origdoublepage}%
}

\let\plainappendixpage\appendixpage
\makeatletter
\renewcommand{\appendixpage}{%
\begingroup
\let\ps@plain\ps@empty
\plainappendixpage
\endgroup}
\makeatother

\tikzcdset{every label/.append style = {font = \small}}

\usepackage{amsmath,amsfonts,amssymb,relsize,geometry}
\usepackage[numbers]{natbib}
\pdfoutput = 1

\usepackage[english]{babel}
\usepackage{amsmath,amssymb,amsthm}
\usepackage{amsfonts}
\usepackage{amsxtra}
\usepackage{array,dcolumn}
\usepackage{graphicx}
\usepackage[font=small,labelfont=bf]{caption}
\usepackage{chngcntr}
 
\counterwithin{figure}{chapter}

\usepackage[hyperfootnotes=true,
        pdffitwindow=true,
        plainpages=false,
        pdfpagelabels=true,
        pdfpagemode=UseOutlines,
        pdfpagelayout=SinglePage,
        pagebackref,
        hyperindex,hidelinks]{hyperref}

\usepackage[activate={true,nocompatibility},spacing,kerning]{microtype}
\usepackage[sc,osf]{mathpazo}
\microtypecontext{spacing=nonfrench}

\newcommand{\mathsym}[1]{{}}
\newcommand{\unicode}[1]{{}}

\theoremstyle{plain}
\newtheorem{theorem}{Theorem}
\numberwithin{theorem}{chapter}
\newtheorem{lemma}{Lemma}
\numberwithin{lemma}{chapter}
\newtheorem{corollary}{Corollary}
\numberwithin{corollary}{chapter}
\newtheorem{proposition}{Proposition}
\numberwithin{proposition}{chapter}

\theoremstyle{definition}
\newtheorem{definition}{Definition}
\numberwithin{definition}{chapter}
\newtheorem{example}{Example}
\numberwithin{example}{chapter}

\theoremstyle{remark}
\newtheorem{remark}{Remark}
\numberwithin{remark}{chapter}
\newtheorem{note}{Note}
\numberwithin{note}{chapter}

\newcommand{\E}{\mathbb{E}}
\newcommand{\+}{\!+\!}
\newcommand{\m}{\!\\!}
\newcommand{\half}{\tfrac{1}{2}}

\renewcommand{\leq}{\leqslant}
\renewcommand{\geq}{\geqslant}
\DeclareMathOperator{\arcsinh}{arcsinh}
\DeclareMathOperator{\Tr}{Tr}
\DeclareMathOperator{\Det}{Det}
\DeclareMathOperator{\Pf}{Pf}
\def\mean#1{\left< #1 \right>}

\clearpage
\begin{document}
\thispagestyle{plain}
\begin{titlepage}
    \begin{center}
        \vspace*{1cm}
        
\LARGE
\textbf{Recursive characterisations of random matrix ensembles and associated combinatorial objects}
        
        \vspace{1cm}
       
       \Large
\textbf{Anas A.~Rahman}\\
       \normalsize
ORCID iD: 0000-0003-2317-6685

\vspace{2.5cm}

        \large      
Doctor of Philosophy -- Science
\\January, 2022
\\The School of Mathematics and Statistics
\\The University of Melbourne
        
\vfill

        \normalsize
Submitted in total fulfilment for the degree of
\\Doctor of Philosophy -- Science
    \end{center}
\end{titlepage}

\clearemptydoublepage
\setcounter{page}{1}
\phantomsection
\addcontentsline{toc}{chapter}{Abstract}
\begin{center}
    \large
    \textbf{Abstract}
\end{center}
\small

Mixed moments and cumulants of random matrices have been studied extensively over the last half-century with applications in a large variety of fields ranging from enumerative geometry to quantum mechanics. It is particularly interesting to expand the cumulants in $1/N$, where $N$ denotes matrix size, and study the coefficients of these expansions along with their generating functions, which we call correlator expansion coefficients. We give an overview of the recursive characterisations of random matrix ensembles that are currently at the forefront of random matrix theory by way of studying two classes of ensembles using two different types of recursive schemes: Established theory on Selberg correlation integrals is used to derive linear differential equations on the eigenvalue densities and resolvents of the classical matrix ensembles, which lead to $1$-point recursions, understood to be analogues of the Harer--Zagier recursion, for the expansion coefficients of the associated $1$-point cumulants, while loop equation analysis is used to recursively compute some leading order correlator expansion coefficients pertaining to certain products of random matrices that have recently come into interest due to their connections to Muttalib--Borodin ensembles and integrals of Harish-Chandra--Itzykson--Zuber type. We also show how the aforementioned differential equations can be used to characterise the large $N$ limiting statistics of the classical matrix ensembles' eigenvalue densities in the global and edge scaling regimes.

A common theme between the two classes of ensembles studied in this thesis is that their representative random matrices can, for the most part, be constructed from Ginibre matrices (matrices whose entries are independently and identically distributed normal variables), which allows for their mixed moments and cumulants to be interpreted as enumerations of particular ribbon graphs, topological and combinatorial maps, and/or topological and combinatorial hypermaps. A major part of this thesis is devoted to a comprehensive review of how the Isserlis--Wick theorem implies these interpretations for the mixed moments and cumulants of the Gaussian and Laguerre ensembles, which leads on to original discussion on how the relevant theory extends to the matrix products mentioned above. Thus, the loop equations derived in this thesis have the added value of solving certain problems in enumerative combinatorics.

In order to make this thesis self-contained and to properly motivate our original contributions, a decent portion of our development constitutes a pedagogically detailed survey of the contiguous literature. It is therefore expected that this thesis will serve as a valuable resource for readers wanting a well-rounded introduction to classical matrix ensembles, matrix product ensembles, $1$-point recursions, loop equations, Selberg correlation integrals, and ribbon graphs.

\normalsize

\newpage
\phantomsection
\addcontentsline{toc}{chapter}{Declaration}
\begin{center}
    \large
    \textbf{Declaration}
\end{center}

This is to certify that:
\begin{enumerate}
\item the thesis comprises only my original work towards the degree of Doctor of Philosophy except where indicated in the preface;
\item due acknowledgement has been made in the text to all other material used; and
\item the thesis is fewer than 100\,000 words in length, exclusive of bibliographies, tables, maps, and appendices.
\end{enumerate}

\vspace{1.5cm}
\begin{flushright}
\includegraphics[width=0.3\textwidth]{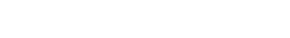}\\ \vspace{-0.5cm}
\rule{12em}{0.6pt}\\
\textbf{Anas A. Rahman}\\
January 22, 2022
\end{flushright}

\newpage
\phantomsection
\addcontentsline{toc}{chapter}{Preface}
\begin{center}
    \large
    \textbf{Preface}
\end{center}
The introductory content of Chapter 1 is wholly original to this thesis. So too are the contents of Chapter 2 preceding \S\ref{s2.1.2}, whereas the content of \S\ref{s2.1.2} is drawn from the work ``Linear differential equations for the resolvents of the classical matrix ensembles'' \citep{RF21} done by the present author in collaboration with Peter Forrester and published by \textit{Random Matrices: Theory and Applications} in 2021. Section \ref{s2.2} contains a mild enhancement of content from the work ``Relations between moments for the Jacobi and Cauchy random matrix ensembles'' \citep{FR21} done by the present author in collaboration with Peter Forrester and published by the \textit{Journal of Mathematical Physics} in 2021. The contribution of the present author towards the works \citep{RF21}, \citep{FR21} is equal to that of Forrester.

Thereafter, the contents of Sections \ref{s2.3}, \ref{s2.4}, and \ref{s3.1} are amalgamations of content from \citep{RF21}, \citep{FR21} reworked slightly for coherence and supplemented with additional, original results for completeness; Appendix \ref{appendixB} is taken from \citep{RF21}. In \S\ref{s3.3.3}, Section \ref{s4.2}, Section~\ref{s4.3}, Appendix \ref{appendixC}, and Appendix \ref{appendixD}, we give a heavily detailed presentation of the ongoing work ``Combinatorics and loop equations for antisymmetrised and Hermitised matrix product ensembles'' \citep{DR22} of the present author and Stephane Dartois. The results pertaining to \citep{DR22} are born of vigorous discussions between the present author and Dartois, but the corresponding outcomes displayed in this thesis are due solely to the present author.

The 2017 article \citep{FRW17} by the present author, Peter Forrester, and Nicholas Witte is briefly reviewed in \S\ref{s4.1.1}, but is work mostly completed prior to commencement of this thesis. Thus, it is cited appropriately where necessary. All other contents of this thesis not addressed in the above are original to this thesis and have not been published elsewhere.

The candidature of the author was financially supported by the Australian Government Research Training Program Scholarship, the ARC Centre of Excellence for Mathematical and Statistical Frontiers, and the ARC Grant DP210102887.

\newpage
\phantomsection
\addcontentsline{toc}{chapter}{Acknowledgements}
\begin{center}
    \large
    \textbf{Acknowledgements}
\end{center}

First and foremost, I would like to thank my supervisors Peter J.~Forrester, Mario Kieburg, and Paul T.~Norbury for the many years of patient guidance through the various topics encountered in this thesis. In particular, Peter's uncanny ability to recall obscure tidbits from his vast knowledgebase and provide insightful feedback on written text combined with his generous willingness to do so upon request at almost any time of day, regardless of frequency, has most certainly been a privilege to experience. It would have been impossible for me to learn the amount of random matrix theory that I did in the last few years without the countless hours of effort put in by Peter. Likewise, I would like to thank my co-supervisors Mario and Paul for introducing me to topics in analysis and geometry, respectively, which have contributed to my growth as a mathematician.

Next, I would like to thank those who would explain things to me for hours on end or just lend me a friendly ear without any obligation to do so: My various interactions with Stephane Dartois, Jesper Ipsen, Shi-Hao Li, Anthony Mays, Norman Do, David Ridout, Nora Ganter, Thomas Quella, Arun Ram, Paul Fijn, and Gufang Zhao have enlightened me on the value and sheer joy of belonging to an academic community. Furthermore, Stephane's skill in explaining intricate topics in a sympathetic and illuminating manner has made it an absolute pleasure to collaborate with him.

As for peers with whom I shared the experience of being a graduate student, I am grateful to Allan Trinh, Jiyuan Zhang, Srivatsa Badariprasad, Wee Chaimanowong, Eric Shen, Campbell Wheeler, and Behrooz Niknami for years of friendship and interesting conversations. It is not unusual to spend a majority of your graduate studies in solitude, especially when working in such a specialised field. Therefore, I consider myself very lucky to have shared an office with two friends, Allan and Jiyuan, with whom I could discuss random matrix theory to my heart's content. On the other hand, conversations with friends from outside random matrix theory have given me opportunities to take a break from thinking about my work while affording me a chance to learn about interesting topics far removed from my field.

Finally on the academic side, I would like to note that my mathematical maturity has benefitted greatly from the seminars, conferences, and summer schools that I have been able to participate in --- I am indebted to the people who make these events possible. In that vein, I am similarly indebted to the academic support staff at our ACEMS node and the School of Mathematics and Statistics for their timely help with all administrative issues, such as sorting out IT access, arranging travel funds, or organising seminars.

In terms of non-academic support, I would like to thank my friends and family for their love and support, without which I could not have completed this journey. I am grateful to my mother for instilling in me the confidence to pursue my passions from a young age. To my family and the friends that I have not been able to see very often, I apologise and thank you for your understanding of my absence. It is in the nature of time for it to grow scarce and it is truly a shame whenever one must spend it alone.

Most of all, my utmost gratitude is reserved for my beautiful fianc\'ee Johanna, who has quelled my worries in my times of woe; who has shown seemingly blind faith in my abilities when I had little faith of my own; who has been understanding at times when I have been unavailable; and who has been the source of my first smile of each day. Thank you.

\newpage~
\thispagestyle{empty}
\microtypesetup{protrusion=false}
\tableofcontents
\listoffigures
\microtypesetup{protrusion=true}

\clearemptydoublepage
\pagenumbering{arabic}
\pagestyle{fancy}
\renewcommand{\sectionmark}[1]{\markright{\thesection.\ #1}}
\renewcommand{\chaptermark}[1]{%
\markboth{\MakeUppercase{%
\chaptername}\ \thechapter.%
\ #1}{}}
\fancyhead{}
\fancyhead[LE]{\slshape \leftmark}
\fancyhead[RO]{\slshape \rightmark}
\fancyfoot{}
\fancyfoot[LE,RO]{\thepage}
\setcounter{page}{1}
\chapter{Introduction}
The overarching theme of this thesis is that of recursive structures within random matrix theory. The focus is on two types of recursions: $\mathit{1}$\textit{-point recursions} for \textit{moment expansion coefficients} and \textit{loop equations} for $n$\textit{-point correlators}. A method of deriving $1$-point recursions from differential equations satisfied by the \textit{eigenvalue densities} of the \textit{classical matrix ensembles} is demonstrated in Chapters 2 and 3, while the method of loop equations is demonstrated through an application to certain matrix product ensembles in Chapter 4. These two classes of random matrix ensembles are respectively introduced in Sections \ref{s1.2} and \ref{s1.3}, along with auxiliary details drawn from the literature so as to give our development sufficient context. Before introducing these ensembles, we review some pertinent fundamentals of random matrix theory.

\setcounter{equation}{0}
\section{Random Matrix Ensembles} \label{s1.1}
Borrowing language from statistical mechanics, an ensemble can be thought of as a virtual collection of all possible states that a given system can be in, with more probable states appearing more often in the collection. Then, if one draws uniformly from the ensemble, one is more likely to retrieve more probable states. One way to obtain such an ensemble is to generate copies of the system of interest ad infinitum, with the final infinite collection being named the ensemble. With this intuition in mind, we henceforth work with the following rigorous definition, in the case that the underlying system is a matrix.

\begin{definition} \label{def1.1}
A \textit{random matrix ensemble} $\mathcal{E}=(\mathcal{S},P)$ is a set $\mathcal{S}$ of matrices of fixed size equipped with a probability density function (p.d.f.) $P:\mathcal{S}\rightarrow[0,\infty)$. To say that a matrix $X$ is drawn from the random matrix ensemble $\mathcal{E}$ is to say that it is a random variable with values in $\mathcal{S}$ and p.d.f. $P$.
\end{definition}
\begin{remark}
It is commonplace to define a random matrix ensemble by prescribing p.d.f.s $P_{ij}$ on the independent entries $X_{ij}$ of a representative random matrix $X\in\mathcal{E}$ along with a specification of how the other entries of $X$ depend on these independent entries. In such cases, the p.d.f. on $\mathcal{S}$ is the joint probability density function (j.p.d.f.) formed by taking the product
\begin{equation}
P(X)=\prod_{i,j\textrm{ s.t.~}\{X_{ij}\}\textrm{ ind.}}P_{ij}(X_{ij}).
\end{equation}
\end{remark}
Definition \ref{def1.1} allows for a broad range of examples: For many random matrix ensembles, the set $\mathcal{S}$ of say $M\times N$ matrices is described by applying constraints to the set of real matrices $\mathbb{M}_{M\times N}(\mathbb{R})$, the set of complex matrices $\mathbb{M}_{M\times N}(\mathbb{C})$, or the set of quaternionic matrices $\mathbb{M}_{M\times N}(\mathbb{H})$ (see Appendix \ref{appendixA} for an introduction to quaternionic matrices), though the definition above does not forbid more exotic matrix entries like octonions \citep{Ni16}, \citep{Fo17} or $p$-adic numbers \citep{Ner13}, to name a few examples --- the Frobenius theorem \citep{Fro77}, \citep{Dys62} directs our attention to matrices with entries in $\mathbb{R},\mathbb{C}$, or $\mathbb{H}$ due to their being the only finite-dimensional associative division algebras over the real numbers, up to isomorphism. The constraints used to construct $\mathcal{S}$ are usually those that require the elements of $\mathcal{S}$ to exhibit features such as orthogonality, unitarity, (skew-)symmetry, Hermiticity, positive definiteness, invariance under complex conjugation, or combinations thereof. Thus, $\mathcal{S}$ is oftentimes simply the set of symmetric, Hermitian, or skew-symmetric matrices, but in physical settings is more likely to be some (possibly trivial) quotient of products of the classical matrix groups $O(N)$, $U(N)$, and $Sp(N)$, of which ten such amalgamations stand out from the viewpoint of modelling Hamiltonians subject to symmetry constraints \citep{AZ97}, \citep{RSFL10}; these are in correspondence with the ten infinite families of matrix Lie algebras. Another type of constraint that has been explored in addition to these is that of requiring entries $X_{ij}$ of the representative random matrix $X\in\mathcal{E}$ to vanish when the $N$-periodic distance between $i$ and $j$ is greater than some threshold called the band width. These so-called \textit{random band matrices} relate to the study of Anderson localisation-delocalisation transitions \citep{Bou18}.

When it comes to determining the p.d.f.~$P$ on $\mathcal{S}$, there are again many options that have garnered interest throughout the literature. There are far too many to cover here in totality, but let us highlight the diversity of topics in random matrix theory through a few examples. A simple yet non-trivial example is that of \textit{Bernoulli matrices} $B$ whose entries $B_{ij}$ are all independently and identically distributed (i.i.d.) Bernoulli random variables $\pm1$: $\mathcal{S}=\mathbb{M}_{N\times N}(\mathbb{R})$ and $P_{ij}(B_{ij})=\tfrac{1}{2}[\delta(B_{ij}-1)+\delta(B_{ij}+1)]$, where $\delta$ is the Dirac delta \citep{Kom67}, \citep{Tik20}. If we instead require that $B$ be symmetric, its diagonal entries be zero, and its independent hence upper triangular entries have p.d.f.
\begin{equation} \label{eq1.1.2}
P_{ij}(B_{ij})=\frac{c}{N}\delta(B_{ij}-1)+\left(1-\frac{c}{N}\right)\delta(B_{ij}),\quad 0<c\equiv c(N)\ll N,
\end{equation}
$B$ belongs to the ensemble of \textit{Erd\H{o}s--R\'enyi matrices} \citep{EKYY13}, \citep{EKYY12}, which are the adjacency matrices of \textit{Erd\H{o}s--R\'enyi random graphs} \citep{ER60}, \citep{Gil59}; the parameter $c$ is the mean connectivity of the nodes. This ensemble is a type of \textit{sparse random matrix ensemble} \citep{RB88}, \citep{Ku08}, \citep{Woo12}, which is an umbrella term for a variety of random matrix ensembles whose shared characteristic is that for each such ensemble, the independent entries $B_{ij}$ of their representative random matrix $B$ (now allowed to be complex and/or have aforementioned constraints relaxed) have p.d.f. $P_{ij}$ of the form given in equation \eqref{eq1.1.2} but with $\delta(B_{ij}-1)$ replaced by any p.d.f. of interest. This brings us to the question of which p.d.f.s are `interesting'.

Moving on from Bernoulli and sparse random matrix ensembles, let us consider a more general applications-based perspective where one would like to model real world systems such as ecologies \citep{May72}, heavy atom spectra \citep{Wig55}, \citep{Dys62b}, \citep{Wig67}, wireless communication channels \citep{TV04}, random quantum states \citep{CN16}, log-gases \citep{Fo10}, or quantum transport \citep{Be97}, among many others.  From this viewpoint, where one chooses to eschew knowledge of the finer details of the real world system at hand in favour of treating them statistically, there are essentially three ideologies that dominate the field. The first is to assume as little as possible about the system and work in full generality, focusing on universal results. In this line of thinking, one studies \textit{Wigner matrix ensembles} \citep{Wig55}, \citep{Wig58}, \citep{TV14} or, more recently, \textit{general Wigner-type matrix ensembles} \citep{AEK17}: Let $X$ be drawn from either the real symmetric or complex Hermitian $N\times N$ Wigner matrix ensemble. Then, its entries $X_{ij}$ are independently distributed for $1\leq i\leq j\leq N$, with mean zero and all moments having an upper bound independent of $i,j$. In addition, the diagonal entries have identical variance while the upper triangular entries have variance one. If we relax the condition on the variances, $X$ is instead of general Wigner-type. Many random matrix ensembles fall into the class of Wigner matrix ensembles due to the freedom enjoyed by the variables $X_{ij}$. For example, results proved for Wigner matrix ensembles hold when the matrix entries are i.i.d. centred normal variables.

The second of the three dominant ideologies is that of studying uniform distributions, which corresponds to a principle often seen in statistical mechanics: In the absence of any relevant information, one should assign equal probability to all states of an ensemble (cf.~the microcanonical ensemble, dice, a deck of cards, etc.). Our discussion up to this point has focused on random matrix ensembles distributed according to p.d.f.s $P$ defined with respect to the flat Lebesgue measure induced by the canonical embedding of $\mathcal{S}$ into Euclidean space; e.g., for the Bernoulli matrix ensemble, we have implicitly taken
\begin{equation*}
\mathrm{d}B=\prod_{i,j=1}^N\mathrm{d}B_{ij},
\end{equation*}
while for the complex Hermitian Wigner matrix ensemble, we have taken
\begin{equation} \label{eq1.1.3}
\mathrm{d}X=\prod_{i=1}^N\mathrm{d}X_{ii}\,\prod_{1\leq j<k\leq N}\mathrm{d}X_{jk}^{(1)}\mathrm{d}X_{jk}^{(2)},\quad X_{jk}=X_{jk}^{(1)}+\mathrm{i}X_{jk}^{(2)}.
\end{equation}
It is not possible to define a uniform probability density with respect to such measures on non-compact domains $\mathcal{S}$. Instead, one must consider compact groups $\mathcal{S}$ endowed with their respective Haar (probability) measures. The prime example is the circular unitary ensemble.
\begin{theorem}[Haar, compact case]
Given a compact group $\mathcal{S}$, let $U\in\mathcal{S}$ be a generic variable. There exists a unique (up to normalisation) measure $\mathrm{d}\mu(U)$ on $\mathcal{S}$, called the \textbf{Haar measure} (see, e.g., \citep[Ch.~11]{Hal50}, \citep[Sec.~2.4, 2.5]{Nac65}), such that
\begin{equation*}
\int_{\mathcal{S}}\mathrm{d}\mu(U)<\infty
\end{equation*}
and for any fixed $U_0\in\mathcal{S}$,
\begin{equation*}
\mathrm{d}\mu(U_0U)=\mathrm{d}\mu(UU_0)=\mathrm{d}\mu(U).
\end{equation*}
If $\mathrm{d}\mu(U)$ integrates to unity on $\mathcal{S}$, it is referred to as the \textbf{Haar probability measure}.
\end{theorem}
The Haar measure is uniform in the sense that for any Borel subset $\mathcal{S}'\subseteq\mathcal{S}$ and fixed $U_0\in\mathcal{S}$, the Haar volume of $\mathcal{S}'$ is invariant under left and right translations of $\mathcal{S}'$ by $U_0$:
\begin{equation*}
\int_{U_0\mathcal{S}'}\mathrm{d}\mu(U)=\int_{\mathcal{S}'U_0}\mathrm{d}\mu(U)=\int_{\mathcal{S}'}\mathrm{d}\mu(U).
\end{equation*}
\begin{definition} \label{def1.2}
The $N\times N$ \textit{circular unitary ensemble} (CUE) is the unitary group $U(N)$ with its Haar measure \citep{Dys62}, \citep{Dys62b}, \citep[Ch.~2]{Fo10} (made explicit in equation \eqref{eq1.2.35}). If $U\in\textrm{CUE}(N)$, then the symmetric unitary matrix $U^T U$ represents the $N\times N$ \textit{circular orthogonal ensemble} (COE); we have $\mathcal{S}\simeq U(N)/O(N)$. If $U\in\textrm{CUE}(2N)$, then $U^DU$ represents the $N\times N$ \textit{circular symplectic ensemble} (CSE), where
\begin{equation} \label{eq1.1.4}
U^D:=J_{2N}\, U^T\, J_{2N}^{-1}
\end{equation}
is the quaternionic adjoint of $U$ (for $X\in\mathbb{M}_{N\times N}(\mathbb{H})$ with the quaternion entries written as $2\times2$ complex matrices, we write $X^\dagger$ for $X^D$; see Appendix \ref{appendixA}), with
\begin{equation} \label{eq1.1.5}
J_{2N}:=I_N\otimes\begin{bmatrix}0&-1\\1&0\end{bmatrix}.
\end{equation}
That is, $J_{2N}$ is the $2N\times2N$ block-diagonal matrix with non-zero blocks given by the $2\times2$ matrix on the right-hand side of equation \eqref{eq1.1.5}. In the CSE case, $\mathcal{S}\simeq U(2N)/Sp(2N)$.
\end{definition}

The third and final ideology is that of working with algebraic combinations of matrices whose entries are i.i.d.~normal variables. By specifying a p.d.f.~on matrix entries and taking such combinations, one is able to model finer details than would be possible with Wigner matrix ensembles. The normal distribution is a natural choice for the entries due to the central limit theorem and the fact that it is the simplest probability distribution on the real line, at least in the sense of moments. More importantly, the normal distribution and, by extension, this class of random matrix ensembles have been investigated extensively throughout the literature. Thus, there are a plethora of tools available for performing concrete computations and, consequently, the study of these types of random matrix ensembles is deeply connected to topics throughout mathematics, sometimes in surprising ways. It is for these reasons that the ensembles we choose to study in this thesis primarily belong to this class. Indeed, connections to Selberg correlation integrals, ribbon graph combinatorics, and loop equations act as the foundation of this thesis. Let us now formally introduce the building block from which most of the ensembles of Sections \ref{s1.2} and \ref{s1.3} are constructed.

\begin{definition} \label{def1.3}
The $M\times N$ real, complex, and quaternionic \textit{Ginibre ensembles} \citep{Gin65} are respectively $\mathbb{M}_{M\times N}(\mathbb{R})$, $\mathbb{M}_{M\times N}(\mathbb{C})$, and $\mathbb{M}_{M\times N}(\mathbb{H})$ with p.d.f.
\begin{equation} \label{eq1.1.6}
P^{(Gin)}(G)=\pi^{-MN\beta/2}\exp\left(-\Tr\,G^\dagger G\right)=\prod_{i=1}^M\prod_{j=1}^N\pi^{-\beta/2}\exp\left(-|G_{ij}|^2\right),
\end{equation}
which is defined with respect to the Lebesgue measure
\begin{equation} \label{eq1.1.7}
\mathrm{d}G=\prod_{i=1}^M\prod_{j=1}^N\prod_{s=1}^\beta\mathrm{d}G_{ij}^{(s)}.
\end{equation}
Here and henceforth, other than in the context of the circular ensembles, the \textit{Dyson index} $\beta$ \citep{Dys62} counts the generic number of real components $G_{ij}^{(s)}$ of each matrix entry $G_{ij}$, so that $\beta=1$ corresponds to real ensembles, $\beta=2$ corresponds to complex ensembles, and $\beta=4$ corresponds to quaternionic ensembles. As seen in equation \eqref{eq1.1.6}, it allows us to treat the real, complex, and quaternionic cases simultaneously.

In similar fashion to the circular ensembles defined earlier, it is customary for the $N\times N$ real, complex, and quaternionic Ginibre ensembles to be referred to as the \textit{Ginibre orthogonal}, \textit{unitary}, and \textit{symplectic ensembles} (GinOE, GinUE, GinSE), respectively. The reasoning behind this terminology is given in \S\ref{s1.2.2}.
\end{definition}
Before using the Ginibre ensembles to construct the classical matrix ensembles and matrix product ensembles that are considered in this thesis, we introduce the observables that will be at the centre of our discussion.

\subsection{Quantities of interest} \label{s1.1.1}
The random matrices studied in this thesis (these are yet to be introduced and do not include those defined above) are diagonalisable and self-adjoint. Thus, they are determined by their eigenvalues and eigenvectors, the former of which are henceforth assumed to be indistinguishable and real. While the eigenvectors of random matrices have been a topic of study \citep{OVW16}, eigenvalue statistics have attracted considerably more attention and will likewise be the subject of our results. Hence, in addition to the p.d.f.s $P(X)$ of say $N\times N$ random matrices $X$, we are also interested in the \textit{eigenvalue j.p.d.f.s} $p(\lambda_1,\ldots,\lambda_N)$ of the eigenvalues $\{\lambda_i\}_{i=1}^N$ of $X$. The eigenvalue j.p.d.f. is defined to be such that
\begin{equation*}
\int_{[a_1,b_1]\times\cdots\times[a_N,b_N]}p(\lambda_1,\ldots,\lambda_N)\,\mathrm{d}\lambda_1\cdots\mathrm{d}\lambda_N
\end{equation*}
is equal to the probability that $a_i\leq\lambda_i\leq b_i$ simultaneously for all $1\leq i\leq N$.
\begin{example} \label{E1.1}
The eigenvalue j.p.d.f. of the $N\times N$ circular ensembles is
\begin{equation} \label{eq1.1.8}
p^{(C)}(e^{\mathrm{i}\theta_1},\ldots,e^{\mathrm{i}\theta_N};\beta)=\frac{1}{\mathcal{N}_{N,\beta}^{(C)}}|\Delta_N(e^{\mathrm{i}\theta})|^\beta,\quad \theta_i\in[0,2\pi),
\end{equation}
where $\mathcal{N}_{N,\beta}^{(C)}=(2\pi)^N\Gamma(\beta N/2+1)/[\Gamma(\beta/2+1)]^N$ is a normalisation constant \citep[I]{Dys62b},
\begin{equation} \label{eq1.1.9}
\Delta_N(\lambda):=\Det\left[\lambda_i^{j-1}\right]_{i,j=1}^N=\prod_{1\leq i<j\leq N}(\lambda_j-\lambda_i)
\end{equation}
is the \textit{Vandermonde determinant}, and $\beta=1,2$, and $4$ correspond to the COE, CUE, and CSE, respectively. From here on out, we will not write out function arguments following semicolons when their values are clear from context.
\end{example}
Related to the eigenvalue j.p.d.f. is the univariate \textit{eigenvalue density} defined by
\begin{equation} \label{eq1.1.10}
\rho(\lambda;N):=N\int_{\mathbb{R}^{N-1}}p(\lambda,\lambda_2,\ldots,\lambda_N)\,\mathrm{d}\lambda_2\cdots\mathrm{d}\lambda_N,
\end{equation}
which is normalised such that the expected number of eigenvalues lying between $a$ and $b$ is given by $\int_a^b\rho(\lambda)\,\mathrm{d}\lambda$. Note that $\rho(\lambda)$ is not a p.d.f. for $N\neq1$.

To enable more concise notation, let us now introduce the ensemble average
\begin{equation} \label{eq1.1.11}
\mean{\mathcal{O}}:=\int_{\mathbb{R}^N}\left\{\mathcal{O}\,p(\lambda_1,\ldots,\lambda_N)\right\}\mathrm{d}\lambda_1\cdots\mathrm{d}\lambda_N.
\end{equation}
It must be understood that $\mathcal{O}$ is to be interpreted as an operator here. For example, if $\mathcal{O}=\frac{\partial}{\partial\lambda_1}\lambda_1$, one should read the above equation \eqref{eq1.1.11} as ``multiply the eigenvalue j.p.d.f. $p(\lambda_1,\ldots,\lambda_N)$ by $\lambda_1$, differentiate the resulting product with respect to $\lambda_1$, then integrate over $\mathbb{R}^N$.'' Using this notation, we can write the eigenvalue density \eqref{eq1.1.10} as the linear statistic
\begin{equation} \label{eq1.1.12}
\rho(\lambda)=\sum_{i=1}^N\mean{\delta(\lambda-\lambda_i)}
\end{equation}
due to the assumed indistinguishability of the eigenvalues $\{\lambda_i\}_{i=1}^N$; if the eigenvalues were distinguishable, equation \eqref{eq1.1.12} would supersede \eqref{eq1.1.10} as the correct definition for the eigenvalue density. While the definition of the ensemble average \eqref{eq1.1.11} is given in terms of the eigenvalue j.p.d.f. $p(\lambda_1,\ldots,\lambda_N)$, we will also utilise this notation for ensemble averages with respect to p.d.f.s $P$ of random matrices. The different uses will be distinguished through the use of subscripts.

For the purposes of this thesis, the eigenvalue densities considered are fully characterised by their \textit{spectral moments} (one may, e.g., check Carleman's condition \citep[Sec.~88]{Wal48})\footnote{A subtlety arises in the case of the Cauchy ensembles defined by equations \eqref{eq1.2.7} and \eqref{eq1.2.9} upcoming. Then, there is a need for analytic continuation to make sense of the moments which otherwise diverge for $k$ large enough; see \S\ref{s1.2.1} and Section \ref{s2.2}.}
\begin{equation} \label{eq1.1.13}
m_k:=\int_{\mathbb{R}}\lambda^k\rho(\lambda)\,\mathrm{d}\lambda,\quad k\in\mathbb{N}.
\end{equation}
Indeed, if we are interested in a linear statistic $\sum_{i=1}^Nf(\lambda_i)$ of the eigenvalues such that $f(\lambda)$ is analytic at zero with Taylor expansion $f(\lambda)=\sum_{k=0}^\infty f_k\,\lambda^k$, we can compute its expectation to be
\begin{equation}
\mean{\sum_{i=1}^Nf(\lambda_i)}=\int_{\mathbb{R}}f(\lambda)\rho(\lambda)\,\mathrm{d}\lambda=\sum_{k=0}^\infty f_k\,m_k.
\end{equation}
The spectral moments can additionally be expressed as an average with respect to the eigenvalue j.p.d.f. $p(\lambda_1,\ldots,\lambda_N)$ and the p.d.f. $P(X)$ of the random matrix $X$ according to
\begin{equation} \label{eq1.1.15}
m_k=\mean{\sum_{i=1}^N\lambda_i^k}=\mean{\Tr\,X^k}_{P(X)},
\end{equation}
where we define
\begin{equation}
\mean{\mathcal{O}}_{P(X)}:=\int_{\mathcal{S}}\left\{\mathcal{O}\,P(X)\right\}\mathrm{d}X
\end{equation}
in analogy with equation \eqref{eq1.1.11}, with $\mathcal{S}$ as given in Definition \ref{def1.1}.

An invaluable tool for studying the spectral moments is the \textit{moment generating function}
\begin{equation} \label{eq1.1.17}
W_1(x):=\sum_{k=0}^\infty\frac{m_k}{x^{k+1}},
\end{equation}
which is also known as the \textit{resolvent} due to it being equal to the \textit{Stieltjes transform} \citep[Sec.~65]{Wal48} of the eigenvalue density:
\begin{equation} \label{eq1.1.18}
W_1(x)=\int_{\textrm{supp}\,\rho}\frac{\rho(\lambda)}{x-\lambda}\mathrm{d}\lambda,\quad x\in\mathbb{C}\setminus \textrm{supp}\,\rho,
\end{equation}
where $\textrm{supp}\,\rho:=\overline{\{\lambda\in\mathbb{R}\,|\,\rho(\lambda)\neq0\}}$ is the \textit{support} of $\rho$ (here, $\overline{\mathcal{S}}$ denotes the closure of $\mathcal{S}$). Obtaining a closed form expression for the resolvent allows one to recover the eigenvalue density via the Sokhotski--Plemelj inversion formula \citep[Sec.~65]{Wal48}
\begin{equation} \label{eq1.1.19}
\rho(\lambda)=\lim_{\varepsilon\to0}\frac{W_1(\lambda-\mathrm{i}\varepsilon)-W_1(\lambda+\mathrm{i}\varepsilon)}{2\pi\mathrm{i}},
\end{equation}
so long as $\rho(\lambda)$ is a priori known to be continuous on its support. Similar to equation \eqref{eq1.1.15}, we can express the resolvent as an ensemble average with respect to the eigenvalue j.p.d.f. $p(\lambda_1,\ldots,\lambda_N)$ and the p.d.f. $P(X)$ on $\mathcal{S}$:
\begin{equation} \label{eq1.1.20}
W_1(x)=\mean{\sum_{i=1}^N\frac{1}{x-\lambda_i}}=\mean{\Tr\,\frac{1}{x-X}}_{P(X)}.
\end{equation}
Being able to choose between three expressions for the resolvent is quite a boon, as lacking a tractable form for the eigenvalue density is no longer a hindrance. This is exactly the case for the matrix product ensembles considered in Chapter 4, and the computations presented there are in terms of the ensemble average $\langle\,\cdot\,\rangle_{P(X)}$. In fact, the resolvent is usually easier to work with than the eigenvalue density, one reason for which being that the former often admits a large $N$ expansion in $1/N$ while the latter does not.
\begin{theorem}[Borot--Guionnet '12] \label{thrm1.1}
Assuming certain hypotheses \citep{BG12} on the eigenvalue j.p.d.f. $p(\lambda_1,\ldots,\lambda_N)$, the resolvent $W_1(x)$ and, more generally, the soon to be defined connected $n$-point correlators $W_n(x_1,\ldots,x_n)$ have large $N$ expansions of the form
\begin{equation} \label{eq1.1.21}
W_n(x_1,\ldots,x_n;N)=N^{2-n}\sum_{l=0}^\infty\frac{W_n^l(x_1,\ldots,x_n)}{N^l},
\end{equation}
where the \textbf{correlator expansion coefficients} $W_n^l(x_1,\ldots,x_n)$ have no dependence on $N$.
\end{theorem}
In Section \ref{s2.4}, we use this theorem to study the large $N$ behaviour of the classical matrix ensembles. However, Theorem \ref{thrm1.1} does not apply directly to a subset of these ensembles due to a technical issue that we must first resolve by scaling the ensembles so that their eigenvalues lie within a compact connected set in the $N\to\infty$ regime. In these cases, we must study the \textit{global scaled eigenvalue density} $\tilde{\rho}(\lambda)$ and \textit{scaled resolvent} $\tilde{W}_1(x)$ defined by
\begin{align}
\tilde{\rho}(\lambda)&:=\frac{c_N}{N}\rho(c_N\lambda), \label{eq1.1.22}
\\\tilde{W}_1(x)&:=c_NW_1(c_Nx)
\\&\;=N\int_{\textrm{supp}\,\tilde{\rho}}\frac{\tilde{\rho}(\lambda)}{x-\lambda}\mathrm{d}\lambda,\quad x\in\mathbb{C}\setminus \textrm{supp}\,\tilde{\rho}, \label{eq1.1.24}
\end{align}
where $c_N$ is a scaling parameter that differs from ensemble to ensemble. (In the case of our matrix product ensembles, we do not check the hypotheses of Theorem \ref{thrm1.1} and instead opt to establish the existence of expansions of the form \eqref{eq1.1.21} via combinatorial arguments given in \S\ref{s3.3.3}.) Now, assuming that the scaled resolvent has an expansion of the form $\tilde{W}_1(x)=\sum_{l=0}^\infty W_1^l(x)N^{1-l}$ as given by equation \eqref{eq1.1.21}, use the Sokhotski--Plemelj inversion formula \eqref{eq1.1.19} to define
\begin{equation} \label{eq1.1.25}
\rho^l(\lambda):=\lim_{\varepsilon\to0}\frac{W_1^l(\lambda-\mathrm{i}\varepsilon)-W_1^l(\lambda+\mathrm{i}\varepsilon)}{2\pi\mathrm{i}}.
\end{equation}
Then, $\tilde{\rho}(\lambda)$ and $\rho^0(\lambda)$ agree in the $N\to\infty$ regime. However, $\tilde{\rho}(\lambda)$ and $\sum_{l=0}^\infty\rho^l(\lambda)N^{-l}$ do not agree at finite $N$, as one might expect. Instead, the latter sum is referred to as the \textit{smoothed eigenvalue density}, so called due to it lacking oscillatory terms that are known to be present in the true global scaled eigenvalue density $\tilde{\rho}(\lambda)$ \citep{GFF05}, \citep{FFG06}, \citep{DF06}. Nonetheless, the moments of the smoothed eigenvalue density, computed via analytic continuation, agree with those of $\tilde{\rho}(\lambda)$, due to the missing oscillatory terms being compensated for by non-integrable singularities at the endpoints of support \citep{WF14}, \citep{FRW17}. A consequence of this is that for $l>0$, the corrections $\rho^l(\lambda)$ integrate to zero on their supports and are thus signed densities. In short, the following diagram does not commute.
\begin{equation*}
\large
\begin{tikzcd}[column sep=7em, row sep=8em]
\rho(\lambda) \arrow[leftrightarrow]{r}{\textrm{scaling }c_N} \arrow[shift left]{d}[anchor=center, rotate=-90,yshift=1.5ex]{\textrm{Stieltjes transform}}& \tilde{\rho}(\lambda) \arrow[shift left]{d}[anchor=center, rotate=-90,yshift=1.5ex]{\textrm{Stieltjes transform}} \arrow[dotted,leftrightarrow]{r}{\textrm{moments}}& \left\{\rho^l(\lambda)\right\}_{l=0}^\infty \arrow[shift left]{d}[anchor=center, rotate=-90,yshift=1.5ex]{\textrm{Stieltjes transform}}\\
W_1(x) \arrow[leftrightarrow]{r}{\textrm{scaling }c_N} \arrow[shift left]{u}[anchor=center, rotate=90,yshift=1.5ex]{\textrm{Sokhotski--Plemelj}} & \tfrac{1}{N}\tilde{W}_1(x) \arrow[shift left]{u}[anchor=center, rotate=90,yshift=1.5ex]{\textrm{Sokhotski--Plemelj}} \arrow{r}{1/N\textrm{ expansion}} & \left\{W_1^l(x)\right\}_{l=0}^\infty \arrow[shift left]{u}[anchor=center, rotate=90,yshift=1.5ex]{\textrm{Sokhotski--Plemelj}}
\end{tikzcd}
\end{equation*}

Let us now turn our attention to Chapter 4, wherein we derive three types of loop equations. The first set of loop equations are satisfied by the \textit{unconnected $n$-point correlators}
\begin{equation} \label{eq1.1.26}
U_n(x_1,\ldots,x_n):=\mean{\prod_{i=1}^n\Tr\,\frac{1}{x_i-X}}_{P(X)},
\end{equation}
which act as generating functions for the \textit{mixed moments}
\begin{equation} \label{eq1.1.27}
m_{k_1,\ldots,k_n}:=\mean{\prod_{i=1}^n\Tr\, X^{k_i}}_{P(X)},\quad k_1,\ldots,k_n\in\mathbb{N}.
\end{equation}
Related to the mixed moments are the \textit{mixed cumulants} $c_{\kappa_i}$, which are defined implicitly by the moment-cumulants relation \citep[Ch.~2]{McC87}
\begin{equation} \label{eq1.1.28}
m_{k_1,\ldots,k_n}=\sum_{K\vdash\{k_1,\ldots,k_n\}}\prod_{\kappa_i\in K}c_{\kappa_i},
\end{equation}
where $K\vdash\{k_1,\ldots,k_n\}$ means that $K$ is a partition of $\{k_1,\ldots,k_n\}$, i.e., $K=\{\kappa_i\}_{i=1}^m$ for some $1\leq m\leq n$ such that the disjoint union $\kappa_1\sqcup\cdots\sqcup\kappa_m$ is equal to $\{k_1,\ldots,k_n\}$. These mixed cumulants are generated by the so-called \textit{connected $n$-point correlators}
\begin{align}
W_n(x_1,\ldots,x_n)&:=\sum_{k_1,\ldots,k_n=0}^{\infty}\frac{c_{k_1,\ldots,k_n}}{x_1^{k_1+1}\cdots x_n^{k_n+1}} \label{eq1.1.29}
\\&\;=\sum_{G\vdash\{x_1,\ldots,x_n\}}(-1)^{\#G-1}(\#G-1)!\prod_{G_i\in G}U_{\#G_i}(G_i) \label{eq1.1.30}
\\&\;=U_n(x_1,J_n)-\sum_{\emptyset\neq J\subseteq J_n}W_{n-\#J}(x_1,J_n\setminus J)U_{\#J}(J),\quad J_n=(x_2,\ldots,x_n), \label{eq1.1.31}
\end{align}
where $\#\mathcal{S}$ denotes the size of the set $\mathcal{S}$ \citep{Smi95}, \citep[pg.~187]{Fo10}, \citep[pp.~8--9]{WF15}\footnote{The otherwise formal series \eqref{eq1.1.29} converges when $|x_1|,\ldots,|x_n|>\underset{\lambda\in\mathrm{supp}\,\rho}{\mathrm{max}}|\lambda|$; cf.~equation \eqref{eq1.1.18}. The characterisations \eqref{eq1.1.30}, \eqref{eq1.1.31} in terms of the $U_n$ follow from (the inverse of) the relation \eqref{eq1.1.28}.}. Setting $n=1,2$ in equation \eqref{eq1.1.30} shows that
\begin{equation*}
W_1(x)=U_1(x),\qquad W_2(x_1,x_2)=\mathrm{Cov}_{P(X)}\left(\Tr\,\frac{1}{x_1-X},\Tr\,\frac{1}{x_2-X}\right),
\end{equation*}
where
\begin{equation*}
\mathrm{Cov}_{P(X)}(f(X),g(X)):=\mean{(f(X)-\mean{f(X)}_{P(X)})(g(X)-\mean{g(X)}_{P(X)})}_{P(X)}
\end{equation*}
denotes the covariance of the linear statistics $f(X)$ and $g(X)$ with respect to $P(X)$; a similar formula is found for $n=3$, but the analogous structures for $n\geq4$ are more complicated. Equation \eqref{eq1.1.31} is used in Chapter 4 to transform the set of loop equations on the unconnected correlators $U_n$ into a second set of loop equations satisfied by the connected correlators $W_n$.

Our interest in the connected correlators $W_n$ over their unconnected counterparts $U_n$ is that the former is of lower order in $N$ due to the cancelling of leading order terms brought on by the inclusion-exclusion structure of equation \eqref{eq1.1.30}. Hence, the connected correlators $W_n$ have the advantage of admitting large $N$ expansions of the form \eqref{eq1.1.21}. This is significant because the loop equations on both the connected and unconnected correlators fail to close and are thus unsolvable, but combining them with the large $N$ expansion \eqref{eq1.1.21} yields a third set of loop equations for the correlator expansion coefficients $W_n^l$ that form a triangular recursive system. Considering $n=1$ for simplicity, this means that $W_1^l$ can be computed through (at most) $\left\lfloor(1+l/2)^2\right\rfloor$ applications of the loop equations (see, e.g., \citep{FRW17}). Although complexity grows strongly with $l$, it is quite feasible to write down $W_1^0,\ldots,W_1^3$ and thus calculate the scaled spectral moments up to a corresponding order in $1/N$. To be precise, let $\tilde{m}_k$ denote the spectral moments of $\tilde{\rho}(\lambda)$ (i.e., the scaled spectral moments of $\rho(\lambda)$) and implicitly define the \textit{moment expansion coefficients} $\tilde{M}_{k,l}$ according to
\begin{equation} \label{eq1.1.32}
\tilde{m}_k=\sum_{l=0}^\infty \tilde{M}_{k,l}\,N^{-l}.
\end{equation}
Then, if $\textrm{Coeff}(x,k)$ denotes the action of expanding in $1/x$ and extracting the coefficient of $1/x^k$, the following diagram commutes.
\begin{equation*}
\large
\begin{tikzcd}[column sep=7em, row sep=7em]
\tilde{W}_1(x) \arrow{r}{\textrm{Coeff}(N,l)} \arrow{d}[anchor=center, rotate=-90,yshift=1.5ex]{\textrm{Coeff}(x,k+1)}& W_1^l(x) \arrow{d}[anchor=center, rotate=-90,yshift=1.5ex]{\textrm{Coeff}(x,k+1)}\\
\tilde{m}_k \arrow{r}{\textrm{Coeff}(N,l)} & \tilde{M}_{k,l}
\end{tikzcd}
\end{equation*}

\begin{remark}
We recognise that the use of the adjectives `unconnected' and `connected' has yet to be motivated. This terminology alludes to the fact that the mixed moments and cumulants often relate to problems of counting topological surfaces that are respectively unconnected or connected; we address the details of this relationship in Chapter 3.
\end{remark}

Having defined the observables of interest in full generality, it is now time to introduce the specific forms that are relevant to the contents of Chapters 2--4.

\setcounter{equation}{0}
\section{Classical Matrix Ensembles} \label{s1.2}
Let us first define the classical matrix ensembles that will be at the focus of Chapter 2.
\begin{definition} \label{def1.4}
Recalling Definition \ref{def1.3}, let $G$ be drawn from the $N\times N$ real, complex, or quaternionic Ginibre ensemble. Then, the random matrix $H=\tfrac{1}{2}(G^\dagger+G)$ represents the corresponding \textit{Gaussian (also known as Hermite) ensemble} \citep{Wig57}, \citep{Dys62b}. If we instead take $G$ to be $M\times N$ Ginibre, then $W=G^\dagger G$ is said to be drawn from the $(M,N)$ \textit{Laguerre (also known as Wishart) ensemble} \citep{Wis28}. We say that $Y$ is an element of the $(M_1,M_2,N)$ real, complex, or quaternionic \textit{Jacobi ensemble} if it is of the form $Y=W_1(W_1+W_2)^{-1}$ where the $W_i$ belong to the corresponding $(M_i,N)$ Laguerre ensembles with $M_i\geq N$ \citep[Ch.~3]{Mui82}.
\end{definition}

The real Gaussian ensemble was first introduced as a statistical model for heavy atom spectra by Wigner \citep{Wig57} in 1957, while its complex and quaternionic counterparts were introduced by Dyson \citep{Dys62b} in 1962 with the same application in mind; in this latter work, Dyson also highlighted, and made application of, analogies with the statistical mechanics of log-gases. These and related papers from the surrounding literature are conveniently collated and reviewed in the 1965 book \citep{Po65} of Porter. The Laguerre (Wishart) ensemble, on the other hand, has been a mainstay of multivariate statistical analysis since the 1928 work \citep{Wis28} of Wishart showing how to change variables from the entries of an $M\times N$ real Ginibre matrix $G$ to the entries of the Wishart--Laguerre matrix $W=G^TG$ defined above. The textbook \citep{Mui82} details applications of the real and complex Laguerre ensembles in multivariate statistics, including its role in estimating covariance matrices and supplying null hypotheses for principal component analysis. Around the turn of the century, the Laguerre ensembles found further applications in the fields of wireless communications \citep{TV04} and quantum transport \citep{Be97} (one such application is briefly discussed early in Section \ref{s3.1}). The review \citep{Be97} also outlines how problems of quantum transport served as early motivation for studying the Jacobi ensembles. Another appreciable motivation was given a decade later in the 2008 work \citep{Jo08} of Johnstone showing how the eigenvalue statistics of the Jacobi ensembles can be used to determine null hypotheses for multivariate analysis of variance.

The matrices given in Definition \ref{def1.4} are all self-adjoint and thus have $N$ real eigenvalues. Taking either the real, complex, or quaternionic case for definiteness, the entries of the Hermite--Gaussian random matrix $H$ are independent (up to the constraint of $H$ being self-adjoint) centred normal variables with the diagonal entries having variance $\frac{1}{2}$ and the real components of the upper triangular entries having variance $\frac{1}{4}$. Thus, it can be observed that the Gaussian ensembles are examples of Wigner matrix ensembles. The diagonal entries of the Wishart--Laguerre random matrix $W$ are i.i.d. chi-squared variables with p.d.f.
\begin{equation*}
P_{ii}^{(L)}(W_{ii})=\frac{1}{\Gamma(\beta M/2)}W_{ii}^{\beta M/2-1}e^{-W_{ii}}.
\end{equation*}
The p.d.f.s of the off-diagonal entries of $W$ and indeed all entries of the Jacobi random matrix $Y$ are too complicated to present here. On the contrary, the p.d.f.s of the random matrices $H,W,$ and $Y$ are themselves quite elegant.

\begin{proposition} \label{prop1.1}
Let $\kappa:=\beta/2$, $a=\kappa(M_1-N+1)-1$, and $b=\kappa(M_2-N+1)-1$. From the p.d.f.~\eqref{eq1.1.6} of the Ginibre ensembles, we can change variables to obtain the p.d.f.s of the $N\times N$ Gaussian, $(M_1,N)$ Laguerre, and $(M_1,M_2,N)$ Jacobi ensembles, respectively \citep[Ch.~1,3]{Fo10}:
\begin{align}
P^{(G)}(H)&=\frac{1}{\mathcal{Z}_{N,\beta}^{(G)}}\exp\left(-\Tr\,H^2\right), \label{eq1.2.1}
\\P^{(L)}(W)&=\frac{1}{\mathcal{Z}_{N,a}^{(L)}}\left(\Det\,W\right)^a\exp\left(-\Tr\,W\right), \label{eq1.2.2}
\\P^{(J)}(Y)&=\frac{1}{\mathcal{Z}_{N,a,b}^{(J)}}\left(\Det\,Y\right)^a\left(\Det(I_N-Y)\right)^b, \label{eq1.2.3}
\end{align}
where the $\mathcal{Z}_N$ are normalisation constants, often referred to as \textbf{partition functions}, whose explicit forms are inconsequential to us but can be obtained from the method described at the end of \S\ref{s2.1.1}.
\end{proposition}

\begin{remark} \label{R1.3}
With $\beta=1,2,4$ corresponding to $\mathbb{F}=\mathbb{R},\mathbb{C},\mathbb{H}$, the random matrix ensembles introduced in Definition \ref{def1.4} are specified by assigning the p.d.f.s \eqref{eq1.2.1}--\eqref{eq1.2.3} to the matrix sets
\begin{align}
\mathcal{S}^{(G)}&=\{H\in\mathbb{M}_{N\times N}(\mathbb{F})\,|\,H=H^\dagger\}, \label{eq1.2.4}
\\ \mathcal{S}^{(L)}&=\{W\in\mathcal{S}^{(G)}\,|\,W\textrm{ positive definite}\},
\\ \mathcal{S}^{(J)}&=\{Y\in\mathcal{S}^{(L)}\,|\,I_N-Y\textrm{ positive definite}\}, \label{eq1.2.6}
\end{align}
respectively.
\end{remark}

\begin{remark} \label{R1.4}
Apart from the matrix $Y$ given in Definition \ref{def1.4}, there are some other important constructions that are known to yield random matrices $Y', Y'', Y'''$, which all have p.d.f.~\eqref{eq1.2.3} with $a,b$ as prescribed in Proposition \ref{prop1.1} \citep[Sec.~3.6]{Fo10}, \citep{ZS00}, \citep{Fo06}.
\begin{enumerate}
\item Let $X$ be a Haar-distributed element of either $O(M_1+M_2)$ ($\beta=1$), $U(M_1+M_2)$ ($\beta=2$), or $Sp(2(M_1+M_2))$ ($\beta=4$) for positive integers $M_1,M_2$. Take $T$ to be the $M_1\times N$ truncation of $X$ with $T_{ij}=X_{ij}$ for $1\leq i\leq M_1$, $1\leq j\leq N$, and $M_1,M_2\geq N$. Define $Y'=T^\dagger T$.
\item For $i=1,2$, let $G_i$ be $M_i\times N$ Ginibre matrices with $M_i\geq N$ and define
\begin{equation*}
X=\begin{bmatrix}G_1\\G_2\end{bmatrix},\quad \tilde{I}=\begin{bmatrix}I_{M_1}&\\&0_{M_2}\end{bmatrix}.
\end{equation*}
Set $Y''=X^\dagger\tilde{I}X(X^\dagger X)^{-1}$.
\item In the notation of Definition \ref{def1.4}, let $Y''':=(W_1+W_2)^{-1/2}W_1(W_1+W_2)^{-1/2}$.
\end{enumerate}
\end{remark}

The term `classical matrix ensemble' originally referred to the Gaussian, Laguerre and Jacobi ensembles due to their relation to the classical orthogonal polynomials, which we briefly review in \S\ref{s1.2.3}. However, the family of classical matrix ensembles has one more constituent according to the contemporary definition.
\begin{definition} \label{def1.5}
A random matrix ensemble is said to be a \textit{classical matrix ensemble} \citep[\S5.4.3]{Fo10}, \citep{FW04} if its eigenvalue j.p.d.f. is of the form
\begin{equation} \label{eq1.2.7}
p^{(w)}(\lambda_1,\ldots,\lambda_N;\beta)=\frac{1}{\mathcal{N}_{N,\beta}^{(w)}}\,\prod_{i=1}^Nw(\lambda_i)\,|\Delta_N(\lambda)|^\beta,\quad \beta=1,2,4,
\end{equation}
where $\mathcal{N}_{N,\beta}^{(w)}$ is a normalisation constant with values given in \citep{DE02}, \citep{FO08} (alternatively, see \S\ref{s2.1.1}), $\Delta_N(\lambda)$ is the Vandermonde determinant, as specified by equation \eqref{eq1.1.9}, and the weight function $w(\lambda)$ is such that
\begin{equation} \label{eq1.2.8}
\frac{\mathrm{d}}{\mathrm{d}\lambda}\log\,w(\lambda)=\frac{w'(\lambda)}{w(\lambda)}=\frac{f(\lambda)}{g(\lambda)},
\end{equation}
with $f$ and $g$ polynomials of degree at most one and two, respectively.
\end{definition}
Up to fractional linear transformations $\lambda\mapsto (a_{11}\lambda+a_{12})/(a_{21}\lambda+a_{22})$ with $[a_{ij}]\in GL_2(\mathbb{R})$, there are precisely four distinct weights satisfying the criterion \eqref{eq1.2.8}:
\begin{equation} \label{eq1.2.9}
w(\lambda)=\begin{cases} e^{-\lambda^2},&\textrm{Gaussian}, \\ \lambda^ae^{-\lambda}\chi_{\lambda>0},&\textrm{Laguerre}, \\ \lambda^a(1-\lambda)^b\chi_{0<\lambda<1},&\textrm{Jacobi}, \\ (1-\mathrm{i}\lambda)^{\eta}(1+\mathrm{i}\lambda)^{\overline{\eta}},&\textrm{generalised Cauchy},\end{cases}
\end{equation}
where the indicator function $\chi_A$ equals one when $A$ is true and zero otherwise; its presence is due to the fact that the Laguerre random matrix $W=G^\dagger G$ is positive definite. Taking $a,b$ as prescribed in Proposition \ref{prop1.1} relates the eigenvalue j.p.d.f.~\eqref{eq1.2.7} with the Gaussian, Laguerre, and Jacobi weights \eqref{eq1.2.9} to the p.d.f.s \eqref{eq1.2.1}--\eqref{eq1.2.3}. More generally, one usually considers $a,b>-1$ real and $\eta=-\kappa(N-1)-1-\alpha$ with $\alpha\in\mathbb{C}$ and $\textrm{Re}(\alpha)>-1/2$ for convergence issues \citep{BO01}, though other values can be considered through analytic continuation --- in this general setting, the function \eqref{eq1.2.7} continues to be referred to as an eigenvalue j.p.d.f.~even if the variables $\{\lambda_i\}_{i=1}^N$ it governs can no longer be interpreted as eigenvalues of an actual random matrix. We will see in \S\ref{s1.2.4} that there is a further generalisation of equation \eqref{eq1.2.7} where $\beta$ is taken to be positive real parameter. Like with the circular ensembles' eigenvalue j.p.d.f.~\eqref{eq1.1.8}, we will suppress the $\beta$ argument of the eigenvalue j.p.d.f.~\eqref{eq1.2.7} when its value is clear from context. Moreover, we will opt to use superscripts $(G),(L),(J),(Cy)$ to distinguish the four cases given in equation \eqref{eq1.2.9}.

The eigenvalue densities corresponding to the eigenvalue j.p.d.f.s of Definition~\ref{def1.5} are given by equation \eqref{eq1.1.10}. Existing derivations of their explicit forms for finite $N$ in terms of (skew-)orthogonal polynomials are presented in \S\ref{s1.2.3}. For now, we review the large $N$ limiting forms of the corresponding global scaled eigenvalue densities $\tilde{\rho}(\lambda)$.
\begin{definition} \label{def1.6}
Recalling that $\kappa=\beta/2$, take $c_N=\sqrt{\kappa N},\kappa N,1$ in equation \eqref{eq1.1.22} for the Gaussian, Laguerre, and Jacobi cases, respectively \citep{WF14}, \citep{FRW17}. Hence, define the \textit{global scaled eigenvalue densities} of the classical matrix ensembles to be
\begin{align}
\tilde{\rho}^{(G)}(\lambda)&:=\sqrt{\frac{\kappa}{N}}\,\rho^{(G)}(\sqrt{\kappa N}\lambda), \label{eq1.2.10}
\\ \tilde{\rho}^{(L)}(\lambda)&:=\kappa\,\rho^{(L)}(\kappa N\lambda),
\\ \tilde{\rho}^{(J)}(\lambda)&:=\frac{1}{N}\,\rho^{(J)}(\lambda),
\\ \tilde{\rho}^{(Cy)}(\lambda)&:=\frac{1}{N}\,\rho^{(Cy)}(\lambda)\Big|_{\eta=-\kappa(N+\hat{\alpha}N-1)-1}, \label{eq1.2.13}
\end{align}
with $\hat{\alpha}$ constant in $N$.
\end{definition}
There is some freedom in choosing the scaling $c_N$, but the required characteristic of the densities \eqref{eq1.2.10}--\eqref{eq1.2.13} is that their large $N$ limiting forms have compact connected support on which they integrate to one. The Jacobi ensembles' eigenvalue densities already have compact connected support $[0,1]$, so to obtain an appropriate expression for $\tilde{\rho}^{(J)}(\lambda)$, one need only renormalise $\rho^{(J)}(\lambda)$ so that it integrates to one on said support. On the other hand, scaling the eigenvalue densities of the (generalised) Cauchy ensembles according to equation \eqref{eq1.1.22} cannot produce a density with compact support as there is no choice of parameter $c_N$ that leads to fast enough decay as $N,|\lambda|\to\infty$. Instead, one needs to take $\eta=-\kappa(N-1)-1-\alpha$ and change variables to relate to the circular Jacobi ensemble, at which point it can be seen that when $\alpha\propto N$, one obtains a compact support in the large $N$ limit. The details of this phenomenon and some other properties that set the Cauchy ensembles apart from the other classical ensembles are discussed in \S\ref{s1.2.1}.

\begin{proposition} \label{prop1.2}
Recall that the $N\to\infty$ limit of $\tilde{\rho}(\lambda)$ is equal to $\rho^0(\lambda)$, the leading order term of the smoothed eigenvalue density. In the Gaussian case, it is given by the celebrated Wigner semi-circle law \citep{Wig58},
\begin{equation} \label{eq1.2.14}
\rho^{(G),0}(\lambda)=\frac{\sqrt{2-\lambda^2}}{\pi}\chi_{|\lambda|<\sqrt{2}}.
\end{equation}
If we suppose that $\gamma_i:=\lim_{N\to\infty}M_i/N$ exists for $i=1,2$, then we have $\rho^{(L),0}(\lambda)$ given by the Mar\v{c}enko-Pastur law \citep{MP67},
\begin{equation} \label{eq1.2.15}
\rho^{(L),0}(\lambda)=(1-\gamma_1)\delta(\lambda)\chi_{\gamma_1<1}+\frac{\sqrt{(\lambda_+^{(MP)}-\lambda)(\lambda-\lambda_-^{(MP)})}}{2\pi\lambda}\chi_{\lambda_-^{(MP)}<\lambda<\lambda_+^{(MP)}},
\end{equation}
and $\rho^{(J),0}(\lambda)$ given by the so-called Wachter law \citep{Wac80},
\begin{equation}
\rho^{(J),0}(\lambda)=(\gamma_1+\gamma_2)\frac{\sqrt{(\lambda_+^{(Wac)}-\lambda)(\lambda-\lambda_-^{(Wac)})}}{2\pi\lambda(1-\lambda)}\chi_{\lambda_-^{(Wac)}<\lambda<\lambda_+^{(Wac)}},
\end{equation}
with
\begin{align}
\lambda_{\pm}^{(MP)}&:=\left(1\pm\sqrt{\gamma_1}\right)^2,
\\ \lambda_{\pm}^{(Wac)}&:=\left(\sqrt{\frac{\gamma_1}{\gamma_1+\gamma_2}\left(1-\frac{1}{\gamma_1+\gamma_2}\right)}\pm\sqrt{\frac{1}{\gamma_1+\gamma_2}\left(1-\frac{\gamma_1}{\gamma_1+\gamma_2}\right)}\right)^2.
\end{align}
Writing $\hat{\alpha}=\hat{\alpha}_1+\mathrm{i}\hat{\alpha}_2$ with $\hat{\alpha}_1,\hat{\alpha}_2$ real and constant in $N$, one surmises from the equivalent result for the circular Jacobi ensemble \citep[Eq.~(5.6)]{BNR09} that in the Cauchy case,
\begin{equation} \label{eq1.2.19}
\rho^{(Cy),0}(\lambda)=\hat{\alpha}_1\frac{\sqrt{(\lambda_+^{(Cy)}-\lambda)(\lambda-\lambda_-^{(Cy)})}}{\pi(1+\lambda^2)}\chi_{\lambda_-^{(Cy)}<\lambda<\lambda_+^{(Cy)}},
\end{equation}
where
\begin{equation}
\lambda_{\pm}^{(Cy)}:=\frac{-\hat{\alpha}_2(\hat{\alpha}_1+1)\pm\sqrt{(\hat{\alpha}_1^2+\hat{\alpha}_2^2)(2\hat{\alpha}_1+1)}}{\hat{\alpha}_1^2}.
\end{equation}
\end{proposition}
\begin{note} \label{N1.1}
Due to how $\kappa$ is involved in the scalings of Definition \ref{def1.6}, all of the large $N$ limiting forms given in Proposition \ref{prop1.2} are independent of $\beta$. It is known that the same does not hold for the $1/N$ correction terms $\rho^1(\lambda)$, as will be seen in \S\ref{s2.4.1}.
\end{note}

The limiting densities $\rho^{(G),0}(\lambda)$, $\rho^{(L),0}(\lambda)$, and $\rho^{(J),0}(\lambda)$ take on parameter-independent forms when $a,b$ are constant in $N$ since then, $\gamma_1=\gamma_2=1$. They display, respectively, two soft edges, a soft and hard edge, and two hard edges. Thus, from a topological viewpoint, all possible combinations of soft and hard edges on a connected interval are represented by the eigenvalue densities of the Gaussian, Laguerre, and Jacobi ensembles in the relatively simple regime $a,b={\rm O}(1)$. Here, a \textit{soft edge} refers to an endpoint $\lambda$ of $\mathrm{supp}\,\rho^0$ with the property that, for all large $N$, $c_N\lambda$ lies within the interior of $\mathrm{supp}\,\rho$, while a \textit{hard edge} refers to such an endpoint when $c_N\lambda$ is also an endpoint of $\mathrm{supp}\,\rho$ \citep{FNH99}, \citep{Fo06a}. In the present setting concerning classical matrix ensembles, $\rho^0(\lambda)$ exhibits square root profiles at its soft edges owing to tails extending past said edges being scaled to negligibility, whereas its hard edges are located at inverse equare root singularities corresponding to strict constraints enforced by the indicator functions seen in equation \eqref{eq1.2.9}; see Figures \ref{fig1.1}--\ref{fig1.3} below.
\begin{figure}[H]
    \centering
    \begin{minipage}{0.32\textwidth}
        \centering
        \includegraphics[width=0.9\textwidth]{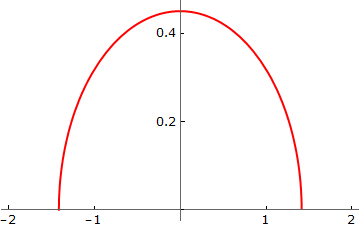}
        \caption[Limiting eigenvalue density of the Gaussian ensembles]{$\rho^{(G),0}(\lambda)$} \label{fig1.1}
    \end{minipage}\hfill
    \begin{minipage}{0.32\textwidth}
        \centering
        \includegraphics[width=0.9\textwidth]{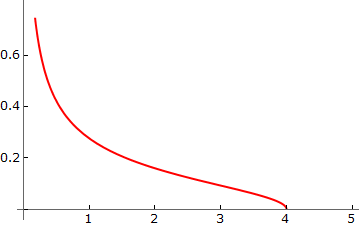}
        \caption[Limiting eigenvalue density of the Laguerre ensembles]{$\rho^{(L),0}(\lambda)|_{\gamma_1=1}$}
    \end{minipage}\hfill
    \begin{minipage}{0.32\textwidth}
        \centering
        \includegraphics[width=0.9\textwidth]{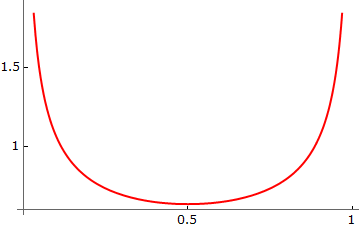}
        \caption[Limiting eigenvalue density of the Jacobi ensembles]{$\rho^{(J),0}(\lambda)|_{\gamma_1=\gamma_2=1}$} \label{fig1.3}
    \end{minipage}
\end{figure}
The other regime of interest is when the parameters $a$ and/or $b$ are proportional to $N$ \citep{Be97}, \citep{BFB97}, \citep{VV08}, \citep{No08}, \citep{LV11}, \citep{MS11}, \citep{MS12}, \citep{CMSV16a}, \citep{CMSV16b}. This situation is a little more complicated, as the limiting forms $\rho^{(L),0}(\lambda)$ and $\rho^{(J),0}(\lambda)$ are no longer parameter-independent. In fact, when $a$ ($b$) is linear in $N$, we have $\gamma_1>1$ ($\gamma_2>1$); consequently, the endpoint $\lambda_-$ ($\lambda_+$) turns from a hard edge to a soft edge, a phenomenon which is discussed further in \S\ref{s2.4.2}. In the Cauchy case, one sees that $\rho^{(Cy),0}(\lambda)$ also has soft edges when we take $\alpha=\hat{\alpha}\kappa N$ with $\hat{\alpha}={\rm O}(1)$, but unlike the $\rho^0(\lambda)$ of the other classical matrix ensembles, this is not a degeneration of a hard edge: When $\alpha$ is constant in $N$, the Cauchy ensembles technically do not have a global scaled eigenvalue density, but the large $N$ limiting form of its equivalent is
\begin{equation}
\lim_{N\to\infty}\tilde{\rho}^{(Cy)}(\lambda)\Big|_{\hat{\alpha}=0}=\rho^{(Cy),0}(\lambda)\Big|_{\hat{\alpha}_1=\hat{\alpha}_2=0}=\frac{1}{\pi(1+\lambda^2)}.
\end{equation}
As this is supported on the real line, we do not see a hard nor soft edge. Instead, one sees a \textit{spectrum singularity} at infinity, which is a term we formally introduce in \S\ref{s1.2.1}.

Another reason for focusing on the Gaussian, Laguerre, and Jacobi ensembles while relegating the Cauchy ensembles to secondary interest is that when considering fractional linear transformations taken from $GL_2(\mathbb{C})$ rather than $GL_2(\mathbb{R})$, the Cauchy weight can be seen to be equivalent to the Jacobi weight. Indeed, a simple change of variables shows that
\begin{equation} \label{eq1.2.22}
w^{(Cy)}(\lambda)=4^{\textrm{Re}(\eta)}\,w^{(J)}\left(\tfrac{1}{2}(1-\mathrm{i}\lambda)\right)\Big|_{a=\overline{b}=\eta}.
\end{equation}
Thus, our results for the Jacobi ensembles translate to the corresponding Cauchy ensembles through analytic continuation in the parameters $a,b$. In this vein, the proofs of Chapter 2 concerning the Jacobi ensembles also transfer to the Gaussian and Laguerre ensembles via Lebesgue's dominated convergence theorem combined with the following limiting procedure.
\begin{lemma} \label{L1.1}
Let $w^{(G)}(\lambda),w^{(L)}(\lambda)$, and $w^{(J)}(\lambda)$ be given by equation \eqref{eq1.2.9} in the Gaussian, Laguerre, and Jacobi cases, respectively. Then,
\begin{equation} \label{eq1.2.23}
\lim_{b\rightarrow\infty}b^a\,w^{(J)}(\lambda/b)=w^{(L)}(\lambda)
\end{equation}
and if we set $a=b=L$,
\begin{equation} \label{eq1.2.24}
\lim_{L\rightarrow\infty}4^{2L}\,w^{(J)}\left(\tfrac{1}{2}(1+\lambda/\sqrt{L})\right)=w^{(G)}(\lambda).
\end{equation}
\end{lemma}
Moreover, it can be observed that $w^{(L)}(\lambda^2)|_{a=0}=w^{(G)}(\lambda)$, so there is a hierarchy placing the Jacobi ensembles above the Laguerre ensembles, which, in turn, sit above the Gaussian ensembles. This corresponds to the loss of parameter $b$ and then $a$.

Some of the tools used in the proofs of Chapter 2 apply to the more general \textit{classical $\beta$ ensembles}, so we defer their introductions to \S\ref{s1.2.4}. Other properties of the classical matrix ensembles that we wish to review follow from their status as \textit{(skew-)orthogonal polynomial ensembles} (SOPEs), which are themselves examples of \textit{invariant matrix ensembles}. These classes of random matrix ensembles are discussed in \S\ref{s1.2.3} and \S\ref{s1.2.2}, respectively. Before doing so, let us first look at the Cauchy ensembles in more depth.

\subsection{The Cauchy and circular Jacobi ensembles} \label{s1.2.1}
For specific values of $a,b$ and with $\beta\in\{1,2,4\}$, we have seen from Definition \ref{def1.4} that $p^{(G)}$, $p^{(L)}$, and $p^{(J)}$ are j.p.d.f.s of the eigenvalues of random matrices that are constructed from Ginibre matrices. In contrast, for fixed $\beta\in\{1,2,4\}$, we can interpret $p^{(Cy)}$ as the eigenvalue j.p.d.f.~of a random matrix related to an appropriate circular ensemble. To see this, let $U$ be drawn from the $N\times N$ circular orthogonal, unitary, or symplectic ensemble. Then, define the Hermitian random matrix $A$ via the \textit{Cayley transformation}
\begin{equation} \label{eq1.2.25}
A:=\mathrm{i}\frac{U+I_N}{U-I_N}=\mathrm{i}(U+I_N)(U-I_N)^{-1}\in\mathbb{M}_{N\times N}(\mathbb{C}).
\end{equation}
The eigenvalues $\{\lambda_i\}_{i=1}^N$ of $A$ are related to the eigenvalues $\{e^{\mathrm{i}\theta_i}\}_{i=1}^N$ of $U$ by the stereographic projection
\begin{equation} \label{eq1.2.26}
\lambda_i=\cot(\theta_i/2).
\end{equation}
Applying the associated inverse mapping $e^{\mathrm{i}\theta_i}=(\lambda_i+\mathrm{i})/(\lambda_i-\mathrm{i})$ to the eigenvalue j.p.d.f. \eqref{eq1.1.8} of the circular ensembles shows that the eigenvalues of $A$ are distributed according to the eigenvalue j.p.d.f.~\eqref{eq1.2.7} with the Cauchy weight \eqref{eq1.2.9}, when $\eta=-\kappa(N-1)-1$ (i.e., when $\alpha=0$).

When studying the eigenvalue j.p.d.f.~\eqref{eq1.2.7}, there is no reason to restrict our parameters to only those that relate the j.p.d.f.~to a realisable random matrix. In fact, for most of this thesis, we consider the Laguerre and Jacobi ensembles with $a,b>-1$ real. The natural way to generalise the Cauchy ensembles' eigenvalue j.p.d.f.s is to introduce the complex parameter $\alpha$ with $\textrm{Re}(\alpha)>-1/2$, and write $\eta=-\kappa(N-1)-1-\alpha$. This ensures that $w^{(Cy)}(\lambda)$ continues to be real-valued on the real line and that
\begin{equation*}
\mathcal{N}_{N,\beta}^{(Cy)}=\int_{\mathbb{R}^N}p^{(Cy)}(\lambda_1,\ldots,\lambda_N;\beta)\,\mathrm{d}\lambda_1\cdots\mathrm{d}\lambda_N
\end{equation*}
is finite (we require $\textrm{Re}(\alpha)>-1/2$ for finiteness due to the non-compact domain of integration). Applying the stereographic projection \eqref{eq1.2.26} to $p^{(Cy)}$ with $\alpha\neq0$ yields the eigenvalue j.p.d.f.~of the circular Jacobi ensemble, which is an extension of the circular ensembles.

\begin{definition}
Taking $\beta>0$ real, let $\alpha\in\mathbb{C}$, $\alpha_1:=\textrm{Re}(\alpha)>-1/2$, and $\alpha_2:=\textrm{Im}(\alpha)$. The size-$N$ \textit{circular Jacobi ensemble} is the system of `eigenvalues'\footnote{Although calling $p^{(cJ)}$ an eigenvalue j.p.d.f.~is a priori artifical in some sense, we will see in \S\ref{s1.2.4} that random matrices have been constructed whose eigenvalues are indeed distributed according to $p^{(cJ)}$ \citep{BNR09}. Similar matrix constructions for the eigenvalue j.p.d.f. \eqref{eq1.2.7} with $\beta>0$ general real are also outlined in \S\ref{s1.2.4}.} on the unit circle $\{e^{\mathrm{i}\theta}\,|\,\theta\in[0,2\pi)\}$ distributed according to the eigenvalue j.p.d.f.
\begin{equation}
p^{(cJ)}(e^{\mathrm{i}\theta_1},\ldots,e^{\mathrm{i}\theta_N};\beta):=\frac{1}{\mathcal{N}_{N,\beta}^{(cJ)}}\prod_{i=1}^Nw^{(cJ)}(e^{\mathrm{i}\theta_i})\,|\Delta_N(e^{\mathrm{i}\theta})|^{\beta},
\end{equation}
where $\mathcal{N}_{N,\beta}^{(cJ)}$ is a normalisation constant, $\Delta_N(e^{\mathrm{i}\theta})$ is the Vandermonde determinant \eqref{eq1.1.9}, and the circular Jacobi weight \citep{BNR09}, \citep{Li17} is given by
\begin{equation} \label{eq1.2.28}
w^{(cJ)}(e^{\mathrm{i}\theta}):=(1-e^{-\mathrm{i}\theta})^\alpha(1-e^{\mathrm{i}\theta})^{\overline{\alpha}}=e^{\alpha_2(\theta-\pi)}\,|1-e^{\mathrm{i}\theta}|^{2\alpha_1}.
\end{equation}
\end{definition}

The weight $w^{(cJ)}$ is said to have a \textit{spectrum singularity} of Fisher-Hartwig type \citep{FH68}, \citep{BM94}, \citep{XZ20} at $\theta=0$; when $\alpha_1\neq0$, it is of root-type (it is actually $\log\,w^{(cJ)}(e^{\mathrm{i}\theta})$ that is singular) and when $\alpha_2\neq0$, it is of jump-type. Accordingly, we say that the Cauchy ensembles display a spectrum singularity in the regime $|\lambda|\to\infty$. Thus, studying the Cauchy ensembles allows us to shine light on aspects of the spectrum singularity scaling regime, supplementing similar studies of the soft and hard edges exhibited by the other classical ensembles. Earlier, we mentioned that the spectrum singularity of the Cauchy ensembles transforms into a pair of soft edges when we take $\alpha\propto N$ --- the same behaviour has been recorded for the circular Jacobi ensemble \citep{Li17}. This mechanism is interesting to us because when we have a spectrum singularity, our methods involving the resolvent $W_1(x)$ cannot be used since not enough spectral moments $m_k^{(Cy)}$ converge. However, this ceases to be an issue when we set $\alpha=\hat{\alpha}\kappa N$ and work with $\tilde{\rho}^{(Cy)}(\lambda)$ as defined in \eqref{eq1.2.13}. To be precise, note that when $\lambda_1,\ldots,\lambda_N\to\infty$,
\begin{equation*}
|\Delta_N(\lambda)|^{\beta}\sim\prod_{i=1}^N\lambda_i^{\beta(N-1)},\qquad w^{(Cy)}(\lambda_i)|_{\eta=-\kappa(N-1)-1-\alpha}\sim\lambda_i^{-\beta(N-1)-2-2\alpha_1},
\end{equation*}
so by equation \eqref{eq1.1.15}, $m_k^{(Cy)}$ is convergent only when $-1<k<2\alpha_1+1$. This upper bound disappears upon setting $\alpha_1=\hat{\alpha}_1\kappa N$, as introduced in Proposition \ref{prop1.2}, and taking $N\to\infty$.

In this thesis, the connection between the Cauchy and circular Jacobi ensembles will not be used to study the Cauchy ensembles, but will rather serve as motivation for doing so. Instead, as mentioned earlier, our results concerning the Cauchy ensembles flow from equivalent results for the Jacobi ensembles. Along the way, we will also formulate certain intermediate results in terms of the \textit{shifted Jacobi ensembles}, which are defined to have eigenvalue j.p.d.f. \eqref{eq1.2.7} with weight
\begin{equation} \label{eq1.2.29}
w^{(sJ)}(\lambda):=(1-\lambda)^a(1+\lambda)^b\chi_{-1<\lambda<1}=w^{(J)}\left(\tfrac{1}{2}(1-\lambda)\right).
\end{equation}
Since results for the Cauchy and shifted Jacobi ensembles descend from those for the (un-shifted) Jacobi ensembles, we present only those results where the relevant transformation gives rise to significant simplifications. Hence, we will report primarily on the \textit{symmetric} Cauchy and shifted Jacobi ensembles, which corresponds to taking $\alpha\in\mathbb{R}$ and $a=b$.

\subsection{Invariant matrix ensembles} \label{s1.2.2}
Given a group $\mathcal{G}$ and a group action $\mathcal{A}:\mathcal{G}\times\mathcal{S}\rightarrow\mathcal{S}$, a random matrix ensemble $\mathcal{E}=(\mathcal{S},P)$ is said to be an $\mathcal{A}$\textit{-invariant matrix ensemble} if the probability measure $\mathrm{d}P(X):=P(X)\,\mathrm{d}X$ (recall that in the case of the circular ensembles, we do not prescribe a p.d.f.~$P(X)$, but work with a probability measure $\mathrm{d}P(X)$ directly) satisfies
\begin{equation}
\mathrm{d}P(\mathcal{A}(g,X))=\mathrm{d}P(X)
\end{equation}
for every $g\in\mathcal{G}$. For example, the CUE of Definition \ref{def1.2} is invariant under left and right multiplication by elements of $U(N)$, while the COE and CSE have invariant structures induced by those of the CUE \citep[I, Thrms.~1,5]{Dys62b}.

The adjoint action is by far the most prolific group action studied in random matrix theory. Indeed, the term \textit{invariant matrix ensemble}, with $\mathcal{A}$ unspecified, refers to invariance under the mapping $X\mapsto gXg^{-1}$ for all $g$ drawn from some appropriate compact group $\mathcal{G}$; when such an invariant matrix ensemble is assumed to be irreducible and the matrix set underlying it is such that $\mathcal{S}\subseteq\mathbb{M}_{N\times N}(\mathbb{R})$, $\mathbb{M}_{N\times N}(\mathbb{C})$, or $\mathbb{M}_{N\times N}(\mathbb{H})$, the group $\mathcal{G}$ must be one of $O(N)$, $U(N)$, or $Sp(2N)$ \citep{Dys62}. Then, the invariant matrix ensemble at hand is more precisely referred to as an \textit{orthogonal ($O(N)$), unitary ($U(N)$), or symplectic ($Sp(2N)$) ensemble}, and if the generic entries of the elements of both $\mathcal{S},\mathcal{G}$ belong to the same number system, the statistics of the ensemble are basis-independent (conjugation by an element of $\mathcal{G}$ is equivalent to a change of basis). The adjectives orthogonal, unitary, and symplectic have already been introduced in the circular and Ginibre cases in Definitions \ref{def1.2} and \ref{def1.3}, where we saw that they correspond to $\beta=1,2$, and $4$, respectively. It can moreover be checked that these invariances extend to the Cauchy matrix $A$ defined in equation \eqref{eq1.2.25} and we show below that the real, complex, and quaternionic matrices of Definition \ref{def1.4} have orthogonal, unitary, and symplectic invariance, respectively. Thus, with G, L, J, Cy labelling the classical weights \eqref{eq1.2.9} and O, U, S the invariance classes, we will for the remainder of this thesis refer, e.g., to the real Gaussian ensemble as the GOE and the $\beta=4$ Cauchy ensemble as the CySE.

\subsubsection{Invariance of the circular, Ginibre, Gaussian, Laguerre, and Jacobi ensembles}
In the CUE case, $\mathrm{d}P(X)$ is the Haar measure on $U(N)$, so it is trivially invariant under the mapping $X\mapsto UXU^\dagger$ for $U\in U(N)$ --- it is a slightly more astute observation that this invariance translates to invariance of the COE (CSE) under the adjoint action of $O(N)$ ($Sp(2N)$) \citep[I]{Dys62b} (recall the forms of $\mathcal{S}$ given in Definition~\ref{def1.2}). Otherwise, the probability measure $\mathrm{d}P(X)$ can be seen to be invariant under the adjoint action of a group $\mathcal{G}$ if one is able to establish the invariance of the p.d.f.~$P(X)$ and the Lebesgue measure $\mathrm{d}X$ separately. It is immediate that $P^{(Gin)}|_{M=N}(X)$, $P^{(G)}(X)$, $P^{(L)}(X)$, and $P^{(J)}(X)$, as specified in Definition~\ref{def1.3} and Proposition \ref{prop1.1}, are invariant under the appropriate adjoint actions (conjugation of $X$ by elements of $O(N),U(N),Sp(2N)$ corresponding to $\beta=1,2,4$) since the functions $\Det\,X$, $\Tr\,X^\dagger X$, and $\Tr\,X^k$ ($k\in\mathbb{N}$) all have this property. More involved is the invariance of the Lebesgue measure $\mathrm{d}X$, where the strategy used depends on whether or not the associated matrix set $\mathcal{S}$ consists purely of self-adjoint matrices. In regards to Remark \ref{R1.3}, one may use the diagonalisation formula $X=U\Lambda U^\dagger$ and some calculus of differential forms to see that the Lebesgue measure on the sets of self-adjoint matrices $\mathcal{S}^{(G)},\mathcal{S}^{(L)},\mathcal{S}^{(J)}$ \eqref{eq1.2.4}--\eqref{eq1.2.6} can be written in a form that is manifestly invariant under the relevant adjoint action.
\begin{proposition} \label{prop1.3}
Given an $N\times N$ real ($\beta=1$), complex ($\beta=2$), or quaternionic ($\beta=4$) self-adjoint matrix variable $X$, diagonalise it as $X=U\Lambda U^\dagger$, where $\Lambda=\mathrm{diag}(\lambda_1,\ldots,\lambda_N)$ is a diagonal matrix of eigenvalues, and $U$ is a corresponding matrix of eigenvectors; the spectral theorem dictates that $U$ is an element of $O(N)$ in the real case, $U(N)$ in the complex case, or $Sp(2N)$ in the quaternionic case. To make the mapping $X\mapsto U\Lambda U^\dagger$ bijective, we insist that the eigenvalues be listed in increasing order and that the first row of $U$ be non-negative real. The canonical Lebesgue~measure
\begin{equation}
\mathrm{d}X=\prod_{i=1}^N\mathrm{d}X_{ii}\prod_{1\leq j<k\leq N}\prod_{s=1}^{\beta}\mathrm{d}X_{jk}^{(s)}
\end{equation}
on the set of self-adjoint matrices (recall that $X_{jk}^{(s)}$ denotes the $s\textsuperscript{th}$ real component of the matrix entry $X_{jk}$, cf.~the measures \eqref{eq1.1.3} and \eqref{eq1.1.7}) decomposes as
\begin{equation} \label{eq1.2.32}
\mathrm{d}X=|\Delta_N(\lambda)|^{\beta}\,\mathrm{d}\Lambda\,\mathrm{d}\mu_{\mathrm{Haar}}(U),
\end{equation}
where $\mathrm{d}\Lambda=\prod_{i=1}^N\mathrm{d}\lambda_i$ is the Lebesgue measure on the \textbf{Weyl chamber} $\{\lambda_1<\cdots<\lambda_N\}\subset\mathbb{R}^N$~and 
\begin{equation} \label{eq1.2.33}
\mathrm{d}\mu_{\mathrm{Haar}}(U)=\bigwedge_{1\leq l<k\leq N}\bigwedge_{s=1}^{\beta}\left(U^\dagger\left[\mathrm{d}U_{ij}\right]_{i,j=1}^N\right)_{lk}^{(s)}
\end{equation}
is the (unnormalised) Haar measure on the quotient space $O(N)/O(1)^N$ ($\beta=1$), $U(N)/U(1)^N$ ($\beta=2$), or $Sp(2N)/Sp(2)^N$ ($\beta=4$) given by the exterior (wedge) product of the independent real components of the entries of the matrix of differentials $U^\dagger\left[\mathrm{d}U_{ij}\right]_{i,j=1}^N$ \citep{Hua63}, \citep[Ch.~1]{Fo10}.
\end{proposition}

\begin{proof}[Proof sketch]
Using Leibniz's rule, take the exterior derivative of the diagonalisation formula $X=U\Lambda U^\dagger$ and then conjugate the result by $U^\dagger$ to obtain
\begin{equation} \label{eq1.2.34}
U^\dagger\left[\mathrm{d}X_{ij}\right]_{i,j=1}^NU=U^\dagger\left[\mathrm{d}U_{ij}\right]_{i,j=1}^N\,\Lambda-\Lambda\,U^\dagger\left[\mathrm{d}U_{ij}\right]_{i,j=1}^N+\textrm{diag}(\mathrm{d}\lambda_1,\ldots,\mathrm{d}\lambda_N).
\end{equation}
Carrying out the necessary matrix multiplications shows that the right-hand side is a self-adjoint matrix with the $(i,j)$ entry given by $\mathrm{d}\lambda_i\chi_{i=j}+(\lambda_j-\lambda_i)\left(U^\dagger[\mathrm{d}U_{kl}]_{k,l=1}^N\right)_{ij}$ for $i\leq j$. Taking the exterior product of the real components of these entries and replacing the Jacobian by its absolute value yields the right-hand side of equation \eqref{eq1.2.32}, while applying the same prescription to the entries of the left-hand side of equation \eqref{eq1.2.34} gives $\mathrm{d}X$, as desired --- note that $\left(U^\dagger[\mathrm{d}U_{kl}]_{k,l=1}^N\right)_{ij}$ contains $\beta$ real components, resulting in as many powers of $|\Delta_N(\lambda)|$ in \eqref{eq1.2.32}. The quotient spaces $O(N)/O(1)^N,U(N)/U(1)^N,Sp(2N)/Sp(2)^N$ are respectively isomorphic to the subsets of $O(N),U(N),Sp(2N)$ that house the matrix $U$.
\end{proof}

\begin{note} \label{N1.2}
The Haar measures on $O(N),U(N),Sp(2N)$ are computed from $U^\dagger[\mathrm{d}U_{ij}]_{i,j=1}^N$ in the same way as $\mathrm{d}\mu_{\mathrm{Haar}}(U)$ above, but the outcome is different due to the extra $N$ parameters. Indeed, if $U\in O(N),U(N),Sp(2N)$ instead of $O(N)/O(1)^N,U(N)/U(1)^N,Sp(2N)/Sp(2)^N$, the relevant procedure returns
\begin{equation} \label{eq1.2.35}
\mathrm{d}\mu'_{\mathrm{Haar}}(U)=\bigwedge_{h=1}^N\left(U^\dagger\left[\mathrm{d}U_{ij}\right]_{i,j=1}^N\right)_{hh}\bigwedge_{1\leq l<k\leq N}\bigwedge_{s=1}^{\beta}\left(U^\dagger\left[\mathrm{d}U_{ij}\right]_{i,j=1}^N\right)_{lk}^{(s)}.
\end{equation}
This is not to be confused with the measures defining the circular ensembles, where equivalence holds only in the CUE case.
\end{note}

It is now straightforward to see that the Lebesgue measure on the set of self-adjoint matrices is invariant under the adjoint action: Taking the complex case for definiteness, write $X\in\mathcal{S}^{(G)}$ as $X=U\Lambda U^\dagger$ according to the prescription given in Propositon \ref{prop1.3}, let $U'\in U(N)$, and define $X'=U'XU'^\dagger$. Then, letting $D$ be a diagonal matrix of phases such that $U'UD$ has first row non-negative real, we have $X'=(U'UD)\Lambda(U'UD)^\dagger$ and hence
\begin{equation}
\mathrm{d}X'=|\Delta_N(\lambda)|^2\,\mathrm{d}\Lambda\,\mathrm{d}\mu_{\mathrm{Haar}}(U'UD)=\mathrm{d}X.
\end{equation}
The second equality follows from $(U'UD)^\dagger\left[\mathrm{d}(U'UD)_{ij}\right]_{i,j=1}^N=D^\dagger U^\dagger\left[\mathrm{d}U_{ij}\right]_{i,j=1}^ND$ and the fact that the exterior product of the independent real components of the entries of this second matrix agrees with $\mathrm{d}\mu_{\mathrm{Haar}}(U)$ \eqref{eq1.2.33}, up to sign \citep[Thrm.~1.20]{Mat97}, \citep[pg.~17]{Fo10}. This latter equivalence of measures can be checked by writing $D=\prod_{l=1}^ND_l$ with $D_l:=[D_{ij}\chi_{i=j=l}]_{i,j=1}^N$ and explicitly showing for a matrix of differentials $\Omega=[\mathrm{d}\omega_{ij}]_{i,j=1}^N$ with only upper triangular entries independent (as is true of $U^\dagger\left[\mathrm{d}U_{ij}\right]_{i,j=1}^N$) that
\begin{equation}
\bigwedge_{1\leq j<k\leq N}(D_l\Omega D_l^\dagger)_{jk}^{(1)}\wedge(D_l\Omega D_l^\dagger)_{jk}^{(2)}=\bigwedge_{1\leq j<k\leq N}\mathrm{d}\omega_{jk}^{(1)}\wedge\mathrm{d}\omega_{jk}^{(2)},\quad1\leq l\leq N.
\end{equation}
Extending these arguments to the real and quaternionic cases verifies that the Gaussian, Laguerre, and Jacobi ensembles, as introduced in Definition \ref{def1.4} and further specified by Proposition \ref{prop1.1} and Remark \ref{R1.3}, are invariant matrix ensembles.

\begin{remark} \label{R1.5}
Similar to the self-adjoint matrices above, elements $X$ of the circular ensembles have unique eigendecompositions of the form $X=U\exp(\mathrm{i}\Theta)U^{\dagger}$, where $U$ is as given in Proposition \ref{prop1.3} and $\Theta=\mathrm{diag}(\theta_1,\ldots,\theta_N)$ is such that the eigenangles $\{\theta_i\}_{i=1}^N$ are listed in increasing order. The analogue of the decomposition \eqref{eq1.2.32} in the circular cases is
\begin{equation} \label{eq1.2.38}
\mathrm{d}P(X)=\frac{1}{\mathcal{Z}_{N,\beta}^{(C)}}|\Delta_N(e^{\mathrm{i}\theta})|^{\beta}\,\mathrm{d}\Theta\,\mathrm{d}\mu_{\mathrm{Haar}}(U),
\end{equation}
where $\mathrm{d}\mu_{\mathrm{Haar}}(U)$ is again as defined in Proposition \ref{prop1.3}, $\mathrm{d}\Theta=\prod_{i=1}^N\mathrm{d}\theta_i$ is the Lebesgue measure on $\{0\leq\theta_1<\cdots<\theta_N<2\pi\}$ \citep[I]{Dys62b}, and $\mathcal{Z}_{N,\beta}^{(C)}$ is a normalisation constant. Note that in both equations \eqref{eq1.2.32} and \eqref{eq1.2.38}, the Jacobian drawn out from the change of variables is the absolute value of the Vandermonde determinant to the $\beta\textsuperscript{th}$ power, making the related invariant matrix ensembles suitable for modelling systems of random particles that interact via pairwise logarithmic repulsion (i.e., log-gases \citep{Fo10}).
\end{remark}

In the situation concerning $N\times N$ Ginibre matrices $G$, things are more complicated. In the 1965 work \citep{Gin65}, Ginibre extended the strategy outlined above by starting with the diagonalisation formula $G=P\Lambda P^{-1}$, where $\Lambda$ is a matrix of eigenvalues and $P$ is a corresponding matrix of eigenvectors (this formula is valid with probability one since the subset of matrices with repeated eigenvalues has Lebesgue measure zero). Since $G$ is not self-adjoint, its eigenvalues are generically complex in the complex and quaternionic cases (when $G$ is quaternionic, it has $N$ pairs of complex conjugate eigenvalues that are thus fully determined by those in the upper half plane), while in the real case, $G$ has a mixture of real and complex conjugate eigenvalues. The eigenvalues of $G$ do not have a natural ordering, but this can be accounted for by a factor of $N!$ or by assigning some arbitrary ordering. More importantly, the diagonalising matrix $P$ is not necessarily orthogonal, unitary, nor symplectic. Nonetheless, the QR decomposition can be used to write $P=UT$ where $U$ is as specified in Proposition \ref{prop1.3}, and $T$ is an upper triangular matrix with diagonal entries real and positive. Thus, it was shown in \citep{Gin65} that the relevant Lebesgue measure \eqref{eq1.1.7} decomposes as
\begin{equation} \label{eq1.2.39}
\mathrm{d}G=J(\lambda)\,\mathrm{d}\Lambda\,\mathrm{d}\mu_{\mathrm{Tri}}(T)\,\mathrm{d}\mu_{\mathrm{Haar}}(U),
\end{equation}
where $\mathrm{d}\Lambda$ is the Lebesgue measure on the space housing the eigenvalues, $\mathrm{d}\mu_{\mathrm{Haar}}(U)$ is the Haar measure defined in Proposition \ref{prop1.3}, $J(\lambda)$ is a Jacobian consisting of products of Vandermonde determinants involving the eigenvalues, and $\mathrm{d}\mu_{\mathrm{Tri}}(T)$ is some measure dependent on the entries of $T$; we leave the latter two unspecified and refer the interested reader to \citep{Gin65}. As was argued for the decomposition \eqref{eq1.2.32}, it follows from the factorisation \eqref{eq1.2.39} that $\mathrm{d}G$ is invariant under the desired adjoint action. Combined with the invariance of $P^{(Gin)}|_{M=N}(G)$ established earlier, we conclude that the $N\times N$ Ginibre ensembles are invariant matrix ensembles.

Soon after, Dyson \citep[App.~A.33]{Meh04} provided a simplification of the above in the complex and quaternionic cases by using the Schur decomposition $G=U\hat{T}U^\dagger$, with $U$ as in Proposition~\ref{prop1.3} and $\hat{T}$ upper triangular with diagonal entries being the eigenvalues of $G$ --- the measure $\mathrm{d}G$ was yet again shown to factorise in a similar manner to that in equation \eqref{eq1.2.39}. Note that one can choose $\hat{T}=T\Lambda T^{-1}$, where $T$ is the upper triangular matrix used by Ginibre. The real case was treated in this manner by Edelman in 1997 \citep{Ede97} in a parallel effort to Lehmann and Sommers' 1991 work \citep{LS91}, with $\hat{T}$ now upper quasi-triangular. The quasi-triangular form is due to the complex conjugate eigenvalue pairs being encoded within appropriate $2\times2$ real matrices, which is necessitated by the requirement that $\hat{T}$ be real.

\subsubsection{Relation to eigenvalue j.p.d.f.s}
To obtain the eigenvalue j.p.d.f.~$p(\lambda_1,\ldots,\lambda_N)$ from the probability measure $\mathrm{d}P(X)$, one changes variables to the eigenvalues and a set of auxiliary variables, then integrates over the latter. In the ensembles discussed above, the auxiliary variables relate to the entries of the matrix $U\in O(N)/O(1)^N$ ($\beta=1$), $U(N)/U(1)^N$ ($\beta=2$), $Sp(2N)/Sp(2)^N$ ($\beta=4$) and, in the Ginibre cases, either the entries of $T$ or the strictly upper triangular entries of $\hat{T}$. Ginibre \citep{Gin65} was able to integrate out the auxiliary variables and obtain the eigenvalue j.p.d.f.~for the complex and quaternionic Ginibre ensembles, but could only address the real Ginibre ensemble in the case that all of the eigenvalues were real. Approximately 25 years later, the works \citep{LS91}, \citep{Ede97} used (a form of) the Schur decomposition to obtain the eigenvalue j.p.d.f.~of the real Ginibre ensemble with any number of real eigenvalues. Although Ginibre's substitution $G=UT\Lambda T^{-1}U^\dagger$ and the Schur decomposition $G=U\hat{T}U^\dagger$ both lead to expressions for $\mathrm{d}G$ that factorise nicely, only the latter transforms $P^{(Gin)}|_{M=N}(G)$ \eqref{eq1.1.6} into an amicable form. Indeed, the benefit of using the Schur decomposition is that
\begin{equation}
P^{(Gin)}|_{M=N}(U\hat{T}U^\dagger)=\hat{P}(\hat{T})\,\prod_{i=1}^Ne^{-\lambda_i^2},
\end{equation}
where $\hat{P}(\hat{T})$ is some function depending purely on the strictly upper triangular entries of $\hat{T}$.

In the interest of brevity, we do not review the Ginibre ensembles in any further detail here, but note that their eigenvalue j.p.d.f.s have vastly different forms for $\beta=1,2,4$, in contrast to the universal structures of the eigenvalue j.p.d.f.s \eqref{eq1.1.8}, \eqref{eq1.2.7} of the circular and classical matrix ensembles. A succinct review of the eigenvalue j.p.d.f.s of the Ginibre ensembles can be found in \citep{FN07}.

\begin{remark} \label{R1.6}
Contrary to the eigenvalues, the squared singular values of the Ginibre matrix $G$ are non-negative real and are well-defined even when $G$ is rectangular. Moreover, they are equivalent to the eigenvalues of the Wishart--Laguerre matrix $G^\dagger G$.
\end{remark}

Moving on to the Gaussian, Laguerre, and Jacobi ensembles, things are simpler than in the Ginibre cases since our diagonalising matrices are now orthogonal, unitary, or symplectic, as appropriate. Combined with the invariance of the p.d.f.~$P(X)$, this means that under the change of variables $X=U\Lambda U^\dagger$ prescribed by Proposition \ref{prop1.3}, we have that $P(X)=P(\Lambda)$. Hence, letting $P(X)$ be one of the p.d.f.s specified in Proposition \ref{prop1.1} and fixing $(\beta,\mathcal{G}(N))=(1,O(N))$, $(2,U(N))$, or $(4,Sp(2N))$, Proposition \ref{prop1.3} implies that
\begin{equation} \label{eq1.2.41}
\int_{\mathcal{G}(N)/\mathcal{G}(1)^N}\mathrm{d}P(X)=P(\Lambda)\,|\Delta_N(\lambda)|^{\beta}\,\mathrm{d}\Lambda\int_{\mathcal{G}(N)/\mathcal{G}(1)^N}\mathrm{d}\mu_{\mathrm{Haar}}(U).
\end{equation}
As the eigenvalue and eigenvector statistics are decoupled, one reads off that the eigenvalue j.p.d.f. is thus given by
\begin{equation} \label{eq1.2.42}
p(\lambda_1,\ldots,\lambda_N)=\frac{\mathrm{vol}(\mathcal{G}(N)/\mathcal{G}(1)^N)}{N!}P(\Lambda)\,|\Delta_N(\lambda)|^{\beta},
\end{equation}
where
\begin{equation}
\mathrm{vol}(\mathcal{G}(N)/\mathcal{G}(1)^N):=\int_{\mathcal{G}(N)/\mathcal{G}(1)^N}\mathrm{d}\mu_{\mathrm{Haar}}(U)
\end{equation}
is the volume of $\mathcal{G}(N)/\mathcal{G}(1)^N$ with respect to the Haar measure defined in Proposition \ref{prop1.3}, while the factor of $N!$ is needed to compensate our removal of the ordering on the eigenvalues; the ordering $\lambda_1<\cdots<\lambda_N$ is implicit in the definition of $\mathrm{d}\Lambda$, but clashes with the definition of $p(\lambda_1,\ldots,\lambda_N)$, which assumes that all eigenvalues are indistinguishable. With $w(\lambda)$ the appropriate classical weight \eqref{eq1.2.9}, it is easy to check that $P(\Lambda)=\prod_{i=1}^Nw(\lambda_i)$, which confirms that equation \eqref{eq1.2.42} is consistent with the definition \eqref{eq1.2.7} of the eigenvalue j.p.d.f. $p^{(w)}(\lambda_1,\ldots,\lambda_N)$. Furthermore, we are led to the observation that the partition functions of Proposition \ref{prop1.1} can be obtained from the computation
\begin{equation} \label{eq1.2.44}
\mathcal{Z}_N=\frac{\mathrm{vol}(\mathcal{G}(N)/\mathcal{G}(1)^N)}{N!}\mathcal{N}_{N,\beta},
\end{equation}
as long as the value of $\mathrm{vol}(\mathcal{G}(N)/\mathcal{G}(1)^N)$ is known. This volume can be computed in a few ways, one of which being a geometric argument of Dyson's given in \citep[I]{Dys62b}, and in fact already known and implemented by Hurwitz \citep{Hur97}. Another approach is to simply compute the ratio $N!\mathcal{Z}_N/\mathcal{N}_{N,\beta}$ in a tractable case. In particular, since $P^{(G)}(X)$ is a product of independent Gaussian p.d.f.s, it is straightforward to see that $\mathcal{Z}_{N,\beta}^{(G)}=(\pi)^{N/2}(\pi/2)^{N(N-1)\beta/4}$, while $\mathcal{N}_{N,\beta}^{(G)}$ can be derived from the Selberg integral theory outlined in \S\ref{s2.1.1} or simply read off from \citep{DE02}. In short, recalling the (unnormalised) Haar measure $\mathrm{d}\mu'_{\mathrm{Haar}}(U)$ \eqref{eq1.2.35},
\begin{equation} \label{eq1.2.45}
\mathrm{vol}(\mathcal{G}(N)/\mathcal{G}(1)^N)=\frac{\int_{\mathcal{G}(N)}\mathrm{d}\mu'_{\mathrm{Haar}}(U)}{\left(\int_{\mathcal{G}(1)}\mathrm{d}\mu'_{\mathrm{Haar}}(U)\right)^N}=N!\pi^{N(N-1)\beta/4}\prod_{j=1}^N\frac{\Gamma(\beta/2+1)}{\Gamma(\beta j/2+1)}.
\end{equation}

\begin{remark}
The above argument translates directly to the circular ensembles, with $\Lambda$ replaced by $\exp(\mathrm{i}\Theta)$ and $\mathrm{d}P(X)$ specified by equation \eqref{eq1.2.38}. The analogue of equation \eqref{eq1.2.42} is then
\begin{equation}
p^{(C)}(e^{\mathrm{i}\theta_1},\ldots,e^{\mathrm{i}\theta_N})=\frac{\mathrm{vol}(\mathcal{G}(N)/\mathcal{G}(1)^N)}{N!}\,|\Delta_N(e^{\mathrm{i}\theta})|^{\beta},
\end{equation}
which is consistent with equation \eqref{eq1.1.8}. Recalling the values of $\mathcal{N}_{N,\beta}^{(C)}$ given in Example \ref{E1.1}, we see that
\begin{equation}
\mathcal{Z}_{N,\beta}^{(C)}=\mathrm{vol}(\mathcal{G}(N)/\mathcal{G}(1)^N)
\end{equation}
exactly when $\beta=2$ and $\mathcal{G}(N)=U(N)$, in keeping with Note \ref{N1.2}.
\end{remark}

\subsubsection{Other invariant matrix ensembles}
When establishing the invariance of the p.d.f.s~$P(X)$ of the Gaussian, Laguerre, and Jacobi ensembles given in Proposition \ref{prop1.1}, we used the fact that they are functions of $\Det\,X$ and $\Tr\,X^k$ for $k\in\mathbb{N}$. This argument extends naturally to the situation where the parameters $a,b$ are continuous, even though their origin in terms of (rectangular) Ginibre matrices restricts them to a discrete set. In fact, it makes sense to endow the matrix sets $\mathcal{S}^{(L)},\mathcal{S}^{(J)}$ with the p.d.f.s $P^{(L)}(W),P^{(J)}(Y)$ with general real parameters $a,b>-1$, seeing as then the eigenvalue j.p.d.f.~induced by equation \eqref{eq1.2.42} is integrable and supported on an interval in the real line. Moreover, we are able to give the Cauchy analogue of these p.d.f.s.

\begin{proposition} \label{prop1.4}
Let $\beta=1,2,4$, equivalently $\kappa=1/2,1,2$, correspond to fixing the field $\mathbb{F}$ as $\mathbb{R},\mathbb{C},\mathbb{H}$. Then, matrices $X$ drawn from $\mathcal{S}^{(G)}$ \eqref{eq1.2.4}, the set of self-adjoint $N\times N$ matrices, with p.d.f.
\begin{equation}
P^{(Cy)}(X):=\frac{1}{\mathcal{Z}_{N,\beta,\alpha}^{(Cy)}}\Det(I_N+X^2)^{-\kappa(N-1)-1-\alpha}
\end{equation}
have eigenvalues distributed according the eigenvalue j.p.d.f.~$p^{(Cy)}$ as defined by equation \eqref{eq1.2.7} with Cauchy weight \eqref{eq1.2.9}. The partition function $\mathcal{Z}_{N,\beta,\alpha}^{(Cy)}$ is given by $\mathcal{N}_{N,\beta}^{(Cy)}$ \eqref{eq2.1.14} multiplied by the appropriate volume \eqref{eq1.2.45} as prescribed by equation \eqref{eq1.2.44}.
\end{proposition}

Thus, there exist invariant matrix ensembles that correspond to the eigenvalue j.p.d.f.~\eqref{eq1.2.7} for every classical weight \eqref{eq1.2.9} and all valid parameters $a,b,\alpha$. All that is lost due to this formulation is that the relevant random matrices can no longer be related to the Ginibre ensembles in a natural way. Rather, we will see in \S\ref{s1.2.4} that there are constructions involving tridiagonal matrices which hold for continuous parameter values.

\begin{remark}
Using equation \eqref{eq1.2.42}, one is able to interpret a given multivariable symmetric function $p(\lambda_1,\ldots,\lambda_N)$ as the eigenvalue j.p.d.f.~of a self-adjoint real, complex, or quaternionic random matrix $X=U\Lambda U^\dagger$, with $\Lambda=\mathrm{diag}(\lambda_1,\ldots,\lambda_N)$ and $U\in O(N),U(N),Sp(2N)$ Haar-distributed according to $\mathrm{d}\mu_{\mathrm{Haar}}(U)$ \eqref{eq1.2.33}. Indeed, it is known from symmetric function theory \citep{Mac79} that the factor $P(\Lambda)$ in the right-hand side of equation \eqref{eq1.2.42} is a function of $\{\Tr\,\Lambda^k\}_{k=0}^N$, hence $\{\Tr\,X^k\}_{k=0}^N$. Thus, the matrix $X$ has a well-defined p.d.f.~$P(X)$ on a subset of the space of self-adjoint matrices, assuming that $p(\lambda_1,\ldots,\lambda_N)$ is sufficiently integrable.
\end{remark}
Since the upcoming (skew-)orthogonal polynomial ensembles and classical $\beta$ ensembles are specified by eigenvalue j.p.d.f.s of the type considered in the above remark, they can be interpreted as invariant matrix ensembles. As we end this subsection, let us remark that the matrix product ensembles discussed in \S\ref{s1.3.1} are also invariant matrix ensembles due to the invariance of the Ginibre and Laguerre ensembles (see, e.g., \citep{IK14} and references therein).

\subsection{Skew-orthogonal polynomial ensembles} \label{s1.2.3}
The subclass of invariant matrix ensembles considered in this subsection are those concerning self-adjoint matrices $X$ with p.d.f.s $P(X)$ such that one has the decomposition
\begin{equation}
P(X)=\prod_{i=1}^Nw(\lambda_i),
\end{equation}
where $\{\lambda_i\}_{i=1}^N$ are the indistinguishable eigenvalues of $X$ and $w(\lambda)$ is some continuous weight function that is real-valued on $\mathbb{R}$ and decays fast enough as $|\lambda|\to\infty$. Equivalently, we are interested in j.p.d.f.s of the form \eqref{eq1.2.7}, but with the weights no longer constrained to be classical.
\begin{definition} \label{def1.8}
Let $\beta=1,2$, or $4$, fix $N\in\mathbb{N}$, and let $w(\lambda)$ be a continuous, non-negative, real-valued function such that $w(\lambda)={\rm o}(\lambda^{-\beta(N-1)-1})$. Then, the j.p.d.f.
\begin{equation} \label{eq1.2.50}
p^{(w)}(\lambda_1,\ldots,\lambda_N;\beta)=\frac{1}{\mathcal{N}_{N,\beta}^{(w)}}\prod_{i=1}^Nw(\lambda_i)\,|\Delta_N(\lambda)|^{\beta}
\end{equation}
describes a size-$N$ \textit{orthogonal polynomial ensemble (OPE)} when $\beta=2$, \textit{skew-orthogonal polynomial ensemble of real type ($\mathbb{R}$-SOPE)} when $\beta=1$, or \textit{skew-orthogonal polynomial ensemble of quaternion type ($\mathbb{H}$-SOPE)} when $\beta=4$ \citep[Ch.~5]{Meh04}. As usual, $\mathcal{N}_{N,\beta}^{(w)}$ is a normalisation constant and $\Delta_N(\lambda)$ is the Vandermonde determinant \eqref{eq1.1.9}.
\end{definition}
\begin{note}
The condition $w(\lambda)={\rm o}(\lambda^{-\beta(N-1)-1})$ ensures that the j.p.d.f.~specified by the above definition has a well-defined (convergent) normalisation constant, but it does not necessarily have convergent spectral moments. As an example, the Cauchy ensembles of \S\ref{s1.2.1} exhibit these features.
\end{note}

To understand the nomenclature introduced in Definition \ref{def1.8}, a few key observations regarding the Vandermonde determinant need to be made.
\begin{lemma} \label{L1.2}
Following \citep[Ch.~5]{Meh04}, \citep[Ch.~6]{Fo10}, let $\{q_j(\lambda)\}_{j=0}^{\infty}$ be a set of monic polynomials such that $q_j(\lambda)$ is degree $j$ for each $j$. Then,
\begin{gather}
\Delta_N(\lambda):=\Det\left[\lambda_i^{j-1}\right]_{i,j=1}^N=\Det\left[q_{j-1}(\lambda_i)\right]_{i,j=1}^N, \label{eq1.2.51}
\\ \Delta_N(\lambda)^4=\Det\begin{bmatrix}\lambda_i^{j-1}\\(j-1)\lambda_i^{j-2}\end{bmatrix}_{\substack{i=1,\ldots,N\\j=1,\ldots,2N}}=\Det\begin{bmatrix}q_{j-1}(\lambda_i)\\q_{j-1}'(\lambda_i)\end{bmatrix}_{\substack{i=1,\ldots,N\\j=1,\ldots,2N}}, \label{eq1.2.52}
\end{gather}
and when $N$ is even ($N$ odd can be treated by setting $\mathrm{sgn}(\lambda_{N+1}-\lambda_i)=1$ in the Pfaffian factor),
\begin{equation} \label{eq1.2.53}
|\Delta_N(\lambda)|=\Delta_N(\lambda)\,\Pf\left[\mathrm{sgn}(\lambda_j-\lambda_i)\right]_{i,j=1}^N,
\end{equation}
where $\mathrm{sgn}(0)=0$ and otherwise, $\mathrm{sgn}(\lambda)=\lambda/|\lambda|$.
\end{lemma}
\begin{proof}[Proof sketch]
To obtain the right-hand side of equation \eqref{eq1.2.51} from the definition of the Vandermonde determinant presented on its immediate left, one simply notes that the $j\textsuperscript{th}$ column of the right-hand side is equal to the $j\textsuperscript{th}$ column of the Vandermonde matrix $[\lambda_i^{j-1}]_{i,j=1}^N$ plus columns to its left. Since adding columns does not change the determinant, one can replace the the columns $[\lambda_i^{j-1}]_{i=1}^N$ with $[q_{j-1}(\lambda_i)]_{i=1}^N$ by successively iterating through $j=2,3,\ldots,N$. The middle expression of equation \eqref{eq1.2.52} can be obtained through induction on $N$, and the right-hand side follows from a column-operation argument similar to that for equation \eqref{eq1.2.51}. Equation \eqref{eq1.2.53} follows from the right-hand side of equation \eqref{eq1.1.9} and the identity \citep{deB55} (valid for even $N$, see Appendix \ref{appendixA})
\begin{equation}
\prod_{1\leq i<j\leq N}\mathrm{sgn}(\lambda_j-\lambda_i)=\Pf\left[\mathrm{sgn}(\lambda_j-\lambda_i)\right]_{i,j=1}^N,
\end{equation}
which can be checked by confirming equality in the case $\lambda_1<\ldots<\lambda_N$ and comparing sign changes upon interchanging variables $\lambda_i\leftrightarrow\lambda_j$.
\end{proof}
If one wishes to integrate the determinantal and Pfaffian structures given in the above lemma against the product $\prod_{i=1}^Nw(\lambda_i)$ occurring in equation \eqref{eq1.2.50}, it is beneficial to use monic polynomials $\{q_j(\lambda)\}_{j=0}^{\infty}$ that are (skew-)orthogonal with respect to the weight $w(\lambda)$.

\begin{definition} \label{def1.9}
With $w(\lambda)$ a suitable weight function (usually taken to have all integer moments finite, but satisfying the hypotheses given in Definition \ref{def1.8} is sufficient for our purposes), define the following inner products on polynomial space \citep[Ch.~5]{Meh04}, \citep[Ch.~6]{Fo10}:
\begin{align}
\langle f,g\rangle^{(1)}&:=\frac{1}{2}\int_{\mathbb{R}^2}f(x)g(y)\mathrm{sgn}(y-x)w(x)w(y)\,\mathrm{d}x\mathrm{d}y,
\\ \langle f,g\rangle^{(2)}&:=\int_{\mathbb{R}}f(x)g(x)w(x)\,\mathrm{d}x,
\\ \langle f,g\rangle^{(4)}&:=\frac{1}{2}\int_{\mathbb{R}}\left(f(x)g'(x)-f'(x)g(x)\right)w(x)\,\mathrm{d}x.
\end{align}
For $\beta=1,2,4$, let $\{q_j^{(\beta)}(\lambda)\}_{j=0}^{\infty}$ be a family of monic polynomials such that each $q_j^{(\beta)}(\lambda)$ is degree $j$, and further posit that there exist finite, positive constants $\{r_j^{(\beta)}\}_{j=0}^{\infty}$ such that
\begin{gather}
\mean{q_{j}^{(2)},q_{k}^{(2)}}^{(2)}=r_j^{(2)}\chi_{j=k},
\\ \mean{q_{2j}^{(\sigma)},q_{2k+1}^{(\sigma)}}^{(\sigma)}=-\mean{q_{2k+1}^{(\sigma)},q_{2j}^{(\sigma)}}^{(\sigma)}=r_j^{(\sigma)}\chi_{j=k},
\\ \mean{q_{2j}^{(\sigma)},q_{2k}^{(\sigma)}}^{(\sigma)}=\mean{q_{2j+1}^{(\sigma)},q_{2k+1}^{(\sigma)}}^{(\sigma)}=0
\end{gather}
for all $j,k\geq0$ and $\sigma=1,4$. Then, we say that the family $\{q_j^{(1)}(\lambda)\}_{j=0}^{\infty}$ is \textit{$\mathbb{R}$-skew-orthogonal}, $\{q_j^{(2)}(\lambda)\}_{j=0}^{\infty}$ is \textit{orthogonal}, and $\{q_j^{(4)}(\lambda)\}_{j=0}^{\infty}$ is \textit{$\mathbb{H}$-skew-orthogonal}, all with respect to $w(\lambda)$.
\end{definition}

\begin{note}
The existence of these polynomials is assured by a Grahm--Schmidt type construction. They are unique for $\beta=2$, but for $\beta=1,4$, the skew-orthogonality remains true upon the replacement of $q_{2j+1}^{(\beta)}(\lambda)$ by $q_{2j+1}^{(\beta)}(\lambda)+\xi_{2j}^{(\beta)}q_{2j}^{(\beta)}(\lambda)$ for arbitrary $\xi_{2j}^{(\beta)}$.
\end{note}
Utilising these (skew-)orthogonal polynomials in the expressions given in Lemma \ref{L1.2}, it is possible to express the eigenvalue j.p.d.f.~$p^{(w)}(\lambda_1,\ldots,\lambda_N;\beta)$ \eqref{eq1.2.50} as either a determinant or a Pfaffian. We present this result without reviewing its derivation, but note that the essential idea is to absorb the factors of $w(\lambda_i)$ into the structures presented in Lemma \ref{L1.2}.

\begin{theorem}[Theorem 5.7.1, \citep{Meh04}] \label{thrm1.2}
Following \citep[Ch.~5]{Meh04}, let $\{q_j^{(\beta)}(x)\}_{j=0}^{\infty}$, $\{r_j^{(\beta)}\}_{j=0}^{\infty}$ be as in the preceding definition, and let $N\in\mathbb{N}$ with $N$ even in the case $\beta=1$ ($N$ odd is not much more complicated, but will not be covered here). Next, introduce the auxiliary functions
\begin{align}
\psi_j(x)&:=\int_{\mathbb{R}}q_j^{(1)}(y)\mathrm{sgn}(x-y)w(y)\,\mathrm{d}y,
\\ S^{(1)}_j(x,y)&:=\psi'_{2j+1}(x)\psi_{2j}(y)-\psi'_{2j}(x)\psi_{2j+1}(y),
\\ S^{(4)}_j(x,y)&:=q_{2j+1}'(x)q_{2j}(y)-q_{2j}'(x)q_{2j+1}(y),
\\ D^{(4)}_j(x,y)&:=q_{2j+1}'(x)q_{2j}'(y)-q_{2j}'(x)q_{2j+1}'(y),
\\ I^{(4)}_j(x,y)&:=q_{2j+1}(x)q_{2j}(y)-q_{2j}(x)q_{2j+1}(y),
\end{align}
and the kernels corresponding respectively to the real, complex, and quaternionic cases,
\begin{align}
K_N^{(1)}(x,y)&:=\sum_{k=0}^{N/2-1}\frac{1}{r_k^{(1)}}\begin{bmatrix}\frac{1}{2}\int_{\mathbb{R}}\mathrm{sgn}(x-z)S_k^{(1)}(z,y)\,\mathrm{d}z &S_k^{(1)}(y,x)\\-S_k^{(1)}(x,y)&\partial_yS_k^{(1)}(x,y)\end{bmatrix} \nonumber
\\&\qquad-\frac{1}{2}\begin{bmatrix}\mathrm{sgn}(x-y)&0\\0&0\end{bmatrix},
\\ K_N^{(2)}(x,y)&:=\sqrt{w(x)w(y)}\sum_{k=0}^{N-1}\frac{1}{r_k^{(2)}}q_k^{(2)}(x)q_k^{(2)}(y), \label{eq1.2.67}
\\K_N^{(4)}(x,y)&:=\sqrt{w(x)w(y)}\sum_{k=0}^{N-1}\frac{1}{r_k^{(4)}}\begin{bmatrix}I^{(4)}_k(x,y)&S^{(4)}_k(y,x)\\-S^{(4)}_k(x,y)&D^{(4)}_k(x,y)\end{bmatrix}.
\end{align}
Then, we have the eigenvalue j.p.d.f.~ of Definition \ref{def1.8} given by
\begin{equation} \label{eq1.2.69}
p^{(w)}(\lambda_1,\ldots,\lambda_N;\beta)=\begin{cases}\Det\left[K_N^{(2)}(\lambda_i,\lambda_j)\right]_{i,j=1}^N,&\beta=2,\\\Pf\left[K_N^{(\beta)}(\lambda_i,\lambda_j)\right]_{i,j=1}^N,&\beta=1,4.\end{cases}
\end{equation}
\end{theorem}
\begin{note}
The right-hand side of equation \eqref{eq1.2.69} can be made uniform by interpreting $K_N^{(1)}(x,y)$ and $K_N^{(4)}(x,y)$ as \textit{complex quaternions} (see Note \ref{NA.1} of Appendix \ref{appendixA}) and replacing the Pfaffians by quaternionic determinants, following equation \eqref{eqA.0.4}.
\end{note}

The significance of the above theorem lies in the fact that for $\beta=1,2,4$, the kernels $K_N^{(\beta)}(x,y)$ have the \textit{reproducing properties} \citep{Dys70}
\begin{align}
\int_{\mathbb{R}}\Det\left[K_N^{(2)}(\lambda_i,\lambda_j)\right]_{i,j=1}^{k+1}\,\mathrm{d}\lambda_{k+1}&=(N-k)\Det\left[K_N^{(2)}(\lambda_i,\lambda_j)\right]_{i,j=1}^k, \label{eq1.2.70}
\\ \int_{\mathbb{R}}\Pf\left[K_N^{(\sigma)}(\lambda_i,\lambda_j)\right]_{i,j=1}^{k+1}\,\mathrm{d}\lambda_{k+1}&=(N-k)\Pf\left[K_N^{(\sigma)}(\lambda_i,\lambda_j)\right]_{i,j=1}^k,
\end{align}
where $\sigma=1,4$ and $1\leq k\leq N-1$. Hence, one may iteratively integrate $p^{(w)}(\lambda_1,\ldots,\lambda_N)$ over $\mathrm{d}\lambda_{k+1}\cdots\mathrm{d}\lambda_N$ ($1\leq k\leq N-1$) to obtain the \textit{$k$-point correlation function}
\begin{align}
\rho_k^{(w)}(\lambda_1,\ldots,\lambda_k;N,\beta)&:=\frac{N!}{(N-k)!}\int_{\mathbb{R}^{N-k}}p^{(w)}(\lambda_1,\ldots,\lambda_N;\beta)\,\mathrm{d}\lambda_{k+1}\cdots\mathrm{d}\lambda_N \label{eq1.2.72}
\\&\;=\begin{cases}\Det\left[K_N^{(2)}(\lambda_i,\lambda_j)\right]_{i,j=1}^k,&\beta=2,\\\Pf\left[K_N^{(\beta)}(\lambda_i,\lambda_j)\right]_{i,j=1}^k,&\beta=1,4.\end{cases} \label{eq1.2.73}
\end{align}
The structures on the right-hand side of equation \eqref{eq1.2.73} characterise the OPEs and SOPEs as \textit{determinantal and Pfaffian point processes}, respectively. They imply, in particular, that the eigenvalue density $\rho^{(w)}(\lambda)=\rho_1^{(w)}(\lambda)$ \eqref{eq1.1.10} is given by the remarkably elegant formula
\begin{equation} \label{eq1.2.74}
\rho^{(w)}(\lambda;\beta)=\begin{cases}K_N^{(2)}(\lambda,\lambda),&\beta=2,\\\Pf\left[K_N^{(\beta)}(\lambda,\lambda)\right],&\beta=1,4.\end{cases}
\end{equation}

The right-hand sides of equations \eqref{eq1.2.73} and \eqref{eq1.2.74} can be simplified further still by using the (generalised) Christoffel--Darboux formulae. Focusing first on the $\beta=2$ case and recalling the notation introduced in Definition \ref{def1.9}, it is well known \citep{Sze75} that the orthogonal polynomials $\{q_j^{(2)}(\lambda)\}_{j=0}^{\infty}$ satisfy the three-term recurrence
\begin{equation}
q_j^{(2)}(\lambda)=(\lambda+a_j)q_{j-1}^{(2)}(\lambda)-\frac{r_{j-1}^{(2)}}{r_{j-2}^{(2)}}q_{j-2}^{(2)}(\lambda),
\end{equation}
where the coefficients $\{a_j\}_{j=1}^{\infty}$ depend on the choice of weight $w(\lambda)$. Using this recurrence, one can show that
\begin{align}
q_{j+1}^{(2)}(x)q_j^{(2)}(y)-q_j^{(2)}(x)q_{j+1}^{(2)}(y)&=(x-y)q_j^{(2)}(x)q_j^{(2)}(y) \nonumber
\\&\qquad+\frac{r_j^{(2)}}{r_{j-1}^{(2)}}\left(q_j^{(2)}(x)q_{j-1}^{(2)}(y)-q_{j-1}^{(2)}(x)q_j^{(2)}(y)\right),
\end{align}
which can then be rearranged to give
\begin{equation}
\frac{q_j^{(2)}(x)q_j^{(2)}(y)}{r_j^{(2)}}=\frac{1}{x-y}\left(\frac{q_{j+1}^{(2)}(x)q_j^{(2)}(y)-q_j^{(2)}(x)q_{j+1}^{(2)}(y)}{r_j^{(2)}}-\frac{q_j^{(2)}(x)q_{j-1}^{(2)}(y)-q_{j-1}^{(2)}(x)q_j^{(2)}(y)}{r_{j-1}^{(2)}}\right).
\end{equation}
Telescopically summing this equation over $j=0,\ldots,N-1$ (with $q_{-1}^{(2)}\equiv0$) reveals the \textit{Christoffel--Darboux formula}:
\begin{proposition}
The $\beta=2$ kernel defined in Theorem \ref{thrm1.2} simplifies as \citep{Sze75}
\begin{equation} \label{eq1.2.78}
K_N^{(2)}(x,y)=\frac{\sqrt{w(x)w(y)}}{r_{N-1}^{(2)}}\frac{q_N^{(2)}(x)q_{N-1}^{(2)}(y)-q_{N-1}^{(2)}(x)q_N^{(2)}(y)}{x-y}.
\end{equation}
\end{proposition}
Hence, the sum \eqref{eq1.2.67} involving the orthogonal polynomials $q_0^{(2)},\ldots,q_{N-1}^{(2)}$ can be seen to depend only on $q_N^{(2)}$ and $q_{N-1}^{(2)}$. Taking the limit $y\to x$ using L'H\^{o}pital's rule shows that for finite $N$, the eigenvalue density has the form
\begin{equation} \label{eq1.2.79}
\rho^{(w)}(\lambda)=K_N^{(2)}(\lambda,\lambda)=\frac{w(\lambda)}{r_{N-1}^{(2)}}\left(q_{N-1}^{(2)}(\lambda)\partial_\lambda q^{(2)}_N(\lambda)-q_N^{(2)}(\lambda)\partial_\lambda q^{(2)}_{N-1}(\lambda)\right).
\end{equation}
All that remains to make $\rho^{(w)}(\lambda)$ fully explicit is knowledge of the orthogonal polynomials $\{q_j^{(2)}(\lambda)\}_{j=0}^{\infty}$. In the Gaussian, Laguerre, and Jacobi cases, these are the Hermite, Laguerre, and Jacobi orthogonal polynomials, respectively. In the Cauchy case, one requires the Routh--Romanovski polynomials \citep{RWAK07}, which are related to the Jacobi polynomials by the change of variables implicit in equation \eqref{eq1.2.22}.

Obtaining the $\beta=1,4$ analogues of equations \eqref{eq1.2.78} and \eqref{eq1.2.79} is a little more involved, as one might expect. The initial challenge was to pin down suitable skew-orthogonal polynomials. The works \citep{NW91}, \citep{NW92}, \citep{AFNM00}, \citep{GP02} outline (progressively simpler) constructions of skew-orthogonal polynomials for the Gaussian, Laguerre, and Jacobi weights from the polynomials that are orthogonal with respect to these weights. In these works, the correlation kernels $K_N^{(\beta)}(x,y)$ and eigenvalue densities
\begin{equation}
\rho^{(w)}(\lambda)=w(\lambda)\sum_{k=0}^{N-1}\frac{1}{r_k^{(\beta)}}S_k^{(\beta)}(\lambda,\lambda),
\end{equation}
with $\beta=1,4$, were not given by Christoffel--Darboux formulae. Instead, $\sum_{k=0}^{N-1}S_k^{(\beta)}(x,y)/r_k^{(\beta)}$ was presented as (essentially, up to factors of two) the Christoffel--Darboux sum \eqref{eq1.2.78} plus a `correction term'. Later on, \citep{Gho06} showed that the skew-orthogonal polynomials satisfy recurrence relations, which were used to derive \textit{generalised Christoffel--Darboux formulae} involving the semi-infinite vectors
\begin{equation*}
\Phi^{(\beta)}(\lambda)=w(\lambda)\begin{bmatrix}(r_{2i}^{(\beta)})^{-1/2}q_{2i}^{(\beta)}(\lambda),&(r_{2i+1}^{(\beta)})^{-1/2}q_{2i+1}^{(\beta)}(\lambda)\end{bmatrix}_{i=0,1,\ldots},
\end{equation*}
$\lambda\Phi^{(\beta)}(\lambda)$, $\partial_{\lambda}\Phi^{(\beta)}(\lambda)$, and $\partial_{\lambda}[\lambda\Phi^{(\beta)}(\lambda)]$, together with the semi-infinite matrices describing the transitions between them. In the Gaussian, Laguerre, and Jacobi cases, the aforementioned recurrence relations involve three terms \citep{Gho08}, so the resulting generalised Christoffel--Darboux formulae are relatively compact.

\subsection{Classical $\beta$ ensembles} \label{s1.2.4}
The (skew-)orthogonal polynomial ensembles of the previous subsection concern the eigenvalue j.p.d.f.~\eqref{eq1.2.7} of the classical matrix ensembles generalised to non-classical weights $w(\lambda)$. In this subsection, we return to $w(\lambda)$ being classical, and instead allow general real $\beta>0$, rather than just $\beta=1,2,4$. Thus, the \textit{classical $\beta$ ensembles} are specified by the eigenvalue j.p.d.f.
\begin{equation} \label{eq1.2.81}
p^{(w)}(\lambda_1,\ldots,\lambda_N;\beta)=\frac{1}{\mathcal{N}_{N,\beta}^{(w)}}\,\prod_{i=1}^Nw(\lambda_i)\,|\Delta_N(\lambda)|^\beta,\quad\beta=2\kappa>0,
\end{equation}
with $w(\lambda)$ a classical weight \eqref{eq1.2.9}. In addition to taking $\beta$ to be positive real, we also consider the classical weights with parameters $a,b>-1$ real and $\eta=-\kappa(N-1)-1-\alpha$, $\alpha\in\mathbb{C}$ with $\mathrm{Re}(\alpha)>-1/2$. At the end of \S\ref{s1.2.2}, we argued that even at this level of generality, these eigenvalue j.p.d.f.s describe the eigenvalues of certain ensembles of self-adjoint random matrices. However, generating matrix representatives of these ensembles is not practical since the p.d.f.s $P(X)$ of the matrices $X$ do not translate to j.p.d.f.s on the entries of $X$. Hence, there was a natural motivation to construct alternative matrix models for the classical $\beta$ ensembles, with a focus on specifying p.d.f.s on matrix entries.
\begin{theorem}[Dumitriu--Edelman '02, Killip--Nenciu '04] \label{thrm1.3}
Recall that random variables $x\sim\chi(k)$ and $y\sim B(s,t)$ are respectively chi and beta distributed if they have p.d.f.s
\begin{equation}
\frac{2^{k/2-1}}{\Gamma(k/2)}x^{k-1}e^{-x^2/2}\chi_{x>0},\qquad \frac{2^{1-s-t}\Gamma(s+t)}{\Gamma(s)\Gamma(t)}(1-y)^{s-1}(1+y)^{t-1}\chi_{-1<y<1},
\end{equation}
and let $N(0,\sigma)$ denote the normal distribution with mean $0$ and variance $\sigma$. Introduce the matrices
\begin{equation} \label{eq1.2.83}
H_{\beta}:=\frac{1}{2}\begin{bmatrix}x_{(G)}(1)&y_{(G)}(1)&&&\\y_{(G)}(1)&x_{(G)}(2)&y_{(G)}(2)&&\\&\ddots&\ddots&\ddots&\\&&y_{(G)}(N-2)&x_{(G)}(N-1)&y_{(G)}(N-1)\\&&&y_{(G)}(N-1)&x_{(G)}(N)\end{bmatrix},
\end{equation}
\begin{equation}
B_{\beta}:=\frac{1}{\sqrt{2}}\begin{bmatrix}x_{(L)}(1)&&&\\y_{(L)}(1)&x_{(L)}(2)&&\\&\ddots&\ddots&\\&&y_{(L)}(N-1)&x_{(L)}(N)\end{bmatrix},
\end{equation}
and
\begin{equation}
Y_{\beta}:=\frac{1}{4}\begin{bmatrix}x_{(J)}(1)&y_{(J)}(1)&&&\\y_{(J)}(1)&x_{(J)}(2)&y_{(J)}(2)&&\\&\ddots&\ddots&\ddots&\\&&y_{(J)}(N-2)&x_{(J)}(N-1)&y_{(J)}(N-1)\\&&&y_{(J)}(N-1)&x_{(J)}(N)\end{bmatrix},
\end{equation}
with the entries independently distributed according to
\begin{align*}
x_{(G)}(k)&\sim N(0,2),
\\x_{(L)}(k)&\sim\chi\left(2a+1+\beta(N-k)\right),
\\y_{(G)}(k),y_{(L)}(k)&\sim\chi\left(\beta(N-k)\right),
\\x_{(J)}(k+1)&=(1-\alpha_{2k-1})\alpha_{2k}-(1+\alpha_{2k-1})\alpha_{2k-2}+2,
\\y_{(J)}(k+1)&=\sqrt{(1-\alpha_{2k-1})(1-\alpha_{2k}^2)(1+\alpha_{2k+1})},
\end{align*}
where $\alpha_{2N-1}=\alpha_{-1}=-1$ and for $0\leq k\leq2N-2$,
\begin{equation*}
\alpha_k\sim\begin{cases}B(\tfrac{2N-k-2}{4}\beta+b+1,\tfrac{2N-k-2}{4}\beta+a+1),&k\textrm{ even},\\B(\tfrac{2N-k-3}{4}\beta+a+b+2,\tfrac{2N-k-1}{4}\beta),&k\textrm{ odd}.\end{cases}
\end{equation*}
Then, the eigenvalues of the tridiagonal real symmetric $N\times N$ matrices $H_{\beta},W_{\beta}:=B_{\beta}B_{\beta}^T$, and $Y_{\beta}$ have eigenvalue j.p.d.f.~$p^{(w)}(\lambda_1,\ldots,\lambda_N)$, as specified by equation \eqref{eq1.2.81}, with the Gaussian, Laguerre, and Jacobi weights \eqref{eq1.2.9}, respectively \citep{DE02}, \citep{KN04}.
\end{theorem}
The proofs given in \citep{DE02} for the Gaussian and Laguerre cases of Theorem \ref{thrm1.3} consist of multiplying together the independent entries of the relevant matrices, changing variables to the eigenvalues and eigenvectors, and then integrating out the eigenvector contributions. These constructions are motivated by viewpoints in matrix reduction from numerical linear algebra. For example, in the GOE and GUE, Householder transformations (see, e.g., \citep{householder}) can be used to systematically reduce the corresponding real, respectively complex Hermitian, matrices to the tridiagonal form \eqref{eq1.2.83}.

Another matrix model for the Gaussian case \citep[Sec.~3]{Fo15}, more in line with the recursive theme of this thesis, involves the iterative construction of the sequence of matrices
\begin{equation} \label{eq1.2.86}
M_{N+1}=\begin{bmatrix}M_N&\mathbf{y}\\\mathbf{y}^T&x\end{bmatrix},\qquad\mathbf{y}=[y(i)]_{i=1}^N
\end{equation}
with $x\sim N(0,2)$ and $y(i)\sim\chi(\beta i)$. Replacing $M_N$ by $\mathrm{diag}(x_1,\ldots,x_N)$, the diagonal matrix of its eigenvalues, does not change the eigenvalue distribution of $M_{N+1}$, and further allows one to read off the recurrence
\begin{equation} \label{eq1.2.87}
\frac{\mathrm{Ch}_{N+1}(\lambda)}{\mathrm{Ch}_N(\lambda)}=\lambda-x-\sum_{i=1}^N\frac{y(i)^2}{\lambda-x_i}
\end{equation}
for the characteristic polynomial $\mathrm{Ch}_N(\lambda)=\Det(\lambda I_N-M_N)$. Taking the residues of both sides of equation \eqref{eq1.2.87} at $\lambda=x_i$ (taking $x=0$ without loss of generality), one may express the $\{y(i)\}_{i=1}^N$ in terms of the $\{x_i\}_{i=1}^N$ and the eigenvalues $\{\lambda_i\}_{i=1}^{N+1}$ of $M_{N+1}$. Substituting such expressions into the j.p.d.f.~for $\mathbf{y}$ and incorporating the necessary Jacobian yields a j.p.d.f.~$\hat{p}(\lambda_1,\ldots,\lambda_{N+1};x_1,\ldots,x_N)$ on the eigenvalues of $M_{N+1}$ conditional on the eigenvalues of $M_N$. We leave this j.p.d.f.~unspecified here (it is fully specified in \citep{Fo15}) but conclude with the fact \citep{Fo15} that, after appropriate scaling, the eigenvalue j.p.d.f. $p^{(G)}(\lambda_1,\ldots,\lambda_N)$ defined by equation \eqref{eq1.2.81} with the Gaussian weight satisfies the recurrence
\begin{equation*}
p^{(G)}(\lambda_1,\ldots,\lambda_{N+1})=\int_{\lambda_{N+1}<x_N<\cdots<\lambda_1}p^{(G)}(\lambda_1,\ldots,\lambda_N)\hat{p}(\lambda_1,\ldots,\lambda_{N+1};x_1,\ldots,x_N)\,\mathrm{d}x_1\cdots\mathrm{d}x_N.
\end{equation*}
That is, the eigenvalues of $M_N$ are distributed according to the j.p.d.f.~of the Gaussian $\beta$ ensemble.

The matrix $Y_{\beta}$ given in Theorem \ref{thrm1.3} was derived in \citep{KN04} by first considering a $\beta$-generalisation of $SO(2N)$ with its Haar probability measure, and then using a stereographic projection so as to map the eigenvalues from the unit circle to an interval in the real line. In addition, \citep{KN04} provided a matrix model for the circular $\beta$ ensemble, which is specified by the eigenvalue j.p.d.f.~\eqref{eq1.1.8} with $\beta$ now positive real --- it is a $\beta$-generalisation of the CUE in the same vein as that considered for $SO(2N)$. Without going into details, \citep{KN04} used theory concerning orthogonal polynomials on the unit circle to construct matrices whose eigenvalues are distributed according to the j.p.d.f.~\eqref{eq1.1.8} of the circular $\beta$ ensemble with general $\beta>0$. These matrices are themselves products of two block-diagonal matrices whose entries have fully specified p.d.f.s, so they can be generated numerically in a straightforward manner. This work was extended in \citep{BNR09} through a certain $\alpha$-deformation to produce matrix models for the circular Jacobi ensemble defined in \S\ref{s1.2.1}. Through the Cayley transformation \eqref{eq1.2.25}, the matrix models for the circular Jacobi ensemble essentially serve as such for the Cauchy $\beta$ ensemble.

\begin{remark}
\begin{enumerate}
\item Applying Householder transformations to Ginibre matrices gives rise to Hessenberg matrices, with the upper triangular entries still normally distributed elements of the respective number system. Thus, without the self-adjoint symmetry of the GOE, etc., the simplicity of \eqref{eq1.2.83} is lost and there is no general-$\beta$ matrix model for the Ginibre ensembles.
\item The eigenvalue densities corresponding to the classical $\beta$ ensembles have the large $N$ limiting forms given in Proposition \ref{prop1.2} with the identifications (see Proposition \ref{prop1.1})
\begin{equation} \label{eq1.2.88}
\gamma_1=1+\lim_{N\to\infty}\frac{a-\kappa+1}{\kappa N},\qquad\gamma_2=1+\lim_{N\to\infty}\frac{b-\kappa+1}{\kappa N}.
\end{equation}
This can be understood from \citep[Eq.~(5.6)]{BNR09} in the Cauchy case, and the loop equation analysis done in \citep{WF14}, \citep{FRW17}, otherwise (see also references therein).
\end{enumerate} \label{R1.9}
\end{remark}

The classical $\beta$ ensembles have been studied extensively throughout the last couple of decades, leading to a vast collection of results: see \citep{CMD07}, \citep{Des09}, \citep{RRV11}, \citep{BEMN11}, \citep{WF14}, \citep{Kum19} and references therein for a diverse but non-exhaustive sample. A result that is of great use to us is that the spectral moments of the classical $\beta$ ensembles can be expanded in either $N$ or $1/N$, in line with equation \eqref{eq1.1.32}.
\begin{lemma}[Dumitriu--Edelman '06, Dumitriu--Paquette '12] \label{L1.3}
Let $m_k$ be the $k\textsuperscript{th}$ spectral moment of the Gaussian, Laguerre, or Jacobi $\beta$ ensemble, as specified by equations \eqref{eq1.1.13} and \eqref{eq1.1.15}. In the first two cases, $m_k$ is a degree-$(k+1)$ polynomial in $N$ (identically zero in the Gaussian case with $k$ odd due to the weight $w^{(G)}(\lambda)$ being an even function) \citep{DE06}, while in the Jacobi case, $m_k/N$ has a large $N$ expansion in $1/N$ \citep{DP12}. For later use, we implicitly define the ($N$-independent) polynomial and series coefficients $M_{k,l}$ as follows:
\begin{align}
m_{2k}^{(G)}&=\sum_{l=0}^kM_{k,l}^{(G)}N^{1+k-l}, \label{eq1.2.89}
\\ m_{k}^{(L)}&=\sum_{l=0}^kM_{k,l}^{(L)}N^{1+k-l}, \label{eq1.2.90}
\\ m_{k}^{(J)}&=\sum_{l=0}^\infty M_{k,l}^{(J)}N^{1-l}. \label{eq1.2.91}
\end{align}
\end{lemma}
The existence of these expansions was established in \citep{DE06}, \citep{DP12} using Jack polynomial theory (see also the later work \citep{MRW17} and the discussion in Section \ref{s3.2}). The Cauchy and shifted Jacobi spectral moments $m_k^{(Cy)}, m_k^{(sJ)}$ can be written in terms of the Jacobi spectral moments $m_k^{(J)}$ (see Section \ref{s2.2} and the proof of Lemma \ref{L1.4} below) so that equation \eqref{eq1.2.91} translates to a similar expansion for the $m_k^{(Cy)},m_k^{(sJ)}$.

\subsubsection{Classical even-$\beta$ ensembles}
The methods used in Chapter 2 are not suitable for treating the classical general-$\beta$ ensembles, but do extend past the $\beta=1,2,4$ paradigm corresponding to the orthogonal, unitary, and symplectic ensembles. Indeed, we are technically able to study a scheme in between these two which we will call the \textit{classical even-$\beta$ ensembles}. These are simply the classical $\beta$ ensembles with $\beta$ constrained to be a positive even integer. The significance of $\beta$ being even is that the Vandermonde factor $|\Delta_N(\lambda)|^{\beta}$ in equation \eqref{eq1.2.81} is then a polynomial in the eigenvalues $\{\lambda_i\}_{i=1}^N$. This was seen to be a critical requirement for some results related to Selberg correlation integrals, the $1/r^2$ quantum many-body system, and the Calogero--Sutherland model (see \citep[Sec.~13.2]{Fo10}, \citep[Sec.~2]{Fo15} for a review). In particular, the works \citep{Fo94}, \citep{DF06}, \citep{Fo12}, \citep{FT19} on various subsets of the classical even-$\beta$ ensembles contain (edge scalings of) expressions for the relevant eigenvalue densities $\rho(\lambda)$ in terms of hypergeometric functions and $\beta$-dimensional integrals, which we use as points of comparison in \S\ref{s2.4.2}.

One issue with studying the classical even-$\beta$ ensembles is that the case $\beta=1$ is not included. As outlined earlier, this case is very important due to it corresponding to real symmetric matrices with orthogonal invariance. It turns out that we are able to extend our results in Chapter 2 to the classical $\beta$ ensembles with $\beta$ such that $4/\beta$ is an even integer; this allows for treatment of the $\beta=1$ case, along with classical $\beta$ ensembles corresponding to some non-integer rational values of $\beta$ such as $\beta=2/3$. This extension is possible because the works \citep{DE06}, \citep{DP12} cited in Lemma \ref{L1.3} tell us that the coefficients $M_{k,l}$ therein are palindromic polynomials in $-1/\kappa=-2/\beta$, upon appropriate scaling of the parameters $a,b$. To be precise, since the even-$\beta$ analysis in Chapter 2 focuses on the spectral moments $m_k$ and their generating functions $W_1(x)$, it is amenable to the following $\beta\leftrightarrow4/\beta$ duality relations.

\begin{lemma} \label{L1.4}
Let us highlight the dependence of the spectral moments $m_k$ \eqref{eq1.1.13} and resolvents $W_1(x)$ \eqref{eq1.1.17} on the parameters $N,\kappa=\beta/2,a,b,\alpha$ by listing them as arguments (recall that we write the parameter $\eta$ in the Cauchy weight \eqref{eq1.2.9} as $\eta=-\kappa(N-1)-1-\alpha$ with $\alpha\in\mathbb{C}$ and $\mathrm{Re}(\alpha)>-1/2$). Then, the $k\textsuperscript{th}$ spectral moments of the classical $\beta$ ensembles satisfy the duality relations \citep{DE06}, \citep{DP12} (see also \citep[App.~A]{FL16} by A. Borodin and V. Gorin)
\begin{align}
m_{2k}^{(G)}(N,\kappa)&=(-\kappa)^{k-1}m_{2k}^{(G)}(-\kappa N,1/\kappa),
\\m_k^{(L)}(N,\kappa,a)&=(-\kappa)^{k-1}m_k^{(L)}(-\kappa N,1/\kappa,-a/\kappa),
\\m_k^{(J)}(N,\kappa,a,b)&=-\kappa^{-1}m_k^{(J)}(-\kappa N,1/\kappa,-a/\kappa,-b/\kappa), \label{eq1.2.94}
\\m_k^{(Cy)}(N,\kappa,\alpha)&=-\kappa^{-1}m_k^{(Cy)}(-\kappa N,1/\kappa,-\alpha/\kappa). \label{eq1.2.95}
\end{align}
These extend to the duality relations \citep{WF14}, \citep{FRW17}
\begin{align}
W_1^{(G)}(x;N,\kappa)&=-\mathrm{i}\kappa^{-3/2}W_1^{(G)}(\mathrm{i}x/\sqrt{\kappa};-\kappa N,1/\kappa),
\\W_1^{(L)}(x;N,\kappa,a)&=\kappa^{-2}W_1^{(L)}(-x/\kappa;-\kappa N,1/\kappa,-a/\kappa),
\\W_1^{(J)}(x;N,\kappa,a,b)&=-\kappa^{-1}W_1^{(J)}(x;-\kappa N,1/\kappa,-a/\kappa,-b/\kappa), \label{eq1.2.98}
\\W_1^{(Cy)}(x;N,\kappa,\alpha)&=-\kappa^{-1}W_1^{(Cy)}(x;-\kappa N,1/\kappa,-\alpha/\kappa). \label{eq1.2.99}
\end{align}
\end{lemma}
\begin{proof}
The references cited above do not study the Cauchy case. However, the duality relations for the Jacobi $\beta$ ensemble translate to the Cauchy $\beta$ ensemble through the identities presented in Section \ref{s2.2}: Identity \eqref{eq2.2.3} expresses the spectral moments $m_k^{(sJ)}$ of the shifted Jacobi $\beta$ ensemble (the classical $\beta$ ensemble with weight $w^{(sJ)}(\lambda)=(1-\lambda)^a(1+\lambda)^b\chi_{-1<\lambda<1}$, as introduced at the end of \S\ref{s1.2.1}) in terms of the $m_k^{(J)}$ above, which shows that $m_k^{(sJ)}$ satisfies the duality \eqref{eq1.2.94}. Then, Proposition \ref{prop2.4} combined with Carlson's theorem (cf.~the proof of Corollary \ref{C2.2} and see Remark \ref{R2.5}) shows that
\begin{equation}
m_k^{(Cy)}(N,\kappa,\alpha)=(-1)^{k/2}m_k^{(sJ)}(N,\kappa,a,b)\Big|_{a=\overline{b}=-\kappa(N-1)-1-\alpha},
\end{equation}
so that equation \eqref{eq1.2.94} implies \eqref{eq1.2.95}. To obtain the duality relation \eqref{eq1.2.99}, one can repeat these arguments with either equation \eqref{eq1.1.17} or \eqref{eq1.1.20} specifying the resolvent.
\end{proof}

\setcounter{equation}{0}
\section{Matrix Product Ensembles} \label{s1.3}
From the viewpoint of using random matrices to model transformations such as quantum operators, scattering channels, or evolution operators for certain systems, it is quite natural to use products of random matrices to model repeated applications of such transformations. Hence, the theory of matrix product ensembles has applications in the study of wireless communications \citep{Mu02}, \citep{TV04}, quantum transport \citep{Be97}, the stability and chaoticity of large dynamical systems \citep{CPV93}, \citep{IF18}, finance \citep{BLMP07}, and the stability of neural networks \citep{PSG17}, \citep{HN20}. There are also interesting connections to Fuss--Catalan and Raney numbers \citep{PZ11}, \citep{FL15}.

Investigations into products of random matrices can be traced back to the 1950s, with many fundamental results and techniques being developed in the works \citep{GN50}, \citep{HC57}, \citep{FK60}, \citep{Os68}, \citep{Ra79}, \citep{New86} (among others) carried out over the following four decades; see \citep{Hel78} (group theory) and \citep{CPV93} (statistical physics) for textbook treatments. There has since been a consistent level of research on products of random matrices (some notable works being \citep{JLJN02}, \citep{JW04} and the applications cited above, along with the inception and success of free probability theory \citep{VDN92}, \citep{MS17}), but the last decade has seen a remarkable surge in interest on this topic. The beginning of this latter era is marked by the derivation of correlation kernels (recall Theorem~\ref{thrm1.2}) encapsulating detailed knowledge of the eigenvalue and singular value statistics of certain matrix product ensembles, where the products contain a finite number of factors and each factor is a finite-sized matrix --- most results from before this era pertain to regimes where either the matrix sizes or number of factors are taken to infinity. The reader is referred to \citep{AI15}, \citep{Ips15} for reviews on this period of time.

At the beginning of this chapter, it was said that two types of random matrix ensembles will be studied in this thesis. The first is the family of classical matrix ensembles reviewed in the previous section, while the second consists of the particular matrix product ensembles that we now introduce.
\begin{definition} \label{def1.10}
Let $m,N_0\in\mathbb{N}$, fix $\nu_0:=0,\nu_1,\ldots,\nu_m\in\mathbb{N}$, and define $N_i:=N_0+\nu_i$ with $N:=N_m$. Recalling Definitions \ref{def1.3} and \ref{def1.4} of the Ginibre and Gaussian ensembles, let $H$ be an $N_0\times N_0$ GUE matrix and for $1\leq i\leq m$, let $G_i$ be drawn independently from the $N_{i-1}\times N_i$ complex Ginibre ensemble. Then, the $N\times N$ product
\begin{equation} \label{eq1.3.1}
\mathcal{H}_m:=G_m^\dagger\cdots G_1^\dagger HG_1\cdots G_m
\end{equation}
represents the $(N_0,\ldots,N_{m-1},N)$ \textit{Hermitised matrix product ensemble} \citep{FIL18}. When $m=1$, we alternatively say that $\mathcal{H}_1$ represents the $(N_0,N)$ \textit{Hermitised Laguerre ensemble}.
\end{definition}
\begin{definition} \label{def1.11}
Let $m,N_0/2\in\mathbb{N}$, fix $\nu_0:=0,\nu_1,\ldots,\nu_m\in\mathbb{N}$, and define $N_i:=N_0+2\nu_i$ with $N:=N_m$. Recalling Definition \ref{def1.3} of the Ginibre ensembles, let $J_{N_0}$ be the elementary antisymmetric matrix defined by equation \eqref{eq1.1.5} and for $1\leq i\leq m$, let $G_i$ be drawn independently from the $N_{i-1}\times N_i$ real Ginibre ensemble. Then, the $N\times N$ product
\begin{equation} \label{eq1.3.2}
\mathcal{J}_m:=G_m^T\cdots G_1^TJ_{N_0}G_1\cdots G_m
\end{equation}
represents the $(N_0,\ldots,N_{m-1},N)$ \textit{antisymmetrised matrix product ensemble} \citep{FILZ19}. When $m=1$, we alternatively say that $\mathcal{J}_1$ represents the $(N_0,N)$ \textit{antisymmetrised Laguerre ensemble}.
\end{definition}
We think of these ensembles as perturbations of the more widely studied Wishart product ensembles \citep{AI15}, \citep{Ips15}, which are defined as follows.
\begin{definition} \label{def1.12}
As in Definition \ref{def1.10}, let $m,N_0\in\mathbb{N}$, fix $\nu_0:=0,\nu_1,\ldots,\nu_m\in\mathbb{N}$, and define $N_i:=N_0+\nu_i$ with $N:=N_m$. For $1\leq i\leq m$, let $G_i$ be drawn independently from the $N_{i-1}\times N_i$ real, complex, or quaternionic Ginibre ensemble, as specified by Definition \ref{def1.3} (fixing the number system across all $G_i$). Then, the $N\times N$ product
\begin{equation}
\mathcal{W}_m:=G_m^\dagger\cdots G_1^\dagger G_1\cdots G_m \label{eq1.3.3}
\end{equation}
represents the $(N_0,\ldots,N_{m-1},N)$ \textit{real, complex, or quaternionic Wishart product ensemble}, respectively. Setting $m=1$ recovers the $(N_0,N)$ Laguerre ensemble, in keeping with Definition~\ref{def1.4}.
\end{definition}

Our present goal in studying the Hermitised and antisymmetrised matrix product ensembles is to better understand the ramifications of perturbing the Wishart product ensembles in the corresponding ways (inserting a GUE matrix $H$ or elementary antisymmetric matrix $J_{N_0}$ in the middle of the relevant products), in terms of both changes to the eigenvalue statistics and changes to the effectiveness of existing techniques for analysing these statistics. In particular, we would like to see how the recent loop equation analysis of Dartois and Forrester \citep{DF20} for the $(N,N,N)$ complex Wishart product ensemble extends to the Hermitised and antisymmetrised matrix product ensembles. Thus, in Chapter 4, we derive loop equations characterising the correlator expansion coefficients $W_n^l$ of the Hermitised and antisymmetrised Laguerre ensembles (i.e., the product ensembles of Definitions~\ref{def1.10} and~\ref{def1.11} with $m=1$) after they have been scaled such that their connected $n$-point correlators $W_n$ \eqref{eq1.1.29} admit large $N$ expansions of the form \eqref{eq1.1.21} --- we argue the validity of these large $N$ expansions in \S\ref{s3.3.3} by giving combinatorial interpretations (which are interesting in their own right) to the mixed cumulants generated by the $W_n$. Our decision to restrict the analysis of Chapter 4 to the $m=1$ ensembles stems from the fact that taking $m$ to be any larger results in loop equations that are too unwieldy to fit within the scope of this thesis. Nonetheless, the loop equations derived in Chapter 4 have rich enough structure for us to make meaningful comparison with the loop equations presented in \citep{DF20}. We are also able to glean insight on the expected structure of the loop equations pertaining to arbitrary $m\in\mathbb{N}$.

Exploring the consequences of perturbing the Wishart product ensembles is not our only motivation for studying the Hermitised and antisymmetrised matrix product ensembles, as this could be done by considering any number of alternative, comparable perturbations of said ensembles. Rather, our secondary motivation for studying the Hermitised and antisymmetrised matrix product ensembles is that they have recently been shown to be related to certain Muttalib--Borodin ensembles and a generalisation, respectively analogue, of the Harish-Chandra--Itzykson--Zuber (HCIZ) integral \citep{FIL18}, \citep{FILZ19}. Hence, we are able to compare the results of Chapter 4 against explicit functional forms for the eigenvalue j.p.d.f.s of the Hermitised and antisymmetrised matrix product ensembles that were obtained in \citep{FIL18}, \citep{FILZ19}. Moreover, it is expected that the results of Chapter 4 will help advance our understanding of the broader connection between products of random matrices and Muttalib--Borodin ensembles that is seen in the literature \citep{KS14}, \citep{FLZ15}, \citep{Ips15}, \citep{WF16}, \citep{FW17}, \citep{FIL18}.

Let us now review, in order, the complex Wishart product ensembles, Muttalib--Borodin ensembles, and HCIZ-type integrals. We keep our review brief by focusing mostly on aspects of these topics that connect them to the ensembles that are studied in this thesis.

\subsection{Complex Wishart product ensembles and their correlation kernels} \label{s1.3.1}
If one wishes to capture the essence of multiplying generic random matrices, it is reasonable to study products of Ginibre matrices: On one hand, prescribing p.d.f.s on the matrix factors makes the situation relatively tractable and allows the use of some powerful analytical tools. On the other hand, there is no benefit in imposing symmetry constraints on the involved matrices, since products of self-adjoint matrices are not necessarily self-adjoint. Thus, a large portion of recent work in the field has been on products of Ginibre matrices; see \citep{BJLNS10}, \citep{AB12}, \citep{KZ14}, \citep{Ips15a} and references therein for a sample. Note that products of truncated Haar-distributed unitary matrices (cf.~the first point of Remark~\ref{R1.4}) and products involving inverse Ginibre matrices have also drawn considerable interest \citep{IK14}, \citep{ABKN14}, \citep{Fo14}, \citep{ARRS16}.

In studying the spectral properties of products of Ginibre matrices, one is interested in either the eigenvalues, which are generally complex, or the singular values, which are non-negative real (cf.~Remark~\ref{R1.6}). Both have been studied in, e.g., \citep{AB12}, \citep{IK14}, \citep{ARRS16} and \citep{AKW13}, \citep{AIK13}, respectively. Due to their relative simplicity, our interest lies in the squared singular values, which are equivalent to the eigenvalues of the Wishart product matrices $\mathcal{W}_m$ \eqref{eq1.3.3}.
\begin{remark}  \label{R1.10}
Like the eigenvalues of $\mathcal{W}_m$, the eigenvalues of the matrix product $\mathcal{H}_m$ \eqref{eq1.3.1} are real, since it is Hermitian. In a similar vein, the eigenvalues of the antisymmetric matrix $\mathcal{J}_m$ \eqref{eq1.3.2} lie on the imaginary axis in complex conjugate pairs.
\end{remark}

Let us now present the analogue of Theorem \ref{thrm1.2} for the complex Wishart product ensembles, which we obtain by combining results of \citep{AIK13} and \citep{KZ14}.
\begin{proposition} \label{prop1.6}
Let $p^{(c\mathcal{W}_m)}(\lambda_1,\ldots,\lambda_{N_0})$ be the j.p.d.f.~of the non-zero eigenvalues of the complex Wishart product matrix $\mathcal{W}_m$ \eqref{eq1.3.3} and let us recall \citep{BE53} that the \textbf{Meijer $G$-function} is defined as
\begin{equation} \label{eq1.3.4}
\meijer{m,n}{p,q}{a_1,\ldots,a_p}{b_1,\ldots,b_q}{z}:=\frac{1}{2\pi\mathrm{i}}\int_{\gamma}\frac{\prod_{j=1}^m\Gamma(b_j+u)\prod_{j=1}^n\Gamma(1-a_j-u)}{\prod_{j=m+1}^q\Gamma(1-b_j-u)\prod_{j=n+1}^p\Gamma(a_j+u)}z^{-u}\,\mathrm{d}u,
\end{equation}
where the choice of integration contour $\gamma$ is too involved to specify here. Then, for $1\leq k\leq N_0$, the $k$-point correlation function (cf.~equation \eqref{eq1.2.72}) associated with $p^{(c\mathcal{W}_m)}(\lambda_1,\ldots,\lambda_{N_0})$ has the form
\begin{align}
\rho_k^{(c\mathcal{W}_m)}(\lambda_1,\ldots,\lambda_k;N_0)&:=\frac{N_0!}{(N_0-k)!}\int_{\mathbb{R}^{N_0-k}}p^{(c\mathcal{W}_m)}(\lambda_1,\ldots,\lambda_{N_0})\,\mathrm{d}\lambda_{k+1}\cdots\mathrm{d}\lambda_{N_0} \label{eq1.3.5}
\\&\;=\Det\left[K_{N_0,\nu_1,\ldots,\nu_m}^{(c\mathcal{W}_m)}(\lambda_i,\lambda_j)\right]_{i,j=1}^k, \label{eq1.3.6}
\end{align}
with correlation kernel given by
\begin{equation} \label{eq1.3.7}
K_{N_0,\nu_1,\ldots,\nu_m}^{(c\mathcal{W}_m)}(x,y)=\int_0^1\meijer{1,0}{1,m+1}{N_0}{0,-\nu_1,\ldots,-\nu_m}{ux}\meijer{m,1}{1,m+1}{-N_0}{\nu_1,\ldots,\nu_m,0}{uy}\,\mathrm{d}u.
\end{equation}
\end{proposition}

Equation \eqref{eq1.3.6} tells us that the complex Wishart product ensembles are determinantal point processes, much like what was seen in \S\ref{s1.2.3} for the $\beta=2$ classical matrix ensembles. However, while this property of the $\beta=2$ classical matrix ensembles is a consequence of their being orthogonal polynomial ensembles, the same is not true for the complex Wishart product ensembles. Instead, the complex Wishart product ensembles, together with the Hermitised and antisymmetrised matrix product ensembles, are \textit{biorthogonal ensembles} \citep{Bo98} (see the next subsection for a formal definition of this term). In fact, all three of these matrix product ensembles belong to a further subclass of biorthogonal ensembles that are simply called \textit{polynomial ensembles}\footnote{In the case of the antisymmetrised matrix product ensembles, it is common practice to study the positive eigenvalues of $\mathrm{i}\mathcal{J}_m$, since these are real and independent, in keeping with Remark \ref{R1.10}. Technically, it is the positive eigenvalues of $\mathrm{i}\mathcal{J}_m$ that constitute a biorthogonal ensemble and it is the squares of these eigenvalues that exhibit the structure \eqref{eq1.3.8} of polynomial ensembles.} \citep{KS14}; these so-called polynomial ensembles are characterised by their eigenvalue j.p.d.f.s having the form
\begin{equation}
p(\lambda_1,\ldots,\lambda_{N_0})=\frac{1}{\mathcal{N}_{N_0}}\Delta_{N_0}(\lambda)\Det\left[w_{j-1}(\lambda_i)\right]_{i,j=1}^{N_0},\quad\lambda_1,\ldots,\lambda_{N_0}\in\mathbb{R}, \label{eq1.3.8}
\end{equation}
where $\mathcal{N}_{N_0}$ is a normalisation constant, $\Delta_{N_0}(\lambda)$ is the Vandermonde determinant \eqref{eq1.1.9}, and the family of weight functions $\{w_{j-1}(\lambda)\}_{j=1}^{N_0}$ is such that the above j.p.d.f.~is well-defined.
\begin{proposition} \label{prop1.7}
The j.p.d.f.s $p^{(c\mathcal{W}_m)}(\lambda_1,\ldots,\lambda_{N_0})$ and $p^{(\mathcal{H}_m)}(\lambda_1,\ldots,\lambda_{N_0})$ governing the non-zero eigenvalues of the complex Wishart product matrix $\mathcal{W}_m$ \eqref{eq1.3.3} and Hermitised matrix product $\mathcal{H}_m$ \eqref{eq1.3.1}, respectively, are given by equation \eqref{eq1.3.8} with weights \citep{AIK13}, \citep{FIL18}
\begin{align}
w_j^{(c\mathcal{W}_m)}(\lambda)&=\meijer{m,0}{0,m}{-}{\nu_1+j,\nu_2,\ldots,\nu_m}{\lambda}, \label{eq1.3.9}
\\ w_j^{(\mathcal{H}_m)}(\lambda)&=(\mathrm{sgn}\,\lambda)^j\prod_{i=1}^m\frac{2^{\nu_i-1}}{\sqrt{\pi}}\,\meijer{2m+1,0}{0,2m+1}{-}{\tfrac{\nu_1}{2},\tfrac{\nu_1+1}{2},\ldots,\tfrac{\nu_m}{2},\tfrac{\nu_m+1}{2},\tfrac{j}{2}}{\frac{\lambda^2}{4^m}},
\end{align}
where we recall the definition of the Meijer $G$-function from Proposition \ref{prop1.6}. Similarly, the positive eigenvalues of $\,\mathrm{i}\mathcal{J}_m$ \eqref{eq1.3.2} are distributed according to the eigenvalue j.p.d.f.~\citep{FILZ19}
\begin{align}
p^{(\mathrm{i}\mathcal{J}_m)}(\lambda_1,\ldots,\lambda_{N_0/2})&=\frac{1}{\mathcal{N}^{(\mathrm{i}\mathcal{J}_m)}_{N_0/2}}\prod_{1\leq k<l\leq N_0/2}(\lambda_l^2-\lambda_k^2)\,\Det\left[w^{(\mathrm{i}\mathcal{J}_m)}_{j-1}(\lambda_i)\right]_{i,j=1}^{N_0/2}, \label{eq1.3.11}
\\ w^{(\mathrm{i}\mathcal{J}_m)}_j(\lambda)&=\meijer{m,0}{0,m}{-}{2\nu_1+j,2\nu_2,\ldots,2\nu_m}{\lambda}. \label{eq1.3.12}
\end{align}
The associated normalisation constants can be found in \citep{AIK13}, \citep{FIL18}, \citep{FILZ19}.
\end{proposition}

The eigenvalue statistics of the matrix products considered in the above proposition are unchanged upon replacing the rectangular real and complex Ginibre matrices $G_i$ ($1\leq i\leq m$) therein by corresponding \textit{induced Ginibre matrices} $\hat{G}_i$ \citep{FBKSZ12}, \citep{AIK13}, \citep{IK14}. These matrices $\hat{G}_i$ are drawn from the space of $N_0\times N_0$ real, respectively complex, matrices with p.d.f.
\begin{equation} \label{eq1.3.13}
P^{(indG)}(\hat{G}_i;N_0,\beta,\nu_i)=\frac{1}{\mathcal{Z}_{N_0,\beta,\nu_i}^{(indG)}}\Det(\hat{G}_i^\dagger\hat{G}_i)^{\beta\nu_i/2}\exp\left(-\Tr(\hat{G}_i^\dagger\hat{G}_i)\right),
\end{equation}
where $\mathcal{Z}_{N_0,\beta,\nu_i}^{(indG)}$ is a normalisation constant and $\beta=1,2$ refers to the real and complex cases, respectively. The fact that making the replacements described above does not affect the eigenvalue statistics of interest was used in the proof of equation \eqref{eq1.3.9} given in \citep{AIK13}, where reformulating the product \eqref{eq1.3.3} in terms of square induced Ginibre matrices enabled the use of the HCIZ integral formula \eqref{eq1.3.51}. If one interprets $p^{(c\mathcal{W}_m)}(\lambda_1,\ldots,\lambda_{N_0})$, $p^{(\mathcal{H}_m)}(\lambda_1,\ldots,\lambda_{N_0})$, and $p^{(\mathrm{i}\mathcal{J}_m)}(\lambda_1,\ldots,\lambda_{N_0/2})$ as eigenvalue j.p.d.f.s of the products $\mathcal{W}_m$ (in the complex case), $\mathcal{H}_m$, and $\mathrm{i}\mathcal{J}_m$ rewritten in terms of the induced Ginibre matrices described above, then Propositions \ref{prop1.6} and \ref{prop1.7} hold true for general real parameters $\nu_1,\ldots,\nu_m>-1$ (cf.~the discussion surrounding Proposition \ref{prop1.4}). Moreover, the results of these propositions are invariant under permutations of the parameters $\nu_1,\ldots,\nu_m$ due to the weak commutation relation established in \citep{IK14}.

A point of intrigue regarding the antisymmetrised matrix product ensembles is that they have recently been shown in \citep{FILZ19} to be closely related to the complex Wishart product ensembles generalised through the induced Ginibre matrices described above.
\begin{proposition} \label{prop1.8}
Let $\mathcal{J}_m$ be as specified in Definition \ref{def1.11} and let $\lambda_1,\ldots,\lambda_{N_0/2}$ denote the positive eigenvalues of $\mathrm{i}\mathcal{J}_m$. Then, by \citep[Cor.~1.2]{FILZ19}, the variables $\lambda_j'=\lambda_j^2/4^m$ ($1\leq j\leq N_0/2$) are statistically equivalent to the eigenvalues of the product $\hat{G}_{2m}^\dagger\cdots\hat{G}_1^\dagger\hat{G}_1\cdots\hat{G}_{2m}$ of $(N_0/2)\times(N_0/2)$ induced Ginibre matrices with p.d.f.s \eqref{eq1.3.13}
\begin{equation*}
P^{(indG)}(\hat{G}_{2i-1};N_0/2,2,\nu_i),\qquad P^{(indG)}(\hat{G}_{2i};N_0/2,2,\nu_i-1/2),\quad 1\leq i\leq m.
\end{equation*}
In short, the j.p.d.f.~of the positive eigenvalues of $\mathrm{i}\mathcal{J}_m$ is given by
\begin{multline}
p^{(\mathrm{i}\mathcal{J}_m)}(\lambda_1,\ldots,\lambda_{N_0/2})=2^{N_0(1/2-m)}\lambda_1\cdots\lambda_{N_0/2}
\\ \times p^{(c\mathcal{W}_{2m})}(\lambda_1^2/4^m,\ldots,\lambda_{N_0/2}^2/4^m)\Big|_{\left\{\substack{\nu_{2i-1}\mapsto\nu_i,\\ \nu_{2i}\mapsto\nu_i-1/2}\right\}_{i=1}^m},
\end{multline}
where the right-hand side can be read off by reparametrising equations \eqref{eq1.3.8} and \eqref{eq1.3.9}.
\end{proposition}

In addition to highlighting a relationship between the weights $w_j^{(c\mathcal{W}_m)}(\lambda)$ \eqref{eq1.3.9} and $w_j^{(\mathrm{i}\mathcal{J}_m)}(\lambda)$ \eqref{eq1.3.12}, the above proposition tells us that the positive eigenvalues of $\mathrm{i}\mathcal{J}_m$ constitute a determinantal point process whose $k$-point correlation functions have the form \eqref{eq1.3.6} with correlation kernel given by
\begin{equation} \label{eq1.3.15}
K_{N_0/2,\nu_1,\ldots,\nu_m}^{(\mathrm{i}\mathcal{J}_m)}(x,y)=2^{1-2m}\sqrt{xy}\,K_{N_0/2,\nu_1,\nu_1-1/2,\ldots,\nu_m,\nu_m-1/2}^{(c\mathcal{W}_{2m})}(x^2/4^m,y^2/4^m),
\end{equation}
where the right-hand side is specified by equation \eqref{eq1.3.7}. Let us mention here that it was shown in \citep{FIL18} that the non-zero eigenvalues of the Hermitised matrix product $\mathcal{H}_m$ also constitute a determinantal point process; we do not present the relevant correlation kernel here, but note that it is made explicit in \citep{FIL18}.

\subsubsection{Scalings of the eigenvalue densities and correlation kernels}
Recall from Proposition \ref{prop1.2} that in the case of the classical matrix ensembles, the $N\to\infty$ limits $\rho^0(\lambda)$ of the global scaled eigenvalue densities $\tilde{\rho}(\lambda)$ \eqref{eq1.1.22} have distinct forms for each classical weight with, e.g., the Gaussian and Laguerre weights corresponding to the Wigner semi-circle and Mar\v cenko--Pastur laws, respectively. It turns out that the analogous limiting densities (now taking $N_0\to\infty$) of the matrix product ensembles considered in this section relate to the Fuss--Catalan distribution \citep{PZ11}, \citep{Ips15} (see also the independent works of Biane \citep[App.~A]{BPB99} and Neuschel \citep{Neu14} for a Plancherel--Rotach-like parametrisation in terms of elementary functions)
\begin{equation} \label{eq1.3.16}
\rho^{(FC_m)}(\lambda)=\frac{1}{\sqrt{2\pi}}\frac{m^{m-3/2}}{(m+1)^{m+1/2}}\meijer{m,0}{m,m}{\tfrac{1}{m},0,-\tfrac{1}{m},\ldots,-\tfrac{m-2}{m}}{-\tfrac{1}{m+1},-\tfrac{2}{m+1},\ldots,-\tfrac{m}{m+1}}{\frac{m^m}{(m+1)^{m+1}}\lambda},
\end{equation}
which is so named due to its spectral moments being the Fuss--Catalan numbers \citep{GKP94}:
\begin{equation} \label{eq1.3.17}
m_k^{(FC_m)}:=\int_{\mathbb{R}}\lambda^k\rho^{(FC_m)}(\lambda)\,\mathrm{d}\lambda=\frac{1}{mk+1}\binom{(m+1)k}{k},\quad k\in\mathbb{N}.
\end{equation}
Note that due to properties of the Meijer $G$-function \eqref{eq1.3.4}, the density $\rho^{(FC_m)}(\lambda)$ has compact support $[0,(m+1)^{m+1}/m^m]$, on which it integrates to unity.

\begin{definition} \label{def1.13}
Let $\rho^{(c\mathcal{W}_m)}(\lambda)$, $\rho^{(\mathcal{H}_m)}(\lambda)$, and $\rho^{(\mathrm{i}\mathcal{J}_m)}(\lambda)$ denote the eigenvalue densities \eqref{eq1.1.10}, equivalently the $1$-point correlation functions \eqref{eq1.3.5}, associated with the eigenvalue j.p.d.f.s $p^{(c\mathcal{W}_m)}(\lambda_1,\ldots,\lambda_{N_0})$, $p^{(\mathcal{H}_m)}(\lambda_1,\ldots,\lambda_{N_0})$, and $p^{(\mathrm{i}\mathcal{J}_m)}(\lambda_1,\ldots,\lambda_{N_0/2})$ specified in Proposition \ref{prop1.7}, respectively. Then, in analogy with Definition \ref{def1.6}, and partially following \citep{Ips15}, \citep{FIL18}, define the corresponding \textit{global scaled eigenvalue densities} \eqref{eq1.1.22} to be
\begin{align}
\tilde{\rho}^{(c\mathcal{W}_m)}(\lambda)&:=N_0^{m-1}\rho^{(c\mathcal{W}_m)}(N_0^m\lambda),
\\ \tilde{\rho}^{(\mathcal{H}_m)}(\lambda)&:=\tfrac{1}{\sqrt{2}}N_0^{m-1/2}\rho^{(\mathcal{H}_m)}(\tfrac{1}{\sqrt{2}}N_0^{m+1/2}\lambda), \label{eq1.3.19}
\\ \tilde{\rho}^{(\mathrm{i}\mathcal{J}_m)}(\lambda)&:=2N_0^{m-1}\rho^{(\mathrm{i}\mathcal{J}_m)}(N_0^m\lambda). \label{eq1.3.20}
\end{align}
Furthermore, recall the notation $\rho^0(\lambda)=\lim_{N_0\to\infty}\tilde{\rho}(\lambda)$ introduced in \S\ref{s1.1.1}.
\end{definition}

It has been shown in \citep{BBCC11}, \citep{PZ11}, \citep{FIL18} that for fixed $\nu_1,\ldots,\nu_m>-1$, the large $N_0$ limits of $\tilde{\rho}^{(c\mathcal{W}_m)}(\lambda)$ and $\tilde{\rho}^{(\mathcal{H}_m)}(\lambda)$ are given in terms of the Fuss--Catalan distribution \eqref{eq1.3.16} by
\begin{align}
\rho^{(c\mathcal{W}_m),0}(\lambda)&=\rho^{(FC_m)}(\lambda), \label{eq1.3.21}
\\ \rho^{(\mathcal{H}_m),0}(\lambda)&=|\lambda|\rho^{(FC_{2m+1})}(\lambda^2). \label{eq1.3.22}
\end{align}
Setting $x=y$ in equation \eqref{eq1.3.15} so that it may be read as a relation between the corresponding eigenvalue densities, and then combining this relation with equation \eqref{eq1.3.21} shows that (see also \citep{FL15} for the $m=1$ case)
\begin{equation} \label{eq1.3.23}
\rho^{(\mathrm{i}\mathcal{J}_m),0}(\lambda)=2\lambda\rho^{(FC_{2m})}(\lambda^2)\chi_{\lambda>0}.
\end{equation}
It can immediately be seen that, aside from subtleties concerning the parameter $m$, $\rho^{(\mathcal{H}_m),0}(\lambda)$ is essentially the symmetric extension of $\rho^{(\mathrm{i}\mathcal{J}_m),0}(\lambda)$ to the negative real axis. Moreover, both of these densities relate to $\rho^{(c\mathcal{W}_m),0}(\lambda)$ via the mapping $\lambda\mapsto\sqrt{\lambda}$, much like how the Wigner semi-circle \eqref{eq1.2.14} and Mar\v cenko--Pastur \eqref{eq1.2.15} laws are related by the equality $\rho^{(G),0}(\lambda)=2|\lambda|\,\rho^{(L),0}(2\lambda^2)$ when the Laguerre parameter $a$ is fixed. In particular, recall that $\rho^{(c\mathcal{W}_1),0}(\lambda)=\rho^{(L),0}(\lambda)$ and $\rho^{(\mathcal{H}_0),0}(\lambda)=\rho^{(G),0}(\lambda/\sqrt{2})$, by definition.

The global scaling regimes just discussed are such that the limiting densities $\rho^0(\lambda)$ have compact connected supports. In contrast, one is also interested in local scaling regimes, where the eigenvalue density $\rho(\lambda)$ is centred on a particular point $\lambda_0\in\textrm{supp}\,\rho$ and then scaled so that the spacing between this eigenvalue and its neighbours is of order unity. The resulting statistics depends on whether $\lambda_0$ is positioned at a soft edge, a hard edge, or within the bulk. In the case of the complex Wishart product ensembles with $\nu_1,\ldots,\nu_m>-1$ fixed, $\lambda_0$ is said to be at the soft edge if $\lim_{N_0\to\infty}\lambda_0/N_0^m=(m+1)^{m+1}/m^m$, the hard edge if $\lim_{N_0\to\infty}\lambda_0/N_0^m$ vanishes, and in the bulk if $\lim_{N_0\to\infty}\lambda_0/N_0^m\in(0,(m+1)^{m+1}/m^m)$. (The analogous scaling regimes of $\rho^{(\mathcal{H}_m)}(\lambda)$ and $\rho^{(\mathrm{i}\mathcal{J}_m)}(\lambda)$ are likewise defined by comparing similar scaled limits, with scalings prescribed by equations \eqref{eq1.3.19}, \eqref{eq1.3.20}, to $\pm(m+1)^{(m+1)/2}/m^{m/2}$.)

\begin{proposition} \label{prop1.9}
Let $m\in\mathbb{N}$ and $\nu_1,\ldots,\nu_m>-1$ be fixed. Furthermore, let
\begin{equation}
\begin{array}{ll}
x_0\in(0,(m+1)^{m+1}/m^m),\quad&\quad c_b=N_0^{m-1}/\rho^{(FC_m)}(x_0),
\\ x_+=\left(\frac{N_0}{m}\right)^m(m+1)^{m+1},\quad&\quad c_+=N_0^{m-2/3}\frac{2^{1/3}m^{m-1}}{(m+1)^{m+2/3}}.
\end{array}
\end{equation}
Then, the hard edge, bulk, and soft edge limiting forms of the correlation kernel \eqref{eq1.3.7} are respectively given by \citep{LWZ16}, \citep{KZ14}
\begin{align}
&\lim_{N_0\to\infty}\frac{1}{N_0}K_{N_0,\nu_1,\ldots,\nu_m}^{(c\mathcal{W}_m)}\left(x/N_0,\,y/N_0\right)=K_{\nu_1,\ldots,\nu_m}^{(Meijer)}(x,y), \label{eq1.3.25}
\\ &\lim_{N_0\to\infty}c_bK_{N_0,\nu_1,\ldots,\nu_m}^{(c\mathcal{W}_m)}\left(c_bx+N_0^mx_0,\,c_by+N_0^mx_0\right)=K^{(sine)}(x,y),
\\ &\lim_{N_0\to\infty}c_+K_{N_0,\nu_1,\ldots,\nu_m}^{(c\mathcal{W}_m)}\left(c_+x+x_+,\,c_+y+x_+\right)=K^{(Airy)}(x,y),
\end{align}
where
\begin{align}
K_{\nu_1,\ldots,\nu_m}^{(Meijer)}(x,y)&=\int_0^1\meijer{1,0}{0,m+1}{-}{0,-\nu_1,\ldots,-\nu_m}{ux}\meijer{m,0}{0,m+1}{-}{\nu_1,\ldots,\nu_m,0}{uy}\,\mathrm{d}u, \label{eq1.3.28}
\\ K^{(sine)}(x,y)&=\frac{\sin\pi(x-y)}{\pi(x-y)}, \label{eq1.3.29}
\\ K^{(Airy)}(x,y)&=\frac{\mathrm{Ai}(x)\mathrm{Ai}'(y)-\mathrm{Ai}'(x)\mathrm{Ai}(y)}{x-y} \label{eq1.3.30}
\end{align}
are the Meijer $G$-, sine, and Airy kernels, respectively (recall that the Meijer $G$-function is partially defined in equation \eqref{eq1.3.4} and that $\mathrm{Ai}(x)$ denotes the Airy function).
\end{proposition}

\begin{remark} \label{R1.11}
Recall from the discussion following Figures \ref{fig1.1}--\ref{fig1.3} that the hard edge of the Laguerre ensembles degenerates to a soft edge upon letting the Laguerre parameter $a$ grow linearly with $N$. A similar phenomenon is exhibited by the complex Wishart product ensembles \citep{AIK13}, whose hard edge ceases to be if all of the parameters $\nu_1,\ldots,\nu_m$ scale with $N_0$. If one writes $\nu_i=\hat{\nu}_i N_0+\delta_i$ with $\hat{\nu}_i,\delta_i={\rm O}(1)$ for $1\leq i\leq m$ and further enforces $\hat{\nu}_1=\cdots=\hat{\nu}_l=0$, but $\hat{\nu}_{l+1},\ldots,\hat{\nu}_m>0$ for some $1\leq l\leq m$, then it is known from \citep{Ips15} that equation \eqref{eq1.3.25} should be replaced with
\begin{equation}
\lim_{N_0\to\infty}c_-K_{N_0,\nu_1,\ldots,\nu_m}^{(c\mathcal{W}_m)}(c_-x,c_-y)=K_{\nu_1,\ldots,\nu_l}^{(Meijer)}(x,y),\quad c_-=\hat{\nu}_{l+1}\cdots\hat{\nu}_mN_0^{m-l-1}.
\end{equation}
\end{remark}

It follows from Propositions \ref{prop1.8} and \ref{prop1.9} that the local correlations of the positive eigenvalues of $\mathrm{i}\mathcal{J}_m$ are characterised by the Meijer $G$-, sine, and Airy kernels, in the appropriate scaling regimes. Likewise, it is mentioned (without proof) in \citep{FIL18} that the bulk and soft edge scaling limits of the Hermitised matrix product ensembles are also described by the sine and Airy kernels. More interesting, however, is the microscopic behaviour of these ensembles at the origin. There, one observes the so-called double-sided hard edge phenomenon.
\begin{proposition} \label{prop1.10}
Let $m\in\mathbb{N}$ and $\nu_1,\ldots,\nu_m>-1$ be fixed. Moreover, let $K_{N_0,\nu_1,\ldots,\nu_m}^{(\mathcal{H}_m)}(x,y)$ denote the correlation kernel corresponding to $p^{(\mathcal{H}_m)}(\lambda_1,\ldots,\lambda_{N_0})$ in analogy with equation \eqref{eq1.3.6}. Then, for $x,y\in\mathbb{R}\setminus\{0\}$, one has the limit \citep{FIL18} 
\begin{multline} \label{eq1.3.32}
\lim_{N_0\to\infty}\frac{1}{\sqrt{N_0/2}}K_{N_0,\nu_1,\ldots,\nu_m}^{(\mathcal{H}_m)}\left(\frac{x}{\sqrt{N_0/2}},\frac{y}{\sqrt{N_0/2}}\right)=\frac{\mathrm{sgn}(xy)|x|}{4^m}K_{\tfrac{1}{2},\tfrac{\nu_1}{2},\tfrac{\nu_1+1}{2},\ldots,\tfrac{\nu_m}{2},\tfrac{\nu_m+1}{2}}^{(Meijer)}\left(\frac{x^2}{4^m},\frac{y^2}{4^m}\right)
\\+\frac{|y|}{4^m}K_{-\tfrac{1}{2},\tfrac{\nu_1}{2},\tfrac{\nu_1-1}{2},\ldots,\tfrac{\nu_m}{2},\tfrac{\nu_m-1}{2}}^{(Meijer)}\left(\frac{x^2}{4^m},\frac{y^2}{4^m}\right).
\end{multline}
\end{proposition}

\begin{remark}
It is important to note that the term `double-sided hard edge' mentioned above merely refers to the fact that the local correlations at the origin of the spectra of the Hermitised matrix product ensembles relate to the hard edge of the complex Wishart product ensembles via the mapping $\lambda\mapsto\lambda^2$ and that said spectra do not actually exhibit an edge at the origin. In fact, it is shown in \citep{FIL18} that setting $m=0$ in the right-hand side of equation \eqref{eq1.3.32} recovers the sine kernel, which corresponds to bulk statistics. Let us also take this opportunity to point out that the limiting densities $\rho^{(c\mathcal{W}_m),0}(\lambda)$ of the complex Wishart product ensembles do not always have inverse square root singularities at the hard edge, in contrast to what is known for the classical matrix ensembles. Rather, in the setting of Remark \ref{R1.11}, $\rho^{(c\mathcal{W}_m),0}(\lambda)$ diverges like $\lambda^{-l/(l+1)}$ as $\lambda\to0$ \citep{BJLNS10}.
\end{remark}

It is well known \citep{Fo93a}, \citep{BO01}, \citep[Ch.~7]{Fo10} that the local bulk, soft edge, and hard edge statistics of the $\beta=2$ classical matrix ensembles are respectively governed by the sine \eqref{eq1.3.29}, Airy \eqref{eq1.3.30}, and Bessel kernels (the required edge scalings are given in \S\ref{s2.4.2}), with the latter defined as
\begin{equation}
K^{(Bessel)}_a(x,y)=\frac{J_a(\sqrt{x})\sqrt{y}J_a'(\sqrt{y})-\sqrt{x}J_a'(\sqrt{x})J_a(\sqrt{y})}{2(x-y)},
\end{equation}
where $J_a(x)$ is the Bessel function and $a$ is the exponent seen in the Laguerre and Jacobi weights \eqref{eq1.2.9}. Thus, the microscopic bulk and soft edge statistics of the matrix product ensembles discussed in this section are within the same universality classes as those pertaining to the $\beta=2$ classical matrix ensembles, but this is not the case at the hard edge (although, the Meijer $G$-kernel \eqref{eq1.3.28} reduces to the Bessel kernel upon setting $m=1$, as one would expect \citep{AI15}). In this way, as well as their being biorthogonal ensembles rather than orthogonal polynomial ensembles, our matrix product ensembles display universal structures that are fundamentally different to those seen in the classical setting.

\subsection{Muttalib--Borodin and biorthogonal ensembles}
In their 1988 work \citep{MPK88} on multichannel disordered wires, Mello, Pereyra, and Kumar (building on the 1982 work \citep{Dor82} of Dorokhov) derived a Fokker--Planck equation that came to be known as the DMPK equation. This equation characterises the eigenvalues of the associated transfer matrix, which encode physical statistics of the wire at hand, such as its quantum conductance and electron localisation length. The DMPK equation was shown in the 1993 work \citep{BR93} of Beenakker and Rejaei to be solved by the eigenvalue j.p.d.f.
\begin{multline} \label{eq1.3.34}
p^{(DMPK)}(\lambda_1,\ldots,\lambda_N)=\frac{1}{\mathcal{N}^{(DMPK)}_{N,s}}\prod_{i=1}^N\exp\left(-V_{DMPK}(\lambda_i;s)\right)\Delta_N(\lambda)
\\ \times\prod_{1\leq j<k\leq N}\left(\arcsinh^2\,\lambda_k^{1/2}-\arcsinh^2\,\lambda_j^{1/2}\right),\quad\lambda_1,\ldots,\lambda_N>0,
\end{multline}
where $\Delta_N(\lambda)$ is the Vandermonde determinant \eqref{eq1.1.9}, $s$ is a length scale, $\mathcal{N}^{(DMPK)}_{N,s}$ is a normalisation constant, and
\begin{equation}
V_{DMPK}(\lambda;s)=\frac{N}{s}\arcsinh^2(\sqrt{\lambda})\left(1+{\rm O}(N^{-1})\right)
\end{equation}
is a background potential. Two years later, Muttalib proposed \citep{Mut95} the study of simplifications of the above model that correspond to eigenvalue j.p.d.f.s of the form
\begin{equation} \label{eq1.3.36}
p(\lambda_1,\ldots,\lambda_N)=\frac{1}{\mathcal{N}_{N,\theta}}\prod_{i=1}^N\exp\left(-V(\lambda_i)\right)\Delta_N(\lambda)\prod_{1\leq j<k\leq N}(\lambda_k^\theta-\lambda_j^\theta),\quad\lambda_1,\ldots,\lambda_N>0,
\end{equation}
where $\mathcal{N}_{N,\theta}$ is a normalisation constant, $\theta\in\mathbb{N}$ is a deformation parameter (the ensembles can be thought of as deformations of the $\theta=1$ cases, which correspond to the OPEs discussed in \S\ref{s1.2.3}), and $V(\lambda)$ is a general background potential such that the j.p.d.f.~\eqref{eq1.3.36} is well-defined --- this structure relates to the j.p.d.f.~\eqref{eq1.3.34} in the limit $\theta\to0^+$ \citep{FW17}.

It turns out that the relation between Muttalib's model and the DMPK equation was not pursued much further by the physics community, but varying forms of equation \eqref{eq1.3.36} appeared nonetheless in a range of works \citep{LSZ06}, \citep{CR14}, \citep{Che18}, \citep{KS14}, \citep{FL15} over the following two decades. A significant motivation for many of these works was that the associated models were shown to be exactly solvable by Muttalib \citep{Mut95} and Borodin \citep{Bo98}: Muttalib used the Konhauser theory of biorthogonal polynomials \citep{Kon65} (this theory was later realised \citep{AV83} to have been studied much earlier in \citep{Did69}, \citep{Der86}) to prove that the ensembles corresponding to equation \eqref{eq1.3.36} are determinantal point processes; Borodin extended this work by introducing the so-called biorthogonal Hermite, Laguerre, and Jacobi ensembles (see below) and giving explicit formulae for their correlation kernels (cf.~equation \eqref{eq1.3.6}). Thus, in the recent work \citep{FW17}, the name \textit{Muttalib--Borodin ensemble} was given to the systems of positive real eigenvalues governed by j.p.d.f.s of the form \eqref{eq1.3.36}, with $\theta$ now a positive real deformation parameter.

\begin{definition}
Having fixed $N\in\mathbb{N}$, a \textit{biorthogonal ensemble} \citep{Bo98} is a system of eigenvalues $\{\lambda_i\}_{i=1}^N$ supported on a possibly infinite interval $I\subseteq\mathbb{R}$ with eigenvalue j.p.d.f.~of the form
\begin{equation} \label{eq1.3.37}
p^{(b,w)}(\lambda_1,\ldots,\lambda_N)=\frac{1}{\mathcal{N}^{(b,w)}_N}\prod_{i=1}^Nw(\lambda_i)\,\Det\left[f_j(\lambda_i)\right]_{i,j=1}^N\,\Det\left[g_j(\lambda_i)\right]_{i,j=1}^N,
\end{equation}
where $\mathcal{N}^{(b,w)}_N$ is a normalisation constant, $w(\lambda)$ is an admissible weight supported on $I$ (that is, $w(\lambda)$ is continuous, non-negative, and real-valued over $I$ \citep{Kon65}; cf.~Definition \ref{def1.8}), and the sequences of real-valued functions $\{f_j(\lambda)\}_{j=1}^N$ and $\{g_j(\lambda)\}_{j=1}^N$, along with $w(\lambda)$, are such that the above j.p.d.f.~is well-defined.
\end{definition}

\begin{remark}
Polynomial ensembles are examples of biorthogonal ensembles since setting
\begin{equation*}
N=N_0,\quad w(\lambda_i)=1,\quad f_j(\lambda_i)=\lambda_i^{j-1},\quad g_j(\lambda_i)=w_{j-1}(\lambda_i)
\end{equation*}
in equation \eqref{eq1.3.37} recovers the structure \eqref{eq1.3.8}. Likewise, Muttalib--Borodin ensembles \eqref{eq1.3.36} can be seen to be biorthogonal ensembles by setting
\begin{equation*}
w(\lambda_i)=\exp(-V(\lambda_i)),\quad f_j(\lambda_i)=\lambda_i^{j-1},\quad g_j(\lambda_i)=\lambda_i^{\theta(j-1)}
\end{equation*}
in equation \eqref{eq1.3.37}. Rewriting equation \eqref{eq1.3.36} by absorbing the factors $\exp(-V(\lambda_i))$ into the determinant $\prod_{1\leq j<k\leq N}(\lambda_k^\theta-\lambda_j^\theta)$ shows that Muttalib--Borodin ensembles are also polynomial ensembles.
\end{remark}

The adjective `biorthogonal' in the above definition refers to the fact that, after reordering the sequences $\{f_j(\lambda)\}_{j=1}^N$ and $\{g_j(\lambda)\}_{j=1}^N$ if necessary (consider the canonical generalisation of \citep[Prop.~2.5]{Bo98} involving LUP decompositions \citep[Cor.~3]{JO97}), there exist further sequences of so-called \textit{biorthogonal functions} \citep{Bo98} (\textit{biorthogonal polynomials} in the case of Muttalib--Borodin ensembles, among others \citep{Kon65}) $\{\xi_j(\lambda)\}_{j=1}^N$ and $\{\eta_j(\lambda)\}_{j=1}^N$, along with a sequence of finite, positive constants $\{r_j\}_{j=1}^N$, such that for each $1\leq i,j\leq N$,
\begin{align*}
\xi_j(\lambda)-f_j(\lambda)&\in\mathrm{Span}\left\{f_1(\lambda),\ldots,f_{j-1}(\lambda)\right\},
\\ \eta_j(\lambda)-g_j(\lambda)&\in\mathrm{Span}\left\{g_1(\lambda),\ldots,g_{j-1}(\lambda)\right\},
\end{align*}
and (recalling that the indicator function $\chi_A$ equals one when $A$ is true and zero otherwise)
\begin{equation} \label{eq1.3.38}
\int_I\xi_i(\lambda)\eta_j(\lambda)w(\lambda)\,\mathrm{d}\lambda=r_j\chi_{i=j}.
\end{equation}

Using the invariance of determinants under row and column operations in a similar fashion to Lemma \ref{L1.2}, equation \eqref{eq1.3.37} can be rewritten as \citep{Mut95}, \citep{Bo98}
\begin{align}
p^{(b,w)}(\lambda_1,\ldots,\lambda_N)&=\frac{1}{\mathcal{N}^{(b,w)}_N}\prod_{i=1}^Nw(\lambda_i)\,\Det\left[\xi_j(\lambda_i)\right]_{i,j=1}^N\,\Det\left[\eta_j(\lambda_i)\right]_{i,j=1}^N \label{eq1.3.39}
\\&=\Det\left[K_N^{(b,w)}(\lambda_i,\lambda_j)\right]_{i,j=1}^N, \label{eq1.3.40}
\end{align}
where $\{\xi_j(\lambda);\eta_j(\lambda)\}_{j=1}^N$ is the system of biorthogonal functions described above and
\begin{equation} \label{eq1.3.41}
K_N^{(b,w)}(x,y):=\sqrt{w(x)w(y)}\sum_{k=0}^{N-1}\frac{1}{r_k}\xi_k(x)\eta_k(y)
\end{equation}
is the correlation kernel obtained by interchanging the indices $i\leftrightarrow j$ in the second determinant of equation \eqref{eq1.3.39} and then combining all factors therein into a single determinant; cf.~equation \eqref{eq1.2.67}. It was shown by Muttalib \citep{Mut95} that the biorthogonality relation \eqref{eq1.3.38} implies that the above correlation kernel has the reproducing property \eqref{eq1.2.70}. Thus, in analogy with equations \eqref{eq1.2.73} and \eqref{eq1.3.6}, the $k$-point correlation function ($1\leq k\leq N$) corresponding to $p^{(b,w)}(\lambda_1,\ldots,\lambda_N)$ is given by
\begin{align}
\rho_k^{(b,w)}(\lambda_1,\ldots,\lambda_k;N)&:=\frac{N!}{(N-k)!}\int_{\mathbb{R}^{N-k}}p^{(b,w)}(\lambda_1,\ldots,\lambda_N)\,\mathrm{d}\lambda_{k+1}\cdots\mathrm{d}\lambda_N
\\&\;=\Det\left[K_N^{(b,w)}(\lambda_i,\lambda_j)\right]_{i,j=1}^k,
\end{align}
so that biorthogonal ensembles fall within the class of determinantal point processes.

As mentioned earlier, the biorthogonal ensembles at the focus of Borodin's work \citep{Bo98} were the biorthogonal Hermite, Laguerre, and Jacobi ensembles. In present day terminology \citep{FI18}, these correspond to the Hermite, Laguerre, and Jacobi Muttalib--Borodin ensembles (the latter with parameter $b$ set to zero), which have eigenvalue j.p.d.f.s of the form
\begin{equation} \label{eq1.3.44}
p^{(w,\theta)}(\lambda_1,\ldots,\lambda_N)=\frac{1}{\mathcal{N}^{(w)}_{N,\theta}}\prod_{i=1}^Nw(\lambda_i)\,\Delta_N(\lambda)\prod_{1\leq j<k\leq N}((\mathrm{sgn}\,\lambda_k)|\lambda_k|^\theta-(\mathrm{sgn}\,\lambda_j)|\lambda_j|^\theta)
\end{equation}
defined on $\mathbb{R}^N$, with respective (semi-)classical weights (cf.~equation \eqref{eq1.2.9})
\begin{equation} \label{eq1.3.45}
w(\lambda)=\begin{cases} |\lambda|^ae^{-\lambda^2},&\textrm{generalised Hermite}, \\ \lambda^ae^{-\lambda}\chi_{\lambda>0},&\textrm{Laguerre}, \\ \lambda^a(1-\lambda)^b\chi_{0<\lambda<1},&\textrm{Jacobi}.\end{cases}
\end{equation}
Here, $\mathcal{N}_{N,\theta}^{(w)}$ is again a normalisation constant and $\theta>0$, $a,b>-1$ are real parameters. As one might expect, the j.p.d.f.~\eqref{eq1.3.44} is a generalisation of the form \eqref{eq1.3.36}; they are equivalent if all eigenvalues are positive and/or if $\theta$ is an odd integer.

The correlation kernels \eqref{eq1.3.41} governing the statistics of the Muttalib--Borodin ensembles with weights \eqref{eq1.3.45} (with $b=0$) were made explicit in \citep{Bo98}, along with their (double-sided) hard edge limiting forms. The associated biorthogonal polynomials were studied earlier in \citep{Kon67}, \citep{MT82}, \citep{TM86}. More recently, Forrester and Ipsen \citep{FI18} have shown how Selberg integral theory can be used to obtain explicit formulae for the biorthogonal polynomials related to the weights \eqref{eq1.3.45}, in addition to the so-called Jacobi prime, generalised symmetric Jacobi, and generalised Cauchy weights. Incidentally, the biorthogonal functions relating to the correlation kernels $K_{N_0,\nu_1,\ldots,\nu_m}^{(c\mathcal{W}_m)}(x,y)$, $K_{N_0/2,\nu_1,\ldots,\nu_m}^{(\mathrm{i}\mathcal{J}_m)}(x,y)$, and $K_{N_0,\nu_1,\ldots,\nu_m}^{(\mathcal{H}_m)}(x,y)$ discussed in the previous subsection have also been made explicit in the works \citep{AIK13}, \citep{FIL18}, \citep{FILZ19}.

\subsubsection{Relations to matrix product ensembles}
Our interest lies in the Laguerre and Hermite Muttalib--Borodin ensembles, as they have been shown \citep{FLZ15}, \citep{Ips15}, \citep{FIL18} to approximate the complex Wishart and Hermitised matrix product ensembles upon applying the asymptotic formula \citep[Sec.~5.7]{Luk69}
\begin{equation} \label{eq1.3.46}
\meijer{m,0}{0,m}{-}{\nu_1,\ldots,\nu_m}{\lambda}\underset{\lambda\rightarrow\infty}{\sim}\frac{1}{\sqrt{m}}\left(\frac{2\pi}{\lambda^{1/m}}\right)^{(m-1)/2}\lambda^{(\nu_1+\cdots+\nu_m)/p}e^{-m\lambda^{1/m}}\left(1+{\rm O}(\lambda^{1/m})\right)
\end{equation}
to the Meijer $G$-functions in Proposition \ref{prop1.7} and appropriately changing variables.
\begin{proposition} \label{prop1.11}
Let $m,N_0\in\mathbb{N}$ and $\nu_1,\ldots,\nu_m>-1$ be fixed (with $N_0$ even when working with $\mathrm{i}\mathcal{J}_m$). Furthermore, let $p^{(H,\theta)}(\lambda_1,\ldots,\lambda_N;a)$ and $p^{(L,\theta)}(\lambda_1,\ldots,\lambda_N;a)$ denote the eigenvalue j.p.d.f.~\eqref{eq1.3.44} with $w(\lambda)$ being the generalised Hermite and Laguerre weights \eqref{eq1.3.45}, respectively. Then, substituting the large $\lambda$ approximation \eqref{eq1.3.46} into the formulae of Proposition \ref{prop1.7} shows that in the $\lambda_1,\ldots,\lambda_{N_0}\to\infty$ limit with $\bar{\nu}:=\nu_1+\cdots+\nu_m$ and $\gamma_i=\lambda_i/\sqrt{2m+1}$ ($1\leq i\leq N_0$),
\begin{align}
p^{(c\mathcal{W}_m)}\left(\left(\tfrac{\lambda_1}{m}\right)^m,\ldots,\left(\tfrac{\lambda_{N_0}}{m}\right)^m\right)&\approx \prod_{i=1}^{N_0}\left(\tfrac{\lambda_i}{m}\right)^{1-m}\,p^{(L,m)}\left(\lambda_1,\ldots,\lambda_{N_0};\,\bar{\nu}+\tfrac{m-1}{2}\right), \label{eq1.3.47}
\\ p^{(\mathrm{i}\mathcal{J}_m)}\left(\left(\tfrac{\lambda_1}{m}\right)^m,\ldots,\left(\tfrac{\lambda_{N_0/2}}{m}\right)^m\right)&\approx \prod_{i=1}^{N_0/2}\left(\tfrac{\lambda_i}{m}\right)^{1-m}\,p^{(L,2m)}\left(\lambda_1,\ldots,\lambda_{N_0/2};\,2\bar{\nu}+\tfrac{m-1}{2}\right), \label{eq1.3.48}
\\ p^{(\mathcal{H}_m)}\left(2^m\gamma_1^{2m+1},\ldots,2^m\gamma_{N_0}^{2m+1}\right)&\approx \prod_{i=1}^{N_0}\frac{\gamma_i^{-2m}}{2^m\sqrt{2m+1}}\,p^{(H,2m+1)}\left(\lambda_1,\ldots,\lambda_{N_0};\,2\bar{\nu}+m\right). \label{eq1.3.49}
\end{align}
\end{proposition}

\begin{remark} \label{R1.14}
It is shown in \citep[Cor.~4.4]{FILZ19} that the approximation \eqref{eq1.3.48} is in fact exact for $m=1$. This complements the fact that the approximations \eqref{eq1.3.47} and \eqref{eq1.3.49} are also exact for $m=1$ and $m=0$, respectively. (Comparing equations \eqref{eq1.2.7}, \eqref{eq1.2.9} to equations \eqref{eq1.3.44}, \eqref{eq1.3.45} reveals that $p^{(L,1)}(\lambda_1,\ldots,\lambda_N;a)$ and $p^{(H,1)}(\lambda_1,\ldots,\lambda_N;0)$ are respectively the eigenvalue j.p.d.f.s of the Laguerre and Gaussian unitary ensembles.)
\end{remark}

The works \citep{FLZ15}, \citep{Ips15} contain formulae similar to the approximation \eqref{eq1.3.47}, while the approximation \eqref{eq1.3.49} is given in \citep{FIL18}. Indeed, the approximation \eqref{eq1.3.49} represents the success of the latter work in relating the Hermite Muttalib--Borodin ensembles to products of random matrices with explicit p.d.f.s on their entries. Thus, \citep{FIL18} completed the task of finding matrix model realisations of the Muttalib--Borodin ensembles with (semi-)classical weights \eqref{eq1.3.45}, supplementing analogous realisations of the Laguerre and Jacobi Muttalib--Borodin ensembles given in the earlier works \citep{AMW13}, \citep{FLZ15}, \citep{FW17}, \citep{Che18}. We note, in particular, that the Jacobi Muttalib--Borodin ensembles relate to products of Ginibre matrices and truncations of Haar-distributed unitary matrices (see \citep{KKS16}, \citep[\S2.2, \S3.2]{FW17} and references therein).

In light of the scalings discussed in \S\ref{s1.3.1}, a question of obvious interest regarding Proposition \ref{prop1.11} is, ``How accurate are the approximations \eqref{eq1.3.47}--\eqref{eq1.3.49} in the large $N_0$ limit and where do they break down?'' In other words, ``When taking $N_0\to\infty$, how do the eigenvalue statistics of the matrix product ensembles discussed in \S\ref{s1.3.1} differ from those of the Muttalib--Borodin ensembles that they relate to via Proposition \ref{prop1.11}?'' It turns out that the global, bulk, and soft edge scaling statistics of the Muttalib--Borodin ensembles pertaining to the right-hand sides of the approximations \eqref{eq1.3.47}--\eqref{eq1.3.49} are equivalent to those of the corresponding matrix product ensembles \citep{Bo98}, \citep{Zha15}, \citep{FL15}, \citep{FW17}, \citep{FIL18} --- that is, the respective statistics are yet again described by the Fuss--Catalan distribution \eqref{eq1.3.16}, the sine kernel \eqref{eq1.3.29}, and the Airy kernel \eqref{eq1.3.30}. This makes sense at a heuristic level since, by the discussion preceding Proposition \ref{prop1.9}, the eigenvalues of the matrix product ensembles considered in Proposition \ref{prop1.11} that are not at the (double-sided) hard edge grow like $N_0^m$ and are thus large enough, in the $N_0\to\infty$ limit, to fall within the scope of said proposition.

As one might expect, Proposition \ref{prop1.11} breaks down at the origin, but not too drastically: Let $K_N^{(L,\theta)}(x,y;a)$ denote the correlation kernel \eqref{eq1.3.41} corresponding via equation \eqref{eq1.3.40} to the eigenvalue j.p.d.f.~$p^{(L,\theta)}(\lambda_1,\ldots,\lambda_N;a)$ specified in Proposition \ref{prop1.11}. Borodin \citep{Bo98} showed that the hard edge scaling limit of this kernel is given by
\begin{align*}
K^{(a,\theta)}(x,y)&:=\lim_{N\to\infty}N^{-1/\theta}K_N^{(L,\theta)}(N^{-1/\theta}x,N^{-1/\theta}y;a)
\\&\;=\theta x^a\int_0^1J_{(a+1)/\theta,1/\theta}(xt)\, J_{a+1,\theta}((yt)^\theta)t^a\,\mathrm{d}t,
\end{align*}
where
\begin{equation*}
J_{a,b}(x)=\sum_{j=0}^\infty\frac{(-x)^j}{j!\Gamma(a+jb)}
\end{equation*}
is Wright's generalised Bessel function \citep{Wr35}. It has been shown \citep[Thrm.~5.1]{KS14} for $\theta\in\mathbb{N}$, as relevant to Proposition \ref{prop1.11} (they also show an analogous result for $1/\theta\in\mathbb{N}$ that we do not display here), that
\begin{equation}
x^{1/\theta-1}K^{(a,\theta)}(\theta x^{1/\theta},\theta y^{1/\theta})=K^{(Meijer)}_{\tfrac{a+1}{\theta}-1,\tfrac{a+2}{\theta}-1,\ldots,\tfrac{a+\theta}{\theta}-1}(y,x),
\end{equation}
where $K^{(Meijer)}_{\nu_1,\ldots,\nu_m}(x,y)$ is the Meijer $G$-kernel \eqref{eq1.3.28} describing the hard edge statistics of the complex Wishart product ensembles via equation \eqref{eq1.3.25}. In the Hermite case, Borodin \citep{Bo98} showed that scaling the correlation kernel corresponding to the eigenvalue j.p.d.f. $p^{(H,\theta)}(\lambda_1,\ldots,\lambda_N;a)$ to the origin results in a limiting kernel that is related to $K^{(a,\theta)}(x,y)$ in a similar fashion to equation \eqref{eq1.3.32}. Thus, in analogy with Proposition \ref{prop1.10}, the Hermite Muttalib--Borodin ensembles exhibit a double-sided hard edge at the origin that is described by the Meijer $G$-kernel (some discussion on this point is also given in \citep{FIL18}).

Let us highlight the remarkable fact that applying the large eigenvalue approximation \eqref{eq1.3.46} to our matrix product ensembles of interest does not disrupt the small eigenvalue statistics enough to change the fact that their (double-sided) hard edge statistics are described by the Meijer $G$-kernel --- one need only reparametrise said kernel. Thus, it is understood in the present day that the matrix product ensembles reviewed in this section, together with their approximations in terms of Muttalib--Borodin ensembles, belong to a universality class of eigenvalue ensembles that are characterised by having their bulk, soft edge, and (double-sided) hard edge statistics described respectively by the sine kernel, the Airy kernel, and the Meijer $G$-kernel with various parametrisations. This universality class has consequently garnered great interest, with recent works showing that it contains other matrix product ensembles \citep{Fo14}, \citep{KS14}, \citep{KKS16}, along with the class of Muttalib--Borodin ensembles with $\theta>0$ real and general weight of the form $w(\lambda)=\lambda^ae^{-NV(\lambda)}\chi_{\lambda>0}$ \citep{KM19}, \citep{Mol21}, \citep{WZ21} (see also \citep{CGS19}, \citep{CLM21} for related results on this latter class of Muttalib--Borodin ensembles).

\subsection{Integrals of Harish-Chandra--Itzykson--Zuber type}
The \textit{Harish-Chandra--Itzykson--Zuber (HCIZ) integral formula} is the evaluation
\begin{equation} \label{eq1.3.51}
\int_{U(N)}e^{\Tr(AUBU^\dagger)}\,\mathrm{d}\tilde{\mu}'_{\mathrm{Haar}}(U)=\prod_{i=1}^{N-1}i!\,\frac{\Det[e^{a_jb_k}]_{j,k=1}^N}{\Delta_N(a)\Delta_N(b)},
\end{equation}
where $A$ and $B$ are $N\times N$ complex Hermitian matrices with eigenvalues $\{a_j\}_{j=1}^N$ and $\{b_j\}_{j=1}^N$, $\Delta_N(\lambda)$ is the Vandermonde determinant \eqref{eq1.1.9}, and $\mathrm{d}\tilde{\mu}'_{\mathrm{Haar}}(U)=\mathrm{d}\mu'_{\mathrm{Haar}}(U)/\mathrm{vol}(U(N))$ is the Haar probability measure on $U(N)$. (The measure $\mathrm{d}\mu'_{\mathrm{Haar}}(U)$ is specified by equation \eqref{eq1.2.35} with $\beta=2$, while $\mathrm{vol}(U(N))=\int_{U(N)}\mathrm{d}\mu'_{\mathrm{Haar}}(U)$ can be obtained by setting $\beta=2$ in equation \eqref{eq1.2.45} and multiplying the result by $[\mathrm{vol}(U(1))]^N=[2\pi]^N$.) The HCIZ integral formula \eqref{eq1.3.51} is so named due to it being a special case of the general group integral studied by Harish-Chandra in the 1957 work \citep{HC57}, and for its independent derivation and introduction to random matrix theory by Itzykson and Zuber in the 1980 work \citep{IZ80}.

The HCIZ integral and its numerous analogues (some of which can be inferred from Harish-Chandra's original group integral \citep{HC57} as shown in, e.g., \citep{FS02}, \citep{FEFZ07}) have seen direct applications in the study of quantum chromodynamics \citep{KM93}, \citep{JSV96}, \citep{AN11}, \citep{KVZ13} and have otherwise been studied more broadly for their connections to Lie theory, combinatorics, harmonic analysis, and probability theory \citep{Hel78}, \citep{Ter88}, \citep{ZJZ03}. In random matrix theory, HCIZ-type integrals were shown in the early to mid-2000s to be powerful tools for studying complex Gaussian sample covariance matrices (i.e., products $\Sigma^{-1}W$ with $W$ a complex Wishart--Laguerre matrix, as specified in Definition~\ref{def1.4}, and $\Sigma$ a fixed covariance matrix) \citep{BK05}, \citep{BBP05}, \citep{SMM06} and sums involving random matrices \citep{Joh01}, \citep{KT04}, \citep{FR05}.

A breakthrough was made in the 2013 work \citep{AKW13} of Akemann et al., who showed how the formula \eqref{eq1.3.51} can be used to obtain the eigenvalue j.p.d.f.~$p^{(c\mathcal{W}_m)}(\lambda_1,\ldots,\lambda_N)$ of the complex Wishart product ensembles (recall Definition \ref{def1.12}) with arbitrary $m\in\mathbb{N}$ and $\nu_1=\cdots=\nu_m=0$ (so that all of the Ginibre matrices $G_1,\ldots,G_m$ in equation \eqref{eq1.3.3} are $N\times N$) --- their method was subsequently extended in \citep{AIK13} to allow for arbitrary $\nu_1,\ldots,\nu_m>-1$ through the use of induced Ginibre matrices (cf.~the discussion below Proposition~\ref{prop1.7}). The success of these works in applying the HCIZ integral formula to arbitrarily large products of random matrices prompted a series of studies on other random matrix products that could likewise be treated using HCIZ-type integrals. Thus, it was soon found that products of two coupled random matrices \citep{AS16}, \citep{Li18} and products of arbitrarily many truncated Haar-distributed unitary matrices \citep{KKS16} can be studied using appropriate generalisations of the HCIZ integral formula (in fact, the authors of this latter work introduced a previously unknown HCIZ-type integral formula as part of their development).

Let us now draw attention to an analogue and a generalisation of the HCIZ integral formula \eqref{eq1.3.51} that are of particular interest to us: For $N$ an even integer, the \textit{orthogonal Harish-Chandra integral formula} is the result
\begin{equation} \label{eq1.3.52}
\int_{O(N)}e^{\tfrac{1}{2}\Tr(AUBU^T)}\,\mathrm{d}\tilde{\mu}'_{\mathrm{Haar}}(U)=2^{-N/2}\prod_{i=1}^{N/2-1}(2i)!\,\frac{\Det[2\cosh(a_jb_k)]_{j,k=1}^{N/2}}{\prod_{1\leq j<k\leq N/2}(a_k^2-a_j^2)(b_k^2-b_j^2)},
\end{equation}
where $A$ and $B$ are $N\times N$ real antisymmetric matrices with imaginary eigenvalues $\{\pm\mathrm{i}a_j\}_{j=1}^{N/2}$ and $\{\pm\mathrm{i}b_j\}_{j=1}^{N/2}$, while $\mathrm{d}\tilde{\mu}'_{\mathrm{Haar}}(U)=\mathrm{d}\mu'_{\mathrm{Haar}}(U)/\mathrm{vol}(O(N))$ is the Haar probability measure on $O(N)$ (as before, set $\beta=1$ in equations \eqref{eq1.2.35}, \eqref{eq1.2.45}, and note that $\mathrm{vol}(O(1))=2$); Fixing $M,N\in\mathbb{N}$ such that $0\leq M\leq N$, letting $\eta=\mathrm{diag}(-I_M,I_{N-M})$ be a pseudo-metric tensor, and taking $A$ and $B$ to be $N\times N$ complex Hermitian matrices with eigenvalues
\begin{align}
&a_1<\cdots<a_M<0<a_{M+1}<\cdots<a_N, \nonumber
\\ &b_1<\cdots<b_M<0<b_{M+1}<\cdots<b_N, \label{eq1.3.53}
\end{align}
the \textit{hyperbolic HCIZ integral} satisfies the relation
\begin{equation} \label{eq1.3.54}
\int_{U(\eta)/U(1)^N}e^{-\Tr(AUBU^{-1})}\,\mathrm{d}\tilde{\mu}_{\mathrm{Haar}}(U)\propto\frac{\Det[e^{-a_jb_k}]_{j,k=1}^M\Det[e^{-a_jb_k}]_{j,k=M+1}^N}{\Delta_N(a)\Delta_N(b)},
\end{equation}
where $U(\eta):=\{U\in GL_N(\mathbb{C})\,|\,U^\dagger\eta U=U\eta U^\dagger=\eta\}$ is the pseudo-unitary group with metric $\eta$ and $\mathrm{d}\tilde{\mu}_{\mathrm{Haar}}(U)$ is the Haar probability measure on the quotient space $U(\eta)/U(1)^N$. The formula \eqref{eq1.3.52} was made explicit in \citep{FEFZ07}, using the fact that it is a special case of Harish-Chandra's group integral \citep{HC57}. The hyperbolic HCIZ integral, on the other hand, was introduced by Fyodorov \citep{Fyo02}, \citep{FS02}, who used the Duistermaat--Heckman localisation principle to prove the relation \eqref{eq1.3.54}.

The formula \eqref{eq1.3.52} is the canonical analogue of the HCIZ integral formula \eqref{eq1.3.51} for the real case, so it stands to reason that it should be able to treat a matrix product ensemble in the same manner as shown in \citep{AKW13}, \citep{AIK13}. One would be forgiven for thinking that the real Wishart product ensembles could be studied using equation \eqref{eq1.3.52}, but this is not the case since the matrices $A,B$ therein are antisymmetric (this being due to the fact that in the general setting of \citep{HC57}, the matrices $A,B$ must belong to the Lie algebra corresponding to the Lie group that is being integrated over). In fact, it was shown in \citep{AIK13} that the real Wishart product ensembles could be studied in the same way as in the complex case if one knew a closed form expression, comparable to the right-hand side of equation \eqref{eq1.3.52}, for the left-hand side of equation \eqref{eq1.3.52} with $A,B$ real symmetric --- such an integral is not of HCIZ-type and is at best known to evaluate to a sum of ratios of zonal polynomials, rather than anything resembling the right-hand side of equation \eqref{eq1.3.52} (see, e.g., \citep[Prop.~1.4]{Wa08}). It turns out that the correct ensembles to consider are the antisymmetrised matrix product ensembles introduced in Definition \ref{def1.11}:

Inspired by Defosseux's work \citep{Def10} studying the antisymmetrised Laguerre ensemble (Definition \ref{def1.11} with $m=1$) and its quaternionic analogue through group theoretic methods, Forrester et~al.~\citep{FILZ19} used equation \eqref{eq1.3.52} to investigate the eigenvalue statistics of the antisymmetrised matrix product ensembles, in effect deriving the results that we presented earlier in Propositions \ref{prop1.7} and \ref{prop1.8}. It was likewise shown in \citep{FIL18} that the eigenvalue j.p.d.f.~$p^{(\mathcal{H}_m)}(\lambda_1,\ldots,\lambda_N)$ of the Hermitised matrix product ensemble of Definition~\ref{def1.10} with $m\in\mathbb{N}$ and $\nu_1=\cdots=\nu_m=0$ can be obtained via repeated applications of the relation \eqref{eq1.3.54}. This marked the first instance of an HCIZ-type integral with non-compact domain of integration being successfully applied to the study of matrix product ensembles; the key distinction to previous studies was that the constraints \eqref{eq1.3.53} now needed to be taken into account, which are necessary for the integral \eqref{eq1.3.54} to converge.

\setcounter{equation}{0}
\section{Outline of the Thesis} \label{s1.4}
Thus far in this chapter, we have introduced the random matrix ensembles and the statistics of those ensembles that are at the focus of this thesis. Namely, we are interested in the classical matrix ensembles and matrix product ensembles reviewed in Sections \ref{s1.2} and~\ref{s1.3}, respectively. More precisely, we will be studying the eigenvalue densities of these ensembles, along with their (mixed) moments and cumulants and the generating functions of these moments and cumulants, which are all defined in \S\ref{s1.1.1}. As mentioned at the beginning of this chapter, our studies will in some sense culminate with the derivation of two types of recursions: $1$-point recursions for the series coefficients $M_{k,l}$ (see Lemma \ref{L1.3}) of the classical matrix ensembles' spectral moments and loop equations for the connected $n$-point correlators $W_n(x_1,\ldots,x_n)$ \eqref{eq1.1.29} of global scalings of the $(N,N)$ Hermitised and antisymmetrised Laguerre ensembles of Definitions \ref{def1.10} and \ref{def1.11}, respectively. Loosely speaking, our motivation for studying these two classes of random matrix ensembles together is that they both relate to the Ginibre ensembles of Definition \ref{def1.3}, while our decision to study $1$-point recursions alongside loop equations stems from the deeper desire to understand the connection, if any, between these two type of recursive schemes --- such a connection can be anticipated and has been conjectured \citep{CD21} to exist due to the facts that these recursions characterise closely related statistics and that ensembles known to admit $1$-point recursions are often governed by loop equations, as well. In addition to the $1$-point recursions and loop equations derived in this thesis, we also review and develop some theory on differential equations and ribbon graph combinatorics, along with some exotic topological hypermaps.

We commence Chapter 2 with a review of some theory on Selberg correlation integrals (Section \ref{s2.1}) and some relations between the Cauchy and Jacobi ensembles (Section~\ref{s2.2}). The theory of Section \ref{s2.1} is then used in \S\ref{s2.3.1} to derive third order ($\beta=2$) and fifth order ($\beta=1,4$) linear differential equations for the eigenvalue densities $\rho^{(J)}(\lambda;N,\beta)$ and resolvents $W_1^{(J)}(x;N,\beta)$ of the Jacobi ensembles. The theory of Section 2.2 and limiting procedures drawn from Lemma \ref{L1.1} are subsequently used in \S\ref{s2.3.2} and \S\ref{s2.3.3} to obtain analogous differential equations for the Laguerre and Cauchy ensembles --- we also present the intermediate differential equations for the shifted Jacobi ensembles corresponding to the weight $w^{(sJ)}(\lambda)$ \eqref{eq1.2.29}. Section \ref{s2.3} concludes with \S\ref{s2.3.4}, where an alternate extension of Lemma~\ref{L1.1} is used to derive seventh order linear differential equations for the eigenvalue densities and resolvents of the $\beta=2/3,6$ Gaussian $\beta$ ensembles. This serves to demonstrate the fact that our method is technically applicable to any classical $\beta$ ensemble with either $\beta\in2\mathbb{N}$ or $4/\beta\in2\mathbb{N}$ (albeit with difficulty growing rapidly for each new member of these sequences of ensembles in their natural order).

The primary contribution of Section \ref{s2.3} is the exposition of how Selberg correlation integral theory provides a unified approach for obtaining the aforementioned differential equations characterising eigenvalue densities and resolvents of classical $\beta$ ensembles with $\beta\in\{\sigma\in\mathbb{R}\,|\,\sigma\in2\mathbb{N}\textrm{ or }4/\sigma\in2\mathbb{N}\}$. In fact, many of these differential equations have been derived earlier in the literature through a variety of methods: The differential equation for the GUE has been given in \citep{LM79}, \citep{GT05}; for the GOE and GSE in \citep{WF14}; for a special case of the shifted JUE (sJUE) in \citep{GKR05}; for the LUE in \citep{GT05}, \citep{ATK11}; and for the symmetric ($\alpha\in\mathbb{R}$) CyUE in \citep{ABGS20}. We confirm that our method recovers these differential equations and additionally derive new analogues for the (un-shifted) Jacobi ensembles, the LOE and LSE, the symmetric ($a=b$) shifted Jacobi ensembles, the symmetric CyOE and CySE, the non-symmetric ($\alpha\in\mathbb{C}$) CyUE, and the Gaussian $\beta$ ensembles with $\beta=2/3,6$. The differential equations of \citep{GT05} were used therein to obtain optimal bounds for the rate of convergence to the limiting semi-circle law $\rho^{(G),0}(\lambda)$ \eqref{eq1.2.14} (for the GUE) and Mar\v cenko--Pastur law $\rho^{(L),0}(\lambda)$ \eqref{eq1.2.15} (for the LUE). More recently, Kopelevitch \citep{Ko18} used the third order differential equation for the GUE eigenvalue density to study the convergence properties of $1/N$ expansions for averages of linear statistics of the GUE. In Section \ref{s2.4}, we modify the results of Section~\ref{s2.3} to give differential equation characterisations of the classical matrix ensembles in the global and edge scaling regimes. These characterisations complement a plethora of existing results on statistics of the classical matrix ensembles in said scaling regimes, such as the Tracy--Widom law \citep{TW94a}, \citep{TW96}, the characterisations of the classical unitary ensembles in terms of the sine, Bessel, and Airy kernels \citep{Fo93a}, \citep{BO01}, \citep[Ch.~7]{Fo10}, the loop equations for the classical $\beta$ ensembles \citep{WF14}, \citep{FRW17}, and much more.

Chapter 3 is concerned with the (mixed) moments $m_{k_1,\ldots,k_n}$ \eqref{eq1.1.27} and cumulants $c_{k_1,\ldots,k_n}$ \eqref{eq1.1.28} of the ensembles introduced in Sections \ref{s1.2} and \ref{s1.3}. As a second application of the contents of Section \ref{s2.3}, we use the differential equations therein to derive linear recurrence relations on the spectral moments $m_k$ of the classical matrix ensembles, which we present in \S\ref{s3.1.1}. To be precise, we give recurrences on the spectral moments $m_k^{(J)}$ and $m_k^{(L)}$ of the Jacobi and Laguerre ensembles, the differences of even moments $\mu_k^{(sJ)}=m_{2k+2}^{(sJ)}-m_{2k}^{(sJ)}$ of the symmetric shifted Jacobi ensembles, the sums of even moments $\mu_k^{(Cy)}=m_{2k+2}^{(Cy)}+m_{2k}^{(Cy)}$ of the symmetric Cauchy ensembles, and the spectral moments $m_k^{(G)}$ of the Gaussian $\beta$ ensembles with $\beta=2/3,6$ (we also confirm that our method recovers the GUE, GOE, and GSE recurrences of \citep{HZ86}, \citep{Le09}). Of these recurrences, those pertaining to the unitary ensembles have been given earlier in \citep{HT03}, \citep{Le04}, \citep{CMSV16b}, \citep{CMOS19}, \citep{ABGS20} --- the work \citep{CMSV16b} also contains inhomogeneous analogues of our LOE and LSE moment recurrences. A benefit of our approach is that we are able to show that the moment recurrences of \S\ref{s3.1.1} are valid for complex $k$ (so long as the involved moments $m_k$ are convergent), thereby generalising the LUE and sJUE moment recurrences given in \citep{HT03}, \citep{Le04}, \citep{CMSV16b}. In \S\ref{s3.1.2}, we substitute the moment expansions of Lemma~\ref{L1.3} into the recurrences of \S\ref{s3.1.1} to extract $1$-point recursions for the moment expansion coefficients $M_{k,l}$ associated with the Laguerre ensembles, the Gaussian $\beta$ ensembles with $\beta=2/3,6$, and the (shifted) Jacobi unitary ensemble. That is, we show that particular finite linear combinations of the $M_{k,l}$, whose coefficients are polynomials in $k$, are equal to zero. Examples of such $1$-point recursions include the celebrated Harer--Zagier recursion \citep{HZ86} for the $M_{k,l}^{(GUE)}$ and the analogous recursions for the GOE and GSE given by Ledoux \citep{Le09} --- we confirm that our method recovers these recursions and we also find that our LUE recursion is a natural extension of \citep[Thrm.~4.1]{ND18} to the $a\neq0$ setting.

The analysis of Section \ref{s3.1} complements a multitude of studies on the spectral moments of the classical matrix ensembles, with many of these works highlighting connections to skew-orthogonal polynomial theory \citep{HT03}, \citep{Le04}, \citep{Le09}, \citep{LV11}, \citep{MS11}, \citep{CMSV16b}, symmetric function theory \citep{Kad97}, \citep{BF97}, \citep{Dum03}, \citep{MOPS}, \citep{No15}, \citep{FL16}, \citep{MRW17}, and hypergeometric orthogonal polynomials from the Askey scheme \citep{CMOS19}, \citep{ABGS20}. For the sake of completeness, we survey these works in Section \ref{s3.2}. In addition to the theory reviewed in Section \ref{s3.2}, a major motivation for establishing $1$-point recursions on moment expansion coefficients $M_{k,l}$ of the classical matrix ensembles is that, in the Gaussian and Laguerre cases, said coefficients are well known to enumerate certain types of ribbon graphs and (hyper)maps \citep{BIZ80}, \citep{Zvo97}, \citep{Fra03}, \citep{BP09}, \citep{LaC09} --- these combinatorial objects have in turn seen applications in algebraic geometry and mathematical physics \citep{FGZ95}, \citep{MP98}, \citep{LZ04}, \citep{Mon09}, \citep{VWZ84}, \citep{BZ94}, \citep{Sil97}, \citep{JNPZ97}. Thus, we begin Section~\ref{s3.3} by giving a pedagogically extensive review of how the Isserlis--Wick theorem \citep{Iss18}, \citep{Wic50} can be used to prove the aforementioned connection between ribbon graphs and the spectral moments of the Gaussian (\S\ref{s3.3.1}) and Laguerre (\S\ref{s3.3.2}) ensembles. In fact, we go further and elucidate the relationship between ribbon graphs and the mixed moments and cumulants of said ensembles. This theory is then extended in \S\ref{s3.3.3} to give combinatorial interpretations for the mixed cumulants of the Hermitised and antisymmetrised matrix product ensembles; in short, we show that these cumulants enumerate ribbon graphs that can be seen as amalgamations of the ribbon graphs discussed in \S\ref{s3.3.1} and \S\ref{s3.3.2}.

In Chapter 4, we transition to an analytic study of the Hermitised and antisymmetrised Laguerre ensembles (i.e., the matrix product ensembles of Definitions \ref{def1.10} and \ref{def1.11} with $m=1$). It turns out that our matrix product ensembles of interest are not (immediately) amenable to the techniques of Chapter 2 and Section \ref{s3.1} due to the relative intractibility (as compared to the orthogonal polynomials relevant to the classical matrix ensembles) of the Meijer $G$-functions displayed in Proposition \ref{prop1.7}. Thus, we take this opportunity to showcase the loop equation formalism, which is well suited to studying the relevant matrix products $\mathcal{H}_1=G_1^\dagger HG_1$ \eqref{eq1.3.1} and $\mathcal{J}_1=G_1^TJ_{N_0}G_1$ \eqref{eq1.3.2} due to their being constructed from Ginibre matrices. Aside from shedding light on the eigenvalue statistics, recently studied in \citep{FIL18}, \citep{FILZ19}, of the Hermitised and antisymmetrised matrix product ensembles, our development provides an avenue for exploring properties of loop equations for matrix product ensembles in general (cf.~\citep{DF20}). Thus, in Section \ref{s4.1}, we give a brief review of loop equations for the classical $\beta$ ensembles and the closely related theory of the topological recursion, before moving onto Sections \ref{s4.2} and \ref{s4.3}, where we derive loop equations for the antisymmetrised and Hermitised Laguerre ensembles, respectively. As mentioned in \S\ref{s1.1.1}, we first derive loop equations for the unconnected $n$-point correlators $U_n(x_1,\ldots,x_n)$ \eqref{eq1.1.26} associated with appropriate global scalings of the antisymmetrised and Hermitised Laguerre ensembles, use those loop equations and equation \eqref{eq1.1.31} to obtain loop equations for the connected analogues $W_n(x_1,\ldots,x_n)$, and then finally show how inserting the large $N$ expansion \eqref{eq1.1.21} into the second set of loop equations enables us to compute the expansion coefficients $W_n^l(x_1,\ldots,x_n)$ in a systematic manner. We round out Sections \ref{s4.2} and \ref{s4.3} with some concrete computations of $W_1^0(x_1)$ and $W_2^0(x_1,x_2)$, which allow us to perform some consistency checks against the combinatorics of \S\ref{s3.3.3}. Finally, Section~\ref{s4.4} contains a brief discussion on some open questions that proved to be outside the scope of this thesis.

\chapter{Differential Equations for the Classical Matrix Ensembles}
The primary goal of this chapter is to establish linear differential equations satisfied by the eigenvalue densities and resolvents of the classical matrix ensembles introduced in Section \ref{s1.2}. With the eigenvalue j.p.d.f.~specified by equation \eqref{eq1.2.7}, replacing $N$ by $N+1$ in equation \eqref{eq1.1.10} shows that the eigenvalue density is given in terms of the absolute value of the $\beta\textsuperscript{th}$ moment of the characteristic polynomial according to
\begin{align}
\rho^{(w)}(\lambda;N+1,\beta)&=\left.\frac{N+1}{\mathcal{N}_{N+1,\beta}^{(w)}}\int_{\mathbb{R}^N}\prod_{i=1}^{N+1}w(\lambda_i)\,|\Delta_{N+1}(\lambda)|^{\beta}\,\mathrm{d}\lambda_1\cdots\mathrm{d}\lambda_N\right|_{\lambda_{N+1}=\lambda} \nonumber
\\&=\frac{(N+1)\mathcal{N}_{N,\beta}^{(w)}}{\mathcal{N}_{N+1,\beta}^{(w)}}\,w(\lambda)\mean{\prod_{i=1}^N|\lambda-\lambda_i|^{\beta}}. \label{eq2.0.1}
\end{align}
The average here is over the eigenvalue j.p.d.f.~$p^{(w)}(\lambda_1,\ldots,\lambda_N;\beta)$ for $N$ eigenvalues, in keeping with equation \eqref{eq1.1.11}. The crux of this chapter is that when $w(\lambda)$ is taken to be the Jacobi weight, this average is an example of a particular class of Selberg correlation integrals. We review these integrals in Section \ref{s2.1}, along with some related theory that is needed to obtain the sought linear differential equations.

In \S\ref{s2.3.1}, we use the contents of Section \ref{s2.1} to establish order three ($\beta=2$) and order five ($\beta=1,4$) linear differential equations that characterise the eigenvalue densities $\rho^{(J)}(\lambda)$ and resolvents $W_1^{(J)}(x)$ of the Jacobi ensembles. From them we derive analogous differential equations for the other classical ensembles in \S\ref{s2.3.2}--\ref{s2.3.4}. In the Cauchy case, these follow from the change of variables suggested by equations \eqref{eq1.2.22}, \eqref{eq1.2.29} and some subtleties that we outline in Section \ref{s2.2}. In the Laguerre case, we simply apply some limiting procedures drawn from Lemma \ref{L1.1} to the differential equations of \S\ref{s2.3.1}. Similar limiting procedures are applicable in the Gaussian case as well, though the corresponding differential equations are already well known \citep{LM79}, \citep{GT05}, \citep{WF14}. Hence, we instead use Lemma \ref{L1.1} to tweak the proofs given in \S\ref{s2.3.1} and treat the Gaussian $\beta$ ensembles with $\beta=6$ and, through the $\beta\leftrightarrow4/\beta$ duality relation displayed in Lemma \ref{L1.4}, $\beta=2/3$. We choose to study these Gaussian $\beta$ ensembles because they are the next easiest classical $\beta$ ensemble to address --- while our techniques are valid for all classical weights and $\beta=\sigma,4/\sigma$ with $\sigma$ a positive even integer, there is a technical increase in complexity as the weight changes from Gaussian, to Laguerre, to Jacobi, and also when $\sigma$ is increased.

For the sake of completeness and to illustrate the expected form of the results in Section~\ref{s2.3}, we present the GOE, GUE, and GSE differential equations here. We use \citep{WF14} as our source and thus take the Gaussian weight to be $w^{(g)}(\lambda):=e^{-\beta N\lambda^2/(4g)}$, noting this change of weight with the superscript $(g)$. (The coupling constant $g$ determines the length scale: To leading order in $N$, the spectrum is supported on $(-2\sqrt{g},2\sqrt{g})$. The weight $w^{(g)}(\lambda)$ is equivalent to the weight $w^{(G)}(\lambda)$ specified in equation \eqref{eq1.2.9} upon setting $g=\beta N/4$.)

\begin{proposition} \label{prop2.1}
Define
\begin{align} \label{eq2.0.2}
\mathcal{D}_{N,\beta}^{(g)} =
\begin{cases}
\left(\frac{g}{\sqrt{\kappa}N}\right)^2\frac{\mathrm{d}^3}{\mathrm{d}x^3}-y_{(g)}^2\frac{\mathrm{d}}{\mathrm{d}x}+x,&\beta=2,
\\-\left(\frac{g}{\sqrt{\kappa}N}\right)^4\frac{\mathrm{d}^5}{\mathrm{d}x^5}+5\left[\half y_{(g)}^2-h\left(\frac{g}{\sqrt{\kappa}N}\right)\right]\left(\frac{g}{\sqrt{\kappa}N}\right)^2\frac{\mathrm{d}^3}{\mathrm{d}x^3}-3\left(\frac{g}{\sqrt{\kappa}N}\right)^2x\frac{\mathrm{d}^2}{\mathrm{d}x^2}&
\\\quad-\left[y_{(g)}^4-4h\left(\frac{g}{\sqrt{\kappa}N}\right)y_{(g)}^2-\left(\frac{g}{\sqrt{\kappa}N}\right)^2\right]\frac{\mathrm{d}}{\mathrm{d}x}+\left[y_{(g)}^2-2h\left(\frac{g}{\sqrt{\kappa}N}\right)\right]x,&\beta=1,4,
\end{cases}
\end{align}
where $\kappa:=\beta/2$, $h:=\sqrt{\kappa}-1/\sqrt{\kappa}$, and $y_{(g)}:=\sqrt{x^2-4g}$. Then, for $\beta=1,2$, and $4$,
\begin{align} \label{eq2.0.3}
\mathcal{D}_{N,\beta}^{(g)}\,\rho^{(g)}(x;N,\beta) = 0
\end{align}
and
\begin{align} \label{eq2.0.4}
\mathcal{D}_{N,\beta}^{(g)}\,\frac{1}{N}W_1^{(g)}(x;N,\beta) =
\begin{cases}
2,&\beta=2,
\\2y_{(g)}^2-10h\left(\frac{g}{\sqrt{\kappa}N}\right),&\beta=1,4.
\end{cases}
\end{align}
\end{proposition}

\begin{remark}
\begin{enumerate}
\item It can be observed that for $\beta=1$ or $4$,
\begin{align}
\mathcal{D}_{N,\beta}^{(g)}&=-\left(\frac{g}{\sqrt{\kappa}N}\right)^2\left[\mathcal{D}_{N,2}^{(g)}+2x\right]\frac{\mathrm{d}^2}{\mathrm{d}x^2}+\left[y_{(g)}^2-5h\left(\frac{g}{\sqrt{\kappa}N}\right)\right]\mathcal{D}_{N,2}^{(g)} \nonumber
\\&\quad+\half y_{(g)}^2\left(\frac{g}{\sqrt{\kappa}N}\right)^2\frac{\mathrm{d}^3}{\mathrm{d}x^3}-\left[hy_{(g)}^2-\left(\frac{g}{\sqrt{\kappa}N}\right)\right]\left(\frac{g}{\sqrt{\kappa}N}\right)\frac{\mathrm{d}}{\mathrm{d}x}+3h\left(\frac{g}{\sqrt{\kappa}N}\right)x. \label{eq2.0.5}
\end{align}
This form of the differential operator should be compared with differential equations for the GOE and GSE eigenvalue densities that were derived recently in \citep{Na18}. The differential equations of \citep{Na18} are linear, but differ from \eqref{eq2.0.3} in that they are third order and have inhomogeneous terms dependent on the GUE eigenvalue density.

\item The operator for $\beta=2$ is even in $N$. In the case of $\beta=1$ and $4$, it is invariant under the mapping $(N,\kappa)\mapsto(-N\kappa,1/\kappa)$, which is consistent with the known duality relations presented in Lemma \ref{L1.4}. As discussed in \S\ref{s1.2.4}, the corresponding duality relations for the Laguerre and Jacobi ensembles given in Lemma \ref{L1.4} are what allow us to obtain differential equations for $\beta=1$ in the upcoming derivations. They also suggest that in certain cases, the LUE and JUE spectral moments $m_k$, when scaled to be ${\rm O}(N)$, are odd functions in $N$ (see \S\ref{s2.4.1} and \S\ref{s3.1.2} for more details).

\item The structure \eqref{eq2.0.1} tells us that $f_{N+1,\beta}(x):=\rho(x;N+1,\beta)/w(x)$ has a large $x$ expansion of the form
\begin{equation} \label{eq2.0.6}
f_{N+1,\beta}(x)\overset{x\rightarrow\infty}{=}\sum_{k=0}^{\infty}a_kx^{\beta N-k}
\end{equation}
with $a_0=(N+1)\mathcal{N}_{N,\beta}/\mathcal{N}_{N+1,\beta}$ fixed and all other constant coefficients $a_k$ yet to be determined. In particular, for $\beta$ an even integer, this series must terminate since the average in \eqref{eq2.0.1} is manifestly a polynomial of degree $\beta N$ in $\lambda$. Substituting $\rho^{(g)}(x;N,2)=w^{(g)}(x)f_{N,2}^{(g)}(x)$ into \eqref{eq2.0.3} shows (choosing the GUE for simplicity),
\begin{equation*}
\left\{\left(\frac{g}{N}\right)^2\frac{\mathrm{d}^3}{\mathrm{d}x^3}-\frac{3g}{N}x\frac{\mathrm{d}^2}{\mathrm{d}x^2}+\left(2x^2+\frac{g}{N}(4N-3)\right)\frac{\mathrm{d}}{\mathrm{d}x}-4(N-1)x\right\}f_{N,2}^{(g)}(x)=0.
\end{equation*}
One can check that this differential equation admits a unique polynomial solution consistent with equation \eqref{eq2.0.6}, and with $\{a_k\}_{k>0}$ completely determined by the choice of $a_0$. This latter feature is true of all the differential equation characterisations we will obtain for $\rho(x;N,\beta)$ in this chapter. It holds true because the underlying differential-difference equation \eqref{eq2.1.17} is effectively a (multi-dimensional) first order recurrence.
\end{enumerate} \label{R2.1}
\end{remark}

The third order differential equation for the GUE eigenvalue density (equation \eqref{eq2.0.3} with $\beta=2$) can be traced back to the work of Lawes and March \citep{LM79}. There, the setting is that of interpreting the GUE eigenvalue density as the squared ground state wave function of spinless non-interacting fermions in one dimension, in the presence of a harmonic confining potential. In the random matrix theory literature, this result appeared in the work of G{\"o}tze and Tikhomirov \citep{GT05}, making use of an earlier result of Haagerup and Thorbj{\o}rnsen \citep{HT03} characterising the two-sided Laplace transform of the density in terms of a hypergeometric function. Differential equations \eqref{eq2.0.3}, \eqref{eq2.0.4} with $\beta=1,4$ were derived in \citep{WF14} using a method based on known evaluations of the eigenvalue densities in terms of Hermite polynomials (recall the discussion in \S\ref{s1.2.3}, particularly that which concerns \citep{AFNM00}). Our approach, using Selberg correlation integrals and duality formulae, is different to those just recounted, and moreover unifies all the classical cases.

\setcounter{equation}{0}
\section{Selberg Correlation Integrals} \label{s2.1}
As mentioned above, our initial interest lies in the Selberg correlation integral that is specified as the average of $\prod_{i=1}^N|\lambda-\lambda_i|^{\beta}$ ($0<\lambda<1$, $\beta>0$) against the j.p.d.f.~ of the Jacobi $\beta$ ensemble, which we recall from equations \eqref{eq1.2.81} and \eqref{eq1.2.9} as
\begin{equation} \label{eq2.1.1}
p^{(J)}(\lambda_1,\ldots,\lambda_N;\beta)=\frac{1}{\mathcal{N}_{N,\beta}^{(J)}}\,\prod_{i=1}^N\lambda_i^a(1-\lambda_i)^b\,|\Delta_N(\lambda)|^\beta,\quad0<\lambda_1,\ldots,\lambda_N<1,\,\beta>0.
\end{equation}
The normalisation constant $\mathcal{N}_{N,\beta}^{(J)}$ is known as the Selberg integral, which, in turn, explains why the average \eqref{eq2.0.1} is referred to as a Selberg correlation integral.

\subsection{The Selberg and Dixon--Anderson integrals} \label{s2.1.1}
Highlighting the dependence on the parameters $a,b$, we denote the \textit{Selberg integral} by
\begin{equation} \label{eq2.1.2}
S_N(a,b,\kappa):=\mathcal{N}_{N,\beta}^{(J)}=\int_{[0,1]^N}\prod_{i=1}^N\lambda_i^a(1-\lambda_i)^b\,|\Delta_N(\lambda)|^{2\kappa}\,\mathrm{d}\lambda_1\cdots\mathrm{d}\lambda_N,
\end{equation}
where we now allow $a,b\in\mathbb{C}$ with $\mathrm{Re}(a),\mathrm{Re}(b)>-1$. This integral was first computed by Selberg \citep{Sel44} to be given by
\begin{equation} \label{eq2.1.3}
S_N(a,b,\kappa)=\prod_{j=0}^{N-1}\frac{\Gamma(a+1+\kappa j)\Gamma(b+1+\kappa j)\Gamma(1+\kappa(j+1))}{\Gamma(a+b+2+\kappa(N+j-1))\Gamma(1+\kappa)},
\end{equation}
with the requirements $\mathrm{Re}(a),\mathrm{Re}(b)>-1$ now seen to be necessary for avoiding the poles in the numerator of the right-hand side of equation \eqref{eq2.1.3}. Since the work of Selberg, there have been a number of proofs produced for the above evaluation \citep{DF85}, \citep{Aom87}, \citep{And91} (see \citep[Ch.~4]{Fo10} for a review). To make connection with the recursive theme of this thesis, we present below the method of Anderson, which consists of setting up a recurrence in $N$.

Following \citep[Sec.~4.2]{Fo10}, we fix parameters $s_1,\ldots,s_{N+1}>0$ and refer to the quantity
\begin{align}
D_N^{(s_1,\ldots,s_{N+1})}(a_1,\ldots,a_{N+1})&:=\int_{a_{N+1}<\lambda_N<\cdots<\lambda_1<a_1}\prod_{i=1}^N\prod_{j=1}^{N+1}|\lambda_i-a_j|^{s_j-1}\,\Delta_N(\lambda)\,\mathrm{d}\lambda_1\cdots\mathrm{d}\lambda_N \nonumber
\\&\;=\frac{\Gamma(s_1)\cdots\Gamma(s_{N+1})}{\Gamma(s_1+\cdots+s_{N+1})}\prod_{1\leq j<k\leq N+1}(a_j-a_k)^{s_j+s_k-1}\label{eq2.1.4}
\end{align}
as the \textit{Dixon--Anderson integral}, so-called because of its independent derivations in 1905 by Dixon \citep{Dix05} and in 1991 by Anderson \citep{And91}. The equality of both sides of equation \eqref{eq2.1.4} is proven in \citep{Fo10} as follows: Let $\{w_i\}_{i=1}^{N+1}$ be drawn from the \textit{Dirichlet distribution}, which we specify by the j.p.d.f.
\begin{equation} \label{eq2.1.5}
\frac{\Gamma(s_1+\cdots+s_{N+1})}{\Gamma(s_1)\cdots\Gamma(s_{N+1})}\prod_{i=1}^{N+1}w_i^{s_i-1}
\end{equation}
supported on the standard $N$-simplex
\begin{equation*}
\left\{(w_1,\ldots,w_{N+1})\in\mathbb{R}^{N+1}\,\Big|\,\sum_{i=1}^{N+1}w_i=1\textrm{ and }w_i>0\textrm{ for all }i=1,\ldots,N+1\right\}.
\end{equation*}
Now, introduce the random rational function
\begin{equation} \label{eq2.1.6}
R(\lambda):=\sum_{i=1}^{N+1}\frac{w_i}{a_i-\lambda}.
\end{equation}
In a similar fashion to the calculation surrounding the ratio of characteristic polynomials \eqref{eq1.2.87} of the matrix model $M_N$ \eqref{eq1.2.86} for the Gaussian $\beta$ ensemble that was outlined in \S\ref{s1.2.4}, taking the residues of both sides of equation \eqref{eq2.1.6} at $\lambda=a_i$ expresses the $w_i$ in terms of the $\{a_i\}_{i=1}^{N+1}$ and the zeroes $\{\lambda_i\}_{i=1}^N$ of $R(\lambda)$. Substituting these expressions into the j.p.d.f.~\eqref{eq2.1.5} and including the Jacobian yields the \textit{Dixon--Anderson j.p.d.f.}
\begin{multline}
p^{(s_1,\ldots,s_{N+1})}(\lambda_1,\ldots,\lambda_N;a_1,\ldots,a_{N+1}):=\frac{\Gamma(s_1+\cdots+s_{N+1})}{\Gamma(s_1)\cdots\Gamma(s_{N+1})}\prod_{i=1}^N\prod_{p=1}^{N+1}|\lambda_i-a_p|^{s_p-1}
\\\times\prod_{1\leq j<k\leq N}(a_j-a_k)^{1-s_j-s_k}\,\Delta_N(\lambda).
\end{multline}
A graphical argument reveals the interlacing condition $a_{N+1}<\lambda_N<\cdots<\lambda_1<a_1$; integrating $p^{s_1,\ldots,s_{N+1}}(\lambda_1,\ldots,\lambda_N;a_1,\ldots,a_{N+1})$ in the variables $\lambda_1,\ldots,\lambda_N$ over this region gives equation \eqref{eq2.1.4}.

To prove the Selberg integral formula \eqref{eq2.1.3}, consider the generalised integral \citep[Sec.~4.2]{Fo10}
\begin{multline} \label{eq2.1.8}
K_N(a,b,\kappa):=\int_{\Omega}\,\prod_{l=1}^{N+1}x_l^a(1-x_l)^b\,\prod_{i=1}^N\prod_{j=1}^{N+1}|y_i-x_j|^{\kappa-1}
\\\times|\Delta_N(y)\Delta_{N+1}(x)|\,\mathrm{d}x_1\cdots\mathrm{d}x_{N+1}\mathrm{d}y_1\cdots\mathrm{d}y_N,
\end{multline}
where $\Omega$ denotes the region $0<x_{N+1}<y_N<\cdots<y_1<x_1<1$. Evaluating this integral over the $\{x_i\}_{i=1}^{N+1}$ first reveals that
\begin{align}
K_N(a,b,\kappa)&=\int_{0<y_N<\cdots<y_1<1}D_{N+1}^{(b+1,(\kappa)^N,a+1)}(1,y_1,\ldots,y_N,0)\,|\Delta_N(y)|\,\mathrm{d}y_1\cdots\mathrm{d}y_N \nonumber
\\&=S_N(a+\kappa,b+\kappa,\kappa)\frac{\Gamma(a+1)\Gamma(b+1)\Gamma(\kappa)^N}{N!\Gamma(a+b+2+N\kappa)}, \label{eq2.1.9}
\end{align}
where $(\kappa)^N$ denotes the $N$-tuple $(\kappa,\ldots,\kappa)$ and $D_{N+1}^{(b+1,(\kappa)^N,a+1)}$ is the Dixon--Anderson integral as defined in equation \eqref{eq2.1.4}. The second line here follows from substituting in the right-hand side of equation \eqref{eq2.1.4}, recalling the definition \eqref{eq2.1.2} of the Selberg integral, and introducing a factor of $N!$ to symmetrise the variables. If one instead evaluates $K_N(a,b,\kappa)$ by integrating over the $\{y_i\}_{i=1}^N$ first, one observes that
\begin{align}
K_N(a,b,\kappa)&=\int_{0<x_{N+1}<\cdots<x_1<1}D_N^{(\kappa)^{N+1}}(x_1,\ldots,x_{N+1})\,|\Delta_{N+1}(x)|\,\mathrm{d}x_1\cdots\mathrm{d}x_{N+1} \nonumber
\\&=S_{N+1}(a,b,\kappa)\frac{\Gamma(\kappa)^{N+1}}{(N+1)!\Gamma((N+1)\kappa)}. \label{eq2.1.10}
\end{align}
Comparing the two expressions \eqref{eq2.1.9} and \eqref{eq2.1.10} for $K_N(a,b,\kappa)$ results in the recurrence
\begin{equation}
S_{N+1}(a,b,\kappa)=S_N(a+\kappa,b+\kappa,\kappa)\frac{(N+1)\Gamma((N+1)\kappa)\Gamma(a+1)\Gamma(b+1)}{\Gamma(a+b+2+N\kappa)\Gamma(\kappa)}.
\end{equation}
Solving this recurrence via induction on $N$, we finally obtain equation \eqref{eq2.1.3}.

\subsubsection{Normalisation constants for the classical $\beta$ ensembles}
By the definition of the Selberg integral, the normalisation constant $\mathcal{N}_{N,\beta}^{(J)}$ in the eigenvalue j.p.d.f.~of the Jacobi $\beta$ ensemble is given by equation \eqref{eq2.1.3}. To obtain the corresponding normalisation constant for the Laguerre $\beta$ ensemble, we extend Lemma \ref{L1.1} by setting $\lambda_i=\gamma_i/b$ in the right-hand side of equation \eqref{eq2.1.2} and taking the limit $b\to\infty$ to see that
\begin{align}
\mathcal{N}_{N,\beta}^{(L)}&:=\int_{[0,\infty)^N}\prod_{i=1}^N\lambda_i^ae^{-\lambda_i}\,|\Delta_N(\lambda)|^{2\kappa}\,\mathrm{d}\lambda_1\cdots\mathrm{d}\lambda_N \nonumber
\\&\;=\lim_{b\to\infty}b^{N(\kappa (N-1)+a+1)}S_N(a,b,\kappa) \nonumber
\\&\;=\prod_{j=0}^{N-1}\frac{\Gamma(1+\kappa(j+1))\Gamma(a+1+\kappa j)}{\Gamma(1+\kappa)}, \label{eq2.1.12}
\end{align}
where the final line is obtained by incorporating Selberg's evaluation \eqref{eq2.1.3} and using Stirling's approximation (recall that $\kappa:=\beta/2$).

In the Gaussian case, one similarly sets $\lambda_i=(1+\gamma_i/\sqrt{L})/2$ and $a=b=L$ in the right-hand side of equation \eqref{eq2.1.2}, so that taking the limit $L\to\infty$ shows that
\begin{align}
\mathcal{N}_{N,\beta}^{(G)}&:=\int_{\mathbb{R}^N}\prod_{i=1}^Ne^{-\lambda_i^2}\,|\Delta_N(\lambda)|^{2\kappa}\,\mathrm{d}\lambda_1\cdots\mathrm{d}\lambda_N \nonumber
\\&\;=\lim_{L\to\infty}2^{-N(2L+\kappa(N-1)+1)}L^{-N(\kappa(N-1)+1)/2}S_N(L,L,\kappa) \nonumber
\\&\;=(2\pi)^{N/2}2^{-N(\kappa(N-1)+1)/2}\prod_{j=0}^{N-1}\frac{\Gamma(1+\kappa(j+1))}{\Gamma(1+\kappa)}.
\end{align}

The normalisation constant for the Cauchy $\beta$ ensemble is given by
\begin{align}
\mathcal{N}_{N,\beta}^{(Cy)}&:=\int_{\mathbb{R}^N}\prod_{i=1}^N(1-\mathrm{i}\lambda)^{\eta}(1+\mathrm{i}\lambda)^{\overline{\eta}}\,|\Delta_N(\lambda)|^{2\kappa}\,\mathrm{d}\lambda_1\cdots\mathrm{d}\lambda_N \nonumber
\\&\;=2^{-N(\kappa(N-1)+2\mathrm{Re}(\alpha))}\pi^N\prod_{j=0}^{N-1}\frac{\Gamma(2\mathrm{Re}(\alpha)+1+\kappa j)\Gamma(1+\kappa(j+1))}{\Gamma(\alpha+1+\kappa j)\Gamma(\overline{\alpha}+1+\kappa j)\Gamma(1+\kappa)}, \label{eq2.1.14}
\end{align}
where we have retained our choice of reparametrisation $\eta=-\kappa(N-1)-1-\alpha$. The above evaluation can be obtained by changing variables in the Selberg integral, but the necessary analytic continuation is a little subtle, so we defer proof of this fact to Section \ref{s2.2}.

We conclude this subsection by noting that Definition \ref{def1.5} is now complete (so too is the definition of the classical $\beta$ ensembles given in \S\ref{s1.2.4}), with the normalisation constants of the classical matrix ($\beta$) ensembles' eigenvalue j.p.d.f.s~ $p^{(w)}(\lambda_1,\ldots,\lambda_N)$ specified by equations \eqref{eq2.1.3} and \eqref{eq2.1.12}--\eqref{eq2.1.14}. From the discussion in \S\ref{s1.2.2}, substituting these values into equation \eqref{eq1.2.44} (using equation \eqref{eq1.2.45}, as well) then completes Propositions \ref{prop1.1} and \ref{prop1.4} by making the partition functions $\mathcal{Z}_N$ explicit.

\subsection{Further preliminaries concerning Selberg correlation integrals} \label{s2.1.2}
As outlined thus far, our initial object of interest is the average displayed in equation \eqref{eq2.0.1} with $w(\lambda)=\lambda^a(1-\lambda)^b$ the Jacobi weight \eqref{eq1.2.9}. For later reference, we denote it
\begin{equation} \label{eq2.1.15}
I_{N,\beta}^{(J)}(\lambda) := \mean{ \prod_{i=1}^N|\lambda-\lambda_i|^{\beta} }_{\text{JE}_{N,\beta}(a,b)}=\frac{\mathcal{N}_{N+1,\beta}^{(J)}\,\rho^{(J)}(\lambda;N+1,\beta)}{(N+1)\mathcal{N}_{N,\beta}^{(J)}\,w^{(J)}(\lambda)},
\end{equation}
where the subscript $\mathrm{JE}_{N,\beta}(a,b)$ indicates that the average is with respect to the eigenvalue j.p.d.f.~$p^{(J)}(\lambda_1,\ldots,\lambda_N;\beta)$ made explicit in equation \eqref{eq2.1.1}. In order to derive differential equations satisfied by the corresponding eigenvalue density \eqref{eq2.0.1}, we must make a detour and first study the auxiliary average
\begin{equation} \label{eq2.1.16}
J_{n,p}^{(N)}(\lambda) := \mean{ \prod_{i=1}^n(\lambda_i-\lambda)^{N+\chi_{i\leq p}} }_{\text{JE}_{n,4/\beta}(a',b')}.
\end{equation}
The significance of $J_{n,p}^{(N)}(\lambda)$ is that it satisfies the differential-difference equation \citep{Fo93}
\begin{multline*}
(n-p)E_pJ_{n,p+1}^{(N)}(\lambda) = -(A_p\lambda+B_p)J_{n,p}^{(N)}(\lambda)+\lambda(\lambda-1)\frac{\mathrm{d}}{\mathrm{d}\lambda}J_{n,p}^{(N)}(\lambda)+D_p\lambda(\lambda-1)J_{n,p-1}^{(N)}(\lambda),
\end{multline*}
\begin{align}
A_p& = (n-p)\left(a'+b'+\tfrac{2}{\kappa}(n-p-1)+2N+2\right), \nonumber
\\B_p& = (p-n)\left(a'+N+1+\tfrac{1}{\kappa}(n-p-1)\right), \nonumber
\\D_p& = p\left(\tfrac{1}{\kappa}(n-p)+N+1\right), \nonumber
\\E_p& = a'+b'+\tfrac{1}{\kappa}(2n-p-2)+N+2, \label{eq2.1.17}
\end{align}
later observed to be equivalent to a particular matrix differential equation \citep{FR12}.

\begin{remark} \label{R2.2}
The order $p$ elementary symmetric polynomial on $N$ variables is
\begin{equation} \label{eq2.1.18}
e_p(x_1,\ldots,x_N)=\sum_{1\leq j_1<j_2<\cdots<j_p\leq N}x_{j_1}\cdots x_{j_p}
\end{equation}
with the convention $e_p(x_1,\ldots,x_N)=0$ if $p>N$. It is known \citep[Ch.~4]{Fo10} that the above differential-difference equation \eqref{eq2.1.17} is satisfied by a broader class of functions
\begin{equation} \label{eq2.1.19}
\tilde{J}_{n,p,q}^{(N)}(\lambda):=\frac{1}{\mathcal{N}_{N,p}}\mean{ \prod_{i=1}^n|\lambda_i-\lambda|^N\,\chi_{\lambda_1,\ldots,\lambda_{N-q}<x}\,e_p(\lambda-\lambda_1,\ldots,\lambda-\lambda_N) }_{\text{JE}_{n,4/\beta}(a',b')},
\end{equation}
where $\mathcal{N}_{N,p}:=\binom{N}{p}$, this being the number of terms in the sum on the right-hand side of equation \eqref{eq2.1.18}. In the case $q=N$, we see that $\tilde{J}_{n,p,q}^{(N)}(\lambda)$ is equal to $J_{n,p}^{(N)}(\lambda)$ due to symmetry of the integrand in the right-hand side of equation \eqref{eq2.1.16}.
\end{remark}

The differential-difference equation \eqref{eq2.1.17} can be seen to be equivalent to a closed system of differential equations upon letting $p=0,1,\ldots,n$. Indeed, appropriately substituting these $n+1$ equations into each other (more details in \S\ref{s2.3.1}) yields an order $n+1$ linear differential equation for $J_{n,0}^{(N)}(\lambda)$. Making the substitutions
\begin{equation} \label{eq2.1.20}
(n,N,\beta,a',b')\mapsto(N,\beta,4/\beta,a,b)
\end{equation}
in equation \eqref{eq2.1.16}, with $\beta$ a positive even integer to remove the absolute value sign, then reveals that the aforementioned differential equation becomes order $N+1$ and is satisfied by $I_{N,\beta}^{(J)}(\lambda)$ \eqref{eq2.1.15}. However, even though our present goal is to obtain a linear differential equation for $I_{N,\beta}^{(J)}(\lambda)$, a differential equation of order $N+1$ is not suitable for the large $N$ analysis that we wish to perform. Thus, an alternative approach must be taken.

Instead of making the substitutions \eqref{eq2.1.20}, we opt to use the duality \citep[Ch.~13]{Fo10}
\begin{equation} \label{eq2.1.21}
\frac{S_N(a,b,\kappa)}{S_N(a+n,b,\kappa)}\mean{ \prod_{i=1}^N(\lambda_i-\lambda)^n }_{\text{JE}_{N,\beta}(a,b)}\, = \,\mean{ \prod_{i=1}^n(1-\lambda\lambda_i)^N }_{\text{JE}_{n,4/\beta}(a',b')},
\end{equation}
where $S_N$ is the Selberg integral as defined in the previous subsection, and we now take
\begin{equation}
a' = \frac{1}{\kappa}(a+b+2)+N-2,\qquad b' = -\frac{1}{\kappa}(b+n)-N.
\end{equation}
This duality relation can be drawn from the fact that both sides of equation \eqref{eq2.1.21} are known to equal \citep{Ka93}
\begin{align} \label{eq2.1.23}
{}_2F_1^{(\kappa)}\left(-N,(a+b+n+1)/\kappa+N-1,(a+n)/\kappa;(x)^n\right),
\end{align}
where ${}_2F_1^{(\kappa)}$ is the generalised multivariate hypergeometric function and $(x)^n$ is again the $n$-tuple $(x,\ldots,x)$. Note that on the right-hand side of equation \eqref{eq2.1.21}, the parameter $b'$ is in general less than $-1$, and so the integral must be understood in the sense of analytic continuation (cf.~Section \ref{s2.2} forthcoming). The left-hand side of equation \eqref{eq2.1.21} is proportional to $I_{N,\beta}^{(J)}(\lambda)$ \eqref{eq2.1.15} upon taking $n=\beta$ a positive even integer, while a simple change of variables shows that the right-hand side is equal to $(-\lambda)^{nN}J_{n,0}^{(N)}(1/\lambda)$. Combining these two facts, we thus have for $\beta$ a positive even integer that
\begin{equation} \label{eq2.1.24}
I_{N,\beta}^{(J)}(\lambda) \propto (-\lambda)^{\beta N}\,J_{\beta,0}^{(N)}(1/\lambda).
\end{equation}
Hence, the order $n+1$ differential equation for $J_{n,0}^{(N)}(x)$ discussed in the paragraph prior translates to an order $\beta+1$ linear differential equation for $I_{N,\beta}^{(J)}(\lambda)$. These differential equations are derived in detail in \S\ref{s2.3.1}.

\setcounter{equation}{0}
\section{Relating the Cauchy and Jacobi Ensembles Through Analytic Continuation} \label{s2.2}
At the end of \S\ref{s1.2.1}, we said that in deriving results for the Cauchy ensembles, we will go through the shifted Jacobi ensembles, which correspond to the weight \eqref{eq1.2.29}
\begin{equation*}
w^{(sJ)}(\lambda)=(1-\lambda)^a(1+\lambda)^b\chi_{-1<\lambda<1}.
\end{equation*}
In doing so, not only will we bring more clarity to our derivations, but we will also have the benefit of making our results explicit in the Jacobi case with the shifted weight. This latter point is mainly one of convenience, both for the reader interested in the shifted Jacobi ensembles, and ourselves when comparing against the existing literature. Relating the un-shifted Jacobi ensembles to their shifted counterparts is very easy: One need only make the change of variables $\lambda_i\mapsto\tfrac{1}{2}(1-\lambda_i)$, where $\{\lambda_i\}_{i=1}^N$ are the eigenvalues --- there are no complications in applying this change of variables in the upcoming calculations. Making this change of variables in equation \eqref{eq2.1.1}, we see that the eigenvalue j.p.d.f.~of the shifted Jacobi $\beta$ ensemble is
\begin{equation}
p^{(sJ)}(\lambda_1,\ldots,\lambda_N;\beta)=\frac{1}{\mathcal{N}_{N,\beta}^{(sJ)}}\,\prod_{i=1}^N(1-\lambda_i)^a(1+\lambda_i)^b\,|\Delta_N(\lambda)|^\beta,
\end{equation}
where $-1<\lambda_1,\ldots,\lambda_N<1$, $\beta=2\kappa>0$, and
\begin{equation} \label{eq2.2.2}
\mathcal{N}_{N,\beta}^{(sJ)}=2^{N(\kappa(N-1)+a+b+1)}\mathcal{N}_{N,\beta}^{(J)}=2^{N(\kappa(N-1)+a+b+1)}S_N(a,b,\kappa),
\end{equation}
with the value of $S_N(a,b,\kappa)$ given in equation \eqref{eq2.1.3}. Denoting an average with respect to this j.p.d.f.~by the subscript $\mathrm{sJE}_{N,\beta}(a,b)$, in keeping with the notation in equation \eqref{eq2.1.15}, the spectral moments \eqref{eq1.1.15} of the shifted and un-shifted Jacobi $\beta$ ensembles are related as follows:
\begin{align}
m_k^{(sJ)}&=\mean{\sum_{i=1}^N\lambda_i^k}_{\mathrm{sJE}_{N,\beta}(a,b)}=\mean{\sum_{i=1}^N(1-2\lambda_i)^k}_{\mathrm{JE}_{N,\beta}(a,b)} \nonumber
\\&=\sum_{s=0}^k\binom{k}{s}(-2)^s\mean{\sum_{i=1}^N\lambda_i^s}_{\mathrm{JE}_{N,\beta}(a,b)}=\sum_{s=0}^k\binom{k}{s}(-2)^sm_s^{(J)}, \label{eq2.2.3}
\end{align}
where the binomial expansion has been used in the second line.

Having established the mechanisms for moving between shifted and un-shifted Jacobi ensembles, the rest of this section is devoted to relating the Cauchy ensembles to the shifted Jacobi ensembles. We first discuss the symmetric case (recall from \S\ref{s1.2.1} that this corresponds to taking $a=b$ and $\alpha\in\mathbb{R}$), where the necessary arguments are easier to follow, before outlining the (non-symmetric) general case.

\subsection{Relating the symmetric Cauchy and shifted Jacobi ensembles} \label{s2.2.1}
When $a=b$ and $\alpha\in\mathbb{R}$, the Cauchy and shifted Jacobi weights $w^{(Cy)}(\lambda)=(1+\lambda^2)^{\eta}$ and $w^{(sJ)}(\lambda)=(1-\lambda^2)^a\chi_{-1<\lambda<1}$ are even functions of $\lambda$ (if $\alpha:=-\kappa(N-1)-1-\eta$ is real, then so too is $\eta$), i.e., symmetric about $\lambda=0$. Integrations of multivariable symmetric functions against these weights can be related via the following proposition.
 
 \begin{proposition}\label{prop2.2}
 Let $f(x_1,\dots,x_N)$ be a multivariable symmetric polynomial of degree $d$ in each
 $x_i$. For $2 \eta < -(d +1)$, define
 \begin{equation}\label{eq2.2.4} 
I_{N,\eta}^{(Cy)}[f(x_1,\ldots,x_N)] := \int_{-\infty}^\infty \mathrm{d}x_1 \, (1 + x_1^2)^\eta \cdots \int_{-\infty}^\infty \mathrm{d}x_N \, (1 + x_N^2)^\eta \,
 f(x_1,\dots,x_N),
 \end{equation}
 and for $\eta$ outside of this range, define $ I_{N,\eta}^{(Cy)}[f(x_1,\ldots,x_N)] $ by its analytic continuation. Also,
 in relation to the shifted Jacobi weight with $a=b > -1$, define
  \begin{equation}\label{eq2.2.5} 
I_{N,a}^{(sJ)}[f(x_1,\ldots,x_N)] := \int_{-1}^1 \mathrm{d}x_1 \, (1 - x_1^2)^a \cdots \int_{-1}^1 \mathrm{d}x_N \, (1 - x_N^2)^a \,
 f(x_1,\dots,x_N),
 \end{equation} 
 and for $a$ outside of this range, define $ I_{N,a}^{(J)}[f(x_1,\ldots,x_N)]$ by its analytic continuation.
 We have
   \begin{equation}\label{eq2.2.6} 
  I_{N,\eta}^{(Cy)}[f(\mathrm{i}x_1,\dots \mathrm{i}x_N)] = (\tan \pi \eta)^N    I_{N,\eta}^{(sJ)}[f(x_1,\dots  x_N)].
  \end{equation} 
  \end{proposition}
  
  \begin{proof}
  Let $p \in \mathbb Z_{\ge 0}$. Suppose $2 \eta < - (p+1)$ and $a>-1$. A simple
  change of variables (for $k$ even)
  and use of the Euler beta integral evaluation (see \citep[Exercises 5.4, q.2]{Fo10})  shows that
  \begin{equation}\label{eq2.2.7}   
  \int_{-\infty}^\infty  (1 + x^2)^\eta x^{k} \, \mathrm{d}x =
  \begin{cases} 0, & k \: {\rm odd}, \\
  (-1)^{k/2} \tan \pi \eta \, {\Gamma(1 + \eta) \Gamma((k+1)/2) \over
  \Gamma((k+3)/2 + \eta)}, & k \: {\rm even},
  \end{cases}
   \end{equation} 
  and
  \begin{equation}\label{eq2.2.8}   
   \int_{-1}^1  (1 - x^2)^a x^{k} \, \mathrm{d}x =
  \begin{cases} 0, & k \: {\rm odd}, \\
   \, {\Gamma(1 + a) \Gamma((k+1)/2) \over
  \Gamma((k+3)/2 + a)}, & k \: {\rm even}.
  \end{cases}
   \end{equation}  
 
 The functions $f(x_1,\ldots,x_N)$ and $f(\mathrm{i}x_1,\ldots,\mathrm{i}x_N)$ are polynomials, so the computation of $ I_{N,\eta}^{(Cy)}[f(\mathrm{i}x_1,\ldots,\mathrm{i}x_N)] $
 and $  I_{N,a}^{(sJ)}[f(x_1,\ldots,x_N)] $ reduces to the above one-dimensional integrals.
 Since, as analytic functions of $\eta$, we read off from the respective evaluations that
 $$
 \int_{-\infty}^\infty  (1 + x^2)^\eta (\mathrm{i}x)^{2k} \, \mathrm{d}x =    \tan \pi \eta   \int_{-1}^1  (1 - x^2)^\eta x^{2k} \, \mathrm{d}x, 
 $$
 the stated result \eqref{eq2.2.6} follows
 \end{proof}
 
 One immediate consequence is a proof for the value of $\mathcal{N}_{N,\beta}^{(Cy)}$ given in equation \eqref{eq2.1.14}, in the symmetric case with $\kappa=\beta/2\in\mathbb{N}$.
 
 \begin{corollary}\label{C2.1}
 Let $\alpha$ be real and let $\eta=-\kappa(N-1)-1-\alpha$. For $\beta$ even, we have
 \begin{equation}\label{eq2.2.9}
(-1)^{\kappa N (N - 1) / 2} \mathcal N_{N,\beta}^{(Cy)} = ( -\tan  \pi \alpha )^N  \mathcal N_{N,\beta}^{(sJ)} \Big |_{a = b = - \kappa (N - 1) - 1 - \alpha},
 \end{equation}
 where both sides are to be interpreted as analytic functions in $\alpha$.
 \end{corollary}
 
 \begin{proof}
 With $\alpha$ and thus $\eta$ real, we have
 $$
 \mathcal N_{N,\beta}^{(Cy)} = \int_{-\infty}^\infty \mathrm{d}x_1 (1 + x_1^2)^{\eta} \cdots
   \int_{-\infty}^\infty \mathrm{d}x_N (1 + x_N^2)^{\eta} \prod_{1 \leq j < k \leq N} | x_k - x_j|^\beta.
   $$
   Also, with $a=b$, we have
   $$
 \mathcal N_{N,\beta}^{(sJ)} =    \int_{-1}^1 \mathrm{d}x_1 \, (1 - x_1^2)^a \cdots   \int_{-1}^1 \mathrm{d}x_N \, (1 - x_N^2)^a 
 \prod_{1 \leq j < k \leq N} | x_k - x_j|^\beta.
 $$
 For $\beta$ even, $ \prod_{1 \leq j < k \leq N} | x_k - x_j|^\beta$ is a multivariable symmetric polynomial, so
 application of Proposition~\ref{prop2.2} gives equation \eqref{eq2.2.9}.
 \end{proof}
 
 \begin{remark}
 The explicit form of the analytic continuations in $\alpha$ of both sides of equation \eqref{eq2.2.9}
 is known from equations \eqref{eq2.1.3}, \eqref{eq2.1.14}, and \eqref{eq2.2.2}. In the notation therein, the equality \eqref{eq2.2.9} requires
  \begin{multline}\label{eq2.2.10}
 S_N(-\kappa(N-1)-1-\alpha,
  -\kappa (N - 1) - 1 - \alpha, \kappa) 
\\= (-1)^{\kappa N(N-1)/2}(-2\pi\cot \pi \alpha)^{N} \prod_{j=0}^{N-1}\frac{\Gamma(2\alpha+1+\kappa j)\Gamma(1+\kappa(j+1))}{\Gamma(\alpha+1+\kappa j)^2\Gamma(1+\kappa)}.
  \end{multline}
  Under the assumption that $\beta$ is even, this can be checked upon the
  manipulation $j \mapsto N - 1 - j$ in the product defining $S_N$, and then use of the
  reflection equation for the gamma functions in that product. Agreement with the right-hand side of equation \eqref{eq2.2.10} is obtained.
   \end{remark}
   
   We can make use of Corollary \ref{C2.1} and further apply Proposition \ref{prop2.2}
   to relate the eigenvalue densities of the Cauchy and shifted Jacobi ensembles.
   
  \begin{proposition}\label{prop2.3} 
  In the setting of Corollary \ref{C2.1},
   \begin{equation}\label{eq2.2.11}
   \rho^{(Cy)}(\mathrm{i}\lambda;N,\beta) = -\cot \pi \alpha \,  \rho^{(sJ)}(\lambda;N,\beta)  \Big |_{a = b = - \kappa (N - 1) - 1 - \alpha}.
  \end{equation}
    \end{proposition}
    
    \begin{remark} \label{R2.4}
A straightforward extension of the above proposition is the relation
\begin{equation} \label{eq2.2.12}
\rho_k^{(Cy)}(\mathrm{i}\lambda_1,\ldots,\mathrm{i}\lambda_k;N,\beta) = (-\cot\pi\alpha)^k \rho_k^{(sJ)}(\lambda_1,\ldots,\lambda_k;N,\beta) \Big |_{a = b = - \kappa (N - 1) - 1 - \alpha}
\end{equation}
on the $k$-point correlation functions defined by equation \eqref{eq1.2.72}. A similar such extension holds true for Proposition \ref{prop2.5} forthcoming, as well.
\end{remark}
    \begin{remark} \label{R2.5}
    Although established for $\beta$ an even integer via reasoning based on  Proposition~\ref{prop2.2},
    an application of Carlson's theorem to the differences of the left- and right-hand sides of equations \eqref{eq2.2.9} and \eqref{eq2.2.11} interpreted as functions of $\kappa-1$ (see, e.g., \citep[Sec.~4.1]{Fo10}) shows that Corollary \ref{C2.1} and Proposition \ref{prop2.3} remain true for general $\beta > 0$; Carlson's theorem states that if a function $f(\kappa-1)$ is analytic for $\mathrm{Re}(\kappa)\geq1$, has the bound $|f(\kappa-1)|=\mathrm{O}(e^{\tau|\kappa-1|})$ for some $\tau<\pi$, and vanishes at all $\kappa\in\mathbb{N}$, then $f\equiv0$.
    \end{remark}
  
  For the particular values of $\beta$ even, $\beta = 2, 4$, we will use the identity \eqref{eq2.2.11} relating the eigenvalue density of the Cauchy ensemble for $\alpha$ real to (an analytic continuation of) the eigenvalue density of the shifted Jacobi ensemble supported on $(-1,1)$ with $a=b$ and the same value of $\beta$ to study properties of the former --- extension to $\beta=1$ is enabled through the above remark or Lemma \ref{L1.4}. This is presented in \S\ref{s2.3.2}, but first we outline a relationship between the Cauchy ensemble when ${\rm Im}(\alpha)\neq0$ (also known as the non-symmetric case or the generalised Cauchy ensemble) and the shifted Jacobi ensemble now requiring $a=\overline{b}$.

\subsection{Relating the non-symmetric Cauchy and shifted Jacobi ensembles} \label{s2.2.2}
A classical result of Cauchy \citep{Ca74} gives
  \begin{equation}\label{eq2.2.13}
  \int_{-\infty}^\infty {\mathrm{d}t \over (1 - \mathrm{i} t)^\gamma (1 + \mathrm{i} t)^\delta} = 2^{2 - \gamma - \delta}\pi {\Gamma(\gamma+\delta-1) \over \Gamma(\gamma) \Gamma(\delta)}
   \end{equation}
   subject to the requirement that ${\rm Re}(\gamma+\delta) > 1$; outside of this range we consider the integral as
   defined by the analytic continuation given by the right-hand side. Use of the reflection equation for the
   gamma function allows this to be rewritten
  \begin{equation}\label{eq2.2.14} 
    \int_{-\infty}^\infty (1 - \mathrm{i}t)^\gamma (1 + \mathrm{i}t)^\delta \, \mathrm{d}t =    2^{\gamma+\delta+2} {\sin \pi \gamma \,\sin \pi \delta \over
    \sin \pi (\gamma+\delta) } {\Gamma(\gamma+1) \Gamma(\delta + 1) \over \Gamma (\gamma+\delta + 2)},
   \end{equation}
   subject now to the requirement  ${\rm Re}(\gamma+\delta) < - 1$ on the left-hand side.
   
   The form \eqref{eq2.2.14} is to be compared against the Euler beta function evaluation
   \begin{equation}\label{eq2.2.15} 
   \int_0^1 t^c ( 1 - t)^d \, \mathrm{d}t = {\Gamma(c+1) \Gamma(d+1) \over \Gamma(c + d + 2)}
     \end{equation}
     or, equivalently,
     \begin{equation}\label{eq2.2.16}
  \int_{-1}^1 (1-t)^c ( 1 + t)^d \, \mathrm{d}t =    2^{c+d+1}    {\Gamma(c+1) \Gamma(d+1) \over \Gamma(c + d + 2)} ,  
   \end{equation}    
     where on the left-hand side, it is required that ${\rm Re}(c),{\rm Re}(d) > -1$.
     The agreement in the gamma function dependence of both integrals allows for a relation between
     multiple integrals analogous to that in Proposition \ref{prop2.2} to be derived.
   
     \begin{proposition}\label{prop2.4} 
     Let $f(x_1,\dots,x_N)$ be a multivariable symmetric polynomial of degree $\tilde{d}$ in each
 $x_i$. For ${\rm Re}(\gamma+\delta) < - \tilde{d} - 1$, define
 \begin{multline}\label{eq2.2.17} 
\tilde{I}_{N,\gamma,\delta}^{(Cy)}[f(x_1,\ldots,x_N)] := \int_{-\infty}^\infty \mathrm{d}x_1 \, (1 -  \mathrm{i} x_1)^\gamma ( 1 + \mathrm{i} x_1)^\delta  \cdots
\\ \cdots \int_{-\infty}^\infty \mathrm{d}x_N \,   (1 -  \mathrm{i} x_N)^\gamma ( 1 + \mathrm{i} x_N)^\delta\,
 f(x_1,\dots,x_N),
 \end{multline}
 and for $\gamma,\delta$ outside of this range, define $ \tilde{I}_{N,\gamma,\delta}^{(Cy)}[f(x_1,\ldots,x_N)] $ by its analytic continuation. Also,
 in relation to the shifted Jacobi weight with ${\rm Re}(c),{\rm Re}(d) > -1$, define
  \begin{multline}\label{eq2.2.18} 
\tilde{I}_{N,c,d}^{(sJ)}[f(x_1,\ldots,x_N)] := \int_{-1}^1 \mathrm{d}x_1 \, (1-x_1)^c (1 + x_1)^d \cdots
\\ \cdots \int_{-1}^1 \mathrm{d}x_N \,  (1 - x_N)^c (1 + x_N)^d \,
 f(x_1,\dots,x_N),
 \end{multline} 
 and for $c,d$ outside of this range, define $ \tilde{I}_{N,c,d}^{(sJ)}[f(x_1,\ldots,x_N)]$ by its analytic continuation.
 We have
   \begin{equation}\label{eq2.2.19} 
  \tilde{I}_{N,\gamma,\delta}^{(Cy)}[f(1-\mathrm{i}x_1,\dots ,1-\mathrm{i} x_N )] = \left ( 2 {\sin \pi \gamma\, \sin \pi \delta \over
    \sin \pi (\gamma+\delta) }  \right )^N
     \tilde{I}_{N,\gamma,\delta}^{(sJ)}[f(1-x_1,\dots  1-x_N)].
  \end{equation} 
  \end{proposition}

  \begin{proof}
  For $p \in \mathbb Z_{\ge 0}$, it follows from equations \eqref{eq2.2.14} and \eqref{eq2.2.16} upon setting $c=\gamma$ and $d=\delta$ that
   \begin{equation}\label{eq2.2.20} 
    \int_{-\infty}^\infty (1 - \mathrm{i}t)^{\gamma+p} (1 + \mathrm{i}t)^\delta \, \mathrm{d}t =    2^{\gamma+\delta+p+2} {\sin \pi \gamma\, \sin \pi \delta \over
    \sin \pi (\gamma+\delta) } {\Gamma(\gamma+p+1) \Gamma(\delta + 1) \over \Gamma (\gamma+\delta+p + 2)},
   \end{equation}
  and
  \begin{equation}\label{eq2.2.21} 
   \int_{-1}^1 (1 - t)^{\gamma+p} ( 1 + t)^\delta \, \mathrm{d}t =  2^{\gamma+\delta+p+1}  {\Gamma(\gamma+p+1) \Gamma(\delta+1) \over \Gamma(\gamma+\delta+p + 2)}.
     \end{equation}
 Hence, in the sense of analytic continuation,
   \begin{equation}\label{eq2.2.22}     
     \int_{-\infty}^\infty (1 - \mathrm{i}t)^{\gamma+p} (1 + \mathrm{i}t)^\delta \, \mathrm{d}t =     2 {\sin \pi \gamma\, \sin \pi \delta \over
    \sin \pi (\gamma+\delta) }   \int_{-1}^1 (1-t)^{\gamma+p} ( 1 + t)^\delta \, \mathrm{d}t.
  \end{equation}    
     
     The stated result now follows from the assumption that $f(x_1,\ldots,x_N)$ in \eqref{eq2.2.17} and \eqref{eq2.2.18}
     is a polynomial and so the evaluation of the multiple integrals reduces to the one-dimensional
     integrals \eqref{eq2.2.20} and \eqref{eq2.2.21}, which are related by equation \eqref{eq2.2.22}.
     
     \end{proof}
     
We can use Proposition \ref{prop2.4} to extend Corollary \ref{C2.1} to the case that $\alpha$, hence $\eta$, are complex.
   \begin{corollary}\label{C2.2} 
   Let $\alpha$ be, in general, complex and recall that we choose $\eta=-\kappa(N-1)-1-\alpha$. For $\kappa=\beta/2\in\mathbb{N}$,
   \begin{equation}\label{eq2.2.23}
(-1)^{\kappa N (N - 1) / 2} \mathcal N_{N,\beta}^{(Cy)} = \left( -2{ \sin  \pi \alpha \,  \sin  \pi \overline{\alpha} \over \sin \pi (\alpha + \overline{\alpha}) }  \right)^N  \mathcal N_{N,\beta}^{(sJ)} \Big |_{\overline{a}= b = - \kappa (N - 1) - 1 - \alpha},
 \end{equation}
 where the right-hand side is to be regarded as defined by its analytic continuation.
  \end{corollary}    
  
  \begin{proof}
  We observe that for $\beta$ even, the product of differences $|\Delta_N(\lambda)|^{\beta}$ in the definition of the
  normalisation constants,
\begin{equation*}
\mathcal{N}_{N,\beta}=\int_{\mathbb{R}^N}p(\lambda_1,\ldots,\lambda_N)\,\mathrm{d}\lambda_1\cdots\mathrm{d}\lambda_N,
\end{equation*}
is a polynomial, and moreover,
  \begin{equation*}
  (\lambda_k - \lambda_j)^\beta =  (-1)^{\kappa}  ((1 - \mathrm{i}\lambda_k) - (1 - \mathrm{i} \lambda_j))^\beta.
  \end{equation*}
  The result now follows from the definition of the normalisation constants
  and the identity \eqref{eq2.2.19}.
  \end{proof}
  
  \begin{remark}
\begin{enumerate}
\item Analogous to equation \eqref{eq2.2.10}, the identity \eqref{eq2.2.23} can be rewritten as
    \begin{multline}\label{eq2.2.24}
 S_N(-\kappa(N-1)-1-\alpha,
  -\kappa (N - 1) - 1 - \overline{\alpha}, \kappa) 
\\= (-1)^{\kappa N(N-1)/2}\left(-\pi\frac{\sin\,\pi(\alpha+\overline{\alpha})}{\sin\,\pi\alpha\,\sin\,\pi\overline{\alpha}}\right)^{N} \prod_{j=0}^{N-1}\frac{\Gamma(2\mathrm{Re}(\alpha)+1+\kappa j)\Gamma(1+\kappa(j+1))}{\Gamma(\alpha+1+\kappa j)\Gamma(\overline{\alpha}+1+\kappa j)\Gamma(1+\kappa)}.
  \end{multline}
 This can be verified using the same steps as for equation \eqref{eq2.2.10}, and is known to be true for $\beta>0$ due yet again to Carlson's theorem (cf.~Remark \ref{R2.5}).
\item In the case $\alpha$ is real, identity \eqref{eq2.2.23} reduces to \eqref{eq2.2.9}.
\end{enumerate}
 \end{remark} 
 
 We can make use of  Corollary \ref{C2.2} and a further application of
 Proposition \ref{prop2.4} in the specification \eqref{eq2.0.1} of the eigenvalue densities $\rho^{(Cy)}(\lambda)$ and $\rho^{(sJ)}(\lambda)$ to deduce the analogue of Proposition
  \ref{prop2.3} in the non-symmetric case.
   
  \begin{proposition}\label{prop2.5} 
  In the setting of Corollary \ref{C2.2},
   \begin{equation}\label{eq2.2.25}
   \rho^{(Cy)}(\mathrm{i}\lambda;N,\beta) = -{\sin \pi (\alpha + \overline{\alpha}) \over 2 \sin \pi \alpha\, \sin \pi \overline{\alpha}}  \,  \rho^{(sJ)}(\lambda;N,\beta)  \Big |_{\overline{a} = b=- \kappa (N - 1) - 1 - \alpha}.
  \end{equation}
    \end{proposition}
    
    \begin{remark}
\begin{enumerate}
\item In the case $\alpha$ is real, equation \eqref{eq2.2.25} reduces to \eqref{eq2.2.11}.
\item As discussed in \S\ref{s1.2.3}, there exist expressions in terms of orthogonal polynomials
  for both sides of equation \eqref{eq2.2.25} when $\beta = 1,2$, or $4$. The expressions for the right-hand side can be found in \citep{AFNM00}, and the left-hand side in \citep{FLT20}. These expressions can be checked
  to be consistent with equation \eqref{eq2.2.25}, using the fact that the orthogonal polynomials
  associated with the shifted Jacobi weight are the Jacobi polynomials $P_N^{(a,b)}(\lambda)$,
  while those associated with the Cauchy weight are the scaled
  Jacobi polynomials $\mathrm{i}^{-N}P_N^{(\eta,\overline{\eta})}(\mathrm{i}\lambda)$.
\end{enumerate}
  \end{remark}

\setcounter{equation}{0}
\section{Linear Differential Equations for the Eigenvalue Densities and Resolvents} \label{s2.3}
The Selberg integral theory in \S\ref{s2.1.2} allows us to derive order $\beta+1$ linear differential equations satisfied by the eigenvalue densities \eqref{eq2.0.1} of the Jacobi $\beta$ ensembles with $\beta=2,4$; the $\beta\leftrightarrow4/\beta$ duality given in Lemma \ref{L1.4} then extends our results to $\beta=1$. Analogous to Proposition \ref{prop2.1}, these differential equations are homogeneous, and they have inhomogeneous counterparts that are solved by the corresponding resolvents (recall that the resolvents have multiple valid definitions, as introduced in \S\ref{s1.1.1}).

After presenting the aforementioned differential equations for the Jacobi ensembles in \S\ref{s2.3.1}, we use the theory of Section \ref{s2.2} to translate our results to the shifted Jacobi and Cauchy ensembles. We treat the $\beta=2$ case in full generality and report on the $\beta=1,4$ cases only in the symmetric case, since the non-symmetric $\beta=1,4$ differential equations have cumbersome presentations with little expository benefit. The differential equations are all of the same orders and simplify greatly in the symmetric case. They are given in \S\ref{s2.3.2}, along with brief discussions regarding connections to the circular Jacobi ensemble.

Extending the limiting procedures in Lemma \ref{L1.1}, in a similar fashion to what was seen for the normalisation constants at the end of \S\ref{s2.1.1}, the Jacobi ensembles' differential equations transform into differential equations satisfied by the Laguerre ensembles' eigenvalue densities and resolvents. We again treat $\beta=1,2$, and $4$, and the differential equations are of the same orders, but with simpler coefficients due to the loss of parameter $b$. Repeating this exercise to scale out both the parameters $a$ and $b$ results in the even simpler differential equations for the Gaussian ensembles displayed in Proposition \ref{prop2.1}. Rather than outlining this calculation, we demonstrate an alternate extension of Lemma \ref{L1.1} that allows us to derive differential equations for the Gaussian ensembles without needing to first make the Jacobi ensemble analogues explicit. Indeed, in \S\ref{s2.3.4}, we derive seventh order linear differential equations that characterise the eigenvalue densities and resolvents of the $\beta=2/3$ and $\beta=6$ Gaussian $\beta$ ensembles, without any knowledge of the Jacobi $\beta$ ensembles for these values of $\beta$. These differential equations serve as a proof of concept: Our method applies for positive integer $\beta$ outside of the classical regime $\beta=1,2,4$, and also allows us to access non-integer $\beta$ values.

As discussed in Section \ref{s1.4}, the interest in the upcoming differential equations is that comparable and/or equivalent (in the GOE, GSE, and classical unitary cases) results in the works \citep{LM79}, \citep{GT05}, \citep{GKR05}, \citep{ATK11}, \citep{WF14}, \citep{Na18}, \citep{ABGS20} have been shown therein and also in, e.g., \citep{Le04}, \citep{Ko18} to be amenable to various forms of analysis. The differential equations of this section will be given two applications in this thesis: In Section \ref{s2.4}, we apply global and edge scalings to the differential equations to give characterisations of the relevant ensembles at these scaling limits, while in Section \ref{s3.1}, we present linear recurrences for the spectral moments $m_k$ of the classical matrix ensembles, supplementing the works \citep{HZ86}, \citep{Le09}, \citep{CMOS19}, among others. Another appeal of our derivations below is that they treat all of the classical matrix ensembles through a unified method, in contrast to the aforementioned works whose results we recover.

\subsection{Differential equations for the Jacobi ensembles} \label{s2.3.1}
Recall from \S\ref{s2.1.2} that the eigenvalue density $\rho^{(J)}(\lambda)$ is a scalar multiple of $w^{(J)}(\lambda)$ times the function $I_{N,\beta}^{(J)}(\lambda)$ defined in equation \eqref{eq2.1.15}. When $\beta$ is an even integer, this latter function is moreover related, via equation \eqref{eq2.1.24}, to the auxiliary function $J_{\beta,0}^{(N)}(\lambda)$ defined in equation \eqref{eq2.1.16}. Using the differential-difference equation \eqref{eq2.1.17} for the $J_{\beta,p}^{(N)}(\lambda)$, now with $p=0,1,\ldots,\beta$, we derive a differential equation for $J_{\beta,0}^{(N)}(\lambda)$. Applying straightforward manipulations according to equations \eqref{eq2.1.15} and \eqref{eq2.1.17} then yields the sought differential equations for $\rho^{(J)}(\lambda;N,\beta)$ with $\beta\in2\mathbb{N}$.

\begin{lemma} \label{L2.1}
The function $J_{2,0}^{(N)}(x)$ \eqref{eq2.1.16} satisfies the differential equation
\begin{multline} \label{eq2.3.1}
0=x^2(x-1)^2\frac{\mathrm{d}^3}{\mathrm{d}x^3}J_{2,0}^{(N)}(x)-\left[3C_2(x)-2(1-2x)\right]x(x-1)\frac{\mathrm{d}^2}{\mathrm{d}x^2}J_{2,0}^{(N)}(x)
\\+\left[\left((a+N)(1+4N)+4+N\right)x(x-1)+\left(2C_2(x)-3(1-2x)\right)C_2(x)\right]\frac{\mathrm{d}}{\mathrm{d}x}J_{2,0}^{(N)}(x)
\\-2N(a+N)\left[2C_2(x)-3(1-2x)\right]J_{2,0}^{(N)}(x),
\end{multline}
where $C_2(x)=(a+2N)(x-1)-b-2x$.
\end{lemma}
\begin{proof}
Setting $n=2$ and taking $p=0,1,2$ in equation \eqref{eq2.1.17}, we obtain the matrix differential equation
\begin{align} \label{eq2.3.2}
\frac{\mathrm{d}}{\mathrm{d}x}\begin{bmatrix}J_{2,0}^{(N)}(x)\\J_{2,1}^{(N)}(x)\\J_{2,2}^{(N)}(x)\end{bmatrix}=\begin{bmatrix}\frac{A_0x+B_0}{x(x-1)}&\frac{2E_0}{x(x-1)}&0\\-D_1&\frac{A_1x+B_1}{x(x-1)}&\frac{E_1}{x(x-1)}\\0&-D_2&0 \end{bmatrix}\begin{bmatrix}J_{2,0}^{(N)}(x)\\J_{2,1}^{(N)}(x)\\J_{2,2}^{(N)}(x)\end{bmatrix}.
\end{align}
The second row gives an expression for $J_{2,2}^{(N)}(x)$ which transforms the third row into a differential equation involving only $J_{2,0}^{(N)}(x)$ and $J_{2,1}^{(N)}(x)$. This equation further transforms into an equation for just $J_{2,0}^{(N)}(x)$ upon substitution of the expression for $J_{2,1}^{(N)}(x)$ drawn from the first row:
\begin{align}
0&=x^2(x-1)^2\frac{\mathrm{d}^3}{\mathrm{d}x^3}J_{2,0}^{(N)}(x)-\left[\tilde{C}_{2,0}(x)+\tilde{C}_{2,1}(x)+1-2x\right]x(x-1)\frac{\mathrm{d}^2}{\mathrm{d}x^2}J_{2,0}^{(N)}(x) \nonumber
\\&\quad+\left[(D_2E_1+2D_1E_0-A_1-2A_0+2)x(x-1)+\tilde{C}_{2,0}(x)\tilde{C}_{2,1}(x)\right]\frac{\mathrm{d}}{\mathrm{d}x}J_{2,0}^{(N)}(x) \nonumber
\\&\quad+\left[A_0\tilde{C}_{2,1}(x)+(A_1-D_2E_1)(A_0x+B_0)-2D_1E_0(1-2x)\right]J_{2,0}^{(N)}(x), \label{eq2.3.3}
\end{align}
where $\tilde{C}_{2,p}(x)=(A_p-2)x+B_p+1$ with $n=\beta=2$. Substituting the appropriate values for the constants $A_p,B_p,D_p$ and $E_p$ gives the claimed result.
\end{proof}

\begin{lemma} \label{L2.2}
Now taking $n=\beta=4$, the function $J_{4,0}^{(N)}(x)$ satisfies the differential equation with polynomial coefficients
\begin{align} \label{eq2.3.4}
0&=4x^4(x-1)^4\frac{\mathrm{d}^5}{\mathrm{d}x^5}J_{4,0}^{(N)}(x)- 20\left[(a+4N)(x-1)-b-2x\right] x^3(x-1)^3\frac{\mathrm{d}^4}{\mathrm{d}x^4}J_{4,0}^{(N)}(x) \nonumber
\\&\quad+\left[5(a+4N)^2-5a(a-2)-12\right] x^3(x-1)^3\frac{\mathrm{d}^3}{\mathrm{d}x^3}J_{4,0}^{(N)}(x) + \cdots
\end{align}
where the (lengthier) specific forms of the coefficients of the lower order derivatives have been suppressed. (One may obtain the full expression by inverting the upcoming proof of Theorem \ref{thrm2.2}. In fact, this is the most efficient method of obtaining the full expression using computer algebra.)
\end{lemma}
\begin{proof}
Like the preceding proof, setting $n=4$ in equation \eqref{eq2.1.17} and taking $p=0,1,\ldots,4$ yields the matrix differential equation
\begin{align} \label{eq2.3.5}
\frac{\mathrm{d}}{\mathrm{d}x}\begin{bmatrix}J_{4,0}^{(N)}(x)\\J_{4,1}^{(N)}(x)\\J_{4,2}^{(N)}(x)\\J_{4,3}^{(N)}(x)\\J_{4,4}^{(N)}(x)\end{bmatrix}=\begin{bmatrix}\frac{A_0x+B_0}{x(x-1)}&\frac{4E_0}{x(x-1)}&0&0&0\\-D_1&\frac{A_1x+B_1}{x(x-1)}&\frac{3E_1}{x(x-1)}&0&0\\0&-D_2&\frac{A_2x+B_2}{x(x-1)}&\frac{2E_2}{x(x-1)}&0\\0&0&-D_3&\frac{A_3x+B_3}{x(x-1)}&\frac{E_3}{x(x-1)}\\0&0&0&-D_4&0\end{bmatrix}\begin{bmatrix}J_{4,0}^{(N)}(x)\\J_{4,1}^{(N)}(x)\\J_{4,2}^{(N)}(x)\\J_{4,3}^{(N)}(x)\\J_{4,4}^{(N)}(x)\end{bmatrix}.
\end{align}
For $1\leq p\leq4$, the $p\textsuperscript{th}$ row gives an expression for $J_{4,p}^{(N)}(x)$ in terms of $\frac{\mathrm{d}}{\mathrm{d}x}J_{4,p-1}^{(N)}(x)$ and $J_{4,k}^{(N)}(x)$ with $k<p$. Substituting these expressions (in the order of decreasing $p$) into the differential equation corresponding to the fifth row yields a fifth order differential equation for $J_{4,0}^{(N)}(x)$ similar to that of Lemma \ref{L2.1}.
\end{proof}

Now, we may easily obtain differential equations for $I_{N,\beta}^{(J)}(\lambda)$ for $\beta=2$ and $4$, which, in turn, give us differential equations for $\rho^{(J)}(\lambda;N,\beta)$ for $\beta=1,2,4$.
\begin{theorem} \label{thrm2.1}
Define
\begin{align} \label{eq2.3.6}
\mathcal{D}_{N,2}^{(J)}&=x^3(1-x)^3\frac{\mathrm{d}^3}{\mathrm{d}x^3}+4(1-2x)x^2(1-x)^2\frac{\mathrm{d}^2}{\mathrm{d}x^2} \nonumber
\\&\quad+\left[(a+b+2N)^2-14\right]x^2(1-x)^2\frac{\mathrm{d}}{\mathrm{d}x}-\left[a^2(1-x)+b^2x-2\right]x(1-x)\frac{\mathrm{d}}{\mathrm{d}x} \nonumber
\\&\quad+\tfrac{1}{2}\left[(a+b+2N)^2-4\right](1-2x)x(1-x)+\tfrac{3}{2}\left[a^2-b^2\right]x(1-x) \nonumber
\\&\quad-a^2(1-x)+b^2x.
\end{align}
Then,
\begin{align} \label{eq2.3.7}
\mathcal{D}_{N,2}^{(J)}\,\rho^{(J)}(x;N,2)=0
\end{align}
and
\begin{align} \label{eq2.3.8}
\mathcal{D}_{N,2}^{(J)}\,\frac{1}{N}W_1^{(J)}(x;N,2)=(a+b+N)(a(1-x)+bx).
\end{align}
\end{theorem}

\begin{proof}
We change variables $x\mapsto1/x$ in equation \eqref{eq2.3.1} to obtain a differential equation for $J_{2,0}^{(N)}(1/x)$ using the fact that $\frac{\mathrm{d}}{\mathrm{d}(1/x)}=-x^2\frac{\mathrm{d}}{\mathrm{d}x}$. Since this differential equation is homogeneous, we can ignore constants of proportionality and substitute in
\begin{align*}
J_{2,0}^{(N)}(1/x)=x^{-a-2N}(1-x)^{-b}\rho^{(J)}(x;N+1,2)
\end{align*}
according to equations \eqref{eq2.1.15} and \eqref{eq2.1.24}. Repeatedly applying the product rule and then replacing $N+1$ by $N$ gives equation \eqref{eq2.3.7}. Applying the Stieltjes transform to this equation term by term (see Appendix \ref{appendixB}) and substituting in the values of the spectral moments $m_1^{(J)}$ and $m_2^{(J)}$ with $\beta=2$ \citep{MRW17} yields equation \eqref{eq2.3.8}.
\end{proof}
Differential equation \eqref{eq2.3.7} lowers the order by one relative to the fourth order differential equation
for $\rho^{(J)}(\lambda;N,2)$ given in the recent work \citep[Prop.~6.7]{CMOS19}.

\begin{remark} \label{R2.7}
With $1\leq k\leq N$, let $\rho_k^{(w)}(\lambda_1,\ldots,\lambda_N)$ denote the $k$-point correlation function of the random matrix ensemble corresponding to the classical weight $w(\lambda)$ \eqref{eq1.2.9}, as specified by equation \eqref{eq1.2.72}. It is well known (see, e.g., \citep[Ch.~9]{Fo10}) that the generating function $E_N^{(w)}((s,\infty);\xi)$ for the probabilities $\{E_{N,k}^{(w)}(s,\infty)\}_{k=0}^N$ of there being exactly $k$ eigenvalues in the interval $(s,\infty)$ can be written in terms of the correlation functions according to
\begin{equation} \label{eq2.3.9}
E_N^{(w)}((s,\infty);\xi)=1+\sum_{k=1}^N\frac{(-\xi)^k}{k!}\int_s^{\infty}\mathrm{d}x_1\cdots\int_s^{\infty}\mathrm{d}x_k\,\rho_k^{(w)}(x_1,\ldots,x_k).
\end{equation}
The logarithmic derivative of these generating functions multiplied by simple polynomials dependent on the weight $w(\lambda)$ are known to be characterised by particular $\sigma$ Painlev\'e equations \citep{TW94}, \citep{FW00}, \citep{WF00}, \citep{FW02}, \citep{FW04}. It has been observed in \citep[\S 3.3]{FT18} that the third order linear differential equations \eqref{eq2.0.3} and \eqref{eq2.3.43} for the eigenvalue densities of the Gaussian and Laguerre unitary ensembles are equivalent to related $\sigma$ Painlev\'e equations (see, e.g., \citep[Ch.~8]{Fo10}). An analogous result holds true for the differential equation \eqref{eq2.3.7} for $\rho^{(J)}(x;N,2)$. Thus, with
$v_1 = v_3 = N + (a+b)/2$, $v_2 = (a+b)/2$, $v_4 = (b-a)/2$, in studying the gap for the interval $(0,s)$ one encounters the nonlinear equation \citep[Eq.~(8.76)]{Fo10}
\begin{eqnarray*}
&&
(t(1-t)f'')^2 - 4t(1-t)(f')^3 + 4(1-2t)(f')^2f + 4f'f^2 - 4 f^2 v_1^2
\nonumber \\
&&
\quad + (f')^2\Big ( 4tv_1^2(1-t) - (v_2 - v_4)^2 - 4tv_2 v_4 \Big )
+ 4ff'(-v_1^2+2tv_1^2+v_2v_4)  = 0,
\end{eqnarray*}
subject to the boundary condition
$
f(t) \underset{t\rightarrow0^+}{\sim} - \xi t(1-t) \rho^{(J)}(t;N,2)$. Substituting this boundary condition for $f$ and equating terms of order $\xi^2$ shows that $u(t):=  t(1-t) \rho^{(J)}(t;N,2)$ satisfies the
second order nonlinear differential equation
\begin{multline} \label{eq2.3.10}
(t(1-t) u''(t))^2 - 4 (u(t))^2 v_1^2 + (u'(t))^2 (4tv_1^2 (1 - t) - (v_2 - v_4)^2 - 4 t v_2 v_4)  \\
+ 4 u(t) u'(t) (- v_1^2 + 2 t v_1^2 + v_2 v_4) = 0.
\end{multline}
Differentiating this and simplifying gives a third order linear differential equation that agrees with equation \eqref{eq2.3.7}.
\end{remark}

\begin{theorem} \label{thrm2.2}
Recalling that $\kappa:=\beta/2$, let
\begin{align*}
a_{\beta}:=\frac{a}{\kappa-1},\quad b_{\beta}:=\frac{b}{\kappa-1},\quad N_{\beta}:=(\kappa-1)N
\end{align*}
so that $(a_{4},b_{4},N_{4})=(a,b,N)$ and $(a_{1},b_{1},N_{1})=(-2a,-2b,-N/2)$. For $\beta=1$ or $4$, define
\begin{align}
\mathcal{D}_{N,\beta}^{(J)}&=4x^5(1-x)^5\frac{\mathrm{d}^5}{\mathrm{d}x^5}+40(1-2x)x^4(1-x)^4\frac{\mathrm{d}^4}{\mathrm{d}x^4}+\left[5\tilde{c}^2-493\right]x^4(1-x)^4\frac{\mathrm{d}^3}{\mathrm{d}x^3}\nonumber
\\&\quad-\left[5f_+(x;\tilde{a},\tilde{b})-88\right]x^3(1-x)^3\frac{\mathrm{d}^3}{\mathrm{d}x^3}+41\left[\tilde{a}-\tilde{b}\right]x^3(1-x)^3\frac{\mathrm{d}^2}{\mathrm{d}x^2}\nonumber
\\&\quad+\left[19\tilde{c}^2-539\right](1-2x)x^3(1-x)^3\frac{\mathrm{d}^2}{\mathrm{d}x^2}-22f_-(x;\tilde{a},\tilde{b})x^2(1-x)^2\frac{\mathrm{d}^2}{\mathrm{d}x^2}\nonumber
\\&\quad+16(1-2x)x^2(1-x)^2\frac{\mathrm{d}^2}{\mathrm{d}x^2}+\left[\tilde{c}^4-64\tilde{c}^2+719\right]x^3(1-x)^3\frac{\mathrm{d}}{\mathrm{d}x}\nonumber
\\&\quad-\left[(\tilde{c}^2-45)(\tilde{a}+\tilde{b}-6)+(\tilde{a}-\tilde{b})^2-248\right]x^2(1-x)^2\frac{\mathrm{d}}{\mathrm{d}x}\nonumber
\\&\quad-\left[(\tilde{c}^2-37)(\tilde{a}-\tilde{b})\right](1-2x)x^2(1-x)^2\frac{\mathrm{d}}{\mathrm{d}x}\nonumber
\\&\quad+\left[f_+(x;\tilde{a}^2,\tilde{b}^2)-14f_+(x;\tilde{a},\tilde{b})-16\right]x(1-x)\frac{\mathrm{d}}{\mathrm{d}x}\nonumber
\\&\quad+\tfrac{1}{2}\left[5(\tilde{c}^2-9)(\tilde{a}-\tilde{b})\right]x^2(1-x)^2+\tfrac{1}{2}(\tilde{c}^2-9)^2(1-2x)x^2(1-x)^2 \nonumber
\\&\quad-\tfrac{1}{2}\left[(3\tilde{c}^2-35)f_-(x;\tilde{a},\tilde{b})+\tfrac{7}{2}(\tilde{a}^2-\tilde{b}^2)+4(\tilde{a}-\tilde{b})\right]x(1-x) \nonumber
\\&\quad-\tfrac{1}{2}\left[4\tilde{c}^2-36+\tfrac{3}{2}(\tilde{a}-\tilde{b})^2\right](1-2x)x(1-x)+f_-(x;\tilde{a}^2,\tilde{b}^2), \label{eq2.3.11}
\end{align}
where
\begin{align}
\tilde{a}=a_{\beta}(a_{\beta}-2),\quad\tilde{b}&=b_{\beta}(b_{\beta}-2),\quad \tilde{c}=a_{\beta}+b_{\beta}+4N_{\beta}-1,\nonumber
\\f_{\pm}(x;\tilde{a},\tilde{b})&=\tilde{a}(1-x)\pm\tilde{b}x. \label{eq2.3.12}
\end{align}
Then, for $\beta=1$ and $4$,
\begin{equation} \label{eq2.3.13}
\mathcal{D}_{N,\beta}^{(J)}\,\rho^{(J)}(x;N,\beta)=0
\end{equation}
and
\begin{multline} \label{eq2.3.14}
\mathcal{D}_{N,\beta}^{(J)}\,\frac{1}{N}W_1^{(J)}(x;N,\beta)=(\tilde{c}-2N_{\beta})x(1-x)\Big[(a_{\beta}+b_{\beta})(a_{\beta}+b_{\beta}-2)(a_{\beta}(1-x)+b_{\beta}x)
\\+4N_{\beta}(\tilde{c}-2N_{\beta})(2a_{\beta}(1-x)+2b_{\beta}x-1)-a_{\beta}b_{\beta}(a_{\beta}+b_{\beta}-6)-4(a_{\beta}+b_{\beta}-1)\Big]
\\-(\tilde{c}-2N_{\beta})\left(a_{\beta}(a_{\beta}-2)^2(1-x)^2+b_{\beta}(b_{\beta}-2)^2x^2\right).
\end{multline}
\end{theorem}

\begin{proof}
The proof is done in four steps. First, to see that equation \eqref{eq2.3.13} holds for $\beta=4$, one undertakes the same steps as in the proof of Theorem \ref{thrm2.1}. That is, change variables $x\mapsto 1/x$ in equation \eqref{eq2.3.4}, substitute in
\begin{align*}
J_{4,0}^{(N)}(1/x)=x^{-a-4N}(1-x)^{-b}\rho^{(J)}(x;N+1,4),
\end{align*}
and then replace $N+1$ by $N$. For the second step, apply the Stieltjes transform according to Appendix \ref{appendixB} and substitute in the values of the spectral moments $m_1^{(J)}$ to $m_4^{(J)}$ \citep{MOPS}, \citep{MRW17} to obtain \eqref{eq2.3.14} for $\beta=4$. The third step proves equation \eqref{eq2.3.14} for $\beta=1$ by employing the $\beta\leftrightarrow4/\beta$ duality relation \eqref{eq1.2.98}, which leads us to formulate equation \eqref{eq2.3.14} in terms of the $(a_{\beta},b_{\beta},N_{\beta})$ parameters. Since this step essentially redefines constants, applying the inverse Stieltjes transform to this result returns equation \eqref{eq2.3.13} with the new constants.
\end{proof}

In the way that equation \eqref{eq2.3.11} is presented, all of its dependencies on $a,b$ and $N$ are captured by $\tilde{a},\tilde{b}$ and $\tilde{c}$. Moreover, it can be seen that this operator is invariant under the symmetry $(x,a,b)\leftrightarrow(1-x,b,a)$, which is a property of $\rho^{(J)}(x;N,\beta)$. Equations \eqref{eq2.3.7} and \eqref{eq2.3.13} have been checked for $N=1,2$ and to hold in the large $N$ limit, while \eqref{eq2.3.8} and \eqref{eq2.3.14} have been checked to be consistent with expressions for the resolvent expansion coefficients $W_1^{(J),0}(x;\beta),\ldots,W_1^{(J),4}(x;\beta)$ (recall Theorem \ref{thrm1.1} in \S\ref{s1.1.1}) generated via the loop equation analysis presented in \citep{FRW17} (see also \S\ref{s4.1.1} for a summary of the methods used in this work). It should be noted that while the moments $m_1^{(J)}$ to $m_4^{(J)}$ of the Jacobi $\beta$ ensemble's eigenvalue density are complicated rational functions of $a,b$, and $N$, the right-hand sides of equations \eqref{eq2.3.8} and \eqref{eq2.3.14} are relatively simple polynomials, even though they are linear combinations of these moments.

\subsection{Differential equations for the shifted Jacobi and Cauchy ensembles} \label{s2.3.2}
Upon making the simple change of variables $x\mapsto(1-x)/2$ in Theorems \ref{thrm2.1} and \ref{thrm2.2}, analogous differential equations for the shifted Jacobi ensembles can be obtained. The outcomes are not different in any essential way, so no insight is gained from explicitly presenting them. Instead, we further enforce the restriction $b=a$ so that significant simplification occurs. Then, the differential equations are for the eigenvalue densities and resolvents of the symmetric shifted Jacobi ensembles.

 \begin{proposition}\label{prop2.6}
 Define
    \begin{multline}\label{eq2.3.15} 
   \mathcal D_{N,2}^{(sJ)} = (1- x^2 )^3 {\mathrm{d}^3 \over \mathrm{d} x^3} - 8 x ( 1 - x^2 )^2 {\mathrm{d}^2 \over \mathrm{d} x^2} \\
   - 2 (1- x^2 ) [3 - 2N^2 - 4 a N + (2 (a + N)^2-7) x^2] {\mathrm{d} \over \mathrm{d}x} \\
   + 4 x ( a^2 + 1  - N^2 - 2 a N + (a + N)^2 x^2 - x^2)
   \end{multline}
   and for $\beta=1$ and $4$,
\begin{multline}\label{eq2.3.16} 
   \mathcal D_{N,\beta}^{(sJ)} = 4(1-x^2)^5{\mathrm{d}^5 \over \mathrm{d} x^5}-80x(1-x^2)^4{\mathrm{d}^4 \over \mathrm{d}x^4}+(5\tilde{c}^2-493)(1-x^2)^4{\mathrm{d}^3 \over \mathrm{d} x^3} \\-4(5\tilde{a}-88)(1-x^2)^3{\mathrm{d}^3 \over \mathrm{d} x^3}+16(11\tilde{a}-8)x(1-x^2)^2{\mathrm{d}^2 \over \mathrm{d} x^2}-2(19\tilde{c}^2-539)x(1-x^2)^3{\mathrm{d}^2 \over \mathrm{d} x^2} \\
+(\tilde{c}^4-64\tilde{c}^2+719)(1-x^2)^3{\mathrm{d}\over \mathrm{d}x}-8\left[(\tilde{c}^2-45)(\tilde{a}-3)-124\right](1-x^2)^2{\mathrm{d} \over\mathrm{d} x} \\
+16\left[(\tilde{a}-7)^2-65\right](1-x^2){\mathrm{d} \over \mathrm{d}x}-(\tilde{c}^2-9)^2x(1-x^2)^2 \\
+4\left[4(\tilde{c}^2-9)+(3\tilde{c}^2-35)\tilde{a}\right]x(1-x^2)-32\tilde{a}^2x,
   \end{multline}
taking $\tilde{a},\tilde{c}$ as defined in Theorem \ref{thrm2.2}, with $b=a$. Then, for $\beta=1,2$, and $4$, we have
\begin{equation} \label{eq2.3.17} 
\mathcal D_{N,\beta}^{(sJ)}\,\rho^{(J)}(x;N,\beta)|_{a=b}=0
\end{equation}
and
\begin{multline} \label{eq2.3.18}
\mathcal{D}^{(sJ)}_{N,\beta}\,\frac{1}{N}W_1^{(sJ)}(x;N,\beta)|_{a=b}
\\=\begin{cases} 4a(2a+N),&\beta=2, \\8(2N_{\beta}+2a_{\beta}-1)\left[2a_{\beta}^3x^2-2(2N_{\beta}+1)(N_{\beta}-1)(x^2-1)\right.&\\ \quad\left.+a_{\beta}^2((8N_{\beta}-3)x^2-8N_{\beta}-5)+8a_{\beta}(N_{\beta}(N_{\beta}-1)(x^2-1)+1)\right] ,&\beta=1,4,\end{cases}
\end{multline}
where $a_{\beta},N_{\beta}$ are also as in Theorem \ref{thrm2.2}.
   \end{proposition}

Differential equation \eqref{eq2.3.17} with $\beta=2$  is a generalisation of that given in \citep[Sec.~4]{GKR05} for the $\beta=2$ Legendre ensemble, which is the sJUE with $a=b=0$. According to Proposition~\ref{prop2.3} and calculations similar to those seen in Appendix \ref{appendixB}, the differential equations satisfied by $\rho^{(Cy)}(x)$ and $W_1^{(Cy)}(x)$ for $\beta$ even and $\alpha>-1/2$ real can be obtained from those satisfied by $\rho^{(sJ)}(x)|_{a=b}$ and $W_1^{(sJ)}(x)|_{a=b}$, respectively. This is done by setting $a=-\kappa(N-1)-1-\alpha$ and replacing $x$ by $\mathrm{i}x$. Doing so in Proposition \ref{prop2.6} gives third and fifth order linear differential equations for $\beta=2$ and $4$, respectively. Furthermore, Remark \ref{R2.5} tells us that the latter differential equations are valid for $\beta=1$ after reparametrising properly.

\begin{proposition}\label{prop2.7}
 Let $\alpha>-1/2$ be real so that we are restricted to the symmetric Cauchy case. Recalling the definition of $N_{\beta}$ in Theorem \ref{thrm2.2}, let
\begin{equation*}
\alpha_{\beta}:=\frac{\alpha}{\kappa-1},\quad\tilde{\alpha}:=\alpha_{\beta}(\alpha_{\beta}-1),\quad\tilde{N}:=(2N_{\beta}+\alpha_{\beta})^2.
\end{equation*}
Define
    \begin{multline}\label{eq2.3.19} 
   \mathcal D_{N,2}^{(Cy)} = (1 + x^2 )^3 {\mathrm{d}^3 \over\mathrm{d}x^3} + 8 x (1 + x^2 )^2 {\mathrm{d}^2 \over \mathrm{d}x^2}
   + 2 (1 + x^2 ) [3 + 2 N (N + 2\alpha) + (7 -2 \alpha^2) x^2] {\mathrm{d}\over \mathrm{d}x} \\
   + 4 x ( 1 + \alpha^2 + 2 N (N + 2 \alpha) + (1 - \alpha^2) x^2)
   \end{multline}
   and for $\beta=1$ and $4$,
\begin{multline}\label{eq2.3.20} 
   \mathcal D_{N,\beta}^{(Cy)} = 4(1+x^2)^5{\mathrm{d}^5 \over \mathrm{d} x^5}+80x(1+x^2)^4{\mathrm{d}^4 \over \mathrm{d} x^4}-4(5\tilde{\alpha}-122)(1+x^2)^4{\mathrm{d}^3 \over\mathrm{d} x^3} \\
+4(5\tilde{N}-93)(1+x^2)^3{\mathrm{d}^3 \over \mathrm{d}x^3}-8(19\tilde{\alpha}-130)x(1+x^2)^3{\mathrm{d}^2 \over \mathrm{d}^2 x}+16(11\tilde{N}-19)x(1+x^2)^2{\mathrm{d}^2 \over \mathrm{d}^2 x} \\
+8(2\tilde{\alpha}^2-31\tilde{\alpha}+82)(1+x^2)^3{\mathrm{d}\over \mathrm{d} x}-32\left[(\tilde{\alpha}-11)(\tilde{N}-4)-31\right](1+x^2)^2{\mathrm{d}\over \mathrm{d}x} \\
+16\left[(\tilde{N}-8)^2-65\right](1+x^2){\mathrm{d}\over\mathrm{d} x}+16(\tilde{\alpha}-2)^2x(1+x^2)^2 \\
-16\left[(3\tilde{\alpha}-8)\tilde{N}+\tilde{\alpha}\right]x(1+x^2)+32(\tilde{N}-1)^2x.
   \end{multline}
Then, for $\beta=1,2$, and $4$, with $\alpha>-1/2$ real, we have
\begin{equation} \label{eq2.3.21} 
\mathcal D_{\beta,N}^{(Cy)}\,\rho^{(Cy)}(x;N,\beta)=0
\end{equation}
and
\begin{multline} \label{eq2.3.22}
\mathcal{D}^{(Cy)}_{N,\beta}\,\frac{1}{N}W_1^{(Cy)}(x;N,\beta)
\\=\begin{cases} 4(N+\alpha)(N+2\alpha),&\beta=2, \\8(2N_{\beta}+2\alpha_{\beta}-1)\left[(2N_{\beta}+1)(8N_{\beta}^2+x^2-1)-2\alpha_{\beta}(1+x^2\alpha_{\beta}^2)\right.&\\ \quad\left.+4N_{\beta}\alpha_{\beta}(6N_{\beta}+x^2+3)+\alpha_{\beta}^2(3x^2-4N_{\beta}(x^2-2)+5)\right] ,&\beta=1,4.\end{cases}
\end{multline}
   \end{proposition}
Equation \eqref{eq2.3.21} with $\beta=2$ was recently derived in \citep{ABGS20} using a method of Ledoux \citep{Le04}.
\begin{remark} \label{R2.8}
Recall from Remark \ref{R2.7} the relationship between the generating function $E_N^{(w)}((s,\infty);\xi)$ and $k$-point correlation functions $\rho_k^{(w)}$ recounted therein. We now supplement Remark \ref{R2.7} with its counterpart for the Cauchy case. In the case $\beta = 2$, it is known \citep{WF00} that
  \begin{equation}\label{eq2.3.23}     
  \sigma(s) := (1 + s^2) {\mathrm{d} \over \mathrm{d} s} \log E_N^{(Cy)}((s,\infty);\xi)
  \end{equation}
  satisfies the nonlinear equation (which can be identified in terms of the
  $\sigma$-P${}_{VI}$ equation \citep{FW04}; see also equation \eqref{eq2.3.35} below)
  \begin{multline}\label{eq2.3.24}
  (1+s^2)^2 (\sigma'')^2 + 4 (1 + s^2) (\sigma')^3 - 8 s \sigma (\sigma')^2 \\
  + 4 \sigma^2 (\sigma' - \alpha^2) + 8 \alpha^2 s \sigma \sigma' +
  4 [ N (N + 2 \alpha) - \alpha^2 s^2] (\sigma')^2 = 0.
    \end{multline}  
Note that equation \eqref{eq2.3.24}  is independent of the parameter $\xi$ in equation \eqref{eq2.3.9}.

According to equation \eqref{eq2.3.9}, to leading order in $\xi$,
  \begin{equation}\label{eq2.3.25} 
  \sigma(s) = \xi  r(s)  + O(\xi^2), \quad r(s) :=  (1 + s^2)\rho_1^{(Cy)}(s)= (1 + s^2)\rho^{(Cy)}(s).
   \end{equation}
Substituting in \eqref{eq2.3.24} and equating terms to the leading order in $\xi$
(which is $O(\xi^2)$) shows
  \begin{equation}\label{eq2.3.26} 
 (1 + s^2 )^2 (r''(s))^2 - 4 \alpha^2 ( r(s))^2 + 8 \alpha^2 s r(s) r'(s) +
 4 [ N (N + 2 \alpha) - \alpha^2 s^2] ( r'(s))^2 = 0.
  \end{equation}
  Upon differentiating with respect to $s$, a factor of $r''(s)$ can be
  cancelled and a third order linear differential equation results,
   \begin{equation}\label{eq2.3.27} 
   (1 + s^2)^2 r'''(s) + 2 s (1 + s^2) r''(s) + 4 [ N (N + 2 \alpha) - \alpha^2 s^2] r'(s)
   + 4 \alpha^2 s r(s) = 0.
  \end{equation}
  Recalling the definition of $r(s)$ in terms of $ \rho^{(Cy)}(s)$ \eqref{eq2.3.25}, we see
  that equation \eqref{eq2.3.27} is equivalent to the third order differential equation given in Proposition
 \ref{prop2.7}.
 \end{remark}  
 
 The $\beta=1,4$ analogue of equation \eqref{eq2.3.27}, obtained by taking $\rho^{(Cy)}(s)=r(s)/(1+s^2)$ in equation \eqref{eq2.3.21}, is
\begin{multline}\label{eq2.3.28}
(1+s^2)^4r^{(5)}(s)+10s(1+s^2)^3r^{(4)}(s)-(5\tilde{\alpha}-22)(1+s^2)^3r'''(s) \\
+(5\tilde{N}-13)(1+s^2)^2r'''(s)-8(\tilde{\alpha}-1)s(1+s^2)^2r''(s) \\
+2(7\tilde{N}+1)s(1+s^2)r''(s)+4\tilde{\alpha}^2(1+s^2)^2r'(s)-2\left[(4\tilde{\alpha}-1)\tilde{N}+1\right](1+s^2)r'(s) \\
+4(\tilde{N}-1)^2r'(s)-4\tilde{\alpha}^2s(1+s^2)r(s)+4\tilde{\alpha}(\tilde{N}-1)sr(s)=0,
\end{multline}
where $\tilde{\alpha}$ and $\tilde{N}$ are as given in Proposition \ref{prop2.7}. In comparing equations \eqref{eq2.3.27} and \eqref{eq2.3.28} to
their counterparts in Proposition \ref{prop2.7}, we see that degree of each of the coefficients (which alternate between
being even and odd in $s$) has been reduced by $2$.

\begin{remark} \label{R2.9}
Recall from \S\ref{s1.2.1} that the the Cauchy and circular Jacobi ensembles are related by the Cayley transformation \eqref{eq1.2.25} and stereographic projection $\lambda_i=\cot(\theta_i/2)$ \eqref{eq1.2.26}. In particular, the eigenvalue densities of the two
ensembles are related by 
\begin{equation}\label{eq2.3.29}
\rho^{(Cy)}(\lambda) = 2\sin^2(\theta/2)\,\rho^{(cJ)}(e^{\mathrm{i}\theta}).
\end{equation}
Equivalently, in the notation $r(s)$ of equation \eqref{eq2.3.25} with $s=\cot(\theta/2)$,
\begin{equation}\label{eq2.3.30}
r(s) = 2  \rho^{(cJ)}(e^{\mathrm{i}\theta}).
\end{equation}
In the case $\alpha=0$, the circular Jacobi weight $w^{(cJ)}(e^{\mathrm{i}\theta})$ \eqref{eq1.2.28} is a constant and so according to equation \eqref{eq2.3.30}, $r(s)$
is then also a constant. We can see immediately that this is consistent with equations \eqref{eq2.3.27} and \eqref{eq2.3.28}.

Suppose now $\theta$ is scaled by writing $\theta = 2 X/N$. Then, in the scaling limit $N\to\infty$, the circular Jacobi ensemble exhibits a spectrum singularity at $\theta=0$, in keeping with the discussion in \S\ref{s1.2.1}. In view of equation \eqref{eq2.3.30}, in the cases $\beta = 1,2$, and $4$, differential equations for the corresponding density
$\rho^{(s.s.)}(X)$ can be obtained by setting $s=N/X$ and $r(s) = \rho^{(s.s.)}(X)$
 in equations \eqref{eq2.3.27} and \eqref{eq2.3.28}, and then equating terms at leading order in $N$.
 Specifically for $\beta = 2$, we therefore have that $R(X) = \rho^{(s.s.)}(X)$
satisfies the third order linear differential equation
\begin{equation} \label{eq2.3.31}
X^2  R'''(X) + 4 X R''(X) + (2 - 4 \alpha^2 + 4 X^2) R'(X) - {4 \alpha^2 \over X} R(X) = 0.
\end{equation}
This is consistent with the known exact formula \citep{NS93}, \cite[Eq.~(7.49) with $\pi \rho = 1$]{Fo10}
\begin{equation} \label{eq2.3.32}
 \rho^{(s.s.)}(x) = {x \over 2} \Big ( ( J_{\alpha - 1/2}(x) )^2 + ( J_{\alpha + 1/2}(x) )^2 - {2 \alpha \over x}
 J_{\alpha - 1/2}(x)   J_{\alpha + 1/2}(x)  \Big ),
 \end{equation}
 as can be checked using computer algebra.
\end{remark}

Though there is no present benefit in presenting the analogue of Proposition \ref{prop2.6} for the non-symmetric shifted Jacobi ensemble, there is merit in deriving the differential equation satisfied by the eigenvalue density of the non-symmetric CyUE, especially considering its relatively compact form. Repeating the process outlined above, we first make the change of variables $x\mapsto(1-x)/2$ in equation \eqref{eq2.3.7} to obtain a differential equation for the non-symmetric shifted Jacobi unitary ensemble. Setting $a=-N-\alpha$, $b=-N-\overline{\alpha}$, and replacing $x$ by $\mathrm{i}x$ in this differential equation, it follows from Proposition \ref{prop2.5} that we obtain the differential equation satisfied by $\rho^{(Cy)}(x;N,2)$.

\begin{proposition}\label{prop2.8}  
 Define the third order differential operator
 \begin{multline} \label{eq2.3.33}
 \tilde{\mathcal{D}}_{N,2}^{(Cy)} = (1 + x^2)^3 {\mathrm{d}^3 \over \mathrm{d} x^3} + 8 x ( 1 + x^2)^2 {\mathrm{d}^2 \over \mathrm{d} x^2} \\
 + (1 + x^2)[6 + 4N (N + \alpha + \overline{\alpha}) + (\alpha-\overline{\alpha})^2  + 2 \mathrm{i} (\alpha - \overline{\alpha}) (2 N + 
 \alpha + \overline{\alpha}) x + (14 - (\alpha + \overline{\alpha})^2) x^2 ] {\mathrm{d} \over \mathrm{d} x} \\
 + [4 + 8 N(N+\alpha+\overline{\alpha}) + 3 \alpha^2 - 2 \alpha \overline{\alpha} + 3 \bar{\alpha}^2 +(4-(\alpha+\overline{\alpha})^2)x^2]x \\
+ \mathrm{i} (\alpha - \overline{\alpha}) ( 2N + \alpha + \overline{\alpha} )(3x^2-1).
 \end{multline}
 For $\beta = 2$ and $\alpha\in\mathbb{C}$ with constraint $\mathrm{Re}(\alpha)>-1/2$, we have
  \begin{equation} \label{eq2.3.34}
   \tilde{\mathcal{D}}_{N,2}^{(Cy)} \, \rho^{(Cy)}(x) = 0.
     \end{equation}
 \end{proposition}

 \begin{remark} \label{R2.10}
After a simple change of variables, the Jimbo--Miwa--Okamoto $\sigma$-form of the
 Painlev\'e differential equation reads \cite[Eq.~(1.32)]{FW04}
  \begin{equation} \label{eq2.3.35}
  h'\Big ( (1 + t^2)h'' \Big )^2 + 4 \Big ( h' ( h - t h') - \mathrm{i}
  b_1b_2b_3b_4 \Big )^2 + 4 \prod_{k=1}^4 (h' + b_k^2) = 0.
 \end{equation}
 Let $\mathbf b = (b_1, b_2, b_3, b_4)$ and define $e_2'[\mathbf b]$, $e_2[\mathbf b]$ as the
 elementary symmetric polynomials of degree two in $\{b_1, b_3, b_4 \}$ and
 $\{b_1, b_2,b_3, b_4 \}$, respectively.
 Set 
   \begin{equation} \label{eq2.3.36}
   U_N^{(Cy)}(t;(\alpha_1,\alpha_2);\xi) =
   (t^2 + 1) {d \over d t} \log \Big (
   (\mathrm{i}t - 1)^{e_2'[\mathbf b] - e_2[\mathbf b]/2} (\mathrm{i}t + 1)^{e_2[\mathbf b]/2}
   E_N^{(Cy)}((t,\infty);\xi)) \Big ),
 \end{equation}
 where $E_N^{(Cy)}$ is specified by equation \eqref{eq2.3.9} in the non-symmetric case with
 $\alpha = \alpha_1 + \mathrm{i} \alpha_2$ ($\alpha_2 \ne 0$). We have from \cite[Prop.~15]{FW04} that
 $U_N^{(Cy)}$ satisfies the transformed $\sigma$-P${}_{VI}$ equation \eqref{eq2.3.35} with parameters
  \begin{equation} \label{eq2.3.37} 
 \mathbf b = ( - \alpha_1, -\mathrm{i} \alpha_2, N + \alpha_1, \alpha_1).
  \end{equation}
 Analogous to equation \eqref{eq2.3.25}, we have from equations \eqref{eq2.3.9}, \eqref{eq2.3.36}, and \eqref{eq2.3.37} that
 \begin{equation} \label{eq2.3.38}  
  U_N^{(Cy)}(t;(\alpha_1,\alpha_2);\xi) = (N + \alpha_1) \alpha_2 - \alpha_1^2 t + \xi r(t) + O(\xi^2), \qquad
  r(t) := (1 + t^2)   \rho^{(Cy)}(t).
  \end{equation}
  
  Substituting \eqref{eq2.3.38} in equation \eqref{eq2.3.35} and equating terms of leading order in $\xi$, which
  as for equation \eqref{eq2.3.25} occurs at order $\xi^2$, then differentiating and cancelling a factor
  of $r''$ shows
  \begin{multline}\label{eq2.3.39}
   (1 + t^2)^2 r''' +  2   t (1 + t^2) r''  + 4[ ( (N + \alpha_1) \alpha_2 + \alpha_1^2 t) (r - t r') \\
 +  ((N + \alpha_1)^2 -  (N + \alpha_1) \alpha_2 t - \alpha_1^2 - \alpha_2^2 ) r']  = 0.
  \end{multline}
  In the case $\alpha_2 = 0$, this agrees with equation \eqref{eq2.3.27}. Now, substituting for $r(t)$ in terms of $ \rho^{(Cy)}(t)$ as specified in equation \eqref{eq2.3.38} reclaims
  equation \eqref{eq2.3.34}.
  \end{remark}

\subsection{Differential equations for the Laguerre ensembles} \label{s2.3.3}
As mentioned at the beginning of this section, the eigenvalue density $\rho^{(L)}(x)$ of the Laguerre $\beta$ ensemble can be obtained from that of the Jacobi $\beta$ ensemble via a limiting procedure: From equation \eqref{eq2.0.1}, upon extending the idea of Lemma \ref{L1.1}, we see (cf.~equation \eqref{eq2.1.12})
\begin{equation} \label{eq2.3.40}
\rho^{(L)}(x;N+1,\beta) = \lim_{b\rightarrow\infty}b^{(N+3)N\kappa+(N+2)a+N+1}\,\rho^{(J)}\Big (\frac{x}{b};N+1,\beta \Big ).
\end{equation}
This fact allows one to transform differential equations satisfied by $\rho^{(J)}(x)$ into analogues satisfied by $\rho^{(L)}(x)$, which will be presented in a moment.

As an aside, suppose that one would like to obtain differential equations for $\rho^{(L)}(x)$ without prior knowledge of the analogous differential equations for $\rho^{(J)}(x)$, i.e., if the results of \S\ref{s2.3.1} were not available, or if one were interested in ensembles with $\beta\notin\{1,2,4\}$. Then, it is actually more efficient to circumvent computation of the differential equations for $\rho^{(J)}(x)$ and use the aforementioned limiting procedure indirectly. That is, let
$$
L_{\beta,p}^{(N)}(x) =\lim_{b\rightarrow\infty}\left(-\tfrac{x}{b}\right)^{\beta N+p}J_{\beta,p}^{(N)}(b/x), \qquad
 \tilde{L}_{\beta,p}^{(N)}(x) =x^ae^{-x}\,L_{\beta,p}^{(N)}(x)
$$
so that by equation \eqref{eq2.0.1}, $\rho^{(L)}(x;N+1,\beta)$ is proportional to $\tilde{L}_{\beta,0}^{(N)}(x)$ when $\beta$ is a positive even integer. Equation \eqref{eq2.1.17} gives a differential-difference equation for $\left(-\tfrac{x}{b}\right)^{\beta N+p}J_{\beta,p}^{(N)}(b/x)$ and taking the limit $b\rightarrow\infty$ thus gives a differential-difference equation for $L_{\beta,p}^{(N)}(x)$. Substituting $L_{\beta,p}^{(N)}(x)=x^{-a}e^x\tilde{L}_{\beta,p}^{(N)}(x)$ then gives a differential-difference equation for $\tilde{L}_{\beta,p}^{(N)}(x)$. These equations with $p=0,1,\ldots,\beta$ are equivalent to matrix differential equations not unlike \eqref{eq2.3.2} and \eqref{eq2.3.5}. However, having taken the limit $b\rightarrow\infty$, we ensure that these new matrix differential equations are simpler, and have the added benefit of simplifying down to scalar differential equations for $\rho^{(L)}(x;N+1,\beta)$ rather than for auxiliary functions. One may then apply the Stieltjes transform to obtain differential equations for the resolvents, and use the $\beta\leftrightarrow4/\beta$ dualities given in Lemma \ref{L1.4} to obtain mirror differential equations like those seen in Theorem \ref{thrm2.2}.

Since we are interested in $\rho^{(L)}(x)$ with $\beta\in\{1,2,4\}$ and we have differential equations for $\rho^{(J)}(x)$ for these $\beta$ values, we use the more direct approach to give the following proposition.

\begin{proposition} \label{prop2.9}
Retaining the definitions of $a_{\beta},N_{\beta}$, and $\tilde{a}$ from Theorem \ref{thrm2.2}, define
\begin{align} \label{eq2.3.41}
\mathcal{D}_{2,N}^{(L)}&=x^3\frac{\mathrm{d}^3}{\mathrm{d}x^3}+4x^2\frac{\mathrm{d}^2}{\mathrm{d}x^2}-\left[x^2-2(a+2N)x+a^2-2\right]x\frac{\mathrm{d}}{\mathrm{d}x}+\left[(a+2N)x-a^2\right]
\end{align}
and for $\beta=1$ or $4$,
\begin{align} \label{eq2.3.42}
\mathcal{D}_{N,\beta}^{(L)}&=4x^5\frac{\mathrm{d}^5}{\mathrm{d}x^5}+40x^4\frac{\mathrm{d}^4}{\mathrm{d}x^4}-\left[5\left(\frac{x}{\kappa-1}\right)^2-10\left(a_{\beta}+4N_{\beta}\right)\frac{x}{\kappa-1}+5\tilde{a}-88\right]x^3\frac{\mathrm{d}^3}{\mathrm{d}x^3}\nonumber
\\&\quad-\left[16\left(\frac{x}{\kappa-1}\right)^2-38\left(a_{\beta}+4N_{\beta}\right)\frac{x}{\kappa-1}+22\tilde{a}-16\right]x^2\frac{\mathrm{d}^2}{\mathrm{d}x^2}\nonumber
\\&\quad+\left[\left(\frac{x}{\kappa-1}\right)^2-4\left(a_{\beta}+4N_{\beta}\right)\left(\frac{x}{\kappa-1}\right)+2\left(2\left(a_{\beta}+4N_{\beta}\right)^2+\tilde{a}-2\right)\right]\frac{x^3}{(\kappa-1)^2}\frac{\mathrm{d}}{\mathrm{d}x}\nonumber
\\&\quad-\left[4(\tilde{a}-3)\left(a_{\beta}+4N_{\beta}\right)\frac{x}{\kappa-1}-\tilde{a}^2+14\tilde{a}+16\right]x\frac{\mathrm{d}}{\mathrm{d}x}-\left(a_{\beta}+4N_{\beta}\right)\left(\frac{x}{\kappa-1}\right)^3\nonumber
\\&\quad+\left(2\left(a_{\beta}+4N_{\beta}\right)^2+\tilde{a}\right)\left(\frac{x}{\kappa-1}\right)^2-(3\tilde{a}+4)\left(a_{\beta}+4N_{\beta}\right)\frac{x}{\kappa-1}+\tilde{a}^2.
\end{align}
Then, for $\beta=1,2$, and $4$,
\begin{equation} \label{eq2.3.43}
\mathcal{D}_{N,\beta}^{(L)}\,\rho^{(L)}(x;N,\beta)=0
\end{equation}
and
\begin{equation} \label{eq2.3.44}
\mathcal{D}_{N,\beta}^{(L)}\,\frac{1}{N}W_1^{(L)}(x;N,\beta)=
\begin{cases}
(x+a),&\beta=2,
\\ \frac{4}{\kappa-1}\left[2\left(\frac{x}{\kappa-1}\right)^2+(2a_{\beta}-1)\frac{x}{\kappa-1}\right]N_{\beta}&
\\\quad-\frac{1}{\kappa-1}\left[\left(\frac{x}{\kappa-1}\right)^3-(a_{\beta}+2)\left(\frac{x}{\kappa-1}\right)^2\right]&
\\\quad+\frac{1}{\kappa-1}\left[(a_{\beta}^2+4a_{\beta}-4)\frac{x}{\kappa-1}-a_{\beta}(a_{\beta}-2)^2\right],&\beta=1,4.
\end{cases}
\end{equation}
\end{proposition}
\begin{proof}
To obtain equation \eqref{eq2.3.43} from \eqref{eq2.3.7} and \eqref{eq2.3.13}, change variables $x\mapsto x/b$, multiply both sides by $b^{(N+2)(N-1)\kappa+(N+1)a+N}$ according to equation \eqref{eq2.3.40}, and then take the limit $b\rightarrow\infty$. This is equivalent to changing variables $x\mapsto x/b$, extracting terms of leading order in $b$, then rewriting $\rho^{(J)}$ as $\rho^{(L)}$.

To obtain equation \eqref{eq2.3.44}, apply the same prescription to equations \eqref{eq2.3.8} and \eqref{eq2.3.14}. Alternatively, one may apply the Stieltjes transform to equation \eqref{eq2.3.43}. This is considerably easier than in the Jacobi case (see Appendix \ref{appendixB}) since
\begin{align*}
\int_0^{\infty}\frac{x^n}{s-x}\frac{\mathrm{d}^n}{\mathrm{d}x^n}\rho^{(L)}(x)\,\mathrm{d}x=s^n\frac{\mathrm{d}^n}{\mathrm{d}s^n}W_1^{(L)}(s)
\end{align*}
can be computed through repeated integration by parts using the fact that the boundary terms vanish at all stages: For $a>-1$ real and $m>n$ non-negative integers, $\frac{x^m}{s-x}\frac{\mathrm{d}^n}{\mathrm{d}x^n}\rho^{(L)}(x)$ has a factor of $x^{a+m-n}$ which dominates at $x=0$ and a factor of $e^{-x}$ which dominates as $x\rightarrow\infty$.
\end{proof}
Equation \eqref{eq2.3.43} with $\beta=2$ is equivalent to that seen in \citep{GT05}, \citep{ATK11}. When $\beta=1,4$, it has been checked for $N=1,2$ using computer algebra. From inspection, it seems that the natural variable of equations \eqref{eq2.3.43} and \eqref{eq2.3.44} is $x/(\kappa-1)$, which is the limit $\lim_{b\rightarrow\infty}b_{\beta}\left(\tfrac{x}{b}\right)$. This is in keeping with the duality relation on the resolvent presented in Lemma \ref{L1.4}. Strictly speaking, changing variables to $y=x/(\kappa-1)$ is not natural and would be counterproductive since the corresponding weight $\left[(\kappa-1)y\right]^ae^{(1-\kappa)y}$ vanishes at $\kappa=1$ and has different support depending on whether $\kappa<1$ or $\kappa>1$.

\subsection{Differential equations for the Gaussian ensembles} \label{s2.3.4}
The previous subsection contains a discussion on how one would obtain differential equations for the densities and resolvents of Laguerre ensembles with even $\beta$ without knowledge of differential equations for the corresponding Jacobi $\beta$ ensembles' densities and resolvents. We now elucidate those ideas by explicitly applying them to the study of the Gaussian ensembles with $\beta=6$ and consequently, by duality, $\beta=2/3$. Indeed, since we do not have differential equations for the $\beta=6$ Jacobi ensemble at hand, we cannot immediately apply the direct limiting approach used in the proof of Proposition \ref{prop2.9}. Thus, for our purposes, it is in fact more efficient to replicate the proofs of \S\ref{s2.3.1} while taking limits at an earlier stage so as to circumvent the need for investigating the $\beta=6$ Jacobi ensemble altogether.

Like in the Jacobi and Laguerre cases, our initial focus is the average
\begin{equation} \label{eq2.3.45}
I_{N,\beta}^{(G)}(\lambda):=\mean{\prod_{i=1}^N|\lambda-\lambda_i|^{\beta}}_{\text{GE}_{N,\beta}},
\end{equation}
where the subscript $\text{GE}_{N,\beta}$ indicates that the average is with respect to the eigenvalue j.p.d.f.~\eqref{eq1.2.7} with weight $w^{(G)}(\lambda)=e^{-\lambda^2}$. The Gaussian analogue of the duality relation \eqref{eq2.1.21} is \citep{BF97}
\begin{equation} \label{eq2.3.46}
\mean{\prod_{i=1}^N\left(\lambda-\kappa^{-1/2}\lambda_i\right)^n}_{\text{GE}_{N,\beta}}=\mean{\prod_{i=1}^n(\lambda-\mathrm{i}\lambda_i)^N}_{\text{GE}_{n,4/\beta}}.
\end{equation}
Replacing $\lambda$ by $\kappa^{-1/2}\lambda$ in the above duality relation and then factoring $(-\mathrm{i})^{nN}$ from the right-hand side shows that for even $\beta$, $I_{N,\beta}^{(G)}(\lambda)$ is proportional to $G_{\beta,0}^{(N)}(\lambda)$, where
\begin{equation} \label{eq2.3.47}
G_{n,p}^{(N)}(\lambda):=(-\mathrm{i})^{nN+p}\mean{\prod_{i=1}^n\left(\lambda_i+\mathrm{i}\kappa^{-1/2}\lambda\right)^{N+\chi_{i\leq p}}}_{\text{GE}_{n,4/\beta}},\quad 0\leq p\leq n.
\end{equation}
Setting $a'=b'=L$ and changing variables $\lambda_i\mapsto\tfrac{1}{2}\left(1+\tfrac{\lambda_i}{\sqrt{L}}\right)$ in $J_{n,p}^{(N)}(\lambda)$ \eqref{eq2.1.16}, we see that
\begin{equation} \label{eq2.3.48}
G_{n,p}^{(N)}(\lambda)=\lim_{L\rightarrow\infty}4^{nL}(-2\mathrm{i}\sqrt{L})^{nN+p}(2\sqrt{L})^{n(n-1)/\kappa+n}\left.J_{n,p}^{(N)}\left(\tfrac{1}{2}\left(1-\mathrm{i}\sqrt{\tfrac{2}{\beta L}}\lambda\right)\right)\right|_{a'=b'=L}.
\end{equation}
Thus, equation \eqref{eq2.1.17} simplifies to a differential-difference equation for $G_{n,p}^{(N)}(\lambda)$,
\begin{multline} \label{eq2.3.49}
(n-p)G_{n,p+1}^{(N)}(\lambda)=\tfrac{(n-p)}{\sqrt{\kappa}}\lambda G_{n,p}^{(N)}(\lambda)-\tfrac{\sqrt{\kappa}}{2}\frac{\mathrm{d}}{\mathrm{d}\lambda}G_{n,p}^{(N)}(x)
\\+\tfrac{p}{2}\left(\tfrac{1}{\kappa}(n-p)+N+1\right)G_{n,p-1}^{(N)}(\lambda)
\end{multline}
(cf.~\cite[Eq.~(5.5)]{FT19}). Taking the $L\rightarrow\infty$ limit early has already yielded a simpler equation than \eqref{eq2.1.17}. However, we can take this one step further by defining
\begin{equation} \label{eq2.3.50}
\tilde{G}_{n,p}^{(N)}(\lambda)=e^{-\lambda^2}G_{n,p}^{(N)}(\lambda)
\end{equation}
so that $\rho^{(G)}(\lambda;N+1,\beta)$ is proportional to $\tilde{G}_{\beta,0}^{(N)}(\lambda)$. It is then easy to obtain the differential-difference equation
\begin{multline} \label{eq2.3.51}
\frac{\mathrm{d}}{\mathrm{d}\lambda}\tilde{G}_{n,p}^{(N)}(\lambda)=\tfrac{p}{\sqrt{\kappa}}\left(\tfrac{1}{\kappa}(n-p)+N+1\right)\tilde{G}_{n,p-1}^{(N)}(\lambda)
\\+2\left(\tfrac{1}{\kappa}(n-p)-1\right)\lambda\tilde{G}_{n,p}^{(N)}(\lambda)-\tfrac{2}{\sqrt{\kappa}}(n-p)\tilde{G}_{n,p+1}^{(N)}(\lambda).
\end{multline}
With $n=\beta\in2\mathbb{N}$ and $p=0,1,\ldots,n$, this is equivalent to a matrix differential equation on the vector $\mathbf{v}=\left[\tilde{G}_{n,0}^{(N)}(x)\,\cdots\,\tilde{G}_{n,n}^{(N)}(x)\right]^T$, which is moreover equivalent to a scalar differential equation for $\rho^{(G)}(\lambda;N+1,\beta)$.

\begin{proposition} \label{prop2.10}
For $\beta=2/3$ or $6$, define
\begin{align} \label{eq2.3.52}
\mathcal{D}_{N,\beta}^{(G)}&=81(\kappa-1)^{7/2}\frac{\mathrm{d}^7}{\mathrm{d}x^7}+1008\left(3N_{\beta}-\frac{2x^2}{\kappa-1}+2\right)(\kappa-1)^{5/2}\frac{\mathrm{d}^5}{\mathrm{d}x^5} \nonumber
\\&\quad+2016x(\kappa-1)^{3/2}\frac{\mathrm{d}^4}{\mathrm{d}x^4} \nonumber
\\&\quad+64\left(21N_{\beta}-\frac{14x^2}{\kappa-1}+5\right)\left(21N_{\beta}-\frac{14x^2}{\kappa-1}+23\right)(\kappa-1)^{3/2}\frac{\mathrm{d}^3}{\mathrm{d}x^3} \nonumber
\\&\quad+9984\left(3N_{\beta}-\frac{2x^2}{\kappa-1}+2\right)x(\kappa-1)^{1/2}\frac{\mathrm{d}^2}{\mathrm{d}x^2} \nonumber
\\&\quad+256\Bigg[54N_{\beta}\left(4N_{\beta}^2+8N_{\beta}+3\right)-(432N_{\beta}^2+576N_{\beta}+57)\frac{x^2}{\kappa-1} \nonumber
\\&\quad+96(3N_{\beta}+2)\frac{x^4}{(\kappa-1)^2}-\frac{64x^6}{(\kappa-1)^3}-20\Bigg](\kappa-1)^{1/2}\frac{\mathrm{d}}{\mathrm{d}x} \nonumber
\\&\quad+256\left(144N_{\beta}^2+192N_{\beta}-64(3N_{\beta}+2)\frac{x^2}{\kappa-1}+\frac{64x^4}{(\kappa-1)^2}+25\right)\frac{x}{(\kappa-1)^{1/2}},
\end{align}
where we retain the definition $N_{\beta}=(\kappa-1)N$. Then, for these same $\beta$ values,
\begin{equation} \label{eq2.3.53}
\mathcal{D}_{N,\beta}^{(G)}\,\rho^{(G)}(x;N,\beta)=0
\end{equation}
and
\begin{align} \label{eq2.3.54}
\mathcal{D}_{N,\beta}^{(G)}\,\frac{1}{N}W_1^{(G)}(x;N,\beta)&=\frac{2^{11}}{\sqrt{\kappa-1}}\left(\frac{4x^2}{\kappa-1}-6N_{\beta}-7\right)^2-3\frac{2^{12}}{\sqrt{\kappa-1}}.
\end{align}
\end{proposition}

\begin{proof}
Take equation \eqref{eq2.3.51} with $n=\beta=6$ and $N$ replaced by $N-1$ to obtain the matrix differential equation
\begin{equation} \label{eq2.3.55}
\frac{\mathrm{d}\mathbf{v}}{\mathrm{d}x}=\begin{bmatrix}2x&-4\sqrt{3}&0&0&0&0&0\\\frac{3N+5}{3\sqrt{3}}&\frac{4x}{3}&-\frac{10}{\sqrt{3}}&0&0&0&0\\0&\frac{6N+8}{3\sqrt{3}}&\frac{2x}{3}&-\frac{8}{\sqrt{3}}&0&0&0\\0&0&\sqrt{3}(N+1)&0&-2\sqrt{3}&0&0\\0&0&0&\frac{12N+8}{3\sqrt{3}}&-\frac{2x}{3}&-\frac{4}{\sqrt{3}}&0\\0&0&0&0&\frac{15N+5}{3\sqrt{3}}&-\frac{4x}{3}&-\frac{2}{\sqrt{3}}\\0&0&0&0&0&2\sqrt{3}N&-2x\end{bmatrix}\mathbf{v}
\end{equation}
for $\mathbf{v}=\left[\tilde{G}_{6,0}^{(N-1)}(x)\,\cdots\,\tilde{G}_{6,6}^{(N-1)}(x)\right]^T$. Like in the proofs of Lemmas \ref{L2.1} and \ref{L2.2}, for $1\leq p\leq6$, the $p\textsuperscript{th}$ row of the above matrix differential equation gives an expression for $\tilde{G}_{6,p}^{(N-1)}(x)$ in terms of $\frac{\mathrm{d}}{\mathrm{d}x}\tilde{G}_{6,p-1}^{(N-1)}(x)$ and $\tilde{G}_{6,k}^{(N-1)}(x)$ with $k<p$. Substituting these expressions into the equation corresponding to the last row in the order of decreasing $p$ then yields a seventh-order differential equation satisfied by $\tilde{G}_{6,0}^{(N-1)}(x)$. Since $\rho^{(G)}(x;N,6)$ is proportional to $\tilde{G}_{6,0}^{(N-1)}(x)$, this equation is equivalent to \eqref{eq2.3.53} for $\beta=6$. Taking the Stieltjes transform of this result and substituting in the spectral moments $m_2^{(G)}$ and $m_4^{(G)}$ from \citep{WF14} then yields equation \eqref{eq2.3.54} for $\beta=6$. Employing the duality relation given in Lemma \ref{L1.4} for the resolvent then shows that equation \eqref{eq2.3.54} also holds for $\beta=2/3$. Finally, taking the inverse Stieltjes transform of this result shows that equation \eqref{eq2.3.53} holds for $\beta=2/3$ as well.
\end{proof}
Equation \eqref{eq2.3.53} has been checked for $N=1,2$ using computer algebra. Similar to the Laguerre case, it seems like $x/\sqrt{\kappa-1}$ is the natural variable in Proposition \ref{prop2.10}. This is evidently due to the duality relation used in the proof of this proposition. Like in the Laguerre case, there is presently no benefit in changing variables to $x/\sqrt{\kappa-1}$.

It has been mentioned that equation \eqref{eq2.3.51} leads to a matrix differential equation which is equivalent to a scalar differential equation for $\rho^{(G)}(x;N+1,\beta)$ when $\beta$ is even. For $\beta=2$ and $4$, these differential equations are respectively
\begin{equation} \label{eq2.3.56}
\frac{\mathrm{d}}{\mathrm{d}x}\begin{bmatrix}\tilde{G}_{2,0}^{(N-1)}(x)\\\tilde{G}_{2,1}^{(N-1)}(x)\\\tilde{G}_{2,2}^{(N-1)}(x)\end{bmatrix}=\begin{bmatrix}2x&-4&0\\N+1&0&-2\\0&2N&-2x\end{bmatrix}\begin{bmatrix}\tilde{G}_{2,0}^{(N-1)}(x)\\\tilde{G}_{2,1}^{(N-1)}(x)\\\tilde{G}_{2,2}^{(N-1)}(x)\end{bmatrix}
\end{equation}
and
\begin{equation} \label{eq2.3.57}
\frac{\mathrm{d}}{\mathrm{d}x}\begin{bmatrix}\tilde{G}_{4,0}^{(N-1)}(x)\\\tilde{G}_{4,1}^{(N-1)}(x)\\\tilde{G}_{4,2}^{(N-1)}(x)\\\tilde{G}_{4,3}^{(N-1)}(x)\\\tilde{G}_{4,4}^{(N-1)}(x)\end{bmatrix}=\begin{bmatrix}2x&-4\sqrt{2}&0&0&0\\\frac{2N+3}{2\sqrt{2}}&x&-3\sqrt{2}&0&0\\0&\sqrt{2}(N+1)&0&-2\sqrt{2}&0\\0&0&\frac{6N+3}{2\sqrt{2}}&-x&-\sqrt{2}\\0&0&0&2\sqrt{2}N&-2x\end{bmatrix}\begin{bmatrix}\tilde{G}_{4,0}^{(N-1)}(x)\\\tilde{G}_{4,1}^{(N-1)}(x)\\\tilde{G}_{4,2}^{(N-1)}(x)\\\tilde{G}_{4,3}^{(N-1)}(x)\\\tilde{G}_{4,4}^{(N-1)}(x)\end{bmatrix}.
\end{equation}
The corresponding scalar differential equations for the respective eigenvalue densities agree with those of Proposition \ref{prop2.1} after scaling $x\mapsto\sqrt{\frac{N\kappa}{2g}}x$. So too does the differential equation for $\rho^{(G)}(x;N,1)$ obtained through the $\beta\leftrightarrow4/\beta$ duality given in Lemma \ref{L1.4}.

\setcounter{equation}{0}
\section{Scalings of the Differential Equations} \label{s2.4}
The differential equations of the previous section hold true not only for finite integers $N$, but also in the large $N$ limit, upon appropriate scaling. In this section we consider two scaling regimes: In \S\ref{s2.4.1}, we study the classical matrix ensembles (and also the Gaussian $\beta$ ensembles with $\beta=2/3,6$) under the global scalings introduced in Definition \ref{def1.6}, while in \S\ref{s2.4.2}, we look at the soft and hard edge scaling regimes (recall the discussion following Proposition \ref{prop1.2}).

In the global scaling regime, we use Theorem \ref{thrm1.1} to expand the scaled resolvents $\tilde{W}_1(x)$, specified by equation \eqref{eq1.1.24}, in $1/N$ and derive first order differential equations which characterise the expansion coefficients. Being first order, these differential equations are straightforward to solve, and using the Sokhotski--Plemelj inversion formula \eqref{eq1.1.25}, we are able to compute expressions for the large $N$ limiting forms $\rho^0(\lambda)$ of the smoothed eigenvalue densities and their $1/N$ corrections $\rho^1(\lambda)$. As expected, we recover the Wigner semi-circle, Mar\v{c}enko--Pastur, and Wachter laws seen in Proposition \ref{prop1.2}, along with the expression for $\rho^{(Cy),0}(\lambda)$ presented therein. Moreover, in keeping with Note \ref{N1.1} following this proposition, we observe that while the large $N$ limiting forms $\rho^0(\lambda)$ are $\beta$-independent, the $1/N$ corrections $\rho^1(\lambda)$ are different for each $\beta$ that we consider. This is in keeping with the general-$\beta$ results known from loop equation analysis (see, e.g., \citep{WF14}, \citep{FRW17} and references therein).

In contrast to the $\beta$-universality seen for the large $N$ limiting forms of the global scaled eigenvalue densities, we see a different type of universality at the soft and hard edges. Namely, it is known \citep{Fo93a}, \citep{NF95} that for fixed $\beta$, a soft edge of a classical $\beta$ ensemble behaves statistically as that of any other classical $\beta$ ensemble, in the large $N$ limit --- likewise for a hard edge. Thus, we may denote the large $N$ limiting forms of the soft and hard edge scaled eigenvalue densities of the classical $\beta$ ensembles as $\rho^{(soft)}(\lambda;\beta)$ and $\rho^{(hard)}(\lambda;\beta)$, respectively. In \S\ref{s2.4.2}, we apply these scalings to the differential equations of Section \ref{s2.3} to derive linear differential equations satisfied by $\rho^{(soft)}(\lambda;\beta)$ for $\beta\in\{2/3,1,2,4,6\}$ and $\rho^{(hard)}(\lambda;\beta)$ for $\beta\in\{1,2,4\}$.

\subsection{Global scaled differential equations} \label{s2.4.1}
Global scaling of the eigenvalue density refers to a choice of scaling which ensures that, in the large $N$ limit, the scaled density is supported on a finite interval on which it integrates to one. Such a transformation is not unique, due to the length of the interval of support being unspecified; for convenience, we follow the conventions outlined in Definition \ref{def1.6}, which are chosen to be consistent with the earlier works \citep{WF14}, \citep{FRW17}. We recall then that the global scaled eigenvalue densities of the classical $\beta$ ensembles are
\begin{align}
\tilde{\rho}^{(G)}(\lambda)&=\sqrt{\frac{\kappa}{N}}\,\rho^{(G)}(\sqrt{\kappa N}\lambda),
\\ \tilde{\rho}^{(L)}(\lambda)&=\kappa\,\rho^{(L)}(\kappa N\lambda),
\\ \tilde{\rho}^{(J)}(\lambda)&=\frac{1}{N}\,\rho^{(J)}(\lambda),
\\ \tilde{\rho}^{(Cy)}(\lambda)&=\frac{1}{N}\,\rho^{(Cy)}(\lambda)\Big|_{\alpha=\hat{\alpha}\kappa N}, \label{eq2.4.4}
\end{align}
with corresponding scaled resolvents \eqref{eq1.1.24},
\begin{align}
\tilde{W}_1^{(G)}(x)&:=\sqrt{\kappa N}\,W_1^{(G)}(\sqrt{\kappa N}x), \label{eq2.4.5}
\\ \tilde{W}_1^{(L)}(x)&:=\kappa N\,W_1^{(L)}(\kappa Nx), \label{eq2.4.6}
\\ \tilde{W}_1^{(Cy)}(x)&:=W_1^{(Cy)}(x)\Big|_{\alpha=\hat{\alpha}\kappa N}, \label{eq2.4.7}
\end{align}
where $\hat{\alpha}$ is constant in $N$; since the spectrum of the Jacobi $\beta$ ensemble is already supported on $[0,1]$, there is no need to scale its resolvent. In \S\ref{s1.1.1}, it was mentioned that the scaled resolvents are more robust than the global scaled eigenvalue densities in the sense that they admit well-defined large $N$ expansions due to Theorem \ref{thrm1.1}. Thus, following \citep{WF14}, \citep{FRW17} again for consistency, we consider the expansions
\begin{align}
\tilde{W}_1^{(G)}(x;N)&=2N\sum_{l=0}^{\infty}\frac{W_1^{(G),l}(x)}{(2\sqrt{\kappa}N)^l}, \label{eq2.4.8}
\\ \tilde{W}_1^{(L)}(x;N)&=N\sum_{l=0}^{\infty}\frac{W_1^{(L),l}(x)}{(\sqrt{\kappa}N)^l}, \label{eq2.4.9}
\\ W_1^{(J)}(x;N)&=N\sum_{l=0}^{\infty}\frac{W_1^{(J),l}(x)}{(\kappa N)^l}, \label{eq2.4.10}
\\ \tilde{W}_1^{(Cy)}(x;N)&=N\sum_{l=0}^{\infty}\frac{W_1^{(Cy),l}(x)}{(\kappa N)^l}. \label{eq2.4.11}
\end{align}
Substituting these expansions into the differential equations of Section \ref{s2.3} (after scaling according to \eqref{eq2.4.5}--\eqref{eq2.4.7}) and equating terms of like order in $N$ results in linear differential equations for the expansion coefficients $W_1^l(x)$. The differential equations for $W_1^l(x)$ are first order with inhomogeneous terms dependent on $\{W_1^k(x)\}_{k=0}^{l-1}$. Hence, they constitute a recursive process for computing the scaled resolvents $\tilde{W}_1(x)$ up to any desired order in $1/N$. The topological recursion or, equivalently, loop equation formalism \citep{EO09} (see also the review given in Section \ref{s4.1}) accomplishes the same task for general $\beta>0$, but our approach, valid for the specific $\beta$ values considered, is more efficient since there is no need to consider the $n$-point correlators $U_n,W_n$ for $n>1$. As a consequence, our work is able to isolate extra structures, such as those studied by Haagerup and Thorbj{\o}rnsen \citep{HT12}: In \citep{HT12}, the authors showed in the GUE case that $W_1^{(G),l}(x)$ is a polynomial in $(x^2-2)^{-1/2}$, with the polynomial coefficients satisfying a certain three-term recurrence.

\subsubsection{Differential equations for $W_1^{(G),l}(x)$}
We begin with the Gaussian $\beta$ ensembles with $\beta\in\{2/3,1,4,6\}$ and refer to \citep{HT12} for the $\beta=2$ case.
\begin{proposition} \label{prop2.11}
We reuse the notation of Proposition \ref{prop2.1} with $g=1/2$ so that $h=\sqrt{\kappa}-1/\sqrt{\kappa}$ and $y_{(G)}:=\sqrt{x^2-2}$. Then, for $\beta=1$ and $4$, the expansion coefficients of $\tilde{W}_1^{(G)}(x;N,\beta)$ \eqref{eq2.4.8} satisfy the differential equations
\begin{equation} \label{eq2.4.12}
y_{(G)}^2\frac{\mathrm{d}}{\mathrm{d}x}W_1^{(G),0}(x)-xW_1^{(G),0}(x)=-1,
\end{equation}
\begin{equation} \label{eq2.4.13}
y_{(G)}^2\frac{\mathrm{d}}{\mathrm{d}x}W_1^{(G),1}(x)-xW_1^{(G),1}(x)=4h\frac{\mathrm{d}}{\mathrm{d}x}W_1^{(G),0}(x)-\frac{h}{y_{(G)}^2}\left[2xW_1^{(G),0}(x)+5\right],
\end{equation}
and for $l\geq2$, the general differential equation
\begin{multline} \label{eq2.4.14}
y_{(G)}^2\frac{\mathrm{d}}{\mathrm{d}x}W_1^{(G),l}(x)-xW_1^{(G),l}(x)=4h\frac{\mathrm{d}}{\mathrm{d}x}W_1^{(G),l-1}(x)-\frac{2hx}{y_{(G)}^2}W_1^{(G),l-1}(x)
\\+\frac{1}{y_{(G)}^2}\left[\frac{5y_{(G)}^2}{2}\frac{\mathrm{d}^3}{\mathrm{d}x^3}-3x\frac{\mathrm{d}^2}{\mathrm{d}x^2}+\frac{\mathrm{d}}{\mathrm{d}x}\right]W_1^{(G),l-2}(x)
\\-\frac{5h}{y_{(G)}^2}\frac{\mathrm{d}^3}{\mathrm{d}x^3}W_1^{(G),l-3}(x)+\frac{1}{y_{(G)}^2}\frac{\mathrm{d}^5}{\mathrm{d}x^5}W_1^{(G),l-4}(x),
\end{multline}
where we set $W_1^{(G),k}:=0$ for $k<0$.
\end{proposition}
\begin{proof}
Beginning with the $\beta=1,4$ case of equation \eqref{eq2.0.4}, set $g=1/2$ and map $x\mapsto \sqrt{\kappa N}x$ to obtain a differential equation for $\tilde{W}_1^{(G)}(x;N,\beta)$. Substitute in the expansion \eqref{eq2.4.8} and equate terms of like order in $N$. The terms of leading order in $N$ correspond to equation \eqref{eq2.4.12}, the next to leading order terms correspond to equation \eqref{eq2.4.13}, and so on.
\end{proof}

Let us recall from \S\ref{s1.1.1} the relationship between the resolvent expansion coefficients $\{W_1^l(x)\}_{l=0}^{\infty}$ and the coefficients $\{\tilde{M}_{k,l}\}_{l=0}^{\infty}$ of the $1/N$ expansions of the spectral moments $\tilde{m}_k$ of $\tilde{\rho}(\lambda)$. The facts that $\tilde{m}_0=m_0/N=1$ and $W_1^l(x)$ has asymptotic expansion $\sum_{k=0}^{\infty}\tilde{M}_{k,l}/x^{k+1}$ tell us that the solutions of the differential equations \eqref{eq2.4.12}--\eqref{eq2.4.14} are fully determined by the boundary conditions at large $x$,
\begin{align}
W_1^0(x)&=\frac{1}{x}+{\rm O}\left(\frac{1}{x^2}\right), \label{eq2.4.15}
\\ W_1^l(x)&={\rm O}\left(\frac{1}{x^2}\right),\quad l>1. \label{eq2.4.16}
\end{align}
This is in fact true for all of the first order differential equations given in this subsection.

Now, applying the above proof to differential equation \eqref{eq2.3.54} gives analogous differential equations for the expansion coefficients $W_1^{(G),l}(x)$ when $\beta=2/3,6$.
\begin{proposition} \label{prop2.12}
Retain the definition of $y_{(G)}$ from Proposition \ref{prop2.11}. Then, for $\beta=2/3$ and $6$, the expansion coefficients of $\tilde{W}_1^{(G)}(x;N,\beta)$ satisfy the differential equations
\begin{equation} \label{eq2.4.17}
y_{(G)}^2\frac{\mathrm{d}}{\mathrm{d}x}W_1^{(G),0}(x)-xW_1^{(G),0}(x)=-1,
\end{equation}
\begin{equation} \label{eq2.4.18}
y_{(G)}^2\frac{\mathrm{d}}{\mathrm{d}x}W_1^{(G),1}(x)-xW_1^{(G),1}(x)=6h\frac{\mathrm{d}}{\mathrm{d}x}W_1^{(G),0}(x)-\frac{h}{y_{(G)}^2}\left[4xW_1^{(G),0}(x)-7\right],
\end{equation}

\begin{multline} \label{eq2.4.19}
y_{(G)}^2\frac{\mathrm{d}}{\mathrm{d}x}W_1^{(G),2}(x)-xW_1^{(G),2}(x)=6h\frac{\mathrm{d}}{\mathrm{d}x}W_1^{(G),1}(x)-\frac{4hx}{y_{(G)}^2}W_1^{(G),1}(x)-\frac{43}{3y_{(G)}^4}
\\+\frac{1}{12y_{(G)}^4}\left[49y_{(G)}^4\frac{\mathrm{d}^3}{\mathrm{d}x^3}-78xy_{(G)}^2\frac{\mathrm{d}^2}{\mathrm{d}x^2}+3(72-19x^2)\frac{\mathrm{d}}{\mathrm{d}x}+25x\right]W_1^{(G),0}(x),
\end{multline}
and for $l\geq3$, the general differential equation
\begin{multline} \label{eq2.4.20}
y_{(G)}^2\frac{\mathrm{d}}{\mathrm{d}x}W_1^{(G),l}(x)-xW_1^{(G),l}(x)=6h\frac{\mathrm{d}}{\mathrm{d}x}W_1^{(G),l-1}(x)-\frac{4hx}{y_{(G)}^2}W_1^{(G),l-1}(x)
\\+\frac{1}{12y_{(G)}^4}\left[49y_{(G)}^4\frac{\mathrm{d}^3}{\mathrm{d}x^3}-78xy_{(G)}^2\frac{\mathrm{d}^2}{\mathrm{d}x^2}+3(72-19x^2)\frac{\mathrm{d}}{\mathrm{d}x}+25x\right]W_1^{(G),l-2}(x)
\\+\frac{h}{3y_{(G)}^4}\left[49y_{(G)}^2\frac{\mathrm{d}^3}{\mathrm{d}x^3}+39x\frac{\mathrm{d}^2}{\mathrm{d}x^2}-10\frac{\mathrm{d}}{\mathrm{d}x}\right]W_1^{(G),l-3}+\frac{7h}{y_{(G)}^4}\frac{\mathrm{d}^5}{\mathrm{d}x^5}W_1^{(G),l-5}
\\+\frac{1}{18y_{(G)}^4}\left[63y_{(G)}^2\frac{\mathrm{d}^5}{\mathrm{d}x^5}+63x\frac{\mathrm{d}^4}{\mathrm{d}x^4}+230\frac{\mathrm{d}^3}{\mathrm{d}x^3}\right]W_1^{(G),l-4}+\frac{3h}{4y_{(G)}^4}\frac{\mathrm{d}^7}{\mathrm{d}x^7}W_1^{(G),l-6},
\end{multline}
where we again set $W_1^{(G),k}:=0$ for $k<0$.
\end{proposition}
This proposition has been checked against \citep{WF14} up to $l=6$, and thus also serves as a check for differential equation \eqref{eq2.3.54} up to order six in $1/N$. Similar checks for consistency, against \citep{FRW17}, have been carried out for Propositions \ref{prop2.13} to \ref{prop2.15} below.

\subsubsection{Differential equations for $W_1^{(L),l}(x)$ and $W_1^{(J),l}(x)$}
Since the differential equations for the Laguerre and Jacobi ensembles' resolvents involve the parameters $a,b$, equating terms of equal order in $N$ requires us to choose how these parameters depend on $N$. To make our results suitable for most known or potential applications  \citep{Be97}, \citep{BFB97}, \citep{VV08}, \citep{No08}, \citep{LV11}, \citep{MS11}, \citep{MS12}, \citep{CMSV16a}, \citep{CMSV16b} and to improve clarity, we take $a=\hat{a}\kappa N$ and $b=\hat{b}\kappa N $ with $\hat{a},\hat{b}$ constant in $N$. Note that our method easily accommodates for more general $N$-expansions of $a$ and $b$.

\begin{proposition} \label{prop2.13}
Set $a=\hat{a}\kappa N$ with $\hat{a}$ constant in $N$, and let $y_{(L)}:=\sqrt{(\hat{a}-x)^2-4x}$. The expansion coefficients of the LUE scaled resolvent $\tilde{W}_1^{(L)}(x;N,2)$ \eqref{eq2.4.9} satisfy the differential equation
\begin{equation} \label{eq2.4.21}
-xy_{(L)}^2\frac{\mathrm{d}}{\mathrm{d}x}W_1^{(L),0}(x)+\left[(\hat{a}+2)x-\hat{a}^2\right]W_1^{(L),0}(x)=x+\hat{a},
\end{equation}
and for $l\geq2$, the general differential equation
\begin{multline} \label{eq2.4.22}
xy_{(L)}^2\frac{\mathrm{d}}{\mathrm{d}x}W_1^{(L),l}(x)-\left[(\hat{a}+2)x-\hat{a}^2\right]W_1^{(L),l}(x)
\\=\left[x^3\frac{\mathrm{d}^3}{\mathrm{d}x^3}+4x^2\frac{\mathrm{d}^2}{\mathrm{d}x^2}+2x\frac{\mathrm{d}}{\mathrm{d}x}\right]W_1^{(L),l-2}(x).
\end{multline}
\end{proposition}
\begin{proof}
Consider equation \eqref{eq2.3.44} with $\beta=2$. In it, set $a=\hat{a}N$ and map $x\mapsto\kappa N x$ to obtain a differential equation for $\tilde{W}_1^{(L)}(x)\big|_{a=\hat{a}N}$. Substitute in the expansion \eqref{eq2.4.9} and equate terms of equal order in $N$ to extract differential equations \eqref{eq2.4.21}, \eqref{eq2.4.22}.
\end{proof}

In the $\kappa=\beta/2=1$, $a=\hat{a}N$ setting considered here, the operator $\mathcal{D}_{N,2}^{(L)}\big|_{x\mapsto \kappa N x}$ \eqref{eq2.3.41} is even in $N$ while the right-hand side of differential equation \eqref{eq2.3.44} under the mapping $x\mapsto\kappa N x$ is odd in $N$. It follows that $\tilde{W}_1^{(L)}(x)$ \eqref{eq2.4.6} is an odd function of $N$, so $W_1^{(L),k}(x)=0$ when $k$ is odd. Thus, differential-difference equation \eqref{eq2.4.22} holds vacuously for odd $l$, and should otherwise be interpreted as a recursion over even $l$. Note also that the vanishing of $W_1^{(L),k}(x)$ for $k$ odd implies that the spectral moments $\tilde{m}_k^{(L)}$ (recall \S\ref{s1.1.1}) of the scaled eigenvalue density $\tilde{\rho}^{(L)}(\lambda;N,2)$ are hence even functions of $N$, in keeping with the second point of Remark \ref{R2.1}. In the $\beta=1,4$ analogue of Proposition \ref{prop2.13} given below, which is obtained by repeating the above proof starting with the $\beta=1,4$ cases of equation \eqref{eq2.3.44} and taking $a=\hat{a}\kappa N$, such phenomena are not seen since $W_1^{(L),k}(x)$ is non-zero for all $k\in\mathbb{N}$.

\begin{proposition} \label{prop2.14}
Retain the definitions of $h$, $\hat{a}$ and $y_{(L)}$ from Propositions \ref{prop2.1} and \ref{prop2.13}. For $\beta=1$ and $4$, the expansion coefficients of the scaled resolvent $\tilde{W}_1^{(L)}(x;N,\beta)$ satisfy the differential equations
\begin{equation} \label{eq2.4.23}
-xy_{(L)}^2\frac{\mathrm{d}}{\mathrm{d}x}W_1^{(L),0}(x)+\left[(\hat{a}+2)x-\hat{a}^2\right]W_1^{(L),0}(x)=x+\hat{a},
\end{equation}

\begin{multline} \label{eq2.4.24}
xy_{(L)}^4\frac{\mathrm{d}}{\mathrm{d}x}W_1^{(L),1}(x)-y_{(L)}^2\left[(\hat{a}+2)x-\hat{a}^2\right]W_1^{(L),1}(x)
\\=4h\hat{a}xy_{(L)}^2\frac{\mathrm{d}}{\mathrm{d}x}W_1^{(L),0}(x)+2h\hat{a}\left[x^2-3(\hat{a}+2)x+2\hat{a}^2\right]W_1^{(L),0}(x)
\\+2h\left[x(x-1)+2\hat{a}x+2\hat{a}^2\right],
\end{multline}
\begin{multline} \label{eq2.4.25}
xy_{(L)}^4\frac{\mathrm{d}}{\mathrm{d}x}W_1^{(L),2}(x)-y_{(L)}^2\left[(\hat{a}+2)x-\hat{a}^2\right]W_1^{(L),2}(x)
\\=4h\hat{a}xy_{(L)}^2\frac{\mathrm{d}}{\mathrm{d}x}W_1^{(L),1}(x)+2h\hat{a}\left[x^2-3(\hat{a}+2)x+2\hat{a}^2\right]W_1^{(L),1}(x)
\\+\tfrac{5}{2}x^3y_{(L)}^2\frac{\mathrm{d}^3}{\mathrm{d}x^3}W_1^{(L),0}(x)+x^2\left[8x^2-19(\hat{a}+2)x+11\hat{a}^2\right]\frac{\mathrm{d}^2}{\mathrm{d}x^2}W_1^{(L),0}(x)
\\+x\left[2x^2-6(\hat{a}+2)x+5\hat{a}^2\right]\frac{\mathrm{d}}{\mathrm{d}x}W_1^{(L),0}(x)
\\+2\left[(\hat{a}+2)x-\hat{a}^2\right]W_1^{(L),0}(x)-2(\hat{a}+x),
\end{multline}
and for $l\geq3$, the general differential equation
\begin{multline} \label{eq2.4.26}
xy_{(L)}^4\frac{\mathrm{d}}{\mathrm{d}x}W_1^{(L),l}(x)-y_{(L)}^2\left[(\hat{a}+2)x-\hat{a}^2\right]W_1^{(L),l}(x)
\\=4h\hat{a}xy_{(L)}^2\frac{\mathrm{d}}{\mathrm{d}x}W_1^{(L),l-1}(x)+2h\hat{a}\left[x^2-3(\hat{a}+2)x+2\hat{a}^2\right]W_1^{(L),l-1}(x)
\\+\tfrac{5}{2}x^3y_{(L)}^2\frac{\mathrm{d}^3}{\mathrm{d}x^3}W_1^{(L),l-2}(x)+x^2\left[8x^2-19(\hat{a}+2)x+11\hat{a}^2\right]\frac{\mathrm{d}^2}{\mathrm{d}x^2}W_1^{(L),l-2}(x)
\\+x\left[2x^2-6(\hat{a}+2)x+5\hat{a}^2\right]\frac{\mathrm{d}}{\mathrm{d}x}W_1^{(L),l-2}(x)+2\left[(\hat{a}+2)x-\hat{a}^2\right]W_1^{(L),l-2}(x)
\\-h\hat{a}x\left[5x^2\frac{\mathrm{d}^3}{\mathrm{d}x^3}+22x\frac{\mathrm{d}^2}{\mathrm{d}x^2}+14\frac{\mathrm{d}}{\mathrm{d}x}\right]W_1^{(L),l-3}(x)
\\-\left[x^5\frac{\mathrm{d}^5}{\mathrm{d}x^5}+10x^4\frac{\mathrm{d}^4}{\mathrm{d}x^4}+22x^3\frac{\mathrm{d}^3}{\mathrm{d}x^3}+4x^2\frac{\mathrm{d}^2}{\mathrm{d}x^2}-4x\frac{\mathrm{d}}{\mathrm{d}x}\right]W_1^{(L),l-4}(x),
\end{multline}
where we set $W_1^{(L),-1}:=0$.
\end{proposition}

Before moving onto the Cauchy ensembles, we present the differential equations characterising the expansion \eqref{eq2.4.10} in the JUE case with $a=\hat{a}\kappa N$, $b=\hat{b}\kappa N$ as discussed immediately prior to Proposition \ref{prop2.13}. We do not display the analogous results in the JOE and JSE cases due to their (relatively speaking) unwieldy forms, but note that they can be derived from equation \eqref{eq2.3.14} following the same proof as below.

\begin{proposition} \label{prop2.15}
Let $y_{(J)}:=\sqrt{(\hat{a}+\hat{b}+2)^2x^2-2(\hat{a}+2)(\hat{a}+\hat{b})x+\hat{a}^2}$ with $\hat{a},\hat{b}$ constant in $N$. The expansion coefficients specified by equation \eqref{eq2.4.10} in the JUE case with $a=\hat{a}\kappa N$ and $b=\hat{b}\kappa N$ satisfy the differential equation
\begin{multline} \label{eq2.4.27}
x(x-1)y_{(J)}^2\frac{\mathrm{d}}{\mathrm{d}x}W_1^{(J),0}(x)
\\-\left[\hat{a}^2(1-x)^3-\hat{a}(\hat{b}+2)x(1-x)(1-2x)-(\hat{b}+2)^2x^3+2(\hat{b}+1)(3x^2+x)\right]W_1^{(J),0}(x)
\\=(\hat{a}+\hat{b}+1)(\hat{a}(1-x)+\hat{b}x),
\end{multline}
and for $l\geq2$, the general differential equation
\begin{multline} \label{eq2.4.28}
x(1-x)y_{(J)}^2\frac{\mathrm{d}}{\mathrm{d}x}W_1^{(J),l}(x)
\\+\left[\hat{a}^2(1-x)^3-\hat{a}(\hat{b}+2)x(1-x)(1-2x)-(\hat{b}+2)^2x^3+2(\hat{b}+1)(3x^2+x)\right]W_1^{(J),l}(x)
\\=x^3(1-x)^3\frac{\mathrm{d}^3}{\mathrm{d}x^3}W_1^{(J),l-2}(x)+4x^2(1-x)^2(1-2x)\frac{\mathrm{d}^2}{\mathrm{d}x^2}W_1^{(J),l-2}(x)
\\-2x(1-x)(7x(1-x)-1)\frac{\mathrm{d}}{\mathrm{d}x}W_1^{(J),l-2}(x)-2x(1-x)(1-2x)W_1^{(J),l-2}(x).
\end{multline}
\end{proposition}
\begin{proof}
Substitute expansion \eqref{eq2.4.10} into equation \eqref{eq2.3.8} and set $a=\hat{a}N$, $b=\hat{b}N$. Equating terms of equal order in $N$ then produces the sought differential equations.
\end{proof}
Similar to Proposition \ref{prop2.13}, in the $\beta=2$, $a=\hat{a}N$, $b=\hat{b}N$ setting, $W_1^{(J),k}(x)$ vanishes for odd $k$, and the differential equations of Proposition \ref{prop2.15} constitute a recursion over even $l$.

\subsubsection{Differential equations for $W_1^{(Cy),l}(x)$ in the symmetric case $\alpha$ real}
The procedure outlined above extends to the Cauchy ensembles in the expected manner. Having already set $\alpha=\hat{\alpha}\kappa N$ (recall that $\eta=-\kappa(N-1)-1-\alpha$) to ensure that the global scaled eigenvalue density $\tilde{\rho}^{(Cy)}(\lambda)$ has the necessary properties, we simplify further by constraining $\hat{\alpha}$ to be a real and positive constant. This decision to consider only the symmetric Cauchy ensembles is simply due to the fact that the equivalent results in the non-symmetric case are cumbersome to display; they can however be derived in the same way. Thus, setting $\alpha=\hat{\alpha}\kappa N$ in equation \eqref{eq2.3.22} with $\hat{\alpha}={\rm O}(1)$ in $N$, substituting in the expansion \eqref{eq2.4.11} for $\tilde{W}_1^{(Cy)}(x)$, and then collecting terms of like order in $N$, we obtain the following two propositions.

\begin{proposition} \label{prop2.16}
In the Cauchy weight, set $\eta=-\kappa(N-1)-1-\alpha$, $\alpha=\hat{\alpha}\kappa N$ with $\hat{\alpha}$ positive real and constant in $N$. Moreover, let $y_{(Cy)}:=\sqrt{\hat{\alpha}^2x^2-2\hat{\alpha}-1}$. Then, the expansion coefficients of the symmetric CyUE scaled resolvent $\tilde{W}_1^{(Cy)}(x;N,2)\big|_{\hat{\alpha}\in\mathbb{R}_+}$ \eqref{eq2.4.11} satisfy the differential equation
\begin{equation} \label{eq2.4.29}
-(1+x^2)y_{(Cy)}^2\frac{\mathrm{d}}{\mathrm{d}x}W_1^{(Cy),0}(x)-\left[\hat{\alpha}^2x^2-(\hat{\alpha}+2)^2+2\right]xW_1^{(Cy),0}(x)=(\hat{\alpha}+1)(2\hat{\alpha}+1),
\end{equation}
and for $l\geq2$, the general differential equation
\begin{multline} \label{eq2.4.30}
4(1+x^2)y_{(Cy)}^2\frac{\mathrm{d}}{\mathrm{d}x}W_1^{(Cy),l}(x)+4\left[\hat{\alpha}^2x^2-(\hat{\alpha}+2)^2+2\right]xW_1^{(Cy),l}(x)
\\=(1+x^2)^3\frac{\mathrm{d}^3}{\mathrm{d}x^3}W_1^{(Cy),l-2}(x)+8x(1+x^2)^2\frac{\mathrm{d}^2}{\mathrm{d}x^2}W_1^{(Cy),l-2}(x)
\\+2(7x^2+3)(1+x^2)\frac{\mathrm{d}}{\mathrm{d}x}W_1^{(Cy),l-2}(x)+4x(1+x^2)W_1^{(Cy),l-2}(x).
\end{multline}
\end{proposition}
As with the other classical unitary ensembles, $\tilde{W}_1^{(Cy)}(x)$ is an odd function of $N$ when $\beta=2$, so $W_1^{(Cy),k}(x)=0$ when $k$ is odd. It can thus be easily checked that differential-difference equation \eqref{eq2.4.30} trivially holds true for odd values of $l$ and is otherwise to be interpreted as a recursion over even $l$.

\begin{proposition} \label{prop2.17}
Retain the notation of Proposition \ref{prop2.16}. For $\beta=1$ and $4$, the expansion coefficients of the scaled resolvent $\tilde{W}_1^{(Cy)}(x;N,\beta)\big|_{\hat{\alpha}\in\mathbb{R}_+}$ satisfy the differential equations
\begin{equation} \label{eq2.4.31}
-(1+x^2)y_{(Cy)}^2\frac{\mathrm{d}}{\mathrm{d}x}W_1^{(Cy),0}(x)-\left[\hat{\alpha}^2x^2-(\hat{\alpha}+2)^2+2\right]xW_1^{(Cy),0}(x)=(\hat{\alpha}+1)(2\hat{\alpha}+1),
\end{equation}

\begin{multline} \label{eq2.4.32}
2(1+x^2)y_{(Cy)}^4\frac{\mathrm{d}}{\mathrm{d}x}W_1^{(Cy),1}(x)+2y_{(Cy)}^2\left[\hat{\alpha}^2x^2-(\hat{\alpha}+2)^2+2\right]xW_1^{(Cy),1}(x)
\\=4(\kappa-1)\hat{\alpha}y_{(Cy)}^2(1+x^2)^2\frac{\mathrm{d}}{\mathrm{d}x}W_1^{(Cy),0}(x)+2(\kappa-1)\hat{\alpha}\left[2\hat{\alpha}^2x^2-(\hat{\alpha}+3)^2+6\right]x(1+x^2)W_1^{(Cy),0}(x)
\\+(\kappa-1)\hat{\alpha}\left[(8\hat{\alpha}^2+9\hat{\alpha}+2)x^2+(2\hat{\alpha}+1)(5\hat{\alpha}+4)\right],
\end{multline}

\begin{multline} \label{eq2.4.33}
8(1+x^2)y_{(Cy)}^4\frac{\mathrm{d}}{\mathrm{d}x}W_1^{(Cy),2}(x)+8y_{(Cy)}^2\left[\hat{\alpha}^2x^2-(\hat{\alpha}+2)^2+2\right]xW_1^{(Cy),2}(x)
\\=16(\kappa-1)\hat{\alpha}y_{(Cy)}^2(1+x^2)^2\frac{\mathrm{d}}{\mathrm{d}x}W_1^{(Cy),1}(x)
\\+8(\kappa-1)\hat{\alpha}\left[2\hat{\alpha}^2x^2-(\hat{\alpha}+3)^2+6\right]x(1+x^2)W_1^{(Cy),1}(x)+10(\kappa-1)^2y_{(Cy)}^2(1+x^2)^3\frac{\mathrm{d}^3}{\mathrm{d}x^3}W_1^{(Cy),0}(x)
\\+4(\kappa-1)^2\left[19\hat{\alpha}^2x^2-3\hat{\alpha}^2-22(2\hat{\alpha}+1)\right]x(1+x^2)^2\frac{\mathrm{d}^2}{\mathrm{d}x^2}W_1^{(Cy),0}(x)
\\+4(\kappa-1)^2\left[29\hat{\alpha}^2x^4-2\hat{\alpha}^2x^2-44(2\hat{\alpha}+1)x^2+\hat{\alpha}^2-12(2\hat{\alpha}+1)\right](1+x^2)\frac{\mathrm{d}}{\mathrm{d}x}W_1^{(Cy),0}(x)
\\+8(\kappa-1)^2\left[3\hat{\alpha}^2x^4-(\hat{\alpha}+8)^2x^2-4(2\hat{\alpha}+1)\right]W_1^{(Cy),0}(x)
\\-4(\kappa-1)^2\left[(3\hat{\alpha}^2-1)x^2+9\hat{\alpha}^2+10\hat{\alpha}+3\right],
\end{multline}
and for $l\geq3$, the general differential equation
\begin{multline} \label{eq2.4.34}
4(1+x^2)y_{(Cy)}^4\frac{\mathrm{d}}{\mathrm{d}x}W_1^{(Cy),l}(x)+4y_{(Cy)}^2\left[\hat{\alpha}^2x^2-(\hat{\alpha}+2)^2+2\right]xW_1^{(Cy),l}(x)
\\=8(\kappa-1)\hat{\alpha}y_{(Cy)}^2(1+x^2)^2\frac{\mathrm{d}}{\mathrm{d}x}W_1^{(Cy),l-1}(x)
\\+4(\kappa-1)\hat{\alpha}\left[2\hat{\alpha}^2x^2-(\hat{\alpha}+3)^2+6\right]x(1+x^2)W_1^{(Cy),l-1}(x)
\\+5(\kappa-1)^2y_{(Cy)}^2(1+x^2)^3\frac{\mathrm{d}^3}{\mathrm{d}x^3}W_1^{(Cy),l-2}(x)
\\+2(\kappa-1)^2\left[19\hat{\alpha}^2x^2-3\hat{\alpha}^2-22(2\hat{\alpha}+1)\right]x(1+x^2)^2\frac{\mathrm{d}^2}{\mathrm{d}x^2}W_1^{(Cy),l-2}(x)
\\+2(\kappa-1)^2\left[29\hat{\alpha}^2x^4-2\hat{\alpha}^2x^2-44(2\hat{\alpha}+1)x^2+\hat{\alpha}^2-12(2\hat{\alpha}+1)\right](1+x^2)\frac{\mathrm{d}}{\mathrm{d}x}W_1^{(Cy),l-2}(x)
\\+4(\kappa-1)^2\left[3\hat{\alpha}^2x^4-(\hat{\alpha}+8)^2x^2-4(2\hat{\alpha}+1)\right]W_1^{(Cy),l-2}(x)
\\-(\kappa-1)^3\hat{\alpha}\left[5(1+x^2)^4\frac{\mathrm{d}^3}{\mathrm{d}x^3}+38x(1+x^2)^3\frac{\mathrm{d}^2}{\mathrm{d}x^2}\right.
\\\left.+2(31x^2+15)(1+x^2)^2\frac{\mathrm{d}}{\mathrm{d}x}+4x(4x^4+9x^2+5)\right]W_1^{(Cy),l-3}(x)
\\-(\kappa-1)^4\bigg[(1+x^2)^5\frac{\mathrm{d}^5}{\mathrm{d}x^5}+20x(1+x^2)^4\frac{\mathrm{d}^4}{\mathrm{d}x^4}+(122x^2+29)(1+x^2)^3\frac{\mathrm{d}^3}{\mathrm{d}x^3}
\\+4x(65x^2+46)(1+x^2)^2\frac{\mathrm{d}^2}{\mathrm{d}x^2}+4(41x^4+56x^2+14)(1+x^2)\frac{\mathrm{d}}{\mathrm{d}x}
\\+8x(2x^4+4x^2+3)\bigg]W_1^{(Cy),l-4}(x)+4(\kappa-1)^3\hat{\alpha}x^2\chi_{l=3}+2(\kappa-1)^4(1-x^2)\chi_{l=4},
\end{multline}
where we have set $W_1^{(Cy),-1}:=0$.
\end{proposition}

\begin{remark} \label{R2.11}
\begin{enumerate}
\item It can be observed that throughout this subsection, the differential equations characterising the leading order terms $W_1^0(x)$ depend on the classical weight $w(\lambda)$, but are otherwise the same for all $\beta$. Thus, $W_1^0(x)$ is $\beta$-independent and use of the Sokhotski--Plemelj inversion formula \eqref{eq1.1.25} tells us that the large $N$ limiting form $\rho^0(\lambda)$ of $\tilde{\rho}(\lambda)$ also has this property. On the other hand, with $a,b={\rm O}(N)$, $W_1^1(x)$ vanishes for $\beta=2$ but can be seen to be non-zero for $\beta\ne2$, in keeping with Note~\ref{N1.1} following Proposition \ref{prop1.2}. In fact, solving the differential equations of Propositions~\ref{prop2.11}--\ref{prop2.15} with the boundary conditions \eqref{eq2.4.15}, \eqref{eq2.4.16} recovers expressions for $W_1^0(x)$ known from the general-$\beta$ works \citep{WF14}, \citep{FRW17}. Applying the Sokhotski--Plemelj formula \eqref{eq1.1.25} then recovers the expressions for $\rho^0(\lambda)$ given in Proposition \ref{prop1.2} and its $1/N,1/N^2$ corrections seen in \citep{WF14}, \citep{FRW17} (see also the earlier work \citep{ABMV12}).

\item According to the second point of Remark \ref{R1.9} (more specifically equation \eqref{eq1.2.88}), taking $a=\hat{a}\kappa N$ and $b=\hat{b}\kappa N$ with $\hat{a},\hat{b}$ constant in $N$ is equivalent to setting $\gamma_1=1+\hat{a}$ and $\gamma_2=1+\hat{b}$. Taking this into account, we see that the auxiliary functions $y_{(w)}$ used in Propositions \ref{prop2.11} to \ref{prop2.17} are related to the large $N$ limiting forms given in Proposition \ref{prop1.2}. To be precise, $y_{(G)}$, $y_{(L)}$, $y_{(J)}$, and $y_{(Cy)}$ are equal to the Stieltjes transform of $\rho^{(G),0}(\lambda)$, $2\lambda\rho^{(L),0}(\lambda)$, $2\lambda(1-\lambda)\rho^{(J),0}(\lambda)$, and $(1+\lambda^2)\rho^{(Cy),0}(\lambda)$, respectively. The significance of their presence in the differential equations of this subsection is that for $l>0$, the singularities of $W_1^l(x)$ lie exactly at the zeroes of $y_{(w)}$, a fact well known from the viewpoint of topological recursion \citep{EO09}.
\end{enumerate}
\end{remark}

We do not make explicit here the expressions for $W_1^l(x)$ nor $\rho^l(x)$ in the Gaussian, Laguerre, or Jacobi cases, since these can be found in the literature \citep{ABMV12}, \citep{WF14}, \citep{FRW17}. Instead, we demonstrate the structures discussed in the above remark in the Cauchy case.
\begin{proposition} \label{prop2.18}
Retain the notation of Proposition \ref{prop2.16}, with $\hat{\alpha}$ positive real and constant in $N$. For $\beta=1,2$, and $4$, we have
\begin{align}
W_1^{(Cy),0}(x)&=\frac{(\hat{\alpha}+1)x-y_{(Cy)}}{1+x^2},
\\ W_1^{(Cy),1}(x)&=\frac{(\kappa-1)\hat{\alpha}}{2}\left[\frac{1}{y_{(Cy)}}-\frac{\hat{\alpha}x}{y_{(Cy)}^2}\right],
\\ W_1^{(Cy),2}(x)&=\frac{(\kappa-1)^2\hat{\alpha}}{2}\left[\frac{\hat{\alpha}^2((4\hat{\alpha}^2+10\hat{\alpha}+5)x^2+2\hat{\alpha}+1)}{4y_{(Cy)}^5}-\frac{\hat{\alpha}(\hat{\alpha}^2+2\hat{\alpha}+1)x}{y_{(Cy)}^4}\right] \nonumber
\\&\qquad+\frac{\kappa}{8}\frac{\hat{\alpha}^2(2\hat{\alpha}+1)(1+x^2)}{y_{(Cy)}^5},
\end{align}
along with the large $N$ asymptotic for the smoothed eigenvalue density \eqref{eq2.4.4}
\begin{multline} \label{eq2.4.38}
\tilde{\rho}^{(Cy)}(\lambda)\Big|_{\hat{\alpha}\in\mathbb{R}_+}=\rho^{(Cy),0}(\lambda)+\frac{\kappa-1}{\kappa N}\left[\frac{\hat{\alpha}}{2\pi\sqrt{1+2\hat{\alpha}-\hat{\alpha}^2\lambda^2}}\chi_{|\lambda|<\sqrt{2\hat{\alpha}+1}/\hat{\alpha}}\right.
\\ \left.-\frac{1}{4}\delta(\lambda-\sqrt{2\hat{\alpha}+1}/\hat{\alpha})-\frac{1}{4}\delta(\lambda+\sqrt{2\hat{\alpha}+1}/\hat{\alpha})\right]+{\rm O}\left(\frac{1}{N^2}\right),
\end{multline}
where $\delta(\lambda)$ is the Dirac delta and $\rho^{(Cy),0}(\lambda)$ is specified by equation \eqref{eq1.2.19} with $\hat{\alpha}=\hat{\alpha}_1$ and $\hat{\alpha}_2=0$.
\end{proposition}
\begin{proof}
The general solution of equation \eqref{eq2.4.29}, equivalently \eqref{eq2.4.31}, is
\begin{equation*}
W_1^{(Cy),0}(x)=\frac{(\hat{\alpha}+1)x+Cy_{(Cy)}}{1+x^2},
\end{equation*}
with the integration constant $C$ forced to equal $-1$ due to the boundary condition \eqref{eq2.4.15}. Similarly, the integration constants present in the general solutions of equations \eqref{eq2.4.30}, \eqref{eq2.4.32}, and \eqref{eq2.4.33} must be zero due to the boundary condition \eqref{eq2.4.16}. Equation \eqref{eq2.4.38} is obtained by applying the Sokhotski--Plemelj formula \eqref{eq1.1.19} to the series \eqref{eq2.4.11} truncated to have two terms; note that the Stieltjes transform \eqref{eq1.1.18} of $\delta(\lambda-\lambda_0)$ is $1/(x-\lambda_0)$.
\end{proof}

Before moving on to the soft and hard edge scaling regimes, let us remark that the results contained in Proposition \ref{prop2.18} are expected to hold for general real $\beta>0$, in line with what is known about the other classical $\beta$ ensembles \citep{WF14}, \citep{FRW17}.

\subsection{Soft and hard edge scaled differential equations} \label{s2.4.2}
In this subsection we study the eigenvalue densities of the classical matrix ensembles after they have been shifted to be centred on either the largest or smallest eigenvalue and scaled such that the mean spacing between this eigenvalue and its neighbour is order unity. The large $N$ limits of the eigenvalue densities recentred and scaled in this manner have one of two universal forms, depending on whether the centring is performed at a soft or hard edge; we denote these large $N$ limiting forms $\rho^{(soft)}(\lambda;\beta)$ and $\rho^{(hard)}(\lambda;\beta)$, respectively. Recalling from the discussion following Proposition \ref{prop1.2}, centring on the largest (smallest) eigenvalue corresponds to a soft edge if $\rho^0(\lambda)$ exhibits a square root profile at the upper (lower) endpoint of its support, and a hard edge if it instead has an inverse square root profile. Thus, some observations can be made from the contents of Proposition \ref{prop1.2} using equation \eqref{eq1.2.88}: When the parameters $a,b,\alpha$ are proportional to $N$, i.e., upon setting $a=\hat{a}\kappa N$, $b=\hat{b}\kappa N$, and $\alpha=\hat{\alpha}\kappa N$ with $\hat{a},\hat{b},\hat{\alpha}$ constant in $N$, all of the edges of the classical $\beta$ ensembles are of the soft type. If one instead takes $a={\rm O}(1)$, the lower edge of the Laguerre and Jacobi $\beta$ ensembles is of the hard type. Likewise, if $b={\rm O}(1)$, $\rho^{(J),0}(\lambda)$ has a hard edge at the upper endpoint of its support. A point of interest we will observe is that while edge scaling is a procedure local to the endpoint of support being centred on, it depends on both parameters $a,b$ where applicable, even though $a$ ($b$) mainly governs the behaviour of the lower (upper) edge.

The particular scalings required to ensure that the large $N$ limits of the eigenvalue densities of the classical matrix ensembles share the same universal forms can be a little complicated, so we (re)introduce some notation. Recall from the definitions of $a,b,\gamma_1,\gamma_2$ in Propositions \ref{prop1.1} and \ref{prop1.2} and equation \eqref{eq1.2.88} contained in Remark \ref{R1.9} that
\begin{align}
\gamma_1&=\lim_{N\to\infty}\frac{M_1}{N}=1+\lim_{N\to\infty}\frac{a-\kappa+1}{\kappa N},
\\ \gamma_2&=\lim_{N\to\infty}\frac{M_2}{N}=1+\lim_{N\to\infty}\frac{b-\kappa+1}{\kappa N},
\end{align}
where $M_1,M_2$ are now interpreted as continuous parameters dependent on $a,b$. Thus in the large $N$ limit, $\gamma_1,\gamma_2=1$ when $a,b={\rm O}(1)$, while if we set $a=\hat{a}\kappa N$ and $b=\hat{b}\kappa N$ with $\hat{a},\hat{b}={\rm O}(1)$, we have $\gamma_1=1+\hat{a}$ and $\gamma_2=1+\hat{b}$. Recall too from Proposition \ref{prop1.2} that $\lambda_{\pm}^{(MP)},\lambda_{\pm}^{(Wac)},\lambda_{\pm}^{(Cy)}$ denote the endpoints of the supports of $\tilde{\rho}^{(L)}(\lambda),\tilde{\rho}^{(J)}(\lambda),\tilde{\rho}^{(Cy)}(\lambda)$, respectively; in the following, we only consider the symmetric Cauchy ensembles, so we set $\hat{\alpha}\in\mathbb{R}$ and $\lambda_{\pm}^{(Cy)}=\pm\sqrt{2\hat{\alpha}+1}/\hat{\alpha}$. Let us also define
\begin{equation}
\gamma_3:=\frac{(\hat{\alpha}+1)^2}{\hat{\alpha}^3} \label{eq2.4.41}
\end{equation}
for convenience and
\begin{align}
q^2&:=\frac{M_1}{M_1+M_2},\qquad r^2:=\frac{M_2}{M_1+M_2}, \label{eq2.4.42}
\\\tilde{q}^2&:=\frac{N}{M_1+M_2},\qquad \tilde{r}^2:=1-\tilde{q}^2,
\\ u_N&:=\left[\frac{qr\tilde{q}\tilde{r}\sqrt{M_1+M_2}}{\tilde{q}\tilde{r}(q^2-r^2)+qr(\tilde{q}^2-\tilde{r}^2)}\right]^{4/3} \label{eq2.4.44}
\end{align}
following \citep{HF12}.

\subsubsection{Differential equations characterising the soft edge}
We now present differential equations satisfied by $\rho^{(soft)}(\lambda)$ for $\beta\in\{2/3,1,2,4,6\}$.
\begin{theorem} \label{thrm2.3}
Define the soft edge limiting forms of the differential operators $\mathcal{D}_{N,\beta}$ introduced in Section~\ref{s2.3} as
\begin{equation} \label{eq2.4.45}
\mathcal{D}_{\beta}^{(soft)} =
\begin{cases}
\frac{\mathrm{d}^3}{\mathrm{d}x^3}-4x\frac{\mathrm{d}}{\mathrm{d}x}+2,&\beta=2,
\\ \frac{\mathrm{d}^5}{\mathrm{d}x^5}-10\kappa x\frac{\mathrm{d}^3}{\mathrm{d}x^3}+6\kappa\frac{\mathrm{d}^2}{\mathrm{d}x^2}+16\kappa^2x^2\frac{\mathrm{d}}{\mathrm{d}x}-8\kappa^2x,&\beta=1,4,
\\ 3\frac{\mathrm{d}^7}{\mathrm{d}x^7}-56\kappa x\frac{\mathrm{d}^5}{\mathrm{d}x^5}+28\kappa\frac{\mathrm{d}^4}{\mathrm{d}x^4}+\frac{784}{3}\kappa^2x^2\frac{\mathrm{d}^3}{\mathrm{d}x^3}
\\ \quad-208\kappa^2x\frac{\mathrm{d}^2}{\mathrm{d}x^2}-4\kappa^2(64\kappa x^3-17)\frac{\mathrm{d}}{\mathrm{d}x}+128\kappa^3x^2,&\beta=2/3,6,
\end{cases}
\end{equation}
where we recall that $\kappa=\beta/2$. Then, for $\beta\in\{2/3,1,2,4,6\}$, the Gaussian, Laguerre, Jacobi, and symmetric Cauchy $\beta$ ensembles' eigenvalue densities satisfy the following differential equation in the soft edge limit:
\begin{equation} \label{eq2.4.46}
\mathcal{D}_{\beta}^{(soft)}\,\rho^{(soft)}(x;\beta) = 0.
\end{equation}
\end{theorem}
\begin{proof}
Make the change of variables \citep{Fo93a} $x\mapsto\sqrt{\kappa}\left(\sqrt{2N}+\frac{x}{\sqrt{2}N^{1/6}}\right)$ in equations \eqref{eq2.0.3} and \eqref{eq2.3.53}. Then, multiply through by $N^{-1/2}$ for $\beta=2$, $N^{-5/6}$ for $\beta=1$ and $4$, or $N^{-7/6}$ for $\beta=2/3$ and $6$. Equating terms of order one then yields equation \eqref{eq2.4.46} above, while all other terms vanish in the $N\rightarrow\infty$ limit.
\end{proof}

In the case $\beta=2$, the differential equation \eqref{eq2.4.46} satisfied by the soft edge density has been isolated in the earlier works
of Brack et al.~\citep[Eq.~(C.2) with $\hbar^2/m=1$,
$\lambda_M - V(r) = - {1 \over 2} r$, $D=1$]{BKMR10} and of Dean et al.~\citep[Eq.~(207) with $d=1$]{DDMS16}. For $\beta=1,2$, and $4$, the above proof can be replicated by instead considering the differential equations \eqref{eq2.3.7}, \eqref{eq2.3.13}, \eqref{eq2.3.21} and \eqref{eq2.3.43} for the eigenvalue densities of the Jacobi, symmetric Cauchy, and Laguerre ensembles, due to universality. Explicitly, differential equation \eqref{eq2.4.46} is satisfied by the leading order term of the following scaled densities in the large $N$ limit:
\begin{itemize}
\item In the regime of the largest eigenvalue \citep{Fo93a}, \citep{BFP98}, \citep{FT18}, \citep{FT19},
\begin{equation} \label{eq2.4.47}
\rho^{(G)}\left(\sqrt{\kappa}\left(\sqrt{2N}+\delta^{(G)}+\frac{x}{\sqrt{2}N^{1/6}}\right);N,\beta\right),
\end{equation}
where $\delta^{(G)}={\rm o}(N^{-1/6})$ is an arbitrary parameter (the regime of the smallest eigenvalue can be treated by exploiting the symmetry $x\leftrightarrow-x$);
\item In the regime of the largest eigenvalue \citep{Fo93a}, \citep{Jo01}, \citep{Fo12a}, \citep{FT18}, \citep{FT19},
\begin{equation} \label{eq2.4.48}
\rho^{(L)}\left(\kappa\left(\lambda_+^{(MP)}N+\delta_+^{(L,soft)}+\frac{(\sqrt{\gamma_1}+1)^{4/3}}{\sqrt{\gamma_1}}N^{1/3}x\right);N,\beta\right),
\end{equation}
where $\delta^{(L,soft)}_\pm={\rm o}(N^{1/3})$ is henceforth an arbitrary parameter. Note that this choice of scaling is effective whether we set $a=\hat{a}\kappa N$ or take $a$ to be constant in $N$. In the latter case, the argument of \eqref{eq2.4.48} simplifies to $\kappa(4N+\delta_+^{(L,soft)}+2^{4/3}N^{1/3}x)$;
\item In the regime of the smallest eigenvalue, having set $a=\hat{a}\kappa N$ with $\hat{a}={\rm O}(1)$ \citep{BFP98}, \citep{Fo12a},
\begin{equation} \label{eq2.4.49}
\rho^{(L)}\left(\kappa\left(\lambda_-^{(MP)}N-\delta^{(L,soft)}_- - (\sqrt{\gamma_1}-1)^{4/3}\left(\frac{N}{\sqrt{\gamma_1}}\right)^{1/3}x\right);N,\beta\right);
\end{equation}
\item In the regime of the largest eigenvalue, with $b=\hat{b}\kappa N$ and $\hat{b}={\rm O}(1)$ \citep{Jo08}, \citep{Fo12a}, \citep{HF12},
\begin{equation}
\rho^{(J)}\left(\lambda_+^{(Wac)}+\delta^{(J,soft)}+\frac{qr\tilde{q}\tilde{r}}{u_N}x;N,\beta\right),
\end{equation}
where $\delta^{(J,soft)}={\rm o}(N^{-2/3})$ is arbitrary and $q,r,\tilde{q},\tilde{r},u_N$ are defined in \eqref{eq2.4.42}--\eqref{eq2.4.44}. Like with \eqref{eq2.4.48}, this scaling is effective regardless of whether $a$ is constant or linear in $N$ (we must take $b={\rm O}(N)$ to ensure that this is actually a soft edge). The regime of the smallest eigenvalue can be treated by using the symmetry $(x,a,b)\leftrightarrow(1-x,b,a)$;
\item In the regime of the largest eigenvalue, having set $\alpha=\hat{\alpha}\kappa N$ with $\hat{\alpha}$ positive real and constant in $N$,
\begin{equation} \label{eq2.4.51}
\rho^{(Cy)}\left(\lambda_+^{(Cy)}+\delta^{(Cy)}+\left(\frac{\gamma_3^2}{2\lambda_+^{(Cy)}N^2}\right)^{1/3}x;N,\beta\right),
\end{equation}
where $\delta^{(Cy)}={\rm o}(N^{-2/3})$ is an arbitrary parameter and $\gamma_3=(\hat{\alpha}+1)^2/\hat{\alpha}^3$ \eqref{eq2.4.41}.
\end{itemize}

\begin{remark}
For even $\beta$, $\rho^{(soft)}(x)$ has an explicit representation as a $\beta$-dimensional integral due to \citep{DF06}, while \citep{FFG06} provides alternate forms for $\beta=1,2$, and $4$. To date, there is no explicit functional form for $\rho^{(soft)}(x)$ when $\beta=2/3$. On the other hand, \citep{DV13} shows for all $\kappa=\beta/2>0$ that for the first two leading orders as $x\rightarrow\infty$,
\begin{equation*}
\rho^{(soft)}(x)\propto\frac{\exp\left(-4\kappa x^{3/2}/3\right)}{x^{3\kappa/2}},
\end{equation*}
which is an extension of the even-$\beta$ result of \citep{Fo12},
\begin{equation} \label{eq2.4.52}
\rho^{(soft)}(x)\underset{x\rightarrow\infty}{\sim}\frac{1}{\pi}\frac{\Gamma(1+\kappa)}{(8\kappa)^{\kappa}}\frac{\exp\left(-4\kappa x^{3/2}/3\right)}{x^{3\kappa/2}}.
\end{equation}
This result is consistent with the differential equations of Theorem \ref{thrm2.3}. So too is the result \citep{DF06},
\begin{equation} \label{eq2.4.53}
\rho^{(soft)}(x)\underset{x\rightarrow-\infty}{\sim}\frac{\sqrt{|x|}}{\pi},\quad\beta\in2\mathbb{N}.
\end{equation}
Likewise, Theorem \ref{thrm2.4} below is consistent with the result \citep{Fo94},
\begin{equation} \label{eq2.4.54}
\rho^{(hard)}(x)\underset{x\rightarrow\infty}{\sim}\frac{1}{2\pi\sqrt{x}},\quad\beta\in2\mathbb{N}.
\end{equation}
Note that the asymptotic forms \eqref{eq2.4.52}, \eqref{eq2.4.53}, and \eqref{eq2.4.54} respectively capture the facts that moving past the soft edge results in exponential decay, moving from the soft edge into the bulk results in a square root profile, and moving from the hard edge into the bulk shows a square root singularity.
\end{remark}

\subsubsection{Differential equations characterising the hard edge}
We now give the hard edge analogue of Theorem \ref{thrm2.3} for $\beta\in\{1,2,4\}$. For uniformity across the Laguerre and Jacobi ensembles, we study the lower edge centred at $x=0$ (the hard edge at $x=1$ for the Jacobi ensemble can be studied by interchanging $x$ with $1-x$ and $a$ with $b$). In contrast to the soft edge scaling, the differential equations characterising the hard edge are seen to depend on the $a$ paramater, which is taken to be constant in $N$.
\begin{theorem} \label{thrm2.4}
Define the hard edge limiting forms of the differential operators $\mathcal{D}_{N,\beta}$ specified in Section~\ref{s2.3} as
\begin{equation} \label{eq2.4.55}
\mathcal{D}_{\beta}^{(hard)} =
\begin{cases}
x^3\frac{\mathrm{d}^3}{\mathrm{d}x^3}+4x^2\frac{\mathrm{d}^2}{\mathrm{d}x^2}+\left[x-a^2+2\right]x\frac{\mathrm{d}}{\mathrm{d}x}+\tfrac{1}{2}x-a^2,&\beta=2,
\\ 4x^5\frac{\mathrm{d}^5}{\mathrm{d}x^5}+40x^4\frac{\mathrm{d}^4}{\mathrm{d}x^4}+\left[10\kappa x-5\tilde{a}+88\right]x^3\frac{\mathrm{d}^3}{\mathrm{d}x^3}&
\\\quad+\left[38\kappa x-22\tilde{a}+16\right]x^2\frac{\mathrm{d}^2}{\mathrm{d}x^2}&
\\\quad+\left[\left(2\kappa x-\tilde{a}\right)^2+12\kappa x-14\tilde{a}-16\right]x\frac{\mathrm{d}}{\mathrm{d}x}&
\\\quad+(2\kappa x-\tilde{a})(\kappa x-\tilde{a})-4\kappa x,&\beta=1,4,
\end{cases}
\end{equation}
where we retain the definition of $\tilde{a}$ given in Theorem \ref{thrm2.2}. Then, for $\beta\in\{1,2,4\}$, the hard edge scaled eigenvalue densities of the Laguerre and Jacobi $\beta$ ensembles satisfy the differential equation
\begin{equation} \label{eq2.4.56}
\mathcal{D}_{\beta}^{(hard)}\,\rho^{(hard)}(x;\beta) = 0.
\end{equation}
\end{theorem}
\begin{proof}
Since the Gaussian ensembles do not exhibit a hard edge, we turn to the Laguerre ensembles as they are the next simplest to work with. Thus, we begin by changing variables \citep{Fo93a}, \citep{NF95} $x\mapsto\kappa x/(4N)$ in equation \eqref{eq2.3.43}. Equating terms of order one yields the differential equation \eqref{eq2.4.56} above, and we note that all other terms are ${\rm O}(\frac{1}{N})$.
\end{proof}

As in the soft edge case, there is a little bit of freedom in the choice of scaling used in the above proof, and the proof itself can be formulated in terms of the differential equations \eqref{eq2.3.7} and \eqref{eq2.3.13} characterising the eigenvalue densities of the Jacobi ensembles. To be precise, the differential equation \eqref{eq2.4.56} above is satisfied by the leading order term of the following scaled densities in the large $N$ limit:
\begin{itemize}
\item With $\delta^{(L,hard)}={\rm o}(N)$ an arbitrary parameter \citep{Fo93a}, \citep{NF95}, \citep{FT19a},
\begin{equation} \label{eq2.4.57}
\rho^{(L)}\left(\frac{\kappa x}{4N+\delta^{(L,hard)}};\beta\right);
\end{equation}
\item With $\delta^{(J,hard)}={\rm o}(N)$ also an arbitrary parameter,
\begin{equation} \label{eq2.4.58}
\rho^{(J)}\left(\frac{x}{N(4\gamma_2N+\delta^{(J,hard)})};\beta\right).
\end{equation}
While $a$ must be constant in $N$, applying this scaling to the differential equations of Section \ref{s2.3} recovers Theorem \ref{thrm2.4} regardless of whether we take $b$ to be proportional to $N$ or constant.
\end{itemize}

\subsubsection{Optimal scaling at the soft and hard edges}
It has previously been observed in \citep{FT18} that the differential equation characterisation of the soft edge scaled eigenvalue density can be extended to similarly characterise the optimal leading order correction term. The latter is obtained by a tuning of the $\delta^{(\,\cdot\,)}$ parameters in equations \eqref{eq2.4.47}--\eqref{eq2.4.51} so as to obtain the fastest possible decay in $N$ of the leading order correction, and thus the fastest possible convergence to the limit. On this latter point, and considering the Gaussian case for definiteness, we know from \cite{FT19} that for $\beta$ even (at least),
\begin{align}
\hat{\rho}_{N,\beta}(x) &:= {\sqrt{\kappa} \over \sqrt{2} N^{7/6}} 
\rho^{(G)}\left(\sqrt{\kappa}\left(\sqrt{2N}+\delta^{(G)}+\frac{x}{\sqrt{2}N^{1/6}}\right);N,\beta\right) \label{eq2.4.59}
\\ &\;=\rho^{(soft)}(x;\beta) + \Big (
\sqrt{2} N^{1/6} \delta^{(G)} - (1 - 1/\kappa)/(2 N^{1/3}) \Big )
{\mathrm{d} \over \mathrm{d}x} \rho^{(soft)}(x;\beta) \nonumber
\\&\qquad+ {\rm O}(N^{-2/3}), \label{eq2.4.60}
\end{align}
where the ${\rm O}(N^{-2/3})$ terms do not depend as simply on $\rho^{(soft)}(x),{\mathrm{d}\over\mathrm{d}x}\rho^{(soft)}(x)$ as those made explicit above. Thus, choosing $\delta^{(G)} =  (1 - 1/\kappa)/(2 \sqrt{2N})$ gives the fastest
convergence to the limit,
$$
\hat{\rho}_{N,\beta}(x) = \rho^{(soft)}(x;\beta) + {1 \over N^{2/3}} \zeta_{\beta}(x) +
{\rm o}(N^{-2/3})
$$
for some $\zeta_{\beta}(x)$ which, for the values of $\beta$ permitting a differential equation
characterisation of $\rho^{(G)}(x)$ and $\rho^{(soft)}(x)$, can itself be characterised as the solution
of a differential equation. The simplest case is $\beta = 2$, when
\begin{equation} \label{eq2.4.61}
\zeta_2'''(x) - 4 x \zeta_2'(x) + 2 \zeta_2(x) = x^2 {\mathrm{d}\over \mathrm{d}x} \rho^{(soft)}(x;2) - x \rho^{(soft)}(x;2),
\end{equation}
which is an inhomogeneous generalisation of equation \eqref{eq2.4.46} for $\beta = 2$. For the particular Laguerre soft edge scaling \eqref{eq2.4.48}, again considering $\beta=2$, the analogue of equation \eqref{eq2.4.61} is given in \cite[Eq.~(4.19)]{FT18}, with $a=\hat{a}\kappa N$ and $\delta^{(L,soft)}_+=0$.

More generally, multiplying the functions \eqref{eq2.4.47}--\eqref{eq2.4.51} by appropriate scalars and powers of $N$ results in a function whose large $N$ asymptotic expansion has leading order term given by $\rho^{(soft)}(x)$ and next to leading order correction some function generically ${\rm O}(N^{-1/3})$ (when $\beta=2$ and not all relevant parameters are constant, the correction term is actually of order $N^{-2/3}$). Likewise, multiplying the functions \eqref{eq2.4.57} and \eqref{eq2.4.58} by the correct pre-factors produces a function whose large $N$ limit is given by $\rho^{(hard)}(x)$ with correction term of generic order $1/N$. In a similar fashion to the computation \eqref{eq2.4.60} above, proper tuning of the $\delta^{(\,\cdot\,)}$ parameters seen in \eqref{eq2.4.47}--\eqref{eq2.4.51}, \eqref{eq2.4.57}, \eqref{eq2.4.58} ensures that the correction terms are ${\rm O}(N^{-2/3})$ in the soft edge case, and ${\rm O}(1/N^2)$ in the hard edge case, which corresponds to the scaled eigenvalue densities experiencing the optimal rate of convergence to the limiting forms $\rho^{(soft)}(x)$ and $\rho^{(hard)}(x)$, respectively.

For $\beta$ an even integer, \citep{FT19} shows that the scaling \eqref{eq2.4.48} becomes optimal when we set
\begin{equation} \label{eq2.4.62}
\delta^{(L,soft)}_+=\begin{cases}\frac{2a}{\kappa},&a={\rm O}(1),\\ \left(1-\frac{1}{\kappa}\right)\frac{\gamma_1-1}{2\sqrt{\gamma_1}},&a={\rm O}(N).\end{cases}
\end{equation}
Likewise, \citep{FT19a} shows for general real $\beta>0$ and $a\in\mathbb{N}$ that the scaling \eqref{eq2.4.57} is optimal upon setting $\delta^{(L,hard)}=2a/\kappa$, which coincides with the value of $\delta^{(L,soft)}_+$ given above for the $a={\rm O}(1)$ setting. We are able to extend these results by utilising Theorems \ref{thrm2.3} and~\ref{thrm2.4} as follows: Letting $\hat{\rho}_{N,\beta}(x)$ denote the analogue of \eqref{eq2.4.59} constructed from either of the functions \eqref{eq2.4.47}--\eqref{eq2.4.51}, \eqref{eq2.4.57}, or \eqref{eq2.4.58}, we have that $\hat{\rho}_{N,\beta}(x)$ is, at leading order, equal to either $\rho^{(soft)}(x;\beta)$ or $\rho^{(hard)}(x;\beta)$. Then, $\hat{\rho}_{N,\beta}(x)$ satisfies an edge-scaled version of one of the differential equations of Section \ref{s2.3}, while $\rho^{(soft)}(x;\beta)$ and $\rho^{(hard)}(x;\beta)$ satisfy the differential equations of Theorems \ref{thrm2.3} and \ref{thrm2.4}. Hence, the correction term $\hat{\rho}_{N,\beta}(x)-\rho^{(soft)}(x;\beta)$, respectively $\hat{\rho}_{N,\beta}(x)-\rho^{(hard)}(x;\beta)$, can be seen to satisfy a certain differential equation. Analysis of this latter differential equation reveals the order in $N$ of the correction term and its dependence on the $\delta^{(\,\cdot\,)}$ parameters. Thus, for $\beta=1,2$, and $4$, we are able to identify the values of $\delta^{(\,\cdot\,)}$ which optimally tune the scalings \eqref{eq2.4.47}--\eqref{eq2.4.49}, \eqref{eq2.4.51}, \eqref{eq2.4.57}, and \eqref{eq2.4.58}. We confirm that the values $\delta^{(G)}=(1-1/\kappa)/(2\sqrt{2N})$, $\delta^{(L,soft)}_+$ \eqref{eq2.4.62}, and $\delta^{(L,hard)}=2a/\kappa$ recounted above from the works \citep{FT18}, \citep{FT19}, \citep{FT19a} remain optimal when $\beta=1$ and/or $a>-1$ is allowed to be general real. Moreover, we find that the scalings \eqref{eq2.4.49}, \eqref{eq2.4.51} and \eqref{eq2.4.58} are optimal when taking
\begin{align}
\delta^{(L,soft)}_-&=\left(1-\frac{1}{\kappa}\right)\frac{\gamma_1-1}{2\sqrt{\gamma_1}},
\\ \delta^{(Cy)}&=\left(1-\frac{1}{\kappa}\right)\frac{\gamma_3}{2\lambda_+^{(Cy)}N},
\\ \delta^{(J,hard)}&=\begin{cases}\frac{4(a+b)}{\kappa}-4\left(1-\frac{1}{\kappa}\right),&b={\rm O}(1),\\ \frac{2a}{\kappa}(\gamma_2+1)-4\left(1-\frac{1}{\kappa}\right),&b={\rm O}(N).\end{cases}
\end{align}
Note that the value of $\delta^{(L,soft)}_-$ given above is equal to that of $\delta^{(L,soft)}_+$ given in equation \eqref{eq2.4.62} in the $a={\rm O}(N)$ setting. Moreover, the value of $\delta^{(J,hard)}$ given above for the $b={\rm O}(1)$ case agrees with computations seen in the recent study \citep[App.~A]{FL20}. Due to computational limitations, we have not been able to identify the optimal value of $\delta^{(J,soft)}$ for $\beta=1,4$, but note that its value in the $\beta=1$ case can be surmised from \citep{Jo08} (which presumably extends to the $\beta=4$ case through the duality principles of Lemma \ref{L1.4}), while for $\beta=2$, it should reduce to zero in the same fashion as $\delta^{(G)},\delta^{(L,soft)}_-,\delta^{(Cy)}$.

\chapter{Characterisations of the Moments and Cumulants}
It was mentioned in \S\ref{s1.1.1} that the eigenvalue densities $\rho(\lambda)$ of the random matrix ensembles studied in this thesis are fully characterised by their spectral moments (though, one must keep in mind the caveat regarding the Cauchy ensembles)
\begin{equation} \label{eq3.0.1}
m_k=\int_{\mathbb{R}}\lambda^k\rho(\lambda)\,\mathrm{d}\lambda=\mean{\sum_{i=1}^N\lambda_i^k},\quad k\in\mathbb{N}.
\end{equation}
In the above, the latter presentation of the spectral moments is to be read as an average with respect to the eigenvalue j.p.d.f.~of the matrix ensemble of interest. We begin this chapter with an application of the differential equations given in Section \ref{s2.3} to the derivation of linear recurrences characterising the spectral moments of the classical matrix ensembles (and the Gaussian $\beta$ ensembles with $\beta=2/3,6$). We recall from Definition \ref{def1.5} that the classical matrix ensembles are exactly those whose eigenvalue j.p.d.f.s are of the form
\begin{equation}
p^{(w)}(\lambda_1,\ldots,\lambda_N;\beta)=\frac{1}{\mathcal{N}_{N,\beta}^{(w)}}\prod_{i=1}^Nw(\lambda_i)\,|\Delta_N(\lambda)|^{\beta},
\end{equation}
where $\beta=1,2$, or $4$, 
\begin{equation*}
\Delta_N(\lambda)=\prod_{1\leq i<j\leq N}(\lambda_j-\lambda_i)
\end{equation*}
is the Vandermonde determinant \eqref{eq1.1.9}, and $w(\lambda)$ is a suitable classical weight. In particular, we are interested in the spectral moments of the Gaussian, Laguerre, Jacobi, symmetric shifted Jacobi, and symmetric Cauchy ensembles, which are specified respectively by the weights \eqref{eq1.2.9}, \eqref{eq1.2.29}
\begin{align}
w^{(G)}(\lambda)&=e^{-\lambda^2},
\\ w^{(L)}(\lambda)&=\lambda^ae^{-\lambda}\chi_{\lambda>0},
\\ w^{(J)}(\lambda)&=\lambda^a(1-\lambda)^b\chi_{0<\lambda<1},
\\ w^{(sJ)}(\lambda)\big|_{a=b}&=(1-\lambda^2)^a\chi_{-1<\lambda<1}, \label{eq3.0.6}
\\ w^{(Cy)}(\lambda)\big|_{\alpha\in\mathbb{R}}&=(1+\lambda^2)^{-\kappa(N-1)-1-\alpha}; \label{eq3.0.7}
\end{align}
we recall that $\kappa:=\beta/2$ and the indicator function $\chi_A$ is defined to equal one when $A$ is true and zero otherwise. The parameters $a,b,\alpha$ are taken to be real throughout this chapter and, for convergence issues, are further constrained such that $a,b>-1$ and $\alpha>-1/2$.

The aforementioned linear recurrences on the spectral moments are given in \S\ref{s3.1.1}, where we also show that these recurrence relations hold true for the negative-integer moments ($m_{-k}$ with $k\in\mathbb{N}$), which have gained interest primarily due to their connection to problems of quantum transport \citep{BFB97}, \citep{MS11}, \citep{MS12}, \citep{CMSV16b} and more recently due to duality principles relating the negative-integer moments $m_{-k}$ to their counterparts $m_{k-1}$ \citep{CMOS19}; in fact, we show moreover that our moment recurrences are satisfied by a further generalisation (see equation \eqref{eq3.1.1} forthcoming) of the $m_k$ concerning complex values of $k$, which have recently been studied in \citep{CMOS19}, \citep{ABGS20}. Having linear recurrence relations on the spectral moments of the classical matrix ensembles at hand means that we are able to derive so-called $1$-point recursions \citep{CD21} characterising the moment expansion coefficients defined in Lemma \ref{L1.3}. To be precise, we proceed in \S\ref{s3.1.2} by substituting the expansions
\begin{align}
m_{2k}^{(G)}&=\sum_{l=0}^kM_{k,l}^{(G)}N^{1+k-l}, \label{eq3.0.8}
\\ m_{k}^{(L)}&=\sum_{l=0}^kM_{k,l}^{(L)}N^{1+k-l}, \label{eq3.0.9}
\\ m_{k}^{(J)}&=\sum_{l=0}^\infty M_{k,l}^{(J)}N^{1-l} \label{eq3.0.10}
\end{align}
into the moment recurrences of \S\ref{s3.1.1} and equating terms of like order in $N$ to obtain $1$-point recursions on the $M_{k,l}^{(w)}$. Following \citep[Defn.~1]{CD21}, this means that we find for each ensemble some integers $i_{\mathrm{max}},j_{\mathrm{max}}$ and a non-trivial set of polynomials $\{q^{(w)}_{ij}(k)\}_{0\leq i\leq i_{\mathrm{max}},0\leq j\leq j_{\mathrm{max}}}$ such that whenever the involved terms are well-defined, we have the equality
\begin{equation} \label{eq3.0.11}
\sum_{i=0}^{i_{\mathrm{max}}}\sum_{j=0}^{j_{\mathrm{max}}}q^{(w)}_{ij}(k)M^{(w)}_{k-i,l-j}=0.
\end{equation}
We present recursions of this type for the moment expansion coefficients of the Gaussian $\beta$ ensembles with $\beta=2/3,6$, the (classical) Laguerre ensembles, and the (symmetric shifted) Jacobi unitary ensemble, choosing to forgo consideration of the remaining classical matrix ensembles in the interest of brevity. However, we confirm here that applying our method to the GUE recovers the celebrated Harer--Zagier recursion \citep{HZ86}
\begin{equation} \label{eq3.0.12}
4(k+1)M_{k,l}^{(GUE)}=4(2k-1)M_{k-1,l}^{(GUE)}+(k-1)(2k-1)(2k-3)M_{k-2,l-2}^{(GUE)}
\end{equation}
subject to the initial conditions $M_{0,0}^{(GUE)}=2M_{1,0}^{(GUE)}=1$, $M_{1,1}^{(GUE)}=0$, and $M_{k,l}^{(GUE)}=0$ for all $k,l<0$, whereas treating instead the GOE and GSE recovers the corresponding recursions of Ledoux given in \citep{Le09}.

Evidently, the problem of computing the spectral moments of the classical matrix ensembles is not a new one. Aside from recurrences on the moments $m_k$ and their expansion coefficients $M_{k,l}$ known from the works \citep{HZ86}, \citep{HT03}, \citep{Le04}, \citep{Le09}, \citep{WF14}, \citep{CMSV16b}, \citep{CMOS19}, \citep{ABGS20} (which the computations of Section \ref{s3.1} recover and/or supplement), the literature contains a range of different approaches for studying the spectral moments of interest. In Section \ref{s3.2}, we review some of these alternative characterisations of the spectral moments of the classical matrix ensembles; in short, we survey parts of the literature that show how the study of these spectral moments relates to skew-orthogonal polynomial theory, symmetric function theory, and hypergeometric orthogonal polynomials from the Askey scheme.

After the review of Section \ref{s3.2}, the remainder of this chapter focuses on elucidating how the spectral moments $m_k$, mixed moments \eqref{eq1.1.27}
\begin{equation} \label{eq3.0.13}
m_{k_1,\ldots,k_n}=\mean{\prod_{i=1}^n\Tr\,X^{k_i}}_{P(X)},\quad k_1,\ldots,k_n\in\mathbb{N},
\end{equation}
and mixed cumulants $c_{k_1,\ldots,k_n}$ of particular random matrix ensembles relate to the enumeration of ribbon graphs satisfying certain constraints. Here, the term `ribbon graph' needs to be formally introduced and it is helpful to reiterate the definition of `mixed cumulants'. The mixed cumulants $c_{\kappa_i}$ are defined implicitly by the moment-cumulants relation \eqref{eq1.1.28}
\begin{equation} \label{eq3.0.14}
m_{k_1,\ldots,k_n}=\sum_{K\vdash\{k_1,\ldots,k_n\}}\prod_{\kappa_i\in K}c_{\kappa_i},
\end{equation}
where we recall that $K\vdash\{k_1,\ldots,k_n\}$ is read as ``$K$ is a partition of $\{k_1,\ldots,k_n\}$'', which means that $K=\{\kappa_i\}_{i=1}^m$ for some $1\leq m\leq n$ such that the disjoint union $\kappa_1\sqcup\cdots\sqcup\kappa_m$ is equal to $\{k_1,\ldots,k_n\}$. For example, $K\vdash\{4,6\}$ if and only if $K$ is equal to $\{\,\{4,6\}\,\}$ or $\{\,\{4\},\{6\}\,\}$, so
\begin{equation*}
m_{4,6}=c_{4,6}+c_4\,c_6.
\end{equation*}
As for ribbon graphs, we give here a pictorial definition that is well suited to our purposes (cf.~\citep{Kon92}, \citep{MP98}, \citep{LaC09} for alternative but near-equivalent definitions).
\begin{definition} \label{def3.1}
Letting $k\in\mathbb{N}$, a $k$-\textit{ribbon graph} is a collection of $n$ polygons ($0<n\leq 2k$) with labelled vertices and $k$ rectangles, referred to as ribbons, satisfying the following properties:
\begin{enumerate}
\item The number of edges of the polygons must sum to $2k$.
\item For each ribbon, designate a pair of opposite edges to be referred to as ends of the ribbon. Then, each of the $2k$ polygon edges must be identified, in a topological sense, with exactly one ribbon-end, so that each ribbon connects two polygon edges together.
\end{enumerate}
The \textit{Euler genus} of an orientable (non-orientable) connected ribbon graph is the smallest $\tilde{g}\in\mathbb{N}$ such that the ribbon graph can be drawn on a sphere with $\tilde{g}/2$ handles ($\tilde{g}$ cross-caps) without intersecting with itself. In the orientable case, it is double the usual genus $g$.
\end{definition}

\begin{note}
Since there are two ways to identify a pair of lines, depending on the relative orientation of the lines, we allow our ribbons to have at most one M\"obius half-twist. We refer to the class of ribbon graphs with untwisted ribbons as `(globally) orientable', while the term `locally orientable' refers to the scenario where ribbons are allowed to be M\"obius half-twisted. In the literature, ribbon graphs are usually restricted to be of the first type, while the latter type are referred to as M\"obius graphs \citep{MW03}, \citep{KK03}, \citep{BP09}.
\end{note}

\begin{figure}[H]
        \centering
\captionsetup{width=.9\linewidth}
        \includegraphics[width=0.7\textwidth]{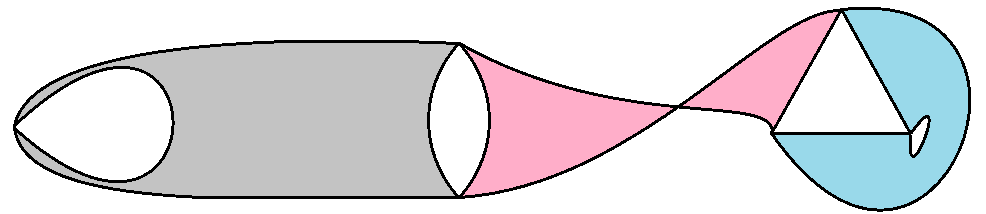}
        \caption[A $3$-ribbon graph with a M\"obius half-twisted ribbon]{A $3$-ribbon graph constructed from a monogon, bigon, and triangle, with a grey ribbon connecting the single edge of the monogon to the left edge of the bigon, a M\"obius half-twisted pink ribbon connecting the right edge of the bigon to an edge of the triangle, and a blue ribbon connecting the remaining two edges of the triangle together.} \label{fig3.1}
\end{figure}

The connection between ribbon graphs and random matrix theory was first explored by Br\'ezin et al.~in their 1978 work \citep{BIPZ78} (see also \citep{BIZ80}), where they were strongly influenced by 't~Hooft's 1974 work \citep{Hoo74} on the large $N$ limit of gauge theories. Closely related ideas were seen independently in \citep{YK83}, \citep{VWZ84}, \citep{HZ86} over the next decade, with Harer and Zagier's 1986 work \citep{HZ86} being of particular interest to us. Technically speaking, Harer and Zagier did not look directly at ribbon graphs, but rather showed and utilised the fact that $2^kM_{k,l}^{(GUE)}$ equals the number of ways of identifying pairs of edges of a $2k$-gon to form a compact orientable surface of genus $l/2$ --- of course, this is equal to the number of orientable genus $l/2$ ribbon graphs that can be built from a single $2k$-gon since the ribbons can be interpreted as identifying the polygon edges at their ends. In 1997, Goulden and Jackson \citep{GJ97} (see also \citep{GHJ01}, \citep{MW03}, \citep{KK03}) gave the GOE analogue of the above statement, which is that $4^kM_{k,l}^{(GOE)}$ equals the number of locally orientable ribbon graphs of Euler genus $l$ that can be built from a single $2k$-gon. Other related works from this time period include \citep{Pen88}, \citep{Kon92}, \citep{Jac94}, \citep{Sil97}.

In \S\ref{s3.3.1}, we give a detailed review of how the spectral moments of the GUE and GOE relate to the ribbon graphs described above. Then, in \S\ref{s3.3.2}, we review the analogous statements for the LUE \citep{Fra03} and LOE \citep{BP09}; the upshot here is that $M_{k,l}^{(LUE)},M_{k,l}^{(LOE)}$ count the same ribbon graphs as $M_{k,l}^{(GUE)},M_{k,l}^{(GOE)}$, respectively, except that the vertices of the underlying $2k$-gons are alternately coloured black and red, and the edge identifications induced by the ribbons respect this bicolouring. Our arguments, which flow from the Isserlis--Wick theorem (see Theorem \ref{thrm3.1} of Section \ref{s3.3}), have been shown to extend to the GSE and LSE \citep{MW03}, \citep{BP09}, but we do not explore this fact in order to keep our discussion concise; $M_{k,l}^{(GSE)},M_{k,l}^{(LSE)}$ can roughly be ascribed the same combinatorial meaning as $M_{k,l}^{(GOE)},M_{k,l}^{(LOE)}$ due to the $\beta\leftrightarrow4/\beta$ duality of Lemma \ref{L1.4}. Finally, in \S\ref{s3.3.3}, we extend the arguments of \S\ref{s3.3.1} and \S\ref{s3.3.2} to produce ribbon graph interpretations for the spectral moments of the Hermitised and antisymmetrised matrix product ensembles defined in Section \ref{s1.3}.

One of our motivations for supplying the review contained in \S\ref{s3.3.2} is that it allows us to interpret the $1$-point recursions of \S\ref{s3.1.2} pertaining to the Laguerre ensembles as solving combinatorial problems similar to that considered by Harer and Zagier --- unfortunately, we are currently unable to find analogous applications for the other recursions given in \S\ref{s3.1.2} due to complications that we outline in Section \ref{s3.3}. As mentioned earlier, the discussion of Section~\ref{s3.3} does not focus solely on the spectral moments of the random matrix ensembles at hand, but also their mixed moments and cumulants. Indeed, we show therein how straightforward generalisations of the ribbon graphs described above, now constructed from multiple polygons, relate to the mixed moments and cumulants of interest. We show moreover that the relevant mixed moments and cumulants are polynomial in $N$. A consequence of this is that, upon appropriate scaling, we are able to assume that the connected $n$-point correlators $W_n$ \eqref{eq1.1.29} of the Hermitised and antisymmetrised matrix product ensembles have large $N$ (actually $N_0$) expansions of the form \eqref{eq1.1.21} without needing to check the hypotheses of Theorem~\ref{thrm1.1}. The existence of these large $N$ expansions is significant because they are crucial to the loop equation analysis presented in Chapter 4.

\setcounter{equation}{0}
\section{Recurrence Relations for the Moments of the Classical Matrix Ensembles} \label{s3.1}
We now derive the aforementioned linear recurrence relations on the spectral moments $m_k$ \eqref{eq3.0.1}, and explain why they hold true when the $m_k$ are instead taken to be the complex-$k$ generalised moments (following Cunden et al.~\citep{CMOS19})
\begin{equation} \label{eq3.1.1}
m_k:=\int_{\mathbb{R}}|\lambda|^k\rho(\lambda)\,\mathrm{d}\lambda,\quad k\in\{k'\in\mathbb{C}\,|\,m_{k'}<\infty\}.
\end{equation}
Note that these generalised moments agree with the spectral moments defined by equation \eqref{eq3.0.1} for $k\in\mathbb{N}$ in the Jacobi and Laguerre cases, and for $k\in2\mathbb{N}$ in the Gaussian, symmetric shifted Jacobi, and symmetric Cauchy cases --- in the latter set of cases, $m_k=0$ for $k$ odd according to equation \eqref{eq3.0.1}, but not equation \eqref{eq3.1.1}. Since the eigenvalue densities studied in this section are either symmetric about the origin or have support contained within the positive real line, the generalised moments \eqref{eq3.1.1} can be interpreted as the spectral moments (we reserve the adjective `spectral' for the moments defined by equation \eqref{eq3.0.1}) of the constrained density $\rho(\lambda)\chi_{\lambda>0}$, with a possible factor of two.

The study of the former type of spectral moments is motivated by the fact that they fully characterise the eigenvalue densities of the classical matrix ensembles, as discussed in \S\ref{s1.1.1}. Generalisation to negative integers $k$ was first considered by Brouwer et al.~\citep{BFB97} in the Laguerre case, where they showed that the reciprocated eigenvalues of the Wigner--Smith time delay matrix $Q$ are distributed according to the Laguerre ensemble with $a=\kappa N$ so that $\mean{\Tr\,Q^k}=m_{-k}^{(L)}|_{a=\kappa N}$ \citep{CMSV16b}. The significance of the Wigner--Smith matrix is that it relates to the study of ballistic chaotic scattering, whereby its eigenvalues, referred to as \textit{proper delay times}, describe the amount of time incident particles spend in the ballistic cavity during a scattering event, among other observables \citep{Be97}, \citep{BFB97}, \citep{CMSV16a}, \citep{CMSV16b}, \citep{LV11}, \citep{MS11}, \citep{MS12}. The more recent extension to the complex-$k$ moments \eqref{eq3.1.1} considered by Cunden et al.~\citep{CMOS19} relates to the study of so-called spectral zeta functions of random matrices $X$, $\zeta_X(k):=\Tr\,(X^\dagger X)^{-k}$ (note that if we take $X$ to be the infinite-dimensional deterministic matrix $\mathrm{diag}(1,\sqrt{2},\sqrt{3},\ldots)$, $\zeta_X(k)$ is then the Riemann zeta function). It is shown in \citep{CMOS19} that spectral zeta functions satisfy a reflection symmetry across the line $\textrm{Re}(k)=1/2$ (much like the Riemann zeta function), which implies reciprocity laws between the moments $m_{-k}$ and $m_{k-1}$ of the LUE and JUE, for $k\in\mathbb{N}$; see Remark \ref{R3.1} within \S\ref{s3.2.3}.

In \S\ref{s3.1.2}, we return to the study of spectral moments \eqref{eq3.0.1} with $k\in\mathbb{N}$ since then, it is known (recall Lemma \ref{L1.3}) that $m_k^{(G)},m_k^{(L)}$ are polynomial in $N$ and $m_k^{(J)}$ is a rational function in $N$. Thus, the expansions \eqref{eq3.0.8}--\eqref{eq3.0.10} are valid and we are able to derive $1$-point recursions \eqref{eq3.0.11} characterising the moment expansion coefficients $M_{k,l}$. As mentioned earlier, $1$-point recursions satisfied by these moment expansion coefficients have been studied in special cases, with particular attention paid to their topological or combinatorial interpretations \citep{HZ86}, \citep{Le09}, \citep{ND18}. These aspects are briefly discussed in \S\ref{s3.1.2}, with the highlighted point being that even though the moment expansion coefficients $M_{k,l}$ may have combinatorial interpretations in certain cases (fully detailed in Section \ref{s3.3}), there are presently no such interpretations for the $1$-point recursions satisfied by them.

\subsection{Recurrences for the spectral moments} \label{s3.1.1}
One may obtain recurrences for the spectral moments $m_k$ of the classical matrix ensembles by recalling from \S\ref{s1.1.1} that the resolvent $W_1(x)$ acts as a generating function for these moments, presuming that the moments are well-defined for all $k\in\mathbb{N}$ or otherwise interpreting the large $x$ expansion of $W_1(x)$ in a formal sense:
\begin{equation}
W_1(x)=\sum_{k=0}^{\infty}\frac{m_k}{x^{k+1}}.
\end{equation}
Substituting this series into the differential equations for $W_1(x)$ given in Section \ref{s2.3} and then equating terms of equal order in $x$ gives relations between the spectral moments. The equations obtained from terms of negative order in $x$ give the upcoming recurrences on the spectral moments, while the terms of order one and positive order in $x$ give the first few moments required to run the recursions (the latter are also available in earlier literature; see, e.g., \citep{FRW17} and references therein).

The recurrence relations derived from the above procedure enable the iterative computation of the spectral moment $m_k$ for any positive integer $k$ starting with knowledge of $m_0,\ldots,m_p$ for some small integer $p$. However, these recurrences are actually valid for the complex-$k$ generalisations defined by equation \eqref{eq3.1.1}, so long as all involved moments $m_k$ are such that in the Gaussian and symmetric shifted Jacobi cases, $\mathrm{Re}(k)>-1$; in the Jacobi and Laguerre cases, $\mathrm{Re}(k)>-a-1$; and in the symmetric Cauchy case, $-1<\mathrm{Re}(k)<2\alpha+1$ (this is a refinement of the constraint $-1<k<2\alpha_1+1$ discussed in \S\ref{s1.2.1}). To ascertain the validity of our moment recurrences in these more general settings, we must provide an alternative proof to that outlined in the previous paragraph. Thus, rather than the differential equations for $W_1(x)$, we begin at the differential equations for the eigenvalue density $\rho(x)$ given in Section \ref{s2.3}. Multiplying both sides of these differential equations by $|x|^k$ and then integrating over the support of the eigenvalue density produces a relation between terms of the form $\int_{\mathrm{supp}\,\rho}x^p|x|^k{\mathrm{d}^n\over\mathrm{d}x^n}\rho(x)\,\mathrm{d}x$ for positive integers $p,n$. The goal then is to use integration by parts to reduce these terms to the form \eqref{eq3.1.1}. This is done in a similar fashion to what is shown in Appendix \ref{appendixB}, with the notable exception being that the domain of integration must be split at the origin due to the absolute value sign in the factor $|x|^k$ --- in any case, all boundary terms arising from integration by parts vanish, owing to our restrictions on $k$.

\begin{proposition} \label{prop3.1}
Recall the definitions of $\tilde{a},\tilde{b}$, and $\tilde{c}$ given in Theorem \ref{thrm2.2}. For $\beta=2$ and $k\in\mathbb{C}$ such that $\mathrm{Re}(k)>2-a$, the moments \eqref{eq3.1.1} of the eigenvalue density of the Jacobi ensemble satisfy the third order linear recurrence
\begin{equation} \label{eq3.1.3}
\sum_{l=0}^3 d_{2,l}^{(J)}m_{k-l}^{(J)}=0,
\end{equation}
where
\begin{align*}
d_{2,0}^{(J)}&=k\left[(a+b+2N)^2-(k-1)^2\right],
\\d_{2,1}^{(J)}&=3k^3-11k^2-k\left[2(a+b+2N)^2+a^2-b^2-14\right]
\\&\quad+3(a+b)(a+2N)+6(N^2-1), \displaybreak
\\d_{2,2}^{(J)}&=(2k-3)\left[2N(a+b+N)+ab\right]-(k-2)\left[3k^2-10k-3a^2+9\right],
\\d_{2,3}^{(J)}&=(k-3)\left[(k-2)^2-a^2\right].
\end{align*}
For $\beta=1,4$ and $k\in\mathbb{C}$ such that $\mathrm{Re}(k)>4-a$, the moments of the Jacobi ensemble's eigenvalue density satisfy the fifth order linear recurrence
\begin{equation} \label{eq3.1.4}
\sum_{l=0}^5 d_{4,l}^{(J)}m_{k-l}^{(J)}=0,
\end{equation}
where
\begin{align*}
d_{4,0}^{(J)}&=k(\tilde{c}^2-(k-2)^2)(\tilde{c}^2-(2k-1)^2),
\\d_{4,1}^{(J)}&=\frac{1}{2}(\tilde{c}^2-9)^2(5-6k)+\frac{1}{2}(\tilde{a}-\tilde{b})\left[(\tilde{c}^2-9)(5-4k)+2k(5(k-1)(k-5)+4k)\right]
\\&\quad+(\tilde{c}^2-9)k\left[5(4k-3)(k-3)+2k\right]-4k^2(k-5)\left[5(k-2)(k-1)-2\right],
\\d_{4,2}^{(J)}&=\tilde{c}^4 (3 k-5)+\tilde{c}^2 \left[\tfrac{1}{2}(\tilde{a}+\tilde{b})(2k-5)+5(\tilde{a}-\tilde{b})(k-2)\right]
\\&\quad-\tilde{c}^2\left[30k^3-171k^2+339k-230\right]-\frac{1}{2} (\tilde{a}+\tilde{b}) \left[5 k^3-44 k^2+129 k-125\right]
\\&\quad-\frac{1}{2} (\tilde{a}-\tilde{b}) \left[35 k^3-246 k^2+581 k-460\right]+\frac{1}{2} (2 k-5) (\tilde{a}-\tilde{b})^2
\\&\quad+40k^4(k-11)+1966k^3-4443k^2+5056k-2305,
\\d_{4,3}^{(J)}&=\frac{1}{2} \tilde{c}^4 (5-2 k)+\tilde{c}^2 \left[\tfrac{1}{4}(\tilde{a}+\tilde{b})(25-8 k)+\tfrac{1}{4}(\tilde{a}-\tilde{b})(45-16k)\right]
\\&\quad+\tilde{c}^2\left[20 k^3-155 k^2+401 k-345\right]+\frac{5}{4}(\tilde{a}+\tilde{b})\left[6k^3-62k^2+216k-253\right]
\\&\quad+\frac{1}{4} (\tilde{a}-\tilde{b})\left[90 k^3-806 k^2+2436 k-2485\right]+\frac{1}{4}(\tilde{a}^2-\tilde{b}^2)(15-4 k)
\\&\quad+\frac{1}{4}(\tilde{a}-\tilde{b})^2(25-8 k)-4k^3(10k^2-140k+789) +8923 k^2-12600 k+\tfrac{14125}{2},
\\d_{4,4}^{(J)}&=(k-4) \left[k^3+k^2-18 k-\left(\tilde{c}^2-\tilde{b}-4 k^2+29 k-51\right)\left(5 k^2-29 k+40\right)\right]
\\&\quad+\frac{1}{2}\tilde{a}^2(6 k-25)+\frac{1}{2}\tilde{a}\left[(4 k-15) \left(\tilde{c}^2-\tilde{b}-10 k^2+76k-147\right)-2 (k-5)\right],
\\d_{4,5}^{(J)}&=(k-5)\left[4(k-5)(k-4)-\tilde{a}\right]\left[\tilde{a}-(k-4)(k-2)\right].
\end{align*}
The sequences of spectral moments $\{m_k^{(J)}\}_{k\in\mathbb{Z},\,k>-a-1}$ are fully determined for $\beta=1,2,4$ by the above recurrence relations and the initial terms $m_0^{(J)},m_1^{(J)},\ldots,m_4^{(J)}$, which can be computed through MOPS \citep{MOPS} or via the methods presented in \citep{MRW17}, \citep{FRW17} --- according to the discussion at the beginning of this subsection, the required initial terms are also encoded within the right-hand sides of differential equations \eqref{eq2.3.8} and \eqref{eq2.3.14}.
\end{proposition}

The recurrence \eqref{eq3.1.3} for the moments $m_k^{(J)}$ of the JUE eigenvalue density was recently given in \citep[Prop.~4.8]{CMOS19}, wherein it is formulated as a recurrence on the differences of moments $\Delta m_k^{(J)}:=m_k^{(J)}-m_{k+1}^{(J)}$. The fact that this recurrence relation can be naturally formulated in terms of the differences $\Delta m_k^{(J)}$ is equivalent to the observation that $\sum_{l=0}^3d_{2,l}^{(J)}=0$. Since $\sum_{l=0}^5d_{4,l}^{(J)}=0$, we see that it is also reasonable to rewrite the recurrence relation \eqref{eq3.1.4} in terms of the $\Delta m_k^{(J)}$. It turns out that the analogous recurrences for the symmetric shifted Jacobi and symmetric Cauchy ensembles are most compactly presented when written in terms of the differences and sums, respectively, of even moments
\begin{align}
\mu_k^{(sJ)}&:=m_{2k+2}^{(sJ)}-m_{2k}^{(sJ)}, \label{eq3.1.5}
\\ \mu_k^{(Cy)}&:=m_{2k+2}^{(Cy)}+m_{2k}^{(Cy)}. \label{eq3.1.6}
\end{align}
Here, the moments $m_k$ are again defined by equation \eqref{eq3.1.1} with $\rho(\lambda)$ being the eigenvalue density of the classical matrix ensemble specified by either the constrained weight \eqref{eq3.0.6} or \eqref{eq3.0.7}, respectively. The fact that the $\mu_k^{(sJ)},\mu_k^{(Cy)}$ recurrences are simpler than the corresponding recurrences on the $m_{2k}^{(sJ)},m_{2k}^{(Cy)}$ is in keeping with the fact that the former are, respectively, the moments of $r^{(sJ)}(x):=(1-x^2)\rho^{(sJ)}(x)|_{a=b}$ and $r^{(Cy)}(x):=(1+x^2)\rho^{(Cy)}(x)|_{\alpha\in\mathbb{R}}$, which are characterised by differential equations whose coefficients are of degree two less than the corresponding coefficients of the differential equations \eqref{eq2.3.17}, \eqref{eq2.3.21} for $\rho^{(sJ)}(x)$ and $\rho^{(Cy)}(x)$; the differential equations satisfied by $r^{(sJ)}(x)$ can be surmised from those satisfied by $r^{(Cy)}(x)$, which are themselves given in \S\ref{s2.3.2} as equations \eqref{eq2.3.27}, \eqref{eq2.3.28}.
\begin{proposition} \label{prop3.2}
Define $\mu_k^{(sJ)}$ by equation \eqref{eq3.1.5} and retain the definitions of $a_{\beta},\tilde{a}$, and $\tilde{c}$ given in Theorem \ref{thrm2.2}. For $\beta=2$ and $k\in\mathbb{C}$ such that $\mathrm{Re}(k)>1/2$, we have the second order linear recurrence
\begin{multline} \label{eq3.1.7}
(2k+4)\left[(2k+3)^2-4(a+N)^2\right]\mu_{k+1}^{(sJ)}-2(2k+1)\left[(2k+2)^2-2N(N+2a)\right]\mu_k^{(sJ)}
\\+(2k+1)(2k)(2k-1)\mu_{k-1}^{(sJ)}=0.
\end{multline}
The initial condition determining the sequence $\{ \mu_k^{(sJ)} |_{\beta = 2} \}_{k\in\mathbb{N}}$ is
\begin{equation} \label{eq3.1.8}
\mu_0^{(sJ)} = {2 N (a + N) (2 a + N)  \over 1 - 4 (a + N)^2}.
\end{equation}
For $\beta=1,4$ and $k\in\mathbb{C}$ such that $\mathrm{Re}(k)>3/2$, we have the fourth order linear recurrence
\begin{equation}
\sum_{l=-2}^2d_{4,l}^{(sJ)}\mu_{k-l}^{(sJ)}=0,
\end{equation}
where
\begin{align*}
d^{(sJ)}_{4,-2}&:=-4(2k+1)(2k)(2k-1)(2k-2)(2k-3),\\
d^{(sJ)}_{4,-1}&:=(2k+1)(2k)(2k-1)\left[20\tilde{a}-5\tilde{c}^2+16k(4k+5)+77\right],\\
d^{(sJ)}_{4,0}&:=(2k+1)\Big[ 8\tilde{a}(5k+8)(2k+3)-\left(\tilde{c}^2-4\tilde{a}-30k^2-67k-42\right)^2 \\
&\qquad+516k^4+2292k^3+3653k^2+2534k+673\Big], \\
d^{(sJ)}_{4,1}&:=\tilde{c}^2\left[(\tilde{c}^2-4\tilde{a})(4k+7)-120k^3-656k^2-1210k-746\right] \\
&\quad+\tilde{a}\left[160k^3+736k^2+1128k+580\right] \\
&\quad+512k^5+4480k^4+16056k^3+29360k^2+27270k+10243, \\
d^{(sJ)}_{4,2}&:=-2(k+3)(\tilde{c}+4k+11)(\tilde{c}+2k+4)(\tilde{c}-2k-4)(\tilde{c}-4k-11).
\end{align*}
The initial conditions determining the sequences $\{\mu_k^{(sJ)} |_{\beta =1,4} \}_{k\in\mathbb{N}}$ are
\begin{align}
\mu_0^{(sJ)}&=\frac{(\tilde{c}-1)\left(\tilde{c}+2a_{\beta}-3\right)\left(\tilde{c}-2a_{\beta}+1\right)}{8\tilde{c}(\tilde{c}-3)(1-\kappa)}, \label{eq3.1.10}
\\ \mu_1^{(sJ)}&=\frac{(\tilde{c}^2-5)(\tilde{c}-7)-4\tilde{a}(\tilde{c}-1)}{4(\tilde{c}^2-4)(\tilde{c}-7)}\mu_0^{(sJ)}. \label{eq3.1.11}
\end{align}
\end{proposition}

We note that the generalisation of recurrence \eqref{eq3.1.7} concerning the non-symmetric shifted JUE corresponding to the weight $w^{(sJ)}(\lambda)=(1-\lambda)^a(1+\lambda)^b\chi_{-1<\lambda<1}$ and $\beta=2$ was first given by Ledoux in \citep{Le04}, albeit consideration was given only to the $\mu^{(sJ)}_k$ with $k\in\mathbb{N}$. Moving on, let us observe that the contents of Section \ref{s2.2}, particularly Proposition \ref{prop2.2}, imply the relation
\begin{equation} \label{eq3.1.12}
\mu_k^{(sJ)}\Big|_{a\mapsto-\kappa(N-1)-1-\alpha}=(-1)^{k-1}\mu_k^{(Cy)},
\end{equation}
where we recall that $\kappa=\beta/2$. A simple consequence of this relation is that Proposition \ref{prop3.2} leads to the following recurrence relations for the sums of moments $\mu_k^{(Cy)}$.
\begin{corollary} \label{C3.1}
Define $\mu_k^{(Cy)}$ by equation \eqref{eq3.1.6} and retain the definitions of $d^{(sJ)}_{4,-2},d^{(sJ)}_{4,-1},\ldots,d^{(sJ)}_{4,2}$ listed in Proposition \ref{prop3.2}. For $\beta=2$ and $k\in\mathbb{C}$ such that $1/2<\mathrm{Re}(k)<\alpha-3/2$, we have the second order linear recurrence (recently given in \citep{ABGS20})
\begin{multline} \label{eq3.1.13}
(2k+4)\left[(2k+3)^2-4\alpha^2\right]\mu_{k+1}^{(Cy)}+2(2k+1)\left[(2k+2)^2+2N(N+2\alpha)\right]\mu_k^{(Cy)} \\
+(2k+1)(2k)(2k-1)\mu_{k-1}^{(Cy)}=0.
\end{multline}
The initial condition determining the sequence $\{ \mu_k^{(Cy)} |_{\beta = 2} \}_{k\in\mathbb{N}}$ is
\begin{equation}
\mu_0^{(Cy)} = {2N \alpha (N + 2 \alpha) \over (2 \alpha - 1) (2 \alpha + 1)},\quad\alpha>1/2.
\end{equation}
For $\beta=1,4$ and $k\in\mathbb{C}$ such that $3/2<\mathrm{Re}(k)<\alpha-5/2$, we have the fourth order linear recurrence
\begin{equation} \label{eq3.1.15}
\sum_{l=-2}^2d_{4,l}^{(Cy)}\mu_{k+l}^{(Cy)}=0,\qquad d_{4,l}^{(Cy)}:=(-1)^{l-1}d_{4,l}^{(sJ)}\Big|_{a\mapsto-\kappa(N-1)-1-\alpha}.
\end{equation}
The initial conditions required to fully characterise the sequences $\{\mu_k^{(Cy)}|_{\beta=1,4}\}_{k\in\mathbb{N}}$ are obtained by parsing equations \eqref{eq3.1.10} and \eqref{eq3.1.11} through the relation \eqref{eq3.1.12}.
\end{corollary}

In a similar vein to how the arguments of Section \ref{s2.2} enable us to simply translate the recurrences of Proposition \ref{prop3.2} for the $\mu_k^{(sJ)}$ into the analogous recurrences on the $\mu_k^{(Cy)}$ given in Corollary \ref{C3.1}, the limiting procedure \eqref{eq1.2.23} described in Lemma \ref{L1.1} can be used to obtain recurrences on the (generalised) spectral moments $m_k^{(L)}$ of the Laguerre ensembles from the recurrences presented in Proposition \ref{prop3.1}. Indeed, to derive the following recurrences on the $m_k^{(L)}$, one may circumvent the strategy outlined at the beginning of this subsection by setting $m_k^{(J)}=b^{-k}m_k^{(L)}$ in Proposition \ref{prop3.1} and then taking the limit $b\to\infty$ (cf.~equation \eqref{eq2.3.40}, in addition to Lemma \ref{L1.1}).
\begin{corollary} \label{C3.2}
Recall the definitions of $a_{\beta},N_{\beta}$, and $\tilde{a}$ given in Theorem \ref{thrm2.2}. For $\beta=2$ and $k\in\mathbb{C}$ such that $\mathrm{Re}(k)>1-a$, the moments of the LUE eigenvalue density satisfy the second order linear recurrence
\begin{equation} \label{eq3.1.16}
(k+1)m_k^{(L)}=(2k-1)(a+2N)m_{k-1}^{(L)}+(k-2)\left[(k-1)^2-a^2\right]m_{k-2}^{(L)}.
\end{equation}
For $\beta=1,4$ and $k\in\mathbb{C}$ such that $\mathrm{Re}(k)>3-a$, we have the fourth order linear recurrence
\begin{equation} \label{eq3.1.17}
\sum_{l=0}^4d_{4,l}^{(L)}(\kappa-1)^lm_{k-l}^{(L)}=0,
\end{equation}
where
\begin{align*}
d_{4,0}^{(L)}&=k+1,
\\d_{4,1}^{(L)}&=(1-4k)(a_{\beta}+4N_{\beta}),
\\d_{4,2}^{(L)}&=(1-k)(5k^2-11k+4)+(2k-3)\left[\tilde{a}+2(a_{\beta}+4N_{\beta})^2\right],
\\d_{4,3}^{(L)}&=(a_{\beta}+4N_{\beta})\left[(11-4k)\tilde{a}+10k^3-68k^2+146k-96\right],
\\d_{4,4}^{(L)}&=(k-4)\left[\tilde{a}-4(k-4)(k-3)\right]\left[\tilde{a}-(k-3)(k-1)\right].
\end{align*}
The initial terms $m_0^{(L)},m_1^{(L)},\ldots,m_3^{(L)}$ required to fully determine the sequences $\{m_k^{(L)}\}_{k\in\mathbb{Z},\,k>-a-1}$ in the orthogonal, unitary, and symplectic cases can be found in \citep{MRW17}, \citep{FRW17}.
\end{corollary}
It should come as no surprise that recurrence \eqref{eq3.1.16} on the moments of the LUE eigenvalue density has already been given in the literature, seeing as this is true of its JUE analogue \eqref{eq3.1.3}, which is more involved. Indeed, it was first obtained by Haagerup and Thorbj{\o}rnsen \citep{HT03}, where they considered only the case $a,k\in\mathbb{N}$. A bit over a decade later, Cunden et al.~\citep{CMSV16b} showed that the LUE moment recurrence \eqref{eq3.1.16} holds for all integers $k>-a-1$ with $a>-1$ a continuous real parameter. On the other hand, recurrence \eqref{eq3.1.17} pertaining to the moments of the LOE and LSE can be seen to be a homogeneous simplification of \citep[Eq.~(43)]{CMSV16b}, wherein the inhomogeneous terms depend on the moments of the LUE.

Another simple observation is that for each of $\beta=1,2$, and $4$, the $m_k^{(L)}$ recurrences displayed in Corollary \ref{C3.2} involve one less term than the corresponding $m_k^{(J)}$ recurrences given in Proposition \ref{prop3.1} --- the terms in the recurrences are themselves drastically simpler, as well. This is in keeping with the loss of parameter $b$. A similar simplification of terms can be observed in the moment recurrences in the Gaussian case, again to be understood as a consequence of the hierarchy implied by Lemma \ref{L1.1}. However, we do not see a reduction in the order of the moment recurrences when moving from the Laguerre to Gaussian ensembles. Thus, the spectral (and generalised) moments of the GUE satisfy a second order linear recurrence first derived by Harer and Zagier in \citep{HZ86}, while the spectral moments of the GOE and GSE are characterised by fourth order linear recurrences given by Ledoux in \citep{Le09}. We do not present them here, but note (as was mentioned earlier) that the strategies discussed in this section recover the recurrences of \citep{HZ86}, \citep{Le09}, with the added insight that they hold for complex $k$. Furthermore, our methods allow us to derive linear recurrence relations for the $\beta=2/3$ and $\beta=6$ Gaussian ensembles' spectral moments, which we now give.
\begin{proposition}
For $\beta=2/3,6$ and $k\in\mathbb{C}$ such that $\mathrm{Re}(k)>11$, the moments of the Gaussian $\beta$ ensemble's eigenvalue density satisfy the sixth order linear recurrence
\begin{equation}
\sum_{l=0}^6d_{6,l}^{(G)}\left(\frac{\kappa-1}{4}\right)^lm_{k-2l}^{(G)}=0,
\end{equation}
where
\begin{align*}
d_{6,0}^{(G)}&=-4(k+2),
\\d_{6,1}^{(G)}&=8(3k-1)(3N_{\beta}+2),
\\d_{6,2}^{(G)}&=48(8-3k)N_{\beta}(3N_{\beta}+4)+49k^3-216k^2+92k+320,
\\d_{6,3}^{(G)}&=4(k-5)(3N_{\beta}+2)\left[24N_{\beta}(3N_{\beta}+4)-49k^2+304k-442\right],
\\d_{6,4}^{(G)}&= 2(k-5)_3\left[294N_{\beta}(3N_{\beta}+4)-63k(k-6)-274\right],
\\d_{6,5}^{(G)}&=252(k-5)_5(3N_{\beta}+2),
\\d_{6,6}^{(G)}&=81(k-5)_7,
\end{align*}
and we retain the definition $N_{\beta}=(\kappa-1)N$. Here, $(x)_n=x(x-1)\cdots(x-n+1)$ is the falling Pochhammer symbol. The initial terms $m_0^{(G)}, m_2^{(G)},\ldots,m_{10}^{(G)}$ needed to determine the sequences $\{m_k^{(G)}|_{\beta=2/3,6}\}_{k\in\mathbb{N}}$ are listed in \cite{WF14}.
\end{proposition}

As we proceed to our discussion on $1$-point recursions, let us conclude this subsection with a couple of general remarks. The recurrences given above describe relations between the complex-$k$ generalised moments $m_k$ defined by equation \eqref{eq3.1.1}, so long as all moments involved are well-defined. However, the recurrences hold true when the $m_k$ are taken to be (integer-$k$) spectral moments as defined by equation \eqref{eq3.0.1}, even though these spectral moments fail to coincide with the generalised moments when considering the Gaussian, symmetric Jacobi, or symmetric Cauchy ensembles, with $k$ an odd integer. In these cases, the definitions \eqref{eq3.0.1} and \eqref{eq3.1.1} disagree only in that, according to the first definition, $m_k=0$ when $k$ is odd --- the relevant recurrences are then seen to hold trivially for odd integers $k$ due to their homogeneous structure and the fact that they run over every second integer-$k$ moment (e.g., recurrence \eqref{eq3.1.15} involves $m_{k+6}^{(Cy)},m_{k+4}^{(Cy)},\ldots,m_{k-4}^{(Cy)}$).

The methods employed in \citep{Le09} manifest a coupling between the GOE and GUE moments, which does not arise naturally from our viewpoint. Likewise for the coupling between the LOE and LUE moments implied by the recurrence \citep[Eq.~(43)]{CMSV16b}. As made clear in \citep{CMSV16b}, both couplings can be traced back to a structural formula expressing the $\beta=1$ eigenvalue densities in terms of their $\beta=2$ counterparts plus what can be regarded as rank one corrections. This inter-relation was discussed briefly in \S\ref{s1.2.3}, where we saw that the context underlying it is that the skew-orthogonal polynomials pertaining to the GOE and LOE can be constructed from the Hermite and Laguerre orthogonal polynomials used to study the GUE and LUE, respectively. At present, there is no evidence implying a relation connecting the moments of the $\beta=6$ Gaussian ensemble to the moments of the GOE and/or GUE.

\subsection{$1$-point recursions for the moment expansion coefficients} \label{s3.1.2}
Following on from the preceding subsection, it is a simple matter to obtain $1$-point recursions of the form \eqref{eq3.0.11} for the coefficients of the moment expansions \eqref{eq3.0.8}--\eqref{eq3.0.10}. In fact, all one needs to do is substitute the moment expansions \eqref{eq3.0.8}--\eqref{eq3.0.10} into the recurrence relations of \S\ref{s3.1.1}, now considering $k$ large enough integers such that all involved moments are well-defined according to definition \eqref{eq3.0.1}, and then equate terms of equal order in $N$. There is however another (instructive) avenue towards the sought $1$-point recursions beginning at the differential-difference equations of \S\ref{s2.4.1}: In the large $N$ limit, the spectral moments $m_{2k}^{(G)},m_k^{(L)}$, and $m^{(J)}_k$ do not converge, so the moment expansions \eqref{eq3.0.8}--\eqref{eq3.0.10} may only be understood in a formal sense. Rectifying this issue by introducing the scalings
\begin{align}
\tilde{m}_{2k}^{(G)}&:=\kappa^{-k}N^{-k-1}\,m_{2k}^{(G)}, \label{eq3.1.19}
\\ \tilde{m}_k^{(L)}&:=\kappa^{-k}N^{-k-1}\,m_{k}^{(L)}, \label{eq3.1.20}
\\ \tilde{m}_k^{(J)}&:=m_k^{(J)}/N, \label{eq3.1.21}
\end{align}
as is compliant with Definition \ref{def1.6} (that is, $\tilde{m}_k=\int_{\mathbb{R}}\lambda^k\tilde{\rho}(\lambda)\,\mathrm{d}\lambda$), we observe that the (scaled) resolvents $\tilde{W}_1^{(G)}(x)$ \eqref{eq2.4.5}, $\tilde{W}_1^{(L)}(x)$ \eqref{eq2.4.6}, and $W_1^{(J)}(x)$ act as generating functions for $\{\tilde{m}_{2k}^{(G)}\}_{k\in\mathbb{N}},\{\tilde{m}_k^{(L)}\}_{k\in\mathbb{N}}$, and $\{\tilde{m}_k^{(J)}\}_{k\in\mathbb{N}}$, respectively. Moreover, the scaled moments have convergent large $N$ expansions (finite sums in the Gaussian and Laguerre cases) of the form $\tilde{m}_k=\sum_{l=0}^{\infty}\tilde{M}_{k,l}\,N^{-l}$ \eqref{eq1.1.32}, with the expansion coefficients $\tilde{M}_{k,l}$ of the scaled moments themselves generated by the resolvent expansion coefficients $W_1^l(x)$ defined implicitly in equations \eqref{eq2.4.8}--\eqref{eq2.4.10}. Thus, substituting the large $x$ expansions
\begin{align*}
W_1^{(G),l}(x)&=2^{l+1}\kappa^{l/2}\sum_{k=0}^{\infty}\tilde{M}^{(G)}_{k,l}\,x^{-k-1},
\\ W_1^{(L),l}(x)&=\kappa^{l/2}\sum_{k=0}^{\infty}\tilde{M}^{(L)}_{k,l}\,x^{-k-1},
\\ W_1^{(J),l}(x)&=\sum_{k=0}^{\infty}\tilde{M}^{(J)}_{k,l}\,x^{-k-1}
\end{align*}
into the differential-difference equations of \S\ref{s2.4.1} and then comparing terms of like order in $x$ yields $1$-point recursions on the $\tilde{M}_{k,l}$ (recall the diagram displayed at the end of \S\ref{s1.1.1}). To obtain equivalent recursions for the unscaled moment expansion coefficients $M_{k,l}$, it is enough to note from equations \eqref{eq3.1.19}--\eqref{eq3.1.21} that
\begin{equation*}
M_{k,l}^{(G)}=\kappa^k\,\tilde{M}_{2k,l}^{(G)},\qquad M_{k,l}^{(L)}=\kappa^k\,\tilde{M}_{k,l}^{(L)}, \qquad M_{k,l}^{(J)}=\tilde{M}_{k,l}^{(J)}.
\end{equation*}

Confirming agreement between the two derivations outlined above, flowing separately from the results of \S\ref{s3.1.1} and \S\ref{s2.4.1}, respectively, serves as a consistency check. That aside, the upcoming $1$-point recursions are themselves motivated by the fact that they enable efficient computation of the expansion coefficients $M_{k,l}$, assuming knowledge of sufficiently many of them for small $k,l\in\mathbb{N}$ --- we do not provide the necessary initial conditions here, but note that they can be computed through the strategy discussed at the beginning of \S\ref{s3.1.1}. Compared to the recurrences of \S\ref{s3.1.1}, which run over the spectral moments $m_k$, the $1$-point recursions are faster at isolating the leading order behaviour of the $m_k$, in the large $N$ limit. On the other hand, changing viewpoint to that of Section \ref{s3.3}, the $1$-point recursions are interesting because they relate enumerations of ribbon graphs in a manner that has no obvious combinatorial interpretation; we make only brief comments on the relevant ribbon graphs here, deferring full discussion to Section \ref{s3.3}.

As mentioned earlier, our $1$-point recursions agree with that of Harer and Zagier \eqref{eq3.0.12} in the GUE case and of Ledoux \citep{Le09} in the GOE and GSE cases. Let us recall that the motivation for studying the former came from the interpretation of $2^kM_{k,l}^{(GUE)}$ as the number of ways of gluing the edges of a $2k$-gon to form a compact orientable surface of genus $l/2$ (cf.~Figure \ref{fig3.5} of \S\ref{s3.3.1}), while $4^kM_{k,l}^{(GOE)}$ is known \citep{GJ97}, \citep{GHJ01}, \citep{MW03}, \citep{KK03} to be equal to the number of such gluings that form a compact locally orientable surface of Euler characteristic $2-l$ (see Figure \ref{fig3.9}); $M_{k,l}^{(GSE)}$ has been given equivalent interpretations in \citep{MW03}, \citep{BP09}, which can be understood through the $\beta\leftrightarrow4/\beta$ duality of Lemma \ref{L1.4}. We do not recount the GOE, GUE, and GSE $1$-point recursions here, for brevity. Likewise, we do not report on the symmetric Cauchy ensembles, nor the (symmetric shifted) Jacobi ensemble when $\beta\ne2$; though these have not been given in the literature before, they are relatively cumbersome to display while being as equally easy to derive from the recurrences of \S\ref{s3.1.1} as the recursions given below. Our first result is hence on the Gaussian $\beta$ ensembles with $\beta=2/3$ and $6$.
\begin{proposition} \label{prop3.4}
Expand the spectral moments of the Gaussian ensemble according to \eqref{eq3.0.8}. Then, for $\beta=2/3,6$ and $k\geq6$,
\begin{equation} \label{eq3.1.22}
(k+1)M_{k,l}^{(G)}=\sum_{i=1}^6\sum_{j=0}^i\frac{(\kappa-1)^{2i-j}}{2^i}f_{i,j}M_{k-i,l-j}^{(G)},
\end{equation}
where
\begin{equation*}
\begin{gathered}
f_{1,0}=3(6k-1),\quad f_{1,1}=2(6k-1),\quad f_{2,0}=36(4-3k),
\\f_{2,1}=48(4-3k),\quad f_{2,2}=49k^3-108k^2+23k+40,\quad f_{3,0}=108(2k-5),
\\f_{3,1}=216(2k-5),\quad f_{3,2}=3(5-2k)(98k^2-304k+189),
\\f_{3,3}=2(5-2k)(98k^2-304k+221),\quad f_{4,2}=\tfrac{441}{2}(2k-5)_3,\quad f_{4,3}=294(2k-5)_3,
\\f_{4,4}=\tfrac{1}{2}(2k-5)_3\left(126k(3-k)-137\right),\quad f_{5,4}=\tfrac{189}{2}(2k-5)_5,
\\f_{5,5}=63(2k-5)_5,\quad f_{6,6}=\tfrac{81}{8}(2k-5)_7,
\end{gathered}
\end{equation*}
and all other $f_{i,j}$ are zero. We also set $M_{k,l}^{(G)}=0$ if $l<0$ or $l>k$.
\end{proposition}
When equation \eqref{eq3.1.22} is used to compute $M_{k,0}^{(G)}$, it reduces to
\begin{multline} \label{eq3.1.23}
16(k+1)M_{k,0}^{(G)}=12(6k-1)(\kappa-1)^2M_{k-1,0}^{(G)}-36(3k-4)(\kappa-1)^4M_{k-2,0}^{(G)}
\\+27(2k-5)(\kappa-1)^6M_{k-3,0}^{(G)},\quad\kappa=1/3\textrm{ or }3.
\end{multline}
This is in keeping with the limiting scaled eigenvalue density $\rho^{(G),0}(\lambda)$ of the Gaussian ensembles equalling the Wigner semi-circle law specified by the density $\sqrt{2-\lambda^2}/\pi$ \eqref{eq1.2.14} supported on $|\lambda|<\sqrt{2}$: up to a scale factor the Catalan numbers are the even moments. Thus, equation \eqref{eq3.1.23} can essentially be seen to be a four term linear recurrence for the Catalan numbers. To the best of the author's knowledge, it does not have a combinatorial interpretation comparable to what is known for the traditional two term recurrence given in equation \eqref{eq3.1.25} below. However, we note that equation \eqref{eq3.1.23} and its three term analogue \eqref{eq3.1.27} are implied by repeated applications of equation \eqref{eq3.1.25} with appropriate scaling.

To expand the Laguerre and Jacobi ensembles' spectral moments in $N$, we need to decide how the parameters $a$ and $b$ scale with $N$. We were faced with this same decision in \S\ref{s2.4.1}, where we opted to set $a=\hat{a}\kappa N$ and $b=\hat{b}\kappa N$ with $\hat{a},\hat{b}$ constant in $N$. We continue with this choice for consistency, with one caveat: Since the $1$-point recursion in the LUE case is relatively simple, we consider the slightly more general parametrisation $a=\hat{a}\kappa N+\delta_a$, where $\hat{a},\delta_a={\rm O}(1)$ --- this gives us an opportunity to demonstrate the insight gained, at the cost of complexity, from implementing this generalisation. Substituting this generalised parametrisation together with the expansion \eqref{eq3.0.9} into recurrence \eqref{eq3.1.16} yields the LUE $1$-point recursion.
\begin{proposition} \label{prop3.5}
For $\beta=2$, $a=\hat{a}N+\delta_a$ with $\hat{a},\delta_a$ constant in $N$, and $k\geq2$,
\begin{multline} \label{eq3.1.24}
(k+1)M_{k,l}^{(LUE)}=(2k-1)\left[(2+\hat{a})M_{k-1,l}^{(LUE)}+\delta_aM_{k-1,l-1}^{(LUE)}\right]
\\-\hat{a}^2(k-2)M_{k-2,l}^{(LUE)}+(k-2)\left((k-1)^2-\delta_a^2\right)M_{k-2,l-2}^{(LUE)},
\end{multline}
where we set $M_{k,l}^{(LUE)}=0$ if $l<0$ or $l>k$.
\end{proposition}
When this is used to compute $M_{k,0}^{(LUE)}$ in the case $a=\delta_a={\rm O}(1)$ and thus $\hat{a}=0$, one recovers the familiar Catalan recursion
\begin{equation} \label{eq3.1.25}
(k+1)M_{k,0}^{(LUE)}=2(2k-1)M_{k-1,0}^{(LUE)}.
\end{equation}
This recurrence is well known (see, e.g., \citep{Sta99}) and can be seen to be consistent with Proposition \ref{prop1.2}: When $a={\rm O}(1)$, we have from equation \eqref{eq1.2.88} that $\gamma_1=1$ so that the Mar\v{c}enko--Pastur density $\rho^{(L),0}(\lambda)$ \eqref{eq1.2.15} simplifies to $\sqrt{4/\lambda-1}/(2\pi)\chi_{0<\lambda<4}$, whose $k\textsuperscript{th}$ spectral moment is exactly the $k\textsuperscript{th}$ Catalan number. Observe also that the Harer--Zagier recursion \eqref{eq3.0.12} reduces to \eqref{eq3.1.25} upon setting $l=0$ and noting that $M_{k,0}^{(LUE)}=2^kM_{k,0}^{(GUE)}$, this equality being implied by the fact that $\rho^{(G),0}(\lambda)=2|\lambda|\,\rho^{(L),0}(2\lambda^2)$.

In contrast to the four term recurrence \eqref{eq3.1.23}, equation \eqref{eq3.1.25} has a famous combinatorial interpretation, which we now present in the language of Section \ref{s3.3}. For $k\in\mathbb{N}$ and $\beta=2$, the expansion coefficients $M_{k,0}^{(LUE)}$ count the number of unique planar ribbon graphs that can be constructed from a $2k$-gon\footnote{In \S\ref{s3.3.2}, we will see that the ribbon graphs pertaining to the Laguerre ensembles must obey a certain bicolouring constraint. This constraint is automatically satisfied when the ribbon graphs are planar, so we ignore it for now.} \citep{Fra03}. To see that they are equal to the $k\textsuperscript{th}$ Catalan number, observe that the set of planar ribbon graphs that can be constructed from a $2k$-gon are in bijection with the set of balanced parenthesisations involving $k$ pairs of parentheses, which are well known to be enumerated by the Catalan numbers \citep{Sta99}; we make the bijection explicit by moving clockwise around the $2k$-gon, starting at the top edge, assigning the label `$1$' (open parenthesis) to the first free edge and `$-1$' (corresponding closed parenthesis) to the edge connected to it by a ribbon, `$2$' and `$-2$' to the next free edge and its ribbon-connected counterpart, and so on. As exemplified in Figure~\ref{fig3.2} below, the edges are then labelled --- in clockwise order starting at the top edge --- by the sequence $(i_1,i_2,\ldots,i_{2k})$ with $i_1=1$ and $j<l$ whenever $0<i_j<i_l$ or $0<i_j=-i_l$.
\begin{figure}[H]
        \centering
\captionsetup{width=.9\linewidth}
        \includegraphics[width=0.9\textwidth]{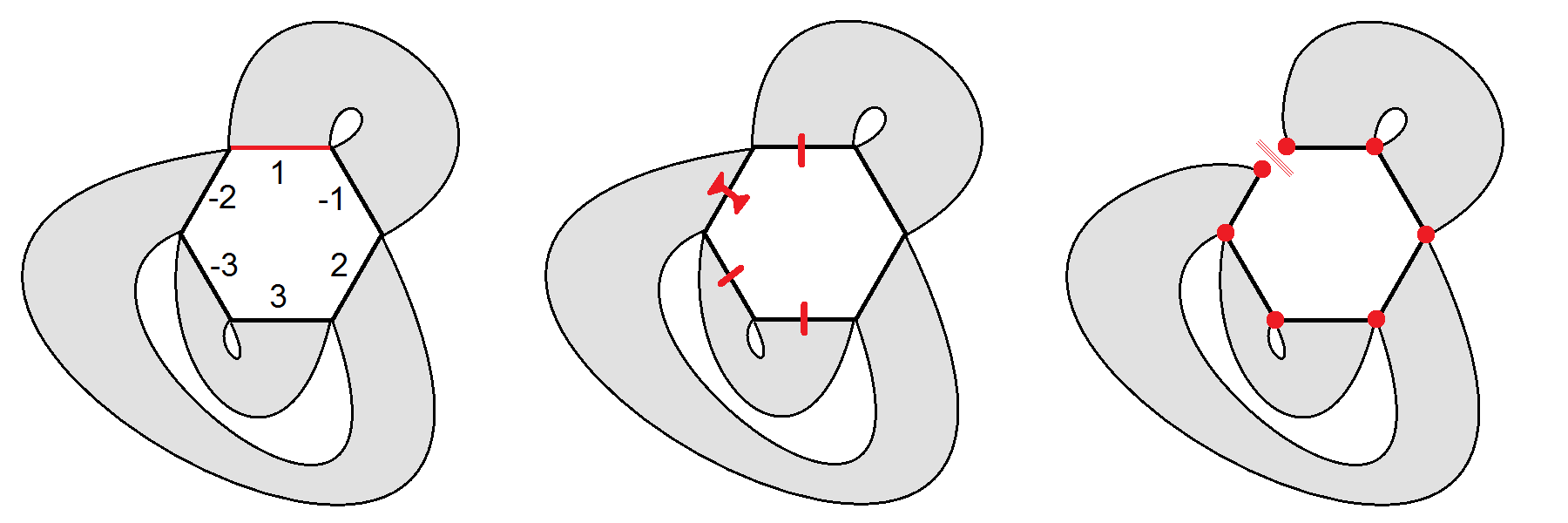}
        \caption[A labelled planar ribbon graph with viable edge and vertex markings]{The first image relates a planar ribbon graph to the balanced parenthesisation $()(())$ by labelling the top edge `$1$' and then, moving clockwise, labelling the tail of the ribbon connected to the top edge `$-1$', the head of the next ribbon `$2$', the head of the final ribbon `$3$', and then the tails of these ribbons `$-3$' and `$-2$'. In the second illustration, we highlight the edges of the hexagon which fit the edge-marking criteria below; in addition to the sixth edge in the clockwise ordering, which is special, an edge is eligible if it precedes an edge that is the tail of some ribbon. In the final image, we show how to split the vertex preceding the top edge in the clockwise ordering so that there are seven vertices available for the vertex-marking procedure outlined below.} \label{fig3.2}
\end{figure}
To elucidate the combinatorics underlying equation \eqref{eq3.1.25}, we take the ribbon graphs labelled in the manner described above and assign them markings in one of two ways. The first marking convention requires that one edge of the $2k$-gon be marked, with eligible edges being those labelled by $i_j$ with either $j=2k$ or $j$ such that $i_{j+1}<0$; that is, we mark either the final edge in the clockwise ordering, or one that immediately precedes a negatively-labelled edge (see the second illustration of Figure \ref{fig3.2} above). There are $(k+1)$ such eligible edges, so that the set of $k$-ribbon graphs edge-marked in this fashion are enumerated by the left-hand side of equation \eqref{eq3.1.25}. For the second marking convention, we split the vertex connecting the first and last edges so that, in clockwise ordering, we have a vertex immediately preceding the first edge, a vertex immediately following the last edge, and $2k-1$ vertices in between. Labelling one of these $2k+1$ vertices by either of $\pm0$ results in a set of $2(2k+1)M_{k,0}^{(LUE)}$ vertex-marked ribbon graphs. Since this number is just the right-hand side of equation \eqref{eq3.1.25} with $k$ replaced by $k+1$, it is unsurprising that there is a bijection between the set of $k$-ribbon graphs edge-marked in the first described convention and the set of $(k-1)$-ribbon graphs vertex-marked in the second convention. To map from the first set to the second, delete the marked edge, the ribbon connected to it, and the edge at the other end of said ribbon, so that each deleted edge collapses to a vertex. Of these two edge-turned-vertices, mark the one corresponding to the deleted edge that was positively-labelled; label it `$-0$' if the positively-labelled edge was marked, or `$0$' otherwise. Finally, we split the vertex connecting the first and last edges. For the reverse mapping, split the marked vertex into two and connect them with an edge. If the vertex was marked with a `$-0$', insert a marked edge immediately anticlockwise to the new edge, and connect the two with a ribbon. If the vertex was instead marked with a `$0$', we need to travel clockwise to a second vertex, turn it into a marked edge, and connect said edge to the first through a ribbon; to uniquely determine the second vertex, we simply pick the furthest one in the clockwise ordering (stopping before reaching the first vertex and edge) that allows us to perform this procedure whilst still producing a planar ribbon graph.
\begin{example} \label{ex3.1}
The planar ribbon graphs displayed in Figure \ref{fig3.3} below map to each other via the bijection detailed above. On the left, we have a planar $3$-ribbon graph labelled by the sequence $(1,-1,2,3,-3,-2)$, with a marking on the final edge `$-2$'. On the right, we have a planar $2$-ribbon graph labelled by the sequence $(1,-1,2,-2)$, with the vertex between edges `$-1$' and `$2$' marked with a `$+0$'. To map the left image to the right, we delete the marked edge, together with its partner `$2$' and the ribbon connecting them, then relabel the edges $3,-3$ as $2,-2$ and split the vertex connecting edges `$1$' and `$-2$'. Since the original edge `$2$' (we focus on the positively-labelled deleted edge) was nestled between edges `$-1$' and `$3$' and was unmarked, we mark the corresponding vertex with a `$+0$' (if the edge marking in the left image was at the edge '$2$', this vertex-marking would have been a `$-0$'). To obtain the left illustration from the right, we expand the vertex labelled `$+0$' into an edge, the vertex $d$ into a marked edge, and connect the two with a ribbon. Then, we join up the new marked edge and the top edge `$1$' and relabel the edges in the appropriate manner. The new marked edge has to be at vertex $d$ because it is the furthest along in the clockwise ordering since vertices $a,b$ appear earlier than vertex `$+0$' in said ordering; note that picking vertex $c$ would result in a non-planar (genus one) ribbon graph.
\end{example}

\begin{figure}[H]
        \centering
\captionsetup{width=.9\linewidth}
        \includegraphics[width=0.7\textwidth]{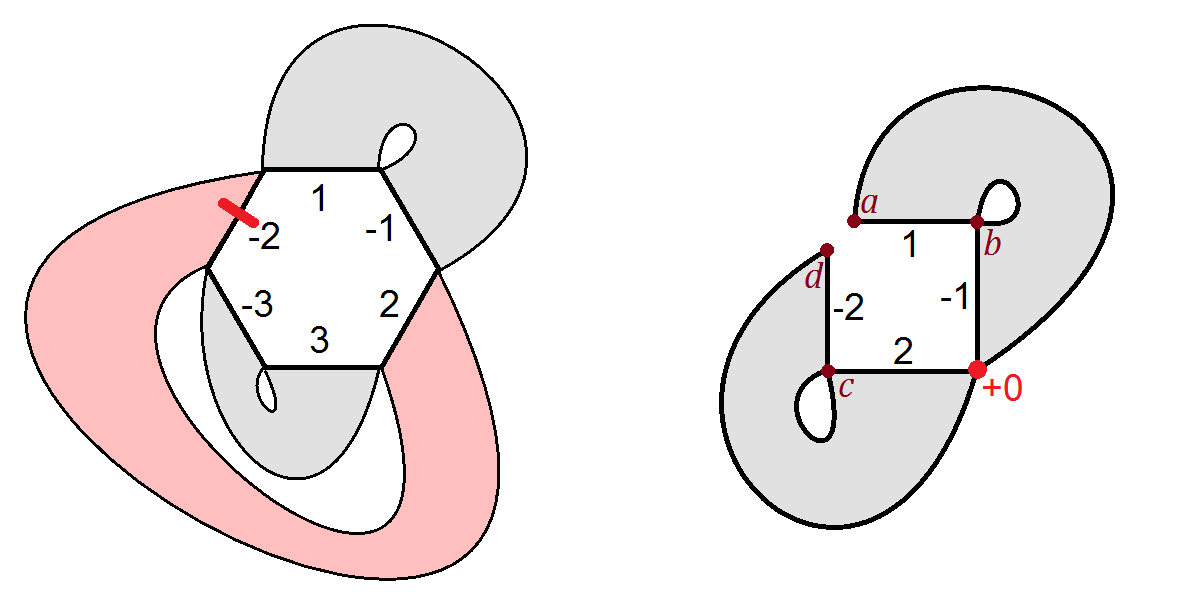}
        \caption[A pair of marked, planar ribbon graphs that correspond to each other]{Representatives of the sets of edge-marked planar $3$-ribbon graphs with a hexagon at their core and vertex-marked planar $2$-ribbon graphs stemming from a square. These sets are in bijection and have cardinality given manifestly by, respectively, the left- and right-hand sides of equation \eqref{eq3.1.25} with $k=3$.} \label{fig3.3}
\end{figure}

When we remove the constraint $l=0$, $M_{k,l}^{(LUE)}$ retains a combinatorial interpretation if we fix $a=\delta_a=0$. Then, as discussed in \S\ref{s3.3.2}, it enumerates ribbon graphs that are now bicoloured and of genus $l/2$ \citep{Fra03} (when $a=0$, $M_{k,l}^{(LUE)}$ vanishes for all odd $l$). Equivalently, $M_{k,l}^{(LUE)}|_{a=0}$ counts the number of ways of forming a compact orientable genus $l/2$ surface by gluing together the edges of a $2k$-gon whose vertices are alternately coloured black and red, with the gluing respecting this bicolouring. When $a=0$, the general-$l$ recursion \eqref{eq3.1.24} reduces to a recursion that runs over even $l$, which was recently derived in \citep[Thrm.~4.1]{ND18} as a means of enumerating the aforementioned gluings. Keeping $\delta_a=0$, but letting $\hat{a}\geq0$ be free, $M_{k,l}^{(LUE)}|_{a=\hat{a}N}$ counts the same ribbon graphs, with the red vertices now weighted by $(\hat{a}+1)$. Thus, $M_{k,l}^{(LUE)}|_{a=\hat{a}N}$ is a polynomial in $(\hat{a}+1)$ with the coefficient of $(\hat{a}+1)^p$ equalling the number of genus $l/2$ bicoloured gluings of the aforementioned $2k$-gon that have $p$ distinct red vertices surviving the gluing procedure (again, we refer to \S\ref{s3.3.2} for details). In this $a=\hat{a}N$ setting, recursion \eqref{eq3.1.24} continues to run over even $l$ so that $m_k^{(LUE)}|_{a=\hat{a}N}$ is manifestly an odd function in $N$, in keeping with the second point of Remark~\ref{R2.1}.

A curious point to note is that even though the moment expansion coefficients $M_{k,l}^{(LUE)}$ have combinatorial interpretations for $k,l\in\mathbb{N}$ and $a=\hat{a}N$ with $\hat{a}\geq0$, the recursion \eqref{eq3.1.24} relating them does not presently have such an interpretation (at least not one comparable to the bijection demonstrated in Example \ref{ex3.1} above). Indeed, it may be the case that this $1$-point recursion describes a relation between the $M_{k,l}^{(LUE)}|_{a=\hat{a}N}$ which simply does not correspond to a topological procedure at the level of the ribbon graphs themselves --- ascertaining the validity of this intriguing proposition remains an open problem. Similar to the LUE case, the moment expansion coefficients $M_{k,l}$ of the GUE, GOE, GSE, LOE, and LSE also have combinatorial meaning, to be detailed in Section \ref{s3.3}, but the $1$-point recursions characterising them are yet to be understood in a combinatorial sense. With the GUE, GOE, and GSE $1$-point recursions known from the works \citep{HZ86}, \citep{Le09}, we now supply the analogous recursions for the LOE and LSE.
\begin{proposition} \label{prop3.6}
For $\beta=1$ and $4$, $a=\hat{a}\kappa N$ with $\hat{a}$ constant in $N$, and $k\geq4$,
\begin{equation} \label{eq3.1.26}
(k+1)M_{k,l}^{(L)}=\sum_{i=1}^4\sum_{j=0}^i(\kappa-1)^jg_{i,j}M_{k-i,l-j}^{(L)},
\end{equation}
where
\begin{equation*}
\begin{gathered}
g_{1,0}=(4k-1)\left(\hat{a}\kappa+4(\kappa-1)\right)(\kappa-1),
\\g_{2,0}=(3-2k)\left(\hat{a}^2\kappa^2+2(\hat{a}\kappa+4(\kappa-1))^2(\kappa-1)^2\right),\quad g_{2,1}=2(2k-3)\hat{a}\kappa,
\\g_{2,2}=(k-1)\left(5k^2-11k+4\right),\quad g_{3,0}=(4k-11)(\hat{a}\kappa+4(\kappa-1))\hat{a}^2\kappa^2(\kappa-1),
\\g_{3,1}=2(11-4k)(\hat{a}\kappa+4(\kappa-1))\hat{a}\kappa(\kappa-1),
\\g_{3,2}=2(3-k)\left(5k^2-19k+16\right)(\hat{a}\kappa+4(\kappa-1))(\kappa-1),\quad g_{4,0}=(4-k)\hat{a}^4\kappa^4,
\\g_{4,1}=4(k-4)\hat{a}^3\kappa^3,\quad g_{4,2}=(k-4)\left(5k^2-32k+47\right)\hat{a}^2\kappa^2,
\\g_{4,3}=2(4-k)(k-3)(5k-17)\hat{a}\kappa,\quad g_{4,4}=4(1-k)\left[(k-4)(k-3)\right]^2,
\end{gathered}
\end{equation*}
and all other $g_{i,j}$ are zero. We also set $M_{k,l}^{(L)}=0$ if $l<0$ or $l>k$.
\end{proposition}

When $a=0$, $M_{k,l}^{(LOE)}$ is shown in \S\ref{s3.3.2} to count the same bicoloured ribbon graphs as $M_{k,l}^{(LUE)}$ except that the ribbons are now allowed to have M\"obius half-twists so that the ribbon graphs are no longer guaranteed to be globally orientable, though they still have Euler genus~$l$ (recall Definition \ref{def3.1} and see Figures \ref{fig3.15} and \ref{fig3.19} for examples). Thus, in keeping with the discussion comparing equations \eqref{eq3.1.23} and \eqref{eq3.1.25}, we see that the three term analogue of these equations obtained by setting $a=l=0$ in Proposition \ref{prop3.6},
\begin{equation} \label{eq3.1.27}
(k+1)M_{k,0}^{(L)}=4(4k-1)(\kappa-1)^2M_{k-1,0}^{(L)}-32(2k-3)(\kappa-1)^4M_{k-2,0}^{(L)},
\end{equation}
is, up to scaling, satisfied by the Catalan numbers when $\kappa=1/2$ or $2$. Indeed, this recursion can be used to verify that $2^kM_{k,0}^{(LOE)}|_{a=0}$, equivalently $2^{-k}M_{k,0}^{(LSE)}|_{a=0}$, is equal to the $k\textsuperscript{th}$ Catalan number. Moreover, we observe that equation \eqref{eq3.1.27} is equivalent to that obtained by setting $l=0$ in the GOE $1$-point recursion given in \citep[Cor.~7]{Le09}.

Now returning to the general-$l$ case, let us compare the moment expansion coefficients $M_{k,l}^{(L)}$ and $M_{k,l}^{(G)}$. We immediately observe that the former is somehow more complicated than the latter: $M_{k,l}^{(L)}$ enumerates the same ribbon graphs as $M_{k,l}^{(G)}$ but with the extra bicolouring constraint described earlier and in \S\ref{s3.3.2}, so that $M_{k,l}^{(L)}|_{a=0}\leq M_{k,l}^{(G)}$ and when $a=\hat{a}\kappa N\neq0$, $(\hat{a}+1)$ acts as a weight that keeps track of the ratio of red to black vertices in the ribbon graphs. Note also that the presence of the parameter $a$ results in the expansion coefficients $M_{k,l}^{(L)}$ and the $1$-point recursions \eqref{eq3.1.24}, \eqref{eq3.1.26} being more complex than their Gaussian counterparts. This increase in complexity is linked to the fact that the Laguerre ensembles sit above the Gaussian ensembles in the hierarchy implied by Lemma \ref{L1.1}. We posit that since the Jacobi ensembles inhabit the tier immediately above the Laguerre ensembles in the aforementioned hierarchy, it is not unreasonable to expect that the $M_{k,l}^{(J)}$ \eqref{eq3.0.10} count some type of ribbon graphs that include, as special cases, the ribbon graphs enumerated by the $M_{k,l}^{(G)}$ or $M_{k,l}^{(L)}$. However, there is no such construction currently known in the literature, with attempts at replicating the constructions of Section \ref{s3.3} thwarted by the fact that the Jacobi matrices of Definition \ref{def1.4} involve inverse matrices whose entries are not compatible with the Isserlis--Wick theorem (see the discussion following Theorem \ref{thrm3.1} of Section \ref{s3.3}). It is nonetheless the author's belief, which we justify below (see also the discussion in \S\ref{s4.4.3} regarding the works \citep{YK83}, \citep{BYK88}, \citep{CDO21}, \citep{GGR20}, \citep{GGR21}), that the $M_{k,l}^{(J)}$ should have combinatorial interpretations in line with what is known for the Gaussian and Laguerre ensembles. 

To motivate the study of the Jacobi ensembles' moment expansion coefficients $M_{k,l}^{(J)}$ with a combinatorial flavour, let us review some properties of the classical matrix ensembles that are relevant in the combinatorial setting. To begin, recall from the paragraph preceding Corollary \ref{C3.2} that $m_k^{(J)}$, consequently $M_{k,l}^{(J)}$, reduces to $m_k^{(L)}$, respectively $M_{k,l}^{(L)}$, when taking $b\to\infty$ (possibly by setting $b=\hat{b}\kappa N$ and working in the large $N$ limit), so that if $M_{k,l}^{(J)}$ were to enumerate a class of ribbon graphs, these ribbon graphs would degenerate to those enumerated by $M_{k,l}^{(L)}$ in the large $b$ limit. This is understood to be a simple consequence of the limit \eqref{eq1.2.23} presented in Lemma \ref{L1.1}, which is key to comparing the Laguerre and Jacobi ensembles. The other limit given in Lemma \ref{L1.1}, relating the Jacobi and Gaussian ensembles, has a cleaner form when written in terms of the symmetric shifted Jacobi ensemble:
\begin{equation*}
\lim_{a\to\infty}w^{(sJ)}(\lambda/\sqrt{a})\big|_{a=b}=w^{(G)}(\lambda).
\end{equation*}
A similar observation can be made by comparing the identities
\begin{align}
\rho^{(G),0}(\lambda)&=\sqrt{2}(1-\lambda^2/2)\rho^{(sJ),0}(\lambda/\sqrt{2})\big|_{a=b=0}, \label{eq3.1.28}
\\ \rho^{(G),0}(\lambda)&=|\lambda|(1-\lambda^2/2)\rho^{(J),0}(\lambda^2/2)\big|_{a=b=0}. \label{eq3.1.29}
\end{align}
Recalling the discussion following equation \eqref{eq3.1.25}, equation \eqref{eq3.1.29} is reminiscent of the analogous relation between $\rho^{(G),0}(\lambda)$ and $\rho^{(L),0}(\lambda)$, suggesting that while the GUE can be compared at leading order to the sJUE by a linear change of variables, the LUE is more closely related to the JUE. If we expand $\mu_k^{(sJ)}$ \eqref{eq3.1.5} as
\begin{equation} \label{eq3.1.30}
\mu_k^{(sJ)}=\sum_{l=0}^{\infty}\Delta M_{k,l}^{(sJ)}N^{1-l}
\end{equation}
and define $\Delta M_{k,l}^{(J)}:=M_{k,l}^{(J)}-M_{k+1,l}^{(J)}$, equations \eqref{eq3.1.28} and \eqref{eq3.1.29} imply that
\begin{equation} \label{eq3.1.31}
2^k\Delta M_{k,0}^{(J)}\big|_{a=b=0}=-2^{k+1}\Delta M_{k,0}^{(sJ)}\big|_{a=0}=M_{k,0}^{(G)},
\end{equation}
which links the differences of the moments of $\rho^{(J),0}(\lambda)|_{a=b=0}$ and $\rho^{(sJ),0}(\lambda)|_{a=b=0}$ to the Catalan numbers. Combining these observations together, it seems that there is merit in studying the moment expansion coefficients of both the un-shifted and symmetric shifted Jacobi ensembles. At this point in time, a good approach for progressing this line of study is to find more ways of comparing the moment expansion coefficients $M_{k,l}$ of  the classical matrix ensembles. To this end, we now derive the $1$-point recursions satisfied by $M_{k,l}^{(JUE)}$, $\Delta M_{k,l}^{(JUE)}$, and $\Delta M_{k,l}^{(sJUE)}$, fixing $\beta=2$ for clarity.

\begin{proposition} \label{prop3.7}
Fix $\beta=2$, $a=\hat{a}N$, and $b=\hat{b}N$, with $\hat{a},\hat{b}={\rm O}(1)$. Then, for $k\geq3$,
\begin{equation} \label{eq3.1.32}
k(\hat{a}+\hat{b}+2)^2M_{k,l}^{(JUE)}=k(k-1)^2M_{k,l-2}^{(JUE)}+\sum_{i=1}^3\left(h_{i,0}M_{k-i,l}^{(JUE)}+h_{i,1}M_{k-i,l-2}^{(JUE)}\right),
\end{equation}
where
\begin{equation*}
\begin{gathered}
h_{1,0}=2(4k-3)(\hat{a}+\hat{b}+1)+\left(3\hat{a}(k-1)+\hat{b}k\right)(\hat{a}+\hat{b}),
\\h_{1,1}=(1-k)(3k^2-8k+6),
\\h_{2,0}=3\hat{a}^2(2-k)+(3-2k)\left((\hat{a}+2)(\hat{b}+2)-2\right),
\\ h_{2,1}=(k-2)(3k^2-10k+9),\quad h_{3,0}=\hat{a}^2(k-3),\quad h_{3,1}=(3-k)(k-2)^2,
\end{gathered}
\end{equation*}
and we set $M_{k,l}^{(JUE)}=0$ if $l<0$.
\end{proposition}
\begin{proof}
The derivation of this recursion is slightly different to that described earlier, so we supply the details. Substitute expansion \eqref{eq3.0.10} into the recurrence \eqref{eq3.1.3}, along with the substitutions $a=\hat{a}N$ and $b=\hat{b}N$, and then equate terms of order $3-l$ in $N$. Note that with our choice of $a$ and $b$, the coefficients $d_{2,0}^{(J)},\ldots,d_{2,3}^{(J)}$ all contain a term of order two and a term of order one in $N$, and no other terms. The ${\rm O}(N^2)$ terms of $d_{2,i}^{(J)}$ correspond to the coefficients $h_{i,0}$ of $M_{k-i,l}^{(JUE)}$ in equation \eqref{eq3.1.32}, while the terms of order one in $N$ correspond to the coefficients $h_{i,1}$ of $M_{k-i,l-2}^{(JUE)}$.
\end{proof}
Proposition \ref{prop3.7} leads directly to the $1$-point recursion on the $\Delta M_{k,l}^{(JUE)}|_{a=\hat{a}N,\,b=\hat{b}N}=M_{k,l}^{(JUE)}|_{a=\hat{a}N,\,b=\hat{b}N}-M_{k+1,l}^{(JUE)}|_{a=\hat{a}N,\,b=\hat{b}N}$, while applying the inverse of equation \eqref{eq2.2.3} yields the equivalent relation on the $\Delta M_{k,l}^{(sJUE)}|_{a=\hat{a}N}$ \eqref{eq3.1.30}. Alternatively, the latter can be derived from equation \eqref{eq3.1.7} by following the prescription at the beginning of this subsection.
\begin{corollary} \label{C3.3}
In the setting of Proposition \ref{prop3.7} with $\Delta M_{k,l}^{(JUE)}=M_{k,l}^{(JUE)}-M_{k+1,l}^{(JUE)}$ and $\Delta M_{k,l}^{(sJUE)}$ defined implicitly by equation \eqref{eq3.1.30}, we have for $k\geq2$ the $1$-point recursions
\begin{multline} \label{eq3.1.33}
(k+1)(\hat{a}+\hat{b}+2)^2\Delta M_{k,l}^{(JUE)}=(2k-1)\left(\hat{a}(\hat{a}+\hat{b}+2)+2(\hat{b}+1)\right)\Delta M_{k-1,l}^{(JUE)}
\\+(2-k)\hat{a}^2\Delta M_{k-2,l}^{(JUE)}+k^2(k+1)\Delta M_{k,l-2}^{(JUE)}
\\-k(k-1)(2k-1)\Delta M_{k-1,l-2}^{(JUE)}+(k-2)(k-1)^2\Delta M_{k-2,l-2}^{(JUE)}
\end{multline}
and
\begin{multline} \label{eq3.1.34}
4(k+1)(\hat{a}+1)^2\Delta M_{k,l}^{(sJUE)}=2(2k-1)(2\hat{a}+1)\Delta M_{k-1,l}^{(sJUE)}+(k+1)(2k+1)^2\Delta M_{k,l-2}^{(sJUE)}
\\+4k^2(2k-1)\Delta M_{k-1,l-2}^{(sJUE)}+(k-1)(2k-1)(2k-3)\Delta M_{k-2,l-2}^{(sJUE)},
\end{multline}
with $\Delta M_{k,l}^{(JUE)},\Delta M_{k,l}^{(sJUE)}$ set to zero for all $l<0$.
\end{corollary}
There are three immediate observations relating to Proposition \ref{prop3.7} and Corollary \ref{C3.3}. Firstly, the recursions therein run over even $l$, similar to Proposition \ref{prop3.5}. Indeed, when the first few moments $m_k^{(JUE)}, m_k^{(sJUE)}$ with $k=0,\ldots,3$ (known from \citep{MOPS}, \citep{MRW17}, \citep{FRW17}) are expanded as series in $1/N$, they do not contain terms of even powers in $N$, so the recursions \eqref{eq3.1.32}--\eqref{eq3.1.34} are trivially satisfied when $l$ is odd. As with Proposition \ref{prop3.5}, this feature is due to our choice of taking $\beta=2$ and $a,b\propto N$. Our second observation is that, unlike the analogous $1$-point recursions for the GUE and LUE moment expansion coefficients, the recursions of Proposition \ref{prop3.7} and Corollary \ref{C3.3} show that $M_{k,l}^{(JUE)},\Delta M_{k,l}^{(JUE)},\Delta M_{k,l}^{(sJUE)}$ depend, respectively, on $M_{k,l-2}^{(JUE)},\Delta M_{k,l-2}^{(JUE)},\Delta M_{k,l-2}^{(sJUE)}$. Thus, in contrast to the Gaussian and Laguerre cases, knowledge of $m_{k'}^{(JUE)},\Delta m_{k'}^{(JUE)},\mu_{k'}^{(sJUE)}$ with $k'<k$ is not enough to determine $M_{k,l}^{(JUE)},\Delta M_{k,l}^{(JUE)},\Delta M_{k,l}^{(sJUE)}$ for $l>0$. Our final observation is that of the three recursions \eqref{eq3.1.32}--\eqref{eq3.1.34}, the final one is the simplest.

If we are to assume that the moment expansion coefficients of the Jacobi ensembles count some type of surface similar to those counted by the GUE and LUE spectral moments, comparing the corresponding $1$-point recursions may shed light on what these surfaces actually are. For example, the fact that the recursions of Proposition \ref{prop3.7} and Corollary \ref{C3.3} run over even $l$ suggests that $l/2$ might play the role of genus in the Jacobi case, as it does in the Gaussian and Laguerre cases. On the other hand, the second observation given above could give a hint as to what sets the JUE apart from the GUE and LUE. Since setting $a=0$ in Proposition \ref{prop3.5} yields a simpler recursion that retains an interpretation in terms of ribbon graphs, we set $a=b=0$ in Proposition \ref{prop3.7} and Corollary \ref{C3.3} for comparison:
\begin{corollary} \label{C3.4}
In the context of Proposition \ref{prop3.7} and Corollary \ref{C3.3}, setting $a=b=0$ reduces recursion \eqref{eq3.1.32} to
\begin{multline} \label{eq3.1.35}
4kM_{k,l}^{(JUE)}=k(k-1)^2M_{k,l-2}^{(JUE)}+2(4k-3)M_{k-1,l}^{(JUE)}
\\+(1-k)(3k^2-8k+6)M_{k-1,l-2}^{(JUE)}+2(3-2k)M_{k-2,l}^{(JUE)}
\\+(k-2)(3k^2-10k+9)M_{k-2,l-2}^{(JUE)}+(3-k)(k-2)^2M_{k-3,l-2}^{(JUE)},
\end{multline}
recursion \eqref{eq3.1.33} to
\begin{multline} \label{eq3.1.36}
4(k+1)\Delta M_{k,l}^{(JUE)}=2(2k-1)\Delta M_{k-1,l}^{(JUE)}+k^2(k+1)\Delta M_{k,l-2}^{(JUE)}
\\-k(k-1)(2k-1)\Delta M_{k-1,l-2}^{(JUE)}+(k-2)(k-1)^2\Delta M_{k-2,l-2}^{(JUE)},
\end{multline}
and recursion \eqref{eq3.1.34} to
\begin{multline} \label{eq3.1.37}
4(k+1)\Delta M_{k,l}^{(sJUE)}=2(2k-1)\Delta M_{k-1,l}^{(sJUE)}+(k+1)(2k+1)^2\Delta M_{k,l-2}^{(sJUE)}
\\+4k^2(2k-1)\Delta M_{k-1,l-2}^{(sJUE)}+(k-1)(2k-1)(2k-3)\Delta M_{k-2,l-2}^{(sJUE)}.
\end{multline}
\end{corollary}
Comparing these forms to the GUE analogue \eqref{eq3.0.12} and the result \citep[Thrm.~4.1]{ND18} of setting $a=0$ in Proposition \ref{prop3.5},
\begin{equation} \label{eq3.1.38}
(k+1)M_{k,l}^{(LUE)}=2(2k-1)M_{k-1,l}^{(LUE)}+(k-2)(k-1)^2M_{k-2,l-2}^{(LUE)},
\end{equation}
we note that equations \eqref{eq3.1.36} and \eqref{eq3.1.37} contain familiar terms, whereas the same is not true for equation \eqref{eq3.1.35}. Setting equation \eqref{eq3.1.35} aside, we see that upon ignoring the terms with indices $(k,l-2)$ and $(k-1,l-2)$, equation \eqref{eq3.1.36} has the same structure as equation \eqref{eq3.1.38}, while equation \eqref{eq3.1.37} is almost equivalent to the Harer--Zagier recursion \eqref{eq3.0.12}. Hence, it seems that the JUE should be compared to the LUE and the sJUE should be compared to the GUE (cf.~the discussion following equations \eqref{eq3.1.28} and \eqref{eq3.1.29}). In addition, the coefficients of the ignored terms contain, respectively, the factors $(k+1)$ and $(2k-1)$, which suggests that we should group them with the terms with indices $(k,l)$ and $(k-1,l)$. As expected, setting $l=0$ in equations \eqref{eq3.1.36} and \eqref{eq3.1.37} shows that $2^k\Delta M_{k,0}^{(JUE)}|_{a=b=0}$ and $-2^{k+1}\Delta M_{k,0}^{(sJUE)}|_{a=0}$ satisfy the two term Catalan recursion \eqref{eq3.1.25}, which is in keeping with equation \eqref{eq3.1.31}. Moving past these observations, the challenge now is to extend the combinatorial interpretation of equation \eqref{eq3.1.25} to the general-$l$ recursions \eqref{eq3.0.12}, \eqref{eq3.1.36}, \eqref{eq3.1.37}, and \eqref{eq3.1.38}. As mentioned earlier, this remains an open problem that we are unable to address here. Nonetheless, some comments on promising directions are given in \S\ref{s4.4.2}. As we move on to the next section, let us make a final remark: Experimentation for $0\leq k\leq 10$, $0\leq l\leq5$ suggests that for each $k,l\geq0$, $-2^{2(k+l)+1}\Delta M_{k,l}^{(sJUE)}|_{a=0}$ is a positive integer, which supports the idea that these scaled differences of moments are equal to the cardinality of some class of combinatorial objects. Moreover, it can be seen that these positive integers grow extremely quickly as $k,l$ are increased, relative to what is seen for their GUE and LUE counterparts; to be more concrete, comparing equations \eqref{eq3.0.12} and \eqref{eq3.1.37} shows that for all $k,l\geq0$,
\begin{equation*}
-2^{2(k+l)+1}\Delta M_{k,l}^{(sJUE)}|_{a=0}\geq M_{k,l}^{(GUE)}\geq M_{k,l}^{(LUE)}|_{a=0}.
\end{equation*}

\setcounter{equation}{0}
\section{Established Results on the Moments of the Classical Matrix Ensembles} \label{s3.2}
In the previous section, we derived recursions characterising the spectral moments of the classical matrix ensembles, along with their complex-$k$ generalisations \eqref{eq3.1.1} and their expansion coefficients $M_{k,l}$. To summarise Sections \ref{s2.3} and \ref{s3.1}, our methodology consisted of using the theory of Selberg correlation integrals to obtain differential equations satisfied by the eigenvalue densities $\rho(x)$ and resolvents $W_1(x)$ of the classical matrix ensembles --- applying integration by parts on the former set of differential equations or inserting the expansion $W_1(x)=\sum_{k=0}^{\infty}m_k/x^{k+1}$ in the latter set resulted in the sought recursions. Some of the recursions derived in Section \ref{s3.1} have already been given in the literature, but our approach involving Selberg correlation integrals is distinct and is able to treat all of the classical matrix ensembles in a unified manner. In this section, we give a select review on recent studies that provide characterisations of the moments of the classical matrix ensembles, focusing particularly on those that are different to the results of Section \ref{s3.1} (classical results on this topic are reviewed in a plethora of texts, including \citep{Meh04}, \citep{Fo10}). Seeing as this gives us an opportunity to showcase connections to various fields of mathematics, we compartmentalise our review by methodology rather than outcome.

\subsection{Results from skew-orthogonal polynomial theory}
A straightforward approach to computing the spectral moments is to simply integrate monomials against closed form expressions for the corresponding eigenvalue densities. This approach was taken by Livan and Vivo in 2011 \citep{LV11}, with their starting point being expressions for the classical Laguerre and Jacobi ensembles' eigenvalue densities known from (skew-)orthogonal polynomial theory. Recalling the discussion of \S\ref{s1.2.3}, the idea then is to take $\rho(\lambda)$ to be given by equation \eqref{eq1.2.67} with $x=y=\lambda$ when $\beta=2$, and by this plus a correction term when $\beta=1,4$. Integrating these expressions against $\lambda^k$ for $k\in\mathbb{N}$, Livan and Vivo gave exact expressions for the moments of the classical Laguerre and Jacobi ensembles in terms of sums of ratios of gamma functions. At this same time, in a parallel but independent work, Mezzadri and Simm \citep{MS11} derived equivalent expressions for the spectral moments $m_k$ of the classical Gaussian, Laguerre, and Jacobi ensembles, moreover showing that their expressions hold true for all $k\in\mathbb{C}$ such that the $m_k$ are well-defined. In order to ease the required integrations against the aforementioned expressions for the eigenvalue densities $\rho^{(w)}(\lambda)$, a key idea of Mezzadri and Simm was to use polynomials that are orthogonal with respect to the perturbations $\lambda^kw(\lambda)$ of the classical weights $w(\lambda)$ \eqref{eq1.2.9}.

The moment recurrences given in \citep{HT03}, \citep{Le04}, \citep{Le09}, \citep{CMSV16b} were also obtained through consideration of (skew-)orthogonal polynomial theory. In the 2003 work \citep{HT03}, Haagerup and Thorbj{\o}rnsen used elementary identities satisfied by the Hermite and Laguerre orthogonal polynomials to express the \textit{exponential moment generating function} (cf.~equation \eqref{eq1.1.18})
\begin{equation*}
R_1(x):=\int_{\mathrm{supp}\,\rho}\exp(x\lambda)\rho(\lambda)\,\mathrm{d}\lambda,
\end{equation*}
for the GUE and LUE cases in terms of (confluent) hypergeometric functions. They then transformed known differential equations satisfied by these hypergeometric functions into differential equations characterising $R_1(x)$. From there, Haagerup and Thorbj{\o}rnsen obtained recurrences for the GUE and LUE (positive-integer) spectral moments through essentially the same strategy as used in \S\ref{s3.1.1}, thereby recovering the recursion of Harer and Zagier \citep{HZ86}, and finding its LUE analogue. Ledoux \citep{Le04} extended these ideas to the JUE case a year later, and to the GOE case \citep{Le09} in 2009 --- clever manipulations were needed to handle the correction term that must be added to the GUE eigenvalue density to obtain its GOE analogue. In 2016, Cunden et al.~\citep{CMSV16b} extended the work of Ledoux to derive an inhomogeneous moment recurrence for the LOE, where the inhomogeneous terms were given in terms of the LUE moments. This recurrence, together with the LUE moment recurrence given in \citep{HT03}, was shown to hold for negative-integer moments ($m_{-k}$ with $0<k<a+1$).

\subsection{Results from symmetric function theory}
Of the classical $\beta$ ensembles (recall \S\ref{s1.2.4}), only the $\beta=1,2$, and $4$ regimes can be studied through (skew-)orthogonal polynomial theory. To compute the spectral moments in the general $\beta$ case, other approaches must be taken. One option is to use loop equation analysis to iteratively compute the resolvent expansion coefficients $W_1^0(x),W_1^1(x),\ldots$ \eqref{eq1.1.21} and then interpret them as generating functions for the moment expansion coefficients $M_{k,l}$. In the Gaussian and Laguerre cases, the spectral moments are polynomials in $N$, so this procedure can be used to compute the spectral moment $m_k$ for any given $k$, assuming enough computation power is available; in the case of the Jacobi and Cauchy $\beta$ ensembles, the best one can hope for is a truncation of the large $N$ expansion of the spectral moments. We do not review the loop equation formalism any further here since it will be revisited in Chapter~4, but note that the Gaussian, Laguerre, and Jacobi $\beta$ ensembles were studied in this manner in the 2014 and 2017 works \citep{WF14}, \citep{FRW17}.

While the loop equation formalism can be used to compute (the leading order behaviour of) spectral moments of the classical $\beta$ ensembles, it does not provide any closed form expressions for these moments. Such closed form expressions can however be obtained in the general $\beta$ setting through symmetric function theory. The relevance of said theory should not be surprising since equation \eqref{eq3.0.1} shows that the $k\textsuperscript{th}$ spectral moment is given by the integral of the symmetric polynomial $\sum_{i=1}^N\lambda_i^k$ multiplied by the appropriate eigenvalue j.p.d.f. Some well known bases for the ring of symmetric polynomials are the sets of monomial symmetric polynomials, Schur polynomials, zonal polynomials, and the Jack polynomials. As an algebra, it is also generated by the sets of power-sum symmetric polynomials and elementary symmetric polynomials. We refer the reader to \citep{Mac79} for definitions of these polynomials, but note that the elementary symmetric polynomials were mentioned earlier in Remark \ref{R2.2}. Of particular interest to us is the set of \textit{Jack polynomials of type ``C''}, which differ from those of type ``J'' and ``P'' by a choice of normalisation (see, e.g., \citep{Dum03}) and are written as
\begin{equation*}
C_{\sigma}^{(\alpha)}(\lambda_1,\ldots,\lambda_N).
\end{equation*}
They are indexed by partitions $\sigma$ of integers (i.e., $\sigma=(\sigma_1,\ldots,\sigma_k)$ with $k,\sigma_1,\ldots,\sigma_k\in\mathbb{N}$ and $\sigma_1\geq\sigma_2\geq\cdots\geq\sigma_k>0$) and a positive real parameter $\alpha$ not to be confused with that associated with the Cauchy ensemble; setting $\alpha=1$ recovers the Schur polynomials while $\alpha=2$ corresponds to the zonal polynomials.

In the 1997 work on Selberg correlation integrals \citep{Kad97}, Kadell gave an explicit closed form expression, in terms of ratios of gamma functions, for the integral
\begin{equation} \label{eq3.2.1}
\int_{[0,1]^N}C_{\sigma}^{(2/\beta)}(\lambda_1,\ldots,\lambda_N)\,p^{(J)}(\lambda_1,\ldots,\lambda_N;\beta)\,\mathrm{d}\lambda_1\cdots\mathrm{d}\lambda_N,
\end{equation}
where $p^{(J)}(\lambda_1,\ldots,\lambda_N;\beta)$ is the eigenvalue j.p.d.f.~\eqref{eq1.2.81} of the Jacobi $\beta$ ensemble and $\sigma$ is any given partition. In the same year, Baker and Forrester \citep{BF97} showed that when $w(\lambda)$ is either the Gaussian or Laguerre weight \eqref{eq1.2.9}, the above integral \eqref{eq3.2.1} with $p^{(J)}$ replaced by $p^{(w)}$ is equal to the generalised Hermite (up to a minus sign), respectively Laguerre, polynomial of index $\sigma$ evaluated at zero, these generalised polynomials being the multivariable symmetric polynomials that are orthogonal with respect to the inner products
\begin{equation*}
\mean{f,g}^{(w)}:=\int_{\mathbb{R}^N}f(\lambda_1,\ldots,\lambda_N)g(\lambda_1,\ldots,\lambda_N)\,p^{(w)}(\lambda_1,\ldots,\lambda_N;\beta)\,\mathrm{d}\lambda_1\cdots\mathrm{d}\lambda_N.
\end{equation*}
In the Jacobi case, one can surmise from \citep{BF97} a result analogous to that of Kadell \citep{Kad97}.

Around 2003 \citep{Dum03}, Dumitriu et al.~used the results of \citep{Kad97}, \citep{BF97} (among others) to devise algorithms, implemented in the MOPS package \citep{MOPS}, for computing the spectral moments of the Gaussian, Laguerre, and Jacobi $\beta$ ensembles. These algorithms essentially boil down to expressing the polynomial $\sum_{i=1}^N\lambda_i^k$ as a linear combination of Jack ``C'' polynomials so that the problem of computing the spectral moments \eqref{eq3.0.1} reduces to the computation of integrals of the form \eqref{eq3.2.1}. One should note that these algorithms are not equivalent to closed form expressions for the spectral moments. However, in the 2017 work \citep{MRW17}, Mezzadri et al.~provide such closed form expressions in the Jacobi and Laguerre cases by way of giving explicit formulae for the coefficients in the aforementioned linear combination of Jack ``C'' polynomials that is equal to $\sum_{i=1}^N\lambda_i^k$. Thus, they express $m_k^{(J)}$ and $m_k^{(L)}$ as sums over partitions of $k$ with each summand being an explicit ratio of gamma functions. Prompted by Fyodorov and Le Doussal, they also show how to use a relation between $\left\{C_{\sigma}^{(\alpha)}(1/\lambda_1,\ldots,1/\lambda_N)\right\}_{\sigma}$ and $\left\{C_{\sigma'}^{(\alpha)}(\lambda_1,\ldots,\lambda_N)\right\}_{\sigma'}$ to extend their expressions for $m_k^{(J)},m_k^{(L)}$ to the case that $k$ is a negative integer. In fact, these expressions for the integer moments of the Jacobi and Laguerre $\beta$ ensembles as sums of ratios of gamma functions were given by Fyodorov and Le Doussal in the independent work \citep{FL16}. Moreover, this latter work contains contour integral representations for $m_k^{(J)},m_k^{(L)}$ ($k\in\mathbb{Z}$ sufficiently large) due to Borodin and Gorin. Before moving on, let us finally mention that Novaes also studied the negative-integer moments of the LUE using Schur polynomials in the 2015 work \citep{No15}.

\subsection{Results in terms of hypergeometric orthogonal polynomials} \label{s3.2.3}
In the recent work \citep{CMOS19}, Cunden at al.~made the novel observation that in the classical Gaussian, Laguerre, and Jacobi cases, simple linear combinations of the complex-$k$ moments $m_k$ \eqref{eq3.1.1}, interpreted as functions of $k$ and renormalised by factors dependent on $N$, are hypergeometric orthogonal polynomials in $k$. To be precise, they showed that when $\beta=2$ and $N$ is fixed, $m_k^{(G)}$ is essentially a Meixner--Pollaczek polynomial in $k$, $m_k^{(L)}$ is a continuous dual Hahn polynomial, and $\Delta m_k^{(J)}=m_k^{(J)}-m_{k+1}^{(J)}$ is a Wilson polynomial. These three polynomials belong to the Askey scheme of hypergeometric orthogonal polynomials, with the first two being degenerations of the latter. Cunden et al.~prove this observation in the Gaussian and Laguerre cases by seeing agreement in the hypergeometric function representations of the moments and the corresponding polynomials when $k\in\mathbb{N}$, and then extending this agreement to $k\in\mathbb{C}$ through an application of Carlson's theorem (cf.~Remark \ref{R2.5}); in the Jacobi case, lacking such a representation in terms of hypergeometric functions, they derive the analogue of recurrence \eqref{eq3.1.3} for $\Delta m_k^{(J)}$ and show that this recurrence, after appropriate scaling, also characterises the Wilson polynomials in a unique sense. By noting that certain linear combinations of the $\beta=1,4$ moments are known to be given by linear combinations of their $\beta=2$ counterparts, Cunden et al.~further show within \citep{CMOS19} how the polynomial characterisations of the $\beta=2$ moments extends to the linear combinations considered in the $\beta=1,4$ cases.

\begin{remark} \label{R3.1}
It was shown in \citep{CMOS19} that the LUE moments and JUE moment differences satisfy the following reciprocity laws:
\begin{align}
m_{-k-1}^{(L)}&=\left(\prod_{j=-k}^k\frac{1}{a-j}\right)m_k^{(L)},
\\\Delta m_{-k-1}^{(J)}&=\left(\prod_{j=-k}^k\frac{a+b+2N-j}{a-j}\right)\Delta m_k^{(J)}.
\end{align}
It is not immediately obvious that there exist similar laws for the moments of the orthogonal or symplectic ensembles; one can experiment using data computable from Proposition \ref{prop3.1} and Corollary \ref{C3.2}. It is believed that this is a consequence of three term recurrences, as seen in the $\beta=2$ cases, being simpler than their five term ($\beta=1,4$) analogues in an integrable sense. Indeed, this is the reason that Cunden et al.~were able to place the $\beta=2$ moments $m_k^{(G)},m_k^{(L)},m_k^{(J)}$ in the Askey scheme of hypergeometric orthogonal polynomials, but were only able to do so for linear combinations of the moments when taking $\beta=1,4$.
\end{remark}

Soon after the work \citep{CMOS19}, Assiotis et al.~showed \citep{ABGS20} that the complex-$k$ sum of moments $\mu_k^{(CyUE)}$ \eqref{eq3.1.6} of the symmetric Cauchy unitary ensemble, again understood as a function of $k$ with fixed $N$ and properly renormalised, is also a hypergeometric orthogonal polynomial. This time, the relevant polynomials are the continuous Hahn polynomials, which are also degenerations of the Wilson polynomials. Recalling that the hypergeometric functions are defined by
\begin{equation*}
\hypergeometric{p}{q}{a_1,\ldots,a_p}{b_1,\ldots,b_q}{z}:=\sum_{n=0}^{\infty}\frac{(a_1)^{(n)}\cdots(a_p)^{(n)}}{(b_1)^{(n)}\cdots(b_q)^{(n)}}\frac{z^n}{n!},
\end{equation*}
with $(x)^{(n)}=x(x+1)\cdots(x+n-1)$ denoting the rising Pochhammer symbol, the \textit{continuous Hahn polynomials} are defined as \citep{KLS10}
\begin{equation} \label{eq3.2.4}
S_n(x;a,b,c,d):=\mathrm{i}^n\frac{(a+c)^{(n)}(a+d)^{(n)}}{n!}\hypergeometric{3}{2}{-n,n+a+b+c+d-1,a+\mathrm{i}x}{a+c,a+d}{1}.
\end{equation}
Assiotis et al.~proceed by deriving a differential equation characterising the eigenvalue density $\rho^{(Cy)}(\lambda;N,2)$ of the CyUE, using this differential equation to obtain recurrence \eqref{eq3.1.13} on the sums of moments $\mu_k^{(Cy)}$ \eqref{eq3.1.6}, and then confirming that this recurrence also characterises the continuous Hahn polynomials. Thus, they show that
\begin{multline}
\mu_k^{(CyUE)}=\frac{\Gamma\left(k+1/2\right)\Gamma\left(\alpha-k-1/2\right)}{\Gamma\left(\alpha+3/2\right)\Gamma\left(-\alpha-1/2\right)\sqrt{\pi}}\frac{\mathrm{i}^{1-N}}{2}\Gamma\left(1/2-\alpha-N\right)\alpha(2\alpha+N)
\\ \times S_{N-1}\left(-\mathrm{i}(k+1);1,\frac{1}{2}+\alpha,1,\frac{1}{2}+\alpha\right),
\end{multline}
with $S_{N-1}$ defined by equation \eqref{eq3.2.4}.

We provide now the analogue of the above result of Assiotis et al.~in the case of the symmetric shifted Jacobi unitary ensemble corresponding to the weight $w^{(sJ)}(\lambda)\big|_{a=b}$ \eqref{eq3.0.6}. Let us first recall from the third point of Remark \ref{R2.1} that $\rho^{(sJUE)}(\lambda)\big|_{a=b}$ is equal to $(1-\lambda^2)^a$ multiplied by a polynomial of order $N-1$. Thus, we see from equations \eqref{eq2.2.8} and \eqref{eq3.1.5} that
\begin{equation} \label{eq3.2.6}
\tilde{\mu}_k:=\frac{\Gamma(k+N+a+3/2)}{\Gamma(k+1/2)}\mu_k^{(sJUE)}\Big|_{a=b}
\end{equation}
is a polynomial in $k$ of order $N-1$. Substituting this scaling into recurrence \eqref{eq3.1.7} results in the simpler recurrence
\begin{multline} \label{eq3.2.7}
(2k+4)[  (2 k + 3) - 2 (a + N)  ] \tilde{\mu}_{k+1} + 2  
 [  ( 2k+2)^2  - 2 N ( N + 2a)] \tilde{\mu}_k \\
 + (2k ) (2 (N + a + k) + 1)  \tilde{\mu}_{k-1} = 0
\end{multline}
valid for $k\in\mathbb{C}$ with $\mathrm{Re}(k)>1/2$. Now, let us constrain the continuous Hahn polynomials \eqref{eq3.2.4} by defining
\begin{equation}
s_n(x;a,b):=S_n(x;a,b,\overline{a},\overline{b})
\end{equation}
and observe that this polynomial satisfies the recurrence \citep[18.22.13--18.22.15]{DLMF}
\begin{multline} \label{eq3.2.9}
A(x)s_n(x+\mathrm{i};a,b)-\left[A(x)+C(x)-n(n+2\mathrm{Re}(a+b)-1)\right]s_n(x;a,b)
\\+C(x)s_n(x-\mathrm{i};a,b)=0,
\end{multline}
with
\begin{equation*}
A(x)=(x+\mathrm{i}\overline{a})(x+\mathrm{i}\overline{b}),\qquad C(x)=(x-\mathrm{i}\overline{a})(x-\mathrm{i}\overline{b}).
\end{equation*}

\begin{proposition} \label{prop3.8}
  For $k\in\mathbb{C}$ with ${\rm Re}(k)> -1/2$ and $\tilde{\mu}_k$ defined by equation \eqref{eq3.2.6}, we have
    \begin{equation}\label{eq3.2.10}
  \tilde{\mu}_k   =   { \tilde{\mu}_0  \mathrm{i}^{1-N} \over N (3/2 - (a + N))^{(N-1)}} 
   s_{N-1} \Big (\mathrm{i}(k+1);1,{1 \over 2} - (a + N) \Big ),
   \end{equation}
with
\begin{equation} \label{eq3.2.11}
\tilde{\mu}_0 =  {\Gamma(N + a + 3/2) \over \Gamma(1/2)}
 \Big ( {2 N (a + N) (2 a + N)  \over 1 - 4 (a + N)^2}  \Big ).
\end{equation}
Thus, the difference of moments $ {\mu}_k^{(sJUE)}$ of the symmetric shifted Jacobi unitary ensemble is given in terms of the continuous Hahn polynomials according to
     \begin{equation}\label{eq3.2.12}
 {\mu}_k^{(sJUE)}    =    {\tilde{\mu}_0 \mathrm{i}^{1-N}  \Gamma(k+1/2) \over  N \Gamma(k +  N  + a + 3/2)  (3/2 - (a + N))^{(N-1)}} 
   s_{N-1} \Big (\mathrm{i}(k+1);1,{1 \over 2} - (a + N) \Big ).
   \end{equation}
   \end{proposition}
\begin{proof}
Equation \eqref{eq3.2.11} is obtained by combining equations \eqref{eq3.1.8} and \eqref{eq3.2.6}. Comparing equations \eqref{eq3.2.7} and \eqref{eq3.2.9} shows that
\begin{equation*}
q(k;a,N):=C_{N,a}\,s_{N-1} \Big (\mathrm{i}(k+1);1,{1 \over 2} - (a + N) \Big ),
\end{equation*}
with $C_{N,a}$ independent of $k$, satisfies the recurrence \eqref{eq3.1.8}. To obtain agreement with $\tilde{\mu}_k$ at $k=0$, we set
\begin{equation*}
    C_{N,a} =  { \tilde{\mu}_0  \over s_{N-1}\Big (\mathrm{i};1,{1 \over 2} - ( a  + N) \Big )} = 
     { \tilde{\mu}_0 \,\mathrm{i}^{1-N} \over N (3/2 - (a + N))^{(N-1)}} ,
  \end{equation*}      
  where the second equality follows from definition \eqref{eq3.2.4} of the continuous Hahn polynomials. With this choice, both $q(k;a,N)$ and $\tilde{\mu}_k$ are polynomials of order $N-1$ in $k$ that coincide at $k=0$. As they both satisfy recurrence \eqref{eq3.1.8}, they also coincide at $k\in\mathbb{Z}$. This agreement is extended to $k\in\mathbb{C}$ via Carlson's theorem (recall Remark \ref{R2.5}). Finally, equation \eqref{eq3.2.12} follows from substituting equation \eqref{eq3.2.6} into equation \eqref{eq3.2.10}.
\end{proof}

\setcounter{equation}{0}
\section{The Isserlis--Wick Theorem and Ribbon Graphs} \label{s3.3}
A major motivation for studying ribbon graphs (recall Definition \ref{def3.1}) is their relation to topological (hyper)maps, which we briefly discuss throughout this section. For now, we note that the term `map' is not synonymous with `function' or `mapping'. Instead, topological maps are so-called because they are reminiscent of the geographic maps of countries that one would find in an atlas --- they are graphs embedded in surfaces. These topological maps have been studied by combinatorialists and geometers since the 1960s \citep{Tut63}, \citep{JS78}, but became prominent in algebraic geometry around 1984 due to Grothendieck's work on \textit{dessins d'enfants} (children's drawings) \citep{Sch94} --- technically, a topological map is equivalent to a `clean' dessin (see Remark \ref{R3.3} at the end of \S\ref{s3.3.1}).

The connection between ribbon graphs (equivalently, topological maps) and random matrix theory is that, through consequences of the Isserlis--Wick theorem that we will soon expound, random matrix theory provides a neat method for enumerating ribbon graphs; the upshot here is that the pairwise polygon-edge identifications represented by the ribbons of a ribbon graph can be interpreted as pairings of random matrix entries. Algebraic geometers and mathematical physicists have used this relationship in a number of intriguing ways, of which we mention a select few:

\begin{enumerate}
\item Let $\mathcal{M}_g^s$ denote the moduli space of genus $g$ compact Riemann surfaces with $s$ marked points (i.e., the $(6g-6+2s)$-dimensional space of parameters that determine the complex structure of such surfaces). Using the uniqueness of the Jenkins--Strebel quadratic differential on compact Riemann surfaces \citep{Jen57}, \citep{Str67}, \citep{Str84}, Harer \citep{Har86} showed how to canonically assign a genus $g$, orientable, connected, metrised ribbon graph built from $s$ polygons to each point of $\mathcal{M}_g^s$ (here, \textit{metrised} means that each ribbon is labelled with a length). This had the consequence of endowing $\mathcal{M}_g^s$ with the structure of a simplicial complex (each $k$-simplex corresponding to a $k$-ribbon graph representative of all metrised ribbon graphs that reduce to it upon `forgetting' ribbon lengths), which enabled Harer and Zagier \citep{HZ86} to reduce the problem of computing the (orbifold) Euler characteristic of $\mathcal{M}_g^s$ to that of enumerating ribbon graphs. This treatment was extended to the moduli space of real algebraic curves in \citep{GJ97}, \citep{GHJ01}, with the relevant ribbon graphs now allowed to be non-orientable.
\item In string theory (see, e.g., \citep{Pol98} for an introduction), one replaces particles with \textit{strings}, which are topologically lines or circles. The trajectory of a string is called a worldsheet, which is a surface with boundaries. Considering only closed (circular) strings, a worldsheet is a surface of, say, genus $g$ with $s$ punctures --- the handles counted by the genus arise from strings splitting and combining, while the punctures are simply contractions of the boundaries that correspond to the initial and final states of strings. In the theory of 2D quantum gravity, one considers a generic worldsheet with its structure, that is, a choice of metric, left undetermined. The partition function is formulated as an integral over the space of all possible worldsheet metrics, with the integrand given in terms of aspects of the worldsheet that depend on the choice of metric. This integral is difficult to make sense of because the space of worldsheet metrics is unwieldy for many reasons. In 1990, progress was made by the independent works \citep{BK90}, \citep{DS90}, \citep{GM90}, where the idea was to discretise the worldsheet and approximate it by triangulations (i.e., topological maps with trivalent vertices or ribbon graphs built purely from triangles). When considering the limit of infinitely many triangles, these discrete approximations improve to the point that each choice of triangulation corresponds to a worldsheet metric. Consequently, the relevant partition function can essentially be written as a sum over orientable, connected ribbon graphs.
\item Another approach to 2D quantum gravity, called topological gravity, has partition function equal to a generating function for certain intersection numbers. These intersection numbers are given by integrals over the compactifications $\overline{\mathcal{M}}_g^s$ of the moduli spaces defined in point 1 above. A 1991 conjecture of Witten \citep{Wit91} is that topological gravity should be equivalent to the formalism outlined in point 2 above. This conjecture was famously proven by Kontsevich the following year \citep{Kon92} using the ideas of Harer that we reviewed in point 1. In short, Kontsevich used the structure of $\mathcal{M}_g^s$ as a simplicial complex (with each simplex corresponding to a specific ribbon graph) in order to relate the aforementioned intersection numbers to explicit sums over ribbon graphs.
\end{enumerate}
For detailed surveys of these topics, see \citep{FGZ95}, \citep{MP98}, \citep{LZ04}, \citep{Mon09}.

The studies listed above share the philosophy of reducing a difficult problem to that of enumerating ribbon graphs, and then using random matrix theory to perform said enumerations. In contrast to this philosophy, the literature also contains many examples of using ribbon graphs (and similar objects) to diagrammatically encode calculations of the (mixed) moments and cumulants defined through equations \eqref{eq3.0.1}, \eqref{eq3.0.13}, \eqref{eq3.0.14}. Naturally, such examples were first seen in the physics literature \citep{VWZ84}, \citep{BZ94}, \citep{Sil97}, \citep{JNPZ97}, since both random matrix theory and diagrammatic theories \citep{Hoo74}, \citep{BIPZ78} were prevalent in the field during the mid-twentieth century. In particular, Wick's theorem \citep{Wic50} on reducing products of creation and annihilation operators to sums of products of creation-annihilation operator pairings was well known to quantum field theorists and easily extends to treatment of similar such products of random matrix entries. In fact, the required generalisation pertaining to products of normally distributed random variables was actually known much earlier to probability theorists as Isserlis' theorem \citep{Iss18}. Let us now state a version of this theorem that is suitable for our purposes.

\begin{theorem}[Isserlis--Wick] \label{thrm3.1}
For $k\in\mathbb{N}$, let $x_1,\ldots,x_k$ be centred normal variables. Then, if $k$ is odd, the expectation $\mean{x_1\cdots x_k}$ is equal to zero, while for $k$ even, it decomposes as
\begin{equation} \label{eq3.3.1}
\mean{x_1\cdots x_k}=\sum_{\sigma\in\mathfrak{P}_k}\mean{x_{\sigma(1)}x_{\sigma(2)}}\cdots\mean{x_{\sigma(k-1)}x_{\sigma(k)}},
\end{equation}
where $\mathfrak{P}_k$ is the set of all permutations of $\{1,\ldots,k\}$ such that for all odd integers $i,j$ with $i<j$, $\sigma(i)<\sigma(i+1)$ and $\sigma(i)<\sigma(j)$.
\end{theorem}

The key idea of this section, in conjunction with the Isserlis--Wick theorem, is that for random matrices $X$ that can be expressed as sums and products of Ginibre matrices (recall Definition \ref{def1.3}), $\Tr\,X^k$ is given by a sum of products of normally distributed random variables. Thus, recalling Definition \ref{def1.4}, the (mixed) moments and cumulants of the GOE and LOE can be simplified using the Isserlis--Wick theorem; graphically, the covariances $\langle x_{\sigma(j)}x_{\sigma(j+1)}\rangle$ in the summand of equation \eqref{eq3.3.1} are to be represented by ribbons connecting edges labelled $x_{\sigma(j)}$ and $x_{\sigma(j+1)}$. By linearity of the expectation, the Isserlis--Wick theorem also holds when the variables $x_1,\ldots,x_k$ are drawn from mean-zero complex normal distributions, so it can be used to study the GUE and LUE, as well. Similar reasoning applies in the GSE and LSE cases, too \citep{MW03}, \citep{BP09}, though we do not discuss these cases for brevity (instead, we appeal to the $\beta\leftrightarrow4/\beta$ duality of Lemma \ref{L1.4}).

Unfortunately, the Isserlis--Wick theorem cannot be used to relate ribbon graphs to any of the other ensembles studied thus far in this thesis. Indeed, the Gaussian $\beta$ ensembles with $\beta=2/3,6$ do not have matrix realisations in terms of Ginibre matrices and must instead be understood through the general-$\beta$ constructions of \S\ref{s1.2.4}. Likewise, there is no relation between the Cauchy and Ginibre ensembles, while matrix realisations of the Jacobi ensembles involve inverse Ginibre matrices (see Definition \ref{def1.4} and Remark \ref{R1.4}). However, it should be noted that the combinatorial approach of \citep{YK83} was used in \citep{BYK88} to express $m_k^{(JOE)}$ in terms of the negative-integer moments of the LOE; the latter do not have any known ribbon graph interpretations (see \S\ref{s4.4.3} for some further discussion on this point).

In \S\ref{s3.3.1}, we show how the Isserlis--Wick theorem leads to interpretations of the (mixed) moments and cumulants of the GUE in terms of (globally) orientable ribbon graphs, and then explain how the GOE analogues are similarly related to locally orientable ribbon graphs. In \S\ref{s3.3.2}, we repeat these exercises for the LUE and LOE, detailing a necessary bicolouring constraint on the vertices of the polygons that are present in the relevant ribbon graphs. Finally, in \S\ref{s3.3.3}, we combine the ideas of \S\ref{s3.3.1} and \S\ref{s3.3.2} to derive ribbon graph representations for the mixed moments and cumulants of the Hermitised and antisymmetrised matrix products introduced in Section \ref{s1.3}, thereby justifying the existence of the large $N$ expansions \eqref{eq1.1.21} for the corresponding connected $n$-point correlators $W_n$ (this latter point being significant in the context of Chapter 4). We will be relaxed with our referencing since our discussion will be an amalgamation of new results, results already cited, and results that are most likely known to the community but unpublished in full detail. That said, let us note that the exposition of the first half of \S\ref{s3.3.1} is equivalent to that of \citep{BIZ80}, \citep{Zvo97}, while the first half of \S\ref{s3.3.2} contains statements that loosely follow from \citep{Fra03}. The second half of both of these subsections discuss concepts that have been studied and reviewed in \citep{BP09}, \citep{LaC09} and references therein. The contents of \S\ref{s3.3.3} are original and based on the ongoing work \citep{DR22}.
\vspace{-0.2cm}

\subsection{Moments of the Gaussian unitary and orthogonal ensembles} \label{s3.3.1}
\vspace{-0.1cm}
Let us now demonstrate how the Isserlis--Wick theorem can be used to express the spectral moments $m_k^{(GUE)}$ as sums over certain $(k/2)$-ribbon graphs. Afterwards, we illustrate how these ideas extend to the mixed moments and cumulants of the GUE and then show how to tweak our arguments to make them suitable for the GOE case. At the end of this subsection, we also discuss connections to topological and combinatorial (hyper)maps.

\subsubsection{Ribbon graphs for the GUE spectral moments}
Let $H$ be drawn from the $N\times N$ Gaussian unitary ensemble. According to the convention established in Section \ref{s1.2}, this means that the entries $\{H_{ij}\}_{i,j=1}^N$ of the complex Hermitian matrix $H$ are centred normal variables with covariances given by
\begin{equation} \label{eq3.3.2}
\mean{H_{ij}H_{kl}}=\frac{1}{2}\chi_{k=j}\chi_{l=i},
\end{equation}
where we recall that the indicator function $\chi_A$ equals one when $A$ is true and zero otherwise. Here and throughout the remainder of this subsection, we suppress the fact that our averages are with respect to the p.d.f.~$P^{(G)}(H)$ given in equation \eqref{eq1.2.1}. Our first order of business is to use the formula \eqref{eq3.3.2} to give a multi-sum expression for the GUE spectral moments.
\begin{lemma} \label{L3.1}
Let $k\in\mathbb{N}$, recall the definition of $\mathfrak{P}_k$ from Theorem \ref{thrm3.1}, and set $i_{k+1}:=i_1$. The spectral moment $m_k^{(GUE)}$ is zero if $k$ is odd, while for even $k$,
\begin{equation} \label{eq3.3.5}
m_k^{(GUE)}=2^{-k/2}\sum_{\sigma\in\mathfrak{P}_k}\sum_{i_1,\ldots,i_k=1}^N(\chi_{i_{\sigma(1)}=i_{\sigma(2)+1}}\chi_{i_{\sigma(2)}=i_{\sigma(1)+1}})\cdots(\chi_{i_{\sigma(k-1)}=i_{\sigma(k)+1}}\chi_{i_{\sigma(k)}=i_{\sigma(k-1)+1}}).
\end{equation}
\end{lemma}
\begin{proof}
Taking equation \eqref{eq1.1.15} as our definition for $m_k^{(GUE)}$, it is well known that
\begin{align}
m_k^{(GUE)}&=\mean{\Tr\,H^k} \nonumber
\\&=\mean{\sum_{i_1,\ldots,i_k=1}^NH_{i_1i_2}H_{i_2i_3}\cdots H_{i_{k-1}i_k}H_{i_ki_1}} \nonumber
\\&=\sum_{i_1,\ldots,i_k=1}^N\mean{H_{i_1i_2}H_{i_2i_3}\cdots H_{i_{k-1}i_k}H_{i_ki_1}}; \label{eq3.3.3}
\end{align}
the second line can be checked via straightforward induction, while the third line follows from linearity of the expectation. The summand on the right-hand side of equation \eqref{eq3.3.3} is the expected value of a product of normally distributed random variables, so the Isserlis--Wick theorem applies. Hence, it vanishes for odd $k$, and for even $k$, we have
\begin{multline} \label{eq3.3.4}
\mean{H_{i_1i_2}H_{i_2i_3}\cdots H_{i_{k-1}i_k}H_{i_ki_1}}
\\=\sum_{\sigma\in\mathfrak{P}_k}\mean{H_{i_{\sigma(1)}i_{\sigma(1)+1}}H_{i_{\sigma(2)}i_{\sigma(2)+1}}}\cdots\mean{H_{i_{\sigma(k-1)}i_{\sigma(k-1)+1}}H_{i_{\sigma(k)}i_{\sigma(k)+1}}}.
\end{multline}
Substituting this expression into equation \eqref{eq3.3.3}, interchanging the order of summation, and using equation \eqref{eq3.3.2} then produces the desired result.
\end{proof}
The inner sum in equation \eqref{eq3.3.5} reduces to $\sum_{j_1,\ldots,j_l=1}^N1=N^l$ for some $1\leq l\leq k/2+1$ since many of the indices $i_1,\ldots,i_k$ must be identified to ensure that the product of indicator functions is non-zero. To better understand this statement, it is convenient to recast it in terms of ribbon graphs. This can be done in a natural way by recognising that each choice of $\sigma\in\mathfrak{P}_k$ determines a partitioning of the set $\{H_{i_1i_2},H_{i_2i_3},\ldots,H_{i_ki_1}\}$ into $k$ disjoint pairs: To begin, we encode the fact that the first index of each element of $\{H_{i_1i_2},H_{i_2i_3},\ldots,H_{i_ki_1}\}$ is equal to the second index of one other element of the same set by drawing a $k$-gon whose vertices are labelled $i_1,i_2,\ldots,i_k$ in clockwise order; each edge of the $k$-gon can be oriented in two ways, with $(i_l\to i_{l+1})$ representing the matrix entry $H_{i_li_{l+1}}$ and $(i_{l+1}\to i_l)$ representing $\overline{H}_{i_li_{l+1}}$. Next, for each $l=1,\ldots,k/2$, we represent the identification of the oriented polygon edges $(i_{\sigma(2l-1)}\to i_{\sigma(2l-1)+1})$ and $(i_{\sigma(2l)+1}\to i_{\sigma(2l)})$, implied by the factor $\chi_{i_{\sigma(2l-1)}=i_{\sigma(2l)+1}}\chi_{i_{\sigma(2l)}=i_{\sigma(2l-1)+1}}$ in the right-hand side of equation \eqref{eq3.3.5}, by connecting them with an untwisted ribbon.
\begin{figure}[H]
        \centering
\captionsetup{width=.9\linewidth}
        \includegraphics[width=0.72\textwidth]{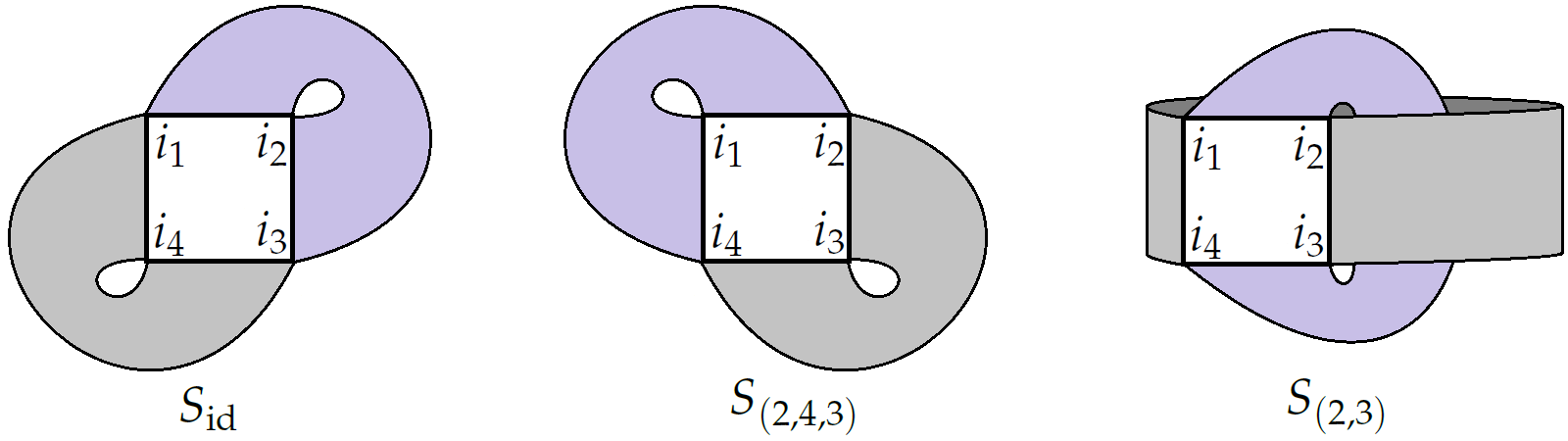}
        \caption[The ribbon graphs that contribute to the value of $m_4^{(GUE)}$]{These are the ribbon graph representations of the terms given in equations \eqref{eq3.3.7}--\eqref{eq3.3.9} below. The purple (grey) ribbons represent the first (second) factor in the summands of the corresponding equations.} \label{fig3.4}
\end{figure}

\begin{example} \label{ex3.2}
Let us illustrate the above formalism in the $k=4$ case. The set $\mathfrak{P}_4$ contains three permutations: the identity, $(2,4,3)$, and $(2,3)$. Hence, equation \eqref{eq3.3.5} tells us that
\begin{equation} \label{eq3.3.6}
m_4^{(GUE)}=\frac{1}{4}\left(S_{\mathrm{id}}+S_{(2,4,3)}+S_{(2,3)}\right)
\end{equation}
with
\begin{align}
S_{\mathrm{id}}&:=4\sum_{i_1,\ldots,i_4=1}^N\mean{H_{i_1i_2}H_{i_2i_3}}\mean{H_{i_3i_4}H_{i_4i_1}}=\sum_{i_1,\ldots,i_4=1}^N(\chi_{i_1=i_3}\chi_{i_2=i_2})(\chi_{i_3=i_1}\chi_{i_4=i_4}), \label{eq3.3.7}
\\S_{(2,4,3)}&:=4\sum_{i_1,\ldots,i_4=1}^N\mean{H_{i_1i_2}H_{i_4i_1}}\mean{H_{i_2i_3}H_{i_3i_4}}=\sum_{i_1,\ldots,i_4=1}^N(\chi_{i_1=i_1}\chi_{i_4=i_2})(\chi_{i_2=i_4}\chi_{i_3=i_3}),
\\S_{(2,3)}&:=4\sum_{i_1,\ldots,i_4=1}^N\mean{H_{i_1i_2}H_{i_3i_4}}\mean{H_{i_2i_3}H_{i_4i_1}}=\sum_{i_1,\ldots,i_4=1}^N(\chi_{i_1=i_4}\chi_{i_3=i_2})(\chi_{i_2=i_1}\chi_{i_4=i_3}). \label{eq3.3.9}
\end{align}
These sums are then represented by the ribbon graphs in Figure \ref{fig3.4}. Seen as two-dimensional surfaces, they have, respectively, three, three, and one boundaries. This reflects the fact that $i_3\equiv i_1$ in the first ribbon graph, $i_4\equiv i_2$ in the second ribbon graph, and $i_4\equiv i_3\equiv i_2\equiv i_1$ in the third ribbon graph. Hence,
\begin{align}
S_{\mathrm{id}}&=\sum_{i_1,i_2,i_4=1}^N1=N^3,
\\ S_{(2,4,3)}&=\sum_{i_1,i_2,i_3=1}^N1=N^3,
\\ S_{(2,3)}&=\sum_{i_1=1}^N1=N,
\end{align}
so that by equation \eqref{eq3.3.6}, $m_4^{(GUE)}=N^3/2+N/4$.
\end{example}

The computation of the above example extends quite simply to the case of $k$ being a general positive even integer: One need only interpret $\mathfrak{P}_k$ within Lemma \ref{L3.1} as the set of all orientable $(k/2)$-ribbon graphs that can be built from a single $k$-gon.
\begin{lemma} \label{L3.2}
Let $k\in2\mathbb{N}$ and $\mathfrak{P}_k$ be as in Theorem \ref{thrm3.1}. Then, we have
\begin{equation} \label{eq3.3.13}
m_k^{(GUE)}=2^{-k/2}\sum_{\sigma\in\mathrm{P}_k}N^{V(\sigma)},
\end{equation}
where $V(\sigma)$ is equal to the number of boundaries of the ribbon graph constructed from $\sigma$ according to the prescription above Figure \ref{fig3.4}.
\end{lemma}
\begin{proof}
Inspection of the ribbon graph corresponding to $\sigma$ shows that each indicator function in the summand of equation \eqref{eq3.3.5} is represented by the side of a ribbon identifying two vertices of the $k$-gon at the centre of the ribbon graph. Thus, the number of unique summation indices $i_1,\ldots,i_k$ is equal to $V(\sigma)$, the number of boundaries of said ribbon graph. Hence, we see the reduction
\begin{equation*}
\sum_{i_1,\ldots,i_k=1}^N(\chi_{i_{\sigma(1)}=i_{\sigma(2)+1}}\chi_{i_{\sigma(2)}=i_{\sigma(1)+1}})\cdots(\chi_{i_{\sigma(k-1)}=i_{\sigma(k)+1}}\chi_{i_{\sigma(k)}=i_{\sigma(k-1)+1}})=\sum_{j_1,\ldots,j_{V(\sigma)}=1}^N1=N^{V(\sigma)}.
\end{equation*}
Substituting this into equation \eqref{eq3.3.5} produces the sought result.
\end{proof}

\begin{remark}
Recalling that ribbons represent edge identifications of the underlying $k$-gons, $V(\sigma)$ is also equal to the number of distinct vertices retained by a $k$-gon after identifying edges according to the ribbon graph corresponding to $\sigma$; see Figure \ref{fig3.5} below.
\end{remark}

\begin{figure}[H]
        \centering
\captionsetup{width=.9\linewidth}
        \includegraphics[width=0.7\textwidth]{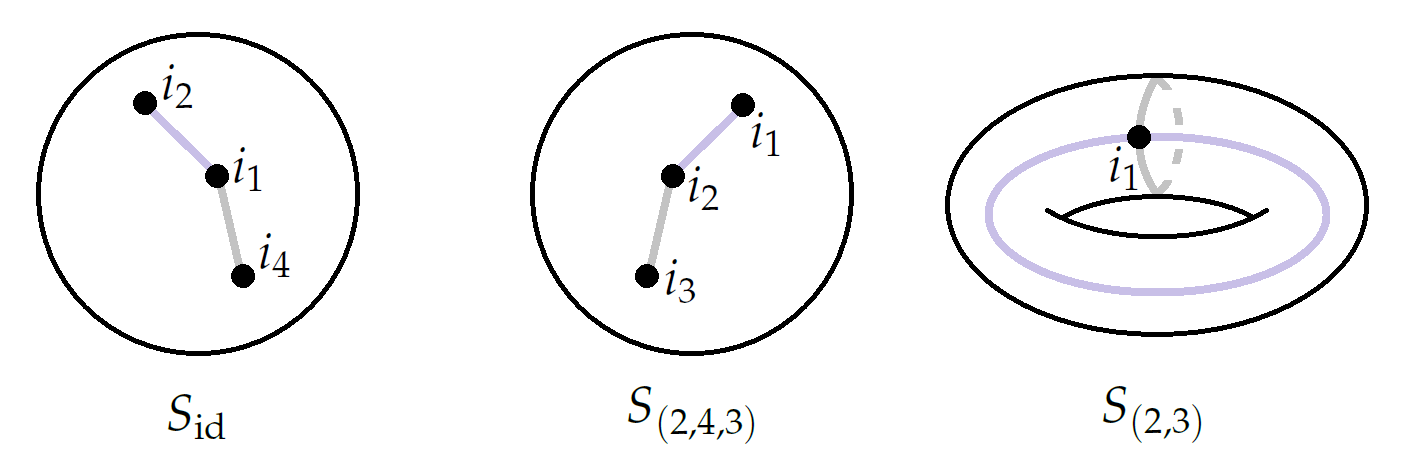}
        \caption[The surfaces corresponding to the ribbon graphs of Figure \ref{fig3.4}]{These compact orientable surfaces (sphere, sphere, and torus) are obtained by enlargening the white squares in Figure \ref{fig3.4} while collapsing the ribbons therein to their now-identified ends. Three, three, and one distinct vertices survive the edge identification procedure, respectively.} \label{fig3.5}
\end{figure}

It is evident from equation \eqref{eq3.3.13} that $m_k^{(GUE)}$ is a polynomial in $N$. In fact, according to the statement (which we are yet to fully justify) following the proof of Lemma \ref{L3.1}, it is a polynomial of degree $k/2+1$. This means that equation \eqref{eq3.3.13} can be rewritten as
\begin{equation} \label{eq3.3.14}
m_k^{(GUE)}=2^{-k/2}\sum_{V=1}^{k/2+1}N^V\,\#\{\sigma\in\mathfrak{P}_k\,|\,V(\sigma)=V\},\quad k\in2\mathbb{N},
\end{equation}
where $\#\mathcal{S}$ denotes the number of elements in the set $\mathcal{S}$. With our current definitions of the function $V(\sigma)$, computing the coefficient of $N^V$ on the right-hand side of equation \eqref{eq3.3.14} requires us to generate all orientable $(k/2)$-ribbon graphs that can be built from a $k$-gon and then count how many of them have exactly $V$ boundaries. However, we are able to give a much neater interpretation of these polynomial coefficients, while also justifying our assumption that the degree of $m_k^{(GUE)}$ in $N$ is $k/2+1$, by classifying our ribbon graphs with respect to the genera of the corresponding compact surfaces.

Let us proceed by first observing that rather than collapsing the ribbons of a ribbon graph (as described in Figure \ref{fig3.5}), another way of obtaining the compact surface represented by a connected ribbon graph is to glue open disks along the boundaries of said ribbon graph. This is seen to be true by noting that contracting the disks in the resulting polygonised surface to points while also collapsing the ribbons to edges is equivalent to just collapsing the ribbons in the original ribbon graph (see Figure~\ref{fig3.6} below). Hence, connected ribbon graphs can be drawn on the compact surfaces that they represent and, moreover, these surfaces are of minimal (Euler) genus such that this is possible without self-intersections (if one were able to draw a ribbon graph on a surface of lower (Euler) genus without self-intersections, then shrinking the ribbons and disks bounded by the ribbons in the described way could not produce the compact surface that the ribbon graph represents). As one would expect, this means that the (Euler) genus of a ribbon graph (recall Definition \ref{def3.1}) is equal to that of the compact surface it represents.

\begin{figure}[H]
        \centering
\captionsetup{width=.9\linewidth}
        \includegraphics[width=0.7\textwidth]{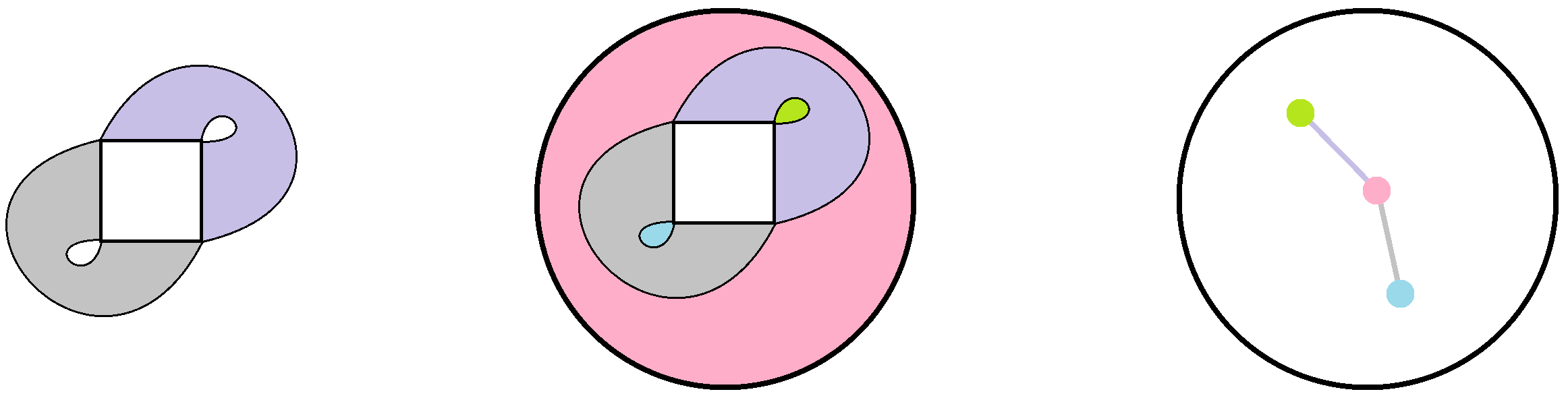}
        \caption[A ribbon graph and the compact orientable surface it represents]{Gluing a pink, green, and blue disk to the boundaries of the ribbon graph illustrated on the left produces the polygonised sphere in the middle. Shrinking the ribbons to edges and (deformed) disks to vertices while enlargening the white square produces the illustration on the right, which is exactly the first surface presented in Figure \ref{fig3.5}.} \label{fig3.6}
\end{figure}

Recall that the Euler characteristic of a compact surface is given by the classical formula 
\begin{equation*}
\chi=F-E+V,
\end{equation*}
where $F,E,V$ are respectively the number of faces, edges, and vertices of a chosen polygonisation of the surface. In our case, the surfaces of interest are constructed by gluing together pairs of edges of a $k$-gon, so an obvious choice of polygonisation for a given surface is the edge-identified $k$-gon that it is equivalent to. Thus, we have $F=1$ and $E=k/2$ since our $k$-gons have one face and $k$ edges which are identified pairwise into $k/2$ distinct edges (cf.~Figure \ref{fig3.5}). Combining this with Euler's formula and noting that the Euler characteristic of a compact orientable surface is also given by $\chi=2-2g$, where $g$ is the genus of the surface, we have
\begin{equation} \label{eq3.3.15}
V=1+k/2-2g.
\end{equation}
Thus, we may reformulate equation \eqref{eq3.3.14} as a \textit{genus expansion}.
\begin{proposition} \label{prop3.9x}
Let $k\in\mathbb{N}$ and let $\mathfrak{P}_k$ be as in Theorem \ref{thrm3.1}. If $k$ is odd, the spectral moment $m_k^{(GUE)}$ is zero, while if $k$ is even, we have that
\begin{equation} \label{eq3.3.16}
m_k^{(GUE)}=2^{-k/2}\sum_{l=0}^{k/2}N^{1+k/2-l}\,\#\{\sigma\in\mathfrak{P}_k\,|\,g(\sigma)=l/2\},
\end{equation}
where $g(\sigma)$ is equal to the genus of the ribbon graph labelled by $\sigma$.
\end{proposition}
\begin{proof}
Lemma \ref{L3.2} implies equation \eqref{eq3.3.14} with the possible requirement that the upper terminal of the sum be changed. Equation \eqref{eq3.3.15} and the discussion above Figure~\ref{fig3.6} tells us that the coefficient of $N^V$ on the right-hand side of equation \eqref{eq3.3.14} is equal to the number of genus $(1+k/2-V)/2$, connected, orientable ribbon graphs that consist of a single $k$-gon and $k/2$ ribbons. Moreover, since $g\geq0$, we see that $V\leq k/2+1$, which establishes that the upper limit of the sum in equation \eqref{eq3.3.14} is indeed $k/2+1$; it is a simple exercise to check that the identity permutation $\sigma=\mathrm{id}\in\mathfrak{P}_k$ corresponds to a (genus zero) sphere, so that $V(\mathrm{id})=k/2+1$. Letting $l=2g$ and changing the summation index in equation \eqref{eq3.3.14} according to equation \eqref{eq3.3.15} gives equation \eqref{eq3.3.16}, as desired.
\end{proof}

Replacing $k$ by $2k$ in equation \eqref{eq3.3.16} gives a proof of equation \eqref{eq3.0.8}, i.e., equation \eqref{eq1.2.89} of Lemma~\ref{L1.3}, in the GUE case and confirms our claim following Figure \ref{fig3.1} that $2^kM_{k,l}^{(GUE)}$ counts the number of orientable genus $l/2$ ribbon graphs that can be constructed from a $2k$-gon. Furthermore, since $g\in\mathbb{N}$ and $l=2g$, $M_{k,l}^{(GUE)}=0$ for $l$ odd and $m_{2k}^{(GUE)}$ is consequently an odd polynomial in $N$, in keeping with the second point of Remark~\ref{R2.1}. Finally, let us point out that the interpretation of $M_{k,l}^{(GUE)}$ in terms of ribbon graphs extends to a representation of the generating functions $W_1^{(GUE),l}(x)$ (recall the discussion at the beginning of \S\ref{s3.1.2}) as genus $l/2$ compact orientable surfaces with one hole --- if one can replace the hole of such a representation of $W_1^{(GUE),l}(x)$ by a $2k$-gon for some positive integer $k$ and connect the edges of the $2k$-gon with ribbons drawn on the surface without self-intersections such that excising the resulting ribbon graph yields a disjoint union of sets homeomorphic to open disks, then that ribbon graph contributes to the value of $M_{k,l}^{(GUE)}$.

\begin{figure}[H]
        \centering
\captionsetup{width=.9\linewidth}
        \includegraphics[width=0.9\textwidth]{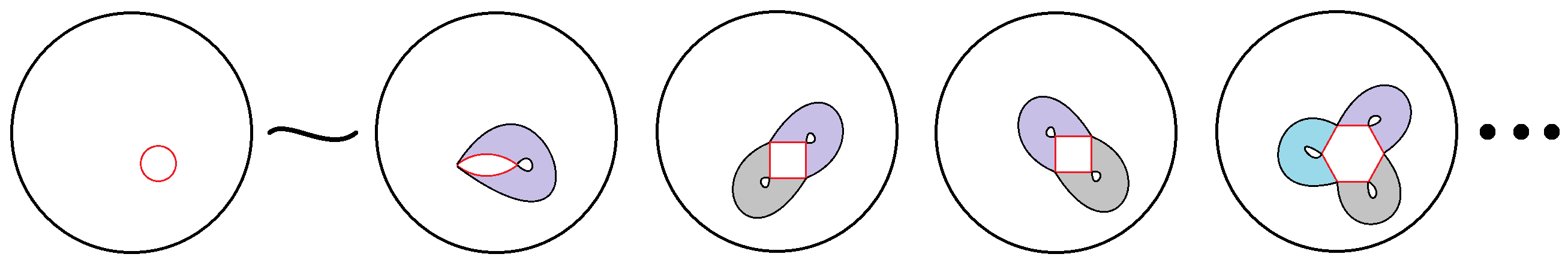}
        \caption[The surface representation of $W_1^{(GUE),0}(x)$]{The sphere with a hole on the left represents $W_1^{(GUE),0}(x)$. On the right, we have a collection of ribbon graphs drawn on this sphere with the hole replaced by a suitable $2k$-gon. The illustrations on the right contribute to the value of $M_{k,0}^{(GUE)}$ with $k=1,2,2,3$, respectively. Therefore, they also contribute to the coefficients of the $1/x$ expansion of $W_1^{(GUE),0}(x)$ since the latter is essentially a generating function for the $M_{k,0}^{(GUE)}$.} \label{fig3.7}
\end{figure}

\begin{remark}
\begin{enumerate}
\item Recalling definition \eqref{eq3.1.19} of $\tilde{m}_{2k}^{(G)}$, Proposition \ref{prop3.9x} tells us that
\begin{equation*}
\tilde{m}_{2k}^{(GUE)}=2^{-k}\sum_{l=0}^kN^{-l}\,\#\{\sigma\in\mathfrak{P}_{2k}\,|\,g(\sigma)=l/2\}.
\end{equation*}
Thus, from equation \eqref{eq1.1.32}, $\tilde{M}_{2k,0}^{(GUE)}=\lim_{N\to\infty}\tilde{m}_{2k}^{(GUE)}=M_{k,0}^{(GUE)}$ is equal to $2^{-k}$ multiplied by the number of genus zero (i.e., planar) ribbon graphs that can be built from a $2k$-gon. Hence, from the discussion contained in the caption of Figure \ref{fig3.2} and the paragraph preceding it, $2^k\tilde{M}_{2k,0}^{(GUE)}$ is equal to the $k\textsuperscript{th}$ Catalan number. Consequently, $W_1^{(GUE),0}(x)$, as described below equation \eqref{eq3.1.19}, is essentially a generating function for the Catalan numbers, which is well known \citep{Sta99} to be given by
\begin{equation*}
W_1^{(GUE),0}(x)=\tfrac{1}{2}(x-\sqrt{x^2-2})
\end{equation*}
after appropriate scaling (see also differential equation \eqref{eq2.4.17}). Of course, applying the Sokhotski--Plemelj inversion formula \eqref{eq1.1.25} yields the Wigner semi-circle law \eqref{eq1.2.14}. This proof of the Wigner semi-circle law is a rewording of Wigner's original proof \citep{Wig55}, which used the so-called \textit{method of moments}.
\item The above proof can be extended to show that the Wigner semi-circle law holds for the Hermitian Wigner matrix ensemble, which is represented by the same matrices as the GUE except that the moments $\langle H_{ij}^{p_1+1}H_{ji}^{p_2+1}\rangle$ ($p_1,p_2\in\mathbb{N}$) no longer need to be zero whenever $p_1,p_2>0$. Setting rigour aside (see, e.g., \citep[Ch.~2]{AGZ09} for a detailed treatment), the idea is that when $H$ is a Hermitian Wigner matrix, the right-hand side of equation \eqref{eq3.3.4} needs to be replaced by a sum over all partitions of $k$, rather than just pair partitions. As an example, the contribution of the terms $\mean{H_{i_1i_2}H_{i_2i_3}}\mean{H_{i_3i_4}H_{i_4i_5}H_{i_5i_6}H_{i_6i_1}}$ and $\mean{H_{i_1i_2}H_{i_2i_3}H_{i_3i_4}}\mean{H_{i_4i_5}H_{i_5i_6}H_{i_6i_1}}$ to the value of $\mean{H_{i_1i_2}H_{i_2i_3}\cdots H_{i_5i_6}H_{i_6i_1}}$ must now be taken into account. However, since these terms are interpreted as products of indicator functions which constrain our summation indices $i_1,\ldots,i_6$, only pair partitions contribute at leading order --- $\mean{H_{i_1i_2}H_{i_2i_3}H_{i_3i_4}H_{i_4i_5}}$ is non-zero only if $i_3,i_4,i_5$ depend on $i_1,i_2$ while $\mean{H_{i_1i_2}H_{i_2i_3}}\mean{H_{i_3i_4}H_{i_4i_5}}$ being non-zero merely constrains $i_3,i_5$ to depend on $i_1,i_2,i_4$. Thus, at leading order in $N$, the spectral moments of the Hermitian Wigner matrix $H$ are given by the Catalan numbers up to some normalisation factor, and the argument given above follows through.
\end{enumerate} \label{R3.2}
\end{remark}

\subsubsection{Ribbon graphs for the GUE mixed moments and cumulants}
Letting $H$ be again drawn from the $N\times N$ GUE, the mixed moments of the GUE \eqref{eq3.0.13} are given by
\begin{equation*}
m_{k_1,\ldots,k_n}^{(GUE)}=\mean{\prod_{i=1}^n\Tr\,H^{k_i}},\quad k_1,\ldots,k_n\in\mathbb{N}.
\end{equation*}
The analogue of equation \eqref{eq3.3.3} is then
\begin{multline} \label{eq3.3.17}
m_{k_1,\ldots,k_n}^{(GUE)}=\sum_{i_1^{(1)},\ldots,i_{k_1}^{(1)}=1}^N\cdots\sum_{i_1^{(n)},\ldots,i_{k_n}^{(n)}=1}^N\left\langle\left(H_{i_1^{(1)}i_2^{(1)}}H_{i_2^{(1)}i_3^{(1)}}\cdots H_{i_{k_1}^{(1)}i_1^{(1)}}\right)\cdots\right.
\\\left.\cdots\left(H_{i_1^{(n)}i_2^{(n)}}H_{i_2^{(n)}i_3^{(n)}}\cdots H_{i_{k_n}^{(n)}i_1^{(n)}}\right)\right\rangle.
\end{multline}
The summand on the right-hand side of this equation is the expected value of a product of centred normal variables, so it can be simplified using the Isserlis--Wick theorem, in a similar manner to equation \eqref{eq3.3.4}. Thus, $m_{k_1,\ldots,k_n}^{(GUE)}$ vanishes if $k_1+\cdots+k_n$ is an odd integer, whereas for $k_1+\cdots+k_n$ an even integer, we can express $m_{k_1,\ldots,k_n}^{(GUE)}$ in terms of ribbon graphs: For each $j=1,\ldots,n$, the set $\{H_{i_1^{(j)}i_2^{(j)}},H_{i_2^{(j)}i_3^{(j)}},\ldots,H_{i_{k_j}^{(j)}i_1^{(j)}}\}$ is represented by a $k_j$-gon with vertices labelled $i_1^{(j)},i_2^{(j)},\ldots,i_{k_j}^{(j)}$ in clockwise order, while the factors $\langle H_{i_p^{(j)}i_{p+1}^{(j)}}H_{i_{q}^{(l)}i_{q+1}^{(l)}}\rangle$ (setting $i_{k_j+1}^{(j)}:=i_1^{(j)}$) in the sum of products of covariances produced by the Isserlis--Wick theorem are represented by untwisted ribbons connecting the appropriate polygon edges.

\begin{figure}[H]
        \centering
\captionsetup{width=.9\linewidth}
        \includegraphics[width=0.70\textwidth]{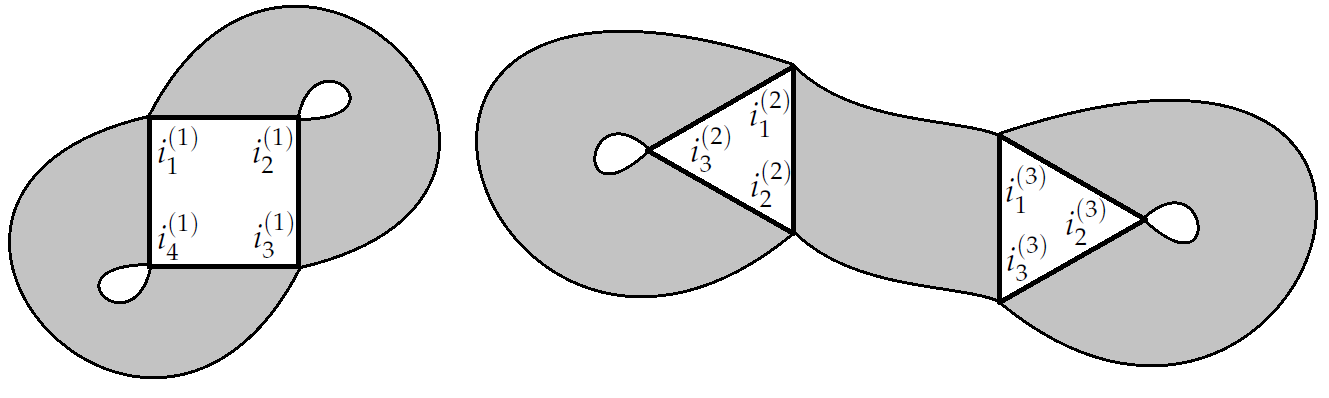}
        \caption[A ribbon graph contributing to the computation of $m_{4,3,3}^{(GUE)}$]{In computing $m_{4,3,3}^{(GUE)}=\mean{\Tr(H^4)\Tr(H^3)\Tr(H^3)}$, one first draws a square and two triangles with vertices labelled as shown. Then, one must list out all possible ways of pairing the polygon edges using untwisted ribbons. The illustrated planar ribbon graph represents $\langle H_{i_1^{(1)}i_2^{(1)}}H_{i_2^{(1)}i_3^{(1)}}\rangle\langle H_{i_3^{(1)}i_4^{(1)}}H_{i_4^{(1)}i_1^{(1)}}\rangle\langle H_{i_2^{(2)}i_3^{(2)}}H_{i_3^{(2)}i_1^{(2)}}\rangle\langle H_{i_1^{(2)}i_2^{(2)}}H_{i_3^{(3)}i_1^{(3)}}\rangle\langle H_{i_1^{(3)}i_2^{(3)}}H_{i_2^{(3)}i_3^{(3)}}\rangle$, which, when summed over the indices, contributes a value of $N^6$ to $m_{4,3,3}^{(GUE)}$.} \label{fig3.8}
\end{figure}

As a slight abuse of notation, let us now define $\mathfrak{P}_{k_1,\ldots,k_n}$, for positive integers $k_1,\ldots,k_n$ such that $k_1+\cdots+k_n$ is even, to be the set of all $\tfrac{1}{2}(k_1+\cdots+k_n)$-ribbon graphs that can be constructed by drawing $n$ polygons with respectively $k_1,k_2,\ldots,k_n$ edges and then connecting said edges with untwisted ribbons. Then, equation \eqref{eq3.3.14} extends to the form
\begin{equation} \label{eq3.3.18}
m_{k_1,\ldots,k_n}^{(GUE)}=2^{-\tfrac{1}{2}(k_1+\cdots+k_n)}\sum_{V=1}^{\tfrac{1}{2}(k_1+\cdots+k_n)+n}N^V\,\#\{\Gamma\in\mathfrak{P}_{k_1,\ldots,k_n}\,|\,V(\Gamma)=V\},
\end{equation}
where $V(\Gamma)$ is equal to the number of boundaries of $\Gamma$, which is equivalent to the number of distinct polygon vertices retained by $\Gamma$ upon collapsing its ribbons in order to identify their ends (e.g., in Figure \ref{fig3.8} above, collapsing the ribbons \`a la Figure \ref{fig3.5} results in there being six distinct vertices labelled by $i_1^{(1)}\equiv i_3^{(1)}$, $i_2^{(1)}$, $i_4^{(1)}$, $i_1^{(2)}\equiv i_2^{(2)}\equiv i_1^{(3)}\equiv i_3^{(3)}$, $i_3^{(2)}$ and $i_2^{(3)}$). When we discussed the $n=1$ case earlier, we saw that the equivalent function $V(\sigma)$ could be expressed in terms of topological invariants of the ribbon graph corresponding to $\sigma$. We would now like to see if such relations can be replicated in the general-$n$ case and moreover justify the upper limit of the sum in equation \eqref{eq3.3.18}.

Let $\Gamma\in\mathfrak{P}_{k_1,\ldots,k_n}$ and let $\Sigma$ denote the surface obtained by collapsing the ribbons of $\Gamma$ such that the ribbon-ends are identified. As in the $n=1$ case, the Euler characteristic of $\Sigma$ is given by the classical formula $\chi=F-E+V$, where $F,E,V$ are again the number of faces, edges, and vertices of any polygonisation of $\Sigma$. Polygonising $\Sigma$ by the polygons of $\Gamma$ with their edges identified according to the ribbon data of $\Gamma$, we have $F=n$, the number of polygons in $\Gamma$, and $E=(k_1+\cdots+k_n)/2$, half the total number of polygon edges in $\Gamma$. Hence, the number of vertices, $V=V(\Gamma)$, is given by
\begin{equation} \label{eq3.3.19}
V(\Gamma)=\chi+\tfrac{1}{2}(k_1+\cdots+k_n)-n.
\end{equation}
Writing $C(\Sigma)$ for the number of connected components of $\Sigma$ and $g_i\geq0$ for the genus of the $i\textsuperscript{th}$ such component, we have by the additive property of the Euler characteristic that $\chi=\sum_{i=1}^{C(\Sigma)}(2-2g_i)$. It follows that $\chi\leq2n$ since $1\leq C(\Sigma)\leq n$. Combined with the above equation, this means that $V(\Gamma)\leq\tfrac{1}{2}(k_1+\cdots+k_n)+n$, which is precisely the upper limit of the sum in equation \eqref{eq3.3.18}. However, unlike in the $n=1$ case, this upper limit cannot always be attained. Indeed, if any of the $k_i$ are odd, the corresponding polygon must be ribbon-connected to at least one other polygon, so $\Gamma$, equivalently $\Sigma$, cannot have $n$ connected components. This slight complication in determining the degree of $m_{k_1,\ldots,k_n}^{(GUE)}$ as a polynomial in $N$, in addition to the reasons listed below, suggests that we should somehow focus on studying connected ribbon graphs.
\begin{enumerate}
\item If we define
\begin{equation*}
\mathfrak{P}_{k_1,\ldots,k_n}^c:=\{\Gamma\in\mathfrak{P}_{k_1,\ldots,k_n}\,|\,\Gamma\textrm{ is connected}\},
\end{equation*}
we have by the arguments above that for all $\Gamma\in\mathfrak{P}_{k_1,\ldots,k_n}^c$,
\begin{equation*}
V(\Gamma)\leq\tfrac{1}{2}(k_1+\cdots+k_n)+2-n.
\end{equation*}
Furthermore, this relation is a strict equality if $\Gamma$ is planar (genus zero) --- it is easy to convince oneself that there exists at least one such $\Gamma$ in each $\mathfrak{P}_{k_1,\ldots,k_n}^c$.
\item Connected surfaces have well-defined genera, so we can rewrite the analogue of equation \eqref{eq3.3.18} with $\mathfrak{P}_{k_1,\ldots,k_n}$ replaced by $\mathfrak{P}_{k_1,\ldots,k_n}^c$ as a genus expansion, in a similar fashion to equation \eqref{eq3.3.16}.
\item No generality is lost by studying connected ribbon graphs as every ribbon graph is a countable disjoint union of its connected components, which is easy to construct.
\end{enumerate}

To expand on the third point given above, it is well known to combinatorialists \citep[Ch.~5]{Sta99} that if the mixed moment $m_{k_1,\ldots,k_n}$ on the left-hand side of the moment-cumulants relation \eqref{eq3.0.14} is the weighted count of some structures of a given type (graphs, marked surfaces, etc.) built from $n$ objects, then the corresponding cumulants $c_{k_1,\ldots,k_n}$ are the analogous counts of the subsets of such structures that are connected. This can be understood in the context of ribbon graphs from the following inductive argument: If we assume for all $p<n$ and $k_1,\ldots,k_p\in\mathbb{N}$ such that $k_1+\cdots+k_p$ is an even integer that the mixed cumulants $c_{k_1,\ldots,k_p}$ enumerate (with suitable weights) the number of connected ribbon graphs that can be built from $p$ polygons with respectively $k_1,\ldots,k_p$ sides, then the product $c_{k_1,\ldots,k_p}c_{l_1,\ldots,l_q}$ ($q<n$) counts the number of ribbon graphs consisting of two connected components with the first being a ribbon graph built from $p$ polygons with $k_1,\ldots,k_p$ sides and the second component being a ribbon graph built from $q$ polygons with $l_1,\ldots,l_q$ sides. Extending this reasoning to general products, we see that the sum
\begin{equation*}
m_{k_1,\ldots,k_n}-c_{k_1,\ldots,k_n}=\sum_{\substack{K\vdash\{k_1,\ldots,k_n\}\\ K\neq\{\{k_1,\ldots,k_n\}\}}}\prod_{\kappa_i\in K}c_{\kappa_i}
\end{equation*}
counts the number of disconnected ribbon graphs that can be constructed from $n$ polygons with $k_1,\ldots,k_n$ sides. Since $m_{k_1,\ldots,k_n}$ enumerates both the disconnected and connected ribbon graphs that can be built from $n$ polygons with $k_1,\ldots,k_n$ sides, it follows that $c_{k_1,\ldots,k_n}$ must enumerate all of the connected ribbon graphs that can be constructed in said manner.

\begin{remark} \label{R3.4x}
If $m_{k_1,\ldots,k_n}$ vanishes whenever $k_1+\cdots+k_n$ is odd, so too does $c_{k_1,\ldots,k_n}$. This can be seen from the inverse of the moment-cumulants relation (cf.~equation \eqref{eq1.1.30}),
\begin{equation} \label{eq3.3.20x}
c_{k_1,\ldots,k_n}=\sum_{K\vdash\{k_1,\ldots,k_n\}}(-1)^{\#K-1}(\#K-1)!\prod_{\kappa_i\in K}m_{\kappa_i}.
\end{equation}
\end{remark}

Having established that we are interested in enumerating connected, orientable ribbon graphs and that these enumerations are given by the mixed cumulants of the GUE, let us now give the general-$n$ analogue of Proposition \ref{prop3.9x}.
\begin{proposition} \label{prop3.10x}
Let $n,k_1,\ldots,k_n\in\mathbb{N}$ be such that $k_1+\cdots+k_n$ is even and let $\mathfrak{P}_{k_1,\ldots,k_n}^c$ be the set of connected, orientable $\tfrac{1}{2}(k_1+\cdots+k_n)$-ribbon graphs built from $n$ polygons with respectively $k_1,\ldots,k_n$ sides. The mixed cumulant $c_{k_1,\ldots,k_n}^{(GUE)}$ is given by
\begin{equation} \label{eq3.3.22}
c_{k_1,\ldots,k_n}^{(GUE)}=\left(\frac{N}{2}\right)^{\tfrac{1}{2}(k_1+\cdots+k_n)}\sum_{l=0}^{\tfrac{1}{2}(k_1+\cdots+k_n)+1-n}N^{2-n-l}\,\#\{\Gamma\in\mathfrak{P}_{k_1,\ldots,k_n}^c\,|\,g(\Gamma)=l/2\},
\end{equation}
where $g(\Gamma)$ is the genus of $\Gamma$.
\end{proposition}
\begin{proof}
First, note that if $\Gamma\in\mathfrak{P}_{k_1,\ldots,k_n}^c$, it is connected with Euler characteristic $\chi=2-2g$, so equation \eqref{eq3.3.19} reads
\begin{equation} \label{eq3.3.20}
V(\Gamma)=\tfrac{1}{2}(k_1+\cdots+k_n)+2-n-2g\leq\tfrac{1}{2}(k_1+\cdots+k_n)+2-n.
\end{equation}
Thus, the analogue of equation \eqref{eq3.3.18} corresponding to connected ribbon graphs is
\begin{equation} \label{eq3.3.22x}
c_{k_1,\ldots,k_n}^{(GUE)}=2^{-\tfrac{1}{2}(k_1+\cdots+k_n)}\sum_{V=1}^{\tfrac{1}{2}(k_1+\cdots+k_n)+2-n}N^V\,\#\{\Gamma\in\mathfrak{P}_{k_1,\ldots,k_n}^c\,|\,V(\Gamma)=V\}.
\end{equation}
Setting $l=2g$ and changing summation index according to equation \eqref{eq3.3.20} in equation \eqref{eq3.3.22x} concludes the proof.
\end{proof}

In comparing equations \eqref{eq3.3.18} and \eqref{eq3.3.22}, we see that a benefit of studying the mixed cumulants is that upon scaling them by a factor of $N^{-(k_1+\cdots+k_n)/2}$ (equivalent to considering the cumulants of the GUE scaled by replacing the matrix $H$ in $P^{(G)}(H)$ \eqref{eq1.2.1} by $\sqrt{N}H$ as is consistent with Definition \ref{def1.6}), they are ${\rm O}(N^{2-n})$, while the mixed moments are ${\rm O}(N^n)$ when scaled in this manner. Thus, the connected $n$-point correlators $\tilde{W}_n^{(GUE)}$ of the global scaled GUE, which are generating functions for the scaled cumulants, have large $N$ expansions of the form given in Theorem \ref{thrm1.1}, whereas the corresponding unconnected $n$-point correlators $\tilde{U}_n^{(GUE)}$ \eqref{eq1.1.26} do not. In keeping with the discussion surrounding Figure \ref{fig3.7}, let us also mention that it is convenient to represent the correlator expansion coefficients $W_n^{(GUE),l}$ defined in Theorem \ref{thrm1.1} as genus $l/2$ compact, connected, orientable surfaces with $n$ holes.

\subsubsection{Ribbon graphs in the GOE case}
The calculations pertaining to the GUE that have been outlined thus far in this subsection follow through in much the same way when $H$ is drawn from the $N\times N$ Gaussian orthogonal ensemble. Indeed, when $H$ is an $N\times N$ GOE matrix, the right-hand side of equation \eqref{eq3.3.3} now equals $m_k^{(GOE)}$ with the summand still simplifying to the form given in equation \eqref{eq3.3.4} when $k$ is a positive even integer (and vanishing if $k$ is odd). The key difference in studying the GOE is that equation \eqref{eq3.3.2} must be replaced with
\begin{equation} \label{eq3.3.23}
\mean{H_{ij}H_{kl}}=\frac{1}{4}\left(\chi_{k=j}\chi_{l=i}+\chi_{k=i}\chi_{l=j}\right),
\end{equation}
which reflects the fact that $H$ is now a real symmetric matrix with off-diagonal entries having variance $1/4$ (when $i\neq j$, $\langle H_{ij}^2\rangle=\mean{H_{ij}H_{ji}}=1/4$) and diagonal entries having variance $1/2$. To ease notation, let us define
\begin{equation}
\xi_{ij}^{kl}(t):=(1-t)\chi_{k=j}\chi_{l=i}+t\chi_{k=i}\chi_{l=j}
\end{equation}
so that the right-hand side of equation \eqref{eq3.3.23} is equal to $\tfrac{1}{4}[\xi_{ij}^{kl}(0)+\xi_{ij}^{kl}(1)]$ (the parameter $t$ can be thought of as keeping track of orientability; cf.~the role of $\gamma$ in \citep{GHJ01}). Then, substituting equation \eqref{eq3.3.23} into equation \eqref{eq3.3.4} shows that for $k$ a positive even integer,
\begin{equation} \label{eq3.3.25}
\mean{H_{i_1i_2}H_{i_2i_3}\cdots H_{i_{k-1}i_k}H_{i_ki_1}}=2^{-k}\sum_{\sigma\in\mathfrak{P}_k}\sum_{t_1,\ldots,t_{k/2}=0}^1\xi_{i_{\sigma(1)}i_{\sigma(1)+1}}^{i_{\sigma(2)}i_{\sigma(2)+1}}(t_1)\cdots\xi_{i_{\sigma(k-1)}i_{\sigma(k-1)+1}}^{i_{\sigma(k)}i_{\sigma(k)+1}}(t_{k/2}).
\end{equation}
As in the GUE case, we now encode the index-matching of the set $\{H_{i_1i_2},H_{i_2i_3},\ldots,H_{i_ki_1}\}$ by drawing a $k$-gon with vertices labelled $i_1,i_2,\ldots,i_k$ in clockwise order, and then represent the edge-matching implied by the summand of equation \eqref{eq3.3.25} by drawing $k/2$ ribbons connecting the edges of said $k$-gon. The factors $\xi_{i_{\sigma(2l-1)}i_{\sigma(2l-1)+1}}^{i_{\sigma(2l)}i_{\sigma(2l)+1}}(t_l)$ ($1\leq l\leq k/2$) in the right-hand side of equation \eqref{eq3.3.25} are equivalent to the products $\chi_{i_{\sigma(2l-1)}=i_{\sigma(2l)+1}}\chi_{i_{\sigma(2l)}=i_{\sigma(2l-1)+1}}$ seen in equation \eqref{eq3.3.5} when we set $t_l=0$. Thus, the term $\xi_{i_{\sigma(2l-1)}i_{\sigma(2l-1)+1}}^{i_{\sigma(2l)}i_{\sigma(2l)+1}}(0)$ is represented by an untwisted ribbon connecting the oriented polygon edges $(i_{\sigma(2l-1)}\to i_{\sigma(2l-1)+1})$ and $(i_{\sigma(2l)+1}\to i_{\sigma(2l)})$. On the other hand, the term $\xi_{i_{\sigma(2l-1)}i_{\sigma(2l-1)+1}}^{i_{\sigma(2l)}i_{\sigma(2l)+1}}(1)$ is non-zero if the edge $(i_{\sigma(2l-1)}\to i_{\sigma(2l-1)+1})$ is identified to $(i_{\sigma(2l)}\to i_{\sigma(2l)+1})$, so we represent it by a M\"obius half-twisted ribbon connecting these two edges.

\begin{figure}[H]
        \centering
\captionsetup{width=.9\linewidth}
        \includegraphics[width=0.7\textwidth]{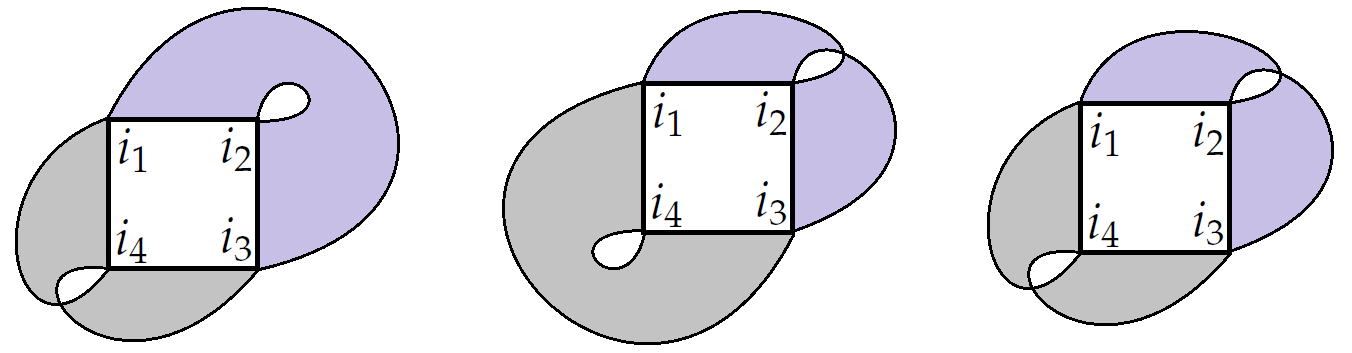}
        \caption[Some ribbon graphs that contribute to the fourth GOE spectral moment]{These ribbon graphs represent the term $\xi_{i_1i_2}^{i_2i_3}(t_1)\xi_{i_3i_4}^{i_4i_1}(t_2)$ for various choices of $t_1,t_2\in\{0,1\}$, with the purple (grey) ribbons representing the first (second) factor. In order, the graphs correspond to setting $(t_1,t_2)=(0,1),(1,0),$ and $(1,1)$; they can be obtained by twisting either or both of the ribbons of the left image of Figure \ref{fig3.4}, which incidentally corresponds to setting $(t_1,t_2)=(0,0)$.} \label{fig3.9}
\end{figure}

Extending the above reasoning to products of the form presented in the summand of the right-hand side of equation \eqref{eq3.3.17} shows that computing the mixed moments and cumulants of the GOE amounts to enumerating locally orientable ribbon graphs of specific types. The necessary arguments for obtaining explicit expressions for these moments and cumulants are the same as in the GUE case, with the key numeric still being the number of boundaries of each ribbon graph. Without reiterating any further details, let us simply state the GOE analogue of Proposition \ref{prop3.10x}.
\begin{proposition}
Let $n,k_1,\ldots,k_n\in\mathbb{N}$ be such that $k_1+\cdots+k_n$ is even and let $\tilde{\mathfrak{P}}_{k_1,\ldots,k_n}^c$ be the set of all connected, locally orientable $\tfrac{1}{2}(k_1+\cdots+k_n)$-ribbon graphs that can be built from $n$ polygons with respectively $k_1,\ldots,k_n$ edges. The mixed cumulant $c_{k_1,\ldots,k_n}^{(GOE)}$ is a degree $\tfrac{1}{2}(k_1+\cdots+k_n)+2-n$ polynomial in $N$ given by the formula
\begin{equation} \label{eq3.3.26}
c_{k_1,\ldots,k_n}^{(GOE)}=\left(\frac{N}{4}\right)^{\tfrac{1}{2}(k_1+\cdots+k_n)}\sum_{l=0}^{\tfrac{1}{2}(k_1+\cdots+k_n)+1-n}N^{2-n-l}\,\#\{\Gamma\in\tilde{\mathfrak{P}}_{k_1,\ldots,k_n}^c\,|\,\tilde{g}(\Gamma)=l\},
\end{equation}
where $\tilde{g}(\Gamma)$ is the Euler genus of $\Gamma$ (recall Definition \ref{def3.1}).
\end{proposition}

\begin{remark}
The set $\tilde{\mathfrak{P}}_{k_1,\ldots,k_n}^c$ defined above is in bijection with $\mathfrak{P}_{k_1,\ldots,k_n}^c\times\{0,1\}^{(k_1+\cdots+k_n)/2}$ since $\Gamma\in\tilde{\mathfrak{P}}_{k_1,\ldots,k_n}^c$ if and only if it is the result of twisting the ribbons of some ribbon graph drawn from $\mathfrak{P}_{k_1,\ldots,k_n}^c$; the tuple $(t_1,t_2,\ldots,t_{(k_1+\cdots+k_n)/2})\in\{0,1\}^{(k_1+\cdots+k_n)/2}$ determines which ribbons are to be twisted.
\end{remark}

Upon setting $n=1$, comparing equation \eqref{eq3.3.26} to equation \eqref{eq3.0.8} confirms that $M_{k,l}^{(GOE)}$ has the combinatorial interpretation described in the paragraph below Figure \ref{fig3.1} (as an aside, observe that the ribbon graph in Figure \ref{fig3.1} is genus zero and thus contributes a value of $N^2$ to the computation of $c_{1,2,3}^{(GOE)}$). Furthermore, the equality
\begin{equation*}
\{\Gamma\in\tilde{\mathfrak{P}}_k^c\,|\,\tilde{g}(\Gamma)=0\}=\{\Gamma\in\mathfrak{P}_k^c\,|\,g(\Gamma)=0\}
\end{equation*}
implies that the moment expansion coefficients $M_{k,0}^{(GUE)}$ and $M_{k,0}^{(GOE)}$ differ only by a factor of $2^{k/2}$, so the proof of Remark \ref{R3.2} can be used to show that the global scaled eigenvalue density of the GOE is given by the Wigner semi-circle law \eqref{eq1.2.14}.

A key difference between $c_{k_1,\ldots,k_n}^{(GUE)}$ and $c_{k_1,\ldots,k_n}^{(GOE)}$ is that, since the (Euler) genus is a non-negative integer, the summand of the expression \eqref{eq3.3.22} for $c_{k_1,\ldots,k_n}^{(GUE)}$ is identically zero for odd values of $l$, but this is not true for the corresponding expression \eqref{eq3.3.26} for $c_{k_1,\ldots,k_n}^{(GOE)}$. Hence, unlike in the GUE case, the correlator expansion coefficients $W_n^{(GOE),l}$ do not vanish for odd values of $l$. Another difference between $W_n^{(GUE),l}$ and $W_n^{(GOE),l}$ is that it does not make sense to represent the latter as a compact surface on which to draw ribbon graphs (like in Figure~\ref{fig3.7}) since it is a generating function for cumulants that count both orientable and non-orientable ribbon graphs.

\subsubsection{Relations to topological and combinatorial maps and hypermaps}
It is oftentimes beneficial, either for visualisation or computational purposes, to consider alternative representations of ribbon graphs. We now outline the connection between ribbon graphs and two such representations: topological and combinatorial maps \citep{Tut84}, \citep{LZ04}, \citep{LaC09}.
\begin{definition} \label{def3.2}
A \textit{topological map} is a graph embedded in a compact surface such that the edges of the graph do not intersect and excising the graph from the surface results in a disjoint union of open sets that are homeomorphic to open disks --- these open sets are referred to as the \textit{faces} of the topological map.
\end{definition}
\begin{note}
The second condition in the above definition requires that for a given embedding to be a valid topological map, each connected component of the graph must live on a separate connected component of the surface. Most authors require topological maps, hence the involved graphs, to be connected. In such settings, it suffices to define a topological map as the embedding of a graph in a connected compact surface of minimal (Euler) genus such that the edges of the graph do not intersect.
\end{note}

There are two natural ways of transforming a ribbon graph into a topological map. In the first formalism, one glues open disks to the boundaries of a given ribbon graph and then contracts the polygons of the ribbon graph to points which become the vertices of a topological map. Shrinking the polygons to vertices also has the consequence of thinning the ribbons to edges connecting said vertices. The disks that were initially glued to the ribbon graph become the faces of the topological map.
\begin{figure}[H]
        \centering
\captionsetup{width=.9\linewidth}
        \includegraphics[width=0.7\textwidth]{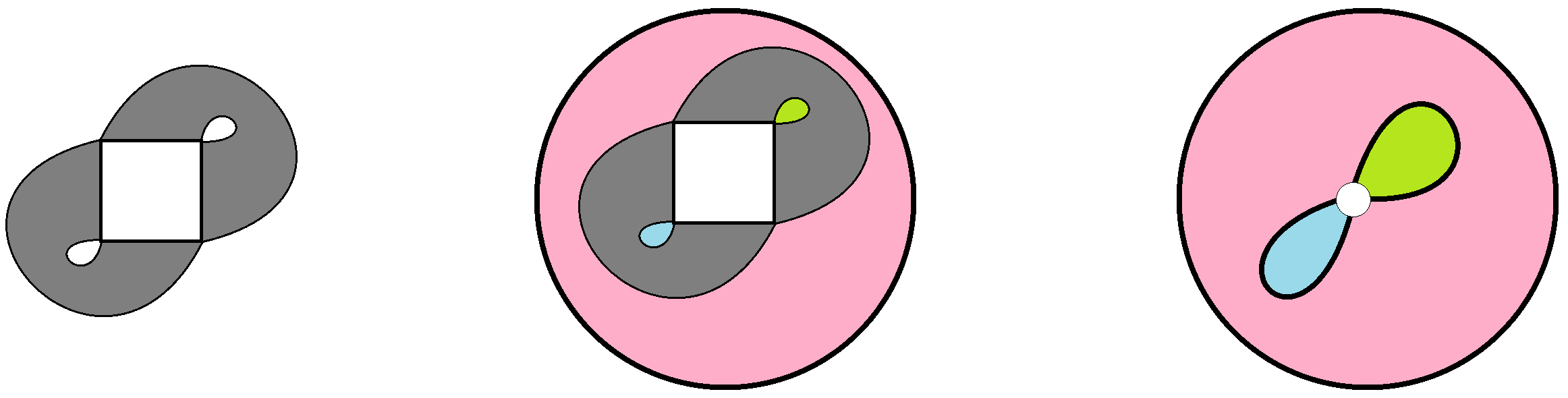}
        \caption[A topological map representation of a ribbon graph]{As in Figure \ref{fig3.6}, we glue three open disks (pink, green, and blue) to the boundaries of the ribbon graph on the left in order to embed it into a sphere (middle image). Shrinking the white square to a vertex and the ribbons to edges incident to said vertex produces the topological map on the right. Our usual convention, which we break here, will be to colour the vertices black and the faces white.} \label{fig3.10}
\end{figure}

In the second formalism, one also glues disks to the boundaries of the given ribbon graph, but then contracts these disks to vertices. The ribbons then collapse to their ends, which become the edges of the topological map, and the polygons of the ribbon graph become the faces of the topological map. This construction was already exemplified in Figure \ref{fig3.6}. The two formalisms just described lead to topological maps that are dual to each other in the sense that canonically follows from the duality principle well known in standard graph theory: one can be obtained from the other by placing a vertex at the centre of every face, drawing an edge between each pair of new vertices whenever the corresponding faces of the original topological map share an edge, and then deleting the edges and vertices of the original topological map. For example, the topological maps of Figures \ref{fig3.6} and \ref{fig3.10} are dual. For the remainder of this subsection, when we refer to the topological map corresponding to a ribbon graph, we will mean the one constructed via the first formalism (i.e., that illustrated in Figure \ref{fig3.10}).

It is important to observe at this point that neither of the above constructions are injective. For example, the first two ribbon graphs displayed in Figure \ref{fig3.4}, which can be distinguished due to the labelling of the polygon vertices (that we often leave implicit), produce the same topological map. In the case of orientable ribbon graphs, we are able to turn the first construction into a bijection by additionally marking a half-edge of each of the $n$ vertices in a systematic manner, thereby forming a so-called $n$-\textit{rooted topological map} \citep{GLS18}. One possible convention is to mark the half-edges corresponding to the ribbon-ends $(i_1^{(l)}\to i_2^{(l)})$ and label them by $l$; see \citep{LaC09}, \citep{PAC18} for more comprehensive reviews of ($1$-)rooted topological maps.
\begin{figure}[H]
        \centering
\captionsetup{width=.9\linewidth}
        \includegraphics[width=0.7\textwidth]{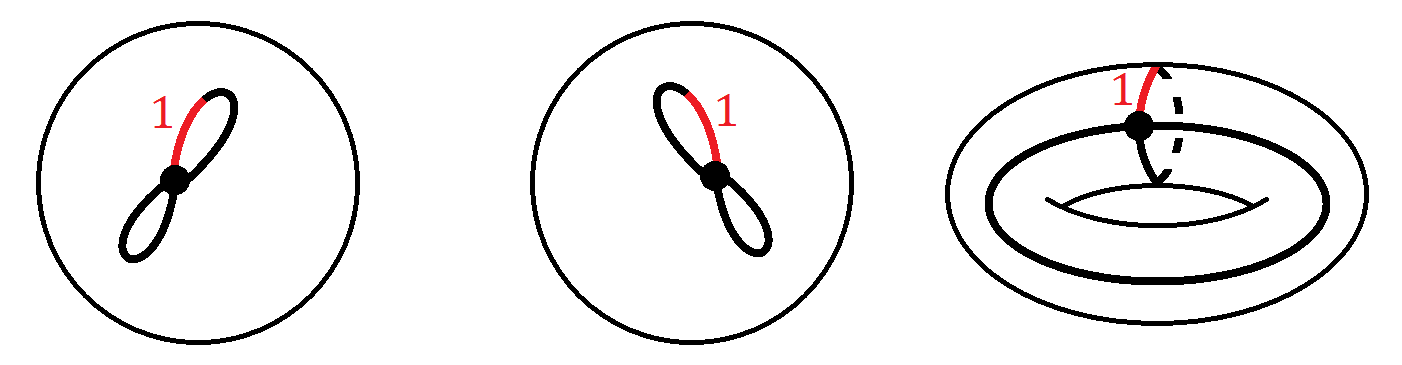}
        \caption[Rooted topological maps for the fourth GUE spectral moment]{These rooted topological maps are constructed from the ribbon graphs of Figure~\ref{fig3.4} using the first prescribed formalism. The half-edges that are root-marked red correspond to the ribbon ends that are glued to the polygon edges $(i_1\to i_2)$.} \label{fig3.11}
\end{figure}

In the case of locally orientable ribbon graphs, one also needs to keep track of local orientation at the vertices and which edges of a given topological map correspond to twisted ribbons. This can be done by assigning orientation to the half-edge root-markings and labelling each edge by, say, $\pm1$, with $+1$ indicating that the edge represents an untwisted ribbon and $-1$ instead referring to a M\"obius half-twisted ribbon (cf.~\citep{MY05}). Seeing as how introducing markings and labellings to give a combinatorial flavour to our topological maps makes the theory more tractable, one might wonder if much more can be gained by introducing even more labels. It turns out that it is possible to transition to a fully combinatorial theory, removing the need for explicitly constructing graphs or surfaces altogether. Thus, we are led to \textit{combinatorial maps}, for which there are multiple definitions.
\begin{definition}[Tutte '84] \label{def3.3}
Following \citep{Tut84}, \citep[\S17.10]{GR01}, a \textit{combinatorial map} is a quadruple $(E_Q,\tau_0,\tau_1,\tau_2)$ where
\begin{enumerate}
\item $E_Q$ is a finite set with the number of elements being divisible by four,
\item $\tau_0,\tau_1,\tau_2$ are fixed-point free involutions on $E_Q$,
\item $\tau_0\tau_1=\tau_1\tau_0$ and $\tau_0\tau_1$ is also fixed-point free,
\item the group $\mean{\tau_0,\tau_1,\tau_2}$ generated by $\tau_0,\tau_1,\tau_2$ is transitive, meaning that it has exactly one orbit, that being all of $E_Q$.
\end{enumerate}
The elements of $E_Q$ are called \textit{quarter-edges}. The orbits of $\tau_0$, $\mean{\tau_0,\tau_1}$, $\mean{\tau_0,\tau_2}$, and $\mean{\tau_1,\tau_2}$ are respectively referred to as the \textit{half-edges}, \textit{edges}, \textit{vertices}, and \textit{faces} of the combinatorial map.
\end{definition}

\begin{figure}[H]
        \centering
\captionsetup{width=.9\linewidth}
        \includegraphics[width=0.6\textwidth]{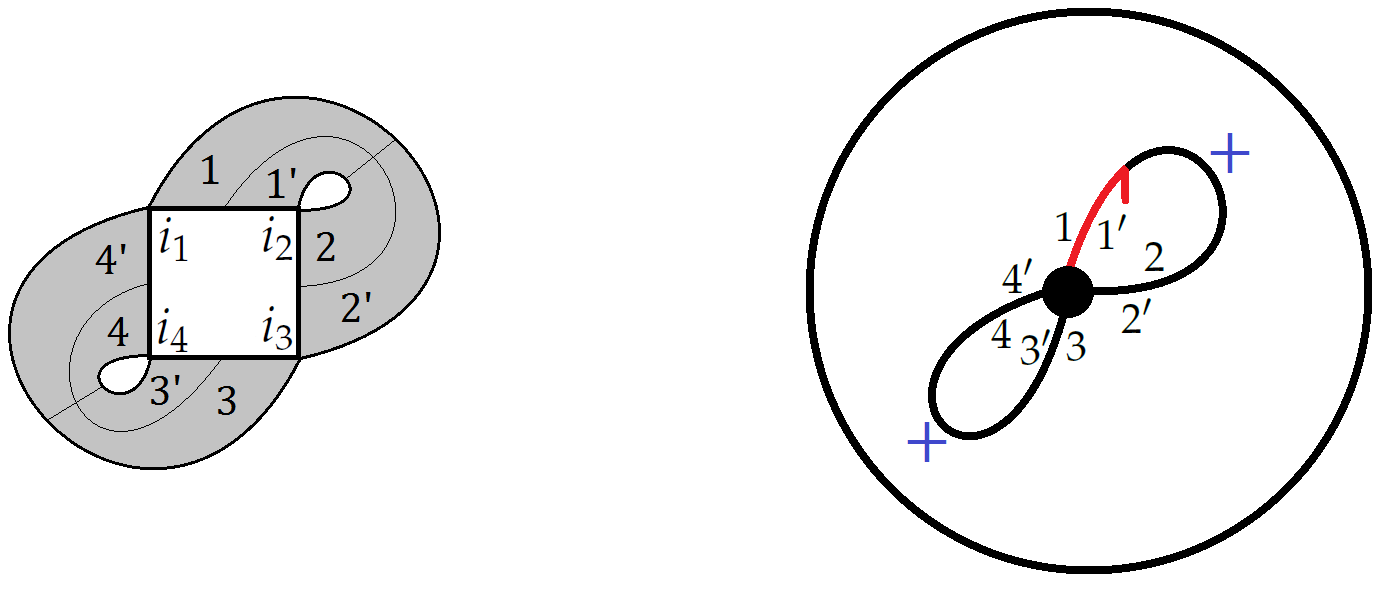}
        \caption[A ribbon graph and topological map with labelled quarter-edges]{We divide the ribbons of a ribbon graph (left) and edges of the corresponding topological map (right) into quarters that are labelled $1,1',2,2',3,3',4,4'$ in clockwise order. The half-edge marked red and oriented clockwise corresponds to the ribbon-end $(i_1\to i_2)$, while the blue plus signs signify that these edges represent untwisted ribbons.} \label{fig3.12}
\end{figure}

Combinatorial maps are in one-to-one correspondence with connected, $n$-rooted topological maps that have been edge-labelled $\pm1$ in the aforementioned manner (the connectedness being enforced by the fourth condition in the above definition). This correspondence follows from the suggestive names given to the elements of $E_Q$ and the various orbits highlighted in Definition \ref{def3.3}: Given a suitable topological map, one must first divide each edge both width- and length-wise to form quarter-edges and then assign these quarter-edges distinct labels. Then, the corresponding combinatorial map has for $E_Q$ the set of said quarter-edges, while the involutions $\tau_0,\tau_1,\tau_2$ on $E_Q$ are such that $\tau_0$ interchanges two quarter-edges if and only if they belong to the same half-edge, $\tau_1$ interchanges two quarter-edges if and only if they belong to the same length-wise half-edge, and $\tau_2$ interchanges two quarter-edges if and only if they are incident to the same vertex without belonging to the same half-edge and are consecutive to each other in the cyclic ordering around that vertex. The two possible choices of (local) orientation at the vertices of the topological map are encoded in the cycles of $\tau_0\tau_2$.
\begin{example}
The combinatorial map corresponding to Figure \ref{fig3.12} below is $(E_Q,\tau_0,\tau_1,\tau_2)$ with
\begin{equation*}
\begin{array}{ll}E_Q=\{1,1',2,2',3,3',4,4'\},&\tau_0=(1,1')(2,2')(3,3')(4,4'),
\\ \tau_1=(1,2')(1',2)(3,4')(3',4),&\tau_2=(1,4')(1',2)(2',3)(3',4).\end{array}
\end{equation*}
Observe that the first cycle of $\tau_0\tau_2=(1,4,3,2)(1',2',3',4')$ describes an anticlockwise ordering of the half-edges incident to the single vertex of the topological map, while the second cycle of $\tau_0\tau_2$ describes the opposite ordering. These cyclic orderings each prescribe a choice of orientation for the topological map. 
\end{example}

A combinatorial map is (globally) orientable if and only if $\mean{\tau_0\tau_1,\tau_0\tau_2}$ has two distinct orbits \citep{Tut84}. If this is the case, one may remove the redundancy of $\tau_0$ by identifying quarter-edges whenever they belong to the same half-edge. In the context of Figure \ref{fig3.12}, this amounts to setting $j'\equiv j$ for $j=1,\ldots,4$. Identifying quarter-edges according to $\tau_0$ induces a choice of orientation and moreover leads to a simpler combinatorial theory on the set of half-edges $E_H\simeq E_Q/\tau_0$, which has half as many elements as that considered in Definition~\ref{def3.3}. This simplification is to be expected since every compact surface has an orientable double cover \citep{Hat02}, so a theory allowing for non-orientable objects should intuitively require twice as many labels as an equivalent theory focusing solely on orientable structures.
\begin{definition}[Edmonds '60] \label{def3.4}
Following \citep{Edm60}, \citep{LZ04}, an \textit{oriented combinatorial map} is a triple $(E_H,\tau_e,\tau_v)$ where
\begin{enumerate}
\item $E_H$ is a finite set with an even number of elements,
\item $\tau_v$ is a permutation of $E_H$ and $\tau_e$ is a fixed-point free involution on $E_H$,
\item the group $\mean{\tau_e,\tau_v}$ is transitive.
\end{enumerate}
The elements of $E_H$ are called \textit{half-edges}. The cycles of $\tau_e$, $\tau_v$, and $\tau_v^{-1}\tau_e^{-1}$ are respectively called the \textit{edges}, \textit{vertices}, and \textit{faces} of the oriented combinatorial map.
\end{definition}

This definition is preferred to Definition \ref{def3.3} in the orientable case because the correspondence between oriented combinatorial maps and orientable, connected, $n$-rooted topological maps is somehow more transparent. Indeed, if one labels the half-edges of such a topological map and settles on a choice of orientation, then the corresponding oriented combinatorial map has $E_H$ being the set of labelled half-edges, $\tau_e$ being the involution that interchanges half-edges that belong to the same edge, and $\tau_v$ being a product of disjoint cycles where each cycle describes the cyclic ordering of the half-edges incident to a vertex that is induced by the orientation of the topological map. Each cycle of the face permutation $\tau_v^{-1}\tau_e^{-1}$ describes a face of the topological map by listing out the first half of each edge that borders the face when traversing in the direction determined by the choice of orientation.
\begin{figure}[H]
        \centering
\captionsetup{width=.9\linewidth}
        \includegraphics[width=0.3\textwidth]{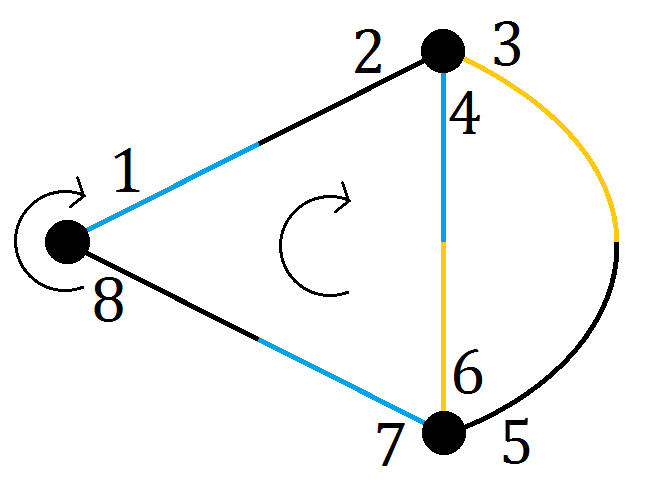}
        \caption[An oriented planar graph with labelled half-edges]{Illustrated is an oriented planar graph with labelled half-edges. It should be thought of as being embedded in the sphere and thus equivalent to an oriented topological map. The corresponding oriented combinatorial map is $(E_H,\tau_e,\tau_v)$ with $E_H=\{1,2,\ldots,8\}$, $\tau_e=(1,2)(3,5)(4,6)(7,8)$, and $\tau_v=(1,8)(2,3,4)(5,7,6)$. Note that the cycles of $\tau_v$ comply with the orientation of the topological map. Likewise, the first, second, and third cycle of the face permutation $\tau_v^{-1}\tau_e^{-1}=(1,4,7)(2,8,5)(3,6)$ respectively list out, in appropriate cyclic order, the blue, black, and yellow halves of the edges bordering the triangular face, the `external face', and the bigonal face.} \label{fig3.13}
\end{figure}

Let us now briefly review the connection between the structures discussed above and their hypermap analogues. A \textit{topological hypermap} is simply a connected topological map whose vertices are coloured black and white (later, we use red) in such a way that each edge connects a black vertex to a white one \citep{LZ04}. In that vein, a \textit{combinatorial hypermap} is the triple of Definition~\ref{def3.4} except that $\tau_e$ is now free to be any permutation of $E_H$; $E_H$ is then the set of edges of the hypermap, while the cycles of $\tau_v$ ($\tau_e$) are interpreted as black vertices (white vertices or so-called hyperedges). Thus, an oriented combinatorial map is a combinatorial hypermap with $\tau_e$ constrained to be a fixed-point free involution. Similarly, a topological hypermap with all white vertices constrained to be bivalent is equivalent to a connected topological map --- the edges of the hypermap are interpreted as half-edges of the topological map so that each white vertex sits at the middle of a map-edge.

\begin{remark} \label{R3.3}
In the literature surrounding Grothendieck's dessins d'enfants \citep{Sch94}, a topological hypermap is referred to as a dessin d'enfant, while a topological map --- seen as a topological hypermap with bivalent white vertices --- is referred to as a \textit{clean} dessin d'enfant.
\end{remark}

\subsection{Moments of the Laguerre unitary and orthogonal ensembles} \label{s3.3.2}
Since the Laguerre unitary and orthogonal ensembles are represented by matrices that can be expressed as products of Ginibre matrices, whose entries are normally distributed, their (mixed) moments and cumulants can also be studied using the Isserlis--Wick theorem. Assuming mastery over the Gaussian case, as outlined in the previous subsection, we first show how the spectral moments $m_k^{(LUE)}$ can be written as sums over certain bicoloured ribbon graphs. Then, as in the Gaussian case, we discuss how the presented ideas extend to the mixed cumulants of the LUE and also the LOE.

\subsubsection{Ribbon graphs for the LUE spectral moments}
Let $G$ be drawn from the $M\times N$ complex Ginibre ensemble and throughout this subsection, let $\mean{\,\cdot\,}$ denote averages with respect to the p.d.f.~$P^{(Gin)}(G)$ given in Definition \ref{def1.3}. According to Definition~\ref{def1.4}, the $N\times N$ Wishart--Laguerre matrix $W=G^\dagger G$ then represents the $(M,N)$ Laguerre unitary ensemble and the spectral moments of this ensemble are given by \eqref{eq1.1.15}
\begin{align}
m_k^{(LUE)}&=\mean{\Tr\,W^k},\quad k\in\mathbb{N} \nonumber
\\&=\mean{\Tr\,(G^\dagger G)^k} \nonumber
\\&=\sum_{i_1,\ldots,i_k=1}^N\sum_{j_1,\ldots,j_k=1}^M\mean{(G^\dagger)_{i_1j_1}G_{j_1i_2}(G^\dagger)_{i_2j_2}G_{j_2i_3}\cdots(G^\dagger)_{i_kj_k}G_{j_ki_1}}, \label{eq3.3.27}
\end{align}
in analogy with equation \eqref{eq3.3.3}. Since the real components of the entries of $G$ are independent, centred, normal variables such that
\begin{equation} \label{eq3.3.28}
\mean{(G^\dagger)_{ij}G_{kl}}=\chi_{i=l}\chi_{j=k},
\end{equation}
the Isserlis--Wick theorem then shows that
\begin{align}
\mean{(G^\dagger)_{i_1j_1}G_{j_1i_2}\cdots(G^\dagger)_{i_kj_k}G_{j_ki_1}}&=\sum_{\sigma\in\mathfrak{S}_k}\prod_{l=1}^k\mean{(G^\dagger)_{i_lj_l}G_{j_{\sigma(l)}i_{\sigma(l)+1}}} \nonumber
\\&=\sum_{\sigma\in\mathfrak{S}_k}\prod_{l=1}^k(\chi_{i_l=i_{\sigma(l)+1}}\chi_{j_l=j_{\sigma(l)}}), \label{eq3.3.29}
\end{align}
where $\mathfrak{S}_k$ is the set of all permutations of $\{1,\ldots,k\}$ and we have defined $i_{k+1}:=i_1$. Hence, combining equations \eqref{eq3.3.27} and \eqref{eq3.3.29} gives the analogue of Lemma \ref{L3.1} for the LUE.
\begin{lemma} \label{L3.3}
Fix $k\in\mathbb{N}$, let $\mathfrak{S}_k$ be the set of permutations of $\{1,\ldots,k\}$ and set $i_{k+1}:=i_1$. We have that the spectral moments of the LUE are given by
\begin{equation} \label{eq3.3.30}
m_k^{(LUE)}=\sum_{\sigma\in\mathfrak{S}_k}\sum_{i_1,\ldots,i_k=1}^N\sum_{j_1,\ldots,j_k=1}^M\prod_{l=1}^k(\chi_{i_l=i_{\sigma(l)+1}}\chi_{j_l=j_{\sigma(l)}}).
\end{equation}
\end{lemma}
As one might expect, the summand of the right-hand side of equation \eqref{eq3.3.30} can be represented by ribbon graphs. To do so, we first draw a $2k$-gon and label its vertices $i_1,j_1,i_2,j_2,\ldots,i_k,j_k$ in clockwise order. Next, to distinguish the two index sets, we colour the vertices labelled by the $i_1,\ldots,i_k$ black and those labelled by the $j_1,\ldots,j_k$ red. Then, for $1\leq l\leq k$, edges of the form $(i_l\to j_l)$ represent $(G^\dagger)_{i_lj_l}$, while those of the form $(j_l\to i_{l+1})$ represent $G_{j_li_{l+1}}$. The factors in the summand of the right-hand side of equation \eqref{eq3.3.30} are thus represented by untwisted ribbons connecting these two types of edges. The condition that the ribbons cannot join edges of the same type is equivalent to that of only allowing ribbon graphs that respect the colouring of the vertices.

\begin{figure}[H]
        \centering
\captionsetup{width=.9\linewidth}
        \includegraphics[width=0.6\textwidth]{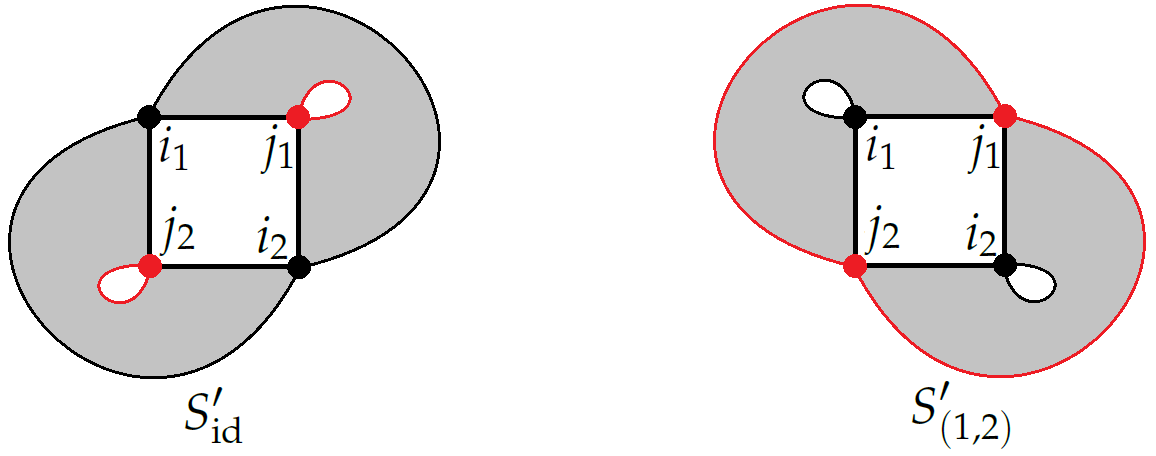}
        \caption[The ribbon graphs that contribute to the second LUE spectral moment]{The illustrated ribbon graphs represent the terms given in equations \eqref{eq3.3.32} and \eqref{eq3.3.33} below. Having alternately coloured the vertices of the squares black and red according to the formalism described above, we see that the vertex colourings induce colours on the boundaries of the ribbon graph. The number of black (red) boundaries is equal to the exponent of $N$ ($M$) in the evaluations of $S_{\mathrm{id}}'$ and $S_{(1,2)}'$.} \label{fig3.14}
\end{figure}

\begin{example} \label{ex3.4}
Let us now give the LUE analogue of Example \ref{ex3.2}, wherein we saw that $m_4^{(GUE)}$ is equal to $(N^3+N^3+N)/4$, with each of the terms $N^3,N$ corresponding to a ribbon graph of Figure \ref{fig3.4}. In the LUE case, we compute the second spectral moment $m_2^{(LUE)}$. By equation \eqref{eq3.3.30}, it is given by
\begin{equation}
m_2^{(LUE)}=S_{\mathrm{id}}'+S_{(1,2)}'
\end{equation}
with
\begin{align}
S_{\mathrm{id}}'&:=\sum_{i_1,i_2=1}^N\sum_{j_1,j_2=1}^M(\chi_{i_1=i_2}\chi_{j_1=j_1})(\chi_{i_2=i_1}\chi_{j_2=j_2})=M^2N, \label{eq3.3.32}
\\S_{(1,2)}'&:=\sum_{i_1,i_2=1}^N\sum_{j_1,j_2=1}^M(\chi_{i_1=i_1}\chi_{j_1=j_2})(\chi_{i_2=i_2}\chi_{j_2=j_1})=MN^2. \label{eq3.3.33}
\end{align}
The evaluation $S_{\mathrm{id}}'=M^2N$ follows from the fact that the product of indicator functions in equation \eqref{eq3.3.32} enforces the equivalence $i_1\equiv i_2$, but otherwise does not constrain the indices $j_1,j_2$ --- similarly for $S_{(1,2)}'$. These equivalences of summation indices can be read off from the ribbon graphs of Figure~\ref{fig3.14} above.
\end{example}

In general, the $k\textsuperscript{th}$ spectral moment for the LUE is given by
\begin{equation} \label{eq3.3.34}
m_k^{(LUE)}=\sum_{\sigma\in\mathfrak{S}_k}M^{V_r(\sigma)}N^{V_b(\sigma)},
\end{equation}
where $V_r(\sigma)$ and $V_b(\sigma)$ are equal to the number of red, respectively black, boundaries of the ribbon graph corresponding to $\sigma$ --- the proof of this fact is the same as for Lemma \ref{L3.2}. As with the GUE case, we are able to reformulate equation \eqref{eq3.3.34} as a genus expansion, so long as the red and black vertices of the involved ribbon graphs are placed on equal footing. 
\begin{proposition}
Let $k\in\mathbb{N}$ and $\mathfrak{S}_k$ be as in Lemma \ref{L3.3}
Setting the Laguerre parameter $a=\hat{a}N$ with $\hat{a}={\rm O}(1)$ so that $M=(\hat{a}+1)N$, as per the convention set out in \S\ref{s2.4.1} and \S\ref{s3.1.2}, we have
\begin{equation} \label{eq3.3.36}
m_k^{(LUE)}=\sum_{l=0}^kN^{1+k-l}\sum_{p=1}^k(\hat{a}+1)^p\,\#\{\sigma\in\mathfrak{S}_k\,|\,g(\sigma)=l/2\textrm{ and }V_r(\sigma)=p\},
\end{equation}
where $g(\sigma)$ is again the genus of the ribbon graph corresponding to $\sigma$ --- equivalently, $g(\sigma)$ is the minimal genus of the surface on which this ribbon graph can be embedded into without self-intersections and it is also the genus of the surface obtained by pairwise identifying the edges of a $2k$-gon according to the factors $(\chi_{i_l=i_{\sigma(l)+1}}\chi_{j_l=j_{\sigma(l)}})$ in the right-hand side of equation \eqref{eq3.3.30}.
\end{proposition}
\begin{proof}
Observe that the construction outlined above Example \ref{ex3.4} describes a bijection between $\mathfrak{S}_k$ and the set of orientable $k$-ribbon graphs built from $2k$-gons with well-defined bicoloured boundaries (i.e., each ribbon has both a black and red side, and black (red) vertices lie on black (red) boundaries). Thus, interpreting $\mathfrak{S}_k$ as this latter set, equation \eqref{eq3.3.34} can be rewritten as
\begin{equation}
m_k^{(LUE)}=\sum_{V_r,V_b=1}^kM^{V_r}N^{V_b}\,\#\{\sigma\in\mathfrak{S}_k\,|\,V_r(\sigma)=V_r\textrm{ and }V_b(\sigma)=V_b\},
\end{equation}
where $\#\mathcal{S}$ has the same meaning as in equation \eqref{eq3.3.14}. Substituting in $M=(\hat{a}+1)N$ and using equation \eqref{eq3.1.15} to relate $V(\sigma)=V_r(\sigma)+V_b(\sigma)$, the total number of boundaries of the ribbon graph labelled by $\sigma$, to $g(\sigma)$ yields equation \eqref{eq3.3.36}.
\end{proof}

Comparing equation \eqref{eq3.3.36} to equation \eqref{eq3.0.9} shows that $M_{k,l}^{(LUE)}$ has precisely the interpretation described below Figure \ref{fig3.3}. Each ribbon graph that contributes to the computation of $m_k^{(LUE)}$ also contributes to the value of $m_{2k}^{(GUE)}$ upon `forgetting' the bicolouring; compare Figures \ref{fig3.4} and \ref{fig3.14}. However, not every LUE ribbon graph arises in this way, since some orientable ribbon graphs have no valid bicolouring. For example, colouring any vertex of the third ribbon graph displayed in Figure \ref{fig3.4} red forces the single boundary and hence all vertices to also be coloured red, which violates the bicolouring requirement. Nonetheless, every planar ribbon graph has a valid bicolouring, so that the genus zero ribbon graphs pertaining to the calculation of $m_{2k}^{(GUE)}$ are in one-to-one correspondence with the genus zero bicoloured ribbon graphs contributing to the value of $m_k^{(LUE)}$. In keeping with the discussion following Proposition \ref{prop3.5}, this implies the very simple relationship $M_{k,0}^{(LUE)}=2^k M_{k,0}^{(GUE)}$ when $\hat{a}=0$. In combination with the proof of the Wigner semi-circle law given in Remark~\ref{R3.2}, we thus have a combinatorial proof of the Mar\v{c}enko--Pastur law \eqref{eq1.2.15} in the regime $a={\rm O}(1)$.

\subsubsection{Ribbon graphs for the LUE and LOE mixed cumulants}
The mixed moments $m_{k_1,\ldots,k_n}^{(LUE)}$ and cumulants $c_{k_1,\ldots,k_n}^{(LUE)}$ of the LUE are respectively defined through the formulae \eqref{eq3.0.13}, \eqref{eq3.0.14}
\begin{align}
m_{k_1,\ldots,k_n}^{(LUE)}&=\mean{\prod_{i=1}^n\Tr\,W^{k_i}},\quad k_1,\ldots,k_n\in\mathbb{N} \label{eq3.3.37}
\\&=\sum_{K\vdash\{k_1,\ldots,k_n\}}\prod_{\kappa_i\in K}c_{\kappa_i}^{(LUE)}. \label{eq3.3.38}
\end{align}
The proofs of equations \eqref{eq3.3.18} and \eqref{eq3.3.22} can be extended to give their LUE analogues.
\begin{proposition} \label{prop3.13x}
Let $n,k_1,\ldots,k_n\in\mathbb{N}$ and define $\mathfrak{S}_{k_1,\ldots,k_n}$ to be the set of $(k_1+\cdots+k_n)$-ribbon graphs that consist of $n$ polygons with respectively $2k_1,2k_2,\ldots,2k_n$ edges and vertices alternately coloured black and red, whose edges are connected by untwisted ribbons that respect the bicolouring of the polygon vertices --- let $\mathfrak{S}_{k_1,\ldots,k_n}^c$ be the subset of connected ribbon graphs in $\mathfrak{S}_{k_1,\ldots,k_n}$. Then, taking $a=\hat{a}N$ with $\hat{a}={\rm O}(1)$, the mixed moments and cumulants of the LUE are given by
\begin{align}
m_{k_1,\ldots,k_n}^{(LUE)}&=\sum_{V=1}^{k_1+\cdots+k_n+n}N^V \nonumber
\\&\qquad\times\sum_{p=1}^{k_1+\cdots+k_n}(\hat{a}+1)^p\,\#\{\Gamma\in\mathfrak{S}_{k_1,\ldots,k_n}\,|\,V_r(\Gamma)+V_b(\Gamma)=V\textrm{ and }V_r(\Gamma)=p\}, \label{eq3.3.39}
\\c_{k_1,\ldots,k_n}^{(LUE)}&=\sum_{l=0}^{k_1+\cdots+k_n+1-n}N^{k_1+\cdots+k_n+2-n-l} \nonumber
\\&\qquad\times\sum_{p=1}^{k_1+\cdots+k_n+1-n}(\hat{a}+1)^p\,\#\{\Gamma\in\mathfrak{S}_{k_1,\ldots,k_n}^c\,|\,g(\Gamma)=l/2\textrm{ and }V_r(\Gamma)=p\}, \label{eq3.3.40}
\end{align}
where $V_r(\Gamma)$ and $V_b(\Gamma)$ are respectively the number of red and black boundaries of the ribbon graph $\Gamma$ and $g(\Gamma)$ is the genus of $\Gamma$.
\end{proposition}
In contrast to what is seen in the GUE case, $m_{k_1,\ldots,k_n}^{(LUE)}$ is always guaranteed to be a polynomial of degree $k_1+\cdots+k_n+n$ in $N$ since there is at least one $\Gamma\in\mathfrak{S}_{k_1,\ldots,k_n}$ such that $V_r(\Gamma)+V_b(\Gamma)$ equals this degree. This $\Gamma$, which has $n$ connected components that are each of genus zero, exists because the polygons present in the ribbon graphs of $\mathfrak{S}_{k_1,\ldots,k_n}$ each have an even number of sides, which allows for polygon edges to be paired without need for any ribbons connecting two separate polygons --- the GUE analogue of $\mathfrak{S}_{k_1,\ldots,k_n}$ is the set $\mathfrak{P}_{k_1,\ldots,k_n}$, which contains no ribbon graphs with $n$ connected components if any of the $k_i$ are odd.

The $(M,N)$ Laguerre orthogonal ensemble is represented by the $N\times N$ Wishart--Laguerre matrix $W=G^TG$, where $G$ is an element of the $M\times N$ real Ginibre ensemble. Its (mixed) moments and cumulants (which are specified by equations \eqref{eq3.3.37}, \eqref{eq3.3.38} with our new definition of $W$) can be computed in the same way as in the LUE case, except that the involved ribbon graphs are now allowed to have M\"obius half-twisted ribbons. This is because $G$ being real means that $(G^T)_{ij}=G_{ji}$, so equation \eqref{eq3.3.28} should be rewritten as
\begin{equation} \label{eq3.3.41}
\mean{(G^T)_{ij}G_{kl}}=\mean{(G^T)_{ij}(G^T)_{lk}}=\mean{G_{ji}G_{kl}}=\frac{1}{2}\chi_{i=l}\chi_{j=k};
\end{equation}
recalling that our ribbon graphs are built from polygons whose edges represent entries of either $G^T$ or $G$, the first average in equation \eqref{eq3.3.41} is represented by an untwisted ribbon connecting these two types of edges, while the second (third) average therein is represented by a M\"obius half-twisted ribbon connecting together two edges of the $G^T$ ($G$) type.
\begin{proposition} \label{prop3.14x}
Let $n,k_1,\ldots,k_n\in\mathbb{N}$ and define $\tilde{\mathfrak{S}}_{k_1,\ldots,k_n}^c$ to be the set of locally orientable, connected $(k_1+\cdots+k_n)$-ribbon graphs built from $n$ polygons with respectively $2k_1,2k_2,\ldots,2k_n$ edges and vertices alternately coloured black and red, whose boundaries have well-defined colours induced by the vertex colouring. Then, in the setting $a=\hat{a}N$ with $\hat{a}={\rm O}(1)$, we have
\begin{multline} \label{eq3.3.42}
c_{k_1,\ldots,k_n}^{(LOE)}=\sum_{l=0}^{k_1+\cdots+k_n+1-n}\left(\frac{N}{2}\right)^{k_1+\cdots+k_n}N^{2-n-l}
\\\times\sum_{p=1}^{k_1+\cdots+k_n+1-n}(\hat{a}+1)^p\,\#\{\Gamma\in\tilde{\mathfrak{S}}_{k_1,\ldots,k_n}^c\,|\,\tilde{g}(\Gamma)=l\textrm{ and }V_r(\Gamma)=p\},
\end{multline}
where $V_r(\Gamma)$ is the number of red boudnaries of $\Gamma$ and $\tilde{g}(\Gamma)$ is the Euler genus of $\Gamma$.
\end{proposition}
Note that the expression \eqref{eq3.3.42} is very similar to that for $c_{k_1,\ldots,k_n}^{(LUE)}$ given in equation \eqref{eq3.3.40}, with a key difference being that the summand of equation \eqref{eq3.3.42} may be non-zero for odd values of $l$, in contrast to what is seen in equation \eqref{eq3.3.40}. Likewise, the mixed moments of the LOE are given by a perturbation of equation \eqref{eq3.3.39} involving locally orientable ribbon graphs, though we do not display it here.
\begin{figure}[H]
        \centering
\captionsetup{width=.9\linewidth}
        \includegraphics[width=0.55\textwidth]{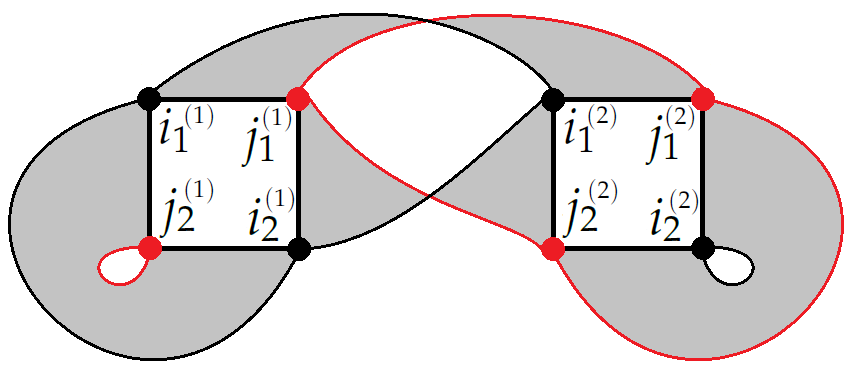}
        \caption[A locally orientable bicoloured ribbon graph]{The illustrated bicoloured, locally orientable ribbon graph represents the term $\langle(G^T)_{i_1^{(1)}j_1^{(1)}}(G^T)_{i_1^{(2)}j_1^{(2)}}\rangle\langle G_{j_1^{(1)}i_2^{(1)}}G_{j_2^{(2)}i_1^{(2)}}\rangle\langle(G^T)_{i_2^{(1)}j_2^{(1)}}G_{j_2^{(1)}i_1^{(1)}}\rangle\langle G_{j_1^{(2)}i_2^{(2)}}(G^T)_{i_2^{(2)}j_2^{(2)}}\rangle$, where we have adopted the vertex labelling convention of Figure \ref{fig3.8}. Interpreting the ribbons as identifying the polygon edges, we have two faces, four edges, and four vertices (equivalently, boundaries). Thus, by the classical formula $\chi=F-E+V=2-\tilde{g}$, this ribbon graph is an element of $\tilde{\mathfrak{S}}_{2,2}^c$ with two red boundaries and Euler genus zero. Hence, by equation \eqref{eq3.3.42}, it contributes a value of $(\hat{a}+1)^2N^4$ to $c_{2,2}^{(LOE)}$.} \label{fig3.15}
\end{figure}

Like in the Gaussian case, the generating functions $\tilde{W}_n^{(LUE)},\tilde{W}_n^{(LOE)}$ of the corresponding mixed cumulants scaled according to Definition~\ref{def1.6} are ${\rm O}(N^{2-n})$, so they have large $N$ expansions of the form given in Theorem \ref{thrm1.1}. Another similarity between the GUE and LUE is that the correlator expansion coefficients $W_n^{(LUE),l}$ (recall their definition in Theorem \ref{thrm1.1} and cf.~the discussion following equation \eqref{eq3.3.22}) can be visualised as genus $l/2$ compact, connected, orientable surfaces with $n$ holes. This represents the fact that for given $(n,l)$, $W_n^{(LUE),l}$ is a generating function for all of the ribbon graphs that can be drawn on such a surface by replacing the holes with even-sided polygons and connecting the edges of these polygons with ribbons such that the resulting (embedded) ribbon graph satisfies the criteria given above Figure \ref{fig3.7} (the ribbon graph does not self-intersect and excising it from the surface results in a disjoint union of sets homeomorphic to open disks), along with the condition that the disks bounded by the ribbons can be coherently bicoloured black and red such that each ribbon borders both a black and red face --- in the LOE case, one needs to be careful and categorise the relevant ribbon graphs by their orientability.

\subsubsection{Relations to topological and combinatorial maps and hypermaps}
As was discussed in \S\ref{s3.3.1}, alternative representations of ribbon graphs are interesting for various reasons. The bicoloured ribbon graphs constructed above can be represented by the topological maps introduced in Definition \ref{def3.2} in a similar fashion to that discussed below said definition, with a few differences: Recall that a true bijection between sets of ribbon graphs and their topological map counterparts requires that the latter be rooted (see Figure \ref{fig3.11}) and that, in the locally orientable (i.e., LOE) case, each vertex be assigned a local orientation. However, in contrast to the GOE case, we no longer need to keep track of which map edges correspond to twisted ribbons since this data can be recovered by checking if the ends of a map edge correspond to polygon edges of the same or different type (recall that the polygon edges of an LOE ribbon graph represent entries of either $G^T$ or $G$) --- this is assuming that the relevant topological maps are constructed using the formalism illustrated in Figure \ref{fig3.10}. (As an aside, observe that there is a bijection between $\tilde{\mathfrak{S}}_{k_1,\ldots,k_n}^c$ and the set $\mathfrak{P}_{2k_1,\ldots,2k_n}^c$ of connected, orientable GUE ribbon graphs: Alternately labelling the edges of each polygon of $\Gamma\in\mathfrak{P}_{2k_1,\ldots,2k_n}^c$ by $G^T$ and $G$, with the `first' or rooted edge of each polygon labelled $G^T$, and then twisting any ribbons connecting polygon edges of the same type results in a ribbon graph that can be uniquely bicoloured in the necessary way.) When constructing a topological map from a bicoloured ribbon graph, the bicolouring of the boundaries of the ribbon graph extends naturally to a face bicolouring of the resulting topological map.

In the combinatorial setting of Definition \ref{def3.3}, the necessary bicolouring constraint is enforced by requiring that $E_Q$ admit a partitioning $E_Q=E_r\cup E_b$ into sets $E_r\simeq E_Q/\tau_0$ and $E_b=E_Q\setminus E_r$ of red and black quarter-edges, respectively, such that $\tau_0(E_x)=E_Q\setminus E_x$ and $\tau_1(E_x)=\tau_2(E_x)=E_x$ for $x=r,b$. The equivalent condition for the oriented combinatorial maps of Definition \ref{def3.4} is that $E_H=E_r'\cup E_b'$ with $E_b'=E_H\setminus E_r'$ and $\tau_e(E_x')=\tau_v(E_x')=E_H\setminus E_x'$, where we again take $x=r,b$. Thus, writing $E_r=\{1',2,3',4\}$ shows that the topological map of Figure \ref{fig3.12} can be bicoloured in a way that relates it to the first ribbon graph displayed in Figure \ref{fig3.14}, while it can be checked that the oriented combinatorial map of Figure \ref{fig3.13} cannot be bicoloured in any valid way.
\begin{figure}[H]
        \centering
\captionsetup{width=.9\linewidth}
        \includegraphics[width=0.45\textwidth]{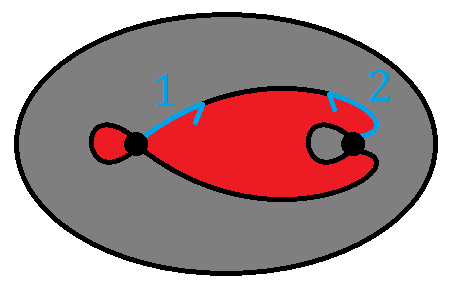}
        \caption[A locally orientable bicoloured topological map]{Pictured is the bicoloured $2$-rooted topological map corresponding to the ribbon graph of Figure \ref{fig3.15}. It has two black (depicted as dark grey) and red faces. Since it is embedded in the sphere, it is orientable. Nonetheless, the notches on the blue root-markings indicate that the local orientations of the vertices are inverse to each other.} \label{fig3.16}
\end{figure}

The topological maps discussed above can be simplified in an interesting way by transforming them into topological hypermaps (whose definition we recall from the end of \S\ref{s3.3.1}). In the LUE case, this is done by contracting red faces to red vertices while identifying half-edges as follows: Traverse clockwise around each vertex, starting at the rooted half-edge, and identify the rooted half-edge with the second visited half-edge (retaining the root-marking), the third visited half-edge with the fourth, and so on. This procedure results in a topological hypermap consisting of red and black vertices that are connected by (possibly rooted) edges. The number of faces and red vertices of the resulting topological hypermap is respectively equal to the number of black and red faces of the original topological map. Moreover, the act of contracting faces to vertices does not affect the genus of the relevant surface, so the topological hypermap obtained through the above procedure has the same genus as the bicoloured topological map (and ribbon graph) that it corresponds to. Thus, we may rewrite equation \eqref{eq3.3.40} in terms of topological hypermaps by interpreting the coefficient
\begin{equation*}
\#\{\Gamma\in\mathfrak{S}_{k_1,\ldots,k_n}^c\,|\,g(\Gamma)=l/2\textrm{ and }V_r(\Gamma)=p\}
\end{equation*}
as the number of genus $l/2$ $n$-rooted topological hypermaps with $p$ red vertices and $n$ black vertices of valency $k_1,\ldots,k_n$ (it is straightforward to see that the above construction is invertible and thus bijective). In terms of combinatorial hypermaps, this is equivalent to counting the number of triples $(E_H,\tau_e,\tau_v)$ such that $E_H$ has $k_1+\ldots+k_n$ edges, $\tau_e$ has $p$ cycles, $\tau_v$ has $n$ cycles of length $k_1,\ldots,k_n$, and $\tau_v^{-1}\tau_e^{-1}$ has $k_1+\cdots+k_n+2-n-p-l$ cycles (this ensures that the number of faces is consistent with the genus equalling $l/2$).
\begin{figure}[H]
        \centering
\captionsetup{width=.9\linewidth}
        \includegraphics[width=0.8\textwidth]{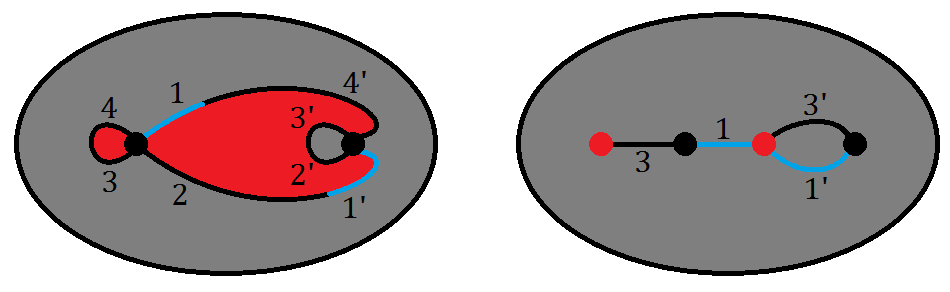}
        \caption[A bicoloured topological map and its hypermap analogue]{The topological hypermap illustrated on the right is obtained from the topological map on the left by contracting the red faces to vertices while making the half-edge identifications $2\equiv1$, $4\equiv3$, $2'\equiv1'$, and $4'\equiv3'$ --- note that we retain the blue root-markings. The corresponding combinatorial map is $(E_H,\tau_e,\tau_v)$ with $E_H=\{1,3,1',3'\}$, $\tau_e=(1,3',1')(3)$, and $\tau_v=(1,3)(1',3')$. Observe that $\tau_v^{-1}\tau_e^{-1}=(1,3',3)(1')$.} \label{fig3.17}
\end{figure}

The procedure described above is also valid in the LOE case, but one needs to decorate the resulting locally orientable topological hypermaps in an appropriate manner to ensure that the mapping is injective. To understand this requirement, let us first recall that each half-edge of a topological map is either of type $G^T$ or $G$, as discussed below equation \eqref{eq3.3.41} (e.g., in the left image of Figure \ref{fig3.17}, the half-edges $1,3,1',3'$ are of type $G^T$, while $2,4,2',4'$ are of type $G$). In the case of LUE topological maps, traversing the boundary of a red face shows that such a boundary is a chain of half-edges with no two consecutive half-edges being of the same type. This is not the case when dealing with LOE topological maps since, e.g., the boundary of the large red face in Figure \ref{fig3.16} has two rooted half-edges (automatically of type $G^T$) meeting each other. To account for this subtlety, we assign $\pm1$ `twist' labellings (cf.~the discussion below Figure \ref{fig3.11}) to the edges of our locally orientable topological hypermaps in such a way that interchanging the type $G^T\leftrightarrow G$ of map half-edges whenever they are represented by twisted hypermap edges (labelled $-1$) results in topological maps whose red faces are bounded by a chain of half-edges of alternating type, as seen in the LUE case. Note that the edges labelled $-1$ are truly twisted in the sense that the faces glued to such edges must switch sides somewhere along them; see Figure \ref{fig3.18} below.
\begin{figure}[H]
        \centering
\captionsetup{width=.9\linewidth}
        \includegraphics[width=0.9\textwidth]{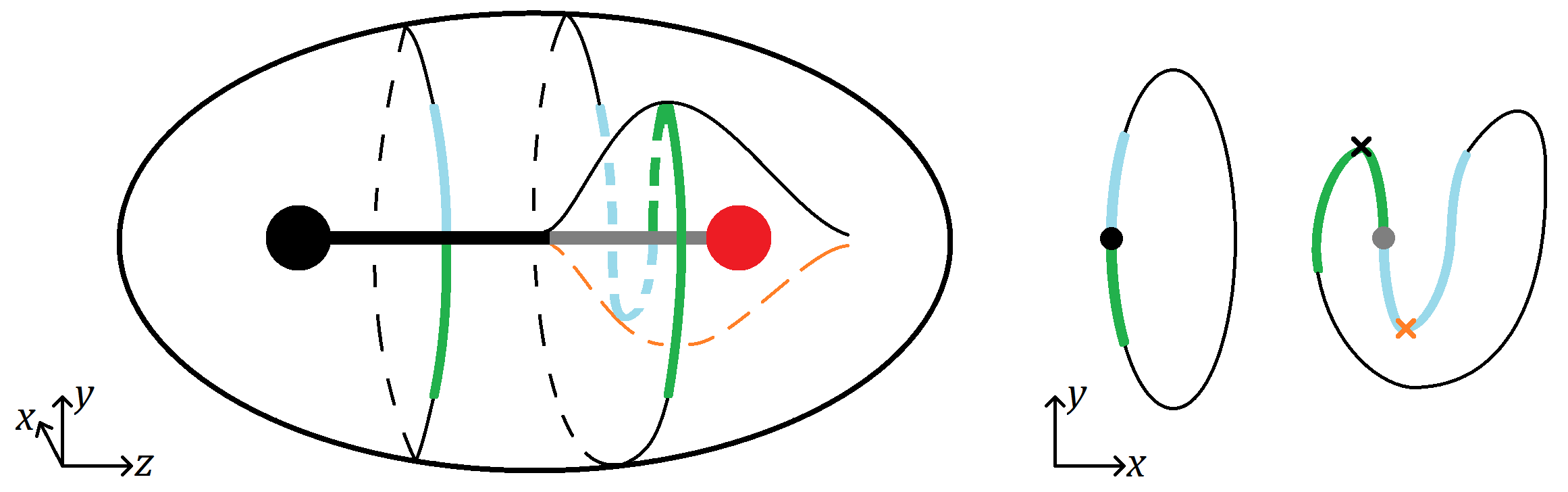}
        \caption[A topological hypermap with a twisted edge]{The left image is the result of gluing a white open disk to a ($-1$-labelled) twisted edge. To make this possible, we have folded a portion of the disk over the grey length of the edge; the green (blue) lines are attached to the boundary of the white disk at points that are relatively close to each other and one should be able to imagine a continuum of similar lines in between them. Folding the disk over the grey length of the edge results in a peak (thin black line) and trough (dotted orange line), which are better visualised through the depicted cross sections --- the first (second) cross section refers to the leftmost (rightmost) green and blue lines in the illustration on the left. In this case, flattening the peak and trough (i.e., untwisting the edge) shows that we simply have a sphere.} \label{fig3.18}
\end{figure}

The alert reader will have noticed that there are multiple ways of assigning twists to the edges of a topological hypermap so that it correctly corresponds to a given locally orientable, bicoloured topological map (e.g., changing the $-1$ label in Figure \ref{fig3.18} to $+1$ is inconsequential). It turns out that each such topological map corresponds to an equivalence class of $\pm1$-labelled topological hypermaps, where two hypermaps are said to be equivalent if one can be obtained from the other by using local surgery to sequentially reverse the local orientations of some vertices while inverting the signs of all edges connected to said vertices.

\begin{figure}[H]
        \centering
\captionsetup{width=.9\linewidth}
        \includegraphics[width=0.6\textwidth]{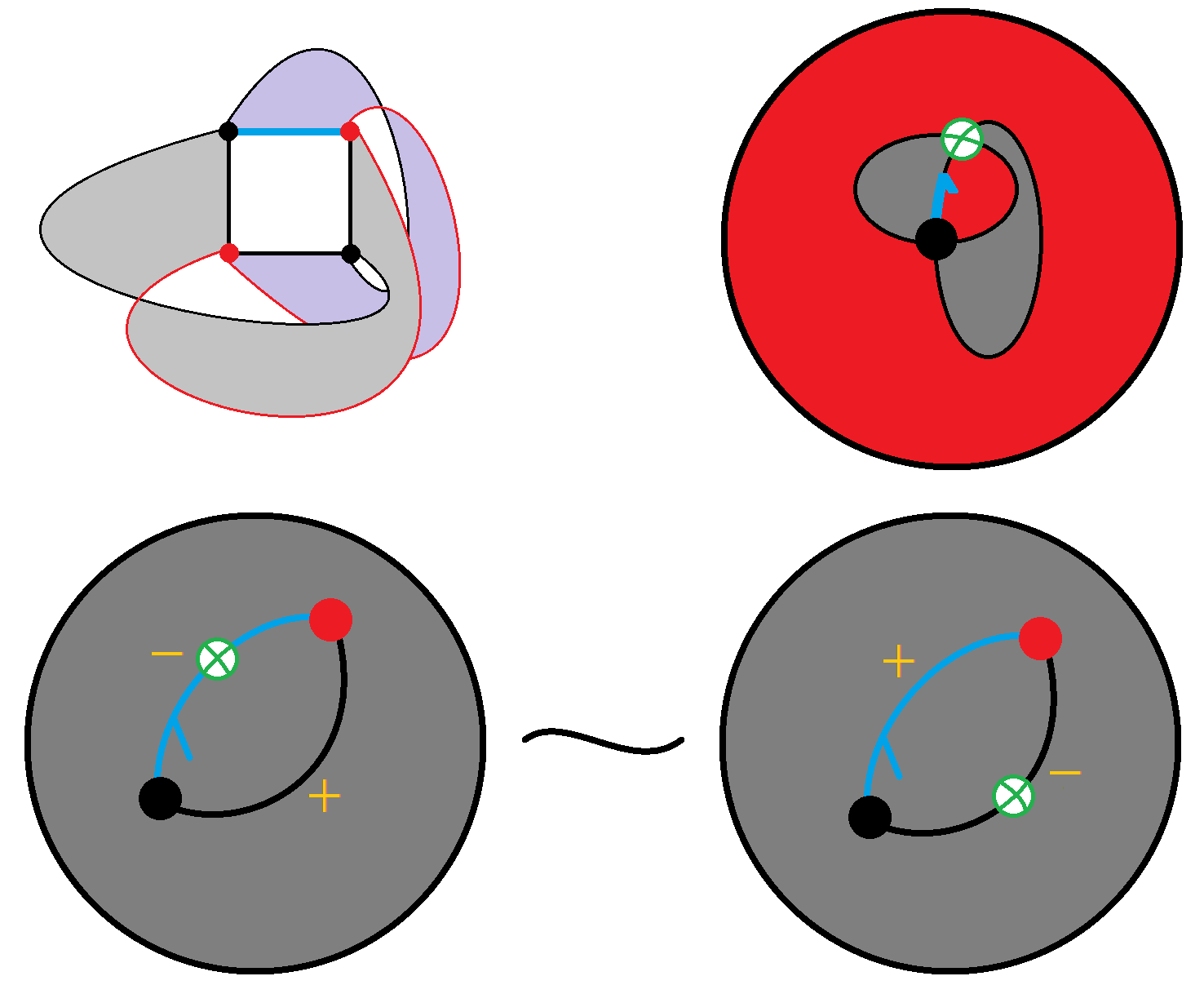}
        \caption[Equivalent topological hypermaps with twisted edges]{On the top left, we have an LOE ribbon graph with two twisted ribbons and a blue root-marking on its top polygon edge. Gluing a red (dark grey) face to the red (black) boundary, then shrinking the ribbons to edges and the white square to a black vertex results in the topological map on the top right; the green $\otimes$ symbol denotes a cross-cap. Contracting the red face to a vertex while identifying half-edges in the necessary manner then results in either of the two equivalent topological hypermaps shown in the bottom.} \label{fig3.19}
\end{figure}

\begin{note}
In comparing Figures \ref{fig3.18} and \ref{fig3.19}, one observes the well known fact that gluing an open disk to a twisted interval (where the twist can be undone without changing its topology) results in a sphere, while doing the same to a loop formed by one twisted and untwisted edge (essentially a M\"obius strip) results in a real projective plane.
\end{note}

Let us now give a reformulation of Proposition \ref{prop3.14x} in terms of hypermaps.
\begin{proposition} \label{prop3.9}
Let $n,k_1,\ldots,k_n\in\mathbb{N}$ and define $\mathfrak{T}_{k_1,\ldots,k_n}$ to be the set of orientable $n$-rooted topological hypermaps with $k_1+\cdots+k_n$ edges, any number of red vertices, and $n$ black vertices of valency $k_1,\ldots,k_n$, with each black vertex having one root edge incident to it; to fix convention, assign all black vertices a clockwise orientation. Next, let $\mathfrak{T}_{k_1,\ldots,k_n}^{*}$ be the set of $\pm1$-labelled locally orientable topological hypermaps generated by taking elements of $\mathfrak{T}_{k_1,\ldots,k_n}$, deleting their faces, labelling each of their edges with either $\pm1$, and then gluing new faces to the boundaries in a manner consistent with that shown in Figure~\ref{fig3.18}. Finally, let $\tilde{\mathfrak{T}}_{k_1,\ldots,k_n}^{*}=\mathfrak{T}_{k_1,\ldots,k_n}^{*}/\sim$, where $\sim$ denotes the equivalence relation described immediately above Figure \ref{fig3.19}. Then, we have that
\begin{multline}
c_{k_1,\ldots,k_n}^{(LOE)}=\sum_{l=0}^{k_1+\cdots+k_n+1-n}\left(\frac{N}{2}\right)^{k_1+\cdots+k_n}N^{2-n-l}
\\\times\sum_{p=1}^{k_1+\ldots+k_n+1-n}(\hat{a}+1)^p\,\#\{\Omega\in\tilde{\mathfrak{T}}_{k_1,\ldots,k_n}^{*}\,|\,\tilde{g}(\Omega)=l\textrm{ and }V_r(\Omega)=p\},
\end{multline}
where $\tilde{g}(\Omega)$ is the Euler genus of $\Omega$ and $V_r(\Omega)$ is the number of red vertices of $\Omega$.
\end{proposition}

As we move on to the next subsection, let us remark that the $\pm1$-labelled topological hypermaps of Proposition \ref{prop3.9} have combinatorial analogues that can be defined by extending Definition~\ref{def3.3} to allow for hyperedges. Without delving into the details, for the sake of brevity, let us simply state that the key ideas are to remove the third condition of Definition~\ref{def3.3} and to interpret the orbits of $\tau_0$, $\mean{\tau_0,\tau_1}$, and $\mean{\tau_0,\tau_2}$ as edges, red (white in the context of \S\ref{s3.3.1}) vertices or hyperedges, and black vertices, respectively.

\subsection{Moments of the Hermitised and antisymmetrised matrix products} \label{s3.3.3}
In this subsection, we amalgamate concepts from the previous two subsections in order to derive ribbon graph interpretations for the mixed moments \eqref{eq3.0.13} and cumulants \eqref{eq3.0.14} of the Hermitised and antisymmetrised matrix product ensembles introduced in Section \ref{s1.3}. We first give some basic results for general products of (possibly) rectangular matrices before highlighting some interesting simplifications in terms of topological hypermaps for the simplest non-trivial products (i.e., the Hermitised and antisymmetrised Laguerre ensembles corresponding to taking $m=1$ and $\nu_1=\nu_0=0$ in Definitions \ref{def1.10} and \ref{def1.11}).

\subsubsection{Ribbon graphs for the mixed moments of the Hermitised matrix product ensembles}
Following Definition \ref{def1.10}, let $m,N_0\in\mathbb{N}$, fix $N_1,\ldots,N_m\in\mathbb{N}$ such that for $1\leq i\leq m$, $N_i\geq N_0$, and write $N:=N_m$. Then, the $(N_0,\ldots,N_{m-1},N)$ Hermitised matrix product ensemble is represented by the $N\times N$ product
\begin{equation} \label{eq3.3.44}
\mathcal{H}_m=G_m^\dagger\cdots G_1^\dagger HG_1\cdots G_m,
\end{equation}
where $H$ is drawn from the $N_0\times N_0$ GUE and for $1\leq i\leq m$, each $G_i$ is drawn independently from the $N_{i-1}\times N_i$ complex Ginibre ensemble. Letting $\mean{\,\cdot\,}$ now, and for the remainder of this subsection, denote averages with respect to the j.p.d.f.
\begin{equation*}P^{(G)}(H)\Big|_{N\mapsto N_0}\,\prod_{i=1}^mP^{(Gin)}(G_i)\Big|_{(M,N)\mapsto(N_{i-1},N_i)},
\end{equation*}
we give a preliminary characterisation of the mixed moments \eqref{eq3.0.13} of the $(N_0,\ldots,N_{m-1},N)$ Hermitised matrix product ensemble.
\begin{lemma} \label{L3.4x}
Fix $m,n,k_1,\ldots,k_n\in\mathbb{N}$ and let $\mathcal{H}_m$ be as in equation \eqref{eq3.3.44}. For $1\leq s\leq n$ and $1\leq t\leq k_s-1$, set $i_{k_s}^{(s;-m-1)}:=i_1^{(s;m+1)}$ and $i_t^{(s;-m-1)}:= i_{t+1}^{(s;m+1)}$. Then, we have that
\begin{align}
m_{k_1,\ldots,k_n}^{(\mathcal{H}_m)}&=\left(\prod_{a=1}^n\prod_{b=1}^{k_a}\prod_{c=1}^{m+1}\sum_{i_b^{(a;\pm c)}=1}^{N_{|c-1|}}\right)\mean{\prod_{s=1}^n\prod_{t=1}^{k_s}H_{i_t^{(s;1)}i_t^{(s;-1)}}} \nonumber
\\&\hspace{12em}\times\prod_{u=1}^m\mean{\prod_{s=1}^n\prod_{t=1}^{k_s}(G^\dagger_u)_{i_t^{(s;u+1)}i_t^{(s;u)}}(G_u)_{i_t^{(s;-u)}i_t^{(s;-u-1)}}}. \label{eq3.3.45}
\end{align}
\end{lemma}
\begin{proof}
Taking equation \eqref{eq3.0.13} as the definition of $m_{k_1,\ldots,k_n}^{(\mathcal{H}_m)}$ and inserting the expansion
\begin{equation*}
\Tr\,\mathcal{H}_m^{k_s}=\sum_{i_1^{(s)},\ldots,i_{k_s}^{(s)}=1}^N(\mathcal{H}_m)_{i_1^{(s)}i_2^{(s)}}(\mathcal{H}_m)_{i_2^{(s)}i_3^{(s)}}\cdots(\mathcal{H}_m)_{i_{k_s}^{(s)}i_1^{(s)}}
\end{equation*}
shows that
\begin{equation*}
m_{k_1,\ldots,k_n}^{(\mathcal{H}_m)}=\mean{\prod_{s=1}^n\Tr\,\mathcal{H}_m^{k_s}}=\left(\prod_{a=1}^n\prod_{b=1}^{k_a}\sum_{i_b^{(a)}=1}^N\right)\mean{\prod_{s=1}^n\prod_{t=1}^{k_s}(\mathcal{H}_m)_{i_t^{(s)}i_{t+1}^{(s)}}},
\end{equation*}
where we have set $i_{k_s+1}^{(s)}:=i_1^{(s)}$ for all $1\leq s\leq n$ --- cf.~equation \eqref{eq3.3.17}. Making the replacement $(\mathcal{H}_m)_{i_t^{(s)}i_{t+1}^{(s)}}\mapsto (\mathcal{H}_m)_{i_t^{(s;m+1)}i_{t+1}^{(s;m+1)}}=(\mathcal{H}_m)_{i_t^{(s;m+1)}i_t^{(s;-m-1)}}$ and using the definition \eqref{eq3.3.44} of $\mathcal{H}_m$ to write
\begin{multline*}
(\mathcal{H}_m)_{i_t^{(s;m+1)}i_t^{(s;-m-1)}}=(G_m^{\dagger})_{i_t^{(s;m+1)}i_t^{(s;m)}}(G_{m-1}^{\dagger})_{i_t^{(s;m)}i_t^{(s;m-1)}}\cdots
\\\cdots(G_1^{\dagger})_{i_t^{(s;2)}i_t^{(s;1)}}H_{i_t^{(s;1)}i_t^{(s;-1)}}(G_1)_{i_t^{(s;-1)}i_t^{(s;-2)}}\cdots(G_m)_{i_t^{(s;-m)}i_t^{(s;-m-1)}}
\end{multline*}
then shows that
\begin{align*}
m_{k_1,\ldots,k_n}^{(\mathcal{H}_m)}&=\left(\prod_{a=1}^n\prod_{b=1}^{k_a}\prod_{c=1}^{m+1}\sum_{i_b^{(a;\pm c)}=1}^{N_{|c-1|}}\right)\left\langle\prod_{s=1}^n\prod_{t=1}^{k_s}(G_m^{\dagger})_{i_t^{(s;m+1)}i_t^{(s;m)}}(G_{m-1}^{\dagger})_{i_t^{(s;m)}i_t^{(s;m-1)}}\cdots\right.
\\&\hspace{11em}\left.\cdots(G_1^{\dagger})_{i_t^{(s;2)}i_t^{(s;1)}}H_{i_t^{(s;1)}i_t^{(s;-1)}}(G_1)_{i_t^{(s;-1)}i_t^{(s;-2)}}\cdots(G_m)_{i_t^{(s;-m)}i_t^{(s;-m-1)}}\right\rangle .
\end{align*}
As the random matrices $H,G_1,\ldots,G_m$ within equation \eqref{eq3.3.44} are assumed to be (mutually) independent, the average in the above can be split into the product of averages shown on the right-hand side of equation \eqref{eq3.3.45}, thereby proving the latter formula.
\end{proof}

As shown in \S\ref{s3.3.1} and \S\ref{s3.3.2}, each average on the right-hand side of equation \eqref{eq3.3.45} can be simplified using the Isserlis--Wick theorem. In particular, the average
\begin{equation*}
\mean{\prod_{s=1}^n\prod_{t=1}^{k_s}H_{i_t^{(s;1)}i_t^{(s;-1)}}}
\end{equation*}
is identically zero if it contains an odd number of entries, so $m_{k_1,\ldots,k_n}^{(\mathcal{H}_m)}$ vanishes if $k_1+\cdots+k_n$ is an odd integer. On the other hand, when $k_1+\cdots+k_n$ is an even integer, we are led to the following ribbon graph construction: For each $s=1,\ldots,n$, draw a $k_s(2m+1)$-gon and label its vertices in clockwise order in such a way that the first $2m+1$ vertices are labelled $i_1^{(s;m+1)},i_1^{(s;m)},\ldots,i_1^{(s;1)},i_1^{(s;-1)},i_1^{(s;-2)},\ldots,i_1^{(s;-m)}$, the next $2m+1$ vertices are labelled $i_2^{(s;m+1)},\ldots,i_2^{(s;1)},i_2^{(s;-1)},\ldots,i_2^{(s;-m)}$, and so on, with the last $2m+1$ vertices being labelled $i_{k_s}^{(s;m+1)},\ldots,i_{k_s}^{(s;1)},i_{k_s}^{(s;-1)},\ldots,i_{k_s}^{(s;-m)}$ --- for $1\leq t\leq k_s$ and $1\leq u\leq m$, edges of the form $(i_t^{(s;1)}\to i_t^{(s;-1)})$ represent $H_{i_t^{(s;1)}i_t^{(s;-1)}}$, while $(i_t^{(s;u+1)}\to i_t^{(s;u)})$ and $(i_t^{(s;-u)}\to i_t^{(s;-u-1)})$ represent $(G^\dagger_u)_{i_t^{(s;u+1)}i_t^{(s;u)}}$ and $(G_u)_{i_t^{(s;-u)}i_t^{(s;-u-1)}}$, respectively.

Applying the Isserlis--Wick theorem to the averages on the right-hand side of equation \eqref{eq3.3.45} results in a product of covariances that are exactly of the type seen in the GUE and LUE cases and can thus be represented by untwisted ribbons connecting polygon edges representing entries from a matrix and its adjoint; that is, for $1\leq u\leq m$, if one end of a ribbon is connected to an edge representing an entry of $G^\dagger_u$ ($H$), then its other end must be connected to an edge representing an entry of $G_u$ ($H$). This latter constraint can be satisfied by assigning, for each $u=1,\ldots,m+1$, a unique colour $\phi_u$ to all polygon vertices labelled by indices of the form $i_t^{(s;\pm u)}$, with $1\leq s\leq n$ and $1\leq t\leq k_s$, and then only allowing ribbon graphs whose boundaries have well-defined colours inherited from the vertices they pass through; see Figure \ref{fig3.20} below.

\begin{figure}[H]
        \centering
\captionsetup{width=.9\linewidth}
        \includegraphics[width=0.8\textwidth]{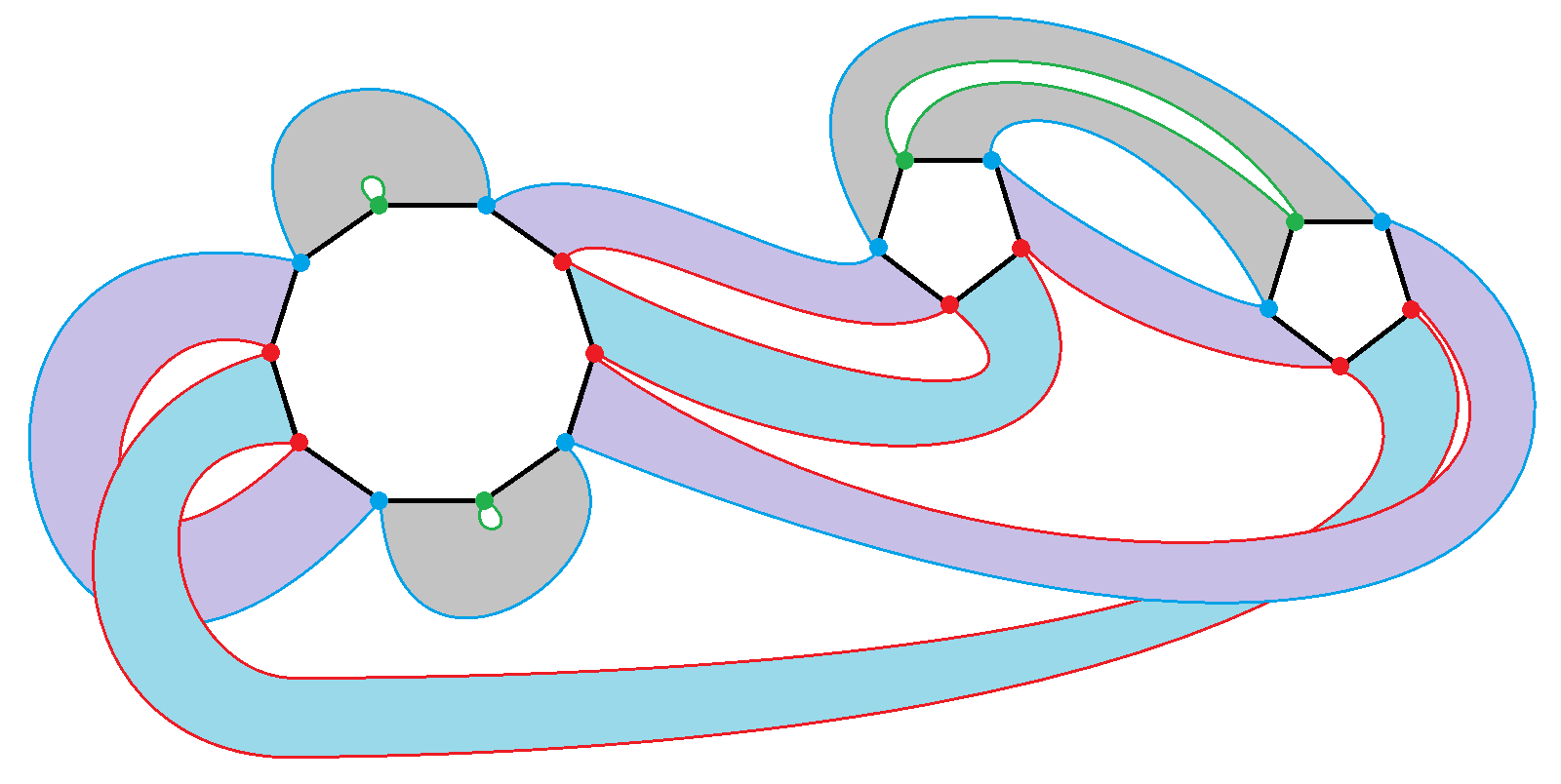}
        \caption[A $3$-coloured ribbon graph contributing to the computation of $m_{2,1,1}^{(\mathcal{H}_2)}$]{In computing $m_{2,1,1}^{(\mathcal{H}_2)}=\langle\Tr(\mathcal{H}_2^2)\Tr(\mathcal{H}_2)^2\rangle$, one first draws a decagon, representing $\Tr(\mathcal{H}_2^2)$, and two pentagons, each representing a factor of $\Tr(\mathcal{H}_2)$, with their vertices coloured as shown (the colours $\phi_1,\phi_2,\phi_3$ are respectively red, blue, and green) --- we do not display the vertex labellings, but mention that the top green vertex of each polygon, moving left to right, is respectively labelled $i_1^{(1;3)}, i_1^{(2;3)}, i_1^{(3;3)}$, while the other vertex labels follow from the prescription given above. In the next step, one simply lists out all possible ways of pairwise identifying the polygon edges using untwisted ribbons whose sides inherit well-defined colours from the vertices; the displayed ribbon graph illustrates one such pairing. For visual clarity, and without any mathematical meaning, we have assigned the colours light blue, purple, and grey to the ribbons with respectively two red sides, a red and blue side, and a blue and green side. These ribbons represent pairings of the form $\langle H_{ij}H_{ji}\rangle$, $\langle (G_1^\dagger)_{ij}(G_1)_{ji}\rangle$, and $\langle (G_2^\dagger)_{ij}(G_2)_{ji}\rangle$ (for suitable $i,j$), respectively.} \label{fig3.20}
\end{figure}

Extending arguments given in the previous two subsections (in particular, the ideas underlying Examples~\ref{ex3.2} and~\ref{ex3.4}) leads to an elegant formula for the mixed moments \eqref{eq3.0.13} and cumulants \eqref{eq3.0.14} of the $(N_0,\ldots,N_{m-1},N)$ Hermitised matrix product ensemble in terms of ribbon graphs, in analogy with equation \eqref{eq3.3.34}.
\begin{lemma} \label{L3.5}
Fix $m,n,k_1,\ldots,k_n\in\mathbb{N}$ and let $\mathcal{H}_m$ be as in equation \eqref{eq3.3.44}. If $k_1+\cdots+k_n$ is odd, we have that
\begin{equation*}
m_{k_1,\ldots,k_n}^{(\mathcal{H}_m)}=c_{k_1,\ldots,k_n}^{(\mathcal{H}_m)}=0.
\end{equation*}
Thus, constraining ourselves to the case that $k_1+\cdots+k_n$ is even, define $\mathfrak{P}^{(m)}_{k_1,\ldots,k_n}$ to be the set of $(m+1)$-coloured orientable $\tfrac{1}{2}(2m+1)(k_1+\cdots+k_n)$-ribbon graphs that can be constructed through the procedure described above Figure \ref{fig3.20} and let $\mathfrak{P}^{(m),c}_{k_1,\ldots,k_n}\subseteq\mathfrak{P}^{(m)}_{k_1,\ldots,k_n}$ be the subset of connected such ribbon graphs. Then,
\begin{align}
m_{k_1,\ldots,k_n}^{(\mathcal{H}_m)}&=2^{-\tfrac{1}{2}(k_1+\cdots+k_n)}\sum_{\Gamma\in\mathfrak{P}^{(m)}_{k_1,\ldots,k_n}}N_0^{V_0(\Gamma)}N_1^{V_1(\Gamma)}\cdots N_m^{V_m(\Gamma)}, \label{eq3.3.46}
\\ c_{k_1,\ldots,k_n}^{(\mathcal{H}_m)}&=2^{-\tfrac{1}{2}(k_1+\cdots+k_n)}\sum_{\Gamma\in\mathfrak{P}^{(m),c}_{k_1,\ldots,k_n}}N_0^{V_0(\Gamma)}N_1^{V_1(\Gamma)}\cdots N_m^{V_m(\Gamma)}, \label{eq3.3.47x}
\end{align}
where, for $0\leq u\leq m$, $V_u(\Gamma)$ is the number of boundaries of $\Gamma$ that are of the colour $\phi_{u+1}$.
\end{lemma}
\begin{proof}
As mentioned above, applying the Isserlis--Wick theorem to the averages in the right-hand side of equation \eqref{eq3.3.45} shows that
\begin{equation*}
\mean{\prod_{s=1}^n\prod_{t=1}^{k_s}H_{i_t^{(s;1)}i_t^{(s;-1)}}}
\end{equation*}
vanishes if $k_1+\cdots+k_n$ is odd and is otherwise given by a sum over $\mathfrak{P}_{k_1,\ldots,k_n}$ (defined above equation \eqref{eq3.3.18}) of a product of indicator functions identifying the summation indices $i_b^{(a;\pm1)}$. Likewise, the averages
\begin{equation*}
\mean{\prod_{s=1}^n\prod_{t=1}^{k_s}(G^\dagger_u)_{i_t^{(s;u+1)}i_t^{(s;u)}}(G_u)_{i_t^{(s;-u)}i_t^{(s;-u-1)}}}
\end{equation*}
are given by a similar sum, but over $\mathfrak{S}_{k_1,\ldots,k_n}$ (recall Proposition \ref{prop3.13x}). Thus, $m_{k_1,\ldots,k_n}^{(\mathcal{H}_m)}=0$ when $k_1+\cdots+k_n$ is odd and, otherwise, the canonical bijection $\mathfrak{P}^{(m)}_{k_1,\ldots,k_n}\simeq\mathfrak{P}_{k_1,\ldots,k_n}\times(\mathfrak{S}_{k_1,\ldots,k_n})^m$ allows us to write the product of averages in the right-hand side of equation \eqref{eq3.3.45} as a sum over $\mathfrak{P}^{(m)}_{k_1,\ldots,k_n}$ of a product of indicator functions.

Interchanging the order of summation and summing over all indices $i_b^{(a;c)}$ then produces equation \eqref{eq3.3.46} --- the product of indicator functions in the summand is such that the number of distinct indices of the form $i_b^{(a;\pm c)}$, with $1\leq a\leq n$ and $1\leq b\leq k_a$ free, but $1\leq c\leq m+1$ fixed, is equal to the number of boundaries of colour $\phi_c$ in the associated ribbon graph.

Finally, we know from Remark \ref{R3.4x} that $c_{k_1,\ldots,k_n}^{(\mathcal{H}_m)}=0$ when $k_1+\cdots+k_n$ is odd, while for $k_1+\cdots+k_n$ even, the discussion above said remark implies equation \eqref{eq3.3.47x}.
\end{proof}

\begin{example}
The ribbon graph of Figure \ref{fig3.20} has two red ($\phi_1$) boundaries, two blue ($\phi_2$) boundaries, and three green ($\phi_3$) boundaries. Thus, by the above lemma, it contributes a value of $N_0^2N_1^2N_2^3$ to both $m_{2,1,1}^{(\mathcal{H}_2)}$ and $c_{2,1,1}^{(\mathcal{H}_2)}$.
\end{example}

In the context of Chapter 4, our interest lies in the mixed cumulants, as opposed to the mixed moments, since the former have (under appropriate conditions) genus expansions with convergence properties that allow their generating functions (after correct scaling) to have large $N_0$ expansions of the form \eqref{eq1.1.21}. Thus, we now give the analogue of the genus expansion \eqref{eq3.3.40} pertaining to the global scaled Hermitised matrix product ensembles, as is relevant to the development of Section \ref{s4.3}.

\begin{proposition} \label{prop3.10}
Let the parameters $m,N_0,\ldots,N_m=N\in\mathbb{N}$ and matrices $H,G_1,\ldots,G_m$ be as in Definition \ref{def1.10}, equivalently equation \eqref{eq3.3.44}, and let us require that the parameters $\nu_i:=N_i-N_0$ be such that for each $i=1,\ldots,m$, there exists a non-negative constant $\hat{\nu}_i:=\nu_i/N_0$ (cf.~Remark~\ref{R1.11}). Furthermore, introduce the scaled GUE matrix $\tilde{H}:=(N_0/2)^{-1/2}H$ and, for each $i=1,\ldots,m$, the scaled complex Ginibre matrices $\tilde{G}_i:=(N_iN_{i-1})^{-1/4}G_i$ (following \citep{BJLNS10}) so that
\begin{equation} \label{eq3.3.47}
\tilde{\mathcal{H}}_m:=\tilde{G}_m^{\dagger}\cdots\tilde{G}_1^{\dagger}\tilde{H}\tilde{G}_1\cdots\tilde{G}_m=\frac{N_0^{-m-1/2}}{(\hat{\nu}_1+1)\cdots(\hat{\nu}_{m-1}+1)}\sqrt{\frac{2}{\hat{\nu}_m+1}}\mathcal{H}_m
\end{equation}
represents the \textbf{global scaled $(N_0,\ldots,N_{m-1},N)$ Hermitised matrix product ensemble} (refining Definition \ref{def1.13}). In this regime, and for $n,k_1,\ldots,k_n\in\mathbb{N}$ such that $k_1+\cdots+k_n$ is even, the mixed cumulants of this ensemble have the genus expansion
\begin{multline} \label{eq3.3.48}
\tilde{c}_{k_1,\ldots,k_n}^{(\mathcal{H}_m)}=\left((\hat{\nu}_1+1)\cdots(\hat{\nu}_{m-1}+1)\sqrt{\hat{\nu}_m+1}\right)^{-k_1-\cdots-k_n}N_0^{2-n}
\\ \times\sum_{l=0}^{\tfrac{1}{2}(2m+1)(k_1+\cdots+k_n)+1-n}\frac{1}{N_0^l}\sum_{p_1,\ldots,p_m=1}^{k_1+\cdots+k_n}(\hat{\nu}_1+1)^{p_1}\cdots(\hat{\nu}_m+1)^{p_m}
\\ \times\#\{\Gamma\in\mathfrak{P}^{(m),c}_{k_1,\ldots,k_n}\,|\,g(\Gamma)=l/2\textrm{ and }V_i(\Gamma)=p_i\textrm{ for each }1\leq i\leq m\},
\end{multline}
where the set $\mathfrak{P}^{(m),c}_{k_1,\ldots,k_n}$ and the functions $V_1(\Gamma),\ldots,V_m(\Gamma)$ are as in Lemma \ref{L3.5} above, and we recall that $g(\Gamma)$ denotes the genus of the ribbon graph $\Gamma$, while $\#\mathcal{S}$ denotes the cardinality of $\mathcal{S}$.
\end{proposition}
\begin{proof}
Substituting $N_i=(\hat{\nu}_i+1)N_0$ ($1\leq i\leq m$) into equation \eqref{eq3.3.47x} shows that
\begin{multline} \label{eq3.3.49}
c_{k_1,\ldots,k_n}^{(\mathcal{H}_m)}=2^{-\tfrac{1}{2}(k_1+\cdots+k_n)}\sum_{V=1}^{\tfrac{1}{2}(2m+1)(k_1+\cdots+k_n)+2-n}N_0^V\sum_{p_1,\ldots,p_m=1}^{k_1+\ldots+k_n}(\hat{\nu}_1+1)^{p_1}\cdots(\hat{\nu}_m+1)^{p_m}
\\ \times\#\{\Gamma\in\mathfrak{P}^{(m),c}_{k_1,\ldots,k_n}\,|\,\sum_{i=0}^mV_i(\Gamma)=V\textrm{ and }V_i(\Gamma)=p_i\textrm{ for each }1\leq i\leq m\},
\end{multline}
where the upper terminal of the sum over $V$, the total number of boundaries of the ribbon graphs $\Gamma$, is obtained through a similar argument to that given in the paragraph containing equation \eqref{eq3.3.19}: In the present setting, the polygonised surface obtained by identifying polygon edges of a given $\Gamma\in\mathfrak{P}^{(m),c}_{k_1,\ldots,k_n}$ by collapsing the ribbons therein to their ends has $F=n$ faces (number of polygons), $E=\tfrac{1}{2}(2m+1)(k_1+\ldots+k_n)$ distinct edges (half the number of polygon edges that are now pairwise identified), $V$ distinct vertices, and non-negative integer genus $g$, so that consideration of the Euler characteristic $\chi=F-E+V=2-2g$ reveals that
\begin{equation*}
V=\tfrac{1}{2}(2m+1)(k_1+\cdots+k_n)+2-2g-n\leq\tfrac{1}{2}(2m+1)(k_1+\cdots+k_n)+2-n.
\end{equation*}
Using this equation, with $l:=2g$, to change variables in equation \eqref{eq3.3.49} then yields the result \eqref{eq3.3.48}, bar a factor of $[(\hat{\nu}_1+1)\cdots(\hat{\nu}_{m-1}+1)\sqrt{(\hat{\nu}_m+1)/2}N_0^{m+1/2}]^{-k_1-\cdots-k_n}$. This factor, which is the $(k_1+\cdots+k_n)\textsuperscript{th}$ power of the scaling factor in equation \eqref{eq3.3.47}, appears as the ratio of scaled and unscaled mixed moments
\begin{align*}
\frac{\tilde{m}_{k_1,\ldots,k_n}^{(\mathcal{H}_m)}}{m_{k_1,\ldots,k_n}^{(\mathcal{H}_m)}}&=\frac{\big\langle \prod_{s=1}^n\Tr\,\tilde{\mathcal{H}}_m^{k_s}\big\rangle}{\big\langle \prod_{s=1}^n\Tr\,\mathcal{H}_m^{k_s}\big\rangle}
\\ \implies m_{k_1,\ldots,k_n}^{(\mathcal{H}_m)}&=\left((\hat{\nu}_1+1)\cdots(\hat{\nu}_{m-1}+1)\sqrt{(\hat{\nu}_m+1)/2}N_0^{m+1/2}\right)^{k_1+\cdots+k_n}\tilde{m}_{k_1,\ldots,k_n}^{(\mathcal{H}_m)};
\end{align*}
inserting this expression into equation \eqref{eq3.3.20x} finally shows that said factor is also equal to the ratio $\tilde{c}_{k_1,\ldots,k_n}^{(\mathcal{H}_m)}/c_{k_1,\ldots,k_n}^{(\mathcal{H}_m)}$, as required.
\end{proof}

Before moving forward, let us make two remarks regarding Proposition~\ref{prop3.10}. The first remark is that, since $\mathfrak{P}_{k_1,\ldots,k_n}^{(m),c}$ consists of orientable ribbon graphs whose genera are non-negative integers, $N_0^{n-2}\tilde{c}_{k_1,\ldots,k_n}^{(\mathcal{H}_m)}$ is an even polynomial in $N_0^{-1}$. Thus, upon setting $\nu_i=\hat{\nu}_iN_0$ with $\hat{\nu}_i={\rm O}(1)$ ($1\leq i\leq m$) and writing the generating functions $\tilde{W}_n^{(\mathcal{H}_m)}$ of the scaled cumulants $\tilde{c}_{k_1,\ldots,k_n}^{(\mathcal{H}_m)}$ in the form given in Theorem \ref{thrm1.1}, one sees that the related expansion coefficients $W_n^{(\mathcal{H}_m),l}$ are identically zero for odd values of $l$, in line with what is known in the GUE and LUE cases. Our second remark relates to the choice of setting $\nu_i=\hat{\nu}_iN_0$ as just discussed, rather than $\nu_i=\hat{\nu}_iN_0+\delta_i$ with $\delta_i={\rm O}(1)$, as in Remark~\ref{R1.11}. This choice, as with the choice of setting $a=\hat{a}N$ with $\hat{a}={\rm O}(1)$ in \S\ref{s3.3.2}, is made so that equations \eqref{eq3.3.40}, \eqref{eq3.3.42}, \eqref{eq3.3.48}, and \eqref{eq3.3.58} upcoming convey their combinatorial content in the most elegant way possible. Nonetheless, the relevant proofs follow through in the more general setting where not all $\delta_i$ are identically zero, albeit the expressions \eqref{eq3.3.40}, \eqref{eq3.3.42}, \eqref{eq3.3.48}, and \eqref{eq3.3.58} cease to hold true past leading order.

\subsubsection{Ribbon graphs for the mixed moments of the antisymmetrised matrix product ensembles}
Let us now take $N_0,\ldots,N_m$ to be positive even integers with $N_1,\ldots,N_m\geq N_0$ and otherwise retain our specifications of the parameters $m,N=N_m\in\mathbb{N}$ and notation $\mean{\,\cdot\,}$. Moreover, for each $i=1,\ldots,m$, let $G_i$ now be drawn independently from the $N_{i-1}\times N_i$ real Ginibre ensemble and recall that $J_{N_0}$ denotes the $N_0\times N_0$ elementary antisymmetric matrix defined in equation \eqref{eq1.1.5} --- for present purposes, it suffices to observe that $J_{N_0}^T=-J_{N_0}$ and $J_{N_0}^2=-J_{N_0}^TJ_{N_0}=-I_{N_0}$. According to Definition \ref{def1.11}, the $N\times N$ product
\begin{equation} \label{eq3.3.51}
\mathcal{J}_m=G_m^T\cdots G_1^TJ_{N_0}G_1\cdots G_m
\end{equation}
represents the $(N_0,\ldots,N_{m-1},N)$ antisymmetrised matrix product ensemble. We begin with the analogue of Lemma \ref{L3.4x} for the mixed moments of this ensemble.
\begin{lemma} \label{L3.6x}
Fix $m,n,k_1,\ldots,k_n\in\mathbb{N}$ and let $\mathcal{J}_m$ be as in equation \eqref{eq3.3.51}. The mixed moment $m_{k_1,\ldots,k_n}^{(\mathcal{J}_m)}$ vanishes if any of the $k_1,\ldots,k_n$ are odd. Otherwise, we have that
\begin{align}
m_{k_1,\ldots,k_n}^{(\mathcal{J}_m)}&=\left(\prod_{a=1}^n\prod_{b=1}^{k_a}\prod_{c=1}^{m+1}\sum_{i_b^{(a;\pm c)}=1}^{N_{|c-1|}}\right)\left(\prod_{s=1}^n\prod_{t=1}^{k_s}(J_{N_0})_{i_t^{(s;1)}i_t^{(s;-1)}}\right) \nonumber
\\&\hspace{12em}\times\prod_{u=1}^m\mean{\prod_{s=1}^n\prod_{t=1}^{k_s}(G^T_u)_{i_t^{(s;u+1)}i_t^{(s;u)}}(G_u)_{i_t^{(s;-u)}i_t^{(s;-u-1)}}}, \label{eq3.3.52}
\end{align}
where we have retained the identifications (introduced in Lemma \ref{L3.4x}) $i_{k_s}^{(s;-m-1)}:=i_1^{(s;m+1)}$ and $i_t^{(s;-m-1)}:= i_{t+1}^{(s;m+1)}$ for $1\leq s\leq n$ and $1\leq t\leq k_s-1$.
\end{lemma}
\begin{proof}
If $k_s$ is odd for some $1\leq s\leq n$, then the antisymmetric nature of $\mathcal{J}_m$ means that $\Tr\,\mathcal{J}_m^{k_s}$ vanishes and, consequently, so too does $m_{k_1,\ldots,k_n}^{(\mathcal{J}_m)}$, by equation \eqref{eq3.0.13}. When all of the $k_1,\ldots,k_n$ are even, the proof is as for Lemma \ref{L3.4x}.
\end{proof}

Replicating the construction preceding Figure \ref{fig3.20} with equation \eqref{eq3.3.41} in mind shows that we are interested in locally orientable ribbon graphs that are built in a similar fashion to those in the Hermitised case: The computation of $m_{k_1,\ldots,k_n}^{(\mathcal{J}_m)}$ requires one to build ribbon graphs from $n$ polygons with respectively $k_1(2m+1),\ldots,k_n(2m+1)$ sides and with vertices labelled and coloured in the exact same manner as in the case of $m_{k_1,\ldots,k_n}^{(\mathcal{H}_m)}$, but the ribbons are now allowed to be M\"obius half-twisted and there are no longer any ribbons connected to the polygon edges of the form $(i_t^{(s;1)}\to i_t^{(s;-1)})$, which now represent $(J_{N_0})_{i_t^{(s;1)}i_t^{(s;-1)}}$ instead of $H_{i_t^{(s;1)}i_t^{(s;-1)}}$. Of course, we still require that the sides of the ribbons connect vertices of like colours. Since the matrices $G_1,\ldots,G_m$ are independent from each other, the collection of locally orientable ribbon graphs just described, which we denote $\tilde{\mathfrak{S}}_{k_1,\ldots,k_n}^{(m)}$, is in bijection with $(\tilde{\mathfrak{S}}_{k_1,\ldots,k_n})^m$, where $\tilde{\mathfrak{S}}_{k_1,\ldots,k_n}$ is the extension of $\tilde{\mathfrak{S}}_{k_1,\ldots,k_n}^c$ (see Proposition \ref{prop3.14x}) allowing for disconnected ribbon graphs. In other words, the ribbon graphs of $\tilde{\mathfrak{S}}_{k_1,\ldots,k_n}^{(m)}$ can be constructed by appropriately combining $m$ distinct LOE ribbon graphs drawn from $\tilde{\mathfrak{S}}_{k_1,\ldots,k_n}$. In that vein, it is helpful to compare the ribbon graphs of $\tilde{\mathfrak{S}}_{k_1,\ldots,k_n}^{(m)}$ to those obtained by deleting the edges representing entries of $J_{N_0}$ while identifying, for $1\leq s\leq n$ and $1\leq t\leq k_s$, the vertices labelled $i_t^{(s;1)}$ to those labelled $i_t^{(s;-1)}$ (i.e., replacing each instance of $(J_{N_0})_{i_t^{(s;1)}i_t^{(s;-1)}}$ in equation \eqref{eq3.3.52} with $\chi_{i_t^{(s;1)}=i_t^{(s;-1)}}$, so that this equation now computes the mixed moments $m_{k_1,\ldots,k_n}^{(r\mathcal{W}_m)}$ of the real Wishart product matrix $\mathcal{W}_m=G_m^T\cdots G_1^TG_1\cdots G_m$; cf.~Definition \ref{def1.12}).
\begin{figure}[H]
        \centering
\captionsetup{width=.9\linewidth}
        \includegraphics[width=0.88\textwidth]{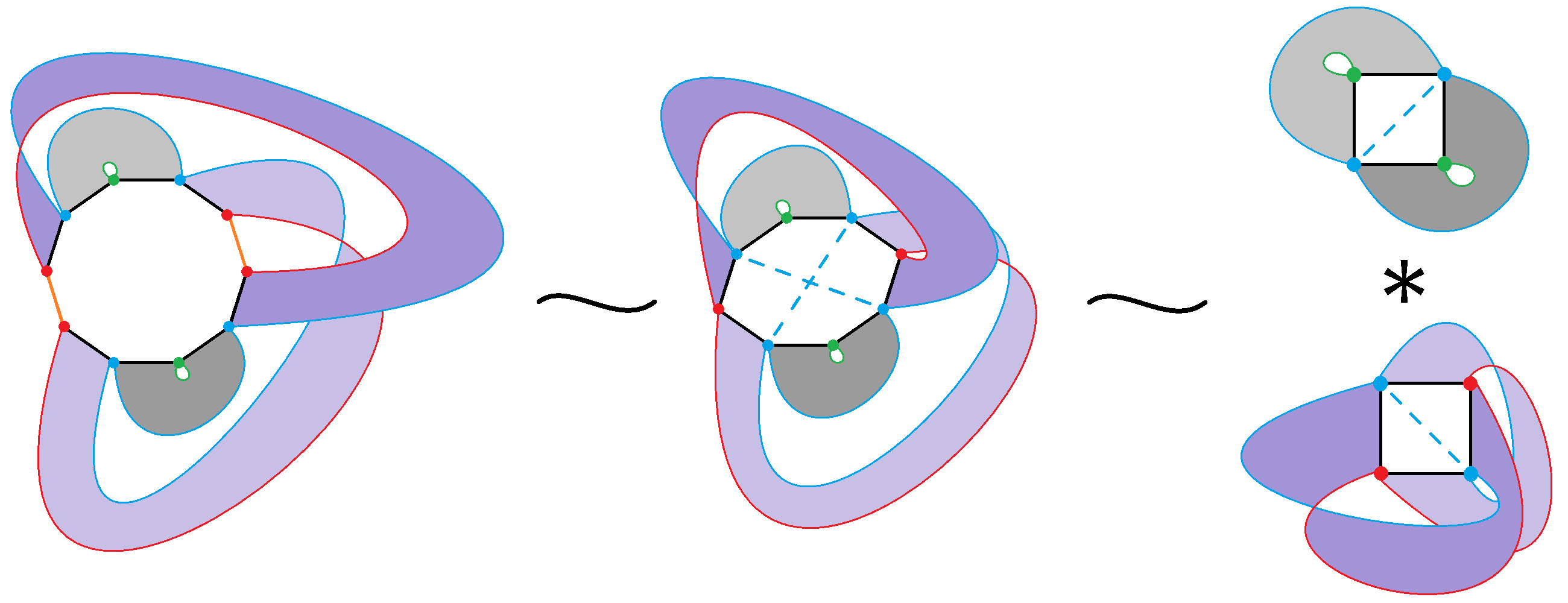}
        \caption[A $3$-coloured ribbon graph contributing to the computation of $m_2^{(\mathcal{J}_2)}$]{The ribbon graph on the left contributes to the computation of $m_2^{(\mathcal{J}_2)}$ and belongs to $\tilde{\mathfrak{S}}_2^{(2)}$. Deleting the orange edges representing entries of $J_{N_0}$ and identifying their ends results in the middle ribbon graph, which we think of as belonging to $(\tilde{\mathfrak{S}}_2)^2$. This latter ribbon graph is a combination of the two ribbon graphs on the right (cut and glue along the dotted blue lines), each of which are drawn from $\tilde{\mathfrak{S}}_2$.} \label{fig3.21}
\end{figure}

Applying the Isserlis--Wick theorem to equation \eqref{eq3.3.52} in the same fashion as in the proof of Lemma \ref{L3.5} produces the following analogue of equation \eqref{eq3.3.46}.
\begin{lemma} \label{L3.7}
Fixing $m,n\in\mathbb{N}$, let $\mathcal{J}_m$ be as in equation \eqref{eq3.3.51}, and take $k_1,\ldots,k_n\in2\mathbb{N}$. Let $\tilde{\mathfrak{S}}^{(m)}_{k_1,\ldots,k_n}$ be the set of $(m+1)$-coloured locally orientable $m(k_1+\cdots+k_n)$-ribbon graphs that can be constructed in the manner described above Figure \ref{fig3.21} and define $f_{\Gamma}(\{i_t^{(s;\pm1)}\}_{1\leq s\leq n,\,1\leq t\leq k_s})$ to be the product of indicator functions that encodes the identifications of the vertices of colour $\phi_1$ that are represented by the ribbon data of $\Gamma$. Then, we have that
\begin{multline} \label{eq3.3.53}
m_{k_1,\ldots,k_n}^{(\mathcal{J}_m)}=2^{-m(k_1+\cdots+k_n)}\sum_{\Gamma\in\tilde{\mathfrak{S}}_{k_1,\ldots,k_n}^{(m)}}N_1^{V_1(\Gamma)}\cdots N_m^{V_m(\Gamma)}\left(\prod_{a=1}^n\prod_{b=1}^{k_a}\sum_{i_b^{(a;\pm1)}=1}^{N_0}\right)\left(\prod_{s=1}^n\prod_{t=1}^{k_s}(J_{N_0})_{i_t^{(s;1)}i_t^{(s;-1)}}\right)
\\ \times f_{\Gamma}\left(\{i_t^{(s;\pm1)}\}_{1\leq s\leq n,\,1\leq t\leq k_s}\right),
\end{multline}
where $V_1(\Gamma),\ldots,V_m(\Gamma)$ are as in Lemma \ref{L3.5}.
\end{lemma}
\begin{example}
Letting $\Gamma$ be the left illustration of Figure~\ref{fig3.21} and following the red sides of the purple ribbons therein shows that
\begin{equation} \label{eq3.3.54}
f_{\Gamma}\left(i_1^{(1;1)},i_1^{(1;-1)},i_2^{(1;1)},i_2^{(1;-1)}\right)=\chi_{i_1^{(1;1)}=i_2^{(1;1)}}\chi_{i_1^{(1;-1)}=i_2^{(1;-1)}}.
\end{equation}
\end{example}

The product $f_{\Gamma}(\{i_t^{(s;\pm1)}\}_{1\leq s\leq n,\,1\leq t\leq k_s})$ is such that if one replaces $(J_{N_0})_{i_t^{(s;1)}i_t^{(s;-1)}}$ by $\chi_{i_t^{(s;1)}=i_t^{(s;-1)}}$ in equation \eqref{eq3.3.53}, as discussed above Figure \ref{fig3.21}, said equation reduces to an expression for the mixed moments of the corresponding real Wishart product ensemble:
\begin{equation} \label{eq3.3.55}
m_{k_1,\ldots,k_n}^{(r\mathcal{W}_m)}=2^{-m(k_1+\cdots+k_n)}\sum_{\Gamma\in\tilde{\mathfrak{S}}_{k_1,\ldots,k_n}^{(m)}}N_0^{V_0(\Gamma)}\cdots N_m^{V_m(\Gamma)},
\end{equation}
where $V_0(\Gamma)$ is the number of boundaries of the ribbon graph $\Gamma$ of colour $\phi_1$ after all of the edges representing entries of $J_{N_0}$ have been given this colour; equivalently, it is the number of such boundaries of the ribbon graph obtained by collapsing said edges of $\Gamma$ to their ends (cf.~the middle illustration of Figure~\ref{fig3.21}, which has $\phi_1$ being red and thus $V_0(\Gamma)=1$). Hence, one can think of the contribution of each ribbon graph $\Gamma\in\tilde{\mathfrak{S}}_{k_1,\ldots,k_n}^{(m)}$ to the value of $m_{k_1,\ldots,k_n}^{(\mathcal{J}_m)}$ as being equal to a perturbation of the value $2^{-m(k_1+\cdots+k_n)}N_0^{V_0(\Gamma)}\cdots N_m^{V_m(\Gamma)}$ contributed by $\Gamma$ to $m_{k_1,\ldots,k_n}^{(r\mathcal{W}_m)}$ via equation \eqref{eq3.3.55}. It turns out that the correct way to perturb this value is to replace the factor $N_0^{V_0(\Gamma)}$ by an appropriate product of traces of monomials in $J_{N_0}$ and $J_{N_0}^T$. To see this, first recall that in the real Wishart product case, the factor $N_0^{V_0(\Gamma)}$ arises from assigning a weight $N_0$ to each distinct boundary, equivalently polygon vertex, of colour $\phi_1$ in $\Gamma$ and that traversing these boundaries tells us how to identify vertex labels so as to arrive at the set of distinct labels. Upon reintroducing the polygon edges representing entries of $J_{N_0}$, traversing these same boundaries instead tells us how to identify indices of these matrix entries. Thus, to obtain the weight assigned to the `first' boundary, i.e., that containing the vertex labelled $i_1^{(1;1)}$, let $\omega_1=(J_{N_0})_{i_1^{(1;1)}i_1^{(1;-1)}}$ and then follow the ribbon border of colour $\phi_1$ connected to the vertex labelled $i_1^{(1;-1)}$ to see which vertex it is identified to. If this latter vertex has label of the form $i_t^{(s;1)}$, identify this label with $i_1^{(1;-1)}$ and rewrite $\omega_1$ as the ordered string $(J_{N_0})_{i_1^{(1;1)}i_1^{(1;-1)}}(J_{N_0})_{i_1^{(1;-1)}i_t^{(s;-1)}}$. On the other hand, if said vertex has label $i_t^{(s;-1)}$, rewrite $\omega_1$ as $(J_{N_0})_{i_1^{(1;1)}i_1^{(1;-1)}}(J_{N_0}^T)_{i_1^{(1;-1)}i_t^{(s;1)}}$. Next, follow the ribbon border connected to the vertex labelled $i_t^{(s;\pm1)}$ and likewise extend $\omega_1$ by concatenating it with a term of the form $(J_{N_0})_{i_t^{(s;\pm1)}i_{t'}^{(s';-1)}}$ or $(J_{N_0}^T)_{i_t^{(s;\pm1)}i_{t'}^{(s';1)}}$, depending on if the ribbon data identifies $i_t^{(s;\pm1)}$ with $i_{t'}^{(s';1)}$ or $i_{t'}^{(s';-1)}$, respectively. Continue on in this manner until the boundary has been fully traversed (that is, the first index of the first matrix entry of $\omega_1$ matches the second index of the last matrix entry of $\omega_1$) and then repeat this exercise for all of the remaining boundaries of colour $\phi_1$ so that we end up with a list of words $\omega_1,\omega_2,\ldots,\omega_{V_0(\Gamma)}$. With this list of words in hand, we may now give a refinement of Lemma \ref{L3.7}.
\begin{lemma} \label{L3.8x}
For each word $\omega_j$ obtained via the procedure described above, let $|\omega_j|$ denote the length of $\omega_j$ and let $\mathrm{sgn}(\omega_j):=(-1)^{\# J^T}$, where $\# J^T$ is the number of entries of $J_{N_0}^T$ in $\omega_j$. In the setting of Lemma \ref{L3.7}, we then have that
\begin{equation} \label{eq3.3.56}
m_{k_1,\ldots,k_n}^{(\mathcal{J}_m)}=2^{-m(k_1+\cdots+k_n)}\sum_{\Gamma\in\tilde{\mathfrak{S}}_{k_1,\ldots,k_n}^{(m)}}N_0^{V_0(\Gamma)}\cdots N_m^{V_m(\Gamma)}\prod_{j=1}^{V_0(\Gamma)}\mathrm{sgn}(\omega_j)\,\mathrm{Re}\left(\mathrm{i}^{|\omega_j|}\right).
\end{equation}
\end{lemma}
\begin{proof}
First, observe that suitably ordering the terms in the summand of equation \eqref{eq3.3.53} and rewriting some factors $(J_{N_0})_{i_t^{(s;1)}i_t^{(s;-1)}}$ as $(J_{N_0}^T)_{i_t^{(s;-1)}i_t^{(s;1)}}$ shows that
\begin{equation*}
\left(\prod_{s=1}^n\prod_{t=1}^{k_s}(J_{N_0})_{i_t^{(s;1)}i_t^{(s;-1)}}\right)\times f_{\Gamma}\left(\{i_t^{(s;\pm1)}\}_{1\leq s\leq n,\,1\leq t\leq k_s}\right)=\omega_1\cdots\omega_{V_0(\Gamma)}.
\end{equation*}
Hence, summing this expression over all indices $i_t^{(s;\pm1)}$ results in a weight of
\begin{equation} \label{eq3.3.58x}
\left(\prod_{a=1}^n\prod_{b=1}^{k_a}\sum_{i_b^{(a;\pm1)}=1}^{N_0}\right)\omega_1\cdots\omega_{V_0(\Gamma)}=\Tr\left(q_\Gamma^{(1)}(J_{N_0},J_{N_0}^T)\right)\cdots\Tr\left(q_\Gamma^{(V_0(\Gamma))}(J_{N_0},J_{N_0}^T)\right),
\end{equation}
where $q_\Gamma^{(j)}$ ($1\leq j\leq V_0(\Gamma)$) is the monomial obtained from the string $\omega_j$ by erasing matrix indices, e.g., the string $(J_{N_0})_{ab}(J_{N_0}^T)_{bc}(J_{N_0})_{ca}$ induces the monomial $J_{N_0}J_{N_0}^TJ_{N_0}$. Note also that
\begin{multline*}
\Tr\left(q_\Gamma^{(j)}(J_{N_0},J_{N_0}^T)\right)=\Tr\left(q_\Gamma^{(j)}(J_{N_0},-J_{N_0})\right)
\\=\Tr\left(q_\Gamma^{(j)}(1,-1)J_{N_0}^{\mathrm{deg}\,q_\Gamma^{(j)}}\right)=q_\Gamma^{(j)}(1,-1)\Tr\left(J_{N_0}^{\mathrm{deg}\,q_\Gamma^{(j)}}\right),
\end{multline*}
which simplifies to either zero, $N_0$, or $-N_0$ (we have abused notation to define $q_\Gamma^{(j)}$ on $\mathbb{R}^2$). Since for $k\in\mathbb{N}$, $J_{N_0}^{2k}=(-I_{N_0})^k=(-1)^kN_0$ and $\Tr\,J_{N_0}^{2k+1}=0$, we see that
\begin{equation*}
q_\Gamma^{(j)}(1,-1)=\mathrm{sgn}(\omega_j),\qquad \Tr\left(J_{N_0}^{\mathrm{deg}\,q_\Gamma^{(j)}}\right)=N_0\,\mathrm{Re}\left(\mathrm{i}^{|\omega_j|}\right).
\end{equation*}
Thus, the weight of the $j\textsuperscript{th}$ boundary is
\begin{equation} \label{eq3.3.59x}
\Tr\left(q_\Gamma^{(j)}(J_{N_0},J_{N_0}^T)\right)=N_0\,\mathrm{sgn}(\omega_j)\,\mathrm{Re}\left(\mathrm{i}^{|\omega_j|}\right).
\end{equation}
Substituting this into equation \eqref{eq3.3.58x} and the result of doing so into equation \eqref{eq3.3.53} completes the proof.
\end{proof}

\begin{example}
Let us briefly revisit the ribbon graph depicted on the left in Figure \ref{fig3.21}. The weight corresponding to the blue and green vertices is $N_1N_2^2$ since there are one blue and two green boundaries. Traversing the red line starting at the vertex corresponding to the label $i_1^{(1;-1)}$ shows us that $i_1^{(1;-1)}\equiv i_2^{(1;-1)}$, in keeping with equation \eqref{eq3.3.54}, so we write $\omega_1=(J_{N_0})_{i_1^{(1;1)}i_1^{(1;-1)}}(J_{N_0}^T)_{i_1^{(1;-1)}i_2^{(1;1)}}$. In a larger ribbon graph, we would then look to extend $\omega_1$ via concatenations, but we see here that $i_2^{(1;1)}\equiv i_1^{(1;1)}$, so we are done (indeed, the red and orange boundary has been fully traversed). We see that $\omega_1$ contains an entry of $J_{N_0}$ followed by an entry of $J_{N_0}^T$, so we have that $\mathrm{sgn}(\omega_1)=(-1)^1=-1$ and $|\omega_1|=2$. Thus, the weight of the single red and orange boundary is \eqref{eq3.3.59x}
\begin{equation*}
\Tr(J_{N_0}J_{N_0}^T)=-N_0\,\mathrm{Re}\left(\mathrm{i}^2\right)=N_0
\end{equation*}
and, hence, our ribbon graph contributes a value of $N_0N_1N_2^2/16$ to the computation of $m_2^{(\mathcal{J}_2)}$.
\end{example}

\begin{remark} \label{R3.4}
One way to think about the formula \eqref{eq3.3.56} is that it is simply the sum \eqref{eq3.3.55} with the summand having each factor of $N_0$ replaced either zero, $N_0$, or $-N_0$. It is possible to apply the arguments given above to the case of the Hermitised product ensembles so that equation \eqref{eq3.3.46} can likewise be interpreted as being given by equation \eqref{eq3.3.55} with the sum being constrained to be over only the orientable ribbon graphs in $\tilde{\mathfrak{S}}_{k_1,\ldots,k_n}^{(m)}$ and with the factor of $N_0^{V_0(\Gamma)}$ in the summand replaced by the GUE mixed moment $m_{k_1',\ldots,k_{V_0(\Gamma)}'}^{(GUE)}|_{N\mapsto N_0}$ \eqref{eq3.3.18}, where $k_1',\ldots,k_{V_0(\Gamma)}'\in\mathbb{N}$ are such that the $i\textsuperscript{th}$ boundary of colour $\phi_1$ seen in $\Gamma$ passes through $k_i'$ vertices of that colour. As an example, we would weight the middle image of Figure \ref{fig3.21} by $m_2^{(GUE)}$ since there is one red boundary passing through two red vertices.
\end{remark}

Returning now to Lemma \ref{L3.8x}, let us give the analogue of Proposition~\ref{prop3.10} in the case of the antisymmetrised matrix product ensembles.
\begin{proposition} \label{prop3.11}
Let the parameters $m,N_0,\ldots,N_m=N\in\mathbb{N}$ and matrices $J_{N_0},G_1,\ldots,G_m$ be as in Definition \ref{def1.11}, equivalently equation \eqref{eq3.3.51}, and let us require that the parameters $\nu_i:=\tfrac{N_i-N_0}{2}$ be such that for each $i=1,\ldots,m$, there exists a non-negative constant $\hat{\nu}_i:=\nu_i/N_0$ (cf.~Remark~\ref{R1.11}). Furthermore, introduce the scaled real Ginibre matrices $\tilde{G}_i:=(\tfrac{N_iN_{i-1}}{4})^{-1/4}G_i$ ($i=1,\ldots,m$) so that
\begin{equation} \label{eq3.3.57}
\tilde{\mathcal{J}}_m:=\tilde{G}_m^T\cdots\tilde{G}_1^TJ_{N_0}\tilde{G}_1\cdots\tilde{G}_m=\frac{2^mN_0^{-m}}{(2\hat{\nu}_1+1)\cdots(2\hat{\nu}_{m-1}+1)\sqrt{2\hat{\nu}_m+1}}\mathcal{J}_m
\end{equation}
represents the \textbf{global scaled $(N_0,\ldots,N_{m-1},N)$ antisymmetrised matrix product ensemble} (again refining Definition \ref{def1.13}). Then, for $n,k_1,\ldots,k_n\in\mathbb{N}$, the mixed cumulants of this ensemble vanish if any of the $k_1,\ldots,k_n$ are odd and are otherwise given by the genus expansion
\begin{multline} \label{eq3.3.58}
\tilde{c}_{k_1,\ldots,k_n}^{(\mathcal{J}_m)}=\left((2\hat{\nu}_1+1)\cdots(2\hat{\nu}_{m-1}+1)\sqrt{2\hat{\nu}_m+1}\right)^{-k_1-\cdots-k_n}N_0^{2-n}
\\ \times\sum_{l=0}^{m(k_1+\cdots+k_n)+1-n}\frac{1}{N_0^l}\sum_{p_1,\ldots,p_m=1}^{k_1+\cdots+k_n}(2\hat{\nu}_1+1)^{p_1}\cdots(2\hat{\nu}_m+1)^{p_m}
\\ \times\sum_{\Gamma\in\tilde{\mathfrak{C}}^{(m),c}_{k_1,\ldots,k_n}}\prod_{j=1}^{p_0}\mathrm{sgn}(\omega_j)\,\mathrm{Re}\left(\mathrm{i}^{|\omega_j|}\right),
\end{multline}
where the strings $\omega_1,\ldots,\omega_{p_0}$ are as introduced above Lemma \ref{L3.8x} and we define
\begin{equation} \label{eq3.3.59}
\tilde{\mathfrak{C}}^{(m),c}_{k_1,\ldots,k_n}:=\{\Gamma\in\tilde{\mathfrak{S}}_{k_1,\ldots,k_n}^{(m),c}\,|\,\tilde{g}(\Gamma)=l\textrm{ and }V_i(\Gamma)=p_i\textrm{ for each }0\leq i\leq m\},
\end{equation}
with $V_0(\Gamma),\ldots,V_m(\Gamma)$ as in Lemma \ref{L3.5} (having assigned the colour $\phi_1$ to the polygon edges representing entries of $J_{N_0}$) and with $\tilde{\mathfrak{S}}^{(m),c}_{k_1,\ldots,k_n}$ being the subset of connected ribbon graphs in $\tilde{\mathfrak{S}}_{k_1,\ldots,k_n}^{(m)}$ (defined in Lemma \ref{L3.7}); recall from Definition \ref{def3.1} that $\tilde{g}(\Gamma)$ denotes the Euler genus of $\Gamma$.
\end{proposition}
\begin{proof}
Observe that $m_{k_1,\ldots,k_n}^{(\mathcal{J}_m)}$ is a weighted count of ribbon graphs with the necessary property for it to be amenable to the inductive argument given in the paragraph above Remark \ref{R3.4x}: the weight of a disconnected ribbon graph is the product of the weights of its connected components (note that the interpretation of $m_{k_1,\ldots,k_n}^{(\mathcal{H}_m)}$ given in Remark \ref{R3.4} does not have this property). Thus, by the aforementioned argument, the mixed cumulants of the antisymmetrised matrix product ensembles are given by equation \eqref{eq3.3.56}, but with $\tilde{\mathfrak{S}}_{k_1,\ldots,k_n}^{(m)}$ replaced by $\tilde{\mathfrak{S}}_{k_1,\ldots,k_n}^{(m),c}$. The fact that the cumulants vanish whenever any of the $k_1,\ldots,k_n$ are odd follows from the same reasoning as in Remark \ref{R3.4x}, while the remainder of this proposition is proved in the same manner as Proposition \ref{prop3.10} with Lemma \ref{L3.8x} as our starting point.
\end{proof}

Since the third line of equation \eqref{eq3.3.58} is manifestly an integer, it is immediate that $N_0^{2-n}\tilde{c}_{k_1,\ldots,k_n}^{(\mathcal{J}_m)}$ is a polynomial in $N_0^{-1}$. However, unlike what was seen in Proposition \ref{prop3.10}, this is not an even polynomial, since we now allow for non-orientable ribbon graphs with odd Euler genus (i.e., terms corresponding to odd values of $l$ now have non-zero contribution; this is exactly what was seen in \S\ref{s3.3.1} and \S\ref{s3.3.2} when comparing the GUE/LUE to the GOE/LOE). Another point of contrast is that the mixed cumulants $\tilde{c}_{k_1,\ldots,k_n}^{(\mathcal{J}_m)}$ can take negative values, unlike $\tilde{c}_{k_1,\ldots,k_n}^{(\mathcal{H}_m)}$ (and, indeed, all other mixed cumulants discussed in this section). We recognise that Proposition~\ref{prop3.11} is a little less intuitive than Proposition \ref{prop3.10} due to the awkward manner in which the words $\omega_1,\ldots,\omega_{p_0}$ must be constructed for each $\Gamma\in\tilde{\mathfrak{C}}^{(m),c}_{k_1,\ldots,k_n}$. It turns out that the third line of equation \eqref{eq3.3.58} can be simplified further, but we do not present this simplification here, opting instead to discuss it in the $m=1$, $\nu_1=\nu_0=0$ setting.

\subsubsection{Topological hypermaps in the setting with $m=1$ and $\nu_1=\nu_0=0$}
We now use the ideas presented at the ends of \S\ref{s3.3.1} and \S\ref{s3.3.2} to derive topological hypermap formulations of Propositions \ref{prop3.10} and \ref{prop3.11}. To further simplify our results, we restrict ourselves to the $m=1$ case concerning products of square matrices (i.e., $N_1=N_0=N$ and $\nu_1=\nu_0=0$). Our arguments are similar to those given briefly in the case of the $m=2$, $(N,N,N)$ complex Wishart product ensemble in \citep{DF20}. It is pertinent that we point out that the upcoming derivations do not have any natural extensions to the $m>1$ cases, but that such cases can be treated combinatorially by introducing the theory of \textit{constellations}. We do not discuss constellations here and instead refer the reader to the textbook \citep{LZ04}.

\begin{figure}[H]
        \centering
\captionsetup{width=.9\linewidth}
        \includegraphics[width=0.8\textwidth]{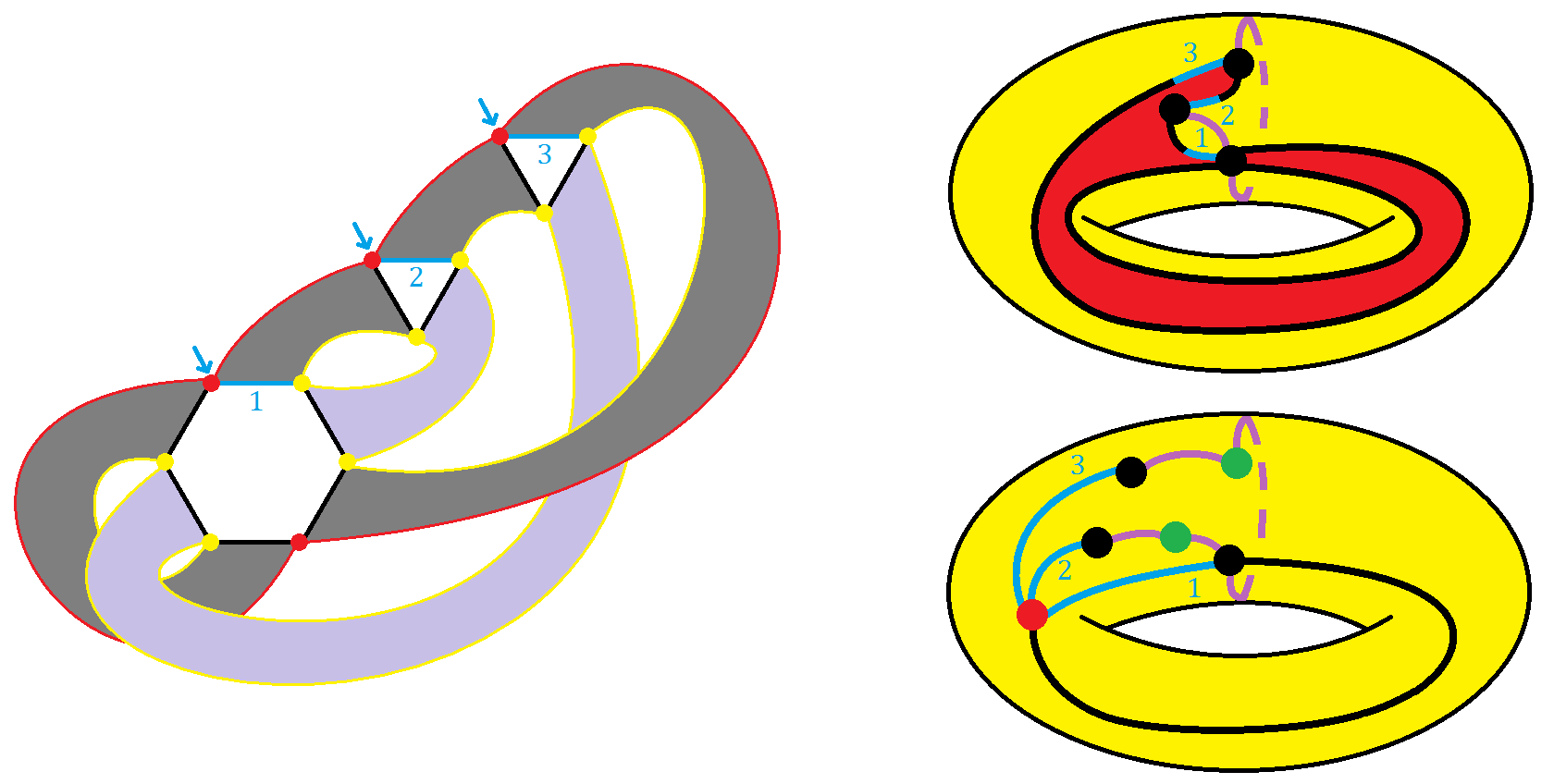}
        \caption[A topological hypermap contributing to the computation of $c_{2,1,1}^{(\mathcal{H}_1)}$]{The ribbon graph on the left contributes to the computation of $c_{2,1,1}^{(\mathcal{H}_1)}$ (obtained by removing grey ribbons from Figure \ref{fig3.20}). The blue arrows point to the rooted vertices and the rooted polygon edges are likewise coloured blue. Gluing faces to the boundaries and shrinking the ribbons and polygons results in the toroidal (genus one) topological map depicted on the top right. Inserting green vertices in the middle of the purple edges and shrinking the red faces in the aforedescribed manner then yields the topological hypermap on the bottom right. The blue edges are actually root-marked black edges.} \label{fig3.22}
\end{figure}

Let us first consider the $(N,N)$ Hermitised Laguerre ensemble represented by the $N\times N$ matrix $\mathcal{H}_1=G^{\dagger}HG$ with $G$ drawn from the $N\times N$ complex Ginibre ensemble and $H$ drawn from the $N\times N$ GUE. From the discussion preceding Figure \ref{fig3.20}, we know that the mixed cumulants $c_{k_1,\ldots,k_n}^{(\mathcal{H}_m)}$ \eqref{eq3.0.13} of this ensemble can be computed by enumerating connected, orientable ribbon graphs constructed in the following manner: Draw $n$ polygons with respectively $3k_1,3k_2,\ldots,3k_n$ sides, pick one vertex of each polygon to be the `first' or rooted vertex of that polygon and then, moving clockwise around the polygon while starting at the first vertex, colour each sequence of three vertices red, yellow, and yellow, in that order (taking $\phi_1,\phi_2$ to be yellow and red, respectively). Next, connect polygon edges with untwisted ribbons whose sides identify vertices of like colour in a way that results in a connected ribbon graph. For visual clarity, we also colour the ribbons connecting edges representing entries of $G^\dagger$ ($H$) to those representing $G$ ($H$) black (purple); see Figure \ref{fig3.22} below for an example. These ribbon graphs can immediately be reformulated as $n$-rooted topological maps by gluing yellow (red) faces to the yellow (red) boundaries, shrinking the polygons to vertices, thinning the ribbons to edges connecting said vertices, and root-marking the half-edges appearing immediately clockwise to rooted vertices (cf.~Figures \ref{fig3.10} and \ref{fig3.11}). To simplify even further to $n$-rooted topological hypermaps (which we recall were introduced at the end of \S\ref{s3.3.1}), we contract the red faces to red vertices while identifying black half-edges in the manner shown in Figure \ref{fig3.17}, and then split the purple edges into two by inserting green vertices in the middle of them; again, see Figure \ref{fig3.22} below.

We now give a reformulation of the $m=1$, $\nu_1=\nu_0=0$ case of Proposition \ref{prop3.10} in terms of these particular topological hypermaps.
\begin{proposition} \label{prop3.12}
For $n,k_1,\ldots,k_n\in\mathbb{N}$ with $k_1+\cdots+k_n$ even, let $\mathfrak{H}_{k_1,\ldots,k_n}$ be the set of orientable $n$-rooted topological hypermaps consisting of $k_1+\cdots+k_n$ black edges and as many purple edges, $\tfrac{1}{2}(k_1+\cdots+k_n)$ green vertices, any number of red vertices, and $n$ black vertices such that 
\begin{enumerate}
\item the green vertices are bivalent with only purple edges attached to them, 
\item the red vertices only have black edges attached to them, 
\item for $i=1,\ldots,n$, the $i\textsuperscript{th}$ black vertex has $k_i$ purple and black edges attached to it in alternating cyclic order,
\item for $i=1,\ldots,n$, one black edge incident to the $i\textsuperscript{th}$ black vertex is root-marked and labelled by $i$.
\end{enumerate}
Then, appropriately simplifying equation \eqref{eq3.3.48} shows that the mixed cumulants of the global scaled $(N,N)$ Hermitised Laguerre ensemble (represented by $\tilde{\mathcal{H}}_1:=\tilde{G}^\dagger \tilde{H}\tilde{G}$ with $\tilde{G}:=G/\sqrt{N}$ and $\tilde{H}:=\sqrt{2/N}H$) is given by
\begin{equation} \label{eq3.3.60}
\tilde{c}_{k_1,\ldots,k_n}^{(\mathcal{H}_1)}=N^{2-n}\sum_{l=0}^{\tfrac{3}{2}(k_1+\cdots+k_n)+1-n}\frac{1}{N^l}\,\#\{\Omega\in\mathfrak{H}_{k_1,\ldots,k_n}\,|\,g(\Omega)=l/2\}.
\end{equation}
\end{proposition}

Moving on, consider now the $(N,N)$ antisymmetrised Laguerre ensemble represented by the $N\times N$ matrix $\mathcal{J}_1=G^TJ_NG$, where $N$ is an even integer, $G$ is drawn from the $N\times N$ real Ginibre ensemble, and $J_N$ is the $N\times N$ elementary antisymmetric matrix defned in equation \eqref{eq1.1.5}. From the discussion leading to Lemma \ref{L3.8x}, we know that the mixed cumulants $c_{k_1,\ldots,k_n}^{(\mathcal{J}_1)}$ can be computed by counting LOE ribbon graphs with appropriate weights --- the weight of each ribbon graph is the product of the weights of its boundaries, with black boundaries being weighted by $N$ and red boundaries being weighted by the trace of a suitable monomial \eqref{eq3.3.59x} in $J_N$ and $J_N^T$. Recall that these monomials can be determined by replacing each red vertex of a given LOE ribbon graph $\Gamma$ by a new polygon edge representing an entry of $J_N$ and then traversing the red boundaries while constructing the strings $\omega_1,\ldots,\omega_{V_r(\Gamma)}$ described above Lemma \ref{L3.8x}, where $V_r(\Gamma)$ is the number of red boundaries of $\Gamma$. Translating this idea to the setting concerning the associated topological maps, which are obtained by gluing red (black) faces to the red (black) boundaries and then shrinking polygons and ribbons to vertices and edges, we need to compute a string $\omega_j$ for each red face of the topological map.

Now, recall from the discussion between Figures \ref{fig3.17} and \ref{fig3.18} that the LOE topological maps have half-edges either of type $G^T$ or $G$ and further observe that half-edges of type $G^T$ ($G$) correspond to ribbon ends that make contact with vertices with labels of the form $i_t^{(s;1)}$ ($i_t^{(s;-1)}$); here, we remember that we are working with ribbon graphs that have been obtained by replacing the red vertices of LOE ribbon graphs with polygon edges representing $J_N$ and we have adopted the vertex labelling convention given above Figure \ref{fig3.21}. Thus, an edge of type $G^T-G^T$ ($G-G$) signifies an identification of the form $i_t^{(s;1)}\equiv i_{t'}^{(s';1)}$ ($i_t^{(s;-1)}\equiv i_{t'}^{(s';-1)}$) so that in the relevant string $\omega_j$, we have either of the substrings $(J_{N}^T)_{i_t^{(s;-1)}i_t^{(s;1)}}(J_{N})_{i_{t'}^{(s';1)}i_{t'}^{(s';-1)}}$ or $(J_{N}^T)_{i_{t'}^{(s';-1)}i_{t'}^{(s';1)}}(J_{N})_{i_t^{(s;1)}i_t^{(s;-1)}}$ ($(J_{N})_{i_t^{(s;1)}i_t^{(s;-1)}}(J_{N}^T)_{i_{t'}^{(s';-1)}i_{t'}^{(s';1)}}$ or $(J_{N})_{i_{t'}^{(s';1)}i_{t'}^{(s';-1)}}(J_{N}^T)_{i_t^{(s;-1)}i_t^{(s;1)}}$) appearing --- edges of type $G^T-G^T$ and $G-G$ correspond to substrings of the form $(J_{N})_{i_t^{(s;1)}i_t^{(s;-1)}}(J_{N})_{i_{t'}^{(s';1)}i_{t'}^{(s';-1)}}$ or $(J_{N}^T)_{i_t^{(s;-1)}i_t^{(s;1)}}(J_{N}^T)_{i_{t'}^{(s';-1)}i_{t'}^{(s';1)}}$. Hence, rewriting a term $(J_{N})_{i_t^{(s;1)}i_t^{(s;-1)}}$ as $(J_{N}^T)_{i_t^{(s;-1)}i_t^{(s;1)}}$ or vice versa in a given string $\omega_j$ is equivalent to interchanging the types $G^T\leftrightarrow G$ of the half-edges representing the ribbon ends that interact with the vertices labelled $i_t^{(s;\pm1)}$. Consequently, the number of $J^T_{N}$ seen in the monomial $q_\Gamma^{(j)}(J_{N},J_{N}^T)$ induced by $\omega_j$ is equal to the minimum number of times such $G^T\leftrightarrow G$ interchanges would need to be applied to the $j\textsuperscript{th}$ red boundary to yield a red boundary made up of a chain of half-edges of alternating type. However, these are the same $G^T\leftrightarrow G$ interchanges discussed between Figures \ref{fig3.17} and \ref{fig3.18}: They can be kept track of by transforming our topological maps into hypermaps through the manner shown in Figure \ref{fig3.17} and then applying a $\pm1$ labelling to the hypermap edges following the prescription given above Figure \ref{fig3.18}. Then, the monomial trace $\Tr\,q_\Gamma^{(j)}(J_{N},J_{N}^T)$ weighting the $j\textsuperscript{th}$ red map face is equal to $\Tr\,J_{N}^{a_j^+}(J_{N}^T)^{a_j^-}$, where $a_j^{\pm}$ is equal to the number of edges of type $\pm1$ incident to the red hypermap vertex corresponding to the $j\textsuperscript{th}$ red map face --- consequently, we have that
\begin{equation} \label{eq3.3.64x}
\mathrm{sgn}(\omega_j)=(-1)^{a_j^-},\qquad |\omega_j|=a_j^-+a_j^+.
\end{equation}

\begin{example}
Although the above prescription is quite longwinded, it is much easier to compute the weight of a given topological hypermap than the ribbon graph it represents. For example, if we want to compute the contribution of the ribbon graph displayed on the top left of Figure \ref{fig3.19} to the value of $c_2^{(\mathcal{J}_1)}$, we first need to replace the red vertices with diagonal edges so that the white square becomes a hexagon. The new edge in the top right represents $(J_{N})_{i_1^{(1;1)}i_1^{(1;-1)}}$, while the new edge in the bottom left represents $(J_{N})_{i_2^{(1;1)}i_1^{(2;-1)}}$. Following the red ribbon border starting at the bottom vertex of the edge representing $(J_{N})_{i_1^{(1;1)}i_1^{(1;-1)}}$ shows that $i_1^{(1;-1)}\equiv i_2^{(1;-1)}$ and continuing to traverse this red boundary by following the other red ribbon border confirms that $i_2^{(1;1)}\equiv i_1^{(1;1)}$. Thus, we have $\omega_1=(J_{N})_{i_1^{(1;1)}i_1^{(1;-1)}}(J_{N}^T)_{i_1^{(1;-1)}i_1^{(1;1)}}$ and so this boundary is weighted $\Tr\,J_{N}J_{N}^T=N\,\mathrm{sgn}(\omega_1)\,\mathrm{Re}(\mathrm{i}^{|\omega_1|})=N$. Since we also have one black boundary, which is automatically weighted $N$, this ribbon graph contributes a value of $N^2/4$ to $c_2^{(\mathcal{J}_1)}$ in our $\nu_1=\nu_0=0$ setting. Now, if we look instead at either of the topological hypermaps given in the bottom of Figure \ref{fig3.19}, we immediately see that the red vertex has one twisted and one untwisted edge adjacent to it, so it is weighted by $N(-1)^{1}\mathrm{Re}(\mathrm{i}^2)=N$; there being one black vertex means that the hypermap is weighted $N^2/4$.
\end{example}

Working with topological hypermaps rather than ribbon graphs allow us to derive the following simplification of the $m=1$, $\nu_1=\nu_0=0$ case of Proposition \ref{prop3.11}.
\begin{proposition} \label{prop3.13}
For $n\in\mathbb{N}$ and $k_1,\ldots,k_n\in2\mathbb{N}$, let $\tilde{\mathfrak{A}}_{k_1,\ldots,k_n}$ be the subset of $\pm1$-labelled locally orientable $n$-rooted topological hypermaps in the set $\tilde{\mathfrak{T}}_{k_1,\ldots,k_n}^{*}$, defined in Proposition \ref{prop3.9}, whose vertices all have even valency. Furthermore, let $\tilde{\mathfrak{A}}_{k_1,\ldots,k_n}^{\tilde{g},\pm}\subseteq\tilde{\mathfrak{A}}_{k_1,\ldots,k_n}$ be such that for each $\Omega\in\tilde{\mathfrak{A}}_{k_1,\ldots,k_n}^{\pm}$, $\Omega$ has Euler genus $\tilde{g}$ and the product of the signs of the edges of $\Omega$ is $\pm1$. Then, the mixed cumulants of the global scaled $(N,N)$ antisymmetrised Laguerre ensemble (represented by $\tilde{\mathcal{J}}_1:=\tilde{G}^TJ_N\tilde{G}$ with $\tilde{G}=\sqrt{2/N}G$) is given by
\begin{equation} \label{eq3.3.61}
\tilde{c}_{k_1,\ldots,k_n}^{(\mathcal{J}_1)}=(-1)^{\tfrac{1}{2}(k_1+\cdots+k_n)}N^{2-n}\sum_{l=0}^{k_1+\cdots+k_n+1-n}\frac{1}{N^l}\left(\#\tilde{\mathfrak{A}}_{k_1,\ldots,k_n}^{l,+}-\#\tilde{\mathfrak{A}}_{k_1,\ldots,k_n}^{l,-}\right),
\end{equation}
where we recall that $\#\mathcal{S}$ denotes the number of elements in $\mathcal{S}$.
\end{proposition}
\begin{proof}
Having converted the ribbon graphs of Proposition \ref{prop3.11} (with $m=1$) into topological hypermaps in the manner prescribed above, we first note that hypermaps with black vertices of odd valency have vanishing weight: Recall from Proposition \ref{prop3.11} that the cumulants $c_{k_1,\ldots,k_n}^{(\mathcal{J}_m)}$ vanish if any of the $k_i$ are odd --- the black vertices of our topological hypermaps have valency $k_1,\ldots,k_n$.

Next, note that the $j\textsuperscript{th}$ red vertex (which corresponds to the $j\textsuperscript{th}$ red boundary of the original ribbon graph) has weight $\mathrm{sgn}(\omega_j)\,\mathrm{Re}(\mathrm{i}^{|\omega_j|})$ according to Proposition \ref{prop3.11}, which simplifies to
\begin{equation} \label{eq3.3.65x}
\mathrm{sgn}(\omega_j)\,\mathrm{Re}(\mathrm{i}^{|\omega_j|})=(-1)^{a_j^-}\mathrm{Re}(\mathrm{i}^{a_j^-+a_j^+})
\end{equation}
by equation \eqref{eq3.3.64x}. This weight also vanishes if $a_j^-+a_j^+$, the valency of the $j\textsuperscript{th}$ red vertex, is odd. Thus, only hypermaps with all vertices of even valency contribute to the value of the mixed cumulants.

Now, using equation \eqref{eq3.3.65x} in the third line of equation \eqref{eq3.3.58} shows that the weight of a topological hypermap due to the weights of the red vertices is the product
\begin{equation*}
\prod_{j=1}^{V_0(\Gamma)}\mathrm{sgn}(\omega_j)\,\mathrm{Re}(\mathrm{i}^{|\omega_j|})=\prod_{j=1}^{V_0(\Gamma)}(-1)^{a_j^-}\mathrm{Re}(\mathrm{i}^{a_j^-+a_j^+}),
\end{equation*}
where, following Lemma \ref{L3.8x} and Proposition \ref{prop3.11}, $V_0(\Gamma)$ is the number of red boundaries of $\Gamma$ and is thus the number of red vertices of the hypermap at hand. As our weight is non-vanishing only when $a_j^-+a_j^+$ is even for all $j$, we see that
\begin{equation*}
\prod_{j=1}^{V_0(\Gamma)}\mathrm{Re}(\mathrm{i}^{a_j^-+a_j^+})=\prod_{j=1}^{V_0(\Gamma)}(-1)^{\tfrac{1}{2}(a_j^-+a_j^+)}=(-1)^{\tfrac{1}{2}(a^-_1+\cdots+a^-_{V_0(\Gamma)}+a^+_1+\cdots+a^+_{V_0(\Gamma)})}=(-1)^{\tfrac{1}{2}(k_1+\cdots+k_n)},
\end{equation*}
where the last line follows from the fact that the sum of the valencies $a_j^-+a_j^+$ of the red vertices is equal to the total number of edges in the hypermap, which is $k_1+\cdots+k_n$. This is the source of the factor seen at the front of the right-hand side of equation \eqref{eq3.3.61}. On the other hand, since each edge is incident to exactly one red vertex, $\prod_{j=1}^{V_0(\Gamma)}(-1)^{a_j^-}$ is simply the product of the signs of all of the edges present in the hypermap, which leads to the definition of $\tilde{\mathfrak{A}}_{k_1,\ldots,k_n}^{\pm}$. Substituting $m=1$, $\nu_1=\nu_0=0$ into equation \eqref{eq3.3.58} while incorporating the simplifications just discussed produces the sought formula \eqref{eq3.3.61}.
\end{proof}

As we move on to the next chapter, let us mention that, although we have not discussed them in full generality, many of the arguments leading to Proposition \ref{prop3.13} can be extended to the general $m\in\mathbb{N}$ setting of Proposition \ref{prop3.11}. Our reason for not presenting these arguments in this general setting is simply that in the related ribbon graphs, the extra ribbons that interact with polygon edges representing entries of $G_2^T, G_2,\ldots, G_m^T,G_m$ are distracting while being completely irrelevant to our discussion, as they do not interact with the polygon edges representing entries of $J_{N_0}$.

\chapter{Loop Equations for the Matrix Product Ensembles}
In this chapter, we take the first step in exploring the feasibility of studying the Hermitised and antisymmetrised matrix product ensembles introduced in Section \ref{s1.3} using the loop equation formalism. Recall from Definition \ref{def1.10} that for $m\in\mathbb{N}$ and $N_0,\ldots, N_m\in\mathbb{N}$ such that $N_1,\ldots,N_m\geq N_0$, the $(N_0,\ldots,N_{m-1},N)$ (writing $N:=N_m$) Hermitised matrix product ensemble is represented by the $N\times N$ random matrix product
\begin{equation} \label{eq4.0.1}
\mathcal{H}_m:=G_m^\dagger\cdots G_1^\dagger HG_1\cdots G_m,
\end{equation}
where $H$ is drawn from the $N_0\times N_0$ GUE and for $1\leq i\leq m$, each $G_i$ is drawn independently from the $N_{i-1}\times N_i$ complex Ginibre ensemble. If we further take $N_0,\ldots,N_m$ to be even and now draw each $G_i$ from the $N_{i-1}\times N_i$ real Ginibre ensemble, we have by Definition \ref{def1.11} that the $(N_0,\ldots,N_{m-1},N)$ antisymmetrised matrix product ensemble is represented by
\begin{equation} \label{eq4.0.2}
\mathcal{J}_m:=G_m^T\cdots G_1^TJ_{N_0}G_1\cdots G_m,
\end{equation}
where $J_{N_0}$ is the elementary antisymmetric matrix defined in equation \eqref{eq1.1.5}. Equivalently, in the notation of Definition \ref{def1.1}, the $(N_0,\ldots,N_{m-1},N)$ Hermitised matrix product ensemble $\mathcal{E}^{(\mathcal{H}_m)}=(\mathcal{S}^{(\mathcal{H}_m)},P^{(\mathcal{H}_m)})$ is the set $\mathcal{S}^{(\mathcal{H}_m)}:=\mathcal{S}^{(G)}\big|_{\mathbb{F}=\mathbb{C}}$ \eqref{eq1.2.4} of $N\times N$ complex Hermitian matrices with the p.d.f.
\begin{equation} \label{eq4.0.3}
P^{(\mathcal{H}_m)}(X):=\delta(X-\mathcal{H}_m)\,P^{(G)}(H)\Big|_{\beta=2,N\mapsto N_0}\,\prod_{i=1}^mP^{(Gin)}(G_i)\Big|_{\beta=2,(M,N)\mapsto(N_{i-1},N_i)},
\end{equation}
where $\delta$ is the Dirac delta and the p.d.f.s $P^{(Gin)}(G)$ of the Ginibre ensemble and $P^{(G)}(H)$ of the Gaussian ensemble are respectively specified in Definition \ref{def1.3} and Proposition \ref{prop1.1}, while the $(N_0,\ldots,N_{m-1},N)$ antisymmetrised matrix product ensemble $\mathcal{E}^{(\mathcal{J}_m)}=(\mathcal{S}^{(\mathcal{J}_m)},P^{(\mathcal{J}_m)})$ is the set $\mathcal{S}^{(\mathcal{J}_m)}$ of $N\times N$ antisymmetric real matrices with the p.d.f.
\begin{equation} \label{eq4.0.4}
P^{(\mathcal{J}_m)}(X):=\delta(X-\mathcal{J}_m)\,\prod_{i=1}^mP^{(Gin)}(G_i)\Big|_{\beta=1,(M,N)\mapsto(N_{i-1},N_i)},
\end{equation}
where the parameters $N_0,\ldots,N_m$ and matrices $G_1,\ldots,G_m$ have been appropriately redefined. What we mean by `taking the first step' is that we will be studying the simplest examples of these ensembles, which correspond to setting $m=1$ and $N_1=N_0=N$ --- as we will soon see, taking $m$ to be any greater would result in loop equations too unwieldy to display in the present setting, while many of the interesting features of such loop equations are already present in the $m=1$ case. In the language of Definitions \ref{def1.10} and \ref{def1.11}, we will thus be deriving loop equations for the Hermitised and antisymmetrised Laguerre ensembles.

One of our motivations for initiating the aforementioned study is that while the eigenvalue j.p.d.f.s of the Hermitised and antisymmetrised matrix product ensembles have recently been made explicit in the works \citep{FIL18}, \citep{FILZ19}, as reviewed in \S\ref{s1.3.1} (technically, we discuss the non-zero real eigenvalues of the matrix $\mathcal{H}_m$ and the positive real eigenvalues of the matrix $\mathrm{i}\mathcal{J}_m$), their moments and associated moment generating functions present challenges relative to the setting of the classical matrix ensembles. Thus, as with our studies of the classical matrix ensembles presented in the earlier parts of this thesis, we would like to derive recursive characterisations of, say, the spectral moments $m_k$ \eqref{eq1.1.13} and resolvents $W_1(x)$ \eqref{eq1.1.17} of the Hermitised and antisymmetrised matrix product ensembles. One might then suggest that we study the ensembles at hand using the methodology laid out in Chapters~2 and~3 to obtain linear differential equations on the resolvent expansion coefficients $W_1^l(x)$ and $1$-point recursions on the moment expansion coefficients $M_{k,l}$ defined implicitly through the expansions \eqref{eq1.1.21} and \eqref{eq1.1.32}. Recall, though, that the development of these chapters is based on the Selberg correlation integral theory reviewed in Section~\ref{s2.1}, which can only be used to study eigenvalue j.p.d.f.s that are (tractably) expressible in terms of the j.p.d.f.~\eqref{eq2.1.1} seen in the Selberg integral \eqref{eq2.1.2}. This is actually the case for the antisymmetrised Laguerre ensemble, since it is an example of a Laguerre Muttalib--Borodin ensemble (recall Proposition~\ref{prop1.11} and Remark \ref{R1.14}) and the work \citep{FI18} relates such Muttalib--Borodin ensembles to the Selberg integral, so it may therefore be possible to apply the techniques of Chapters 2 and 3 to the antisymmetrised Laguerre ensemble, but we do not pursue this line of research since it is quite unlikely that this approach could be extended effectively to the general $m>1$ ensembles $\mathcal{E}^{(\mathcal{J}_m)}$ and $\mathcal{E}^{(\mathcal{H}_m)}$, as we currently have no way of relating the relevant eigenvalue j.p.d.f.s (recall their specifications in terms of the relatively complicated Meijer $G$-functions given in Proposition~\ref{prop1.7}) to the Selberg integral. Instead, we are led to considering the loop equation formalism, which is well known to be applicable to a large variety of random matrix ensembles.

The (final sets of) loop equations derived in this chapter constitute triangular recursive systems on the coefficients $W_n^l$ of the large $N$ expansions \eqref{eq1.1.21}
\begin{equation} \label{eq4.0.5}
W_n(x_1,\ldots,x_n)=N^{2-n}\sum_{l=0}^\infty\frac{W_n^l(x_1,\ldots,x_n)}{N^l}
\end{equation}
of the connected $n$-point correlators $W_n$ \eqref{eq1.1.29}, meaning that they facilitate the systematic, recursive computation of said coefficients. Although their forming triangular recursive systems means that these loop equations are not as efficient as corresponding $1$-point recursions (if they exist) for computing the spectral moments $m_k^{(\mathcal{H}_1)}$ and $m_k^{(\mathcal{J}_1)}$, they have the added benefit of producing the coefficients $W_n^{(\mathcal{H}_1),l}$ and $W_n^{(\mathcal{J}_1),l}$ for $n\geq1$ and $l\geq0$, from which we can extract coefficients of the genus expansions \eqref{eq3.3.60} and \eqref{eq3.3.61} of the corresponding mixed cumulants $\tilde{c}_{k_1,\ldots,k_n}$, thereby giving us a method of enumerating the ribbon graphs and topological hypermaps constructed in \S\ref{s3.3.3}.

Another motivation for deriving loop equations for the Hermitised and antisymmetrised Laguerre ensembles is that they are of higher order than the equivalent loop equations for the random matrix ensembles that are more commonly studied in the literature (see, e.g., \S\ref{s4.1.1} forthcoming). Indeed, we will see in \S\ref{s4.2.2} and \S\ref{s4.3.2} that the loop equations specifying $W_1^0(x)$, also know as the \textit{spectral curves} \citep{EO07}, for the antisymmetrised (Hermitised) Laguerre ensembles are third (fourth) order polynomials in $W_1^{(\mathcal{J}_1),0}(x)$ ($W_1^{(\mathcal{H}_1),0}(x)$), whereas the spectral curves of the classical matrix ensembles reviewed in \S\ref{s4.1.1} are quadratic polynomials in $W_1^0(x)$. Spectral curves of order higher than two have garnered interest in the abstract setting \citep{EO09}, \citep{BE13}, \citep{BHLMR14}, \citep{Ora15}, but there have not been many concrete examples of such higher order spectral curves and, more broadly, loop equations (here, higher order means higher than usual) arising from random matrix theory (see, however, the studies on so-called multi-matrix models \citep{EO09}). As mentioned above, we produce such examples through the analysis carried out in Sections \ref{s4.2} and \ref{s4.3} below, complementing the higher order loop equations given in the recent work \citep{DF20} on the $(N,N,N)$ complex Wishart product ensemble (see Definition~\ref{def1.12}). Note that even though the Hermitised and antisymmetrised Laguerre ensembles are closely related to the complex Wishart product ensemble studied in \citep{DF20}, the associated loop equations are structurally quite different (bar the similarities one expects from universality principles). These differences can be intuitively understood from the discussion of \S\ref{s3.3.3}: At the level of ribbon graphs and topological hypermaps, all three of these ensembles are distinct variations of the Laguerre ensemble.

In Section \ref{s4.1}, we introduce the loop equation formalism by means of giving a general outline of the key ideas behind it. We cannot hope to do much better than this since there are a variety of techniques for obtaining loop equations, as evidenced by the term `loop equations' being somewhat interchangeable with `Ward identities', `Virasoro constraints', `Tutte's equations' and `Schwinger--Dyson equations' (see, e.g., \citep{Mig83} and Chapters 8, 10, 16, 17, 26, and 30 of the handbook \citep{Oxf15}), as well as the method of constructing and recursively solving loop equations being related to the topological recursion \citep{EO09}. This relation to the topological recursion is briefly reviewed in \S\ref{s4.1.2}, while \S\ref{s4.1.1} contains a survey of how loop equations for the Gaussian, Laguerre, and Jacobi $\beta$ ensembles (defined in \S\ref{s1.2.4}) were derived in \citep{EM09}, \citep{FRW17}. Following the introductory content of Section \ref{s4.1}, we demonstrate the loop equation formalism in more detail, using an alternative approach to that discussed in \S\ref{s4.1.1}, by applying it to the $(N,N)$ antisymmetrised and Hermitised Laguerre ensembles in Sections~\ref{s4.2} and \ref{s4.3}, respectively.

\setcounter{equation}{0}
\section{A Brief Introduction to Loop Equations} \label{s4.1}
As with many tools in mathematics, the loop equation formalism is essentially a sophisticated application of integration by parts. Following the notation of Section \ref{s1.1}, let $X$ be an $N\times N$ random matrix drawn from the ensemble $\mathcal{E}=(\mathcal{S},P)$, let $p(\lambda_1,\ldots,\lambda_N)$ be the j.p.d.f.~of the eigenvalues $\{\lambda_i\}_{i=1}^N$ of $X$, let $\mean{\,\cdot\,}$ denote an average with respect to this eigenvalue j.p.d.f., and let $\mean{\,\cdot\,}_{P(X)}$ denote an average with respect to the matrix p.d.f.~$P(X)$. The main idea behind the loop equation formalism \citep{Mig83}, \citep{ACKM93} is to integrate the total derivative of the product of $P(X)$ and a well-chosen product of functions, $F(X)=\prod_i\,F_i(X)$ say, in two different ways: the first is to observe that the integral vanishes by the fundamental theorem of calculus due to the function $F(X)$ having been chosen such that its product with $P(X)$ is zero on the boundary $\partial\mathcal{S}$ of the domain of integration; the second is to expand the derivative under the integral sign using the Leibniz product rule and then integrate each of the terms separately so as to obtain the identity
\begin{equation} \label{eq4.1.1}
0=\int_{\mathcal{S}}\left\{\frac{\mathrm{d}}{\mathrm{d}X}P(X)\right\}F(X)\,\mathrm{d}X+\sum_i\mean{\left\{\frac{\mathrm{d}}{\mathrm{d}X}F_i(X)\right\}\frac{F(X)}{F_i(X)}}_{P(X)}.
\end{equation}
For this identity to be relevant to our goal, one of the integrals on the right-hand side must be equal to either the mixed moment \eqref{eq1.1.27}
\begin{equation} \label{eq4.1.2}
m_{k_1,\ldots,k_n}:=\mean{\prod_{i=1}^N\Tr\,X^{k_i}}_{P(X)}=\mean{\sum_{i_1,\ldots,i_n=1}^N\lambda_{i_1}^{k_1}\lambda_{i_2}^{k_2}\cdots\lambda_{i_n}^{k_n}},
\end{equation}
with $n,k_1,\ldots,k_n\in\mathbb{N}$ generic, or the unconnected $n$-point correlator \eqref{eq1.1.26}
\begin{equation} \label{eq4.1.3}
U_n(x_1,\ldots,x_n):=\mean{\prod_{i=1}^N\Tr\,\frac{1}{x_i-X}}_{P(X)}=\mean{\sum_{i_1,\ldots,i_n=1}^N\frac{1}{(x_1-\lambda_{i_1})\cdots(x_n-\lambda_{i_n})}},
\end{equation}
with generic $n\in\mathbb{N}$.

It turns out that good choices of $F(X)$ are those such that the average $\mean{\frac{\mathrm{d}}{\mathrm{d}X}F(X)}_{P(X)}$ (recall from \S\ref{s1.1.1} that the average of an operator with respect to a p.d.f.~$P(X)$ is the integral of that operator applied to said p.d.f.) can be seen to be slight perturbations of either the average \eqref{eq4.1.2} or \eqref{eq4.1.3}. If, after choosing such an $F(X)$, all of the integrals in the identity \eqref{eq4.1.1} can be written in terms of mixed moments $m_{k_1',\ldots,k_{n'}'}$ with at least one such integral being $m_{k_1,\ldots,k_n}$ exactly, alternatively unconnected correlators $U_{n'}(x_1,\ldots,x_{n'})$ with at least one integral being $U_n(x_1,\ldots,x_n)$, we say that this identity is a loop equation on the mixed moments, respectively unconnected correlators. In practice, at least one of the integrals in the identity \eqref{eq4.1.1} cannot be expressed in the necessary way, so one must repeat this exercise to find simpler complementary identities expressing these undesirable integrals in terms of mixed moments or unconnected correlators, as required.

Assuming we succeed in finding the identities described above, we would then have a set of loop equations on the unconnected correlators (one for each $n\in\mathbb{N}$); note that if one instead has a set of loop equations on the mixed moments, multiplying the $n\textsuperscript{th}$ loop equation by $x_1^{-k_1-1}\cdots x_n^{-k_n-1}$ and summing over $k_1,\ldots,k_n\geq0$ results in the associated loop equation on the unconnected correlators. Such a set of loop equations on the unconnected correlators can be transformed into a corresponding set of loop equations on the connected correlators through the identity \eqref{eq1.1.31}, which we repeat here for convenience,
\begin{equation} \label{eq4.1.4}
W_n(x_1,\ldots,x_n)=U_n(x_1,J_n)-\sum_{\emptyset\neq J\subseteq J_n}W_{n-\#J}(x_1,J_n\setminus J)U_{\#J}(J),\quad J_n=(x_2,\ldots,x_n).
\end{equation}
If we also have a large $N$ expansion (henceforth also referred to as a topological or genus expansion) of the form \eqref{eq4.0.5} available, we may substitute this expansion into the loop equations for the $W_n$ and then equate terms of like order in $N$ to extract loop equations on the $W_n^l$. As we will see in \S\ref{s4.1.1}, \S\ref{s4.2.3}, and \S\ref{s4.3.3}, it is this final set of loop equations that are exactly solvable through a suitable recursive procedure.

Up until now, we have formulated our discussion in terms of averages with respect to the matrix p.d.f.~$P(X)$. However, we can see from equations \eqref{eq4.1.2} and \eqref{eq4.1.3} above that we have the option to work with either the eigenvalue j.p.d.f.~$p(\lambda_1,\ldots,\lambda_N)$ or the matrix p.d.f.~$P(X)$, which is equivalent to working with univariate p.d.f.s on matrix entries. (Note that in the latter case, we do not necessarily mean p.d.f.s on the entries of $X$; if $X$ is a product of simpler matrices, it is often beneficial to work with the entries of the factors of $X$.) Both have their advantages and disadvantages: In the first convention, one has the benefit of only needing to work with the $N$ variables $\lambda_1,\ldots,\lambda_N$, but may have to deal with unwieldy eigenvalue j.p.d.f.s --- this is certainly the case for our matrix product ensembles (at least when taking $m>1$, as we hope to eventually do); see Proposition~\ref{prop1.7}. On the other hand, the second convention requires us to work with many more variables, but each of these variables may be drawn from relatively simple distributions.

Since the Hermitised and antisymmetrised Laguerre ensembles are specified in terms of Ginibre matrices, whose entries are normally distributed, we opt to study these ensembles using averages with respect to the p.d.f.s. \eqref{eq4.0.3} and \eqref{eq4.0.4}. To supplement our development, we review in \S\ref{s4.1.1} how averages with respect to the eigenvalue j.p.d.f.~\eqref{eq1.2.81} have been used to study the Gaussian, Laguerre, and Jacobi $\beta$ ensembles --- recall that these ensembles do not have amenable matrix p.d.f.s outside of the $\beta=1,2,4$ regimes. We continue this review in \S\ref{s4.1.2}, where we discuss how the topological recursion \citep{EO09} results from using residue calculus to simplify the GUE loop equations to their most aesthetic forms.

\subsection{Loop equations for the classical $\beta$ ensembles} \label{s4.1.1}
Loop equations for the \textit{general $\beta$ ensembles}, which correspond to the general-$\beta$ eigenvalue j.p.d.f.~\eqref{eq1.2.81} with $w(\lambda)=:e^{-\kappa NV(\lambda)}$ no longer constrained to be classical (recall that $\kappa=\beta/2$), were first obtained in the 2006 work \citep{CE06} of Chekhov and Eynard. Writing the potential $V(\lambda)$ as the formal power series
\begin{equation} \label{eq4.1.5}
V(\lambda)=1+\sum_{k=1}^{\infty}t_k\lambda^k,
\end{equation}
the loop equations of \citep{CE06} were formulated in terms of the linear integral operator $\hat{K}[\,\cdot\,]$ and functional derivative $\tfrac{\partial}{\partial V(x)}$ defined by (see also \citep{ACKM93})
\begin{equation} \label{eq4.1.6}
\hat{K}\left[f(x)\right]=\oint_{\gamma}\frac{\mathrm{d}\xi}{2\pi\mathrm{i}}\frac{V'(\xi)}{x-\xi}f(\xi),\qquad \frac{\partial}{\partial V(x)}=-\sum_{k=1}^{\infty}\frac{1}{x^{k+1}}\frac{\partial}{\partial t_k},
\end{equation}
where $x$ lies outside the integration contour $\gamma$, which circles the branch cuts of the resolvent $W_1(\xi)$ in a positive direction, and the so-called \textit{loop insertion operator} $\tfrac{\partial}{\partial V(x)}$ is such that the $n$-point connected correlator \eqref{eq4.1.4} can be (formally) written as
\begin{equation*}
W_n(x_1,\ldots,x_n)=\frac{\partial}{\partial V(x_n)}\frac{\partial}{\partial V(x_{n-1})}\cdots\frac{\partial}{\partial V(x_2)}W_1(x_1).
\end{equation*}
These operators posed subtle challenges when it came to solving the relevant loop equations in practice, so about three years later, Eynard and Marchal \citep{EM09} derived a more tractable reformulation of said loop equations. There, it was shown through a consideration of the infinitesimal change of variables 
\begin{equation} \label{eq4.1.7}
\lambda_i\mapsto\lambda_i+\epsilon\frac{1}{x_1-\lambda_i}+{\rm O}(\epsilon^2)
\end{equation}
in the average \eqref{eq4.1.3} that the relevant connected correlators \eqref{eq4.1.4} satisfy, for each $n\geq1$, the loop equation (see also \citep{BMS11}, \citep{BEMN11}, \citep{WF14})
\begin{multline} \label{eq4.1.8}
\kappa\sum_{J\subseteq J_n}W_{\#J+1}(x_1,J)W_{n-\#J}(x_1,J_n\setminus J)+\kappa W_{n+1}(x_1,x_1,J_n)+(\kappa-1)\frac{\partial}{\partial x_1}W_n(x_1,J_n)
\\=\kappa N\left[V'(x_1)W_n(x_1,J_n)-P_n(x_1,J_n)\right]-\sum_{i=2}^n\frac{\partial}{\partial x_i}\left\{\frac{W_{n-1}(x_1,J_n\setminus\{x_i\})-W_{n-1}(J_n)}{x_1-x_i}\right\},
\end{multline}
where $J_n$ is as in equation \eqref{eq4.1.4}, $\#\mathcal{S}$ denotes the size of $\mathcal{S}$, and
\begin{equation} \label{eq4.1.9}
P_n(x_1,J_n):=\mean{\sum_{i_1=1}^N\frac{V'(x_1)-V'(\lambda_{i_1})}{x_1-\lambda_{i_1}}\sum_{i_2,\ldots,i_n=1}^N\frac{1}{(x_2-\lambda_{i_2})\cdots(x_n-\lambda_{i_n})}}_{\mathrm{conn}}.
\end{equation}
Here, the \textit{connected average} $\mean{\,\cdot\,}_{\mathrm{conn}}$ relates to the usual average $\mean{\,\cdot\,}$ in the same way that the connected and unconnected correlators relate to each other via equation \eqref{eq4.1.4}.

Now, it is straightforward to specialise the loop equation \eqref{eq4.1.8} to the case of the Gaussian $\beta$ ensemble: one need only observe that taking $V(\lambda)$ to be the quadratic potential $\lambda^2$ in equation \eqref{eq4.1.9} shows that (see, e.g., \citep[Thrm.~1]{WF14})
\begin{equation} \label{eq4.1.10}
P_n^{(G)}(x_1,J_n)=\mean{2N\sum_{i_2,\ldots,i_n=1}^N\frac{1}{(x_2-\lambda_{i_2})\cdots(x_n-\lambda_{i_n})}}_{\mathrm{conn}}=2N\chi_{n=1},
\end{equation}
where we recall that the indicator function $\chi_A$ equals one when $A$ is true and is otherwise zero. Hence, if we let $\tilde{W}_n^{(G)}(x_1,\ldots,x_n)$ denote the connected $n$-point correlator of the global scaled Gaussian $\beta$ ensemble with eigenvalue j.p.d.f.~(apply the scaling $\lambda_i\mapsto\sqrt{\kappa N}\lambda_i$, in keeping with Definition \ref{def1.6}, to the j.p.d.f.~\eqref{eq1.2.81} with $w(\lambda)=e^{-\lambda^2}$)
\begin{equation}
\tilde{p}^{(G)}(\lambda_1,\ldots,\lambda_N;\beta)=\frac{(\kappa N)^{\tfrac{N}{2}(\kappa(N-1)+1)}}{\mathcal{N}_{N,\beta}^{(G)}}\prod_{i=1}^Ne^{-\kappa N\lambda_i^2}|\Delta_N(\lambda)|^{\beta},
\end{equation}
substituting equation \eqref{eq4.1.10} into the loop equation \eqref{eq4.1.8} shows that
\begin{multline} \label{eq4.1.12}
\kappa\sum_{J\subseteq J_n}\tilde{W}^{(G)}_{\#J+1}(x_1,J)\tilde{W}^{(G)}_{n-\#J}(x_1,J_n\setminus J)+\kappa \tilde{W}^{(G)}_{n+1}(x_1,x_1,J_n)+(\kappa-1)\frac{\partial}{\partial x_1}\tilde{W}^{(G)}_n(x_1,J_n)
\\=2\kappa N\left[x_1\tilde{W}^{(G)}_n(x_1,J_n)-N\chi_{n=1}\right]-\sum_{i=2}^n\frac{\partial}{\partial x_i}\left\{\frac{\tilde{W}^{(G)}_{n-1}(x_1,J_n\setminus\{x_i\})-\tilde{W}^{(G)}_{n-1}(J_n)}{x_1-x_i}\right\}.
\end{multline}
Moreover, invoking Theorem \ref{thrm1.1} to further substitute the expansion \eqref{eq4.0.5}
\begin{equation} \label{eq4.1.13}
\tilde{W}^{(G)}_n(x_1,\ldots,x_n)=N^{2-n}\sum_{l=0}^\infty\frac{W_n^{(G),l}(x_1,\ldots,x_n)}{N^l}
\end{equation}
into equation \eqref{eq4.1.12} and equating terms of equal order in $N$ yields, for $n\geq1$ and $l\geq0$, the $(n,l)$ loop equation on the correlator expansion coefficients,
\begin{multline} \label{eq4.1.14}
\kappa\sum_{J\subseteq J_n}\sum_{k=0}^lW^{(G),k}_{\#J+1}(x_1,J)W^{(G),l-k}_{n-\#J}(x_1,J_n\setminus J)+\kappa W^{(G),l-2}_{n+1}(x_1,x_1,J_n)
\\=(1-\kappa)\frac{\partial}{\partial x_1}W^{(G),l-1}_n(x_1,J_n)+2\kappa x_1W^{(G),l}_n(x_1,J_n)-2\kappa\chi_{n=1,l=0}
\\-\sum_{i=2}^n\frac{\partial}{\partial x_i}\left\{\frac{W^{(G),l}_{n-1}(x_1,J_n\setminus\{x_i\})-W^{(G),l}_{n-1}(J_n)}{x_1-x_i}\right\},
\end{multline}
with $W_n^{(G),-2},W_n^{(G),-1}:=0$ for all $n\geq1$. It can readily be seen that for any given $m\in\mathbb{N}$, the set of loop equations obtained by setting $n=1,\ldots,m$ in equation \eqref{eq4.1.12} involves more than $m$ correlators and is thus unsolvable, but the same is not true for equation \eqref{eq4.1.14}. Indeed, setting $(n,l)=(1,0)$ in equation \eqref{eq4.1.14} produces the so-called spectral curve \citep{EO07}
\begin{equation} \label{eq4.1.15}
\left(W_1^{(G),0}(x_1)\right)^2-2x_1W_1^{(G),0}(x_1)+2=0,
\end{equation}
which can be easily solved to recover the Stieltjes transform $W_1^{(G),0}(x_1)=x_1-\sqrt{x_1^2-2}$ of the Wigner semi-circle law \eqref{eq1.2.14}. Having this expression at hand then enables one to solve equation \eqref{eq4.1.14} with $(n,l)=(1,1)$ to obtain $W_1^{(G),1}(x_1)$, which then allows for the computation of $W_2^{(G),0}(x_1,x_2)$ through the $(2,0)$ loop equation \eqref{eq4.1.14}, and so on. In fact, as mentioned in \S\ref{s1.1.1}, the $(n,l)$ loop equations \eqref{eq4.1.14} can be solved in a triangular recursive manner and this is moreover true for all analogous $W_n^l$ loop equations discussed in this chapter. The entries of the table below specify the order in which said loop equations must be solved to eventually compute $W_1^{10}$ --- given enough computation power and time, $W_n^l$ can, in theory, be computed for any $n\geq1$ and $l\geq0$.

\begin{center}
\small
\begin{tabular}{c c|c c c c c c c c c c c}
&$l$&$0$&$1$&$2$&$3$&$4$&$5$&$6$&$7$&$8$&$9$&$10$
\\$n$&$W_n^l$
\\ \hline $1$&&$1$&$2$&$4$&$6$&$9$&$12$&$16$&$20$&$25$&$30$&$36$
\\$2$&&$3$&$5$&$8$&$11$&$15$&$19$&$24$&$29$&$35$&&
\\$3$&&$7$&$10$&$14$&$18$&$23$&$28$&$34$&&&&
\\$4$&&$13$&$17$&$22$&$27$&$33$&&&&&&
\\$5$&&$21$&$26$&$32$&&&&&&&&
\\$6$&&$31$&
\end{tabular}
\end{center}

It turns out that obtaining $W_n^l$ loop equations of the form \eqref{eq4.1.14} for the Laguerre and Jacobi $\beta$ ensembles is more involved than in the Gaussian case, especially if one wishes to allow the Laguerre and Jacobi parameters $a,b$ to vary with $N$. This is because the potentials $V(\lambda)$ corresponding to the global scaled Laguerre and Jacobi $\beta$ ensembles (compare the weight $w(\lambda)=e^{-\kappa NV(\lambda)}$ to the results of applying the scalings of Definition \ref{def1.6} to the weights given in equation \eqref{eq1.2.9}) are respectively
\begin{equation} \label{eq4.1.16}
V^{(L)}(\lambda)=\lambda-\frac{a}{\kappa N}\log(\lambda),\qquad V^{(J)}(\lambda)=-\frac{a}{\kappa N}\log(\lambda)-\frac{b}{\kappa N}\log(1-\lambda),
\end{equation}
which means that the auxiliary function $P_n(x_1,J_n)$ specified by equation \eqref{eq4.1.9} is noticeably more complicated than seen in equation \eqref{eq4.1.10}. Thus, loop equations for the Laguerre and Jacobi $\beta$ ensembles were first made explicit in the 2017 work \citep{FRW17} of Forrester, the present author, and Witte. The proofs therein used an adaptation of  Aomoto's method \citep{Aom87} for proving the Selberg integral formula \eqref{eq2.1.3}, which differs slightly in spirit from the derivation of \citep{EM09} reviewed above. (Note that the same adaptation of Aomoto's method was also used by Witte and Forrester in 2015 to obtain loop equations for the circular $\beta$ ensembles \citep{WF15}.) Let us now review how Aomoto's method was used in \citep{FRW17} to tackle the problem of specifying $P_n(x_1,J_n)$.

In keeping with the general formalism prescribed at the beginning of this section, we begin by considering the average
\begin{equation} \label{eq4.1.17}
\mean{\sum_{i_1=1}^N\frac{\partial}{\partial \lambda_{i_1}}\frac{1}{x_1-\lambda_{i_1}}\sum_{i_2,\ldots,i_n=1}^N\frac{1}{(x_2-\lambda_{i_2})\cdots(x_n-\lambda_{i_n})}}
\end{equation}
with respect to the global scaled eigenvalue j.p.d.f.~$\tilde{p}(\lambda_1,\ldots,\lambda_N)$ of either the Gaussian, Laguerre, or Jacobi $\beta$ ensemble; recall that the differential operator $\frac{\partial}{\partial \lambda_{i_1}}$ acts on the product of all terms on its right in the integral formulation of this average, including $\tilde{p}(\lambda_1,\ldots,\lambda_N)$. The equivalent of making the change of variables \eqref{eq4.1.7} in the definition of the unconnected correlators is to apply integration by parts to the above average in order to obtain an identity of the form \eqref{eq4.1.1}. Temporarily taking $a,b>0$ so that the weights $w^{(L)}(\lambda),w^{(J)}(\lambda)$ equal (or limit to) zero at the boundary of the domain of integration (and later relaxing this requirement to $a,b>-1$ using analytic continuation), we see that this average must vanish by the fundamental theorem of calculus. On the other hand, using the Leibniz product rule to expand the integrand shows that
\begin{align}
0&=\mean{\sum_{i_1=1}^N\frac{\partial}{\partial \lambda_{i_1}}\frac{1}{x_1-\lambda_{i_1}}\sum_{i_2,\ldots,i_n=1}^N\frac{1}{(x_2-\lambda_{i_2})\cdots(x_n-\lambda_{i_n})}} \nonumber
\\&=\kappa \tilde{U}_{n+1}(x_1,x_1,J_n)+(\kappa-1)\frac{\partial}{\partial x_1}\tilde{U}_n(x_1,J_n)-\sum_{i=2}^n\frac{\partial}{\partial x_i}\left\{\frac{\tilde{U}_{n-1}(x_1,J_n\setminus\{x_i\})-\tilde{U}_{n-1}(J_n)}{x_i-x_1}\right\} \nonumber
\\&\qquad-\kappa N\mean{\sum_{i_1,\ldots,i_n=1}^N\frac{V'(\lambda_{i_1})}{(x_1-\lambda_{i_1})\cdots(x_n-\lambda_{i_n})}}, \label{eq4.1.18}
\end{align}
where $\tilde{U}_n$ denotes the average \eqref{eq4.1.3} with respect to the eigenvalue j.p.d.f.~$\tilde{p}(\lambda_1,\ldots,\lambda_N)$ of the global scaled classical $\beta$ ensemble at hand. In the Gaussian case, $V'(\lambda_{i_1})=2\lambda_{i_1}=2x_1-2(x_1-\lambda_{i_1})$, so the average on the bottom line of equation \eqref{eq4.1.18} simplifies to $2x_1\tilde{U}^{(G)}_n(x_1,J_n)-2N\tilde{U}^{(G)}_{n-1}(J_n)$ and thus equation \eqref{eq4.1.18} can be seen to be a loop equation on the $\tilde{U}_n^{(G)}$ which, upon the use of an inductive argument based on equation \eqref{eq4.1.4} (see \citep[App.~A]{FRW17}), is equivalent to the loop equation \eqref{eq4.1.12} on the $\tilde{W}_n^{(G)}$. In the Laguerre case, we have from equation \eqref{eq4.1.16} that $V'(\lambda_{i_1})=1-a/(\kappa N\lambda_{i_1})$, so the aforementioned average on the bottom line of equation \eqref{eq4.1.18} instead simplifies to
\begin{equation} \label{eq4.1.19}
\left(1-\frac{a}{\kappa Nx_1}\right)\tilde{U}_n^{(L)}(x_1,J_n)-\frac{a}{\kappa Nx_1}\mean{\sum_{i_1,\ldots,i_n=1}^N\frac{1}{\lambda_{i_1}(x_2-\lambda_{i_2})\cdots(x_n-\lambda_{i_n})}},
\end{equation}
which is not immediately expressible in terms of the $\tilde{U}_n^{(L)}$. Thus, we search for a companion to the average \eqref{eq4.1.17} whose expansion via Aomoto's method produces an identity of the form \eqref{eq4.1.1} that is simpler than equation \eqref{eq4.1.18}, but contains the troublesome average seen in equation \eqref{eq4.1.19} above. It was shown in \citep{FRW17} that it is sufficient to consider the average
\begin{equation} \label{eq4.1.20}
\mean{\sum_{i_1=1}^N\frac{\partial}{\partial \lambda_{i_1}}\sum_{i_2,\ldots,i_n=1}^N\frac{1}{(x_2-\lambda_{i_2})\cdots(x_n-\lambda_{i_n})}},
\end{equation}
which, by a replication of the above arguments, is equal to both zero and
\begin{equation} \label{eq4.1.21}
a\mean{\sum_{i_1,\ldots,i_n=1}^N\frac{1}{\lambda_{i_1}(x_2-\lambda_{i_2})\cdots(x_n-\lambda_{i_n})}}-N\tilde{U}_{n-1}^{(L)}(J_n)-\sum_{i=2}^n\frac{\partial}{\partial x_i}\tilde{U}_{n-1}^{(L)}(J_n).
\end{equation}
Thus, combining equation \eqref{eq4.1.18} with the fact that the expression \eqref{eq4.1.21} equals zero results in a loop equation on the $\tilde{U}_n^{(L)}$ that further transforms via equation \eqref{eq4.1.4} into a loop equation on the $\tilde{W}_n^{(L)}$ \citep[Eq.~(3.13)]{FRW17}. This loop equation is the same as equation \eqref{eq4.1.8}, except with each $W_{n'}$ rewritten as $\tilde{W}_{n'}^{(L)}$ and with the replacement
\begin{multline}
\kappa N\left[V'(x_1)W_n(x_1,J_n)-P_n(x_1,J_n)\right]\mapsto\left(\kappa N-\frac{a}{x_1}\right)\tilde{W}_n^{(L)}(x_1,J_n)
\\-\chi_{n=1}\frac{\kappa N^2}{x_1}-\frac{1}{x_1}\sum_{i=2}^n\frac{\partial}{\partial x_i}\tilde{W}_{n-1}^{(L)}(J_n).
\end{multline}
Note that the second line of the right-hand side of the above is exactly $P_n(x_1,J_n)$ \eqref{eq4.1.9} in the Laguerre case.

Finally, to obtain loop equations for the Jacobi $\beta$ ensemble, one repeats the above exercise of applying integration by parts to the averages \eqref{eq4.1.17} and \eqref{eq4.1.20}, now with respect to the eigenvalue j.p.d.f.~$p^{(J)}(\lambda_1,\ldots,\lambda_N)$ \eqref{eq1.2.81}. However, it is once again observed \citep{FRW17} that this does not result in an equation that can be written fully in terms of the $U_n^{(J)}$ (we do not write $\tilde{U}_n^{(J)}$ nor $\tilde{W}_n^{(J)}$ since no scaling is required in the Jacobi case; see the discussion below Definition \ref{def1.6}) --- one must also consider the third average
\begin{equation} \label{eq4.1.23}
\mean{\sum_{i_1=1}^N\frac{\partial}{\partial \lambda_{i_1}}\lambda_{i_1}\sum_{i_2,\ldots,i_n=1}^N\frac{1}{(x_2-\lambda_{i_2})\cdots(x_n-\lambda_{i_n})}}.
\end{equation}
Combining the identity resulting from applying integration by parts to this average with the Jacobi analogues of the identity \eqref{eq4.1.18} and that obtained by setting the expression \eqref{eq4.1.21} to zero then gives a loop equation on the $U_n^{(J)}$, which is again equivalent to a loop equation on the $W_n^{(J)}$ due to the aforementioned inductive argument involving the relation \eqref{eq4.1.4}. Like in the Gaussian and Laguerre cases, this latter loop equation has the form \eqref{eq4.1.8} with each $W_{n'}$ rewritten as $W_{n'}^{(J)}$ and with the replacement \citep[Prop.~4.4]{FRW17}
\begin{align}
\kappa N\left[V'(x_1)W_n(x_1,J_n)-P_n(x_1,J_n)\right]&\mapsto\left(\frac{b}{1-x_1}-\frac{a}{x_1}\right)W_n^{(J)}(x_1,J_n)+\frac{n-1}{x_1(1-x_1)}W_{n-1}^{(J)}(J_n) \nonumber
\\&\qquad-\frac{\chi_{n=1}}{x_1(1-x_1)}[(a+b+1)N+\kappa N(N-1)] \nonumber
\\&\qquad+\frac{1}{x_1(1-x_1)}\sum_{i=2}^nx_i\frac{\partial}{\partial x_i}W_{n-1}^{(J)}(J_n). \label{eq4.1.24}
\end{align}

As we move on, let us mention that in contrast to the Gaussian case, the $W_n^l$ loop equations for the Laguerre and Jacobi $\beta$ ensembles depend on how $a,b$ vary with $N$. Nonetheless, the spectral curves (i.e., the analogues of equation \eqref{eq4.1.15}) are quadratic in $W_1^0(x_1)$ \citep{FRW17}.

\subsection{From loop equations to the topological recursion} \label{s4.1.2}
The topological recursion was first given in its present form in the 2009 work \citep{EO09} of Eynard and Orantin, where it arose as a refinement of the loop equations derived about five years earlier \citep{Eyn04}, \citep{EO07} for the 1-Hermitian matrix model with polynomial potential. This so-called \textit{1-Hermitian matrix model} is the ensemble of eigenvalues with j.p.d.f.
\begin{equation} \label{eq4.1.25}
p^{(1HMM)}(\lambda_1,\ldots,\lambda_N)=\frac{1}{\mathcal{N}_N^{(1HMM)}}\prod_{i=1}^Ne^{-NV(\lambda_i)}|\Delta_N(\lambda)|^2,
\end{equation}
where the potential $V(\lambda)$ is as in equation \eqref{eq4.1.5} \citep{ACKM93}; note that this ensemble is the same as the general $\beta$ ensembles discussed in the previous subsection, but with $\beta=2$. Let us now sketch the derivation of the topological recursion for the 1-Hermitian matrix model (with $V(\lambda)$ constrained to be polynomial) given in the work \citep{EO09}.

Setting $\kappa=\beta/2=1$ in equation \eqref{eq4.1.8} produces the loop equation on the connected $n$-point correlators of the 1-Hermitian matrix model,
\begin{multline} \label{eq4.1.26}
\sum_{J\subseteq J_n}W_{\#J+1}(x_1,J)W_{n-\#J}(x_1,J_n\setminus J)+W_{n+1}(x_1,x_1,J_n)
\\=N\left[V'(x_1)W_n(x_1,J_n)-P_n(x_1,J_n)\right]
\\-\sum_{i=2}^n\frac{\partial}{\partial x_i}\left\{\frac{W_{n-1}(x_1,J_n\setminus\{x_i\})-W_{n-1}(J_n)}{x_1-x_i}\right\},
\end{multline}
where $J_n$ is as in equation \eqref{eq4.1.4} and $P_n(x_1,J_n)$ is as specified by equation \eqref{eq4.1.9}. Let us now restrict the potential $V(\lambda)$ to be independent of $N$ so that substituting the topological expansion \eqref{eq4.0.5} into the above loop equation shows that $P_n(x_1,J_n)$ has an analogous expansion of the form
\begin{equation} \label{eq4.1.27}
P_n(x_1,J_n)=N^{2-n}\sum_{l=0}^{\infty}\frac{P_n^l(x_1,J_n)}{N^l}.
\end{equation}
Thus, substituting both expansions \eqref{eq4.0.5}, \eqref{eq4.1.27} into equation \eqref{eq4.1.26} and equating terms of like order in $N$ then produces the loop equation on the correlator expansion coefficients,
\begin{multline} \label{eq4.1.28}
\sum_{J\subseteq J_n}\sum_{k=0}^lW_{\#J+1}^k(x_1,J)W_{n-\#J}^{l-k}(x_1,J_n\setminus J)+W_{n+1}^{l-2}(x_1,x_1,J_n)
\\=V'(x_1)W_n^l(x_1,J_n)-P_n^l(x_1,J_n)-\sum_{i=2}^n\frac{\partial}{\partial x_i}\left\{\frac{W_{n-1}^l(x_1,J_n\setminus\{x_i\})-W_{n-1}^l(J_n)}{x_1-x_i}\right\}.
\end{multline}
In particular, taking $(n,l)=(1,0)$ yields the spectral curve (cf.~equation \eqref{eq4.1.15}):
\begin{equation} \label{eq4.1.29}
\left(W_1^0(x_1)\right)^2-V'(x_1)W_1^0(x_1)+P_1^0(x_1)=0.
\end{equation}
Having assumed $V(\lambda)$ to be polynomial, equation \eqref{eq4.1.9} tells us that $P_1(x_1)$ must also be polynomial in $x_1$ of degree one less than $V'(x_1)$. Thus, equation \eqref{eq4.1.29} can be shown (see, e.g., \citep{EKR18}) to have solution of the form
\begin{equation} \label{eq4.1.30}
W_1^0(x_1)=\frac{1}{2}\left(V'(x_1)\pm M(x_1)\sqrt{\sigma(x_1)}\right),
\end{equation}
where $M(x_1)$ and $\sigma(x_1)$ are polynomials such that the latter is of even degree, say $2s$, with all roots distinct and $M(x_1)^2\sigma(x_1)=\left(V'(x_1)\right)^2-4P_1^0(x_1)$. From the perspective of random matrix theory, one must take the negative sign in equation \eqref{eq4.1.30} so that $W_1^0(x_1)\sim1/x_1$ as $x_1\to\infty$, in line with, say, equation \eqref{eq2.4.15}.

The first step in simplifying the above loop equations is to circumvent the multivalued nature of $W_1^0(x_1)$ by viewing it as a single-valued holomorphic function on an appropriate two-sheeted cover of the complex plane. This cover can be constructed by taking two copies of the complex plane, gluing them along the $s$ branch cuts of $W_1^0(x_1)$ (which are the $s$ intervals along which the degree-$2s$ polynomial $\sigma(x_1)$ is negative) and then adding a point at infinity to each of the two complex planes so that the resulting surface is a genus $s-1$ compact Riemann surface $\Sigma$ \citep{EKR18}; see Figure \ref{fig4.1} below. Such a covering space $\Sigma$ comes equipped with a holomorphic projection $x:\Sigma\to\mathbb{C}\cup\{\infty\}$ such that for all $x_1\in\mathbb{C}\cup\{\infty\}$ not a root of $\sigma(x_1)$, there exist two distinct points $z,z'\in\Sigma$ such that $x(z)=x(z')=x_1$. More importantly, there exists a single-valued meromorphic function $y:\Sigma\to\mathbb{C}$ such that whenever $z,z'\in\Sigma$ are such that $z\neq z'$, but $x(z)=x(z')$, $y(z)$ and $y(z')$ are the two values of $W_1^0(x(z))$ given by equation \eqref{eq4.1.30}. Here on out, let us work in the one-cut case, which corresponds to assuming that $\sigma(x_1)$ is quadratic with two distinct roots, $u<v$ say, and thus $\Sigma$ is the Riemann sphere $\mathbb{CP}^1$ (being genus zero).

\begin{figure}[H]
        \centering
\captionsetup{width=.9\linewidth}
        \includegraphics[width=0.88\textwidth]{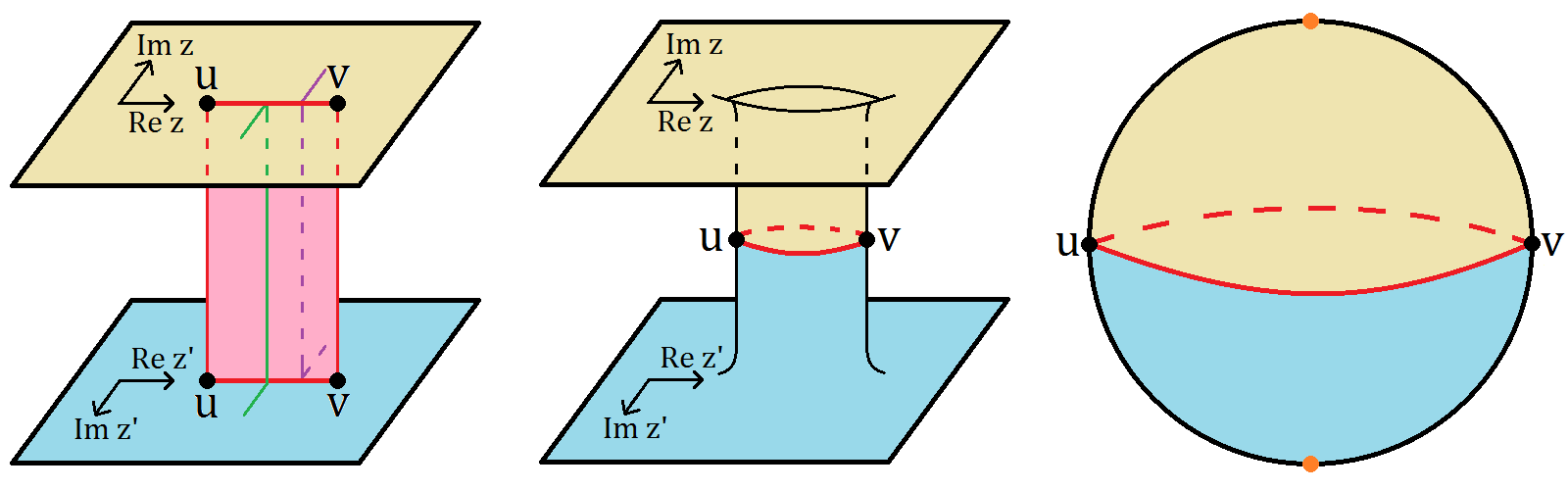}
        \caption[A construction of a two-sheeted covering of the complex plane]{On the left, we have two copies of the complex plane with their branch cuts $[u,v]$ identified; note that the blue plane is obtained from the beige plane through a reflection in the real axis to ensure that if one approaches the branch cut from the beige positive (negative) half plane, one leaves the branch cut by entering the blue negative (positive) half plane. In the middle image, we have opened up the branch cuts into circles and pulled them together so as to form a tube. Compactifying by adding (orange) points at infinity to the beige and blue planes results in the Riemann sphere on the right. Note that if there were more cuts, there would be more tubes which would end up becoming handles in the final surface. For each $x_1\in\mathbb{C}\cup\{\infty\}\setminus\{u,v\}$, there exists a $z$ in the beige hemisphere and $z'$ in the blue hemisphere such that $x(z)=x(z')=x_1$ and $y(z),y(z')$ are the two values of $W_1^0(x_1)$ given in equation \eqref{eq4.1.30}.} \label{fig4.1}
\end{figure}

In the one-cut case, one can use the Joukowsky transform to see that when the endpoints of our branch cut are $u<v$,
\begin{equation} \label{eq4.1.31}
x(z)=\frac{v-u}{4}\left(z+\frac{1}{z}\right)+\frac{u+v}{2}
\end{equation}
has the properties required of the projection map described above: this mapping has exactly $u,v$ as branch points and every $x_1\in\mathbb{C}\cup\{\infty\}\setminus\{u,v\}$ has two preimages under $x$ (note that $x(-1)=u$ and $x(1)=v$). Furthermore, we can choose a sign in equation \eqref{eq4.1.30} to write
\begin{equation} \label{eq4.1.32}
y(z)=\frac{1}{2}\left(V'(x(z))-\frac{v-u}{4}M(x(z))\left(z-\frac{1}{z}\right)\right);
\end{equation}
observe that for all $z\in\Sigma=\mathbb{CP}^1$, $x(z)=x(1/z)$, so if $z\neq\pm1$ is a preimage of $x_1$, $y(z)$ and $y(1/z)$ are the two values of $W_1^0(x_1)$ given in equation \eqref{eq4.1.30}. This reformulation of $(x_1,W_1^0(x_1))$ as the coordinates $(x(z),y(z))$ on $\Sigma$ extends naturally to the spectral curve \eqref{eq4.1.29} and, indeed, to the general $(n,l)$ loop equation \eqref{eq4.1.28}. Moreover, since the $(n,l)$ loop equation specifying $W_n^l(x_1,\ldots,x_n)$ is linear in this correlator expansion coefficient (when $(n,l)\neq(1,0)$), we see by induction that each of these correlators have the same branching structure as $W_1^0(x_1)$ and so can be viewed as single-valued holomorphic functions on $\Sigma^n$. Thus, we are led to rewrite equation \eqref{eq4.1.28} in terms of the coordinates $(x(z),y(z))$ on $\Sigma$. In fact, to facilitate upcoming residue calculus, we are encouraged to introduce differential forms on $\Sigma$, as well. Hence, let us define for $z_1,\ldots,z_n\in\Sigma$ and $(n,l)\neq(1,0),(2,0)$,
\begin{align}
\omega_1^0(z_1)&:=y(z_1)\mathrm{d}x(z_1), \label{eq4.1.33}
\\ \omega_2^0(z_1,z_2)&:=\left[W_2^0(x(z_1),x(z_2))+\frac{1}{(x(z_1)-x(z_2))^2}\right]\mathrm{d}x(z_1)\mathrm{d}x(z_2), \label{eq4.1.34}
\\ \omega_n^l(z_1,\ldots,z_n)&:=W_n^l(x(z_1),\ldots,x(z_n))\mathrm{d}x(z_1)\cdots\mathrm{d}x(z_n), \label{eq4.1.35}
\end{align}
where equation \eqref{eq4.1.33} is a specialisation of equation \eqref{eq4.1.35} based on the replacement of equation \eqref{eq4.1.30} with equation \eqref{eq4.1.32}, equation \eqref{eq4.1.34} is chosen to differ from equation \eqref{eq4.1.35} in a way that ensures that $\omega_2^0(z_1,z_2)$ coincides with the Bergman kernel (also known as the fundamental differential of the second kind; see, e.g., \citep{EO09}, \citep{EKR18}) $\mathrm{d}z_1\mathrm{d}z_2/(z_1-z_2)^2$ on $\mathbb{CP}^1$\footnote{It was first shown in \citep{AJM90}, \citep{ACKM93} that for the one-cut 1-Hermitian matrix model, $W_2^0$ has the universal form implied by the right-hand side of equation \eqref{eq4.1.34} being equal to the Bergman kernel on $\mathbb{CP}^1$. This was also observed \citep{WF14}, \citep{FRW17} for the classical $\beta$ ensembles discussed in \S\ref{s4.1.1}.}, and the $W_n^l(x(z_1),\ldots,x(z_n))$ within equation \eqref{eq4.1.35} is to be interpreted as the single-valued function on $(\mathbb{CP}^1)^n$ obtained from the loop equation \eqref{eq4.1.28} after it has been rewritten in terms of the coordinates $(x(z),y(z))$.

With definitions \eqref{eq4.1.31}--\eqref{eq4.1.35} in hand and noting that $V'(x_1)-2W_1^0(x_1)$ should be replaced by $y(1/z_1)-y(z_1)$ to be consistent with our choice of replacing $W_1^0(x_1)$ with $y(z_1)$, we rewrite equation \eqref{eq4.1.28} for $(n,l)\neq(1,0),(2,0)$ as (cf.~\citep[Sec.~3.3]{Eyn16})
\begin{align}
\omega_n^l(z_1,J_n')&=\frac{1}{\omega_1^0(1/z_1)-\omega_1^0(z_1)}\Bigg[\omega_{n+1}^{l-2}(z_1,z_1,J_n')+\sum_{\substack{J\subseteq J_n'\\0\leq k\leq l}}^{\circ}\omega_{\#J+1}^k(z_1,J)\omega_{n-\#J}^{l-k}(z_1,J_n'\setminus J)\Bigg] \nonumber
\\&\quad-\frac{1}{\omega_1^0(1/z_1)-\omega_1^0(z_1)}\Bigg[\chi_{n=1,l=2}\lim_{\xi\to z_1}\frac{\mathrm{d}x(z_1)\mathrm{d}x(\xi)}{(x(z_1)-x(\xi))^2} \nonumber
\\&\hspace{11em}+2\sum_{i=2}^n\frac{\mathrm{d}x(z_1)\mathrm{d}x(z_i)\omega_{n-1}^l(z_1,J_n'\setminus\{z_i\})}{(x(z_1)-x(z_i))^2}\Bigg] \nonumber
\\&\quad+\frac{\mathrm{d}x(z_1)\cdots\mathrm{d}x(z_n)}{y(1/z_1)-y(z_1)}\Bigg[P_n^l(x(z_1),J_n) +\sum_{i=2}^n\frac{4}{v-u}\frac{z_i^2}{z_i^2-1}\nonumber
\\&\hspace{11em}\times\frac{\partial}{\partial z_i}\left\{\frac{W_{n-1}^l(x(z_1),J_n\setminus\{x(z_i)\})-W_{n-1}^l(J_n)}{x(z_1)-x(z_i)}\right\}\Bigg], \label{eq4.1.36}
\end{align}
where we now have $J_n=(x(z_2),\ldots,x(z_n))$, $J_n'=(z_2,\ldots,z_n)$, and the sum $\sum^{\circ}$ excludes the terms corresponding to $(k,J)=(0,\{z_i\}),(l,J_n'\setminus\{z_i\})$; the terms of the second and third line arise from the fact that occurrences of $\omega_2^0$ in the top line must be accompanied by the extra term shown in equation \eqref{eq4.1.34}, as compared to equation \eqref{eq4.1.35}. Now, replace $z_1$ by $\xi$ in equation \eqref{eq4.1.36}, multiply both sides by a half times the meromorphic differential \citep{EO09}
\begin{equation}
\mathrm{d}S_{\xi,1/\xi}(z_1):=\int_{z=1/\xi}^{z=\xi}\omega_2^0(z_1,z)=\left(\frac{1}{z_1-\xi}-\frac{1}{z_1-1/\xi}\right)\mathrm{d}z_1,\quad\xi\in\mathbb{CP}^1\setminus\{\pm1\},
\end{equation}
and take the sum of the residues at $\xi=\pm1$ on both sides of the resulting equation. This shows that the left-hand side simply reduces as
\begin{multline} \label{eq4.1.38}
\frac{1}{2}\sum_{a=\pm1}\underset{\xi=a}{\mathrm{Res}}\,\mathrm{d}S_{\xi,1/\xi}(z_1)\omega_n^l(\xi,J_n')=\frac{1}{2}\sum_{a=\pm1}\underset{\xi=a}{\mathrm{Res}}\,\left(\frac{1}{z_1-\xi}-\frac{1}{z_1-1/\xi}\right)\omega_n^l(\xi,J_n')\,\mathrm{d}z_1
\\=\sum_{a=\pm1}\underset{\xi=a}{\mathrm{Res}}\,\frac{1}{z_1-\xi}\omega_n^l(\xi,J_n')\,\mathrm{d}z_1=\underset{\xi=z_1}{\mathrm{Res}}\,\frac{1}{\xi-z_1}\omega_n^l(\xi,J_n')\,\mathrm{d}z_1=\omega_n^l(z_1,J_n');
\end{multline}
the second equality is obtained by changing variables $\xi\mapsto1/\xi$ in $\underset{\xi=a}{\mathrm{Res}}\omega_n^l(\xi,J_n')/(z_1-1/\xi)$ and using the fact that $\omega_n^l(1/\xi,J_n')=-\omega_n^l(\xi,J_n')$ (this can be checked inductively through equation \eqref{eq4.1.36} \citep{EO09}, \citep[Sec.~3.3]{Eyn16}), while the third equality follows from the fact that the sum of all residues of a meromorphic form on $\mathbb{CP}^1$ must vanish (again, an inductive argument via equation \eqref{eq4.1.36} shows that $\omega_n^l(\xi,J_n')$ only has residues at $\xi=\pm1$). On the other hand, the right-hand side simplifies remarkably to
\begin{equation} \label{eq4.1.39}
\sum_{a=\pm1}\underset{\xi=a}{\mathrm{Res}}\,K(\xi,z_1)\Bigg[\omega_{n+1}^{l-2}(\xi,1/\xi,J_n')+\sum_{\substack{J\subseteq J_n'\\0\leq k\leq l}}^{\circ}\omega_{\#J+1}^k(\xi,J)\omega_{n-\#J}^{l-k}(1/\xi,J_n'\setminus J)\Bigg],
\end{equation}
where we define the \textit{recursion kernel} as
\begin{equation} \label{eq4.1.40}
K(\xi,z_1):=\frac{\mathrm{d}S_{\xi,1/\xi}(z_1)}{2(\omega_1^0(\xi)-\omega_1^0(1/\xi))}.
\end{equation}
The simplified form \eqref{eq4.1.39} results from the fact that all of the terms below the first line of equation \eqref{eq4.1.36} either do not have residue at $z_1=\pm1$ or the sum of their non-zero residues vanish by symmetry arguments \citep[Sec.~3.3]{Eyn16}. Finally, setting the left- and right-hand sides \eqref{eq4.1.38} and \eqref{eq4.1.39} equal to each other gives us the (Eynard--Orantin) \textit{topological recursion} for the one-cut 1-Hermitian matrix model with polynomial potential:
\begin{multline} \label{eq4.1.41}
\omega_n^l(z_1,J_n')=\sum_{a=\pm1}\underset{\xi=a}{\mathrm{Res}}\,K(\xi,z_1)\Bigg[\omega_{n+1}^{l-2}(\xi,1/\xi,J_n')+\sum_{\substack{J\subseteq J_n'\\0\leq k\leq l}}^{\circ}\omega_{\#J+1}^k(\xi,J)\omega_{n-\#J}^{l-k}(1/\xi,J_n'\setminus J)\Bigg].
\end{multline}
Observe that this topological recursion formula can be seen to be a simplification of the loop equation \eqref{eq4.1.28} obtained by essentially replacing the right-hand side by $W_n^l(x_1,J_n)$. Note that this means that knowledge of the $P_n^l(x_1,J_n)$ is not needed to compute the $\omega_n^l$ --- recall from \S\ref{s4.1.1} that without the topological recursion formula available, other techniques, such as Aomoto's method, must be used to deal with $P_n^l(x_1,J_n)$.

Let us now explain why the topological recursion is so named. Assume for simplicity that the potential $V(\lambda)$ is the Gaussian potential $\lambda^2$ (it has however been known since the 1970s \citep{Hoo74}, \citep{BIPZ78}, \citep{BIZ80} that the following argument applies formally to more general potentials; see, e.g., \citep[Ch.~2]{Eyn16}). Then, by the discussion of \S\ref{s3.3.1}, the $W_n^l$ are zero for odd values of $l$ and are otherwise generating functions for the compact, connected, orientable ribbon graphs of genus $l/2$ that can be built from $n$ polygons. Thus, we should take $l$ to be even in equation \eqref{eq4.1.41} and discard terms in the sum involving odd values of $k$. Moreover, in keeping with the discussion involving Figure \ref{fig3.7} and that following Proposition \ref{prop3.10x}, it is convenient to visualise $\omega_n^l$ as a genus $l/2$ surface with $n$ holes or punctures. We can then interpret the topological recursion formula \eqref{eq4.1.41} in the following diagrammatic fashion \citep{EO07}, \citep{EO09}: The process of multiplying by the recursion kernel $K(\xi,z_1)$ and then taking the sum of the residues at $\xi=\pm1$ is represented by the gluing of a pair of pants (i.e., a sphere with three holes) to either two holes of a genus $l/2-1$ surface with $n+1$ holes to increase the genus and lower the number of holes by one each, or to one hole each of two separate surfaces in order to form a genus $l/2$ surface with $n$ holes; see Figure \ref{fig4.2} below. The recursion thus runs over the Euler characteristic $\chi=2-l+n$.

\begin{figure}[H]
        \centering
\captionsetup{width=.9\linewidth}
        \includegraphics[width=0.88\textwidth]{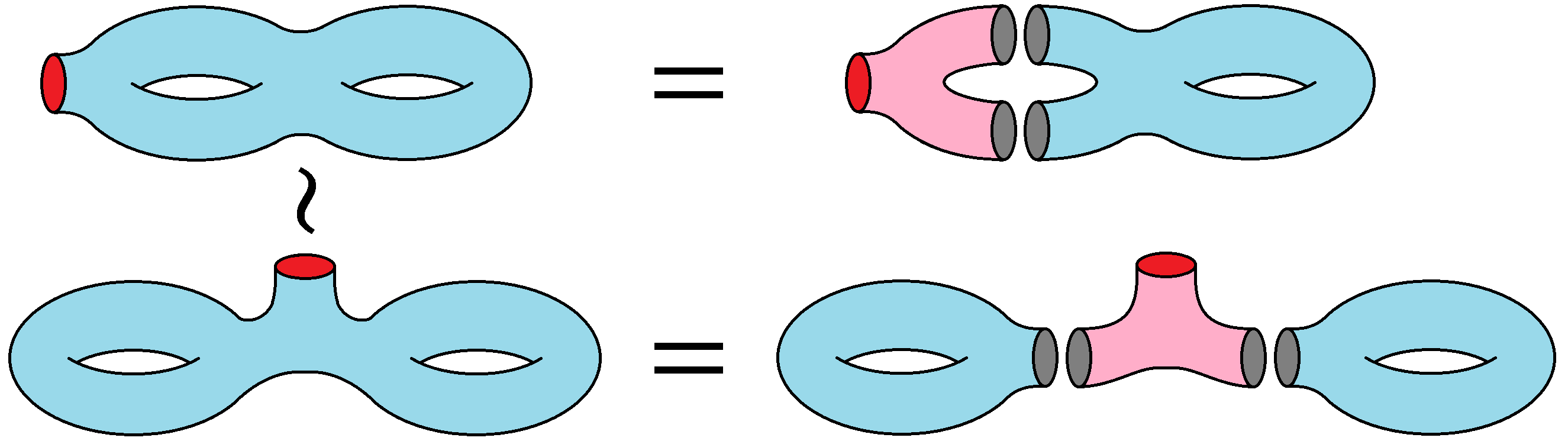}
        \caption[A diagrammatic interpretation of the topological recursion]{On the left, we have two homeomorphic genus $2$ surfaces with $1$ hole representing the term $\omega_1^4(z_1)$ in the left-hand side of the topological recursion formula \eqref{eq4.1.41} with $(n,l)=(1,4)$ (having taken $n$ to be the number of holes and $l/2$ to be the genus). In the top row, we see that gluing a pair of pants (in pink) representing the operator $\sum_{a=\pm1}\underset{\xi=a}{\mathrm{Res}}K(\xi,z_1)$ to a genus $1$ surface with $2$ holes representing $\omega_2^2(\xi,1/\xi)$ produces the surface representing $\omega_1^4(z_1)$. In the bottom row, we see that gluing such a pair of pants to two genus $1$ surfaces with $1$ hole each produces the surface on the bottom left --- the two blue surfaces on the bottom right represent $\omega_1^2(\xi)$ and $\omega_1^2(1/\xi)$.} \label{fig4.2}
\end{figure}

The diagrammatic interpretation of the topological recursion described above can be traced back to Tutte's equation and recursion \citep{Tut63}, \citep{WL72} (see also \citep[Sec.~1.3]{Eyn16} for a textbook treatment) on the coefficients of the genus expansions \eqref{eq3.3.22} of the GUE mixed cumulants $c_{k_1,\ldots,k_n}^{(GUE)}$. In the language of ribbon graphs (rather than topological maps, as considered by Tutte), the idea behind Tutte's recursion is to simply observe that a given connected, orientable ribbon graph with $n$ polygons can be obtained from a similar such ribbon graph with $n+1$ polygons by shrinking a particular ribbon to identify its ends so that the two relevant polygons merge into one --- if instead deleting said ribbon results in a connected (disconnected) ribbon graph with two unpaired polygon edges, then we have a ribbon graph analogue of the decomposition displayed in the top (bottom) row of Figure \ref{fig4.2}.

The key insight of the seminal work \citep{EO09} was that the mechanism underlying the topological recursion formula \eqref{eq4.1.41} is common to many enumerative problems in random matrix theory and algebraic geometry. In the algebro-geometric setting, one abstracts by taking the spectral curve to be a triple $(\Sigma,x,y)$ of a Riemann surface $\Sigma$ and two meromorphic functions $x,y:\Sigma\to\mathbb{C}\cup\{\infty\}$, defines $\omega_1^0(z_1)=y(z_1)\mathrm{d}x(z_1)$, defines $\omega_2^0(z_1,z_2)$ to be the Bergman kernel on $\Sigma^2$, and then recursively defines all other $\omega_n^l$ through equation \eqref{eq4.1.41} reformulated in terms of the genus $g=l/2$ (i.e., setting $\omega_n^{l'}=0$ whenever $l'$ is odd) and with the sum over $a=\pm1$ replaced by a sum over all branch points, equivalently zeroes of the differential form $\mathrm{d}x$. Thus, for particular choices of initial data $(\Sigma,x,y,\omega_2^0)$, the topological recursion has been shown to govern, for example, Gromov--Witten invariants of $\mathbb{CP}^1$, various Hurwitz numbers, dessins d'enfants, the intersection numbers mentioned at the beginning of Section \ref{s3.3}, and Weil--Petersson volumes; see, e.g., \citep{EO09}, \citep{DMSS13}, \citep{NS14}, \citep{CD21} and references therein. Indeed, Mirzakhani's groundbreaking recursion \citep{Mirz07}, \citep{Mirz07a} for Weil--Petersson volumes has been shown to be an instance of the topological recursion \citep{Eyn11}.

Since the work of Eynard and Orantin \citep{EO09}, the topological recursion has been generalised and reformulated in a variety of ways: Eynard and Marchal \citep{EM09} showed that the topological recursion formula \eqref{eq4.1.41} can be applied to general $\beta$ ensembles upon redefining the recursion kernel $K(\xi,z_1)$ \eqref{eq4.1.40} in a suitable $\beta$-dependent way; Bouchard et al.~ \citep{BE13}, \citep{BHLMR14} studied generalisations of the topological recursion involving higher order spectral curves by introducing terms on the right-hand side of equation \eqref{eq4.1.41}; Do and Norbury \citep{ND18} and Chekhov and Norbury \citep{CN19} have studied aspects of the topological recursion when the spectral curve is irregular (i.e., the eigenvalue density has hard edges) --- in particular, they showed that the connected correlators of the LUE with $a=0$ and the Legendre unitary ensemble, equivalently the sJUE with $a=b=0$, obey the topological recursion (like \citep{FRW17}, their results provide an alternative to Proposition \ref{prop3.5} and Corollary \ref{C3.4} for computing the moment expansion coefficients of the LUE with $a=0$ and the Legendre unitary ensemble); Kontsevich and Soibelman \citep{KS18} have recently reformulated the topological recursion by replacing the initial data with a certain set of tensors derived from $\omega_1^0$ and $\omega_2^0$.

\setcounter{equation}{0}
\section{Loop Equations for the Antisymmetrised Laguerre Ensemble} \label{s4.2}
In this section, we derive loop equations for the mixed moments $\tilde{m}^{(\mathcal{J}_1)}_{k_1,\ldots,k_n}$ \eqref{eq4.1.2}, the unconnected $n$-point correlators $\tilde{U}_n^{(\mathcal{J}_1)}(x_1,\ldots,x_n)$ \eqref{eq4.1.3}, the corresponding connected correlators $\tilde{W}_n^{(\mathcal{J}_1)}(x_1,\ldots,x_n)$ \eqref{eq4.1.4}, and finally some correlator expansion coefficients $W_n^{(\mathcal{J}_1),l}(x_1,\ldots,x_n)$ \eqref{eq4.0.5} of the global scaled $(N,N)$ antisymmetrised Laguerre ensemble, whose definition we recall from Proposition~\ref{prop3.11}. To improve clarity, we begin by introducing some new notation (local to this section).

Thus, take $N$ to be an even integer and introduce the scaled real $N\times N$ Ginibre matrix $X\in\mathbb{M}_{N\times N}(\mathbb{R})$ with p.d.f.
\begin{equation} \label{eq4.2.1}
P(X)=\left(\frac{N}{2\pi}\right)^{N^2/2}\exp\left(-\tfrac{N}{2}\Tr(X^TX)\right).
\end{equation}
Moreover, denote the elementary antisymmetric matrix $J_N$ \eqref{eq1.1.5} by simply the letter $J$ and define $B:=X^TJX$. Then, $X$ is statistically equal to the scaled real Ginibre matrix $\tilde{G}=\tilde{G}_1$ of Propositions \ref{prop3.11} and \ref{prop3.13}, and therefore relates to the real Ginibre matrix $G$ of Definition \ref{def1.3} through the equation $X=\sqrt{2/N}G$. Likewise, the antisymmetric matrix product $B=X^TJX$ equals the matrix $\tilde{\mathcal{J}}_1$ \eqref{eq3.3.57} of Proposition \ref{prop3.11} and is thus related to the unscaled matrix $\mathcal{J}_1$ \eqref{eq4.0.2} according to the equality $B=2\mathcal{J}_1/N$.

Next, let $\mean{\,\cdot\,}$ denote averages with respect to the p.d.f.~$P(X)$ \eqref{eq4.2.1} and recall that, e.g., $\mean{\partial_{X_{ab}}f(X)}$ should be read as ``take the partial derivative of the product $f(X)P(X)$ with respect to $X_{ab}$ and integrate the result over $M_{N\times N}(\mathbb{R})$", while $\mean{\{\partial_{X_{ab}}f(X)\}}$ reads ``multiply the partial derivative of $f(X)$ with respect to $X_{ab}$ by $P(X)$ and then integrate the result over $M_{N\times N}(\mathbb{R})$". For $k_1,\ldots,k_n\in\mathbb{N}$, the mixed moments and unconnected $n$-point correlators of $B$ are respectively specified by (recall equations \eqref{eq4.1.2}, \eqref{eq4.1.3} and cf.~equation \eqref{eq1.1.17})
\begin{align}
m_{k_1,\ldots,k_n}&=\mean{\prod_{i=1}^n\Tr\,B^{k_i}}, \label{eq4.2.2}
\\ U_n(x_1,\ldots,x_n)&=\sum_{k_1,\ldots,k_n=0}^{\infty}\frac{m_{k_1,\ldots,k_n}}{x_1^{k_1+1}\cdots x_n^{k_n+1}}; \label{eq4.2.3}
\end{align}
note that whenever any of the $k_i$ are odd, the trace of the antisymmetric matrix $B^{k_i}$ is zero, so $m_{k_1,\ldots,k_n}$ vanishes. The connected $n$-point correlator $W_n(x_1,\ldots,x_n)$ is most easily specified in terms of the above through equation \eqref{eq4.1.4}, but has an alternative charcterisation as a generating function of the mixed cumulants $c_{k_1,\ldots,k_n}$ --- which are defined implicitly via the moment-cumulants relation \eqref{eq1.1.28} --- through the expansion \eqref{eq1.1.29}
\begin{equation} \label{eq4.2.4}
W_n(x_1,\ldots,x_n)=\sum_{k_1,\ldots,k_n=0}^{\infty}\frac{c_{k_1,\ldots,k_n}}{x_1^{k_1+1}\cdots x_n^{k_n+1}}.
\end{equation}
These quantities are the same as the scaled analogues discussed in \S\ref{s3.3.3}; we have only simplified notation by omitting the tilde and the superscript $(\mathcal{J}_1)$ so that, e.g., $\tilde{c}_{k_1,\ldots,k_n}^{(\mathcal{J}_1)}$ is now written simply as $c_{k_1,\ldots,k_n}$. If we use the superscript $(\mathcal{J}_1)$, without a tilde, to denote statistics of the unscaled ensemble represented by the matrix $\mathcal{J}_1$ \eqref{eq4.0.2}, we have
\begin{align}
m_{k_1,\ldots,k_n}&=\left(\frac{2}{N}\right)^{k_1+\cdots+k_n}m_{k_1,\ldots,k_n}^{(\mathcal{J}_1)},
\\ c_{k_1,\ldots,k_n}&=\left(\frac{2}{N}\right)^{k_1+\cdots+k_n}c_{k_1,\ldots,k_n}^{(\mathcal{J}_1)},
\\ U_n(x_1,\ldots,x_n)&=\left(\frac{N}{2}\right)^nU_n^{(\mathcal{J}_1)}(Nx_1/2,\ldots,Nx_n/2),
\\ W_n(x_1,\ldots,x_n)&=\left(\frac{N}{2}\right)^nW_n^{(\mathcal{J}_1)}(Nx_1/2,\ldots,Nx_n/2).
\end{align}
Now, inserting the genus expansion \eqref{eq3.3.61} for the mixed cumulants $c_{k_1,\ldots,k_n}=\tilde{c}_{k_1,\ldots,k_n}^{(\mathcal{J}_1)}$ into equation \eqref{eq4.2.4} shows that $W_n(x_1,\ldots,x_n)$ admits a large $N$ expansion of the form \eqref{eq4.0.5} and we may thus implicitly define the correlator expansion coefficients $W_n^l(x_1,\ldots,x_n)$ accordingly.

Finally, let us adopt the Einstein summation convention, meaning that repeated matrix indices in a product of matrix entries are to be summed over $1,\ldots,N$. For example, $A_{ab}B_{bc}$ denotes the sum over $b=1,\ldots,N$ of the product of the $(a,b)$ entry of $A$ and $(b,c)$ entry of $B$. Thus, $A_{ab}B_{bc}=(AB)_{ac}$, the $(a,c)$ entry of the matrix product $AB$. In particular, we must take care to remember that $J_{aa}=\Tr(J)=0$ and $\partial_{X_{ab}}X_{ab}=\sum_{a,b=1}^N1=N^2$.

As discussed at the beginning of Section \ref{s4.1}, the first step in deriving our loop equations is to use integration by parts on a suitable perturbation of the average \eqref{eq4.2.2} (recall that a perturbation of the average \eqref{eq4.2.3} can also serve as an alternative starting point) in order to produce an identity of the form \eqref{eq4.1.1}. Let us recall from the discussion of Section \ref{s3.3} (particularly that preceding Example \ref{ex3.2}) that traces $\Tr\,B^k$ can be conveniently represented by $k$-gons formed by chaining together edges labelled by entries of $B$ into a closed loop. The intuition behind the first average we consider in our development is that we would like to delete a term in the average \eqref{eq4.2.2} specifying $m_{k_1+1,k_2,\ldots,k_n}$ to cut open the loop representing $\Tr\,B^{k_1+1}$ and then reintroduce that same term to join the loop back up, except that the reintroduced term is now extracted from the p.d.f.~$P(X)$ \eqref{eq4.2.1} by way of acting on said p.d.f.~with an appropriate differential operator\footnote{This intuition is essentially the spirit behind the terminology `loop equations' and is also the reason why loop equations are sometimes referred to as cut-and-join equations.}.

\begin{lemma} \label{L4.1}
Let us write $\Tr\,B^{k_1+1}=(B^{k_1})_{ad}B_{da}$ (using the Einstein summation convention) and then erase the last term so that we are left with $(B^{k_1})_{ad}$. Moreover, let us define $I_n:=(k_2,\ldots,k_n)$ and use the shorthand $\Tr\,B^{I_n}:=\Tr\,B^{k_2}\Tr\,B^{k_3}\cdots\Tr\,B^{k_n}$. We have that
\begin{equation} \label{eq4.2.9}
\partial_{(X^T)_{ab}}(J^T)_{bc}\partial_{X_{cd}}\exp\left(-\tfrac{N}{2}\Tr(X^TX)\right)=N^2B_{da}\exp\left(-\tfrac{N}{2}\Tr(X^TX)\right)
\end{equation}
and, consequently,
\begin{equation} \label{eq4.2.10}
\mean{(B^{k_1})_{ad}\Tr\,B^{I_n}\partial_{(X^T)_{ab}}(J^T)_{bc}\partial_{X_{cd}}}=N^2m_{k_1+1,I_n}.
\end{equation}
\end{lemma}
\begin{proof}
We give a short proof as a warm-up. First note by the Leibniz product rule that
\begin{multline*}
\partial_{(X^T)_{ab}}\partial_{X_{cd}}\exp\left(-\tfrac{N}{2}\Tr(X^TX)\right)=-N\partial_{(X^T)_{ab}}X_{cd}\exp\left(-\tfrac{N}{2}\Tr(X^TX)\right)
\\=\left(N^2(X^T)_{ab}X_{cd}-N\chi_{a=d,b=c}\right)\exp\left(-\tfrac{N}{2}\Tr(X^TX)\right).
\end{multline*}
Multiplying this result by $(J^T)_{bc}$ gives
\begin{equation*}
\left(N^2(X^TJ^TX)_{ad}-NJ_{bb}\chi_{a=d,b=c}\right)\exp\left(-\tfrac{N}{2}\Tr(X^TX)\right),
\end{equation*}
which reduces to the right-hand side of equation \eqref{eq4.2.9} upon recalling that $J_{bb}=\Tr\,J=0$ and observing that $(X^TJ^TX)^T=X^TJX$, so $(X^TJ^TX)_{ad}=B_{da}$. Pre-multiplying both sides of equation \eqref{eq4.2.9} by $(N/2\pi)^{N^2/2}(B^{k_1})_{ad}\Tr\,B^{I_n}$ and then integrating over $\mathbb{M}_{N\times N}(\mathbb{R})$ while keeping track of notation yields equation \eqref{eq4.2.10}.
\end{proof}

From the above, one might suggest that our desired starting point be the average $\mean{\partial_{(X^T)_{ab}}(J^T)_{bc}\partial_{X_{cd}}(B^{k_1})_{ad}\Tr\,B^{I_n}}$, as starting at the average of a full derivative enables us to use the fundamental theorem of calculus to immediately show that said average is zero --- this is indeed very similar to the initial average \eqref{eq4.1.17} seen in the loop equation analysis of the Laguerre and Jacobi $\beta$ ensembles given in \citep{FRW17}, as reviewed in \S\ref{s4.1.1}. However, one soon sees that this path requires us to consider averages of the form $\mean{\Tr(B^pX^TX)\Tr\,B^q}$, which are not immediately expressible in terms of the mixed moments. As seen in \S\ref{s4.1.1}, this inevitably leads to a game of finding identities that can be combined with the initial analogue of equation \eqref{eq4.1.1} in order to obtain a valid loop equation on the mixed moments. Such games can at times prove to be a tedious matter of trial and error, especially for large products of matrices, and if one does not have much experience in the exercise.

It turns out that, for reasons soon to be made clear, a lot of guesswork can be avoided by beginning our analysis at the average
\begin{equation} \label{eq4.2.11}
\mean{\left(\partial_{(X^T)_{ab}}+NX_{ba}\right)(J^T)_{bc}\left(\partial_{X_{cd}}+N(X^T)_{dc}\right)(B^{k_1})_{ad}\Tr\,B^{I_n}}.
\end{equation}
Thus, in \S\ref{s4.2.1} forthcoming, we apply integration by parts to this average to obtain loop equations on the mixed moments $m_{k_1,\ldots,k_n}$ in a relatively straightforward manner. These loop equations are then easily transformed into loop equations on the unconnected correlators $U_n(x_1,\ldots,x_n)$. Then, in \S\ref{s4.2.2}, we use equations \eqref{eq4.1.4} and \eqref{eq4.0.5} to obtain loop equations for the connected correlators $W_n(x_1,\ldots,x_n)$ and their expansion coefficients $W_n^l(x_1,\ldots,x_n)$ from the $U_n$ loop equations of \S\ref{s4.2.1}. We then conclude this section with some explicit computations and discussions in relation to $W_1^0(x_1)$ and $W_2^0(x_1,x_2)$.

\subsection{Loop equations on the $\tilde{U}_n^{(\mathcal{J}_1)}$} \label{s4.2.1}
Before unpacking the average \eqref{eq4.2.11}, let us first give the companion to Lemma \ref{L4.1}.
\begin{lemma} \label{L4.2}
The partial derivatives of $(B^{k_1})_{ef}$ with respect to $(X^T)_{ab}$ and $X_{cd}$ (taking $e,f$ to be arbitrary and possibly equal to each other or one of $a,b,c,d$) are given by
\begin{align}
\partial_{(X^T)_{ab}}(B^{k_1})_{ef}&=\sum_{p_1+p_2=k_1-1}\left((B^{p_1})_{ea}(JXB^{p_2})_{bf}+(B^{p_1}X^TJ)_{eb}(B^{p_2})_{af}\right), \label{eq4.2.12}
\\ \partial_{X_{cd}}(B^{k_1})_{ef}&=\sum_{p_1+p_2=k_1-1}\left((B^{p_1})_{ed}(JXB^{p_2})_{cf}+(B^{p_1}X^TJ)_{ec}(B^{p_2})_{df}\right), \label{eq4.2.13}
\end{align}
where the sums over $p_1+p_2=k_1-1$ are over the range $p_1=0,1,\ldots,k_1-1$ with $0\leq p_2\leq k_1-1$ set to $k_1-p_1-1$; these sums are empty for $k_1=0$. For later convenience, let us also mention that
\begin{multline} \label{eq4.2.14}
\partial_{(X^T)_{ab}}(B^{k_1}X^T)_{ef}=\sum_{p_1+p_2=k_1-1}\left((B^{p_1})_{ea}(JXB^{p_2}X^T)_{bf}+(B^{p_1}X^TJ)_{eb}(B^{p_2}X^T)_{af}\right)
\\+\chi_{f=b}(B^{k_1})_{ea}.
\end{multline}
\end{lemma}
\begin{proof}
First note that by the Leibniz product rule,
\begin{multline*}
\partial_{(X^T)_{ab}}B_{gh}=\left\{\partial_{(X^T)_{ab}}(X^T)_{gl}\right\}(JX)_{lh}+(X^TJ)_{gl}\left\{\partial_{(X^T)_{ab}}X_{lh}\right\}
\\=\chi_{g=a,l=b}(JX)_{lh}+(X^TJ)_{gl}\chi_{l=b,h=a}=\chi_{g=a}(JX)_{bh}+\chi_{h=a}(X^TJ)_{gb}.
\end{multline*}
Thus, the left-hand side of equation \eqref{eq4.2.13} decomposes as
\begin{multline*}
\partial_{(X^T)_{ab}}(B^{k_1})_{ef}=\sum_{p_1=0}^{k_1-1}(B^{p_1})_{eg}\left\{\partial_{(X^T)_{ab}}B_{gh}\right\}(B^{k_1-p_1-1})_{hf}
\\=\sum_{p_1=0}^{k_1-1}\left((B^{p_1})_{ea}(JX)_{bh}(B^{k_1-p_1-1})_{hf}+(B^{p_1})_{eg}(X^TJ)_{gb}(B^{k_1-p_1-1})_{af}\right),
\end{multline*}
which simplifies to the right-hand side of equation \eqref{eq4.2.12} upon summing over the repeated indices $g,h$. To obtain equation \eqref{eq4.2.13} from equation \eqref{eq4.2.12}, simply observe that setting $(c,d)=(b,a)$ shows that $X_{cd}=(X^T)_{ab}$. For equation \eqref{eq4.2.14}, note that
\begin{align*}
\partial_{(X^T)_{ab}}(B^{k_1}X^T)_{ef}&=\left\{\partial_{(X^T)_{ab}}(B^{k_1})_{eg}\right\}(X^T)_{gf}+(B^{k_1})_{eg}\left\{\partial_{(X^T)_{ab}}(X^T)_{gf}\right\}
\\&=\left\{\partial_{(X^T)_{ab}}(B^{k_1})_{eg}\right\}(X^T)_{gf}+(B^{k_1})_{eg}\chi_{g=a,f=b}
\\&=\left\{\partial_{(X^T)_{ab}}(B^{k_1})_{eg}\right\}(X^T)_{gf}+(B^{k_1})_{ea}\chi_{f=b};
\end{align*}
substituting equation \eqref{eq4.2.12} with $f\mapsto g$ into the braces then completes the proof.
\end{proof}

With these identities in hand, we can now apply integration by parts to the average \eqref{eq4.2.11} in two different ways to obtain a precursor to the loop equation on the $U_n$.
\begin{proposition} \label{prop4.1}
Recalling from Lemma \ref{L4.1} that $I_n=(k_2,\ldots,k_n)$ and $\Tr\,B^{I_n}=\Tr\,B^{k_2}\cdots\Tr\,B^{k_n}$, the average \eqref{eq4.2.11} simplifies to both the left- and right-hand sides of the following equation:
\begin{equation} \label{eq4.2.15}
N^2m_{k_1+1,I_n}=\mean{\left\{\partial_{(X^T)_{ab}}(J^T)_{bc}\partial_{X_{cd}}(B^{k_1})_{ad}\Tr\,B^{I_n}\right\}}.
\end{equation}
\end{proposition}
\begin{proof}
Expanding the argument of the average \eqref{eq4.2.11} shows that it is given by
\begin{align}
&\mean{\left(\partial_{(X^T)_{ab}}+NX_{ba}\right)(J^T)_{bc}\left(\partial_{X_{cd}}+N(X^T)_{dc}\right)(B^{k_1})_{ad}\Tr\,B^{I_n}} \nonumber
\\&=\mean{\partial_{(X^T)_{ab}}(J^T)_{bc}\partial_{X_{cd}}(B^{k_1})_{ad}\Tr\,B^{I_n}}+N\mean{\partial_{(X^T)_{ab}}(J^T)_{bc}(X^T)_{dc}(B^{k_1})_{ad}\Tr\,B^{I_n}} \nonumber
\\&\quad+N\mean{X_{ba}(J^T)_{bc}\partial_{X_{cd}}(B^{k_1})_{ad}\Tr\,B^{I_n}}+N^2\mean{X_{ba}(J^T)_{bc}(X^T)_{dc}(B^{k_1})_{ad}\Tr\,B^{I_n}}.  \label{eq4.2.16}
\end{align}
The first two terms on the right-hand side of this equation are manifestly zero by the fundamental theorem of calculus (they are integrals of derivatives of the exponentially decaying p.d.f.~$P(X)$ \eqref{eq4.2.1} multiplied by polynomials in entries of $X$, which must converge to zero at the boundary of the domain of integration). Moreover, contracting indices in the fourth term on the right-hand side of equation \eqref{eq4.2.16} while rewriting $(J^T)_{bc}$ as $J_{cb}$ shows that this term is equal to the left-hand side of equation \eqref{eq4.2.15}. Finally, to see that the third term on the right-hand side of equation \eqref{eq4.2.16} vanishes, note that for $f(X)$ a function of the entries of $X$, integration by parts shows that
\begin{align*}
\mean{X_{ba}(J^T)_{bc}\partial_{X_{cd}}f(X)}&=\mean{\partial_{X_{cd}}X_{ba}(J^T)_{bc}f(X)}-\mean{\left\{\partial_{X_{cd}}X_{ba}\right\}(J^T)_{bc}f(X)}
\\&=\mean{\partial_{X_{cd}}X_{ba}(J^T)_{bc}f(X)}-\mean{\chi_{a=d,b=c}(J^T)_{bc}f(X)}
\\&=\mean{\partial_{X_{cd}}X_{ba}(J^T)_{bc}f(X)}-\mean{\chi_{a=d}(J^T)_{bb} f(X)},
\end{align*}
which reduces to zero by application of the fundamental theorem of calculus to the first term and by making the substitution $(J^T)_{bb}=\Tr\,J^T=0$ in the second term.

Now, for $f(X)$ again a function of the entries of $X$, observe that
\begin{multline*}
\left(\partial_{(X^T)_{ab}}+NX_{ba}\right)f(X)\exp\left(-\tfrac{N}{2}\Tr(X^TX)\right)
\\=\left\{\partial_{(X^T)_{ab}}f(X)\right\}\exp\left(-\tfrac{N}{2}\Tr(X^TX)\right)+f(X)\left\{\partial_{(X^T)_{ab}}\exp\left(-\tfrac{N}{2}\Tr(X^TX)\right)\right\}
\\+NX_{ba}f(X)\exp\left(-\tfrac{N}{2}\Tr(X^TX)\right)
\\=\left\{\partial_{(X^T)_{ab}}f(X)\right\}\exp\left(-\tfrac{N}{2}\Tr(X^TX)\right)
\end{multline*}
since $\partial_{(X^T)_{ab}}\exp\left(-\tfrac{N}{2}\Tr(X^TX)\right)$ is equal to $-NX_{ba}\exp\left(-\tfrac{N}{2}\Tr(X^TX)\right)$. We similarly have
\begin{equation*}
\left(\partial_{X_{cd}}+N(X^T)_{dc}\right)f(X)\exp\left(-\tfrac{N}{2}\Tr(X^TX)\right)=\left\{\partial_{X_{cd}}f(X)\right\}\exp\left(-\tfrac{N}{2}\Tr(X^TX)\right).
\end{equation*}
Thus, the average \eqref{eq4.2.11} simplifies as
\begin{multline*}
\mean{\left(\partial_{(X^T)_{ab}}+NX_{ba}\right)(J^T)_{bc}\left(\partial_{X_{cd}}+N(X^T)_{dc}\right)(B^{k_1})_{ad}\Tr\,B^{I_n}}
\\=\mean{\left(\partial_{(X^T)_{ab}}+NX_{ba}\right)(J^T)_{bc}\left\{\partial_{X_{cd}}(B^{k_1})_{ad}\Tr\,B^{I_n}\right\}},
\end{multline*}
which can be seen to be precisely the right-hand side of equation \eqref{eq4.2.15} upon writing
\begin{equation*}
f(X)=(J^T)_{bc}\left\{\partial_{X_{cd}}(B^{k_1})_{ad}\Tr\,B^{I_n}\right\}.
\end{equation*}
\end{proof}

Our claim is that using the Leibniz product rule to expand the derivative in the right-hand side of equation \eqref{eq4.2.15} and then properly simplifying each of the averages in the resulting sum will produce our sought loop equation. To demonstrate this claim, let us now write the right-hand side of equation \eqref{eq4.2.15} as
\begin{equation} \label{eq4.2.17}
\mean{\left\{\partial_{(X^T)_{ab}}(J^T)_{bc}\partial_{X_{cd}}(B^{k_1})_{ad}\Tr\,B^{I_n}\right\}}=\mathcal{A}+\sum_{i=2}^n\left(\mathcal{A}_i+\overline{\mathcal{A}}_i+\mathcal{B}_i\right)+\sum_{\substack{2\leq i,j\leq n,\\i\neq j}}\mathcal{C}_{ij},
\end{equation}
where
\begin{align}
\mathcal{A}&=\mean{\left\{\partial_{(X^T)_{ab}}(J^T)_{bc}\partial_{X_{cd}}(B^{k_1})_{ad}\right\}\Tr\,B^{I_n}}, \label{eq4.2.18}
\\ \mathcal{A}_i&=\mean{\left\{\partial_{(X^T)_{ab}}(B^{k_1})_{ad}\right\}(J^T)_{bc}\left\{\partial_{X_{cd}}\Tr\,B^{k_i}\right\}\Tr\,B^{I_n\setminus\{k_i\}}},
\\ \overline{\mathcal{A}}_i&=\mean{\left\{\partial_{(X^T)_{ab}}\Tr\,B^{k_i}\right\}(J^T)_{bc}\left\{\partial_{X_{cd}}(B^{k_1})_{ad}\right\}\Tr\,B^{I_n\setminus\{k_i\}}},
\\ \mathcal{B}_i&=\mean{\left\{\partial_{(X^T)_{ab}}(J^T)_{bc}\partial_{X_{cd}}\Tr\,B^{k_i}\right\}(B^{k_1})_{ad}\Tr\,B^{I_n\setminus\{k_i\}}},
\\ \mathcal{C}_{ij}&=\mean{\left\{\partial_{(X^T)_{ab}}\Tr\,B^{k_i}\right\}(J^T)_{bc}\left\{\partial_{X_{cd}}\Tr\,B^{k_j}\right\}(B^{k_1})_{ad}\Tr\,B^{I_n\setminus\{k_i,k_j\}}}. \label{eq4.2.22}
\end{align}
We show in the following lemma that each of these terms can be expressed in terms of the mixed moments, so combining equations \eqref{eq4.2.15} and \eqref{eq4.2.17} will result in a loop equation on said moments, as claimed.
\begin{lemma} \label{L4.3}
Let us retain the definitions $I_n=(k_1,\ldots,k_n)$ and $\Tr\,B^{I_n}=\Tr\,B^{k_2}\cdots\Tr\,B^{k_n}$ from Lemma \ref{L4.1}. Moreover, for $k\in\mathbb{N}$, let $[k]_{\mathrm{mod}\, 2}$ equal zero when $k$ is even and one when $k$ is odd. Then, the quantities \eqref{eq4.2.18}--\eqref{eq4.2.22} above are given by
\begin{align}
\mathcal{A}&=\sum_{p_1+p_2=k_1-1}m_{p_1,p_2,I_n}-\sum_{p_1+p_2+p_3=k_1-1}m_{p_1,p_2,p_3,I_n}+\frac{1-k_1^2}{2}m_{k_1-1,I_n}, \label{eq4.2.23}
\\ \mathcal{A}_i=\overline{\mathcal{A}}_i&=2k_i[k_i+1]_{\mathrm{mod}\,2}\left(m_{k_1+k_i-1,I_n\setminus\{k_i\}}-\sum_{p_1+p_2=k_1-1}m_{k_i+p_1,p_2,I_n\setminus\{k_i\}}\right), \label{eq4.2.24}
\\ \mathcal{B}_i&=-2k_i[k_i+1]_{\mathrm{mod}\,2}\left(m_{k_1+k_i-1,I_n\setminus\{k_i\}}+\sum_{p_1+p_2=k_i-1}m_{k_1+p_1,p_2,I_n\setminus\{k_i\}}\right), \label{eq4.2.25}
\\ \mathcal{C}_{ij}=\mathcal{C}_{ji}&=-4k_ik_j[k_i+1]_{\mathrm{mod}\,2}[k_j+1]_{\mathrm{mod}\,2}\,m_{k_1+k_i+k_j-1,I_n\setminus\{k_i,k_j\}}. \label{eq4.2.26}
\end{align}
\end{lemma}
\begin{proof}
We present only the computations of $\mathcal{A}$ and $\mathcal{C}_{ij}$ since their derivations contain all of the ideas needed to prove equations \eqref{eq4.2.24} and \eqref{eq4.2.25}. We first give the proof of equation \eqref{eq4.2.26}, as it is simpler than that of equation \eqref{eq4.2.23}. Thus, we begin by using equation \eqref{eq4.2.12} with $(k_1,f)\mapsto(k_i,e)$ to see that
\begin{multline} \label{eq4.2.27}
\partial_{(X^T)_{ab}}\Tr\,B^{k_i}=\partial_{(X^T)_{ab}}(B^{k_i})_{ee}=\sum_{p_1+p_2=k_i-1}\left((JXB^{k_i-1})_{ba}+(B^{k_i-1}X^TJ)_{ab}\right)
\\=(JXB^{k_i-1})_{ba}\sum_{p_1+p_2=k_i-1}\left(1+(-1)^{k_i}\right)=2k_i[k_i+1]_{\mathrm{mod}\,2}(JXB^{k_i-1})_{ba}.
\end{multline}
Here, we have used the fact that $(B^{k_i-1}X^TJ)^T=J^TX(B^T)^{k_i-1}=(-1)^{k_i}JXB^{k_i-1}$ and that $(1+(-1)^{k_i})$ equals two if $k_i$ is even and vanishes, otherwise. Repeating this argument with equation \eqref{eq4.2.13} likewise shows that
\begin{equation*}
\partial_{X_{cd}}\Tr\,B^{k_j}=2k_j[k_j+1]_{\mathrm{mod}\,2}(B^{k_j-1}X^TJ)_{dc}.
\end{equation*}
Multiplying these two results together, in combination with $(J^T)_{bc}$, then reveals the identity
\begin{equation*}
\left\{\partial_{(X^T)_{ab}}\Tr\,B^{k_i}\right\}(J^T)_{bc}\left\{\partial_{X_{cd}}\Tr\,B^{k_j}\right\}=-4k_ik_j[k_i+1]_{\mathrm{mod}\,2}[k_j+1]_{\mathrm{mod}\,2}(B^{k_i+k_j-1})_{da}.
\end{equation*}
Further multiplying this by $(B^{k_1})_{ad}\Tr\,B^{I_n\setminus\{k_i,k_j\}}$, summing over the repeated indices $a,d$, and then taking the average yields the right-hand side of equation \eqref{eq4.2.26}, as required.

Similar to above, we begin the proof of equation \eqref{eq4.2.23} by noting that equation \eqref{eq4.2.13} with $(e,f)=(a,d)$ tells us that
\begin{multline} \label{eq4.2.28}
(J^T)_{bc}\partial_{X_{cd}}(B^{k_1})_{ad}=\sum_{p_1+p_2=k_1-1}\left((-1)^{p_2}(B^{k_1-1}X^T)_{ab}-(B^{p_1}X^T)_{ab}\Tr\,B^{p_2}\right)
\\=[k_1]_{\mathrm{mod}\,2}(B^{k_1-1}X^T)_{ab}-\sum_{p_1+p_2=k_1-1}(B^{p_1}X^T)_{ab}\Tr\,B^{p_2},
\end{multline}
where we have used the fact that $\sum_{p_1+p_2=k_1-1}(-1)^{p_2}=\sum_{p_2=0}^{k_1-1}(-1)^{p_2}$ equals one if $k_1$ is odd and zero, otherwise. Now, taking $(k_1,e,f)\mapsto(k_1-1,a,b)$ in equation \eqref{eq4.2.14} shows that
\begin{align}
\partial_{(X^T)_{ab}}(B^{k_1-1}X^T)_{ab}&=\sum_{p_1+p_2=k_1-2}\left(\Tr\,B^{p_1}\Tr\,B^{p_2+1}+(-1)^{p_2}\Tr\,B^{k_1-1}\right)+N\Tr\,B^{k_1-1} \nonumber
\\&=\sum_{p_1+p_2=k_1-1}\Tr\,B^{p_1}\Tr\,B^{p_2}+[k_1+1]_{\mathrm{mod}\,2}\Tr\,B^{k_1-1} \nonumber
\\&=\sum_{p_1+p_2=k_1-1}\Tr\,B^{p_1}\Tr\,B^{p_2}; \label{eq4.2.29}
\end{align}
the second line follows from making the replacement $p_2\mapsto p_2-1$ in the sum and noting that the new $(p_1,p_2)=(k_1-1,0)$ term this introduces is equivalent to the term $N\Tr\,B^{k_1-1}$ that already appears at the end of the first line, while the final line follows from the fact that $[k_1+1]_{\mathrm{mod}\,2}$ vanishes whenever $k_1$ is odd, but $\Tr\,B^{k_1-1}$ vanishes whenever $k_1$ is even. Moving on to the second term in equation \eqref{eq4.2.28}, using equation \eqref{eq4.2.27} with the replacement $k_i\mapsto p_2$ and equation \eqref{eq4.2.29} with $k_1\mapsto p_1$ shows that
\begin{multline} \label{eq4.2.30}
\partial_{(X^T)_{ab}}(B^{p_1}X^T)_{ab}\Tr\,B^{p_2}=\left\{\partial_{(X^T)_{ab}}(B^{p_1}X^T)_{ab}\right\}\Tr\,B^{p_2}+(B^{p_1}X^T)_{ab}\left\{\partial_{(X^T)_{ab}}\Tr\,B^{p_2}\right\}
\\=\sum_{q_1+q_2=p_1}\Tr\,B^{q_1}\Tr\,B^{q_2}\Tr\,B^{p_2}+[p_1]_{\mathrm{mod}\,2}\Tr\,B^{p_1}\Tr\,B^{p_2}+2p_2[p_2+1]_{\mathrm{mod}\,2}\Tr\,B^{p_1+p_2}
\\=\sum_{q_1+q_2=p_1}\Tr\,B^{q_1}\Tr\,B^{q_2}\Tr\,B^{p_2}+2p_2[p_2+1]_{\mathrm{mod}\,2}\Tr\,B^{p_1+p_2},
\end{multline}
where the last line follows from the fact that $[p_1]_{\mathrm{mod}\,2}$ vanishes if $p_1$ is even, but $\Tr\,B^{p_1}$ vanishes if $p_1$ is odd. Applying the operator $\partial_{(X^T)_{ab}}$ to both sides of equation \eqref{eq4.2.28} and then substituting in equations \eqref{eq4.2.29} and \eqref{eq4.2.30} shows that
\begin{multline} \label{eq4.2.31}
\partial_{(X^T)_{ab}}(J^T)_{bc}\partial_{X_{cd}}(B^{k_1})_{ad}=[k_1]_{\mathrm{mod}\,2}\sum_{p_1+p_2=k_1-1}\Tr\,B^{p_1}\Tr\,B^{p_2}
\\-\sum_{p_1+p_2=k_1-1}\left(2p_2[p_2+1]_{\mathrm{mod}\,2}\Tr\,B^{p_1+p_2}+\sum_{q_1+q_2=p_1}\Tr\,B^{q_1}\Tr\,B^{q_2}\Tr\,B^{p_2}\right).
\end{multline}
The factor of $[k_1]_{\mathrm{mod}\,2}$ in the first term can be removed since it is superfluous: for $\Tr\,B^{p_1},\Tr\,B^{p_2}$ to be non-zero, each of $p_1,p_2$ must be even and, consequently, the sum over $p_1+p_2=k_1-1$ is non-zero only if $k_1$ is odd. Likewise, the second term is non-zero for $k_1$ odd and it simplifies as
\begin{equation*}
-\sum_{p_1+p_2=k_1-1}2p_2[p_2+1]_{\mathrm{mod}\,2}\Tr\,B^{p_1+p_2}=-2\Tr\,B^{k_1-1}(2+4+\cdots+k_1-1)=\frac{1-k_1^2}{2}\Tr\,B^{k_1-1},
\end{equation*}
while replacing $p_1$ in the outer sum of the third term by $q_1+q_2$ and then renaming the summation indices shows that
\begin{multline*}
-\sum_{p_1+p_2=k_1-1}\sum_{q_1+q_2=p_1}\Tr\,B^{q_1}\Tr\,B^{q_2}\Tr\,B^{p_2}=-\sum_{q_1+q_2+p_2=k_1-1}\Tr\,B^{q_1}\Tr\,B^{q_2}\Tr\,B^{p_2}
\\=-\sum_{p_1+p_2+p_3=k_1-1}\Tr\,B^{p_1}\Tr\,B^{p_2}\Tr\,B^{p_3}.
\end{multline*}
Combining these observations together, we thus see that equation \eqref{eq4.2.31} reduces to
\begin{multline*}
\partial_{(X^T)_{ab}}(J^T)_{bc}\partial_{X_{cd}}(B^{k_1})_{ad}=\sum_{p_1+p_2=k_1-1}\Tr\,B^{p_1}\Tr\,B^{p_2}+\frac{1-k_1^2}{2}\Tr\,B^{k_1-1}
\\-\sum_{p_1+p_2+p_3=k_1-1}\Tr\,B^{p_1}\Tr\,B^{p_2}\Tr\,B^{p_3}.
\end{multline*}
Finally, multiplying the right-hand side of this equation by $\Tr\,B^{I_n}$ and taking the average produces the right-hand side of equation \eqref{eq4.2.23}, as required.
\end{proof}

It is now a simple matter of substituting the expressions \eqref{eq4.2.23}--\eqref{eq4.2.26} into the right-hand side of equation \eqref{eq4.2.17} while replacing the left-hand side by $N^2m_{k_1+1,I_n}$, as prescribed by Propositon \ref{prop4.1}, to obtain the desired loop equation on the mixed moments $m_{k_1,\ldots,k_n}$. This loop equation is presented as equation \eqref{eqC.1.1} in Appendix \ref{appendixC.1}.

At this point, it is possible to proceed by substituting the moment-cumulants relation \eqref{eq1.1.28} into the loop equation on the moments to obtain an analogous equation on the cumulants $c_{k_1,\ldots,k_n}$. Simplifying this equation on the $c_{k_1,\ldots,k_n}$ in an appropriate manner (using a similar argument to that given in the proof of Proposition \ref{prop4.3} in \S\ref{s4.2.2} upcoming), multiplying the result by $x_1^{-k_1-1}\cdots x_n^{-k_n-1}$, and summing over $k_1,\ldots,k_n\geq0$ would then yield the loop equation on the $W_n(x_1,\ldots,x_n)$ (cf.~equation \eqref{eq4.2.4}). We do not go down this path, instead opting to first derive the loop equation on the unconnected correlators $U_n(x_1,\ldots,x_n)$ \eqref{eq4.2.3} and then transforming it, through the identity \eqref{eq4.1.4}, into the corresponding loop equation on the $W_n(x_1,\ldots,x_n)$ --- if necessary, the loop equation on the mixed cumulants $c_{k_1,\ldots,k_n}$ can be extracted from the loop equation on the $W_n(x_1,\ldots,x_n)$ by equating coefficients of $x_1^{-k_1-1}\cdots x_n^{-k_n-1}$, possibly by taking suitable residues. (Note that we could have alternatively skipped the derivation of the loop equation on the mixed moments altogether by starting our analysis at a perturbation of the average \eqref{eq4.1.3}, rather than \eqref{eq4.1.2}; our choice of starting point leads to the cleaner presentation.)

\begin{proposition} \label{prop4.2}
Recall that $J_n=(x_2,\ldots,x_n)$, set $U_0:=1$, and define $U_{n'}:=0$ for $n'<0$. Furthermore, define the auxiliary function
\begin{multline}
A_i(x_1,J_n)=x_i^2\left(2U_n(x_1,x_1,J_n\setminus\{x_i\})-U_n(x_i,J_n)\right)-x_1x_iU_n(x_1,J_n)
\\+\frac{1}{x_1}\left(x_1x_iU_{n-1}(J_n)-x_i^2U_{n-1}(x_1,J_n\setminus\{x_i\})\right).
\end{multline}
Then, for $n\geq1$, the unconnected correlators specified by equation \eqref{eq4.2.3} satisfy the loop equation
\begin{multline} \label{eq4.2.33}
0=x_1U_{n+2}(x_1,x_1,x_1,J_n)-U_{n+1}(x_1,x_1,J_n)+N^2x_1U_n(x_1,J_n)-N^3U_{n-1}(J_n)
\\+\frac{1}{2x_1}\left(x_1^2\frac{\partial^2}{\partial x_1^2}+x_1\frac{\partial}{\partial x_1}-1\right)U_n(x_1,J_n)+2\sum_{i=2}^n\frac{\partial}{\partial x_i}\left\{\frac{A_i(x_1,J_n)}{x_1^2-x_i^2}\right\}
\\+8\sum_{2\leq i<j\leq n}\frac{1}{x_1}\frac{\partial^2}{\partial x_i\partial x_j}\Bigg\{\frac{x_1x_i^2x_jU_{n-2}(J_n\setminus\{x_i\})}{(x_1^2-x_j^2)(x_i^2-x_j^2)}-\frac{x_1x_ix_j^2U_{n-2}(J_n\setminus\{x_j\})}{(x_1^2-x_i^2)(x_i^2-x_j^2)}
\\+\frac{x_i^2x_j^2U_{n-2}(x_1,J_n\setminus\{x_i,x_j\})}{(x_1^2-x_i^2)(x_1^2-x_j^2)}\Bigg\}.
\end{multline}
\end{proposition}
\begin{proof}
As mentioned above, inserting the equalities \eqref{eq4.2.23}--\eqref{eq4.2.26} into equation \eqref{eq4.2.17} and combining the result with equation \eqref{eq4.2.15} gives the loop equation \eqref{eqC.1.1} on the mixed moments $m_{k_1,\ldots,k_n}$. Multiplying both sides of this loop equation by $x_1^{-k_1-1}\cdots x_n^{-k_n-1}$ and summing over $k_1,\ldots,k_n\geq0$ then produces the loop equation \eqref{eq4.2.33} above. We detail this transformation term by term in Appendix \ref{appendixC.1}.
\end{proof}

\subsection{Loop equations on the $\tilde{W}_n^{(\mathcal{J}_1)}$ and $W_n^{(\mathcal{J}_1),l}$} \label{s4.2.2}
Our main goal in this subsection is to transform the loop equation \eqref{eq4.2.33} on the $U_n(x_1,\ldots,x_n)$ into a loop equation on the $W_n(x_1,\ldots,x_n)$ through the use of the identity \eqref{eq4.1.4}. Actually, we use higher order analogues of this identity, which we present below.
\begin{lemma} \label{L4.4}
Recall that $J_n=(x_2,\ldots,x_n)$ and let $\mu\vdash(x_1,\ldots,x_n)$ indicate that $\mu$ is a partition of the ordered set $(x_1,\ldots,x_n)$, i.e., $\mu=\{\mu_t\}_{t=1}^m$ for some $1\leq m\leq n$ such that the disjoint union $\sqcup_{t=1}^m\mu_t=\{x_1,\ldots,x_n\}$. Writing $\#\mu$ for the size of $\mu$ ($m$ in the preceding sentence), we have that
\begin{align}
U_n(x_1,J_n)&=\sum_{K_1\sqcup K_2=J_n}U_{\#K_1}(K_1)W_{\#K_2+1}(x_1,K_2), \label{eq4.2.34}
\\ U_{n+1}(x_1,x_1,J_n)&=\sum_{\mu\vdash(x_1,x_1)}\sum_{\sqcup_{t=1}^{\#\mu+1}K_t=J_n}U_{\#K_1}(K_1)\prod_{\mu_t\in\mu}W_{\#K_{t+1}+\#\mu_t}(\mu_t,K_{t+1}), \label{eq4.2.35}
\\ U_{n+2}(x_1,x_1,x_1,J_n)&=\sum_{\mu\vdash(x_1,x_1,x_1)}\sum_{\sqcup_{t=1}^{\#\mu+1}K_t=J_n}U_{\#K_1}(K_1)\prod_{\mu_t\in\mu}W_{\#K_{t+1}+\#\mu_t}(\mu_t,K_{t+1}). \label{eq4.2.36}
\end{align}
\end{lemma}
\begin{proof}
Equation \eqref{eq4.2.34} is a simple rewriting of equation \eqref{eq4.1.4}. This equation tells us, upon increasing $n$ by one and replacing $J_n$ with $(x_1,J_n)$, that
\begin{equation*}
U_{n+1}(x_1,x_1,J_n)=\sum_{K_1\sqcup K_2=(x_1,J_n)}U_{\#K_1}(K_1)W_{\#K_2+1}(x_1,K_2).
\end{equation*}
Splitting this sum based on whether or not $K_1$ contains $x_1$ gives
\begin{equation*}
U_{n+1}(x_1,x_1,J_n)=\sum_{K_1\sqcup K_2=J_n}\left[U_{\#K_1+1}(x_1,K_1)W_{\#K_2+1}(x_1,K_2)+U_{\#K_1}(K_1)W_{\#K_2+2}(x_1,x_2,K_2)\right].
\end{equation*}
Substituting in an appropriate rewriting of equation \eqref{eq4.2.34} for $U_{\#K_1+1}(x_1,K_1)$ then simplifies this expression to equation \eqref{eq4.2.35}. Likewise, increasing $n$ by one in equation \eqref{eq4.2.35} and replacing $J_n$ with $(x_1,J_n)$ shows that
\begin{equation*}
U_{n+2}(x_1,x_1,x_1,J_n)=\sum_{\mu\vdash(x_1,x_1)}\sum_{\sqcup_{t=1}^{\#\mu+1}K_t=(x_1,J_n)}U_{\#K_1}(K_1)\prod_{\mu_t\in\mu}W_{\#K_{t+1}+\#\mu_t}(\mu_t,K_{t+1}).
\end{equation*}
Repeating the exercise described above then produces equation \eqref{eq4.2.36}.
\end{proof}
\begin{note}
Let us remark that the notation in the above is drawn from the work \citep{DF20}; see also \citep{BE13}, \citep{BHLMR14}. We write $\mu=\{\mu_1,\ldots,\mu_{\#\mu}\}$, which implicitly defines the sets $\mu_1,\ldots,\mu_{\#\mu}$. Observe that since we are dealing with partitions of ordered sets, the partition $\mu=\{\,\{x_1\},\,\{x_1,x_1\}\,\}$ is counted three times: the $x_1$ in $\mu_1=\{x_1\}$ could have been the first, second, or third entry in $(x_1,x_1,x_1)$. Although the $\mu_t$ must be non-empty, the $K_t$ are allowed to be empty.
\end{note}

With Lemma \ref{L4.4} in hand, we can now start computing loop equations for the connected correlators $W_n(x_1,\ldots,x_n)$. Thus, setting $n=1$ in equation \eqref{eq4.2.33} and then replacing $U_0$, $U_1(x_1)$, $U_2(x_1,x_1)$, and $U_3(x_1,x_1,x_1)$ by $1$, $W_1(x_1)$, the right-hand side of equation \eqref{eq4.2.35} with $n=1$, and the right-hand side of equation \eqref{eq4.2.36} with $n=1$, respectively, yields the \textit{$n=1$ loop equation on the connected correlators},
\begin{multline} \label{eq4.2.37}
0=x_1\left[W_3(x_1,x_1,x_1)+3W_2(x_1,x_1)W_1(x_1)+W_1(x_1)^3\right]-W_2(x_1,x_1)-W_1(x_1)^2
\\+N^2x_1W_1(x_1)-N^3+\frac{1}{2x_1}\left(x_1^2\frac{\mathrm{d}^2}{\mathrm{d}x_1^2}+x_1\frac{\mathrm{d}}{\mathrm{d}x_1}-1\right)W_1(x_1).
\end{multline}

It can be immediately seen that, similar to the loop equations reviewed in \S\ref{s4.1.1}, this loop equation has too many unknowns ($W_1,W_2$, and $W_3$) to be solvable. However, in analogy with equation \eqref{eq4.1.14}, substituting the large $N$ expansion \eqref{eq4.0.5}, repeated here for convenience,
\begin{equation} \label{eq4.2.38}
W_n(x_1,\ldots,x_n)=N^{2-n}\sum_{l=0}^\infty\frac{W_n^l(x_1,\ldots,x_n)}{N^l}
\end{equation}
into equation \eqref{eq4.2.37} and equating terms of like order in $N$ produces, for each $l\geq0$, the $(1,l)$ loop equation used to compute $W_1^l(x_1)$ (recall from the discussion below equation \eqref{eq4.1.15} that iterating through $(n,l)\in\mathbb{N}^2$ in the correct order allows us to solve the $(n,l)$ loop equation for the correlator expansion coefficient $W_n^l$),
\begin{multline} \label{eq4.2.39}
0=x_1\Big[W_3^{l-4}(x_1,x_1,x_1)+3\sum_{l_1+l_2=l-2}W_2^{l_1}(x_1,x_1)W_1^{l_2}(x_1)+\sum_{l_1+l_2+l_3=l}W_1^{l_1}(x_1)W_1^{l_2}(x_1)W_1^{l_3}(x_1)\Big]
\\-W_2^{l-3}(x_1,x_1)-\sum_{l_1+l_2=l-1}W_1^{l_1}(x_1)W_1^{l_2}(x_1)+x_1W_1^l(x_1)-\chi_{l=0}
\\+\frac{1}{2x_1}\left(x_1^2\frac{\mathrm{d}^2}{\mathrm{d}x_1^2}+x_1\frac{\mathrm{d}}{\mathrm{d}x_1}-1\right)W_1^{l-2}(x_1).
\end{multline}
(As usual, we have set $W_n^{l'}:=0$ for all $n\geq1$ and $l'<0$.) In particular, setting $l=0$ in the above gives the spectral curve
\begin{equation} \label{eq4.2.40}
0=x_1\left(W_1^0(x_1)\right)^3+x_1W_1^0(x_1)-1.
\end{equation}
Being a cubic polynomial in $W_1^0(x_1)$, this equation can be solved to express $W_1^0(x_1)$ as a cube root (although equation \eqref{eq4.2.40} has three solutions, only one obeys the requirement \eqref{eq2.4.15} and is thus relevant to our setting). We do not make $W_1^0(x_1)$ explicit here, but refer to \citep{FL15}, \citep{DF20} and references therein for techniques on obtaining such an expression --- in \S\ref{s4.2.3}, we instead find a rational parametrisation for the spectral curve \eqref{eq4.2.40}, in analogy with the parametrisations $x=x(z)$ \eqref{eq4.1.31} and $W_1^0(x)=y(z)$ \eqref{eq4.1.32} reviewed in \S\ref{s4.1.2}.

Moving on, setting $n=2$ in equation \eqref{eq4.2.33} and then replacing $U_1$ by $W_1$, $U_2$ by the right-hand side of equation \eqref{eq4.2.34} with $n=2$ and suitable variables, $U_3(x_1,x_1,x_2)$ by the right-hand side of equation \eqref{eq4.2.35} with $n=2$, and $U_4(x_1,x_1,x_1,x_2)$ by the right-hand side of equation \eqref{eq4.2.36} with $n=2$ shows that
\begin{multline} \label{eq4.2.41}
0=x_1\sum_{\mu\vdash(x_1,x_1,x_1)}\sum_{\sqcup_{t=1}^{\#\mu+1}K_t=\{x_2\}}U_{\#K_1}(K_1)\prod_{\mu_t\in\mu}W_{\#K_{t+1}+\#\mu_t}(\mu_t,K_{t+1})
\\-\sum_{\mu\vdash(x_1,x_1)}\sum_{\sqcup_{t=1}^{\#\mu+1}K_t=\{x_2\}}U_{\#K_1}(K_1)\prod_{\mu_t\in\mu}W_{\#K_{t+1}+\#\mu_t}(\mu_t,K_{t+1})-N^3U_1(x_2)
\\+\left[N^2x_1+\frac{1}{2x_1}\left(x_1^2\frac{\partial^2}{\partial x_1^2}+x_1\frac{\partial}{\partial x_1}-1\right)\right]\left(W_2(x_1,x_2)+W_1(x_1)U_1(x_2)\right)
\\+\frac{2}{x_1}\frac{\partial}{\partial x_2}\frac{1}{x_1^2-x_2^2}\Big[x_1x_2^2\left(2W_2(x_1,x_1)+2W_1(x_1)^2-W_2(x_2,x_2)-W_1(x_2)^2\right)
\\\hspace{6em}-x_1^2x_2\left(W_2(x_1,x_2)+W_1(x_1)W_1(x_2)\right)
\\+x_1x_2W_1(x_2)-x_2^2W_1(x_1)\Big].
\end{multline}
This is not yet our sought loop equation, as we can simplify it one step further. Indeed, subtracting the product of the right-hand side of equation \eqref{eq4.2.37} by $U_1(x_2)$, which is zero, from equation \eqref{eq4.2.41} reveals the \textit{$n=2$ loop equation on the connected correlators},
\begin{multline} \label{eq4.2.42}
0=x_1\Big[W_4(x_1,x_1,x_1,x_2)+3W_3(x_1,x_1,x_2)W_1(x_1)
\\+3W_2(x_1,x_1)W_2(x_1,x_2)+3W_2(x_1,x_2)W_1(x_1)^2\Big]-W_3(x_1,x_1,x_2)
\\-2W_2(x_1,x_2)W_1(x_1)+N^2x_1W_2(x_1,x_2)+\frac{1}{2x_1}\left(x_1^2\frac{\partial^2}{\partial x_1^2}+x_1\frac{\partial}{\partial x_1}-1\right)W_2(x_1,x_2)
\\+\frac{2}{x_1}\frac{\partial}{\partial x_2}\frac{1}{x_1^2-x_2^2}\Big[x_1x_2^2\left(2W_2(x_1,x_1)+2W_1(x_1)^2-W_2(x_2,x_2)-W_1(x_2)^2\right)
\\-x_1^2x_2\left(W_2(x_1,x_2)+W_1(x_1)W_1(x_2)\right)+x_1x_2W_1(x_2)-x_2^2W_1(x_1)\Big].
\end{multline}

As with equation \eqref{eq4.2.37}, equation \eqref{eq4.2.42} is unsolvable, but substituting in the large $N$ expansion \eqref{eq4.2.38} results in a set of solvable loop equations on the correlator expansion coefficients. Thus, we have for each $l\geq0$, the $(2,l)$ loop equation used to compute the expansion coefficient $W_2^l(x_1,x_2)$,
\begin{align}
0&=x_1\Big[W_4^{l-4}(x_1,x_1,x_2,x_2)+3\sum_{l_1+l_2=l-2}W_3^{l_1}(x_1,x_1,x_2)W_1^{l_2}(x_1) \nonumber
\\&\hspace{6em}+3\sum_{l_1+l_2=l-2}W_2^{l_1}(x_1,x_1)W_2^{l_2}(x_1,x_2)+3\sum_{l_1+l_2+l_3=l}W_2^{l_1}(x_1,x_2)W_1^{l_2}(x_1)W_1^{l_3}(x_1)\Big] \nonumber
\\&\quad-W_3^{l-3}(x_1,x_1,x_2)-2\sum_{l_1+l_2=l-1}W_2^{l_1}(x_1,x_2)W_1^{l_2}(x_1)+x_1W_2^l(x_1,x_2) \nonumber
\\&\quad+\frac{1}{2x_1}\left(x_1^2\frac{\partial^2}{\partial x_1^2}+x_1\frac{\partial}{\partial x_1}-1\right)W_2^{l-2}(x_1,x_2) \nonumber
\\&\quad+\frac{2}{x_1}\frac{\partial}{\partial x_2}\frac{1}{x_1^2-x_2^2}\Big[2x_1x_2^2W_2^{l-2}(x_1,x_1)-x_1x_2^2W_2^{l-2}(x_2,x_2)-x_1^2x_2W_2^{l-2}(x_1,x_2) \nonumber
\\&\hspace{10em}+x_1x_2^2\sum_{l_1+l_2=l}\left(2W_1^{l_1}(x_1)W_1^{l_2}(x_1)-W_1^{l_1}(x_2)W_1^{l_2}(x_2)\right)+x_1x_2W_1^{l-1}(x_2) \nonumber
\\&\hspace{12em}-x_1^2x_2\sum_{l_1+l_2=l}W_1^{l_1}(x_1)W_1^{l_2}(x_2)-x_2^2W_1^{l-1}(x_1)\Big]. \label{eq4.2.43}
\end{align}
Setting $l=0$ then specifies $W_2^0(x_1,x_2)$ as a rational function of $W_1^0$ and its derivative:
\begin{multline} \label{eq4.2.44}
W_2^0(x_1,x_2)=\frac{2}{x_1}\frac{1}{3(W_1^0(x_1))^2+1}\frac{\partial}{\partial x_2}\frac{x_2}{x_2^2-x_1^2}\Big[2x_2(W_1^0(x_1))^2-x_2(W_1^0(x_2))^2
\\-x_1W_1^0(x_1)W_1^0(x_2)\Big].
\end{multline}
We solve this equation using the rational parametrisation mentioned below equation \eqref{eq4.2.40} in \S\ref{s4.2.3}. As expected, we will recover the structure of the Bergman kernel (cf.~equation \eqref{eq4.1.34} and the discussion below it).

Let us now highlight that in obtaining equation \eqref{eq4.2.42} from equation \eqref{eq4.2.41}, we have simply subtracted off terms from the first three lines of equation \eqref{eq4.2.41} that contain a factor of $U_1(x_2)$ while using equation \eqref{eq4.2.37} to argue that this does not change the left-hand side. This idea extends to the general $n\geq3$ case:
\begin{proposition} \label{prop4.3}
Let us retain the notation used in Lemma \ref{L4.4}. For $n\geq3$, the connected correlators \eqref{eq4.2.4} of the global scaled $(N,N)$ antisymmetrised Laguerre ensemble satisfy the loop equation
\begin{align}
0&=\left(x_1\sum_{\mu\vdash(x_1,x_1,x_1)}-\sum_{\mu\vdash(x_1,x_1)}\right)\sum_{\sqcup_{t=1}^{\#\mu}K_t=J_n}\prod_{\mu_t\in\mu}W_{\#K_t+\#\mu_t}(\mu_t,K_t) \nonumber
\\&\quad+N^2x_1W_n(x_1,J_n)+\frac{1}{2x_1}\left(x_1^2\frac{\partial^2}{\partial x_1^2}+x_1\frac{\partial}{\partial x_1}-1\right)W_n(x_1,J_n) \nonumber
\\&\quad+\frac{2}{x_1}\sum_{i=2}^n\frac{\partial}{\partial x_i}\frac{1}{x_1^2-x_i^2}\Bigg[\bigg(2x_1x_i^2\sum_{\mu\vdash(x_1,x_1)}-x_1x_i^2\sum_{\mu\vdash(x_i,x_i)}-x_1^2x_i\sum_{\mu\vdash(x_1,x_i)}\bigg) \nonumber
\\&\hspace{16em}\times\sum_{\sqcup_{t=1}^{\#\mu}K_t=J_n\setminus\{x_i\}}\prod_{\mu_t\in\mu}W_{\#K_t+\#\mu_t}(\mu_t,K_t) \nonumber
\\&\hspace{20em}+x_1x_iW_{n-1}(J_n)-x_i^2W_{n-1}(x_1,J_n\setminus\{x_i\})\Bigg] \nonumber
\\&\quad+8\sum_{2\leq i<j\leq n}\frac{1}{x_1}\frac{\partial^2}{\partial x_i\partial x_j}\Bigg\{\frac{x_1x_i^2x_jW_{n-2}(J_n\setminus\{x_i\})}{(x_1^2-x_j^2)(x_i^2-x_j^2)}-\frac{x_1x_ix_j^2W_{n-2}(J_n\setminus\{x_j\})}{(x_1^2-x_i^2)(x_i^2-x_j^2)} \nonumber
\\&\hspace{20em}+\frac{x_i^2x_j^2W_{n-2}(x_1,J_n\setminus\{x_i,x_j\})}{(x_1^2-x_i^2)(x_1^2-x_j^2)}\Bigg\}. \label{eq4.2.45}
\end{align}
\end{proposition}
\begin{proof}
We give a modification of the inductive argument detailed in \citep[App.~A]{FRW17} (see also \citep{DF20}). To proceed, let us define $E_1(x_1)$ to be the right-hand side of the $n=1$ loop equation \eqref{eq4.2.37}, $E_2(x_1,x_2)$ to be the right-hand side of the $n=2$ loop equation \eqref{eq4.2.42}, and for $n\geq3$, define $E_n(x_1,\ldots,x_n)$ to be the right-hand side of the alleged loop equation \eqref{eq4.2.45} above. Our induction hypothesis is that $E_m(x_1,\ldots,x_m)=0$ for all $1\leq m\leq n$ --- this has already been shown through equations \eqref{eq4.2.37}, \eqref{eq4.2.42} to be true for the base cases $m=1,2$. In a similar manner to the derivation of equation \eqref{eq4.2.41}, use Lemma \ref{L4.4} to write the $U_n$ loop equation \eqref{eq4.2.33} in terms of the $W_n$. Then, check term by term that the right-hand side of this equation agrees with the sum
\begin{equation*}
\sum_{K_1\sqcup K_2=J_n}U_{\#K_1}(K_1)E_{\#K_2+1}(x_1,K_2)=E_n(x_1,J_n)+\sum_{\substack{K_1\sqcup K_2=J_n,\\K_1\neq\emptyset}}U_{\#K_1}(K_1)E_{\#K_2+1}(x_1,K_2).
\end{equation*}
Since the left-hand side of the equation obtained from using Lemma \ref{L4.4} to rewrite the $U_n$ loop equation \eqref{eq4.2.33} is zero, the above sum must also vanish. Hence, we have
\begin{equation*}
0=E_n(x_1,J_n)+\sum_{\substack{K_1\sqcup K_2=J_n,\\K_1\neq\emptyset}}U_{\#K_1}(K_1)E_{\#K_2+1}(x_1,K_2)=E_n(x_1,\ldots,x_n),
\end{equation*}
where the second equality follows from our induction hypothesis. As $E_n(x_1,\ldots,x_n)$ is the right-hand side of equation \eqref{eq4.2.45}, the proposition is proved.
\end{proof}

Like with the loop equations given earlier in this subsection, to extract information from the above loop equation, one must substitute the large $N$ expansion \eqref{eq4.2.38} into said equation and collect terms of equal order in $N$ to obtain the $(n,l)$ loop equations on the correlator expansion coefficients $W_n^l$ --- it is the $(n,l)$ loop equations that can be solved. We do not display the $W_n^l$ loop equations here since it is straightforward to extract them from equation \eqref{eq4.2.34}: one need only replace products of the form $W_{n_1}(x_1,\ldots,x_{n_1})\cdots W_{n_k}(x_1,\ldots,x_{n_k})$ with sums of the form
\begin{equation*}
\sum_{l_1+\cdots+l_k=l-r}W_{n_1}^{l_1}(x_1,\ldots,x_{n_1})\cdots W_{n_k}^{l_k}(x_1,\ldots,x_{n_k}),
\end{equation*}
where $r$ depends on what power of $N$ the original product is multiplied by; we refer to the derivations of equations \eqref{eq4.2.39} and \eqref{eq4.2.43} for examples.

As for the loop equation \eqref{eq4.2.45} itself, some brief comparisons to results in the existing literature are in order. First, let us observe that for each $n\geq3$, the loop equation \eqref{eq4.2.45} is indeed higher order than the equivalent loop equations seen in the classical and general $\beta$ ensembles reviewed in \S\ref{s4.1.1}. By this, we mean that the equivalent of the term $W_{n+1}(x_1,x_1,J_n)$ in equation \eqref{eq4.1.8} is the term $W_{n+2}(x_1,x_1,x_1,J_n)$ corresponding to the $\mu=\{\,\{x_1,x_1,x_1\}\,\}$ term in first sum displayed in equation \eqref{eq4.2.45}. In fact, the fact that this sum is over $\mu\vdash(x_1,x_1,x_1)$, as opposed to $\mu\vdash(x_1,x_1)$, classifies this loop equation as `higher order' than in the classical case. Another way of quantifying the complexity of our loop equation is to note that we have sums over partitions of various size-two ordered sets appearing in many of the terms and our largest sum contains products of three connected correlators --- this is not seen in more traditional settings. Thus, if we were to simplify the $W_n^l$ loop equation corresponding to the $W_n$ loop equation \eqref{eq4.2.45} using residue calculus in a similar manner to that presented in \S\ref{s4.1.2}, our analogue of equation \eqref{eq4.1.41}, if it exists, would need at least an extra sum of a product of three differential forms $\omega_{n'}^{l'}$ on the right-hand side.

Moving past comparisons to loop equations for $\beta$ ensembles, let us recall that even though the above observations show that the loop equation \eqref{eq4.2.45} is structurally quite different from most of the loop equations studied in the literature, there still exist some works studying such higher order loop equations. Namely, multi-matrix models and chains of matrices have been shown to be characterised by such higher order loop equations \citep{EO09}, \citep{Ora15}, while loop equations involving large sums of the form $\sum_{\mu\vdash(x_1,\ldots,x_1)}$ have been studied in the abstract setting for their own sake \citep{BE13}, \citep{BHLMR14}. We also remind the reader that loop equations for the $(N,N,N)$ complex Wishart product ensemble were derived in \citep{DF20}, where they were shown to be of the same order as equation \eqref{eq4.2.45}. Nonetheless, we highlight the fact that the antisymmetric nature of the matrix $B=\tilde{\mathcal{J}}_1$ \eqref{eq3.3.57} shows itself in the loop equation \eqref{eq4.2.45} in a way that has not been seen in previously studied loop equations: Comparing the last line of equation \eqref{eq4.1.12} to the third line of equation \eqref{eq4.2.45}, we see that the denominator $x_1-x_i$ has been replaced by $x_1^2-x_i^2$, which is unchanged when mapping $x_1\mapsto-x_1$ or $x_i\mapsto-x_i$ (this structure is also seen in the last two lines of our loop equation). It can be surmised from the computations of Appendix \ref{appendixC.1} that these differences of squares arise from the fact that the mixed moments $m_{k_1,\ldots,k_n}$ vanish whenever any of the $k_i$ are odd.

\subsection{Discussion on $W_1^{(\mathcal{J}_1),0}$ and $W_2^{(\mathcal{J}_1),0}$} \label{s4.2.3}
Let us now focus our discussion on the $(1,0)$ and $(2,0)$ loop equations \eqref{eq4.2.40}, \eqref{eq4.2.44} so that we may comment on the large $N$ limiting behaviour of the global scaled $(N,N)$ antisymmetrised Laguerre ensemble. Taking $n,l$ to be such small integers also makes it feasible for us to round out this section with some concrete computations that make connections with the content of \S\ref{s3.3.3}.

We commence our discussion by noting that $W_1^0(x_1)$ has the large $x_1$ expansion
\begin{equation*}
W_1^0(x_1)=W_1^{(\mathcal{J}_1),0}(x_1)=\sum_{k=0}^{\infty}\frac{\tilde{M}_{k,0}^{(\mathcal{J}_1)}}{x_1^{k+1}},
\end{equation*}
in keeping with equation \eqref{eq4.2.4} and having adopted the notation of equation \eqref{eq1.1.32} to write $\tilde{m}_k^{(\mathcal{J}_1)}=\sum_{l=0}^\infty\tilde{M}_{k,l}^{(\mathcal{J}_1)}N^{-l}$ (as allowed by Proposition \ref{prop3.13}). Letting $\tilde{M}_{k,0}^{(\mathrm{i}\mathcal{J}_1)}$ denote the analogous scaled moments of the matrix $\mathrm{i}\mathcal{J}_1$, we have that
\begin{equation} \label{eq4.2.46}
\tilde{M}_{k,0}^{(\mathcal{J}_1)}=\lim_{N\to\infty}\mean{\Tr\,\tilde{\mathcal{J}}_1^k}=(-\mathrm{i})^k\lim_{N\to\infty}\mean{\Tr\,(\mathrm{i}\tilde{\mathcal{J}}_1)^k}=(-\mathrm{i})^k\tilde{M}_{k,0}^{(\mathrm{i}\mathcal{J}_1)}.
\end{equation}
As these latter moments are also characterised as the spectral moments of the large $N=N_0$ limit $\rho^{(\mathrm{i}\mathcal{J}_1),0}(\lambda)$ of the scaled density $\tilde{\rho}^{(\mathrm{i}\mathcal{J}_1)}(\lambda)$ \eqref{eq1.3.20} specified in Definition \ref{def1.13} of \S\ref{s1.3.1}, equation \eqref{eq1.3.23} tells us that
\begin{equation} \label{eq4.2.47}
\tilde{M}_{2k,0}^{(\mathcal{J}_1)}=(-1)^km_k^{(FC_2)},
\end{equation}
where $m_k^{(FC_2)}$ is the $m=2$ Fuss--Catalan number specified by equation \eqref{eq1.3.17} (recall that $\tilde{M}_{k,0}^{(\mathcal{J}_1)}$ vanishes for odd values of $k$ due to the fact that $\mathcal{J}_1$ is antisymmetric). Indeed, one can check using computer algebra that the solution of equation \eqref{eq4.2.40} has large $x_1$ expansion
\begin{equation} \label{eq4.2.48}
W_1^0(x_1)=\frac{1}{x_1}-\frac{1}{x_1^3}+\frac{3}{x_1^5}-\frac{12}{x_1^7}+\frac{55}{x_1^9}-\frac{273}{x_1^{11}}+{\rm O}(x_1^{-13});
\end{equation}
observe that for integer $k$, the coefficient of $x_1^{-2k-1}$ is precisely $(-1)^km_k^{(FC_2)}=(-1)^k\frac{1}{2k+1}\binom{3k}{k}$, while there are no terms of the form $x_1^{-2k}$.

Rewriting equation \eqref{eq4.2.46} in terms of the moment generating functions defined by equation \eqref{eq1.1.17} and using equation \eqref{eq4.2.47}, we see that
\begin{equation*}
W_1^{(\mathcal{J}_1),0}(x_1)=\sum_{k=0}^{\infty}\frac{\tilde{M}_{2k,0}^{(\mathcal{J}_1)}}{x_1^{2k+1}}=-x_1\sum_{k=0}^{\infty}\frac{m_k^{(FC_2)}}{(-x_1^2)^{k+1}}=-x_1W_1^{(FC_2)}(-x_1^2),
\end{equation*}
where $W_1^{(FC_2)}(x):=\sum_{k=0}^\infty m_k^{(FC_2)}/x^{k+1}$ is the generating function of the relevant Fuss--Catalan numbers; equivalently, it is the Stieltjes transform \eqref{eq1.1.18} of the $m=2$ Fuss--Catalan distribution $\rho^{(FC_2)}(\lambda)$ \eqref{eq1.3.16}. In combination with equation \eqref{eq4.2.40}, this suggests that upon writing $u=-x_1^2$, the resolvent $W_1^{(FC_2)}(u)$ satisfies the functional relation
\begin{equation} \label{eq4.2.49}
0=u^2\left(W_1^{(FC_2)}(u)\right)^3-uW_1^{(FC_2)}(u)+1.
\end{equation}
This is known to be true (see, e.g., \citep{FL15}, whence equation \eqref{eq1.3.23} could be obtained), and thus serves as confirmation that the loop equation analysis presented in \S\ref{s4.2.2} has produced the correct spectral curve for the global scaled $(N,N)$ antisymmetrised Laguerre ensemble.

\begin{remark} \label{R4.1}
For another consistency check of our spectral curve \eqref{eq4.2.40}, we can compute it using techniques from free probability theory (see, e.g. \citep{MS17}). However, such techniques are designed to probe the macroscopic statistics of products of random matrices and are thus ill suited for accessing statistics that are not of leading order in $N$. These lower order statistics are of interest to us since, by Proposition \ref{prop3.13}, our $1/N$ expansion \eqref{eq4.0.5} has combinatorial meaning at all orders in $N$. Moreover, these combinatorial interpretations are valid for finite $N$, while free probability theory is concerned with the $N\to\infty$ limit. Let us thus draw attention to the fact that our loop equation analysis allows us to compute the mixed cumulants $\tilde{c}_{k_1,\ldots,k_n}^{(\mathcal{J}_1)}$ to any desired order in $N$, assuming the availability of enough computation power and time (recall the discussion below equation \eqref{eq4.1.15}).
\end{remark}

In keeping with the discussion at the end of \S\ref{s4.2.2}, let us reiterate that, as foreshadowed at the beginning of this chapter, our spectral curve \eqref{eq4.2.40} is a third order polynomial in $W_1^0(x_1)=W_1^{(\mathcal{J}_1),0}(x_1)$. Thus, finding a rational parametrisation of our spectral curve is not as straightforward as, say, in the case of the one-cut 1-Hermitian matrix model reviewed in \S\ref{s4.1.2}. However, we are quite fortuitous in that the spectral curve \eqref{eq4.2.40} is a linear polynomial in $x_1$, so we may rearrange it to write
\begin{equation*}
x_1=\frac{1}{\left(W_1^0(x_1)\right)^3+W_1^0(x_1)}.
\end{equation*}
Thus, in analogy with the parametrisations \eqref{eq4.1.31}, \eqref{eq4.1.32}, a natural parametrisation of the above equation presents itself to us: Simply define
\begin{align}
x(z)&=\frac{1}{z^3+z}, \label{eq4.2.50}
\\ y(z)&=z. \label{eq4.2.51}
\end{align}
Then, replacing $x_1$ by $x(z_1)$ and $W_1^0(x_1)=W_1^0(x(z_1))$ by its analytic continuation $y(z_1)$ in the right-hand side of equation \eqref{eq4.2.40} returns zero, showing that this is a valid parametrisation of our spectral curve. Using the Newton polygon method (see \citep{BP00} and references therein) on the complex algebraic curve $0=xy^3+xy-1$ shows that our spectral curve is the genus zero Riemann sphere with viable coordinates $x(z),y(z)$ as given above. Hence, $W_1^0(x(z_1))$ is a single-valued function of $z_1\in\mathbb{CP}^1$ whose value agrees with $W_1^0(x_1)$ when $z_1\in x^{-1}(x_1)$ is chosen to be close to $z_1=0$ (this ensures that we choose the solution of equation \eqref{eq4.2.40} obeying the asymptotic requirement \eqref{eq2.4.15} that $W_1^0(x_1)\sim1/x_1$ as $x_1\to\infty$).

Now, taking $z_1,z_2\in\mathbb{CP}^1$ and substituting $x_1=x(z_1)$ and $x_2=x(z_2)$ \eqref{eq4.2.50} into equation \eqref{eq4.2.44} shows that
\begin{align}
W_2^0(x(z_1),x(z_2))&=\frac{1}{x'(z_1)x'(z_2)}\frac{1}{(z_1-z_2)^2}-\frac{1}{(x(z_1)-x(z_2))^2} \nonumber
\\&\quad-\frac{1}{x'(z_1)x'(z_2)}\frac{1}{(z_1+z_2)^2}+\frac{}{(x(z_1)+x(z_2))^2}. \label{eq4.2.52}
\end{align}
The first line of the right-hand side is the value of $W_2^0(x(z_1),x(z_2))$ obtained in the case of the one-cut 1-Hermitian matrix model upon rearranging equation \eqref{eq4.1.34}, while the second line ensures that $W_2^0(x(z_1),x(z_2))$ is antisymmetric in $z_1,z_2$. Thus, we have a double pole at $z_1=-z_2$ in addition to the double pole at $z_1=z_2$ seen in the ensembles reviewed in Section~\ref{s4.1}. This phenomenon is consistent with the appearance of the denominators of the form $x_a^2-x_b^2$ in equation \eqref{eq4.2.45}, which was briefly discussed at the end of \S\ref{s4.2.2}. In keeping with Remark~\ref{R4.1} above, we show that our loop equation analysis can determine $W_1^1(x_1)$, which is the $1/N$ correction to $W_1^0(x_1)$. This is a simple matter of setting $l=1$ in equation \eqref{eq4.2.39} with $x_1$ replaced by $x(z_1)$ and $W_1^0(x_1)$ replaced by $y(z_1)=z_1$ \eqref{eq4.2.51} to see that
\begin{equation} \label{eq4.2.53}
W_1^1(x(z_1))=\frac{z_1^2}{x(z_1)(3z_1^2+1)}=\frac{z_1^5+z_1^3}{3z_1^2+1}.
\end{equation}

From Proposition \ref{prop3.13}, we know that the mixed cumulants $\tilde{c}_{k_1,\ldots,k_n}^{(\mathcal{J}_1)}$ have a genus expansion of the form
\begin{equation} \label{eq4.2.54}
\tilde{c}_{k_1,\ldots,k_n}^{(\mathcal{J}_1)}=N^{2-n}\sum_{l=0}^{k_1+\cdots+k_n+1-n}\frac{c_{k_1,\ldots,k_n}^{(\mathcal{J}_1),l}}{N^l}.
\end{equation}
Combining this with the large $N$ expansion \eqref{eq4.0.5} and large $x_1,\ldots,x_n$ expansion \eqref{eq4.2.4} of the connected correlators $W_n(x_1,\ldots,x_n)$ then shows that
\begin{equation} \label{eq4.2.55}
W_n^l(x_1,\ldots,x_n)=\sum_{k_1,\ldots,k_n=0}^{\infty}\frac{c_{k_1,\ldots,k_n}^{(\mathcal{J}_1),l}}{x_1^{k_1+1}\cdots x_n^{k_n+1}}.
\end{equation}
Hence, noting that these interpretations as generating functions are valid only in the large $x_1,\ldots,x_n$ regime, we have that the coefficients of the cumulant genus expansions can be extracted from the correlator expansion coefficients using the residue formula
\begin{equation} \label{eq4.2.56}
c_{k_1,\ldots,k_n}^{(\mathcal{J}_1),l}=(-1)^n\underset{x_1,\ldots,x_n=\infty}{\mathrm{Res}}x_1^{k_1}\cdots x_n^{k_n}W_n^l(x_1,\ldots,x_n)\,\mathrm{d}x_1\cdots\mathrm{d}x_n.
\end{equation}
As our formalism produces expressions for the correlator expansion coefficients in terms of the variables $z_1,\ldots,z_n\in\mathbb{CP}^1$ through the map \eqref{eq4.2.50}, this residue formula is only useful to us upon reformulating it in terms of the $z_1,\ldots,z_n$ variables, as well. Thus, observe that $x_i=x(z_i)=\infty$ corresponds to $z_i=0,\pm\mathrm{i}$. Of these three solutions, we are interested in $z_i=0$, as it coincides with the requirement that $y(z_i)=z_i\to0$ as $x(z_i)\to\infty$, which follows from the fact that the branch of $W_1^0(x_1)$ that behaves as a moment generating function is the one that exhibits the asymptotic $W_1^0(x_1)\sim1/x_1$ in the $x_1\to\infty$ limit. Noting also that setting $x_i=x(z_i)$ means that $\mathrm{d}x_i=x'(z_i)\mathrm{d}z_i$, equation \eqref{eq4.2.56} therefore implies that
\begin{equation} \label{eq4.2.57}
c_{k_1,\ldots,k_n}^{(\mathcal{J}_1),l}=(-1)^n\underset{z_1,\ldots,z_n=0}{\mathrm{Res}}W_n^l(x(z_1),\ldots,x(z_n))\prod_{i=1}^nx(z_i)^{k_i}x'(z_i)\,\mathrm{d}z_i.
\end{equation}

For example, inserting $W_1^0(x(z_1))=z_1$ into the above shows that $c_4^{(\mathcal{J}_1),0}=3$, which is exactly the coefficient of $1/x_1^5$ in the large $x_1$ expansion \eqref{eq4.2.48}. Likewise, substituting equations \eqref{eq4.2.52} and \eqref{eq4.2.53} into the formula \eqref{eq4.2.57} shows that $c_2^{(\mathcal{J}_1),1}=1$ and $c_{2,2}^{(\mathcal{J}_1),0}=12$ --- see Appendix \ref{appendixD.1} for complementary data. From Proposition \ref{prop3.13}, these computations imply that, up to the topological equivalence relation described above Proposition \ref{prop3.9}, there are three distinct genus zero rooted topological hypermaps with a degree four black vertex, one or two red vertices of even valency, and an even number of twisted edges; there is one rooted topological hypermap of Euler genus one with one bivalent black vertex, one bivalent red vertex, and one twisted edge; and there are twelve genus zero $2$-rooted topological hypermaps with two degree two black vertices, one or two red vertices of even valency, and an even number of twisted edges. The one topological hypermap counted by $c_2^{(\mathcal{J}_1),1}$ has already been displayed in Figure~\ref{fig3.19}, while the fifteen topological hypermaps counted by $c_4^{(\mathcal{J}_1),0}$ and $c_{2,2}^{(\mathcal{J}_1),0}$ are shown in Figures~\ref{fig4.3} and~\ref{fig4.4} below. Note that in general, $|c_{k_1,\ldots,k_n}^{(\mathcal{J}_1),l}|$ may take a value which is less than the number of hypermaps it corresponds to, due to the cancelling in equation \eqref{eq3.3.61} --- we do not see this here due to $n,l$ being too low.
\begin{figure}[H]
        \centering
\captionsetup{width=.9\linewidth}
        \includegraphics[width=0.7\textwidth]{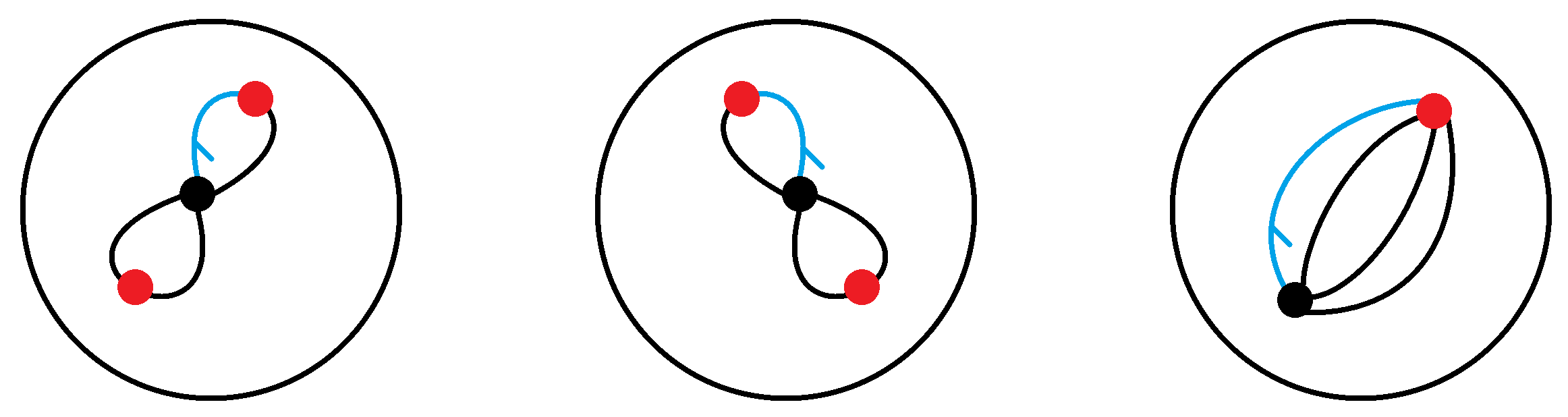}
        \caption[The topological hypermaps corresponding to $\lim_{N\to\infty}\tilde{c}_4^{(\mathcal{J}_1)}/N$]{These three topological hypermaps are counted by $c_4^{(\mathcal{J}_1),0}=3$. The root edges are coloured blue with notches indicating local orientation of the black vertices. All the edges are untwisted, so we do not need the $\pm1$ labelling discussed in Proposition \ref{prop3.13}.} \label{fig4.3}
\end{figure}

\begin{figure}[H]
        \centering
\captionsetup{width=.9\linewidth}
        \includegraphics[width=0.88\textwidth]{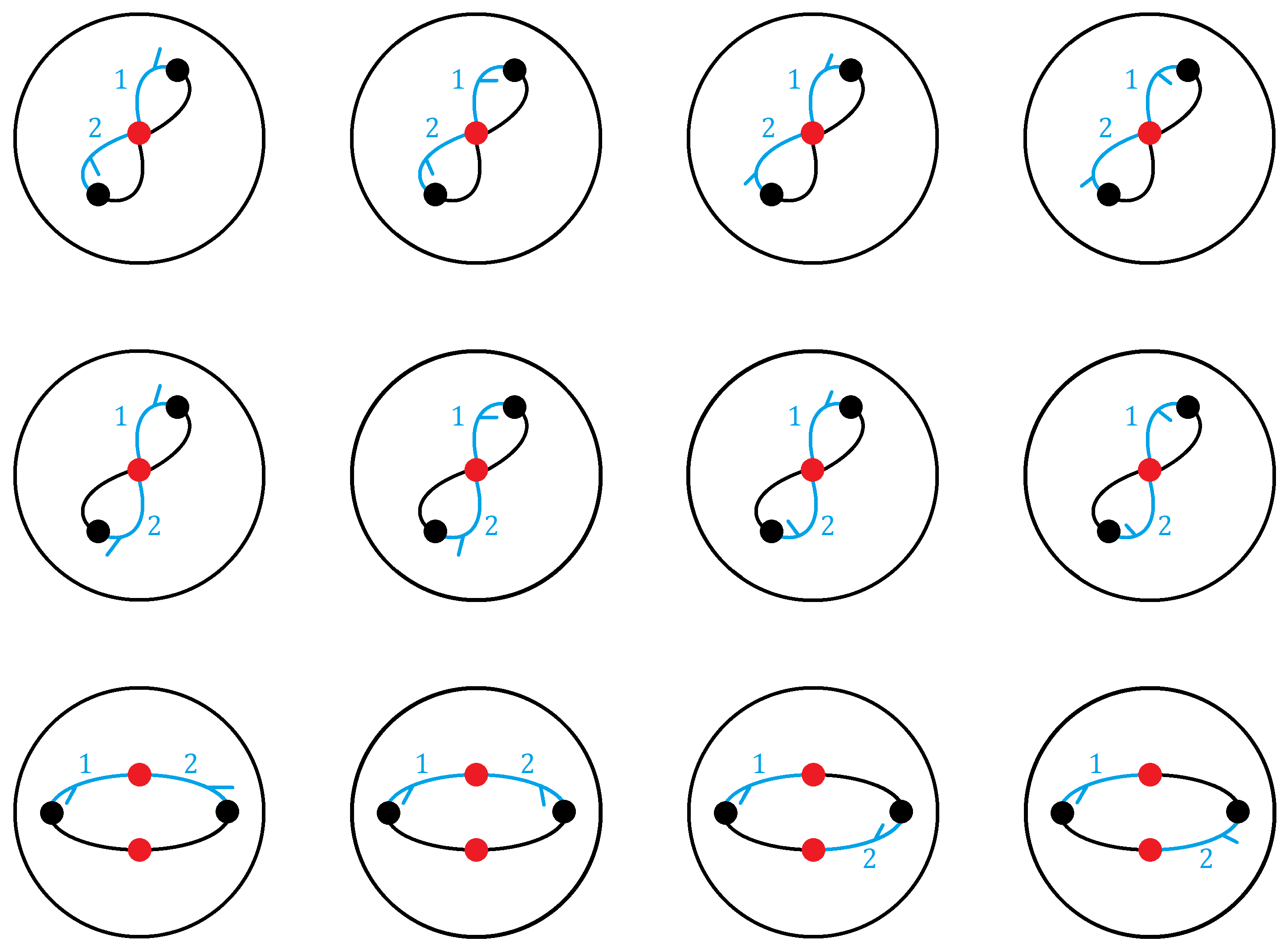}
        \caption[The topological hypermaps corresponding to $\lim_{N\to\infty}\tilde{c}_{2,2}^{(\mathcal{J}_1)}$]{These twelve topological hypermaps are counted by $c_{2,2}^{(\mathcal{J}_1),0}=12$. We have again coloured the two root edges blue, distinguished them with the labels $1,2$, and used notches to mark the local orientation of each black vertex. We have avoided drawing complicated deformations of the sphere, like in Figure \ref{fig3.18}, by having all our edges be untwisted and instead letting our black vertices have possibly inconsistent local orientations.} \label{fig4.4}
\end{figure}

\setcounter{equation}{0}
\section{Loop Equations for the Hermitised Laguerre Ensemble} \label{s4.3}
This section is concerned with the derivation of loop equations for the global scaled $(N,N)$ Hermitised Laguerre ensemble defined in Proposition \ref{prop3.10}. In parallel to the development of the previous section, we derive loop equations for the unconnected $n$-point correlators $\tilde{U}_n^{(\mathcal{H}_1)}(x_1,\ldots,x_n)$ \eqref{eq4.1.3} in \S\ref{s4.3.1} and the connected $n$-point correlators $\tilde{W}_n^{(\mathcal{H}_1)}(x_1,\ldots,x_n)$ \eqref{eq4.1.4}, along with the coefficients $W_n^{(\mathcal{H}_1),l}(x_1,\ldots,x_n)$ of their large $N$ expansions \eqref{eq4.0.5}, in \S\ref{s4.3.2}. These loop equations are used to compute $W_1^{(\mathcal{H}_1),0}(x_1)$ and $W_2^{(\mathcal{H}_1),0}(x_1,x_2)$ in \S\ref{s4.3.3}. In contrast to Section \ref{s4.2}, the discussion of \S\ref{s4.3.1}--\S\ref{s4.3.3} will be relatively succinct: we focus on presenting analogues of the results given in Section \ref{s4.2}, providing technical details of their derivations only when they differ to those shown earlier. As one might expect, the mechanisms underlying the results of Sections \ref{s4.2} and \ref{s4.3} are quite similar.

To begin, let us redefine $X$ to be the scaled complex $N\times N$ Ginibre matrix with p.d.f.
\begin{equation}
\tilde{P}^{(GinUE)}(X)=\left(\frac{N}{\pi}\right)^{N^2}\exp\left(-N\Tr(X^\dagger X)\right)
\end{equation}
and $B$ to be the product $B:=X^\dagger\tilde{H}X$, where $\tilde{H}$ is the scaled $N\times N$ GUE matrix with p.d.f.
\begin{equation}
\tilde{P}^{(GUE)}(\tilde{H})=2^{N(N-1)/2}\left(\frac{N}{2\pi}\right)^{N^2/2}\exp\left(-\tfrac{N}{2}\Tr(\tilde{H}^2)\right).
\end{equation}
Note that $X=\tilde{G}$ in distribution, with $\tilde{G}=\tilde{G}_1$ as in Propositions \ref{prop3.10} and \ref{prop3.12}, $\tilde{H}$ is as in said propositions (this matrix represents a slightly different global scaling of the GUE to the scaling specified in Definition \ref{def1.6}), and therefore $B=\tilde{\mathcal{H}}_1$. With regards to the quantities of Definition \ref{def1.10}, equivalently equation \eqref{eq4.0.1}, we have that $\tilde{H}=\sqrt{2/N}H$, $X=G_1/\sqrt{N}$, and $B=\sqrt{2}N^{-3/2}\mathcal{H}_1$; cf.~equation \eqref{eq3.3.47}. Letting $\mean{\,\cdot\,}$ now denote averages with respect to the product $\tilde{P}^{(GinUE)}(X)\tilde{P}^{(GUE)}(\tilde{H})$, we again simplify notation by omitting tildes and the superscript $(\mathcal{H}_1)$ to write
\begin{align}
m_{k_1,\ldots,k_n}&=\mean{\prod_{i=1}^n\Tr\,B^{k_i}},\quad k_1,\ldots,k_n\in\mathbb{N}, \label{eq4.3.3}
\\ U_n(x_1,\ldots,x_n)&=\sum_{k_1,\ldots,k_n=0}^{\infty}\frac{m_{k_1,\ldots,k_n}}{x_1^{k_1+1}\cdots x_n^{k_n+1}}, \label{eq4.3.4}
\end{align}
as in equations \eqref{eq4.2.2}, \eqref{eq4.2.3} --- recall from \S\ref{s3.3.3} that $m_{k_1,\ldots,k_n}=0$ when $k_1+\cdots+k_n$ is odd. The associated mixed cumulants $c_{k_1,\ldots,k_n}$ and connected correlators $W_n(x_1,\ldots,x_n)$ are specified by equations \eqref{eq1.1.28} and \eqref{eq1.1.29}, respectively (we also recall equation \eqref{eq4.1.4}). Compared to the statistics of the unscaled matrix $\mathcal{H}_1$ \eqref{eq4.0.1}, indicated by the superscript $(\mathcal{H}_1)$, but without any tildes, the statistical equivalence $B=\sqrt{2}N^{-3/2}\mathcal{H}_1$ implies that
\begin{align}
m_{k_1,\ldots,k_n}&=\left(\frac{\sqrt{2}}{N^{3/2}}\right)^{k_1+\cdots+k_n}m_{k_1,\ldots,k_n}^{(\mathcal{H}_1)},
\\ c_{k_1,\ldots,k_n}&=\left(\frac{\sqrt{2}}{N^{3/2}}\right)^{k_1+\cdots+k_n}c_{k_1,\ldots,k_n}^{(\mathcal{H}_1)},
\\ U_n(x_1,\ldots,x_n)&=\left(\frac{N^{3/2}}{\sqrt{2}}\right)^nU_n^{(\mathcal{H}_1)}(N^{3/2}x_1/\sqrt{2},\ldots,N^{3/2}x_n/\sqrt{2}),
\\ W_n(x_1,\ldots,x_n)&=\left(\frac{N^{3/2}}{\sqrt{2}}\right)^nW_n^{(\mathcal{H}_1)}(N^{3/2}x_1/\sqrt{2},\ldots,N^{3/2}x_n/\sqrt{2}).
\end{align}
As in the previous section, the mixed cumulants $c_{k_1,\ldots,k_n}$ having the genus expansion \eqref{eq3.3.60} means that $W_n(x_1,\ldots,x_n)$ admits the large $N$ expansion \eqref{eq4.0.5}. We retain our convention of denoting the associated expansion coefficients as $W_n^l(x_1,\ldots,x_n)$ and note that, by the discussion following Proposition \ref{prop3.10}, $W_n^l=0$ whenever $l$ is odd.

Emulating the key idea of Section \ref{s4.2}, we commence our analysis of the global scaled $(N,N)$ Hermitised Laguerre ensemble by considering the average
\begin{equation} \label{eq4.3.9}
\mean{\left(\partial_{(X^\dagger)_{ab}}+NX_{ba}\right)\left(\partial_{(\tilde{H}^\dagger)_{bc}}+N\tilde{H}_{cb}\right)\left(\partial_{X_{cd}}+N(X^\dagger)_{dc}\right)(B^{k_1})_{ad}\Tr\,B^{I_n}},
\end{equation}
in analogy with the average \eqref{eq4.2.11} ($I_n$ and $\Tr\,B^{I_n}$ are as in Lemma \ref{L4.1}). This choice of starting point is inspired by the following computations:
\begin{align}
\partial_{(X^\dagger)_{ab}}\exp\left(-N\Tr(X^\dagger X)\right)&=-N\left\{\partial_{(X^\dagger)_{ab}}\sum_{e,f=1}^N(X^\dagger)_{ef}X_{fe}\right\}\exp\left(-N\Tr(X^\dagger X)\right) \nonumber
\\&=-N\sum_{e,f=1}^N\Big[\left\{\partial_{(X^\dagger)_{ab}}(X^\dagger)_{ef}\right\}X_{fe} \nonumber
\\&\hspace{6em}+(X^\dagger)_{ef}\left\{\partial_{(X^\dagger)_{ab}}X_{fe}\right\}\Big]\exp\left(-N\Tr(X^\dagger X)\right) \nonumber
\\&=-N\sum_{e,f=1}^N\left[\chi_{e=a,f=b}X_{fe}+0\right]\exp\left(-N\Tr(X^\dagger X)\right) \nonumber
\\&=-NX_{ba}\exp\left(-N\Tr(X^\dagger X)\right), \label{eq4.3.10}
\\\partial_{(\tilde{H}^\dagger)_{bc}}\exp\left(-\tfrac{N}{2}\Tr(\tilde{H}^2)\right)&=-\frac{N}{2}\sum_{e,f=1}^N\Big[\left\{\partial_{(\tilde{H}^\dagger)_{bc}}\tilde{H}_{ef}\right\}\tilde{H}_{fe} \nonumber
\\&\hspace{6em}+\tilde{H}_{ef}\left\{\partial_{(\tilde{H}^\dagger)_{bc}}\tilde{H}_{fe}\right\}\Big]\exp\left(-\tfrac{N}{2}\Tr(\tilde{H}^2)\right) \nonumber
\\&=-N\tilde{H}_{cb}\exp\left(-\tfrac{N}{2}\Tr(\tilde{H}^2)\right), \label{eq4.3.11}
\\ \partial_{X_{cd}}\exp\left(-N\Tr(X^\dagger X)\right)&=-N(X^\dagger)_{dc}\exp\left(-N\Tr(X^\dagger X)\right). \label{eq4.3.12}
\end{align}
The first two equalities in the above follow from the chain and Leibniz product rules and the third equality is due to the fact that, as a complex variable, $X_{fe}$ is independent of $(X^\dagger)_{ab}=\overline{X_{ba}}$ for any choice of $e,f$. Similar reasoning yields equations \eqref{eq4.3.11} and \eqref{eq4.3.12} upon recalling that $\tilde{H}$ being Hermitian implies that $\tilde{H}^\dagger=\tilde{H}$. 

As in Section \ref{s4.2}, let us now proceed by computing the average \eqref{eq4.3.9} in two different ways, thereby deriving loop equations on the mixed moments $m_{k_1,\ldots,k_n}$ and, consequently, the unconnected correlators $U_n(x_1,\ldots,x_n)$.

\subsection{Loop equations on the $\tilde{U}_n^{(\mathcal{H}_1)}$} \label{s4.3.1}
Since $(X^\dagger)_{ab}=\overline{X_{ba}}$ and $X_{cd}$ are independent of each other for all choices of $a,b,c,d$, the analogue of Lemma \ref{L4.2} relevant to the present setting is comparatively simple:
\begin{lemma} \label{L4.5}
Having set $B=X^\dagger\tilde{H}X$ as above, the partial derivatives of $(B^{k_1})_{ef}$ with respect to $(X^\dagger)_{ab}$, $\tilde{H}_{bc}$, and $X_{cd}$ are respectively given by
\begin{align}
\partial_{(X^\dagger)_{ab}}(B^{k_1})_{ef}&=\sum_{p_1+p_2=k_1-1}(B^{p_1})_{ea}(\tilde{H}XB^{p_2})_{bf},
\\ \partial_{\tilde{H}_{bc}}(B^{k_1})_{ef}&=\sum_{p_1+p_2=k_1-1}(B^{p_1}X^\dagger)_{eb}(XB^{p_2})_{cf},
\\ \partial_{X_{cd}}(B^{k_1})_{ef}&=\sum_{p_1+p_2=k_1-1}(B^{p_1}X^\dagger\tilde{H})_{ec}(B^{p_2})_{df}.
\end{align}
Similar to Lemma \ref{L4.2}, these sums are empty when $k_1=0$ and are otherwise taken over the range $p_1=0,1,\ldots,k_1-1$. The indices $e,f$ are arbitrary and the above equations hold true upon setting $e,f$ equal to each other or to one of $a,b,c,d$.
\end{lemma}

Moving on, the analogue of Proposition \ref{prop4.1} corresponding to the global scaled $(N,N)$ Hermitised Laguerre ensemble is as follows:
\begin{proposition} \label{prop4.4}
Writing $I_n=(k_2,\ldots,k_n)$ and $\Tr\,B^{I_n}=\Tr\,B^{k_2}\cdots\Tr\,B^{k_n}$ as usual, the average \eqref{eq4.3.9} simplifies to both the left- and right-hand sides of the equation
\begin{multline} \label{eq4.3.16}
\mean{\left\{\partial_{(X^\dagger)_{ab}}\partial_{(\tilde{H}^\dagger)_{bc}}\partial_{X_{cd}}(B^{k_1})_{ad}\Tr\,B^{I_n}\right\}}
\\=N^3m_{k_1+1,I_n}-k_1\sum_{p_1+p_2=k_1-1}m_{p_1,p_2,I_n}-2\sum_{2\leq i<j\leq n}k_ik_jm_{k_1,k_i+k_j-1,I_n\setminus\{k_i,k_j\}}
\\-\sum_{i=2}^nk_i\left[2k_1m_{k_1+k_i-1,I_n\setminus\{k_i\}}+\sum_{p_1+p_2=k_i-1}m_{k_1,p_1,p_2,I_n\setminus\{k_i\}}\right].
\end{multline}
\end{proposition}
\begin{proof}
Expanding the first two brackets in the argument of the average \eqref{eq4.3.9} and using the independence of $\tilde{H}$ from $X^\dagger,X$ shows that
\begin{multline*}
\mean{\left(\partial_{(X^\dagger)_{ab}}+NX_{ba}\right)\left(\partial_{(\tilde{H}^\dagger)_{bc}}+N\tilde{H}_{cb}\right)\left(\partial_{X_{cd}}+N(X^\dagger)_{dc}\right)(B^{k_1})_{ad}\Tr\,B^{I_n}}
\\=\mean{\partial_{(X^\dagger)_{ab}}\left(\partial_{(\tilde{H}^\dagger)_{bc}}+N\tilde{H}_{cb}\right)\left(\partial_{X_{cd}}+N(X^\dagger)_{dc}\right)(B^{k_1})_{ad}\Tr\,B^{I_n}}
\\+N\mean{\partial_{(\tilde{H}^\dagger)_{bc}}X_{ba}\left(\partial_{X_{cd}}+N(X^\dagger)_{dc}\right)(B^{k_1})_{ad}\Tr\,B^{I_n}}
\\+N^2\mean{X_{ba}\tilde{H}_{cb}\left(\partial_{X_{cd}}+N(X^\dagger)_{dc}\right)(B^{k_1})_{ad}\Tr\,B^{I_n}}.
\end{multline*}
The first two averages on the right-hand side vanish by the fundamental theorem of calculus, while the third average can be simplified using integration by parts:
\begin{align}
N^2&\mean{X_{ba}\tilde{H}_{cb}\left(\partial_{X_{cd}}+N(X^\dagger)_{dc}\right)(B^{k_1})_{ad}\Tr\,B^{I_n}} \nonumber
\\&=N^3\mean{\Tr\,B^{k_1+1}\Tr\,B^{I_n}}+N^2\mean{X_{ba}\tilde{H}_{cb}\partial_{X_{cd}}(B^{k_1})_{ad}\Tr\,B^{I_n}} \nonumber
\\&=N^3\mean{\Tr\,B^{k_1+1}\Tr\,B^{I_n}}+N^2\mean{\partial_{X_{cd}}X_{ba}\tilde{H}_{cb}(B^{k_1})_{ad}\Tr\,B^{I_n}} \nonumber
\\&\quad-N^2\mean{\left\{\partial_{X_{cd}}X_{ba}\right\}\tilde{H}_{cb}(B^{k_1})_{ad}\Tr\,B^{I_n}} \nonumber
\\&=N^3\mean{\Tr\,B^{k_1+1}\Tr\,B^{I_n}}-N^2\mean{\chi_{c=b,d=a}\tilde{H}_{cb}(B^{k_1})_{ad}\Tr\,B^{I_n}} \nonumber
\\&=N^3m_{k_1+1,I_n}-N^2\mean{\Tr\,\tilde{H}\Tr\,B^{k_1}\Tr\,B^{I_n}}; \label{eq4.3.17}
\end{align}
the average $\mean{\partial_{X_{cd}}X_{ba}\tilde{H}_{cb}(B^{k_1})_{ad}\Tr\,B^{I_n}}$ in the third line vanishes due to the fundamental theorem of calculus, yet again. Now, the average $\mean{\Tr\,\tilde{H}\Tr\,B^{k_1}\Tr\,B^{I_n}}$ poses a challenge not seen in the proof of Proposition \ref{prop4.1}, wherein the analogous average $\mean{\Tr\,J\Tr\,B^{k_1}\Tr\,B^{I_n}}$ was zero due to the antisymmetry of $J$. Here, we must derive auxiliary identities to deal with this average, in a similar fashion to how identities resulting from applying integration by parts to the averages \eqref{eq4.1.20} and \eqref{eq4.1.23} were needed to derive loop equations for the Jacobi $\beta$ ensemble. The fundamental theorem of calculus (left-hand side) and the Leibniz product rule combined with Lemma \ref{L4.5} (right-hand side) shows that
\begin{multline} \label{eq4.3.18}
0=\mean{\partial_{(\tilde{H}^\dagger)_{aa}}\Tr\,B^{k_1}\Tr\,B^{I_n}}
\\=k_1\mean{\Tr(B^{k_1-1}X^\dagger X)\Tr\,B^{I_n}}+\sum_{i=2}^nk_i\mean{\Tr(B^{k_i-1}X^\dagger X)\Tr\,B^{k_1}\Tr\,B^{I_n\setminus\{k_i\}}}
\\-N\mean{\Tr\,\tilde{H}\Tr\,B^{k_1}\Tr\,B^{I_n}}.
\end{multline}
The averages involving traces of the form $\Tr(B^{k_j-1}X^\dagger X)$ can be expressed in terms of the mixed moments by likewise showing that for $i=1,\ldots,n$,
\begin{multline*}
0=\mean{\partial_{(X^\dagger)_{ab}}(B^{k_i-1}X^\dagger)_{ab}\Tr\,B^{\{k_1\}\cup I_n\setminus\{k_i\}}}
\\=-N\mean{\Tr(B^{k_i-1}X^\dagger X)\Tr\,B^{\{k_1\}\cup I_n\setminus\{k_i\}}}+\sum_{p_1+p_2=k_i-1}m_{p_1,p_2,\{k_1\}\cup I_n\setminus\{k_i\}}
\\+\sum_{\substack{1\leq j\leq n,\\j\neq i}}k_jm_{k_i+k_j-1,\{k_1\}\cup I_n\setminus\{k_i,k_j\}}.
\end{multline*}
Setting $i=1,\ldots,n$ into this equation and substituting the resulting set of identities into equation \eqref{eq4.3.18} multiplied by $N$ shows that
\begin{align*}
-N^2&\mean{\Tr\,\tilde{H}\Tr\,B^{k_1}\Tr\,B^{I_n}}
\\&=-Nk_1\mean{\Tr(B^{k_1-1}X^\dagger X)\Tr\,B^{I_n}}-N\sum_{i=2}^nk_i\mean{\Tr(B^{k_i-1}X^\dagger X)\Tr\,B^{k_1}\Tr\,B^{I_n\setminus\{k_i\}}}
\\&=-k_1\sum_{p_1+p_2=k_1-1}m_{p_1,p_2,I_n}-k_1\sum_{j=2}^nk_jm_{k_1+k_j-1,I_n\setminus\{k_j\}}-\sum_{i=2}^nk_i\sum_{p_1+p_2=k_i-1}m_{k_1,p_1,p_2,I_n\setminus\{k_i\}}
\\&\qquad-\sum_{i=2}^nk_i\left(k_1m_{k_1+k_i-1,I_n\setminus\{k_i\}}+\sum_{\substack{2\leq j\leq n,\\j\neq i}}k_jm_{k_1,k_i+k_j-1,I_n\setminus\{k_i,k_j\}}\right).
\end{align*}
Substituting this result into equation \eqref{eq4.3.17} then gives the right-hand side of equation \eqref{eq4.3.16}, as required.

To show that the average \eqref{eq4.3.9} simplifies to the left-hand side of equation \eqref{eq4.3.16}, one need only repeat the relevant arguments given in the proof of Proposition \ref{prop4.1}, with the identities \eqref{eq4.3.10}--\eqref{eq4.3.12} in mind.
\end{proof}

The next step in obtaining loop equations on the mixed moments is to expand the right-hand side of equation \eqref{eq4.3.16} using the Leibniz product rule. Letting
\begin{equation*}
f_i(B)=\begin{cases}(B^{k_1})_{ad},&i=1,\\ \Tr\,B^{k_i},&i=2,\ldots,n\end{cases}
\end{equation*}
and writing $\mathcal{A}_{i,j,h}$ for the average obtained by replacing $f_i(B)$ by $\{\partial_{(X^\dagger)_{ab}}f_i(B)\}$, then $f_j(B)$ by $\{\partial_{(\tilde{H}^\dagger)_{bc}}f_j(B)\}$, and finally $f_h(B)$ by $\{\partial_{X_{cd}}f_h(B)\}$ in $\langle f_1(B)\cdots f_n(B)\rangle$ so that, e.g.,
\begin{equation*}
\mathcal{A}_{3,3,1}=\mean{\left\{\partial_{(X^\dagger)_{ab}}\partial_{(\tilde{H}^\dagger)_{bc}}\Tr\,B^{k_3}\right\}\left\{\partial_{X_{cd}}(B^{k_1})_{ad}\right\}\Tr\,B^{I_n\setminus\{k_3\}}},
\end{equation*}
we have that the left-hand side of equation \eqref{eq4.3.16} is given by
\begin{multline} \label{eq4.3.19}
\mean{\left\{\partial_{(X^\dagger)_{ab}}\partial_{(\tilde{H}^\dagger)_{bc}}\partial_{X_{cd}}(B^{k_1})_{ad}\Tr\,B^{I_n}\right\}}=\sum_{i,j,h=1}^n\mathcal{A}_{i,j,h}
\\\hspace{-2em}=\mathcal{A}_{1,1,1}+\sum_{i=2}^n\left(\mathcal{A}_{1,1,i}+\mathcal{A}_{1,i,1}+\mathcal{A}_{i,1,1}+\mathcal{A}_{1,i,i}+\mathcal{A}_{i,1,i}+\mathcal{A}_{i,i,1}+\mathcal{A}_{i,i,i}\right)
\\+\sum_{\substack{2\leq i,j\leq n,\\i\neq j}}\left(\mathcal{A}_{1,i,j}+\mathcal{A}_{i,1,j}+\mathcal{A}_{i,j,1}+\mathcal{A}_{i,i,j}+\mathcal{A}_{i,j,i}+\mathcal{A}_{j,i,i}\right)+\sum_{\substack{2\leq i,j,h\leq n,\\i\neq j\neq h\neq i}}\mathcal{A}_{i,j,h}.
\end{multline}
Every term on the right-hand side of this equation can be expressed in terms of the mixed moments using similar methods to those detailed in the proof of Lemma~\ref{L4.3}, with Lemma \ref{L4.5} supplanting the role of Lemma \ref{L4.2}. Hence, substituting this equation into equation \eqref{eq4.3.16} produces the aforementioned loop equation on the mixed moments --- this loop equation and the required expressions for the $\mathcal{A}_{i',j',h'}$ are displayed in Appendix \ref{appendixC.2}. Multiplying both sides of said loop equation by $x_1^{-k_1-1}\cdots x_n^{-k_n-1}$ and summing over $k_1,\ldots,k_n\geq0$ then gives the following loop equation on the unconnected correlators --- this transformation is carried out term by term in Appendix \ref{appendixC.2}, as well.

\begin{proposition} \label{prop4.5}
Recall that $J_n=(x_2,\ldots,x_n)$, set $U_0:=1$, and define $U_{n'}:=0$ for $n'<0$. Furthermore, define the auxiliary functions (not to be confused with that given in Proposition \ref{prop4.2})
\begin{multline} \label{eq4.3.20}
A_i(x_1,J_n)=\frac{3x_1x_iU_{n+1}(x_1,x_1,x_1,J_n\setminus\{x_i\})-x_1x_iU_{n+1}(x_1,x_1,J_n)}{x_1-x_i}
\\-\frac{x_1x_iU_{n+1}(x_1,x_i,J_n)+x_i^2U_{n+1}(x_i,x_i,J_n)}{x_1-x_i}
\\+\left(\frac{3x_1}{2}\frac{\partial^2}{\partial x_1^2}+\frac{\partial}{\partial x_1}+x_i\frac{\partial^2}{\partial x_1\partial x_i}+\frac{x_i}{2x_1}\frac{\partial^2}{\partial x_i^2}x_i\right)
\\ \times\left\{\frac{x_iU_{n-1}(x_1,J_n\setminus\{x_i\})-x_1U_{n-1}(J_n)}{x_1-x_i}\right\},
\end{multline}
\begin{multline} \label{eq4.3.21}
A_{i,j}(x_1,J_n)=\frac{2x_iU_{n-1}(x_1,J_n\setminus\{x_i\})-2x_jU_{n-1}(x_1,J_n\setminus\{x_j\})}{x_i-x_j}
\\\hspace{-6em}+x_ix_j\Bigg\{\frac{6U_{n-1}(x_1,x_1,J_n\setminus\{x_i,x_j\})+U_{n-1}(J_n)}{(x_1-x_i)(x_1-x_j)}
\\\hspace{4em}-\frac{3U_{n-1}(x_1,J_n\setminus\{x_j\})+3U_{n-1}(x_1,,J_n\setminus\{x_i\})}{(x_1-x_i)(x_1-x_j)}
\\\hspace{4em}-\frac{3U_{n-1}(x_i,J_n\setminus\{x_j\})-2U_{n-1}(x_1,J_n\setminus\{x_i\})}{(x_1-x_i)(x_i-x_j)}
\\+\frac{3U_{n-1}(x_j,J_n\setminus\{x_i\})-2U_{n-1}(x_1,J_n\setminus\{x_j\})}{(x_1-x_j)(x_i-x_j)}\Bigg\},
\end{multline}
\begin{multline} \label{eq4.3.22}
A_{i,j,h}(x_1,J_n)=\frac{x_ix_jx_hU_{n-3}(x_1,J_n\setminus\{x_i,x_j,x_h\})}{(x_1-x_i)(x_1-x_j)(x_1-x_h)}-\frac{x_1x_jx_hU_{n-3}(J_n\setminus\{x_j,x_h\})}{(x_1-x_i)(x_i-x_j)(x_i-x_h)}
\\+\frac{x_1x_ix_hU_{n-3}(J_n\setminus\{x_i,x_h\})}{(x_1-x_j)(x_i-x_j)(x_j-x_h)}-\frac{x_1x_ix_jU_{n-3}(J_n\setminus\{x_i,x_j\})}{(x_1-x_h)(x_i-x_h)(x_j-x_h)}.
\end{multline}
Then, for $n\geq1$, the unconnected correlators specified by equation \eqref{eq4.3.4} satisfy the loop equation
\begin{multline} \label{eq4.3.23}
0=x_1^2U_{n+3}(x_1,x_1,x_1,x_1,J_n)-N^3x_1U_n(x_1,J_n)+N^4U_{n-1}(J_n)
\\\hspace{2em}+\lim_{\xi\to x_1}\left(x_1^2\frac{\partial^2}{\partial x_1^2}+2x_1\frac{\partial}{\partial x_1}\right)\left\{U_{n+1}(\xi,x_1,J_n)+\frac{1}{2}U_{n+1}(x_1,x_1,J_n)\right\}+\sum_{i=2}^n\frac{\partial}{\partial x_i}A_i(x_1,J_n)
\\+\sum_{2\leq i<j\leq n}\frac{\partial^2}{\partial x_i\partial x_j}A_{i,j}(x_1,J_n)+\frac{6}{x_1}\sum_{2\leq i<j<h\leq n}\frac{\partial^3}{\partial x_i\partial x_j\partial x_h}A_{i,j,h}(x_1,J_n).
\end{multline}
\end{proposition}

\subsection{Loop equations on the $\tilde{W}_n^{(\mathcal{H}_1)}$ and $W_n^{(\mathcal{H}_1),l}$} \label{s4.3.2}
Setting $n=1$ in equation \eqref{eq4.3.23} above and using the canonical extension of Lemma \ref{L4.4} specifying $U_{n+3}(x_1,x_1,x_1,x_1,J_n)$ as a suitable sum over $\mu\vdash(x_1,x_1,x_1,x_1)$ produces the $n=1$ \textit{loop equation on the connected correlators} $W_n(x_1,\ldots,x_n)$ \eqref{eq1.1.29},
\begin{multline} \label{eq4.3.24}
0=x_1^2\sum_{\mu\vdash(x_1,x_1,x_1,x_1)}\prod_{\mu_t\in\mu}W_{\#\mu_t}(\mu_t)-N^3x_1W_1(x_1)+N^4
\\+\lim_{\xi\to x_1}\left(x_1^2\frac{\partial^2}{\partial x_1^2}+2x_1\frac{\partial}{\partial x_1}\right)\left\{W_2(\xi,x_1)+\frac{1}{2}W_2(x_1,x_1)\right.
\\+\left.W_1(\xi)W_1(x_1)+\frac{1}{2}\left(W_1(x_1)\right)^2\right\}.
\end{multline}
Inserting the $1/N$ expansion \eqref{eq4.0.5} into this equation and collecting terms of order $N^{4-l}$ ($l\geq0$) yields the $(1,l)$ loop equation
\begin{multline} \label{eq4.3.25}
0=x_1^2\sum_{\mu\vdash(x_1,x_1,x_1,x_1)}\sum_{\substack{l_1+\cdots+l_{\#\mu}\\=l+2\#\mu-8}}\prod_{\mu_t\in\mu}W_{\#\mu_t}^{l_t}(\mu_t)-x_1W_1^l(x_1)+\chi_{l=0}
\\\hspace{-8em}+\lim_{\xi\to x_1}\left(x_1^2\frac{\partial^2}{\partial x_1^2}+2x_1\frac{\partial}{\partial x_1}\right)\left\{W_2^{l-4}(\xi,x_1)+\frac{1}{2}W_2^{l-4}(x_1,x_1)\right.
\\+\left.\sum_{l_1+l_2=l-2}W_1^{l_1}(x_1)\left(W_1^{l_2}(\xi)+\frac{1}{2}W_1^{l_2}(x_1)\right)\right\}.
\end{multline}
(Recall our convention of setting $W_n^{l'}:=0$ for all $n\geq1$ and $l'<0$.) Setting $l=0$ in the above gives the spectral curve
\begin{equation} \label{eq4.3.26}
0=x_1^2\left(W_1^0(x_1)\right)^4-x_1W_1^0(x_1)+1,
\end{equation}
while setting $l=1$ shows that
\begin{equation} \label{eq4.3.27}
0=\left[4x_1\left(W_1^0(x_1)\right)^3-1\right]W_1^1(x_1).
\end{equation}
Equation \eqref{eq4.3.26} has been given in \citep{FIL18}, where it was derived through techniques from free probability theory and also through the relation \eqref{eq1.3.22} between the limiting density $\rho^{(\mathcal{H}_m),0}(\lambda)$ of the Hermitised matrix product ensemble (see Definition \ref{def1.13}) and the Fuss--Catalan distribution $\rho^{(FC_{2m+1})}(\lambda)$ \eqref{eq1.3.16} --- we have that $W_1^{(\mathcal{H}_m),0}(x_1)=x_1W_1^{(FC_{2m+1})}(x_1^2)$ \citep{FIL18} and that equation \eqref{eq4.2.49} is known \citep{FL15} to have the generalisation
\begin{equation} \label{eq4.3.28}
0=u^m\left(W_1^{(FC_m)}(u)\right)^{m+1}-uW_1^{(FC_m)}(u)+1,\quad m\in\mathbb{N}.
\end{equation}
Since $4x_1\left(W_1^0(x_1)\right)^3=1$ is in contradiction with equation \eqref{eq4.3.26}, equation \eqref{eq4.3.27} tells us that $W_1^1(x_1)$ must be identically zero --- this is perfectly in line with our earlier observation (see below Proposition \ref{prop3.10}) that $W_n^l=0$ whenever $l$ is odd.

\begin{definition} \label{def4.1}
For notational convenience, define $\overline{A}_i(x_1,J_n)$ to be $A_i(x_1,J_n)$, as specified by equation \eqref{eq4.3.20}, but with $U_{n-1}(x_1,J_n\setminus\{x_i\})$ replaced by $W_{n-1}(x_1,J_n\setminus\{x_i\})$, $U_{n-1}(J_n)$ replaced by $W_{n-1}(J_n)$, and each instance of $U_{n+1}(K)$ ($K$ an ordered set of variables drawn from $\{x_1,\ldots,x_n\}$) replaced by
\begin{equation*}
\sum_{\mu\vdash K'}\sum_{\sqcup_{t=1}^{\#\mu}K_t=J_n\setminus\{x_i\}}\prod_{\mu_t\in\mu}W_{\#K_t+\#\mu_t}(\mu_t,K_t),
\end{equation*}
where $K'$ is the ordered set obtained by removing $J_n\setminus\{x_i\}$ from $K$. Likewise, define $\overline{A}_{i,j}(x_1,J_n)$ to be $A_{i,j}(x_1,J_n)$ \eqref{eq4.3.21}, but with each instance of $U_{n-1}(K)$ replaced by
\begin{equation*}
\sum_{\mu\vdash K''}\sum_{\sqcup_{t=1}^{\#\mu}K_t=J_n\setminus\{x_i,x_j\}}\prod_{\mu_t\in\mu}W_{\#K_t+\#\mu_t}(\mu_t,K_t),
\end{equation*}
where $K''$ is now the result of removing $J_n\setminus\{x_i,x_j\}$ from $K$. Finally, let $\overline{A}_{i,j,h}(x_1,J_n)$ be the function $A_{i,j,h}(x_1,J_n)$ \eqref{eq4.3.22} with each instance of $U_{n-3}(K)$ replaced by $W_{n-3}(K)$.
\end{definition}

Setting $n=2$ in equation \eqref{eq4.3.23} and subtracting equation \eqref{eq4.3.25} multiplied by $U_1(x_2)$ from the result yields the $n=2$ \textit{loop equation on the connected correlators},
\begin{multline} \label{eq4.3.29}
0=x_1^2\sum_{\mu\vdash(x_1,x_1,x_1,x_1)}\sum_{\sqcup_{t=1}^{\#\mu}K_t=\{x_2\}}\prod_{\mu_t\in\mu}W_{\#K_t+\#\mu_t}(\mu_t,K_t)-N^3x_1W_2(x_1,x_2)+\frac{\partial}{\partial x_2}\overline{A}_2(x_1,x_2)
\\\hspace{-6em}+\lim_{\xi\to x_1}\left(x_1^2\frac{\partial^2}{\partial x_1^2}+2x_1\frac{\partial}{\partial x_1}\right)\Bigg\{\sum_{\mu\vdash(\xi,x_1)}\sum_{\sqcup_{t=1}^{\#\mu}K_t=\{x_2\}}\prod_{\mu_t\in\mu}W_{\#K_t+\#\mu_t}(\mu_t,K_t)
\\+\frac{1}{2}\sum_{\mu\vdash(x_1,x_1)}\sum_{\sqcup_{t=1}^{\#\mu}K_t=\{x_2\}}\prod_{\mu_t\in\mu}W_{\#K_t+\#\mu_t}(\mu_t,K_t)\Bigg\};
\end{multline}
the auxiliary function $\overline{A}_2(x_1,x_2)$ is as in Definition \ref{def4.1} above. Substituting the large $N$ expansion \eqref{eq4.0.5} into this equation, multiplying the result by $N^{-3}$, and taking the limit $N\to\infty$ gives the $(n,l)=(2,0)$ loop equation
\begin{multline} \label{eq4.3.30}
0=4x_1^2\left(W_1^0(x_1)\right)^3W_2^0(x_1,x_2)-x_1W_2^0(x_1,x_2)-\frac{\partial}{\partial x_2}\frac{x_2^2\left(W_1^0(x_2)\right)^3}{x_1-x_2}
\\+\frac{\partial}{\partial x_2}\frac{x_1x_2\left[3\left(W_1^0(x_1)\right)^3-\left(W_1^0(x_1)\right)^2W_1^0(x_2)-W_1^0(x_1)\left(W_1^0(x_2)\right)^2\right]}{x_1-x_2}.
\end{multline}

As seen in \S\ref{s4.2.2}, the above method for deriving equation \eqref{eq4.3.29} extends to the general $n\geq3$ case. Thus, let us now give the analogue of Proposition \ref{prop4.3} for the global scaled $(N,N)$ Hermitised Laguerre ensemble --- we omit the proof, as it is exactly the same as that given for Proposition \ref{prop4.3}, with equation \eqref{eq4.3.23} now assuming the role of equation \eqref{eq4.2.33}.
\begin{proposition} \label{prop4.6}
Let $\overline{A}_i(x_1,J_n)$, $\overline{A}_{i,j}(x_1,J_n)$, and $\overline{A}_{i,j,h}(x_1,J_n)$ be as introduced in Definition \ref{def4.1} above. For $n\geq3$, the connected correlators of the global scaled $(N,N)$ Hermitised Laguerre ensemble specified by equations \eqref{eq4.1.4} and \eqref{eq4.3.4} satisfy the loop equation
\begin{multline} \label{eq4.3.31}
0=x_1^2\sum_{\mu\vdash(x_1,x_1,x_1,x_1)}\sum_{\sqcup_{t=1}^{\#\mu}K_t=J_n}\prod_{\mu_t\in\mu}W_{\#K_t+\#\mu_t}(\mu_t,K_t)-N^3x_1W_n(x_1,J_n)+\sum_{i=2}^n\frac{\partial}{\partial x_2}\overline{A}_i(x_1,J_n)
\\+\sum_{2\leq i<j\leq n}\frac{\partial^2}{\partial x_i\partial x_j}\overline{A}_{i,j}(x_1,J_n)+\frac{6}{x_1}\sum_{2\leq i<j<h\leq n}\frac{\partial^3}{\partial x_i\partial x_j\partial x_h}\overline{A}_{i,j,h}(x_1,J_n)
\\\hspace{-6em}+\lim_{\xi\to x_1}\left(x_1^2\frac{\partial^2}{\partial x_1^2}+2x_1\frac{\partial}{\partial x_1}\right)\Bigg\{\sum_{\mu\vdash(\xi,x_1)}\sum_{\sqcup_{t=1}^{\#\mu}K_t=J_n}\prod_{\mu_t\in\mu}W_{\#K_t+\#\mu_t}(\mu_t,K_t)
\\+\frac{1}{2}\sum_{\mu\vdash(x_1,x_1)}\sum_{\sqcup_{t=1}^{\#\mu}K_t=J_n}\prod_{\mu_t\in\mu}W_{\#K_t+\#\mu_t}(\mu_t,K_t)\Bigg\}.
\end{multline}
\end{proposition}

The above loop equation can be used to derive loop equations characterising the correlator expansion coefficients $W_n^l$ using the method described below Proposition \ref{prop4.3}. Like in \S\ref{s4.2.2}, we do not display these loop equations here, but note that they are straightforward rewritings of equation \eqref{eq4.3.31} in a similar fashion to how equation \eqref{eq4.3.25} is just equation \eqref{eq4.3.24} with products of connected correlators $W_{n'}$ replaced by suitable sums of products of correlator expansion coefficients $W_{n'}^{l'}$.

The loop equation \eqref{eq4.3.31} above is similar to the analogous equation \eqref{eq4.2.45} for the $(N,N)$ antisymmetrised Laguerre ensemble in that, in the same sense as described below Proposition~\ref{prop4.3}, it is of higher order than the loop equations for the classical and general $\beta$ ensembles reviewed in \S\ref{s4.1.1}. Points of difference include the occurrence of the limiting term in the last two lines of equation \eqref{eq4.3.31} and the lack of denominators involving differences of squares, which is a special feature of the antisymmetrised Laguerre ensemble. Moreover, the loop equation \eqref{eq4.3.31} is in fact higher order than equation \eqref{eq4.2.45} (and also the loop equation of \citep{DF20}), this being indicated by the spectral curve \eqref{eq4.3.26} being a quartic polynomial in $W_1^0(x_1)$ instead of cubic, the first sum in the right-hand side of equation \eqref{eq4.3.31} containing a product of four correlators instead of at most three, and there being a sum over $2\leq i<j<h\leq n$ of third order partial derivatives $\partial_{x_i}\partial_{x_j}\partial_{x_h}$ of the simple functions $\overline{A}_{i,j,h}(x_1,J_n)$ involving $W_{n-3}$ in said loop equation; cf.~the last two lines of equation \eqref{eq4.2.45}.

\subsection{Discussion on $W_1^{(\mathcal{H}_1),0}$ and $W_2^{(\mathcal{H}_1),0}$} \label{s4.3.3}
In parallel with \S\ref{s4.2.3}, we now find a rational parametrisation for the spectral curve \eqref{eq4.3.26} of the global scaled $(N,N)$ Hermitised Laguerre ensemble, evaluate $W_1^0(x_1)$, $W_2^0(x_1,x_2)$, and $W_1^2(x_1)$ in terms of this parametrisation, and then use residue calculus to compute some coefficients of the genus expansion \eqref{eq3.3.60}
\begin{equation} \label{eq4.3.32}
\tilde{c}_{k_1,\ldots,k_n}^{(\mathcal{H}_1)}=N^{2-n}\sum_{l=0}^{\tfrac{3}{2}(k_1+\cdots+k_n)+1-n}\frac{c_{k_1,\ldots,k_n}^{(\mathcal{H}_1),l}}{N^l}
\end{equation}
for particular values of $n,k_1,\ldots,k_n$. Thus, our immediate goal is to find rational functions $x(z),y(z)$ such that equation \eqref{eq4.3.26} holds true for all $z$ upon replacing $W_1^0(x_1)$ by $y(z)$ and $x_1$ by $x(z)$. By using the Newton polygon method \citep{BP00}, we know that the complex algebraic curve $0=x^2y^4-xy+1$ is of genus zero and is thus the Riemann sphere $\mathbb{CP}^1$. Hence, our desired parametrisation must exist (see, e.g., \citep[Thrm.~4.63]{SWP08}; an algorithm guaranteed to produce such a parametrisation is given in \citep[Sec.~4.7]{SWP08}). Writing $u(z)=x(z)y(z)^2$ shows that our spectral curve (written in the $x(z),y(z)$ variables) is equivalent to
\begin{equation*}
0=u(z)^2-\frac{u(z)}{y(z)}+1.
\end{equation*}
As this is a quadratic equation in $u(z)$ of the same form as equation \eqref{eq4.1.29}, we may take inspiration from the parametrisation \eqref{eq4.1.31} and let
\begin{equation} \label{eq4.3.33}
\frac{1}{y(z)}=z+\frac{1}{z}\implies y(z)=\frac{z}{z^2+1}.
\end{equation}
Then, $u(z)=z$ (or $u(z)=1/z$, but we ignore this choice without loss of generality) and, consequently,
\begin{equation} \label{eq4.3.34}
x(z)=\frac{u(z)}{y(z)^2}=\frac{(z^2+1)^2}{z}.
\end{equation}
It can immediately be seen that the parametrisations \eqref{eq4.3.33}, \eqref{eq4.3.34} satisfy the equation
\begin{equation*}
0=x(z)^2y(z)^4-x(z)y(z)+1,
\end{equation*}
so $y(z)$ is the single-valued analytic continuation of $W_1^0(x(z))$ to $\mathbb{CP}^1$. That is, $W_1^0(x_1)$ is equal to $y(z)$ when $z\in x^{-1}(x_1)$ is chosen to be close to $z=0$. To see why we must make this choice, note that our desired solution of equation \eqref{eq4.3.26} has the property that $W_1^0(x_1)\sim1/x_1$ as $x_1\to\infty$, which translates to the requirement that $y(z)\sim x(z)y(z)^2=u(z)=z\to0$ as $x(z)\to\infty$ (cf.~the discussion below equations \eqref{eq4.2.50} and \eqref{eq4.2.51}).

Moving on, taking $z_1,z_2\in\mathbb{CP}^1$ and making the substitutions $x_1=x(z_1)$ and $x_2=x(z_2)$ (with $W_1^0(x_i)=y(z_i)$) into equation \eqref{eq4.3.30} shows that
\begin{equation} \label{eq4.3.35}
W_2^0(x(z_1),x(z_2))=\frac{1}{x'(z_1)x'(z_2)}\frac{1}{(z_1-z_2)^2}-\frac{1}{(x(z_1)-x(z_2))^2}.
\end{equation}
This is exactly the expected universal form for $W_2^0$ when working with a genus zero spectral curve; see equation \eqref{eq4.1.34} together with the discussion below it and cf.~the first line of equation \eqref{eq4.2.52}.

Per the comment below equation \eqref{eq4.3.28}, $W_1^1(x_1)$ is identically zero, so the next-to-leading order term in the large $N$ expansion \eqref{eq4.0.5} of $W_1(x_1)$ is $W_1^2(x_1)$, which is equivalently the $1/N^2$ correction to $W_1^0(x_1)$. With the evaluation \eqref{eq4.3.35} in hand, we may set $l=2$ in the loop equation \eqref{eq4.3.25} with $x_1$ replaced by $x(z_1)$ \eqref{eq4.3.34}, $W_1^0(x_1)$ replaced by $y(z_1)$ \eqref{eq4.3.33}, and $W_2^0(x_1,x_1)$ replaced by the limit
\begin{equation*}
\lim_{z_2\to z_1}W_2^0(x(z_1),x(z_2))=\frac{6z_1^6-2z_1^4+10z_1^2+2}{z_1^4\,x'(z_1)^4}
\end{equation*}
to compute
\begin{equation} \label{eq4.3.36}
W_1^2(x(z_1))=-\frac{z_1^3(9z_1^8+38z_1^4+16z_1^2+1)}{(3z_1^2-1)^5(z_1^2+1)^3}.
\end{equation}

As in the case of the antisymmetrised Laguerre ensemble, comparing the genus expansion \eqref{eq4.3.32} of $\tilde{c}_{k_1,\ldots,k_n}^{(\mathcal{H}_1)}$ to the large $x_1,\ldots,x_n$ expansion \eqref{eq1.1.29} and large $N$ expansion \eqref{eq4.0.5} of the connected correlators $W_n(x_1,\ldots,x_n)=\tilde{W}_n^{(\mathcal{H}_1)}(x_1,\ldots,x_n)$ shows that
\begin{equation} \label{eq4.3.37}
W_n^l(x_1,\ldots,x_n)=\sum_{k_1,\ldots,k_n=0}^{\infty}\frac{c_{k_1,\ldots,k_n}^{(\mathcal{H}_1),l}}{x_1^{k_1+1}\cdots x_n^{k_n+1}}.
\end{equation}
Hence, in analogy with equations \eqref{eq4.2.56}, \eqref{eq4.2.57} and recalling that $x(z_1),\ldots,x(z_n)\to\infty$ corresponds to $z_1,\ldots,z_n\to0$ in the regime where equation \eqref{eq1.1.29}, consequently \eqref{eq4.3.37}, is valid, we may use the residue formulae
\begin{align}
c_{k_1,\ldots,k_n}^{(\mathcal{H}_1),l}&=(-1)^n\underset{x_1,\ldots,x_n=\infty}{\mathrm{Res}}x_1^{k_1}\cdots x_n^{k_n}W_n^l(x_1,\ldots,x_n)\,\mathrm{d}x_1\cdots\mathrm{d}x_n \nonumber
\\&=(-1)^n\underset{z_1,\ldots,z_n=0}{\mathrm{Res}}W_n^l(x(z_1),\ldots,x(z_n))\prod_{i=1}^nx(z_i)^{k_i}x'(z_i)\,\mathrm{d}z_i \label{eq4.3.38}
\end{align}
to compute the coefficients of the genus expansion \eqref{eq4.3.32}. Substituting equations \eqref{eq4.3.33}, \eqref{eq4.3.35}, and \eqref{eq4.3.36} into the latter residue formula, we present the evaluation of $c_{k_1,\ldots,k_n}^{(\mathcal{H}_1),l}$ for $(n,l)=(1,0),(1,2),(2,0)$ and some low values of $k_1,k_2$ in Appendix \ref{appendixD.2}. Let us mention here the particular values $c_4^{(\mathcal{H}_1),0}=4$, $c_2^{(\mathcal{H}_1),2}=1$, and $c_{1,1}^{(\mathcal{H}_1),0}=2$, which tell us that there exist four distinct genus zero rooted topological hypermaps with one degree four black vertex satisfying the criteria of Proposition~\ref{prop3.12}; there is one genus one orientable rooted topological hypermap with a  degree four black vertex satisfying said criteria; and there are two genus zero $2$-rooted topological hypermaps with two bivalent black vertices of the type defined in Proposition~\ref{prop3.12}. We display these topological hypermaps in Figure \ref{fig4.5} below.
\begin{figure}[H]
        \centering
\captionsetup{width=.9\linewidth}
        \includegraphics[width=0.95\textwidth]{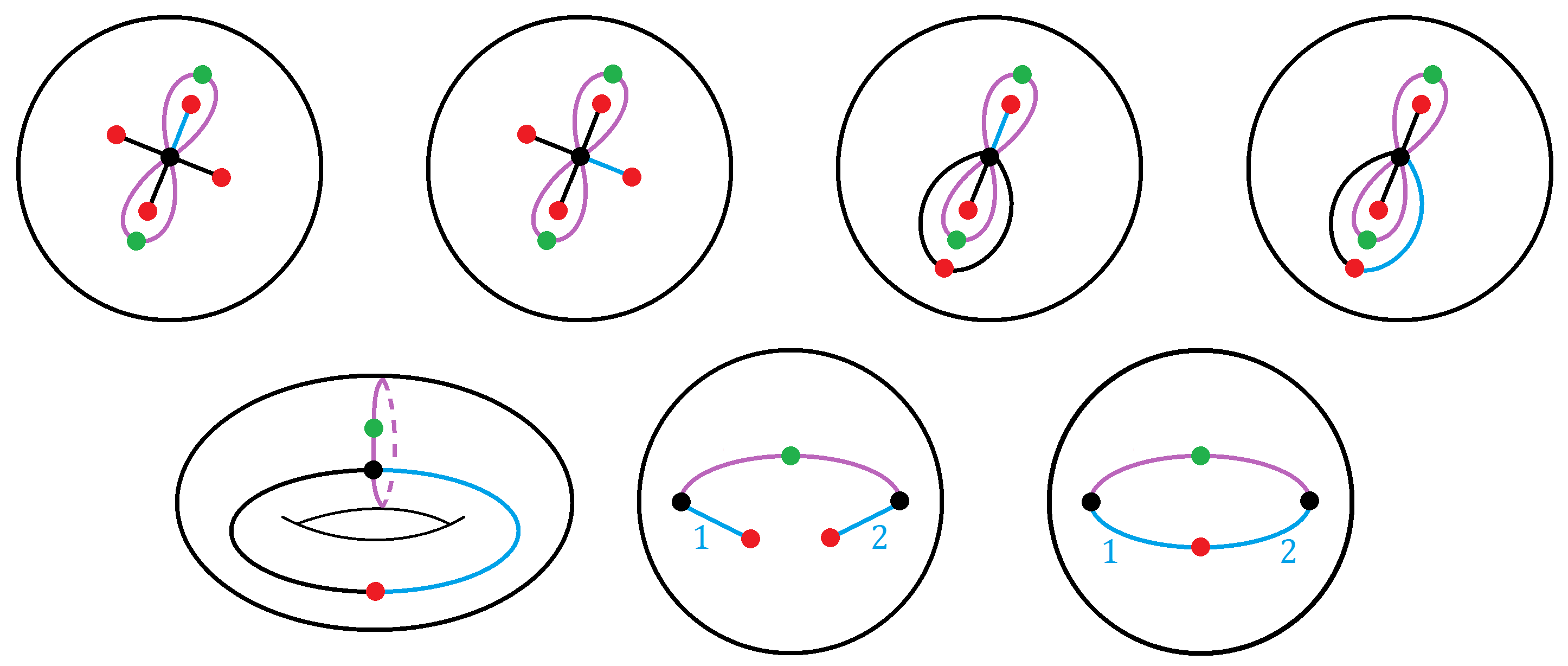}
        \caption[The topological hypermaps corresponding to $c_4^{(\mathcal{H}_1),0}$, $c_2^{(\mathcal{H}_1),2}$, and $c_{1,1}^{(\mathcal{H}_1),0}$]{The four topological hypermaps in the top row are counted by $c_4^{(\mathcal{H}_1),0}$, the toroidal hypermap in the bottom left corresponds to $c_2^{(\mathcal{H}_1),2}=1$, and the remaining two topological hypermaps are enumerated by $c_{1,1}^{(\mathcal{H}_1),0}$. Colours are as in Figure \ref{fig3.22}.} \label{fig4.5}
\end{figure}

\setcounter{equation}{0}
\section{Concluding Remarks and Outlook} \label{s4.4}
In this thesis, we have added onto the well established theory of classical matrix ensembles (some of which is reviewed in Section \ref{s1.2}), elaborated on the relatively under-reported but known combinatorial theory of said ensembles, and then constructed some ribbon graphs and loop equations for the Hermitised and antisymmetrised Laguerre ensembles. Recall from Section \ref{s1.4} that our studies have been in part motivated by the desire to better understand $1$-point recursions, loop equations, and the conjectured \citep{CD21} relationship between these two recursive schemes. This is a broad and difficult problem that we have partially addressed by broadening the range of examples of $1$-point recursions and loop equations. While this is far from bringing said problem to resolution, our examples provide a source of open questions that should stimulate meaningful progress in this endeavour. Thus, as we come to the end of this thesis, let us tie it together with some suggestions on directions for future research.

\subsection{Larger matrix products and topological recursion} \label{s4.4.1}
Recall that one of our original motivations for undertaking the loop equation analysis of the global scaled $(N,N)$ antisymmetrised and Hermitised Laguerre ensembles presented in Sections \ref{s4.2} and \ref{s4.3} was to explore the feasibility of treating the general $m$ antisymmetrised and Hermitised matrix products $\mathcal{J}_m$ \eqref{eq4.0.2} and $\mathcal{H}_m$ \eqref{eq4.0.1} in the same way. We posit that the general procedure outlined in Sections \ref{s4.2} and \ref{s4.3} should be able to produce loop equations for these matrix product ensembles, but note that the outcomes are expected to be prohibitively convoluted for large values of $m$. Indeed, the $m>1$ extensions of equations \eqref{eq4.2.11} and \eqref{eq4.3.9} require us to consider the action of differential operators of the form
\begin{equation*}
\partial_{(X_m^T)_{i_mi_{m-1}}}\cdots\,\partial_{(X_1^T)_{i_1i_0}}(J^T)_{i_0j_0}\partial_{(X_1)_{j_0j_1}}\cdots\,\partial_{(X_m)_{j_{m-1}j_m}}
\end{equation*}
with $X_i,J$ defined analogously to the $X,J$ of Section \ref{s4.2} and
\begin{equation*}
\partial_{(X_m^\dagger)_{i_mi_{m-1}}}\cdots\,\partial_{(X_1^\dagger)_{i_1i_0}}\partial_{(\tilde{H}^\dagger)_{i_0j_0}}\partial_{(X_1)_{j_0j_1}}\cdots\,\partial_{(X_m)_{j_{m-1}j_m}}
\end{equation*}
with $X_i,\tilde{H}$ defined analogously to the $X,\tilde{H}$ of Section \ref{s4.3}, respectively. Consequently, we anticipate that the analogue of equation \eqref{eq4.2.45} pertaining to some fixed $m>1$ should contain a sum over partitions $\mu$ of the $(2m+1)$-tuple $(x_1,\ldots,x_1)$ of the form seen in the first line of equation \eqref{eq4.2.45} and sums of up to $(2m)\textsuperscript{th}$ order partial derivatives of the form seen in the final two lines of equation \eqref{eq4.2.45}; the $m>1$ analogue of equation \eqref{eq4.3.31} should likewise contain a sum over partitions $\mu$ of the $(2m+2)$-tuple $(x_1,\ldots,x_1)$ in addition to sums of up to $(2m+1)\textsuperscript{th}$ order partial derivatives of the canonical extensions of $A_{i,j,h}(x_1,J_n)$ \eqref{eq4.3.22} having up to $(2m+1)$ indices.

It was mentioned in \S\ref{s4.1.2} that applying residue calculus to the loop equation \eqref{eq4.1.28} for the correlator expansion coefficients $W_n^l$ of the one-cut 1-Hermitian matrix model to obtain the simpler topological recursion formula \eqref{eq4.1.41} for the associated differential forms $\omega_n^l$ essentially removes the need for considering the terms $P_n^l(x_1,J_n)$ and
\begin{equation*}
\sum_{i=2}^n\frac{\partial}{\partial x_i}\left\{\frac{W_{n-1}^l(x_1,J_n\setminus\{x_i\})-W_{n-1}^l(J_n)}{x_1-x_i}\right\}
\end{equation*}
seen on the right-hand side of equation \eqref{eq4.1.28}. Likewise, it stands to reason that one option for alleviating the aforementioned complexities associated with the loop equations for the general $m$ antisymmetrised and Hermitised matrix product ensembles is to use the methodology reviewed in \S\ref{s4.1.2} to instead derive topological recursion formulae for these ensembles. This would very likely either replace the aforementioned general $m$ analogues of $A_{i,j,h}(x_1,J_n)$ with drastically simpler versions of themselves or possibly remove the need for considering them altogether --- it was commented in \citep{DF20} that the loop equations therein for the $(N,N,N)$ complex Wishart product ensemble are also expected to reduce to relatively simple topological recursion formulae upon applying the techniques of \S\ref{s4.1.2}. It would thus be very interesting to see if our loop equations (and those of \citep{DF20}) can be simplified to elegant topological recursion formulae in the proposed manner and, if so, to compare the latter to the higher order topological recursion formulae of \citep{BE13}, \citep{BHLMR14}. Although this would still be far off from obtaining loop equations or topological recursion formulae for general $m$, it should at the very least be possible to take $N\to\infty$ and observe a uniform structure across all $m\in\mathbb{N}$, thereby enabling us to recover the spectral curves known from equations \eqref{eq1.3.22}, \eqref{eq1.3.23}, and \eqref{eq4.3.28} \citep{BBCC11}, \citep{PZ11}, \citep{FL15}, \citep{FIL18}, \citep{FILZ19},
\begin{align}
0&=x_1^{2m-1}\left(W_1^{(\mathcal{J}_m),0}(x_1)\right)^{2m+1}+x_1W_1^{(\mathcal{J}_m),0}(x_1)-1,
\\0&=x_1^{2m}\left(W_1^{(\mathcal{H}_m),0}(x_1)\right)^{2m+2}-x_1W_1^{(\mathcal{J}_m),0}(x_1)+1.
\end{align}

As an aside, a related endeavour that may be worth pursuing is to see if the theory of \citep{ND18}, \citep{CN19} concerning spectral curves corresponding to eigenvalue densities with hard edges and that of \citep{EM09} relating to topological recursion for general $\beta$ ensembles can be used to derive explicit topological recursion formulae from the loop equations \citep{FRW17} for the Laguerre and Jacobi $\beta$ ensembles discussed in \S\ref{s4.1.1}. Such explicit formulae should be attainable as these ensembles fall within the abstract setting considered in \citep{BG12}.

\subsection{On $1$-point recursions} \label{s4.4.2}
As the loop equation formalism and topological recursion fully determine the resolvent expansion coefficients $W_1^l$, which act as generating functions for the moment expansion coefficients $\tilde{M}_{k,l}$ defined implicitly through equation \eqref{eq1.1.32}, it is reasonable to say that the data produced by $1$-point recursions on the $\tilde{M}_{k,l}$ is a subset of that produced by the loop equation formalism --- the benefit in considering $1$-point recursions is that they are more efficient at generating this data. Thus, it should come as no surprise that many ensembles that can be studied using $1$-point recursions are also known to be susceptible to loop equation analysis (see \citep{CD21} and references therein for a brief review). Seeing as how the $1$-point recursions given in \S\ref{s3.1.2} characterise random matrix ensembles that have also been shown to be governed by the loop equations reviewed in \S\ref{s4.1.1}, it is natural to ask if there exist $1$-point recursions on the $\tilde{M}_{k,l}=c_k^l$ that are determined by the loop equation analysis of \S\ref{s4.2.2} and \S\ref{s4.3.2}. We suggest two approaches for obtaining such $1$-point recursions: 
\begin{enumerate}
\item It may be possible to derive $1$-point recursions for the antisymmetrised Laguerre ensemble by using the techniques of Chapters 2 and 3 in addition to the theory of \citep{FI18} relating the Laguerre Muttalib--Borodin ensemble to the Selberg integral. (Incidentally, let us mention here that it remains to be seen how far the theory of Chapter 2 can be pushed to uncover features of the classical $\beta$ ensembles with general $\beta\in2\mathbb{N}$.)
\item It should be checked if Chekhov's application \citep{Che16} of an adaptation of the Br\'ezin--Hikami replica method \citep{BH08} to derive the $1$-point recursion \eqref{eq3.1.38} for the LUE with $a=0$ can be repurposed to treat the antisymmetrised and Hermitised Laguerre ensembles. It would likewise be interesting to see if the $1$-point recursions of \S\ref{s3.1.2} can be obtained in this way, as well.
\end{enumerate}

Let us recall from the discussion at the end of \S\ref{s3.1.2} that while the moment expansion coefficients $M_{k,l}$ of the Gaussian and Laguerre ensembles have combinatorial interpretations due to the theory reviewed in \S\ref{s3.3.1} and \S\ref{s3.3.2}, these interpretations are not known to translate to combinatorial interpretations for the $1$-point recursions on said expansion coefficients. However, preliminary experimentation by the present author suggests that dressing the bijective proof of the Catalan recursion \eqref{eq3.1.25} sketched below Figure \ref{fig3.2} (and demonstrated in Example \ref{ex3.1}) with suitable additional rules may result in natural combinatorial proofs of both the Harer--Zagier recursion \eqref{eq3.0.12} and the $1$-point recursion \eqref{eq3.1.24} for the LUE. Confirming this conjecture to be true and comparing the proposed bijections to those mentioned in \citep[Sec.~1]{CD21} would shed light on how the $1$-point recursions of \S\ref{s3.1.2} relate to each other. This is of particular interest to us since the $1$-point recursions for the (shifted) JUE presented in Corollary \ref{C3.4} are very similar to those for the GUE and LUE --- it is hoped that a combinatorial understanding of the GUE and LUE $1$-point recursions  in terms of ribbon graphs could extend to a similar understanding of the $1$-point recursions and, in turn, the spectral moments of the (shifted) JUE. 

\subsection{Combinatorics associated with the Jacobi unitary ensemble} \label{s4.4.3}
In contrast to the approach described above for obtaining combinatorial interpretations for the JUE moments through associated interpretations of its $1$-point recursion, we now show what happens if one tries to obtain such combinatorial interpretations by instead modifying the arguments of Section \ref{s3.3}. First, let $G_1,G_2$ be drawn from the $M_1\times N$, respectively $M_2\times N$, complex Ginibre ensemble of Definition \ref{def1.3} so that for $i=1,2$, $W_i=G_i^\dagger G_i$ is drawn from the $(M_i,N)$ LUE, in accordance with Definition \ref{def1.4}. Then, according to this same definition, $Y=W_1(W_1+W_2)^{-1}$ represents the $(M_1,M_2,N)$ JUE. Letting $\mean{\,\cdot\,}$ denote averages with respect to the product $P^{(L)}(W_1)P^{(L)}(W_2)$, with $P^{(L)}(W)$ specified by equation \eqref{eq1.2.2}, equation \eqref{eq1.1.15} then tells us that the spectral moments of the JUE are given by
\begin{align}
m_k^{(JUE)}&=\mean{\Tr\,Y^k},\quad k\in\mathbb{N} \nonumber
\\&=\mean{\Tr\,\left(I_N+W_2W_1^{-1}\right)^{-k}} \nonumber
\\&=\sum_{l=0}^{\infty}\binom{l+k-1}{l}(-1)^l\mean{\Tr\,(W_2W_1^{-1})^l},
\end{align}
where we have put aside issues of convergence and used the generalised binomial theorem. This computation reduces the problem at hand to that of finding a combinatorial interpretation for the spectral moments of $W_2W_1^{-1}$. Using the independence of $W_1,W_2$, it can be shown \citep{YK83}, \citep{BYK88} that said spectral moments are given by particular sums of products of mixed moments of $W_2$ and $W_1^{-1}$. Alternatively, one may use the cyclic property of the trace to show that these same spectral moments are given by the averages $\langle\Tr\,(G_2W_1^{-1}G_2^\dagger)^l\rangle$, which can be interpreted as the weighted count of LUE ribbon graphs, with the weights being certain mixed moments of the inverse Wishart ensemble (cf.~Remark~\ref{R3.4} preceding Proposition~\ref{prop3.11}). The bottom line is that, since the mixed moments of the inverse Wishart ensemble are currently not known to have any interpretations in terms of ribbon graphs, the argument just given cannot produce such interpretations for the JUE spectral moments.

If one is content with combinatorial interpretations of JUE spectral moments that are not related to ribbon graphs, then the situation is more fortuitous. Indeed, the above construction can be made fully combinatorial by observing that the mixed moments of the inverse Wishart ensemble have recently been characterised in terms of monotone double Hurwitz numbers \citep{CDO21}, \citep{GGR20}. More recently still, the authors of \citep{GGR20} have characterised the mixed moments of the JUE itself in terms of monotone triple Hurwitz numbers \citep{GGR21}. These works complement the earlier work \citep{BG20}, which characterised the mixed moments of the GUE in terms of so-called $2$-orbifold strictly monotone Hurwitz numbers. A curious feature of these works is that the proofs therein of the relationships between mixed moments and Hurwitz numbers do not have a combinatorial flavour of the sort discussed in Section~\ref{s3.3}, even though Hurwitz numbers are very closely related to combinatorial (hyper)maps and constellations (see \citep{LZ04}, \citep{ACEH18}). This gap has been addressed in the GUE case \citep{BCDG19}, but it remains to be seen how the Hurwitz number combinatorics of the LUE and JUE could be related to ribbon graphs. It would be interesting to see if the results of \citep{CDO21}, \citep{GGR20}, \citep{GGR21} can be reverse engineered to determine classes of ribbon graphs that are enumerated by the mixed moments of the JUE. In that vein, the theory of \S\ref{s3.3.3} could benefit from being reformulated in terms of combinatorial hypermaps and/or constellations; this would enable the use of computer algebra for computing the relevant mixed cumulants and could also reveal a connection between said cumulants and new classes of Hurwitz numbers.

\newpage~
\thispagestyle{empty}
\newpage
\phantomsection
\addcontentsline{toc}{chapter}{Bibliography}
\small
\bibliography{Thesis}

\begin{thebibliography}{330}
\providecommand{\natexlab}[1]{#1}
\providecommand{\url}[1]{\texttt{#1}}
\expandafter\ifx\csname urlstyle\endcsname\relax
  \providecommand{\doi}[1]{doi: #1}\else
  \providecommand{\doi}{doi: \begingroup \urlstyle{rm}\Url}\fi

\bibitem[DLM()]{DLMF}
{Digital Library of Mathematical Functions}.
\newblock URL \url{https://dlmf.nist.gov/18}.

\bibitem[Adachi et~al.(2011)Adachi, Toda, and Kubotani]{ATK11}
S.~Adachi, M.~Toda, and H.~Kubotani.
\newblock Asymptotic analysis of singular values of rectangular complex
  matrices in the {Laguerre} and fixed-trace ensembles.
\newblock \emph{J. Phys. A}, 44\penalty0 (29):\penalty0 292002, 2011.
\newblock \doi{10.1088/1751-8113/44/29/292002}.

\bibitem[Adhikari et~al.(2016)Adhikari, Reddy, Reddy, and Saha]{ARRS16}
K.~Adhikari, N.~K. Reddy, T.~R. Reddy, and K.~Saha.
\newblock Determinantal point processes in the plane from products of random
  matrices.
\newblock \emph{Ann. Inst. H. Poincar\'e Probab. Statist.}, 52\penalty0
  (1):\penalty0 16--46, 2016.
\newblock \doi{10.1214/14-AIHP632}.

\bibitem[Adler et~al.(2000)Adler, Forrester, Nagao, and van Moerbeke]{AFNM00}
M.~Adler, P.~J. Forrester, T.~Nagao, and P.~van Moerbeke.
\newblock {Classical Skew Orthogonal Polynomials and Random Matrices}.
\newblock \emph{J. Stat. Phys.}, 99\penalty0 (1--2):\penalty0 141--170, 2000.
\newblock \doi{10.1023/A:1018644606835}.

\bibitem[Adler et~al.(2013)Adler, van Moerbeke, and Wang]{AMW13}
M.~Adler, P.~van Moerbeke, and D.~Wang.
\newblock Random matrix minor processes related to percolation theory.
\newblock \emph{Random Matrices: Theory Appl.}, 2\penalty0 (4):\penalty0
  1350008, 2013.
\newblock \doi{10.1142/S2010326313500081}.

\bibitem[Ajanki et~al.(2017)Ajanki, Erd\H{o}s, and Kr\"uger]{AEK17}
O.~H. Ajanki, L.~Erd\H{o}s, and T.~Kr\"uger.
\newblock Universality for general {Wigner-type} matrices.
\newblock \emph{Probab. Theory Relat. Fields}, 169\penalty0 (3--4):\penalty0
  667--727, 2017.
\newblock \doi{10.1007/s00440-016-0740-2}.

\bibitem[Akemann and Burda(2012)]{AB12}
G.~Akemann and Z.~Burda.
\newblock Universal microscopic correlation functions for products of
  independent {Ginibre} matrices.
\newblock \emph{J. Phys. A}, 45\penalty0 (46):\penalty0 465201, 2012.
\newblock \doi{10.1088/1751-8113/45/46/465201}.

\bibitem[Akemann and Ipsen(2015)]{AI15}
G.~Akemann and J.~R. Ipsen.
\newblock {Recent Exact and Asymptotic Results for Products of Independent
  Random Matrices}.
\newblock \emph{Acta Phys. Pol. B}, 46\penalty0 (9):\penalty0 1747--1784, 2015.
\newblock \doi{10.5506/APhysPolB.46.1747}.

\bibitem[Akemann and Nagao(2011)]{AN11}
G.~Akemann and T.~Nagao.
\newblock {Random matrix theory for the Hermitian Wilson Dirac operator and the
  chGUE-GUE transition}.
\newblock \emph{J. High Energ. Phys.}, 2011\penalty0 (10):\penalty0
  JHEP10(2011)060, 2011.
\newblock \doi{10.1007/JHEP10(2011)060}.

\bibitem[Akemann and Strahov(2016)]{AS16}
G.~Akemann and E.~Strahov.
\newblock {Dropping the Independence: Singular Values for Products of Two
  Coupled Random Matrices}.
\newblock \emph{Commun. Math. Phys.}, 345\penalty0 (1):\penalty0 101--140,
  2016.
\newblock \doi{10.1007/s00220-016-2653-4}.

\bibitem[Akemann et~al.(2013{\natexlab{a}})Akemann, Ipsen, and Kieburg]{AIK13}
G.~Akemann, J.~R. Ipsen, and M.~Kieburg.
\newblock {Products of rectangular random matrices: Singular values and
  progressive scattering}.
\newblock \emph{Phys. Rev. E}, 88\penalty0 (5):\penalty0 052118,
  2013{\natexlab{a}}.
\newblock \doi{10.1103/PhysRevE.88.052118}.

\bibitem[Akemann et~al.(2013{\natexlab{b}})Akemann, Kieburg, and Wei]{AKW13}
G.~Akemann, M.~Kieburg, and L.~Wei.
\newblock {Singular value correlation functions for products of Wishart random
  matrices}.
\newblock \emph{J. Phys. A}, 46\penalty0 (27):\penalty0 275205,
  2013{\natexlab{b}}.
\newblock \doi{10.1088/1751-8113/46/27/275205}.

\bibitem[Akemann et~al.(2014)Akemann, Burda, Kieburg, and Nagao]{ABKN14}
G.~Akemann, Z.~Burda, M.~Kieburg, and T.~Nagao.
\newblock Universal microscopic correlation functions for products of truncated
  unitary matrices.
\newblock \emph{J. Phys. A}, 47\penalty0 (25):\penalty0 255202, 2014.
\newblock \doi{10.1088/1751-8113/47/25/255202}.

\bibitem[Akemann et~al.(2015)Akemann, Baik, and Francesco]{Oxf15}
G.~Akemann, J.~Baik, and P.~Di Francesco, editors.
\newblock \emph{{The Oxford Handbook of Random Matrix Theory}}.
\newblock Oxford University Press, Oxford, 2015.
\newblock \doi{10.1093/oxfordhb/9780198744191.001.0001}.

\bibitem[Al-Salam and Verma(1983)]{AV83}
W.~A. Al-Salam and A.~Verma.
\newblock {$q$-Konhauser polynomials}.
\newblock \emph{Pac. J. Math.}, 108\penalty0 (1):\penalty0 1--7, 1983.

\bibitem[Alexandrov et~al.(2018)Alexandrov, Chapuy, Eynard, and Harnad]{ACEH18}
A.~Alexandrov, G.~Chapuy, B.~Eynard, and J.~Harnad.
\newblock {Weighted Hurwitz numbers and topological recursion: An overview}.
\newblock \emph{J. Math. Phys.}, 59\penalty0 (8):\penalty0 081102, 2018.
\newblock \doi{10.1063/1.5013201}.

\bibitem[Allez et~al.(2012)Allez, Bouchaud, Majumdar, and Vivo]{ABMV12}
R.~Allez, J.-P. Bouchaud, S.~N. Majumdar, and P.~Vivo.
\newblock {Invariant $\beta$-Wishart ensembles, crossover densities and
  asymptotic corrections to the Mar\v{c}enko-Pastur law}.
\newblock \emph{J. Phys. A}, 46\penalty0 (1):\penalty0 015001, 2012.
\newblock \doi{10.1088/1751-8113/46/1/015001}.

\bibitem[Ambj{\o}rn et~al.(1990)Ambj{\o}rn, Jurkiewicz, and Makeenko]{AJM90}
J.~Ambj{\o}rn, J.~Jurkiewicz, and Yu.~M. Makeenko.
\newblock Multiloop correlators for two-dimensional quantum gravity.
\newblock \emph{Phys. Lett. B}, 251\penalty0 (4):\penalty0 517--524, 1990.
\newblock \doi{10.1016/0370-2693(90)90790-D}.

\bibitem[Ambj{\o}rn et~al.(1993)Ambj{\o}rn, Chekhov, Kristjansen, and
  Makeenko]{ACKM93}
J.~Ambj{\o}rn, L.~Chekhov, C.~F. Kristjansen, and Yu. Makeenko.
\newblock Matrix model calculations beyond the spherical limit.
\newblock \emph{Nucl. Phys. B}, 404\penalty0 (1--2):\penalty0 127--172, 1993.
\newblock \doi{10.1016/0550-3213(93)90476-6}.
\newblock Erratum ibid. 449(3):681, 1995. doi: 10.1016/0550-3213(95)00391-5.

\bibitem[Anderson(1991)]{And91}
G.~W. Anderson.
\newblock {A Short Proof of Selberg's Generalized Beta Formula}.
\newblock \emph{Forum Math.}, 3\penalty0 (4):\penalty0 415--417, 1991.
\newblock \doi{10.1515/form.1991.3.415}.

\bibitem[Anderson et~al.(2009)Anderson, Guionnet, and Zeitouni]{AGZ09}
G.~W. Anderson, A.~Guionnet, and O.~Zeitouni.
\newblock \emph{{An Introduction to Random Matrices}}.
\newblock Cambridge University Press, 2009.

\bibitem[Aomoto(1987)]{Aom87}
K.~Aomoto.
\newblock {Jacobi Polynomials Associated with Selberg Integrals}.
\newblock \emph{SIAM J. Math. Anal.}, 18\penalty0 (2):\penalty0 545--549, 1987.
\newblock \doi{10.1137/0518042}.

\bibitem[Assiotis et~al.(2021)Assiotis, Bedert, Gunes, and Soor]{ABGS20}
T.~Assiotis, B.~Bedert, M.~A. Gunes, and A.~Soor.
\newblock {Moments of the Generalized Cauchy Random Matrices and
  continuous-Hahn Polynomials}.
\newblock \emph{Nonlinearity}, 34\penalty0 (7):\penalty0 4923, 2021.
\newblock \doi{10.1088/1361-6544/abfeac}.

\bibitem[Atland and Zirnbauer(1997)]{AZ97}
A.~Atland and M.~R. Zirnbauer.
\newblock Nonstandard symmetry classes in mesoscopic normal-superconducting
  hybrid structures.
\newblock \emph{Phys. Rev. B}, 55\penalty0 (2):\penalty0 1142, 1997.
\newblock \doi{10.1103/PhysRevB.55.1142}.

\bibitem[Bai et~al.(1988)Bai, Yin, and Krishnaiah]{BYK88}
Z.~D. Bai, Y.~Q. Yin, and P.~R. Krishnaiah.
\newblock {On the Limiting Empirical Distribution Function of the Eigenvalues
  of a Multivariate $F$ Matrix}.
\newblock \emph{Theory Probab. Appl.}, 32\penalty0 (3):\penalty0 490--500,
  1988.
\newblock \doi{10.1137/1132067}.

\bibitem[Baik et~al.(2005)Baik, Ben~Arous, and P\'ech\'e]{BBP05}
J.~Baik, G.~Ben~Arous, and S.~P\'ech\'e.
\newblock Phase transition of the largest eigenvalue for nonnull complex sample
  covariance matrices.
\newblock \emph{Ann. Probab.}, 33\penalty0 (5):\penalty0 1643--1697, 2005.
\newblock \doi{10.1214/009117905000000233}.

\bibitem[Baker and Forrester(1997)]{BF97}
T.~H. Baker and P.~J. Forrester.
\newblock {The Calogero--Sutherland Model and Generalized Classical
  Polynomials}.
\newblock \emph{Commun. Math. Phys.}, 188\penalty0 (1):\penalty0 175--216,
  1997.
\newblock \doi{10.1007/s002200050161}.

\bibitem[Baker et~al.(1998)Baker, Forrester, and Pearce]{BFP98}
T.~H. Baker, P.~Forrester, and P.~Pearce.
\newblock {Random matrix ensembles with an effective extensive external
  charge}.
\newblock \emph{J. Phys. A}, 31\penalty0 (29):\penalty0 6087, 1998.
\newblock \doi{10.1088/0305-4470/31/29/002}.

\bibitem[Banica et~al.(2011)Banica, Belinschi, Capitaine, and Collins]{BBCC11}
T.~Banica, S.~T. Belinschi, M.~Capitaine, and B.~Collins.
\newblock {Free Bessel Laws}.
\newblock \emph{Can. J. Math.}, 63\penalty0 (1):\penalty0 3--37, 2011.
\newblock \doi{10.4153/CJM-2010-060-6}.

\bibitem[Basor and Morrison(1994)]{BM94}
E.~L. Basor and K.~E. Morrison.
\newblock {The Fisher--Hartwig conjecture and Toeplitz eigenvalues}.
\newblock \emph{Lin. Alg. and its Appl.}, 202:\penalty0 129--142, 1994.
\newblock \doi{10.1016/0024-3795(94)90187-2}.

\bibitem[Bateman and Erd\'elyi(1953)]{BE53}
H.~Bateman and A.~Erd\'elyi.
\newblock \emph{{Higher Transcendental Functions}}.
\newblock McGraw--Hill, 1953.

\bibitem[Beelen and Pellikaan(2000)]{BP00}
P.~Beelen and R.~Pellikaan.
\newblock {The Newton Polygon of Plane Curves with Many Rational Points}.
\newblock \emph{Des. Codes, Cryptogr.}, 21\penalty0 (1--3):\penalty0 41--67,
  2000.
\newblock \doi{10.1023/A:1008323208670}.

\bibitem[Beenakker(1997)]{Be97}
C.~W.~J. Beenakker.
\newblock Random-matrix theory of quantum transport.
\newblock \emph{Rev. Mod. Phys.}, 69\penalty0 (3):\penalty0 731--808, 1997.
\newblock \doi{10.1103/RevModPhys.69.731}.

\bibitem[Beenakker and Rejaei(1993)]{BR93}
C.~W.~J. Beenakker and B.~Rejaei.
\newblock {Nonlogarithmic Repulsion of Transmission Eigenvalues in a Disordered
  Wire}.
\newblock \emph{Phys. Rev. Lett.}, 71\penalty0 (22):\penalty0 3689--3692, 1993.
\newblock \doi{10.1103/PhysRevLett.71.3689}.

\bibitem[Bercovici and Pata(1999)]{BPB99}
H.~Bercovici and V.~Pata.
\newblock {Stable laws and domains of attraction in free probability theory
  (with an Appendix by Philippe Biane)}.
\newblock \emph{Ann. Math.}, 149\penalty0 (3):\penalty0 1023--1060, 1999.
\newblock \doi{10.2307/121080}.

\bibitem[Bessis et~al.(1980)Bessis, Itzykson, and Zuber]{BIZ80}
D.~Bessis, C.~Itzykson, and J.~B. Zuber.
\newblock {Quantum Field Theory Techniques in Graphical Enumeration}.
\newblock \emph{Adv. Appl. Math.}, 1\penalty0 (2):\penalty0 109--157, 1980.
\newblock \doi{10.1016/0196-8858(80)90008-1}.

\bibitem[Bleher and Kuijlaars(2005)]{BK05}
P.~M. Bleher and A.~B.~J. Kuijlaars.
\newblock {Integral representations for multiple Hermite and multiple Laguerre
  polynomials}.
\newblock \emph{Ann. Inst. Fourier}, 55\penalty0 (6):\penalty0 2001--2014,
  2005.
\newblock \doi{10.5802/aif.2148}.

\bibitem[Borodin(1998)]{Bo98}
A.~Borodin.
\newblock Biorthogonal ensembles.
\newblock \emph{Nucl. Phys. B}, 536\penalty0 (3):\penalty0 704--732, 1998.
\newblock \doi{10.1016/S0550-3213(98)00642-7}.

\bibitem[Borodin and Olshanski(2001)]{BO01}
A.~Borodin and G.~Olshanski.
\newblock {Infinite Random Matrices and Ergodic Measures}.
\newblock \emph{Commun. Math. Phys.}, 223\penalty0 (1):\penalty0 87--123, 2001.
\newblock \doi{10.1007/s002200100529}.

\bibitem[Borot and Garcia-Failde(2020)]{BG20}
G.~Borot and E.~Garcia-Failde.
\newblock {Simple Maps, Hurwitz Numbers, and Topological Recursion}.
\newblock \emph{Commun. Math. Phys.}, 380\penalty0 (2):\penalty0 581--654,
  2020.
\newblock \doi{10.1007/s00220-020-03867-1}.

\bibitem[Borot and Guionnet(2012{\natexlab{a}})]{BG12}
G.~Borot and A.~Guionnet.
\newblock {Asymptotic Expansion of $\beta$ Matrix Models in the One-cut
  Regime}.
\newblock \emph{Commun. Math. Phys.}, 317\penalty0 (2):\penalty0 447--483,
  2012{\natexlab{a}}.
\newblock \doi{10.1007/s00220-012-1619-4}.

\bibitem[Borot and Guionnet(2012{\natexlab{b}})]{FR12}
G.~Borot and A.~Guionnet.
\newblock {A Fuchsian Matrix Differential Equation for Selberg Correlation
  Integrals}.
\newblock \emph{Commun. Math. Phys.}, 309\penalty0 (3):\penalty0 771--792,
  2012{\natexlab{b}}.
\newblock \doi{10.1007/s00220-011-1305-y}.

\bibitem[Borot et~al.(2011)Borot, Eynard, Majumdar, and Nadal]{BEMN11}
G.~Borot, B.~Eynard, S.~N. Majumdar, and C.~Nadal.
\newblock Large deviations of the maximal eigenvalue of random matrices.
\newblock \emph{J. Stat. Mech.: Theory Exp.}, 2011\penalty0 (11):\penalty0
  P11024, 2011.
\newblock \doi{10.1088/1742-5468/2011/11/P11024}.

\bibitem[Borot et~al.(2019)Borot, Charbonnier, Do, and Garcia-Failde]{BCDG19}
G.~Borot, S.~Charbonnier, N.~Do, and E.~Garcia-Failde.
\newblock {Relating Ordinary and Fully Simple Maps via Monotone Hurwitz
  Numbers}.
\newblock \emph{Electron. J. Comb.}, 26\penalty0 (3):\penalty0 P3.43, 2019.
\newblock \doi{10.37236/8634}.

\bibitem[Bouchard and Eynard(2013)]{BE13}
V.~Bouchard and B.~Eynard.
\newblock Think globally, compute locally.
\newblock \emph{J. High Energ. Phys.}, 2013\penalty0 (2):\penalty0 143, 2013.
\newblock \doi{10.1007/JHEP02(2013)143}.

\bibitem[Bouchard et~al.(2014)Bouchard, Hutchinson, Loliencar, Meiers, and
  Rupert]{BHLMR14}
V.~Bouchard, J.~Hutchinson, P.~Loliencar, M.~Meiers, and M.~Rupert.
\newblock {A Generalized Topological Recursion for Arbitrary Ramification}.
\newblock \emph{Ann. Henri Poincar\'e}, 15\penalty0 (1):\penalty0 143--169,
  2014.
\newblock \doi{10.1007/s00023-013-0233-0}.

\bibitem[Bouchaud et~al.(2007)Bouchaud, Laloux, Miceli, and Potters]{BLMP07}
J.~P. Bouchaud, L.~Laloux, M.~A. Miceli, and M.~Potters.
\newblock Large dimension forecasting models and random singular value spectra.
\newblock \emph{Euro. Phys. J. B}, 55\penalty0 (2):\penalty0 201--207, 2007.
\newblock \doi{10.1140/epjb/e2006-00204-0}.

\bibitem[Bourgade(2018)]{Bou18}
P.~Bourgade.
\newblock Random band matrices.
\newblock In \emph{Proc. Int. Cong. of Math., Rio de Janeiro}, volume~3, pages
  2745--2770, 2018.
\newblock \doi{10.1142/9789813272880_0159}.

\bibitem[Bourgade et~al.(2009)Bourgade, Nikeghbali, and Rouault]{BNR09}
P.~Bourgade, A.~Nikeghbali, and A.~Rouault.
\newblock {Circular Jacobi Ensembles and Deformed Verblunsky Coefficients}.
\newblock \emph{Int. Math. Res. Not.}, 2009\penalty0 (23):\penalty0 4357--4394,
  2009.
\newblock \doi{10.1093/imrn/rnp092}.

\bibitem[Brack et~al.(2010)Brack, Koch, Murthy, and Roccia]{BKMR10}
M.~Brack, A.~Koch, M.~V.~N. Murthy, and J.~Roccia.
\newblock Exact and asymptotic local virial theorems for finite fermionic
  systems.
\newblock \emph{J. Phys. A}, 43\penalty0 (25):\penalty0 255204, 2010.
\newblock \doi{10.1088/1751-8113/43/25/255204}.

\bibitem[Br\'ezin and Hikami(2008)]{BH08}
E.~Br\'ezin and S.~Hikami.
\newblock {Intersection Theory from Duality and Replica}.
\newblock \emph{Commun. Math. Phys.}, 283\penalty0 (2):\penalty0 507--521,
  2008.
\newblock \doi{10.1007/s00220-008-0519-0}.

\bibitem[Br\'ezin and Kazakov(1990)]{BK90}
E.~Br\'ezin and V.~A. Kazakov.
\newblock Exactly solvable field theories of closed strings.
\newblock \emph{Phys. Lett. B}, 236\penalty0 (2):\penalty0 144--150, 1990.
\newblock \doi{10.1016/0370-2693(90)90818-Q}.

\bibitem[Br\'ezin and Zee(1994)]{BZ94}
E.~Br\'ezin and A.~Zee.
\newblock Correlation functions in disordered systems.
\newblock \emph{Phys. Rev. E}, 49\penalty0 (4):\penalty0 2588, 1994.
\newblock \doi{10.1103/PhysRevE.49.2588}.

\bibitem[Br\'ezin et~al.(1978)Br\'ezin, Itzykson, Parisi, and Zuber]{BIPZ78}
E.~Br\'ezin, C.~Itzykson, G.~Parisi, and J.~B. Zuber.
\newblock {Planar Diagrams}.
\newblock \emph{Commun. Math. Phys.}, 59\penalty0 (1):\penalty0 35--51, 1978.
\newblock \doi{10.1007/BF01614153}.

\bibitem[Brini et~al.(2011)Brini, Mari\~no, and Stevan]{BMS11}
A.~Brini, M.~Mari\~no, and S.~Stevan.
\newblock The uses of the refined matrix model recursion.
\newblock \emph{J. Math. Phys.}, 52\penalty0 (5):\penalty0 052305, 2011.
\newblock \doi{10.1063/1.3587063}.

\bibitem[Brouwer et~al.(1997)Brouwer, Frahm, and Beenakker]{BFB97}
P.~W. Brouwer, K.~M. Frahm, and C.~W.~J. Beenakker.
\newblock Quantum mechanical time-delay matrix in chaotic scattering.
\newblock \emph{Phys. Rev. Lett.}, 78\penalty0 (25):\penalty0 4737--4740, 1997.
\newblock \doi{10.1103/PhysRevLett.78.4737}.

\bibitem[Bryc and Pierce(2009)]{BP09}
W.~Bryc and V.~Pierce.
\newblock Duality of real and quaternionic random matrices.
\newblock \emph{Electron. J. Probab.}, 14:\penalty0 452--476, 2009.
\newblock \doi{10.1214/EJP.v14-606}.

\bibitem[Burda et~al.(2010)Burda, Jarosz, Livan, Nowak, and Swiech]{BJLNS10}
Z.~Burda, A.~Jarosz, G.~Livan, M.~A. Nowak, and A.~Swiech.
\newblock {Eigenvalues and singular values of products of rectangular Gaussian
  random matrices}.
\newblock \emph{Phys. Rev. E}, 82\penalty0 (6):\penalty0 061114, 2010.
\newblock \doi{10.1103/PhysRevE.82.061114}.

\bibitem[Cauchy(1974)]{Ca74}
A.-L. Cauchy.
\newblock M\'emoire sur les int\'egrales d\'efinies, prises entre des limites
  imaginaires.
\newblock In \emph{{Oeuvres de Cauchy, 2${}^e$ s\'erie, T.~XV}}, pages 59--89.
  Gauthier-Villiars, Paris, 1974.

\bibitem[Charlier et~al.(2021)Charlier, Lenells, and Mauersberger]{CLM21}
C.~Charlier, J.~Lenells, and J.~Mauersberger.
\newblock {Higher Order Large Gap Asymptotics at the Hard Edge for
  Muttalib--Borodin Ensembles}.
\newblock \emph{Commun. Math. Phys.}, 384\penalty0 (5):\penalty0 829--907,
  2021.
\newblock \doi{10.1007/s00220-021-04059-1}.

\bibitem[Chaudhuri and Do(2021)]{CD21}
A.~Chaudhuri and N.~Do.
\newblock {Generalisations of the Harer--Zagier recursion for 1-point
  functions}.
\newblock \emph{J. Algebr. Comb.}, 2021.
\newblock \doi{10.1007/s10801-020-01003-9}.

\bibitem[Chekhov and Eynard(2006)]{CE06}
L.~Chekhov and B.~Eynard.
\newblock {Matrix eigenvalue model: Feynman graph technique for all genera}.
\newblock \emph{J. High Energ. Phys.}, 2006\penalty0 (12):\penalty0 026, 2006.
\newblock \doi{10.1088/1126-6708/2006/12/026}.

\bibitem[Chekhov and Norbury(2019)]{CN19}
L.~Chekhov and P.~Norbury.
\newblock Topological recursion with hard edges.
\newblock \emph{Int. J. Math.}, 30\penalty0 (3):\penalty0 1950014, 2019.
\newblock \doi{10.1142/S0129167X19500149}.

\bibitem[Chekhov(2016)]{Che16}
L.~O. Chekhov.
\newblock {The Harer--Zagier recursion for an irregular spectral curve}.
\newblock \emph{J. Geom. Phys.}, 110:\penalty0 30--43, 2016.
\newblock \doi{10.1016/j.geomphys.2016.07.007}.

\bibitem[Cheliotis(2018)]{Che18}
D.~Cheliotis.
\newblock Triangular random matrices and biorthogonal ensembles.
\newblock \emph{Statist. Probab. Lett.}, 134:\penalty0 36--44, 2018.
\newblock \doi{10.1016/j.spl.2017.10.010}.

\bibitem[Claeys and Romano(2014)]{CR14}
T.~Claeys and S.~Romano.
\newblock Biorthogonal ensembles with two-particle interactions.
\newblock \emph{Nonlinearity}, 27\penalty0 (10):\penalty0 2419, 2014.
\newblock \doi{10.1088/0951-7715/27/10/2419}.

\bibitem[Claeys et~al.(2019)Claeys, Girotti, and Stivigny]{CGS19}
T.~Claeys, M.~Girotti, and D.~Stivigny.
\newblock {Large Gap Asymptotics at the Hard Edge for Product Random Matrices
  and Muttalib--Borodin Ensembles}.
\newblock \emph{Int. Math. Res. Not.}, 2019\penalty0 (9):\penalty0 2800--2847,
  2019.
\newblock \doi{10.1093/imrn/rnx202}.

\bibitem[Collins and Nechita(2016)]{CN16}
B.~Collins and I.~Nechita.
\newblock {Random matrix techniques in quantum information theory}.
\newblock \emph{J. Math. Phys.}, 57\penalty0 (1):\penalty0 015215, 2016.
\newblock \doi{10.1063/1.4936880}.

\bibitem[Crisanti et~al.(1993)Crisanti, Paladin, and Vulpiani]{CPV93}
A.~Crisanti, G.~Paladin, and A.~Vulpiani.
\newblock \emph{{Products of Random Matrices}}.
\newblock Springer-Verlag, Berlin Heidelberg, 1993.
\newblock \doi{10.1007/978-3-642-84942-8}.

\bibitem[Cunden et~al.(2016{\natexlab{a}})Cunden, Mezzadri, Simm, and
  Vivo]{CMSV16a}
F.~D. Cunden, F.~Mezzadri, N.~Simm, and P.~Vivo.
\newblock Correlators for the {Wigner--Smith} time-delay matrix of chaotic
  cavities.
\newblock \emph{J. Phys. A}, 49\penalty0 (18):\penalty0 18LT01,
  2016{\natexlab{a}}.
\newblock \doi{10.1088/1751-8113/49/18/18LT01}.

\bibitem[Cunden et~al.(2016{\natexlab{b}})Cunden, Mezzadri, Simm, and
  Vivo]{CMSV16b}
F.~D. Cunden, F.~Mezzadri, N.~Simm, and P.~Vivo.
\newblock Large-{$N$} expansion for the time-delay matrix of ballistic chaotic
  cavities.
\newblock \emph{J. Math. Phys.}, 57\penalty0 (11):\penalty0 111901,
  2016{\natexlab{b}}.
\newblock \doi{10.1063/1.4966642}.

\bibitem[Cunden et~al.(2019)Cunden, Mezzadri, O'Connell, and Simm]{CMOS19}
F.~D. Cunden, F.~Mezzadri, N.~O'Connell, and N.~Simm.
\newblock {Moments of Random Matrices and Hypergeometric Orthogonal
  Polynomials}.
\newblock \emph{Commun. Math. Phys.}, 369\penalty0 (3):\penalty0 1091--1145,
  2019.
\newblock \doi{10.1007/s00220-019-03323-9}.

\bibitem[Cunden et~al.(2021)Cunden, Dahlqvist, and O'Connell]{CDO21}
F.~D. Cunden, A.~Dahlqvist, and N.~O'Connell.
\newblock {Integer moments of complex Wishart matrices and Hurwitz numbers}.
\newblock \emph{Ann. Inst. H. Poincar\'e D}, 8\penalty0 (2):\penalty0 243--268,
  2021.
\newblock \doi{10.4171/AIHPD/103}.

\bibitem[Dartois and Forrester(2020)]{DF20}
S.~Dartois and P.~J. Forrester.
\newblock {Schwinger--Dyson and loop equations for a product of square Ginibre
  random matrices}.
\newblock \emph{J. Phys. A}, 53\penalty0 (17):\penalty0 175201, 2020.
\newblock \doi{10.1088/1751-8121/ab6fc4}.

\bibitem[Dartois and Rahman()]{DR22}
S.~Dartois and A.~A. Rahman.
\newblock {Combinatorics and loop equations for antisymmetrised and Hermitised
  matrix product ensembles}.
\newblock In preparation.

\bibitem[de~Bruijn(1955)]{deB55}
N.~G. de~Bruijn.
\newblock On some multiple integrals involving determinants.
\newblock \emph{J. Indian Math. Soc.}, 19\penalty0 (3--4):\penalty0 133--151,
  1955.
\newblock \doi{10.18311/jims/1955/17010}.

\bibitem[Dean et~al.(2016)Dean, Le~Doussal, Majumdar, and Schehr]{DDMS16}
D.~S. Dean, P.~Le~Doussal, S.~N. Majumdar, and G.~Schehr.
\newblock {Noninteracting fermions at finite temperature in a $d$-dimensional
  trap: Universal correlations}.
\newblock \emph{Phys. Rev. A}, 94\penalty0 (6):\penalty0 063622, 2016.
\newblock \doi{10.1103/PhysRevA.94.063622}.

\bibitem[Defosseux(2010)]{Def10}
M.~Defosseux.
\newblock Orbit measures, random matrix theory and interlaced determinantal
  processes.
\newblock \emph{Ann. Inst. H. Poincar\'e Probab. Statist.}, 46\penalty0
  (1):\penalty0 209--249, 2010.
\newblock \doi{10.1214/09-AIHP314}.

\bibitem[Deruyts(1886)]{Der86}
J.~Deruyts.
\newblock Sur une class de polyn\^omes conjug\'es.
\newblock \emph{M\'em. Cour. et M\'em. des Savants \'Etr., Acad. Royale des
  Sci., des Lettres et des Beaux-Arts de Belgique}, 48, 1886.

\bibitem[Desrosiers(2009)]{Des09}
P.~Desrosiers.
\newblock Duality in random matrix ensembles for all $\beta$.
\newblock \emph{Nucl. Phys. B}, 817\penalty0 (3):\penalty0 224--251, 2009.
\newblock \doi{10.1016/j.nuclphysb.2009.02.019}.

\bibitem[Desrosiers and Forrester(2006)]{DF06}
P.~Desrosiers and P.~J. Forrester.
\newblock {Hermite and Laguerre} $\beta$-ensembles: {Asymptotic} corrections to
  the eigenvalue density.
\newblock \emph{Nucl. Phys. B}, 743\penalty0 (3):\penalty0 307--332, 2006.
\newblock \doi{10.1016/j.nuclphysb.2006.03.002}.

\bibitem[Di~Francesco(2003)]{Fra03}
P.~Di~Francesco.
\newblock Rectangular matrix models and combinatorics of colored graphs.
\newblock \emph{Nucl. Phys. B}, 648\penalty0 (3):\penalty0 461--496, 2003.
\newblock \doi{10.1016/s0550-3213(02)00900-8}.

\bibitem[Di~Francesco et~al.(1995)Di~Francesco, Ginsparg, and
  Zinn-Justin]{FGZ95}
P.~Di~Francesco, P.~Ginsparg, and J.~Zinn-Justin.
\newblock {2D gravity and random matrices}.
\newblock \emph{Phys. Rep.}, 254\penalty0 (1--2):\penalty0 1--133, 1995.
\newblock \doi{10.1016/0370-1573(94)00084-G}.

\bibitem[Didon(1869)]{Did69}
M.~F. Didon.
\newblock Sur certains syst\`emes de polyn\^omes associ\'es.
\newblock \emph{Ann. Sci. \'Ec. Norm. Sup\'er.}, 6:\penalty0 111--125, 1869.

\bibitem[Dixon(1905)]{Dix05}
A.~L. Dixon.
\newblock {Generalisations of Legendre's Formula $KE'-(K-E)K'=\frac{1}{2}\pi$}.
\newblock \emph{Proc. London Math. Soc.}, s2--3\penalty0 (1):\penalty0
  206--224, 1905.
\newblock \doi{10.1112/plms/s2-3.1.206}.

\bibitem[Do and Norbury(2018)]{ND18}
N.~Do and P.~Norbury.
\newblock Topological recursion for irregular spectral curves.
\newblock \emph{J. London Math. Soc.}, 97\penalty0 (3):\penalty0 398--426,
  2018.
\newblock \doi{10.1112/jlms.12112}.

\bibitem[Dorokhov(1982)]{Dor82}
O.~N. Dorokhov.
\newblock {Transmission coefficient and the localization length of an electron
  in $N$ bound disordered chains}.
\newblock \emph{Pis'ma Zh. Eksp. Teor. Fiz.}, 36\penalty0 (7):\penalty0 259,
  1982.

\bibitem[Dotsenko and Fateev(1985)]{DF85}
V.~S. Dotsenko and V.~A. Fateev.
\newblock {Four-point correlation functions and the operator algebra in 2D
  conformal invariant theories with central charge $C\le1$}.
\newblock \emph{Nucl. Phys. B}, 251:\penalty0 691--734, 1985.
\newblock \doi{10.1016/S0550-3213(85)80004-3}.

\bibitem[Douglas and Shenker(1990)]{DS90}
M.~R. Douglas and S.~H. Shenker.
\newblock Strings in less than one dimension.
\newblock \emph{Nucl. Phys. B}, 335\penalty0 (3):\penalty0 635--654, 1990.
\newblock \doi{10.1016/0550-3213(90)90522-F}.

\bibitem[Dumaz and Vir\'ag(2013)]{DV13}
L.~Dumaz and B.~Vir\'ag.
\newblock {The right tail exponent of the Tracy--Widom $\beta$ distribution}.
\newblock \emph{Ann. Inst. H. Poincar\'e Probab. Statist.}, 49\penalty0
  (4):\penalty0 915--933, 2013.
\newblock \doi{10.1214/11-AIHP475}.

\bibitem[Dumitrescu et~al.(2013)Dumitrescu, Mulase, Safnuk, and Sorkin]{DMSS13}
O.~Dumitrescu, M.~Mulase, B.~Safnuk, and A.~Sorkin.
\newblock {The spectral curve of the Eynard--Orantin recursion via the Laplace
  transform}.
\newblock In A.~Dzhamay, K.~Maruno, and V.~U. Pierce, editors, \emph{{Algebraic
  and Geometric Aspects of Integrable Systems and Random Matrices}}, pages
  263--315. Amer. Math. Soc., Providence, R.I., 2013.
\newblock \doi{10.1090/conm/593/11867}.

\bibitem[Dumitriu(2003)]{Dum03}
I.~Dumitriu.
\newblock \emph{{Eigenvalue Statistics for Beta-Ensembles}}.
\newblock PhD thesis, Massachusetts Institute of Technology, June 2003.

\bibitem[Dumitriu and Edelman(2002)]{DE02}
I.~Dumitriu and A.~Edelman.
\newblock Matrix models for beta ensembles.
\newblock \emph{J. Math. Phys.}, 43\penalty0 (11):\penalty0 5830--5847, 2002.
\newblock \doi{10.1063/1.1507823}.

\bibitem[Dumitriu and Edelman(2006)]{DE06}
I.~Dumitriu and A.~Edelman.
\newblock {Global spectrum fluctuations for the $\beta$-Hermite and
  $\beta$-Laguerre ensembles via matrix models}.
\newblock \emph{J. Math. Phys.}, 47\penalty0 (6):\penalty0 063302, 2006.
\newblock \doi{10.1063/1.2200144}.

\bibitem[Dumitriu and Paquette(2012)]{DP12}
I.~Dumitriu and E.~Paquette.
\newblock {Global fluctuations for linear statistics of $\beta$-Jacobi
  ensembles}.
\newblock \emph{Random Matrices: Theory Appl.}, 1\penalty0 (4):\penalty0
  1250013, 2012.
\newblock \doi{10.1142/S201032631250013X}.

\bibitem[Dumitriu et~al.(2007)Dumitriu, Edelman, and Shuman]{MOPS}
I.~Dumitriu, A.~Edelman, and G.~Shuman.
\newblock {MOPS: Multivariate orthogonal polynomials (symbolically)}.
\newblock \emph{J. Symb. Comput.}, 42\penalty0 (6):\penalty0 587--620, 2007.
\newblock \doi{10.1016/j.jsc.2007.01.005}.

\bibitem[Dyson(1962{\natexlab{a}})]{Dys62}
F.~J. Dyson.
\newblock {The Threefold Way. Algebraic Structure of Symmetry Groups and
  Ensembles in Quantum Mechanics}.
\newblock \emph{J. Math. Phys.}, 3\penalty0 (6):\penalty0 1199--1215,
  1962{\natexlab{a}}.
\newblock \doi{10.1063/1.1703863}.

\bibitem[Dyson(1962{\natexlab{b}})]{Dys62b}
F.~J. Dyson.
\newblock {Statistical Theory of the Energy Levels of Complex Systems. I--III}.
\newblock \emph{J. Math. Phys.}, 3\penalty0 (1):\penalty0 140--175,
  1962{\natexlab{b}}.
\newblock \doi{10.1063/1.1703773}.

\bibitem[Dyson(1970)]{Dys70}
F.~J. Dyson.
\newblock {Correlations between Eigenvalues of a Random Matrix}.
\newblock \emph{Commun. Math. Phys.}, 19\penalty0 (3):\penalty0 235--250, 1970.
\newblock \doi{10.1007/BF01646824}.

\bibitem[Edelman(1997)]{Ede97}
A.~Edelman.
\newblock {The Probability that a Random Real Gaussian Matrix has $k$ Real
  Eigenvalues, Related Distributions, and the Circular Law}.
\newblock \emph{J. Multivariate Anal.}, 60\penalty0 (2):\penalty0 203--232,
  1997.
\newblock \doi{10.1006/jmva.1996.1653}.

\bibitem[Edmonds(1960)]{Edm60}
J.~R. Edmonds.
\newblock {A combinatorial representation for polyhedral surfaces}.
\newblock Master's thesis, University of Maryland, 1960.

\bibitem[Erd\H{o}s et~al.(2012)Erd\H{o}s, Knowles, Yau, and Yin]{EKYY12}
L.~Erd\H{o}s, A.~Knowles, H.-T. Yau, and J.~Yin.
\newblock Spectral {Statistics of Erd\H{o}s--R\'enyi Graphs II}: {Eigenvalue
  Spacing and the Extreme Eigenvalues}.
\newblock \emph{Commun. Math. Phys.}, 314\penalty0 (3):\penalty0 587--640,
  2012.
\newblock \doi{10.1007/s00220-012-1527-7}.

\bibitem[Erd\H{o}s et~al.(2013)Erd\H{o}s, Knowles, Yau, and Yin]{EKYY13}
L.~Erd\H{o}s, A.~Knowles, H.-T. Yau, and J.~Yin.
\newblock Spectral statistics of {Erd\H{o}s--R\'enyi } graphs {I}: {Local}
  semicircle law.
\newblock \emph{Ann. Probab.}, 41\penalty0 (3B):\penalty0 2279--2375, 2013.
\newblock \doi{10.1214/11-AOP734}.

\bibitem[Erd\H{o}s and R\'enyi(1960)]{ER60}
P.~Erd\H{o}s and A.~R\'enyi.
\newblock {On the evolution of random graphs}.
\newblock \emph{Magyar Tud. Akad. Mat. Kutat\'o Int. K\"ozl.}, 5:\penalty0
  17--61, 1960.

\bibitem[Eynard(2004)]{Eyn04}
B.~Eynard.
\newblock Topological expansion for the 1-hermitian matrix model correlation
  functions.
\newblock \emph{J. High Energ. Phys.}, 2004\penalty0 (11):\penalty0 031, 2004.
\newblock \doi{10.1088/1126-6708/2004/11/031}.

\bibitem[Eynard(2011)]{Eyn11}
B.~Eynard.
\newblock {Recursion Between Mumford Volumes of Moduli Spaces}.
\newblock \emph{Ann. Henri Poincar\'e}, 12\penalty0 (8):\penalty0 1431--1447,
  2011.
\newblock \doi{10.1007/s00023-011-0113-4}.

\bibitem[Eynard(2016)]{Eyn16}
B.~Eynard.
\newblock \emph{{Counting Surfaces}}.
\newblock Birkh\"auser, Switzerland, 2016.
\newblock \doi{10.1007/978-3-7643-8797-6}.

\bibitem[Eynard and Marchal(2009)]{EM09}
B.~Eynard and O.~Marchal.
\newblock {Topological expansion of the Bethe ansatz, and non-commutative
  algebraic geometry}.
\newblock \emph{J. High Energ. Phys.}, 2009\penalty0 (3):\penalty0 094, 2009.
\newblock \doi{10.1088/1126-6708/2009/03/094}.

\bibitem[Eynard and Orantin(2008)]{EO07}
B.~Eynard and N.~Orantin.
\newblock Invariants of algebraic curves and topological expansion.
\newblock \emph{Commun. Number Theory Phys.}, 1\penalty0 (2):\penalty0
  347--452, 2008.
\newblock \doi{10.4310/CNTP.2007.v1.n2.a4}.

\bibitem[Eynard and Orantin(2009)]{EO09}
B.~Eynard and N.~Orantin.
\newblock Topological recursion in enumerative geometry and random matrices.
\newblock \emph{J. Phys. A}, 42\penalty0 (29):\penalty0 293001, 2009.
\newblock \doi{10.1088/1751-8113/42/29/293001}.

\bibitem[Eynard et~al.(2018)Eynard, Kimura, and Ribault]{EKR18}
B.~Eynard, T.~Kimura, and S.~Ribault.
\newblock Random matrices.
\newblock \emph{arXiv:1510.04430}, 2018.

\bibitem[Fischmann et~al.(2012)Fischmann, Bruza, Khoruzhenko, Sommers, and
  \.Zyczkowski]{FBKSZ12}
J.~Fischmann, W.~Bruza, B.~A. Khoruzhenko, H.-J. Sommers, and K.~\.Zyczkowski.
\newblock {Induced Ginibre ensemble of random matrices and quantum operations}.
\newblock \emph{J. Phys. A}, 45\penalty0 (7):\penalty0 075203, 2012.
\newblock \doi{10.1088/1751-8113/45/7/075203}.

\bibitem[Fisher and Hartwig(1968)]{FH68}
M.~E. Fisher and R.~E. Hartwig.
\newblock {Toeplitz Determinants: Some Applications, Theorems, and
  Conjectures}.
\newblock \emph{Adv. Chem. Phys.}, 15:\penalty0 333--353, 1968.
\newblock \doi{10.1002/9780470143605.ch18}.

\bibitem[Forrester(1993{\natexlab{a}})]{Fo93}
P.~J. Forrester.
\newblock Recurrence equations for the computation of correlations in the
  $1/r^2$ quantum many-body system.
\newblock \emph{J. Stat. Phys.}, 72\penalty0 (1--2):\penalty0 39--50,
  1993{\natexlab{a}}.
\newblock \doi{10.1007/BF01048039}.

\bibitem[Forrester(1993{\natexlab{b}})]{Fo93a}
P.~J. Forrester.
\newblock The spectrum edge of random matrix ensembles.
\newblock \emph{Nucl. Phys. B}, 402\penalty0 (3):\penalty0 709--728,
  1993{\natexlab{b}}.
\newblock \doi{10.1016/0550-3213(93)90126-A}.

\bibitem[Forrester(1994)]{Fo94}
P.~J. Forrester.
\newblock Exact results and universal asymptotics in the {Laguerre} random
  matrix ensemble.
\newblock \emph{J. Math. Phys.}, 35\penalty0 (5):\penalty0 2539--2551, 1994.
\newblock \doi{10.1063/1.530883}.

\bibitem[Forrester(2006{\natexlab{a}})]{Fo06}
P.~J. Forrester.
\newblock Evenness symmetry and inter-relationships between gap probabilities
  in random matrix theory.
\newblock \emph{Forum Math.}, 18\penalty0 (5):\penalty0 711--743,
  2006{\natexlab{a}}.
\newblock \doi{10.1515/FORUM.2006.036}.

\bibitem[Forrester(2006{\natexlab{b}})]{Fo06a}
P.~J. Forrester.
\newblock {Hard and soft edge spacing distributions for random matrix ensembles
  with orthogonal and symplectic symmetry}.
\newblock \emph{Nonlinearity}, 19\penalty0 (12):\penalty0 2989,
  2006{\natexlab{b}}.
\newblock \doi{10.1088/0951-7715/19/12/015}.

\bibitem[Forrester(2010)]{Fo10}
P.~J. Forrester.
\newblock \emph{Log-gases and random matrices}.
\newblock Princeton University Press, Princeton, 2010.

\bibitem[Forrester(2012{\natexlab{a}})]{Fo12}
P.~J. Forrester.
\newblock {Spectral density asymptotics for Gaussian and Laguerre
  $\beta$-ensembles in the exponentially small region}.
\newblock \emph{J. Phys. A}, 45\penalty0 (7):\penalty0 075206,
  2012{\natexlab{a}}.
\newblock \doi{10.1088/1751-8113/45/7/075206}.

\bibitem[Forrester(2012{\natexlab{b}})]{Fo12a}
P.~J. Forrester.
\newblock {Large deviation eigenvalue density for the soft edge Laguerre and
  Jacobi $\beta$-ensembles}.
\newblock \emph{J. Phys. A}, 45\penalty0 (14):\penalty0 145201,
  2012{\natexlab{b}}.
\newblock \doi{10.1088/1751-8113/45/14/145201}.

\bibitem[Forrester(2014)]{Fo14}
P.~J. Forrester.
\newblock {Eigenvalue statistics for product complex Wishart matrices}.
\newblock \emph{J. Phys. A}, 47\penalty0 (34):\penalty0 345202, 2014.
\newblock \doi{10.1088/1751-8113/47/34/345202}.

\bibitem[Forrester(2015)]{Fo15}
P.~J. Forrester.
\newblock Beta ensembles.
\newblock In G.~Akemann, J.~Baik, and P.~Di Francesco, editors, \emph{{The
  Oxford Handbook of Random Matrix Theory}}, pages 415--432. Oxford University
  Press, Oxford, 2015.
\newblock \doi{10.1093/oxfordhb/9780198744191.013.20}.

\bibitem[Forrester(2017)]{Fo17}
P.~J. Forrester.
\newblock Octonions in random matrix theory.
\newblock \emph{Proc. R. Soc. A}, 473\penalty0 (2200):\penalty0 20160800, 2017.
\newblock \doi{10.1098/rspa.2016.0800}.

\bibitem[Forrester and Ipsen(2018)]{FI18}
P.~J. Forrester and J.~R. Ipsen.
\newblock {Selberg integral theory and Muttalib--Borodin ensembles}.
\newblock \emph{Adv. Appl. Math.}, 95:\penalty0 152--176, 2018.
\newblock \doi{10.1016/j.aam.2017.11.004}.

\bibitem[Forrester and Li(2021)]{FL20}
P.~J. Forrester and S.-H. Li.
\newblock {Rate of convergence at the hard edge for various P\'olya ensembles
  of positive definite matrices}.
\newblock \emph{Integral Transforms Spec. Funct.}, 2021.
\newblock \doi{10.1080/10652469.2021.1952200}.

\bibitem[Forrester and Liu(2015)]{FL15}
P.~J. Forrester and D.-Z. Liu.
\newblock {Raney Distributions and Random Matrix Theory}.
\newblock \emph{J. Stat. Phys.}, 158\penalty0 (5):\penalty0 1051--1082, 2015.
\newblock \doi{10.1007/s10955-014-1150-4}.

\bibitem[Forrester and Nagao(2007)]{FN07}
P.~J. Forrester and T.~Nagao.
\newblock {Eigenvalue Statistics of the Real Ginibre Ensemble}.
\newblock \emph{Phys. Rev. Lett.}, 99\penalty0 (5):\penalty0 050603, 2007.
\newblock \doi{10.1103/PhysRevLett.99.050603}.

\bibitem[Forrester and Ole~Warnaar(2008)]{FO08}
P.~J. Forrester and S.~Ole~Warnaar.
\newblock The importance of the {Selbeg} integral.
\newblock \emph{Bull. Amer. Math. Soc. (N.N.)}, 45\penalty0 (4):\penalty0
  489--534, 2008.
\newblock \doi{10.1090/S0273-0979-08-01221-4}.

\bibitem[Forrester and Rahman(2021)]{FR21}
P.~J. Forrester and A.~A. Rahman.
\newblock {Relations between moments for the Jacobi and Cauchy random matrix
  ensembles}.
\newblock \emph{J. Math. Phys.}, 62\penalty0 (7):\penalty0 073302, 2021.
\newblock \doi{10.1063/5.0039887}.

\bibitem[Forrester and Rains(2005)]{FR05}
P.~J. Forrester and E.~M. Rains.
\newblock Interpretations of some parameter dependent generalizations of
  classical matrix ensembles.
\newblock \emph{Probab. Theory Relat. Fields}, 131\penalty0 (1):\penalty0
  1--61, 2005.
\newblock \doi{10.1007/s00440-004-0375-6}.

\bibitem[Forrester and Trinh(2018)]{FT18}
P.~J. Forrester and A.~K. Trinh.
\newblock {Functional form for the leading correction to the distribution of
  the largest eigenvalue in the GUE and LUE}.
\newblock \emph{J. Math. Phys.}, 59\penalty0 (5):\penalty0 053302, 2018.
\newblock \doi{10.1063/1.5016347}.

\bibitem[Forrester and Trinh(2019{\natexlab{a}})]{FT19}
P.~J. Forrester and A.~K. Trinh.
\newblock {Optimal soft edge scaling variables for the Gaussian and Laguerre
  even $\beta$ ensembles}.
\newblock \emph{Nucl. Phys. B}, 938:\penalty0 621--639, 2019{\natexlab{a}}.
\newblock \doi{10.1016/j.nuclphysb.2018.12.006}.

\bibitem[Forrester and Trinh(2019{\natexlab{b}})]{FT19a}
P.~J. Forrester and A.~K. Trinh.
\newblock {Finite-size corrections at the hard edge for the Laguerre $\beta$
  ensemble}.
\newblock \emph{Studies Appl. Math.}, 143\penalty0 (3):\penalty0 315--336,
  2019{\natexlab{b}}.
\newblock \doi{10.1111/sapm.12279}.

\bibitem[Forrester and Wang(2017)]{FW17}
P.~J. Forrester and D.~Wang.
\newblock {Muttalib--Borodin ensembles in random matrix theory --- realisations
  and correlation functions}.
\newblock \emph{Electron. J. Probab.}, 22:\penalty0 1--43, 2017.
\newblock \doi{10.1214/17-EJP62}.

\bibitem[Forrester and Witte(2000)]{FW00}
P.~J. Forrester and N.~S. Witte.
\newblock {Application of the $\tau$-Function Theory of Painlev\'e Equations to
  Random Matrices: $\textrm{P}_{\textrm{IV}}$, $\textrm{P}_{\textrm{II}}$ and
  the GUE}.
\newblock \emph{Commun. Math. Phys.}, 219\penalty0 (2):\penalty0 357--398,
  2000.
\newblock \doi{10.1007/s002200100422}.

\bibitem[Forrester and Witte(2002)]{FW02}
P.~J. Forrester and N.~S. Witte.
\newblock Application of the $\tau$-function theory of {Painlev\'e} equations
  to random matrices: $\textrm{P}_{\textrm{v}}$, $\textrm{P}_{\textrm{iii}}$,
  the {LUE, JUE, and CUE}.
\newblock \emph{Commun. Pure Appl. Math.}, 55\penalty0 (6):\penalty0 679--727,
  2002.
\newblock \doi{10.1002/cpa.3021}.

\bibitem[Forrester and Witte(2004)]{FW04}
P.~J. Forrester and N.~S. Witte.
\newblock Application of the $\tau$-function theory of {Painlev\'e} equations
  to random matrices: $\textrm{P}_{\textrm{vi}}$, the {JUE, CyUE, cJUE} and
  scaled limits.
\newblock \emph{Nagoya Math. J.}, 174:\penalty0 29--114, 2004.
\newblock \doi{10.1017/S0027763000008801}.

\bibitem[Forrester et~al.(1999)Forrester, Nagao, and Honner]{FNH99}
P.~J. Forrester, T.~Nagao, and G.~Honner.
\newblock Correlations for the orthogonal-unitary and symplectic-unitary
  transitions at the hard and soft edges.
\newblock \emph{Nucl. Phys. B}, 553\penalty0 (3):\penalty0 601--643, 1999.
\newblock \doi{10.1016/S0550-3213(99)00272-2}.

\bibitem[Forrester et~al.(2006)Forrester, Frankel, and Garoni]{FFG06}
P.~J. Forrester, N.~E. Frankel, and T.~M. Garoni.
\newblock Asymptotic form of the density profile for {Gaussian and Laguerre}
  random matrix ensembles with orthogonal and symplectic symmetry.
\newblock \emph{J. Math. Phys.}, 47\penalty0 (2):\penalty0 023301, 2006.
\newblock \doi{10.1063/1.2165254}.

\bibitem[Forrester et~al.(2015)Forrester, Liu, and Zinn-Justin]{FLZ15}
P.~J. Forrester, D.-Z. Liu, and P.~Zinn-Justin.
\newblock {Equilibrium problems for Raney densities}.
\newblock \emph{Nonlinearity}, 28\penalty0 (7):\penalty0 2265, 2015.
\newblock \doi{10.1088/0951-7715/28/7/2265}.

\bibitem[Forrester et~al.(2017)Forrester, Rahman, and Witte]{FRW17}
P.~J. Forrester, A.~A. Rahman, and N.~S. Witte.
\newblock Large {$N$} expansions for the {Laguerre and Jacobi}
  $\beta$-ensembles from the loop equations.
\newblock \emph{J. Math. Phys.}, 58\penalty0 (11):\penalty0 113303, 2017.
\newblock \doi{10.1063/1.4997778}.

\bibitem[Forrester et~al.(2018)Forrester, Ipsen, and Liu]{FIL18}
P.~J. Forrester, J.~R. Ipsen, and D.-Z. Liu.
\newblock {Matrix Product Ensembles of Hermite Type and the Hyperbolic
  Harish-Chandra--Itzykson--Zuber Integral}.
\newblock \emph{Ann. Henri Poincar\'e}, 19\penalty0 (5):\penalty0 1307--1348,
  2018.
\newblock \doi{10.1007/s00023-018-0654-x}.

\bibitem[Forrester et~al.(2019)Forrester, Ipsen, Liu, and Zhang]{FILZ19}
P.~J. Forrester, J.~R. Ipsen, D.-Z. Liu, and L.~Zhang.
\newblock {Orthogonal and symplectic Harish-Chandra integrals and matrix
  product ensembles}.
\newblock \emph{Random Matrices: Theory Appl.}, 8\penalty0 (4):\penalty0
  1950015, 2019.
\newblock \doi{10.1142/S2010326319500151}.

\bibitem[Forrester et~al.(2021)Forrester, Li, and Trinh]{FLT20}
P.~J. Forrester, S.-H. Li, and A.~K. Trinh.
\newblock {Asymptotic correlations with correction for the circular Jacobi
  $\beta$-ensemble}.
\newblock \emph{J. Approx. Theory}, 271:\penalty0 105633, 2021.
\newblock \doi{10.1016/j.jat.2021.105633}.

\bibitem[Frobenius(1877)]{Fro77}
G.~Frobenius.
\newblock {\"Uber lineare Substitutionen und bilineare Formen}.
\newblock \emph{Journal f\"ur die reine und angewandte Mathematik},
  84:\penalty0 1--63, 1877.

\bibitem[Furstenberg and Kesten(1960)]{FK60}
H.~Furstenberg and H.~Kesten.
\newblock {Products of Random Matrices}.
\newblock \emph{Ann. Math. Statist.}, 31\penalty0 (2):\penalty0 457--469, 1960.
\newblock \doi{10.1214/aoms/1177705909}.

\bibitem[Fyodorov(2002)]{Fyo02}
Y.~V. Fyodorov.
\newblock {Negative moments of characteristic polynomials of random matrices:
  Ingham--Siegel integral as an alternative to Hubbard--Stratonovich
  transformation}.
\newblock \emph{Nucl. Phys. B}, 621\penalty0 (3):\penalty0 643--674, 2002.
\newblock \doi{10.1016/S0550-3213(01)00508-9}.

\bibitem[Fyodorov and Le~Doussal(2016)]{FL16}
Y.~V. Fyodorov and P.~Le~Doussal.
\newblock {Moments of the Position of the Maximum for GUE Characteristic
  Polynomials and for Log-Correlated Gaussian Processes}.
\newblock \emph{J. Stat. Phys.}, 164\penalty0 (1):\penalty0 190--240, 2016.
\newblock \doi{10.1007/s10955-016-1536-6}.

\bibitem[Fyodorov and Strahov(2002)]{FS02}
Y.~V. Fyodorov and E.~Strahov.
\newblock {Characteristic polynomials of random Hermitian matrices and
  Duistermaat--Heckman localisation on non-compact K\"ahler manifolds}.
\newblock \emph{Nucl. Phys. B}, 630\penalty0 (3):\penalty0 453--491, 2002.
\newblock \doi{10.1016/S0550-3213(02)00185-2}.

\bibitem[Gaberdiel et~al.(2005)Gaberdiel, Klemm, and Runkel]{GKR05}
M.~R. Gaberdiel, A.~Klemm, and I.~Runkel.
\newblock {Matrix model eigenvalue integrals and twist fields in the
  $su(2)$-WZW model}.
\newblock \emph{J. High Energ. Phys.}, 2005\penalty0 (10):\penalty0
  JHEP10(2005)107, 2005.
\newblock \doi{10.1088/1126-6708/2005/10/107}.

\bibitem[Garoni et~al.(2005)Garoni, Forrester, and Frankel]{GFF05}
T.~M. Garoni, P.~J. Forrester, and N.~E. Frankel.
\newblock Asymptotic corrections to the eigenvalue density of the {GUE and
  LUE}.
\newblock \emph{J. Math. Phys.}, 46\penalty0 (10):\penalty0 103301, 2005.
\newblock \doi{10.1063/1.2035028}.

\bibitem[Gel'fand and Naimark(1950)]{GN50}
I.~M. Gel'fand and M.~A. Naimark.
\newblock Unitary representations of the classical groups.
\newblock \emph{Trudy Mat. Inst. Steklov}, 36:\penalty0 3--288, 1950.

\bibitem[Ghosh(2006)]{Gho06}
S.~Ghosh.
\newblock {Generalized Christoffel--Darboux formula for skew-orthogonal
  polynomials and random matrix theory}.
\newblock \emph{J. Phys. A}, 39\penalty0 (28):\penalty0 8775, 2006.
\newblock \doi{10.1088/0305-4470/39/28/S02}.

\bibitem[Ghosh(2008)]{Gho08}
S.~Ghosh.
\newblock {Generalized Christoffel--Darboux formula for classical
  skew-orthogonal polynomials}.
\newblock \emph{J. Phys. A}, 41\penalty0 (43):\penalty0 435204, 2008.
\newblock \doi{10.1088/1751-8113/41/43/435204}.

\bibitem[Ghosh and Pandey(2002)]{GP02}
S.~Ghosh and A.~Pandey.
\newblock Skew-orthogonal polynomials and random-matrix ensembles.
\newblock \emph{Phys. Rev. E}, 65\penalty0 (4):\penalty0 046221, 2002.
\newblock \doi{10.1103/PhysRevE.65.046221}.

\bibitem[Gilbert(1959)]{Gil59}
E.~N. Gilbert.
\newblock {Random Graphs}.
\newblock \emph{Ann. Math. Statist.}, 30\penalty0 (4):\penalty0 1141--1144,
  1959.
\newblock \doi{10.1214/aoms/1177706098}.

\bibitem[Ginibre(1965)]{Gin65}
J.~Ginibre.
\newblock {Statistical Ensembles of Complex, Quaternion, and Real Matrices}.
\newblock \emph{J. Math. Phys.}, 6\penalty0 (3):\penalty0 440--449, 1965.
\newblock \doi{10.1063/1.1704292}.

\bibitem[Gisonni et~al.(2020)Gisonni, Grava, and Ruzza]{GGR20}
M.~Gisonni, T.~Grava, and G.~Ruzza.
\newblock {Laguerre Ensemble: Correlators, Hurwitz Numbers and Hodge
  Integrals}.
\newblock \emph{Ann. Henri Poincar\'e}, 21\penalty0 (10):\penalty0 3285--3339,
  2020.
\newblock \doi{10.1007/s00023-020-00922-4}.

\bibitem[Gisonni et~al.(2021)Gisonni, Grava, and Ruzza]{GGR21}
M.~Gisonni, T.~Grava, and G.~Ruzza.
\newblock {Jacobi Ensemble, Hurwitz Numbers and Wilson Polynomials}.
\newblock \emph{Lett. Math. Phys.}, 111\penalty0 (3):\penalty0 67, 2021.
\newblock \doi{10.1007/s11005-021-01396-z}.

\bibitem[Godsil and Royle(2001)]{GR01}
C.~Godsil and G.~Royle.
\newblock \emph{{Algebraic Graph Theory}}.
\newblock Springer-Verlag, New York, 2001.
\newblock \doi{10.1007/978-1-4613-0163-9}.

\bibitem[Golub and Van~Loan(1996)]{householder}
G.~H. Golub and C.~F. Van~Loan.
\newblock \emph{{Matrix Computations}}.
\newblock Johns Hopkins University Press, Baltimore, 3rd edition, 1996.

\bibitem[Gopalakrishna et~al.(2018)Gopalakrishna, Labelle, and
  Shramchenko]{GLS18}
K.~Gopalakrishna, P.~Labelle, and V.~Shramchenko.
\newblock {Feynman diagrams, ribbon graphs, and topological recursion of
  Eynard--Orantin}.
\newblock \emph{J. High Energ. Phys.}, 2018\penalty0 (6):\penalty0 162, 2018.
\newblock \doi{10.1007/JHEP06(2018)162}.

\bibitem[G\"{o}tze and Tikhomirov(2005)]{GT05}
F.~G\"{o}tze and A.~Tikhomirov.
\newblock {The rate of convergence for spectra of GUE and LUE matrix
  ensembles}.
\newblock \emph{Centr. Eur. J. Math.}, 3\penalty0 (4):\penalty0 666--704, 2005.
\newblock \doi{10.2478/BF02475626}.

\bibitem[Goulden and Jackson(1997)]{GJ97}
I.~P. Goulden and D.~M. Jackson.
\newblock {Maps in Locally Orientable Surfaces and Integrals Over Real
  Symmetric Surfaces}.
\newblock \emph{Can. J. Math.}, 49\penalty0 (5):\penalty0 865--882, 1997.
\newblock \doi{10.4153/CJM-1997-045-9}.

\bibitem[Goulden et~al.(2001)Goulden, Harer, and Jackson]{GHJ01}
I.~P. Goulden, J.~L. Harer, and D.~M. Jackson.
\newblock {A geometric parametrization for the virtual Euler characteristics of
  the moduli spaces of real and complex algebraic curves}.
\newblock \emph{Trans. Amer. Math. Soc.}, 353\penalty0 (11):\penalty0
  4405--4427, 2001.
\newblock \doi{10.1090/S0002-9947-01-02876-8}.

\bibitem[Graham et~al.(1994)Graham, Knuth, and Patashnik]{GKP94}
R.~L. Graham, D.~E. Knuth, and O.~Patashnik.
\newblock \emph{{Concrete Mathematics}}.
\newblock Addison--Wesley, Reading, Massachusetts, 1994.

\bibitem[Gross and Migdal(1990)]{GM90}
D.~J. Gross and A.~A. Migdal.
\newblock Nonperturbative two-dimensional quantum gravity.
\newblock \emph{Phys. Rev. Lett.}, 64\penalty0 (2):\penalty0 127, 1990.
\newblock \doi{10.1103/PhysRevLett.64.127}.

\bibitem[Haagerup and Thorbj{\o}rnsen(2003)]{HT03}
U.~Haagerup and S.~Thorbj{\o}rnsen.
\newblock {Random matrices with complex Gaussian entries}.
\newblock \emph{Expo. Math.}, 21\penalty0 (4):\penalty0 293--337, 2003.
\newblock \doi{10.1016/S0723-0869(03)80036-1}.

\bibitem[Haagerup and Thorbj{\o}rnsen(2012)]{HT12}
U.~Haagerup and S.~Thorbj{\o}rnsen.
\newblock {Asymptotic Expansions for the Gaussian Unitary Ensemble}.
\newblock \emph{Infin. Dimens. Anal. Quantum Probab. Relat. Top.}, 15\penalty0
  (1):\penalty0 1250003, 2012.
\newblock \doi{10.1142/S0219025712500038}.

\bibitem[Halmos(1950)]{Hal50}
P.~R. Halmos.
\newblock \emph{{Measure Theory}}.
\newblock Springer-Verlag, New York, 1950.
\newblock \doi{10.1007/978-1-4684-9440-2}.

\bibitem[Hanin and Nica(2020)]{HN20}
B.~Hanin and M.~Nica.
\newblock {Products of Many Large Random Matrices and Gradients in Deep Neural
  Networks}.
\newblock \emph{Commun. Math. Phys.}, 376\penalty0 (1):\penalty0 287--322,
  2020.
\newblock \doi{10.1007/s00220-019-03624-z}.

\bibitem[Harer(1986)]{Har86}
J.~L. Harer.
\newblock The virtual cohomological dimension of the mapping class group of an
  orientable surface.
\newblock \emph{Invent. Math.}, 84\penalty0 (1):\penalty0 157--176, 1986.
\newblock \doi{10.1007/BF01390325}.

\bibitem[Harer and Zagier(1986)]{HZ86}
J.~L. Harer and D.~Zagier.
\newblock {The Euler characteristic of the moduli space of curves}.
\newblock \emph{Invent. Math.}, 85\penalty0 (3):\penalty0 457--485, 1986.
\newblock \doi{10.1007/BF01390325}.

\bibitem[Harish-Chandra(1957)]{HC57}
Harish-Chandra.
\newblock {Differential Operators on a Semisimple Lie Algebra}.
\newblock \emph{Amer. J. Math.}, 79\penalty0 (1):\penalty0 87--120, 1957.
\newblock \doi{10.2307/2372387}.

\bibitem[Hatcher(2002)]{Hat02}
A.~Hatcher.
\newblock \emph{{Algebraic Topology}}.
\newblock Cambridge University Press, 2002.

\bibitem[Helgason(1978)]{Hel78}
S.~Helgason.
\newblock \emph{{Differential Geometry, Lie Groups and Symmetric Spaces}}.
\newblock Academic Press, 1978.

\bibitem[Holcomb and Flores(2012)]{HF12}
D.~Holcomb and G.~R.~M. Flores.
\newblock {Edge Scaling of the $\beta$-Jacobi Ensemble}.
\newblock \emph{J. Stat. Phys.}, 149\penalty0 (6):\penalty0 1136--1160, 2012.
\newblock \doi{10.1007/s10955-012-0634-3}.

\bibitem[Hooft(1974)]{Hoo74}
G.~'t Hooft.
\newblock A planar diagram theory for strong interactions.
\newblock \emph{Nucl. Phys. B}, 72\penalty0 (3):\penalty0 461--473, 1974.
\newblock \doi{10.1016/0550-3213(74)90154-0}.

\bibitem[Hua(1963)]{Hua63}
L.~K. Hua.
\newblock \emph{{Harmonic Analysis of Functions of Several Complex Variables in
  the Classical Domains}}.
\newblock Amer. Math. Soc., Providence, R.I., 1963.

\bibitem[Hurwitz(1897)]{Hur97}
A.~Hurwitz.
\newblock {\"uber die Erzeugung der Invarianten durch Integration}.
\newblock \emph{Nachr. Ges. Wiss. G\"ottingen}, 1897:\penalty0 71--90, 1897.

\bibitem[Ipsen(2015{\natexlab{a}})]{Ips15}
J.~R. Ipsen.
\newblock \emph{{Products of Independent Gaussian Random Matrices}}.
\newblock PhD thesis, Bielefeld University, August 2015{\natexlab{a}}.

\bibitem[Ipsen(2015{\natexlab{b}})]{Ips15a}
J.~R. Ipsen.
\newblock {Lyapunov exponents for products of rectangular real, complex and
  quaternionic Ginibre matrices}.
\newblock \emph{J. Phys. A}, 48\penalty0 (15):\penalty0 155204,
  2015{\natexlab{b}}.
\newblock \doi{10.1088/1751-8113/48/15/155204}.

\bibitem[Ipsen and Forrester(2018)]{IF18}
J.~R. Ipsen and P.~J. Forrester.
\newblock {Kac--Rice fixed point analysis for single- and multi-layered complex
  systems}.
\newblock \emph{J. Phys. A}, 51\penalty0 (47):\penalty0 474003, 2018.
\newblock \doi{10.1088/1751-8121/aae76d}.

\bibitem[Ipsen and Kieburg(2014)]{IK14}
J.~R. Ipsen and M.~Kieburg.
\newblock Weak commutation relations and eigenvalue statistics for products of
  rectangular random matrices.
\newblock \emph{Phys. Rev. E}, 89\penalty0 (3):\penalty0 032106, 2014.
\newblock \doi{10.1103/PhysRevE.89.032106}.

\bibitem[Isserlis(1918)]{Iss18}
L.~Isserlis.
\newblock {On a Formula for the Product-Moment Coefficient of any Order of a
  Normal Frequency Distribution in any Number of Variables}.
\newblock \emph{Biometrika}, 12\penalty0 (1/2):\penalty0 134--139, 1918.
\newblock \doi{10.2307/2331932}.

\bibitem[Itzykson and Zuber(1980)]{IZ80}
C.~Itzykson and J.-B. Zuber.
\newblock {The planar approximation. II}.
\newblock \emph{J. Math. Phys.}, 21\penalty0 (3):\penalty0 411--421, 1980.
\newblock \doi{10.1063/1.524438}.

\bibitem[Jackson et~al.(1996)Jackson, \c{S}ener, and Verbaarschot]{JSV96}
A.~D. Jackson, M.~K. \c{S}ener, and J.~J.~M. Verbaarschot.
\newblock {Finite volume partition functions and Itzykson--Zuber integrals}.
\newblock \emph{Phys. Lett. B}, 387\penalty0 (2):\penalty0 355--360, 1996.
\newblock \doi{10.1016/0370-2693(96)00993-8}.

\bibitem[Jackson et~al.(2002)Jackson, Lautrup, Johansen, and Nielsen]{JLJN02}
A.~D. Jackson, B.~Lautrup, P.~Johansen, and M.~Nielsen.
\newblock Products of random matrices.
\newblock \emph{Phys. Rev. E}, 66\penalty0 (6):\penalty0 066124, 2002.
\newblock \doi{10.1103/PhysRevE.66.066124}.

\bibitem[Jackson(1994)]{Jac94}
D.~M. Jackson.
\newblock {On an Integral Representation for the Genus Series for 2-Cell
  Embeddings}.
\newblock \emph{Trans. Amer. Math. Soc.}, 344\penalty0 (2):\penalty0 755--772,
  1994.
\newblock \doi{10.1090/S0002-9947-1994-1236224-5}.

\bibitem[Janik and Wieczorek(2004)]{JW04}
R.~A. Janik and W.~Wieczorek.
\newblock Multiplying unitary random matrices---universality and spectral
  properties.
\newblock \emph{J. Phys. A}, 37\penalty0 (25):\penalty0 6521, 2004.
\newblock \doi{10.1088/0305-4470/37/25/007}.

\bibitem[Janik et~al.(1997)Janik, Nowak, Papp, and Zahed]{JNPZ97}
R.~A. Janik, M.~A. Nowak, G.~Papp, and I.~Zahed.
\newblock Non-hermitian random matrix models.
\newblock \emph{Nucl. Phys. B}, 501\penalty0 (3):\penalty0 603--642, 1997.
\newblock \doi{10.1016/S0550-3213(97)00418-5}.

\bibitem[Jenkins(1957)]{Jen57}
J.~A. Jenkins.
\newblock {On the Existence of Certain General Extremal Metrics}.
\newblock \emph{Ann. Math.}, 66\penalty0 (3):\penalty0 440--453, 1957.
\newblock \doi{10.2307/1969901}.

\bibitem[Johansson(2001)]{Joh01}
K.~Johansson.
\newblock {Universality of the Local Spacing Distribution in Certain Ensembles
  of Hermitian Wigner Matrices}.
\newblock \emph{Commun. Math. Phys.}, 215\penalty0 (3):\penalty0 683--705,
  2001.
\newblock \doi{10.1007/s002200000328}.

\bibitem[Johnson and Okunev(1997)]{JO97}
C.~R. Johnson and P.~Okunev.
\newblock {Necessary And Sufficient Conditions For Existence of the LU
  Factorization of an Arbitrary Matrix}.
\newblock \emph{arXiv:0506382}, 1997.

\bibitem[Johnstone(2001)]{Jo01}
I.~M. Johnstone.
\newblock On the distribution of the largest eigenvalue in principal components
  analysis.
\newblock \emph{Ann. Statist.}, 29\penalty0 (2):\penalty0 295--327, 2001.
\newblock \doi{10.1214/aos/1009210544}.

\bibitem[Johnstone(2008)]{Jo08}
I.~M. Johnstone.
\newblock {Multivariate analysis and Jacobi ensembles: Largest eigenvalue,
  Tracy--Widom limits and rates of convergence}.
\newblock \emph{Ann. Statist.}, 36\penalty0 (6):\penalty0 2638--2716, 2008.
\newblock \doi{10.1214/08-AOS605}.

\bibitem[Jones and Singerman(1978)]{JS78}
G.~A. Jones and D.~Singerman.
\newblock {Theory of Maps on Orientable Surfaces}.
\newblock \emph{Proc. London Math. Soc.}, s3--37\penalty0 (2):\penalty0
  273--307, 1978.
\newblock \doi{10.1112/plms/s3-37.2.273}.

\bibitem[Kadell(1997)]{Kad97}
K.~W.~J. Kadell.
\newblock {The Selberg--Jack Symmetric Functions}.
\newblock \emph{Adv. Math.}, 130\penalty0 (1):\penalty0 33--102, 1997.
\newblock \doi{10.1006/aima.1997.1642}.

\bibitem[Kaneko(1993)]{Ka93}
J.~Kaneko.
\newblock {Selberg Integrals and Hypergeometric Functions Associated with Jack
  Polynomials}.
\newblock \emph{SIAM J. Math. Anal.}, 24\penalty0 (4):\penalty0 1086--1110,
  1993.
\newblock \doi{10.1137/0524064}.

\bibitem[Katori and Komatsuda(2003)]{KK03}
M.~Katori and N.~Komatsuda.
\newblock Moments of vicious walkers and {M\"obius} graph expansions.
\newblock \emph{Phys. Rev. E}, 67\penalty0 (5):\penalty0 051110, 2003.
\newblock \doi{10.1103/PhysRevE.67.051110}.

\bibitem[Katori and Tanemura(2004)]{KT04}
M.~Katori and H.~Tanemura.
\newblock Symmetry of matrix-valued stochastic processes and noncolliding
  diffusion particle systems.
\newblock \emph{J. Math. Phys.}, 45\penalty0 (8):\penalty0 3058, 2004.
\newblock \doi{10.1063/1.1765215}.

\bibitem[Kazakov and Migdal(1993)]{KM93}
V.~A. Kazakov and A.~A. Migdal.
\newblock {Induced gauge theory at large $N$}.
\newblock \emph{Nucl. Phys. B}, 397\penalty0 (1--2):\penalty0 214--238, 1993.
\newblock \doi{10.1016/0550-3213(93)90342-M}.

\bibitem[Kieburg et~al.(2013)Kieburg, Verbaarschot, and Zafeiropoulos]{KVZ13}
M.~Kieburg, J.~J.~M. Verbaarschot, and S.~Zafeiropoulos.
\newblock {Spectral properties of the Wilson--Dirac operator and random matrix
  theory}.
\newblock \emph{Phys. Rev. D}, 88\penalty0 (9):\penalty0 094502, 2013.
\newblock \doi{10.1103/PhysRevD.88.094502}.

\bibitem[Kieburg et~al.(2016)Kieburg, Kuijlaars, and Stivigny]{KKS16}
M.~Kieburg, A.~B.~J. Kuijlaars, and D.~Stivigny.
\newblock {Singular Value Statistics of Matrix Products with Truncated Unitary
  Matrices}.
\newblock \emph{Int. Math. Res. Not.}, 2016\penalty0 (11):\penalty0 3392--3424,
  2016.
\newblock \doi{10.1093/imrn/rnv242}.

\bibitem[Killip and Nenciu(2004)]{KN04}
R.~Killip and I.~Nenciu.
\newblock Matrix models for circular ensembles.
\newblock \emph{Int. Math. Res. Not.}, 2004\penalty0 (50):\penalty0 2665--2701,
  2004.
\newblock \doi{10.1155/S1073792804141597}.

\bibitem[Koekoek et~al.(2010)Koekoek, Lesky, and Swarttouw]{KLS10}
R.~Koekoek, P.~A. Lesky, and R.~F. Swarttouw.
\newblock \emph{{Hypergeometric Orthogonal Polynomials and Their q-Analogues}}.
\newblock Springer-Verlag, Berlin Heidelberg, 2010.
\newblock \doi{10.1007/978-3-642-05014-5}.

\bibitem[Koml\'os(1967)]{Kom67}
J.~Koml\'os.
\newblock On the determinant of $(0,\,1)$ matrices.
\newblock \emph{Studia Sci. Math. Hungar.}, 2:\penalty0 7--21, 1967.

\bibitem[Konhauser(1965)]{Kon65}
J.~D.~E. Konhauser.
\newblock {Some Properties of Biorthogonal Polynomials}.
\newblock \emph{J. Math. Anal. Appl.}, 11:\penalty0 242--260, 1965.
\newblock \doi{10.1016/0022-247X(65)90085-5}.

\bibitem[Konhauser(1967)]{Kon67}
J.~D.~E. Konhauser.
\newblock {Biorthogonal polynomials suggested by the Laguerre polynomials}.
\newblock \emph{Pac. J. Math.}, 21\penalty0 (2):\penalty0 303--314, 1967.

\bibitem[Kontsevich(1992)]{Kon92}
M.~Kontsevich.
\newblock {Intersection Theory on the Moduli Space of Curves and the Matrix
  Airy Function}.
\newblock \emph{Commun. Math. Phys.}, 147\penalty0 (1):\penalty0 1--23, 1992.
\newblock \doi{10.1007/BF02099526}.

\bibitem[Kontsevich and Soibelman(2018)]{KS18}
M.~Kontsevich and Y.~Soibelman.
\newblock Airy structures and symplectic geometry of topological recursion.
\newblock In C.-C.~M. Liu and M.~Mulase, editors, \emph{{Topological Recursion
  and its Influence in Analysis, Geometry, and Topology}}, pages 433--490.
  Amer. Math. Soc., Providence, R.I., 2018.
\newblock \doi{10.1090/pspum/100}.

\bibitem[Kopelevitch(2018)]{Ko18}
O.~Kopelevitch.
\newblock {A Convergent $\frac{1}{N}$ expansion for the GUE}.
\newblock \emph{Ann. Henri Poincar\'e}, 19\penalty0 (12):\penalty0 3883--3899,
  2018.
\newblock \doi{10.1007/s00023-018-0727-x}.

\bibitem[K\"uhn(2008)]{Ku08}
R.~K\"uhn.
\newblock Spectra of sparse random matrices.
\newblock \emph{J. Phys. A}, 41\penalty0 (29):\penalty0 295002, 2008.
\newblock \doi{10.1088/1751-8113/41/29/295002}.

\bibitem[Kuijlaars and Molag(2019)]{KM19}
A.~B.~J. Kuijlaars and L.~D. Molag.
\newblock {The local universality of Muttalib--Borodin biorthogonal ensembles
  with parameter $\theta=\tfrac{1}{2}$}.
\newblock \emph{Nonlinearity}, 32\penalty0 (8):\penalty0 3023, 2019.
\newblock \doi{10.1088/1361-6544/ab247c}.

\bibitem[Kuijlaars and Stivigny(2014)]{KS14}
A.~B.~J. Kuijlaars and D.~Stivigny.
\newblock Singular values of products of random matrices and polynomial
  ensembles.
\newblock \emph{Random Matrices: Theory Appl.}, 3\penalty0 (3):\penalty0
  1450011, 2014.
\newblock \doi{10.1142/S2010326314500117}.

\bibitem[Kuijlaars and Zhang(2014)]{KZ14}
A.~B.~J. Kuijlaars and L.~Zhang.
\newblock {Singular Values of Products of Ginibre Random Matrices, Multiple
  Orthogonal Polynomials and Hard Edge Scaling Limits}.
\newblock \emph{Commun. Math. Phys.}, 332\penalty0 (2):\penalty0 759--781,
  2014.
\newblock \doi{10.1007/s00220-014-2064-3}.

\bibitem[Kumar(2019)]{Kum19}
S.~Kumar.
\newblock {Recursion for the Smallest Eigenvalue Density of
  $\beta$-Wishart--Laguerre Ensemble}.
\newblock \emph{J. Stat. Phys.}, 175\penalty0 (1):\penalty0 126--149, 2019.
\newblock \doi{10.1007/s10955-019-02245-z}.

\bibitem[La~Croix(2009)]{LaC09}
M.~A. La~Croix.
\newblock \emph{{The combinatorics of the Jack parameter and the genus series
  for topological maps}}.
\newblock PhD thesis, University of Waterloo, August 2009.

\bibitem[Lando and Zvonkin(2004)]{LZ04}
S.~K. Lando and A.~K. Zvonkin.
\newblock \emph{{Graphs on Surfaces and Their Applications}}.
\newblock Springer-Verlag, Berlin Heidelberg, 2004.
\newblock \doi{10.1007/978-3-540-38361-1}.

\bibitem[Lawes and March(1979)]{LM79}
G.~P. Lawes and N.~H. March.
\newblock Exact local density method for linear harmonic oscillator.
\newblock \emph{J. Chem. Phys.}, 71\penalty0 (2):\penalty0 1007--1009, 1979.
\newblock \doi{10.1063/1.438398}.

\bibitem[Le~Ca\"{e}r et~al.(2007)Le~Ca\"{e}r, Male, and Delannay]{CMD07}
G.~Le~Ca\"{e}r, C.~Male, and R.~Delannay.
\newblock {Nearest-neighbour spacing distributions of the $\beta$-Hermite
  ensemble of random matrices}.
\newblock \emph{Physica A}, 383\penalty0 (2):\penalty0 190--208, 2007.
\newblock \doi{10.1016/j.physa.2007.04.057}.

\bibitem[Ledoux(2004)]{Le04}
M.~Ledoux.
\newblock {Differential Operators and Spectral Distributions of Invariant
  Ensembles from the Classical Orthogonal Polynomials. The Continuous Case}.
\newblock \emph{Electron. J. Probab.}, 9:\penalty0 177--208, 2004.
\newblock \doi{10.1214/EJP.v9-191}.

\bibitem[Ledoux(2009)]{Le09}
M.~Ledoux.
\newblock {A recursion formula for the moments of the Gaussian orthogonal
  ensemble}.
\newblock \emph{Ann. Inst. H. Poincar\'e Probab. Statist.}, 45\penalty0
  (3):\penalty0 754--769, 2009.
\newblock \doi{10.1214/08-AIHP184}.

\bibitem[Lehmann and Sommers(1991)]{LS91}
N.~Lehmann and H.-J. Sommers.
\newblock {Eigenvalue Statistics of Random Real Matrices}.
\newblock \emph{Phys. Rev. Lett.}, 67\penalty0 (8):\penalty0 941--944, 1991.
\newblock \doi{10.1103/PhysRevLett.67.941}.

\bibitem[Liu(2017)]{Li17}
D.-Z. Liu.
\newblock Limits for circular {Jacobi} beta-ensembles.
\newblock \emph{J. Approx. Theory}, 215:\penalty0 40--67, 2017.
\newblock \doi{10.1016/j.jat.2016.11.005}.

\bibitem[Liu(2018)]{Li18}
D.-Z. Liu.
\newblock {Singular Values for Products of Two Coupled Random Matrices: Hard
  Edge Phase Transition}.
\newblock \emph{Constr. Approx.}, 47:\penalty0 487--528, 2018.
\newblock \doi{10.1007/s00365-017-9389-z}.

\bibitem[Liu et~al.(2016)Liu, Wang, and Zhang]{LWZ16}
D.-Z. Liu, D.~Wang, and L.~Zhang.
\newblock {Bulk and soft-edge universality for singular values of products of
  Ginibre random matrices}.
\newblock \emph{Ann. Inst. H. Poincar\'e Probab. Statist.}, 52\penalty0
  (4):\penalty0 1734--1762, 2016.
\newblock \doi{10.1214/15-AIHP696}.

\bibitem[Livan and Vivo(2011)]{LV11}
G.~Livan and P.~Vivo.
\newblock Moments of {Wishart--Laguerre} and {Jacobi} ensembles of random
  matrices: application to the quantum transport problem in chaotic cavities.
\newblock \emph{Acta Phys. Pol. B}, 42\penalty0 (5):\penalty0 1081--1104, 2011.
\newblock \doi{10.5506/APhysPolB.42.1081}.

\bibitem[Lueck et~al.(2006)Lueck, Sommers, and Zirnbauer]{LSZ06}
T.~Lueck, H.-J. Sommers, and M.~R. Zirnbauer.
\newblock Energy correlations for a random matrix model of disordered bosons.
\newblock \emph{J. Math. Phys.}, 47\penalty0 (10):\penalty0 103304, 2006.
\newblock \doi{10.1063/1.2356798}.

\bibitem[Luke(1969)]{Luk69}
Y.~L. Luke.
\newblock \emph{{The Special Functions and Their Approximations}}.
\newblock Academic Press, 1969.

\bibitem[Macdonald(1979)]{Mac79}
I.~G. Macdonald.
\newblock \emph{{Symmetric Functions and Hall Polynomials}}.
\newblock Clarendon Press, Oxford, 1979.

\bibitem[Madhekar and Thakare(1982)]{MT82}
H.~C. Madhekar and N.~K. Thakare.
\newblock {Biorthogonal polynomials suggested by the Jacobi polynomials}.
\newblock \emph{Pac. J. Math.}, 100\penalty0 (2):\penalty0 417--424, 1982.

\bibitem[Mar\v{c}enko and Pastur(1967)]{MP67}
V.~A. Mar\v{c}enko and L.~A. Pastur.
\newblock Distribution of eigenvalues for some sets of random matrices.
\newblock \emph{Math. USSR-Sbornik}, 1\penalty0 (4):\penalty0 457--483, 1967.
\newblock \doi{10.1070/SM1967v001n04ABEH001994}.

\bibitem[Mathai(1997)]{Mat97}
A.~M. Mathai.
\newblock \emph{{Jacobians of Matrix Transformation and Functions of Matrix
  Arguments}}.
\newblock World Scientific, Singapore, 1997.
\newblock \doi{10.1142/3438}.

\bibitem[May(1972)]{May72}
R.~M. May.
\newblock {Will a Large Complex System be Stable?}
\newblock \emph{Nature}, 238:\penalty0 413--414, 1972.
\newblock \doi{10.1038/238413a0}.

\bibitem[McCullagh(1987)]{McC87}
P.~McCullagh.
\newblock \emph{{Tensor Methods in Statistics}}.
\newblock Chapman \& Hall, 1987.

\bibitem[Mehta(2004)]{Meh04}
M.~L. Mehta.
\newblock \emph{{Random Matrices}}.
\newblock Elsevier, Amsterdam, 3rd edition, 2004.

\bibitem[Mello et~al.(1984)Mello, Pereyra, and Kumar]{MPK88}
P.~A. Mello, P.~Pereyra, and N.~Kumar.
\newblock Macroscopic approach to multichannel disordered conductors.
\newblock \emph{Ann. Phys. (N.~Y.)}, 181\penalty0 (2):\penalty0 290--317, 1984.
\newblock \doi{10.1016/0003-4916(88)90169-8}.

\bibitem[Mezzadri and Simm(2011)]{MS11}
F.~Mezzadri and N.~J. Simm.
\newblock Moments of the transmission eigenvalues, proper delay times and
  random matrix theory {I}.
\newblock \emph{J. Math. Phys.}, 52\penalty0 (10):\penalty0 103511, 2011.
\newblock \doi{10.1063/1.3644378}.

\bibitem[Mezzadri and Simm(2012)]{MS12}
F.~Mezzadri and N.~J. Simm.
\newblock Moments of the transmission eigenvalues, proper delay times and
  random matrix theory {II}.
\newblock \emph{J. Math. Phys.}, 53\penalty0 (5):\penalty0 053504, 2012.
\newblock \doi{10.1063/1.4708623}.

\bibitem[Mezzadri et~al.(2017)Mezzadri, Reynolds, and Winn]{MRW17}
F.~Mezzadri, A.~K. Reynolds, and B.~Winn.
\newblock Moments of the eigenvalue densities and of the secular coefficients
  of $\beta$-ensembles.
\newblock \emph{Nonlinearity}, 30\penalty0 (3):\penalty0 1034--1057, 2017.
\newblock \doi{10.1088/1361-6544/aa518c}.

\bibitem[Migdal(1983)]{Mig83}
A.~A. Migdal.
\newblock {Loop equations and $1/N$ expansion}.
\newblock \emph{Phys. Rep.}, 102\penalty0 (4):\penalty0 199--290, 1983.
\newblock \doi{10.1016/0370-1573(83)90076-5}.

\bibitem[Mingo and Speicher(2017)]{MS17}
J.~A. Mingo and R.~Speicher.
\newblock \emph{{Free Probability and Random Matrices}}.
\newblock Springer-Verlag, New York, 2017.
\newblock \doi{10.1007/978-1-4939-6942-5}.

\bibitem[Mirzakhani(2007{\natexlab{a}})]{Mirz07}
M.~Mirzakhani.
\newblock {Simple geodesics and Weil--Petersson volumes of moduli spaces of
  bordered Riemann surfaces}.
\newblock \emph{Invent. Math.}, 167\penalty0 (1):\penalty0 179--222,
  2007{\natexlab{a}}.
\newblock \doi{10.1007/s00222-006-0013-2}.

\bibitem[Mirzakhani(2007{\natexlab{b}})]{Mirz07a}
M.~Mirzakhani.
\newblock {Weil--Petersson volumes and intersection theory on the moduli space
  of curves}.
\newblock \emph{J. Amer. Math. Soc.}, 20\penalty0 (1):\penalty0 1--23,
  2007{\natexlab{b}}.
\newblock \doi{10.1090/S0894-0347-06-00526-1}.

\bibitem[Molag(2021)]{Mol21}
L.~D. Molag.
\newblock {The local universality of Muttalib--Borodin ensembles when the
  parameter $\theta$ is the reciprocal of an integer}.
\newblock \emph{Nonlinearity}, 34\penalty0 (5):\penalty0 3485, 2021.
\newblock \doi{10.1088/1361-6544/abeab6}.

\bibitem[Mondello(2009)]{Mon09}
G.~Mondello.
\newblock Riemann surfaces, ribbon graphs and combinatorial classes.
\newblock In A.~Papadopoulos, editor, \emph{{Handbook of Teichm\"uller Theory,
  Volume II}}, pages 151--215. Eur. Math. Soc., Germany, 2009.
\newblock \doi{10.4171/055-1/6}.

\bibitem[Muirhead(1982)]{Mui82}
R.~J. Muirhead.
\newblock \emph{Aspects of multivariate statistical theory}.
\newblock Wiley, New York, 1982.
\newblock \doi{10.1002/9780470316559}.

\bibitem[Mulase and Penkava(1998)]{MP98}
M.~Mulase and M.~Penkava.
\newblock {Ribbon graphs, quadratic differentials on Riemann surfaces, and
  algebraic curves defined over $\overline{\mathbb{Q}}$}.
\newblock \emph{Asian J. Math.}, 2\penalty0 (4):\penalty0 875--919, 1998.
\newblock \doi{10.4310/AJM.1998.v2.n4.a11}.

\bibitem[Mulase and Waldron(2003)]{MW03}
M.~Mulase and A.~Waldron.
\newblock {Duality of Orthogonal and Symplectic Matrix Integrals and
  Quaternionic Feynman Graphs}.
\newblock \emph{Commun. Math. Phys.}, 240\penalty0 (3):\penalty0 553--586,
  2003.
\newblock \doi{10.1007/s00220-003-0918-1}.

\bibitem[Mulase and Yu(2005)]{MY05}
M.~Mulase and J.~Yu.
\newblock Non-commutative matrix integrals and representation varieties of
  surface groups in a finite group.
\newblock \emph{Ann. Inst. Fourier}, 55\penalty0 (6):\penalty0 2161--2196,
  2005.
\newblock \doi{10.5802/aif.2157}.

\bibitem[M\"uller(2002)]{Mu02}
R.~R. M\"uller.
\newblock {On the Asymptotic Eigenvalue Distribution of Concatenated
  Vector-Valued Fading Channels}.
\newblock \emph{IEEE Trans. Inf. Theory}, 48\penalty0 (7):\penalty0 2086--2091,
  2002.
\newblock \doi{10.1109/TIT.2002.1013149}.

\bibitem[Muttalib(1995)]{Mut95}
K.~A. Muttalib.
\newblock Random matrix models with additional interactions.
\newblock \emph{J. Phys. A}, 28\penalty0 (5):\penalty0 L159--L164, 1995.
\newblock \doi{10.1088/0305-4470/28/5/003}.

\bibitem[Nachbin(1965)]{Nac65}
L.~Nachbin.
\newblock \emph{{The Haar Integral}}.
\newblock D. Van Nostrand Co., Princeton, New Jersey, 1965.

\bibitem[Nagao and Forrester(1995)]{NF95}
T.~Nagao and P.~J. Forrester.
\newblock Asymptotic correlations at the spectrum edge of random matrices.
\newblock \emph{Nucl. Phys. B}, 435\penalty0 (3):\penalty0 401--420, 1995.
\newblock \doi{10.1016/0550-3213(94)00545-P}.

\bibitem[Nagao and Slevin(1993)]{NS93}
T.~Nagao and K.~Slevin.
\newblock Laguerre ensembles of random matrices: nonuniversal correlation
  functions.
\newblock \emph{J. Math. Phys.}, 34\penalty0 (6):\penalty0 2317--2330, 1993.
\newblock \doi{10.1063/1.530118}.

\bibitem[Nagao and Wadati(1991)]{NW91}
T.~Nagao and M.~Wadati.
\newblock {Correlation Functions of Random Matrix Ensembles Related to
  Classical Orthogonal Polynomials}.
\newblock \emph{J. Phys. Soc. Jpn.}, 60\penalty0 (10):\penalty0 3298--3322,
  1991.
\newblock \doi{10.1143/JPSJ.60.3298}.

\bibitem[Nagao and Wadati(1992)]{NW92}
T.~Nagao and M.~Wadati.
\newblock {Correlation Functions of Random Matrix Ensembles Related to
  Classical Orthogonal Polynomials. II}.
\newblock \emph{J. Phys. Soc. Jpn.}, 61\penalty0 (1):\penalty0 78--88, 1992.
\newblock \doi{10.1143/JPSJ.61.78}.

\bibitem[Naprienko(2018)]{Na18}
Y.~Naprienko.
\newblock {A convergent $1/n$-expansion for GSE and GOE}.
\newblock \emph{arXiv:1801.02359}, 2018.

\bibitem[Neretin(2013)]{Ner13}
Yu.~A. Neretin.
\newblock Hua measures on the space of $p$-adic matrices and inverse limits of
  {Grassmannians}.
\newblock \emph{Izvestiya: Math.}, 77\penalty0 (5):\penalty0 941--953, 2013.
\newblock \doi{10.1070/IM2013v077n05ABEH002665}.

\bibitem[Neuschel(2014)]{Neu14}
T.~Neuschel.
\newblock {Plancherel--Rotach formulae for average characteristic polynomials
  of products of Ginibre random matrices and the Fuss--Catalan distribution}.
\newblock \emph{Random Matrices: Theory Appl.}, 3\penalty0 (1):\penalty0
  1450003, 2014.
\newblock \doi{10.1142/S2010326314500038}.

\bibitem[Newman(1986)]{New86}
C.~M. Newman.
\newblock {The Distribution of Lyapunov Exponents: Exact Results for Random
  Matrices}.
\newblock \emph{Commun. Math. Phys.}, 103\penalty0 (1):\penalty0 121--126,
  1986.
\newblock \doi{10.1007/BF01464284}.

\bibitem[Nieminen(2016)]{Ni16}
J.~M. Nieminen.
\newblock {$N$-by-$N$ Random Matrix Theory with Matrix Representations of
  Octonions}.
\newblock \emph{Acta Phys. Pol. B}, 47\penalty0 (4):\penalty0 1113, 2016.
\newblock \doi{10.5506/APhysPolB.47.1113}.

\bibitem[Norbury and Scott(2014)]{NS14}
P.~Norbury and N.~Scott.
\newblock {Gromov--Witten invariants of $\mathbb{P}^1$ and Eynard--Orantin
  invariants}.
\newblock \emph{Geom. Topol.}, 18\penalty0 (4):\penalty0 1865--1910, 2014.
\newblock \doi{10.2140/gt.2014.18.1865}.

\bibitem[Novaes(2008)]{No08}
M.~Novaes.
\newblock Statistics of quantum transport in chaotic cavities with broken time
  reversal symmetry.
\newblock \emph{Phys. Rev. B}, 78\penalty0 (3):\penalty0 035337, 2008.
\newblock \doi{10.1103/PhysRevB.78.035337}.

\bibitem[Novaes(2015)]{No15}
M.~Novaes.
\newblock {Statistics of time delay and scattering correlation functions in
  chaotic systems. I. Random matrix theory}.
\newblock \emph{J. Math. Phys.}, 56\penalty0 (6):\penalty0 062110, 2015.
\newblock \doi{10.1063/1.4922746}.

\bibitem[Orantin(2015)]{Ora15}
N.~Orantin.
\newblock Chain of matrices, loop equations, and topological recursion.
\newblock In G.~Akemann, J.~Baik, and P.~Di Francesco, editors, \emph{{The
  Oxford Handbook of Random Matrix Theory}}, pages 329--352. Oxford University
  Press, Oxford, 2015.
\newblock \doi{10.1093/oxfordhb/9780198744191.013.16}.

\bibitem[O'Rourke et~al.(2016)O'Rourke, Vu, and Wang]{OVW16}
S.~O'Rourke, V.~Vu, and K.~Wang.
\newblock Eigenvectors of random matrices: {A} survey.
\newblock \emph{J. Combin. Theory Ser. A}, 144:\penalty0 361--442, 2016.
\newblock \doi{10.1016/j.jcta.2016.06.008}.

\bibitem[Oseledec(1968)]{Os68}
V.~I. Oseledec.
\newblock {A multiplicative ergodic theorem. Lyapunov characteristic numbers
  for dynamical systems}.
\newblock \emph{Trudy Moskov. Mat. Ob\v{s}\v{c}.}, 19:\penalty0 179--210, 1968.

\bibitem[Penner(1988)]{Pen88}
R.~C. Penner.
\newblock {Perturbative series and the moduli space of Riemann surfaces}.
\newblock \emph{J. Differ. Geom.}, 27\penalty0 (1):\penalty0 35--53, 1988.
\newblock \doi{10.4310/jdg/1214441648}.

\bibitem[Pennington et~al.(2017)Pennington, Schoenholz, and Ganguli]{PSG17}
J.~Pennington, S.~S. Schoenholz, and S.~Ganguli.
\newblock Resurrecting the sigmoid in deep learning through dynamical isometry:
  theory and practice.
\newblock In \emph{Advances in Neural Information Processing Systems},
  volume~30, pages 4788--4798, 2017.

\bibitem[Penson and \.Zyczkowski(2011)]{PZ11}
K.~A. Penson and K.~\.Zyczkowski.
\newblock {Product of Ginibre matrices: Fuss--Catalan and Raney distributions}.
\newblock \emph{Phys. Rev. E}, 83\penalty0 (6):\penalty0 061118, 2011.
\newblock \doi{10.1103/PhysRevE.83.061118}.

\bibitem[Polchinski(1998)]{Pol98}
J.~Polchinski.
\newblock \emph{{String Theory}}.
\newblock Cambridge University Press, 1998.

\bibitem[Porter(1965)]{Po65}
C.~E. Porter.
\newblock \emph{{Statistical Theories of Spectra: Fluctuations}}.
\newblock Academic Press, 1965.

\bibitem[Prats~Ferrer et~al.(2007)Prats~Ferrer, Eynard, Di~Francesco, and
  Zuber]{FEFZ07}
A.~Prats~Ferrer, B.~Eynard, P.~Di~Francesco, and J.-B. Zuber.
\newblock {Correlation Functions of Harish-Chandra Integrals over the
  Orthogonal and the Symplectic Groups}.
\newblock \emph{J. Stat. Phys.}, 129\penalty0 (5--6):\penalty0 885--935, 2007.
\newblock \doi{10.1007/s10955-007-9350-9}.

\bibitem[Prunotto et~al.(2018)Prunotto, Alberico, and Czerski]{PAC18}
A.~Prunotto, W.~M. Alberico, and P.~Czerski.
\newblock Feynman diagrams and rooted maps.
\newblock \emph{Open Phys.}, 16\penalty0 (1):\penalty0 149--167, 2018.
\newblock \doi{10.1515/phys-2018-0023}.

\bibitem[Raghunathan(1979)]{Ra79}
M.~S. Raghunathan.
\newblock {A proof of Oseledec's multiplicative ergodic theorem}.
\newblock \emph{Israel J. Math.}, 32\penalty0 (4):\penalty0 356--362, 1979.
\newblock \doi{10.1007/BF02760464}.

\bibitem[Rahman and Forrester(2021)]{RF21}
A.~A. Rahman and P.~J. Forrester.
\newblock Linear differential equations for the resolvents of the classical
  matrix ensembles.
\newblock \emph{Random Matrices: Theory Appl.}, 10\penalty0 (3):\penalty0
  2250003, 2021.
\newblock \doi{10.1142/S2010326322500034}.

\bibitem[Ram\'{i}rez et~al.(2011)Ram\'{i}rez, Rider, and Vir\'{a}g]{RRV11}
J.~A. Ram\'{i}rez, B.~Rider, and B.~Vir\'{a}g.
\newblock {Beta ensembles, stochastic Airy spectrum, and a diffusion}.
\newblock \emph{J. Amer. Math. Soc.}, 24\penalty0 (4):\penalty0 919--944, 2011.
\newblock \doi{10.1090/S0894-0347-2011-00703-0}.

\bibitem[Raposo et~al.(2007)Raposo, Weber, Alvarez, and Kirchbach]{RWAK07}
A.~P. Raposo, H.~J. Weber, D.~E. Alvarez, and M.~Kirchbach.
\newblock Romanovski polynomials in selected physics problems.
\newblock \emph{Centr. Eur. J. Phys.}, 5\penalty0 (3):\penalty0 253--284, 2007.
\newblock \doi{10.2478/s11534-007-0018-5}.

\bibitem[Rodgers and Bray(1988)]{RB88}
G.~J. Rodgers and A.~J. Bray.
\newblock Density of states of a sparse random matrix.
\newblock \emph{Phys. Rev. B}, 37\penalty0 (7):\penalty0 3557, 1988.
\newblock \doi{10.1103/PhysRevB.37.3557}.

\bibitem[Ryu et~al.(2010)Ryu, Schnyder, Furusaki, and Ludwig]{RSFL10}
S.~Ryu, A.~P. Schnyder, A.~Furusaki, and A.~W.~W. Ludwig.
\newblock Topological insulators and superconductors: tenfold way and
  dimensional hierarchy.
\newblock \emph{New J. Phys.}, 12\penalty0 (6):\penalty0 065010, 2010.
\newblock \doi{10.1088/1367-2630/12/6/065010}.

\bibitem[Schneps(1994)]{Sch94}
L.~Schneps.
\newblock \emph{{The Grothendieck Theory of Dessins d'Enfants}}.
\newblock Cambridge University Press, 1994.

\bibitem[Selberg(1944)]{Sel44}
A.~Selberg.
\newblock Bemerkninger om et multiplet integral.
\newblock \emph{Norsk Mat. Tidsskr.}, 26:\penalty0 71--78, 1944.

\bibitem[Sendra et~al.(2008)Sendra, Winkler, and P\'erez-Diaz]{SWP08}
J.~R. Sendra, F.~Winkler, and S.~P\'erez-Diaz.
\newblock \emph{{Rational Algebraic Curves: A Computer Algebra Approach}}.
\newblock Springer-Verlag, Berlin Heidelberg, 2008.
\newblock \doi{10.1007/978-3-540-73725-4}.

\bibitem[Silvestrov(1997)]{Sil97}
P.~G. Silvestrov.
\newblock Summing graphs for random band matrices.
\newblock \emph{Phys. Rev. E}, 55\penalty0 (6):\penalty0 6419, 1997.
\newblock \doi{10.1103/PhysRevE.55.6419}.

\bibitem[Simon et~al.(2006)Simon, Moustakas, and Marinelli]{SMM06}
S.~H. Simon, A.~L. Moustakas, and L.~Marinelli.
\newblock {Capacity and Character Expansions: Moment-Generating Function and
  Other Exact Results for MIMO Correlated Channels}.
\newblock \emph{IEEE Trans. Inf. Theory}, 52\penalty0 (12):\penalty0
  5336--5351, 2006.
\newblock \doi{10.1109/TIT.2006.885519}.

\bibitem[Smith(1995)]{Smi95}
P.~J. Smith.
\newblock {A Recursive Formulation of the Old Problem of Obtaining Moments from
  Cumulants and Vice Versa}.
\newblock \emph{The American Statistician}, 49\penalty0 (2):\penalty0 217--218,
  1995.
\newblock \doi{10.2307/2684642}.

\bibitem[Stanley(1999)]{Sta99}
R.~P. Stanley.
\newblock \emph{{Enumerative Combinatorics: Volume 2}}.
\newblock Cambridge University Press, 1999.

\bibitem[Strebel(1967)]{Str67}
K.~Strebel.
\newblock {On quadratic differentials with closed trajectories and second order
  poles}.
\newblock \emph{J. Analyse Math.}, 19\penalty0 (1):\penalty0 373--382, 1967.
\newblock \doi{10.1007/BF02788726}.

\bibitem[Strebel(1984)]{Str84}
K.~Strebel.
\newblock \emph{{Quadratic Differentials}}.
\newblock Springer-Verlag, Berlin Heidelberg, 1984.
\newblock \doi{10.1007/978-3-662-02414-0}.

\bibitem[Szeg\H{o}(1975)]{Sze75}
G.~Szeg\H{o}.
\newblock \emph{{Orthogonal Polynomials}}.
\newblock Amer. Math. Soc., Providence, R.I., 4th edition, 1975.

\bibitem[Tao and Vu(2014)]{TV14}
T.~Tao and V.~Vu.
\newblock Random matrices: {The universality phenomenon for Wigner ensembles}.
\newblock In Van~H. Vu, editor, \emph{{Modern Aspects of Random Matrix
  Theory}}, pages 121--172. Amer. Math. Soc., Providence, R.I., 2014.
\newblock \doi{10.1090/psapm/072}.

\bibitem[Terras(1988)]{Ter88}
A.~Terras.
\newblock \emph{{Harmonic Analysis on Symmetric Spaces and Applications II}}.
\newblock Springer-Verlag, New York, 1988.

\bibitem[Thakare and Madhekar(1986)]{TM86}
N.~K. Thakare and H.~C. Madhekar.
\newblock {Biorthogonal polynomials suggested by the Hermite polynomials}.
\newblock \emph{Indian J. Pure Appl. Math.}, 17\penalty0 (8):\penalty0
  1031--1041, 1986.

\bibitem[Tikhomirov(2020)]{Tik20}
K.~Tikhomirov.
\newblock {Singularity of random Bernoulli matrices}.
\newblock \emph{Ann. Math.}, 191\penalty0 (2):\penalty0 593--634, 2020.
\newblock \doi{10.4007/annals.2020.191.2.6}.

\bibitem[Tracy and Widom(1994{\natexlab{a}})]{TW94}
C.~A. Tracy and H.~Widom.
\newblock {Fredholm Determinants, Differential Equations and Matrix Models}.
\newblock \emph{Commun. Math. Phys.}, 163\penalty0 (1):\penalty0 33--72,
  1994{\natexlab{a}}.
\newblock \doi{10.1007/BF02101734}.

\bibitem[Tracy and Widom(1994{\natexlab{b}})]{TW94a}
C.~A. Tracy and H.~Widom.
\newblock {Level-Spacing Distributions and the Airy Kernel}.
\newblock \emph{Commun. Math. Phys.}, 159\penalty0 (1):\penalty0 151--174,
  1994{\natexlab{b}}.
\newblock \doi{10.1007/BF02100489}.

\bibitem[Tracy and Widom(1996)]{TW96}
C.~A. Tracy and H.~Widom.
\newblock {On Orthogonal and Symplectic Matrix Ensembles}.
\newblock \emph{Commun. Math. Phys.}, 177\penalty0 (3):\penalty0 727--754,
  1996.
\newblock \doi{10.1007/BF02099545}.

\bibitem[Tulino and Verd\'u(2004)]{TV04}
A.~M. Tulino and S.~Verd\'u.
\newblock {Random Matrix Theory and Wireless Communications}.
\newblock \emph{Foundations and Trends in Communications and Information
  Theory}, 1\penalty0 (1):\penalty0 1--182, 2004.
\newblock \doi{10.1561/0100000001}.

\bibitem[Tutte(1963)]{Tut63}
W.~T. Tutte.
\newblock {A Census of Planar Maps}.
\newblock \emph{Can. J. Math.}, 15:\penalty0 249--271, 1963.
\newblock \doi{10.4153/CJM-1963-029-x}.

\bibitem[Tutte(1984)]{Tut84}
W.~T. Tutte.
\newblock Graph theory.
\newblock In Gian-Carlo Rota, editor, \emph{{Encyclopedia of Mathematics and
  its Applications}}. Addison--Wesley, Reading, Massachusetts, 1984.

\bibitem[Verbaarschot et~al.(1984)Verbaarschot, Weidenm\"uller, and
  Zirnbauer]{VWZ84}
J.~Verbaarschot, H.~A. Weidenm\"uller, and M.~Zirnbauer.
\newblock {Evaluation of ensemble averages for simple Hamiltonians perturbed by
  a GOE interaction}.
\newblock \emph{Ann. Phys. (N. Y.)}, 153\penalty0 (2):\penalty0 367--388, 1984.
\newblock \doi{10.1016/0003-4916(84)90023-X}.

\bibitem[Vivo and Vivo(2008)]{VV08}
P.~Vivo and E.~Vivo.
\newblock Transmission eigenvalue densities and moments in chaotic cavities
  from random matrix theory.
\newblock \emph{J. Phys. A}, 41\penalty0 (12):\penalty0 122004, 2008.
\newblock \doi{10.1088/1751-8113/41/12/122004}.

\bibitem[Voiculescu et~al.(1992)Voiculescu, Dykema, and Nica]{VDN92}
D.~V. Voiculescu, K.~J. Dykema, and A.~Nica.
\newblock \emph{{Free Random Variables}}.
\newblock Amer. Math. Soc., Providence, R.I., 1992.

\bibitem[Wachter(1980)]{Wac80}
K.~W. Wachter.
\newblock {The Limiting Empirical Measure of Multiple Discriminant Ratios}.
\newblock \emph{Ann. Statist.}, 8\penalty0 (5):\penalty0 937--957, 1980.
\newblock \doi{10.1214/aos/1176345134}.

\bibitem[Wall(1948)]{Wal48}
H.~S. Wall.
\newblock \emph{Analytic theory of continued fractions}.
\newblock D. Van Nostrand Co., New York, 1948.

\bibitem[Walsh and Lehman(1972)]{WL72}
T.~Walsh and A.~B. Lehman.
\newblock {Counting rooted maps by genus II}.
\newblock \emph{J. Combin. Theory Ser. B}, 13\penalty0 (2):\penalty0 122--141,
  1972.
\newblock \doi{10.1016/0095-8956(72)90049-4}.

\bibitem[Wang(2008)]{Wa08}
D.~Wang.
\newblock \emph{{Spiked Models in Wishart Ensemble}}.
\newblock PhD thesis, Brandeis University, May 2008.

\bibitem[Wang and Zhang(2022)]{WZ21}
D.~Wang and L.~Zhang.
\newblock {A vector Riemann--Hilbert approach to the Muttalib--Borodin
  ensembles}.
\newblock \emph{J. Funct. Anal.}, 282\penalty0 (7):\penalty0 109380, 2022.
\newblock \doi{10.1016/j.jfa.2021.109380}.

\bibitem[Wick(1950)]{Wic50}
G.~C. Wick.
\newblock {The Evaluation of the Collision Matrix}.
\newblock \emph{Phys. Rev.}, 80\penalty0 (2):\penalty0 268--273, 1950.
\newblock \doi{10.1103/PhysRev.80.268}.

\bibitem[Wigner(1955)]{Wig55}
E.~P. Wigner.
\newblock {Characteristic Vectors of Bordered Matrices With Infinite
  Dimensions}.
\newblock \emph{Ann. Math.}, 62\penalty0 (3):\penalty0 548--564, 1955.
\newblock \doi{10.2307/1970079}.

\bibitem[Wigner(1957)]{Wig57}
E.~P. Wigner.
\newblock {Statistical Properties of Real Symmetric Matrices with Many
  Dimensions}.
\newblock In \emph{Can. Math. Congr. Proc., Toronto}, volume~4, page 174, 1957.

\bibitem[Wigner(1958)]{Wig58}
E.~P. Wigner.
\newblock {On the Distribution of the Roots of Certain Symmetric Matrices}.
\newblock \emph{Ann. Math.}, 67\penalty0 (2):\penalty0 325--327, 1958.
\newblock \doi{10.2307/1970008}.

\bibitem[Wigner(1967)]{Wig67}
E.~P. Wigner.
\newblock {Random Matrices in Physics}.
\newblock \emph{SIAM Rev.}, 9\penalty0 (1):\penalty0 1--23, 1967.
\newblock \doi{10.1137/1009001}.

\bibitem[Wishart(1928)]{Wis28}
J.~Wishart.
\newblock {The Generalised Product Moment Distribution in Samples from a Normal
  Multivariate Population}.
\newblock \emph{Biometrika}, 20A\penalty0 (1/2):\penalty0 32--52, 1928.
\newblock \doi{10.2307/2331939}.

\bibitem[Witte and Forrester(2000)]{WF00}
N.~S. Witte and P.~J. Forrester.
\newblock {Gap probabilities in the finite and scaled Cauchy random matrix
  ensembles}.
\newblock \emph{Nonlinearity}, 13\penalty0 (6):\penalty0 1965--1986, 2000.
\newblock \doi{10.1088/0951-7715/13/6/305}.

\bibitem[Witte and Forrester(2014)]{WF14}
N.~S. Witte and P.~J. Forrester.
\newblock Moments of the {Gaussian} $\beta$ ensembles and the {large-$N$}
  expansion of the densities.
\newblock \emph{J. Math. Phys.}, 55\penalty0 (8):\penalty0 083302, 2014.
\newblock \doi{10.1063/1.4886477}.

\bibitem[Witte and Forrester(2015)]{WF15}
N.~S. Witte and P.~J. Forrester.
\newblock Loop equation analysis of the circular $\beta$ ensembles.
\newblock \emph{J. High Energ. Phys.}, 2015\penalty0 (2):\penalty0 173, 2015.
\newblock \doi{10.1007/jhep02(2015)173}.

\bibitem[Witte and Forrester(2016)]{WF16}
N.~S. Witte and P.~J. Forrester.
\newblock {Singular Values of Products of Ginibre Random Matrices}.
\newblock \emph{Studies Appl. Math.}, 138\penalty0 (2):\penalty0 135--184,
  2016.
\newblock \doi{10.1111/sapm.12147}.

\bibitem[Witten(1991)]{Wit91}
E.~Witten.
\newblock Two dimensional gravity and intersection theory on moduli space.
\newblock \emph{Surveys Diff. Geom.}, 1:\penalty0 243--310, 1991.
\newblock \doi{10.4310/SDG.1990.v1.n1.a5}.

\bibitem[Wood(2012)]{Woo12}
P.~M. Wood.
\newblock Universality and the circular law for sparse random matrices.
\newblock \emph{Ann. Appl. Probab.}, 22\penalty0 (3):\penalty0 1266--1300,
  2012.
\newblock \doi{10.1214/11-AAP789}.

\bibitem[Wright(1935)]{Wr35}
E.~M. Wright.
\newblock {The Asymptotic Expansion of the Generalized Bessel Function}.
\newblock \emph{Proc. London Math. Soc.}, s2--38\penalty0 (1):\penalty0
  257--270, 1935.
\newblock \doi{10.1112/plms/s2-38.1.257}.

\bibitem[Xu and Zhao(2020)]{XZ20}
S.~Xu and Y.~Zhao.
\newblock {Gap Probability of the Circular Unitary Ensemble with a
  Fisher--Hartwig Singularity and the Coupled Painlev\'e V System}.
\newblock \emph{Commun. Math. Phys.}, 377\penalty0 (2):\penalty0 1545--1596,
  2020.
\newblock \doi{10.1007/s00220-020-03776-3}.

\bibitem[Yin and Krishnaiah(1983)]{YK83}
Y.~Q. Yin and P.~R. Krishnaiah.
\newblock {A Limit Theorem for the Eigenvalues of Product of Two Random
  Matrices}.
\newblock \emph{J. Multivariate Anal.}, 13\penalty0 (4):\penalty0 489--507,
  1983.
\newblock \doi{10.1016/0047-259X(83)90035-0}.

\bibitem[Zhang(2015)]{Zha15}
L.~Zhang.
\newblock {Local Universality in Biorthogonal Laguerre Ensembles}.
\newblock \emph{J. Stat. Phys.}, 161\penalty0 (3):\penalty0 688--711, 2015.
\newblock \doi{10.1007/s10955-015-1353-3}.

\bibitem[Zinn-Justin and Zuber(2003)]{ZJZ03}
P.~Zinn-Justin and J.-B. Zuber.
\newblock {On some integrals over the $U(N)$ unitary group and their large $N$
  limit}.
\newblock \emph{J. Phys. A}, 36\penalty0 (12):\penalty0 3173--3194, 2003.
\newblock \doi{10.1088/0305-4470/36/12/318}.

\bibitem[Zvonkin(1997)]{Zvo97}
A.~Zvonkin.
\newblock {Matrix integrals and map enumeration: An accessible introduction}.
\newblock \emph{Math. Comput. Modelling}, 26\penalty0 (8-10):\penalty0
  281--304, 1997.
\newblock \doi{10.1016/S0895-7177(97)00210-0}.

\bibitem[\.Zyczkowski and Sommers(2000)]{ZS00}
K.~\.Zyczkowski and H.-J. Sommers.
\newblock Truncations of random unitary matrices.
\newblock \emph{J. Phys. A}, 33\penalty0 (10):\penalty0 2045--2057, 2000.
\newblock \doi{10.1088/0305-4470/33/10/307}.

\end{thebibliography}
\normalsize

\begin{appendices}
\renewcommand\chaptername{Appendix}
\fancyhead[RO]{\slshape \leftmark}
\chapter{Quaternionic Matrices and Pfaffians}\label{appendixA}
The quaternions $\mathbb{H}$ are a number system that contain the real and complex numbers $\mathbb{R},\mathbb{C}$. They are a finite-dimensional associative division algebra over $\mathbb{R}$, of which there are exactly three up to isomorphism (the other two being $\mathbb{R},\mathbb{C}$) \citep{Fro77}. They are also referred to as a \textit{skew-field} since the only field axiom not satisfied by $\mathbb{H}$ is that of commutative multiplication. The quaternions are best realised as the Clifford algebra $Cl_{0,2}(\mathbb{R})$, which is an $\mathbb{R}$-algebra generated by two elements that we label $\mathrm{i}$ and $\mathrm{j}$, each squaring to negative one and together satisfying the anti-commutation relation $\mathrm{ij}+\mathrm{ji}=0$. Defining $\mathrm{k}:=\mathrm{ij}$, one has the isomorphism
\begin{equation} \label{eqA.0.1}
\mathbb{H}\simeq\{a+b\mathrm{i}+c\mathrm{j}+d\mathrm{k}\,|\,a,b,c,d\in\mathbb{R}\}
\end{equation}
with products induced by the following multiplication table:
\begin{center}
\begin{tabular}{ c | c c c c }
$\times$&$1$&i&j&k \\\hline
$1$&$1$&i&j&k\\
i&i&$-1$&k&$-$j\\
j&j&$-$k&$-1$&i\\
k&k&j&$-$i&$-1$
\end{tabular}
\end{center}
\begin{note} \label{NA.1}
Some authors refer to the right-hand side of \eqref{eqA.0.1} as the `real quaternions' due to the coefficients $a,b,c,d$ being real, in contrast to the `complex quaternions' defined by \eqref{eqA.0.1} with $a,b,c,d\in\mathbb{C}$. Throughout this thesis, $\mathbb{H}$ refers strictly to the real quaternions.
\end{note}
We can realise the well known embedding $\mathbb{M}_{M\times N}(\mathbb{H})\hookrightarrow\mathbb{M}_{2M\times2N}(\mathbb{C})$ by mapping each matrix entry according to
\begin{equation} \label{eqA.0.2}
a+b\mathrm{i}+c\mathrm{j}+d\mathrm{k}\mapsto\begin{bmatrix}a+b\mathrm{i}&c+d\mathrm{i}\\-c+d\mathrm{i}&a-b\mathrm{i}\end{bmatrix}.
\end{equation}
In fact, we identify $\mathbb{M}_{M\times N}(\mathbb{H})$ with the set of $M\times N$ matrices whose entries are themselves $2\times2$ matrices of the form given on the right-hand side of \eqref{eqA.0.2}.

The quaternionic conjugate of $q=a+b\mathrm{i}+c\mathrm{j}+d\mathrm{k}$ is $\overline{q}=a-b\mathrm{i}-c\mathrm{j}-d\mathrm{k}$ and the squared norm is $|q|^2=q\overline{q}=\overline{q}q=a^2+b^2+c^2+d^2$. The quaternionic dual of a $2N\times 2M$ complex matrix $X$ is given by
\begin{equation*}
X^D:=-J_{2M}X^TJ_{2N}
\end{equation*}
where
\begin{equation*}
J_{2N}:=I_N\otimes\begin{bmatrix}0&-1\\1&0\end{bmatrix}
\end{equation*}
is the $2N\times2N$ block-diagonal matrix with the diagonal blocks given by
\begin{equation*}
J_2=\begin{bmatrix}0&-1\\1&0\end{bmatrix}.
\end{equation*}
If $Q$ is the matrix representation of the quaternion $q$, then $Q^D$ is the matrix representation of $\overline{q}$. More generally, if $X\in\mathbb{M}_{M\times N}(\mathbb{H})$, i.e., if $X$ is a $2M\times2N$ complex matrix with $2\times2$ block structure such that each block can be interpreted as a quaternion, then $X^D$ is the quaternionic adjoint $X^\dagger$ of $X$. Here, the quaternionic adjoint $X^\dagger$ is equivalent to the transpose of the quaternionic conjugate of $X$. With this in mind, the \textit{(unitary) symplectic group} is defined by an obvious generalisation of the orthogonal and unitary groups,
\begin{equation}
Sp(2N):=\{U\in\mathbb{M}_{N\times N}(\mathbb{H})\,|\,U^\dagger U=I_{2N}\}.
\end{equation}

Due to the fact that $\mathbb{H}$ is not commutative, the determinant is not well-defined for quaternionic matrices. When deciding on a definition, one needs to choose which properties of the determinant must be retained and which can be jettisoned. For our purposes, it is enough that the determinant correspond to the product of eigenvalues. Thus, we adopt a definition of Dyson's for self-adjoint quaternionic matrices \citep{Dys70}.
\begin{definition}
Let $X\in\mathbb{M}_{N\times N}(\mathbb{H})$ be self-adjoint and for the sake of clarity, let $\tilde{X}$ be its $2N\times2N$ complex-matrix representation. Then, define the \textit{quaternionic determinant} as
\begin{equation} \label{eqA.0.4}
\Det\,X:=\Pf\,J_{2N}^{-1}\tilde{X},
\end{equation}
where $\Pf\,Y$ is the Pfaffian of $Y$. Likewise, to ensure that the trace of a quaternionic matrix is the sum of its eigenvalues, we define the \textit{quaterionic trace} as
\begin{equation*}
\Tr\,X:=\sum_{i=1}^N\textrm{Re}\,(X_{ii})=\frac{1}{2}\Tr\,\tilde{X}.
\end{equation*}
\end{definition}
The Pfaffian of an anti-symmetric matrix can be thought of as the square root of its determinant. Strictly speaking, given an anti-symmetric matrix $X\in\mathbb{M}_{2N\times 2N}(\mathbb{C})$,
\begin{equation*}
\Pf\,X=\frac{1}{2^NN!}\sum_{\sigma\in S_{2N}}\textrm{sgn}(\sigma)X_{\sigma(1),\sigma(2)}X_{\sigma(3),\sigma(4)}\cdots X_{\sigma(2N-1),\sigma(2N)},
\end{equation*}
where $S_{2N}$ is the symmetric group of order $(2N)!$ and $\textrm{sgn}(\sigma)$ is the sign of the permutation $\sigma$. Similar to the determinant, the Pfaffian can alternatively be computed through the following cofactor expansion (expanding along the first row):
\begin{equation*}
\Pf\,X=\sum_{j=2}^{2N}(-1)^jX_{1j}\,\Pf\,X(1,j),
\end{equation*}
where the minor $X(1,j)$ is obtained from $X$ by deleting the first and $j\textsuperscript{th}$ rows and columns; if one were instead computing the determinant, only the first row and $j\textsuperscript{th}$ column of $X$ would need to be deleted to produce the desired minor $X(1,j)$. The Pfaffian has an interpretation as a square root because of the fact that $(\Pf\,X)^2=\Det\,X$. Note that this means that the square of the right-hand side of equation \eqref{eqA.0.4} is $\Det\,\tilde{X}$, so that eigenvalues are not double-counted when computing the quaternionic determinant.

\chapter{Particular Stieltjes Transforms}\label{appendixB}
To take the Stieltjes transforms of equations \eqref{eq2.3.7} and \eqref{eq2.3.12}, we need to compute terms of the form
\begin{align*}
\int_0^1\frac{x^p(1-x)^q}{s-x}\frac{\mathrm{d}^n}{\mathrm{d}x^n}\rho^{(J)}(x;N,\beta)\,\mathrm{d}x
\end{align*}
for $\beta=2$ or $4$, $0\leq n\leq5$, $n\leq q\leq n+2$, and $q\leq p\leq q+1$ all integers. To this end, we define
\begin{equation}\label{eqB.0.1}
\mathcal{I}_{\beta}(s;p,q,n,k):=\int_0^1\frac{x^p(1-x)^q}{(s-x)^k}\frac{\mathrm{d}^n}{\mathrm{d}x^n}\rho^{(J)}(x;N,\beta)\,\mathrm{d}x
\end{equation}
for integers $0\leq n\leq q\leq p$ and $k\geq0$. Then, integration by parts gives the identity
\begin{align}\label{eqB.0.2}
\mathcal{I}_{\beta}(s;p,q,n,k)&=(p+q)\mathcal{I}_{\beta}(s;p,q-1,n-1,k)-p\mathcal{I}_{\beta}(s;p-1,q-1,n-1,k)\nonumber
\\&\quad-k\mathcal{I}_{\beta}(s;p,q,n-1,k+1).
\end{align}
Applying this identity $n$ times allows us to reduce $\mathcal{I}_{\beta}(s;p,q,n,k)$ to an expression involving terms of the form $\mathcal{I}_{\beta}(s;p,q,0,k)$. Then, considering $(s-x)^{-k-1}=(-1)^k/k!\,\partial_s^k(s-x)^{-1}$ for $k\geq0$ gives us
\begin{align}
\mathcal{I}_{\beta}(s;p,q,0,k+1)=\frac{(-1)^k}{k!}\frac{\mathrm{d}^k}{\mathrm{d}s^k}\mathcal{I}_{\beta}(s;p,q,0,1),\quad k\geq0,
\end{align}
which allows us to further reduce to an expression involving terms of the form $\mathcal{I}_{\beta}(s;p,q,0,1)$. Finally, factorisation of $x^p-s^p$ for positive integer $p$ yields
\begin{align} \label{eqB.0.4}
\mathcal{I}_{\beta}(s;p,q,0,1)=s^p\mathcal{I}_{\beta}(s;0,q,0,1)-\sum_{l=0}^{p-1}s^{p-l-1}\mathcal{I}_{\beta}(s;l,q,0,0),\quad p\geq1
\end{align}
and likewise
\begin{align}
\mathcal{I}_{\beta}(s;0,q,0,1)=\sum_{m=0}^q\binom{q}{m}(-1)^m\left[s^mW_1^{(J)}(s;N,\beta)-\sum_{l=0}^{m-1}s^{m-l-1}m_l^{(J)}\right],\quad q\geq1,
\end{align}
where $m_l^{(J)}$ is the $l\textsuperscript{th}$ spectral moment of $\rho^{(J)}(x;N,\beta)$. These last two equations thus reduce our expression to one involving powers of $s$, derivatives of $W_1^{(J)}(s;N,\beta)$, and spectral moments of $\rho^{(J)}(x;N,\beta)$.

To begin, we list
\begin{align*}
\mathcal{I}_{\beta}(s;1,1,1,1)&=s(1-s)\frac{\mathrm{d}}{\mathrm{d}s}W_1^{(J)}(s;N,\beta)-N
\\\mathcal{I}_{\beta}(s;2,2,2,1)&=s^2(1-s)^2\frac{\mathrm{d}^2}{\mathrm{d}s^2}W_1^{(J)}(s;N,\beta)-2N(s-2)-6m_1^{(J)}
\\\mathcal{I}_{\beta}(s;3,3,3,1)&=s^3(1-s)^3\frac{\mathrm{d}^3}{\mathrm{d}s^3}W_1^{(J)}(s;N,\beta)-6N\left(s^2-3s+3\right)-24m_1^{(J)}(s-3)-60m_2^{(J)}
\\\mathcal{I}_{\beta}(s;4,4,4,1)&=s^4(1-s)^4\frac{\mathrm{d}^4}{\mathrm{d}s^4}W_1^{(J)}(s;N,\beta)-24N(s^3-4s^2+6s-4)
\\&\qquad-120m_1^{(J)}\left(s^2-4s+6\right)-360m_2^{(J)}(s-4)-840m_3^{(J)}
\\\mathcal{I}_{\beta}(s;5,5,5,1)&=s^5(1-s)^5\frac{\mathrm{d}^5}{\mathrm{d}s^5}W_1^{(J)}(s;N,\beta)-120N(s^4-5s^3+10s^2-10s+5)
\\&\qquad-720m_1^{(J)}(s^3-5s^2+10s-10)-2520m_2^{(J)}(s^2-5s+10)
\\&\qquad-6720m_3^{(J)}(s-5)-15120m_4^{(J)}
\end{align*}
All of the necessary $\mathcal{I}_{\beta}(s;p,q,n,k)$ can be obtained from these through variants of identity \eqref{eqB.0.4} and integration by parts. For example,
\begin{align*}
\mathcal{I}_{\beta}(s;n+1,n,n,1)&=s\mathcal{I}_{\beta}(s;n,n,n,1)-\mathcal{I}_{\beta}(s;n,n,n,0)
\\&=s\mathcal{I}_{\beta}(s;n,n,n,1)+(-1)^{n+1}\int_0^1\frac{\mathrm{d}^n}{\mathrm{d}x^n}(x^n(1-x)^n)\rho^{(J)}(x;N,\beta)\,\mathrm{d}x,
\end{align*}
which only requires knowledge of the spectral moments $m_0^{(J)}$ to $m_n^{(J)}$ of $\rho^{(J)}(x;N,\beta)$, in addition to a term from the above list. It should be noted that applying the Stieltjes transform to $x^p(1-x)^q\frac{\mathrm{d}^n}{\mathrm{d}x^n}\rho^{(J)}(x)$ produces $s^p(1-s)^q\frac{\mathrm{d}^n}{\mathrm{d}s^n}W_1^{(J)}(s)$ plus terms that do not involve $W_1^{(J)}(s)$. It follows that the differential equations for the resolvents will be the same as those for the eigenvalue densities, with additional inhomogeneous terms.

\chapter{Loop Equations on Moments}\label{appendixC}
\section{Loop Equations on the $\tilde{m}_{k_1,\ldots,k_n}^{(\mathcal{J}_1)}$} \label{appendixC.1}
Let $m_{k_1,\ldots,k_n}=\tilde{m}_{k_1,\ldots,k_n}^{(\mathcal{J}_1)}$ be as defined in equation \eqref{eq4.2.2}. Substituting equations \eqref{eq4.2.23}--\eqref{eq4.2.26} into equation \eqref{eq4.2.17} and then substituting the result into equation \eqref{eq4.2.15} yields the following loop equation (recall that $I_n=(k_2,\ldots,k_n)$ and $\Tr\,B^{I_n}=\Tr\,B^{k_2}\cdots\Tr\,B^{k_n}$):
\begin{multline} \label{eqC.1.1}
0=N^2m_{k_1+1,I_n}+\sum_{p_1+p_2+p_3=k_1-1}m_{p_1,p_2,p_3,I_n}-\sum_{p_1+p_2=k_1-1}m_{p_1,p_2,I_n}+\frac{k_1^2-1}{2}m_{k_1-1,I_n}
\\+8\sum_{2\leq i<j\leq n}k_ik_j[k_i+1]_{\mathrm{mod}\,2}[k_j+1]_{\mathrm{mod}\,2}m_{k_1+k_i+k_j-1,I_n\setminus\{k_i,k_j\}}
\\\hspace{-2em}+2\sum_{i=2}^nk_i[k_i+1]_{\mathrm{mod}\,2}\Bigg(2\sum_{p_1+p_2=k_1-1}m_{k_i+p_1,p_2,I_n\setminus\{k_i\}}+\sum_{p_1+p_2=k_i-1}m_{k_1+p_1,p_2,I_n\setminus\{k_i\}}
\\-m_{k_1+k_i-1,I_n\setminus\{k_i\}}\Bigg).
\end{multline}

To convert this equation into the loop equation \eqref{eq4.2.33} on the unconnected correlators $U_n(x_1,\ldots,x_n)=\tilde{U}^{(\mathcal{J}_1)}_n(x_1,\ldots,x_n)$ \eqref{eq4.2.3}, we multiply both sides of equation \eqref{eqC.1.1} by $x_1^{-k_1-1}\cdots x_n^{-k_n-1}$ and sum over $k_1,\ldots,k_n\geq0$. Applying this prescription term by term requires the following glossary (recall that $J_n=(x_2,\ldots,x_n)$):
\begin{equation} \label{eqC.1.2}
\sum_{k_1,\ldots,k_n\geq0}\sum_{p_1+p_2+p_3=k_1-1}\frac{m_{p_1,p_2,p_3,I_n}}{x_1^{k_1+1}\cdots x_n^{k_n+1}}=x_1U_{n+2}(x_1,x_1,x_1,J_n),
\end{equation}
\begin{equation} \label{eqC.1.3}
\sum_{k_1,\ldots,k_n\geq0}\sum_{p_1+p_2=k_1-1}\frac{m_{p_1,p_2,I_n}}{x_1^{k_1+1}\cdots x_n^{k_n+1}}=U_{n+1}(x_1,x_1,J_n),
\end{equation}
\begin{equation} \label{eqC.1.4}
\sum_{k_1,\ldots,k_n\geq0}\frac{m_{k_1+1,I_n}}{x_1^{k_1+1}\cdots x_n^{k_n+1}}=x_1U_n(x_1,J_n)-NU_{n-1}(J_n),
\end{equation}
\begin{equation} \label{eqC.1.5}
\sum_{k_1,\ldots,k_n\geq0}\frac{k_1^2-1}{2}\frac{m_{k_1-1,I_n}}{x_1^{k_1+1}\cdots x_n^{k_n+1}}=\frac{1}{2x_1}\left(x_1^2\frac{\partial^2}{\partial x_1^2}+x_1\frac{\partial}{\partial x_1}-1\right)U_n(x_1,J_n),
\end{equation}
\begin{multline} \label{eqC.1.6}
\sum_{k_1,\ldots,k_n\geq0}k_i\left[k_i+1\right]_{\textrm{mod 2}}\frac{m_{k_1+k_i-1,I_n\setminus\{k_i\}}}{x_1^{k_1+1}\cdots x_n^{k_n+1}}
\\=\frac{1}{x_1}\frac{\partial}{\partial x_i}\left\{\frac{x_i^2U_{n-1}(x_1,J_n\setminus\{x_i\})-x_1x_iU_{n-1}(J_n)}{x_1^2-x_i^2}\right\},
\end{multline}
\begin{multline} \label{eqC.1.7}
\sum_{k_1,\ldots,k_n\geq0}\sum_{p_1+p_2=k_i-1}k_i\left[k_i+1\right]_{\textrm{mod 2}}\frac{m_{k_1+p_1,p_2,I_n\setminus\{k_i\}}}{x_1^{k_1+1}\cdots x_n^{k_n+1}}
\\=\frac{\partial}{\partial x_i}\left\{\frac{x_1x_iU_{n}(x_1,J_n)-x_i^2U_{n}(x_i,J_n)}{x_1^2-x_i^2}\right\},
\end{multline}
\begin{multline} \label{eqC.1.8}
\sum_{k_1,\ldots,k_n\geq0}\sum_{p_1+p_2=k_1-1}k_i\left[k_i+1\right]_{\textrm{mod 2}}\frac{m_{k_i+p_1,p_2,I_n\setminus\{k_i\}}}{x_1^{k_1+1}\cdots x_n^{k_n+1}}
\\=\frac{\partial}{\partial x_i}\left\{\frac{x_i^2U_n(x_1,x_1,J_n\setminus\{x_i\})-x_1x_iU_n(x_1,J_n)}{x_1^2-x_i^2}\right\},
\end{multline}
\begin{multline} \label{eqC.1.9}
\sum_{k_1,\ldots,k_n\geq0}k_ik_j\left[k_i+1\right]_{\textrm{mod 2}}\left[k_j+1\right]_{\textrm{mod 2}}\frac{m_{k_1+k_i+k_j-1,I_n\setminus\{k_i,k_j\}}}{x_1^{k_1+1}\cdots x_n^{k_n+1}}
\\=\frac{1}{x_1}\frac{\partial^2}{\partial x_i \partial x_j}\Bigg\{\frac{x_1x_i^2x_jU_{n-2}(J_n\setminus\{x_i\})}{(x_1^2-x_j^2)(x_i^2-x_j^2)}-\frac{x_1x_ix_j^2U_{n-2}(J_n\setminus\{x_j\})}{(x_1^2-x_i^2)(x_i^2-x_j^2)}
\\+\frac{x_i^2x_j^2U_{n-2}(x_1,J_n\setminus\{x_i,x_j\})}{(x_1^2-x_i^2)(x_1^2-x_j^2)}\Bigg\}.
\end{multline}

The proofs of equations \eqref{eqC.1.2}--\eqref{eqC.1.9} are a little more involved than one might expect, so we compute a few illustrative examples for the reader's convenience:

To prove equation \eqref{eqC.1.2}, one needs to appropriately split up the factor $x_1^{k_1+1}$ and interchange the order of summation to see that
\begin{align*}
\sum_{k_1\geq0}\sum_{p_1+p_2+p_3=k_1-1}\frac{m_{p_1,p_2,p_3,I_n}}{x_1^{k_1+1}}&=x_1\sum_{k_1=0}^\infty\sum_{p_1=0}^{k_1-1}\sum_{p_2=0}^{k_1-p_1-1}\frac{m_{p_1,p_2,k_1-p_1-p_2-1,I_n}}{x_1^{p_1+1}x_1^{p_2+1}x_1^{k_1-p_1-p_2}}
\\&=x_1\sum_{p_1=0}^\infty\sum_{k_1=p_1+1}^\infty\sum_{p_2=0}^{k_1-p_1-1}\frac{m_{p_1,p_2,k_1-p_1-p_2-1,I_n}}{x_1^{p_1+1}x_1^{p_2+1}x_1^{k_1-p_1-p_2}}
\\&=x_1\sum_{p_1=0}^\infty\sum_{k_1=0}^\infty\sum_{p_2=0}^{k_1}\frac{m_{p_1,p_2,k_1-p_2,I_n}}{x_1^{p_1+1}x_1^{p_2+1}x_1^{k_1-p_2+1}} \displaybreak
\\&=x_1\sum_{p_1=0}^\infty\sum_{p_2=0}^\infty\sum_{k_1=p_2}^\infty\frac{m_{p_1,p_2,k_1-p_2,I_n}}{x_1^{p_1+1}x_1^{p_2+1}x_1^{k_1-p_2+1}}
\\&=x_1\sum_{p_1=0}^\infty\sum_{p_2=0}^\infty\sum_{k_1=0}^\infty\frac{m_{p_1,p_2,k_1,I_n}}{x_1^{p_1+1}x_1^{p_2+1}x_1^{k_1+1}}.
\end{align*}
Multiplying this result by $x_2^{-k_2-1}\cdots x_n^{-k_n-1}$ and further summing over $k_2,\ldots,k_n\geq0$ then gives the right-hand side of equation \eqref{eqC.1.2}.

Equation \eqref{eqC.1.4} has a simple but comparatively unique proof:
\begin{multline*}
\sum_{k_1=0}^{\infty}\frac{m_{k_1+1,I_n}}{x_1^{k_1+1}}=\sum_{k_1=-1}^{\infty}\frac{m_{k_1+1,I_n}}{x_1^{k_1+1}}-m_{0,I_n}=x_1\sum_{k_1=-1}^{\infty}\frac{m_{k_1+1,I_n}}{x_1^{k_1+2}}-m_{0,I_n}
\\=x_1\sum_{k_1=0}^{\infty}\frac{m_{k_1,I_n}}{x_1^{k_1+1}}-m_{0,I_n}=x_1\sum_{k_1=0}^{\infty}\frac{m_{k_1,I_n}}{x_1^{k_1+1}}-Nm_{I_n},
\end{multline*}
using the fact that $m_{0,I_n}=\mean{\Tr\,B^0\Tr\,B^{I_n}}=N\mean{\Tr\,B^{I_n}}=Nm_{I_n}$. Multiplying again by $x_2^{-k_2-1}\cdots x_n^{-k_n-1}$ and summing over $k_2,\ldots,k_n\geq0$ produces the desired result.

To prove equation \eqref{eqC.1.6}, we first note that since $[k_i+1]_{\mathrm{mod}\,2}$ vanishes if $k_i$ is odd, we can replace $k_i$ by $2k_i$ in the summand of the left-hand side of equation \eqref{eqC.1.6}. However, since $m_{k_1+2k_i-1,I_n\setminus\{k_i\}}=\mean{\Tr\,B^{k_1+2k_i-1}\Tr\,B^{I_n\setminus\{k_i\}}}$ vanishes whenever $k_1$ is even, we must also replace $k_1$ by $2k_1+1$. Thus, the left-hand side of equation \eqref{eqC.1.6} simplifies as
\begin{align*}
\sum_{k_1,\ldots,k_n\geq0}&k_i\left[k_i+1\right]_{\textrm{mod 2}}\frac{m_{k_1+k_i-1,I_n\setminus\{k_i\}}}{x_1^{k_1+1}\cdots x_n^{k_n+1}}
\\&=-\frac{\partial}{\partial x_i}x_i\sum_{k_1,\ldots,k_n\geq0}\frac{m_{2k_1+2k_i,I_n\setminus\{k_i\}}}{x_1^{2k_1+2}x_2^{k_2+1}\cdots x_{i-1}^{k_{i-1}+1}x_i^{2k_i+1}x_{i+1}^{k_{i+1}+1}\cdots x_n^{k_n+1}}
\\&=-\frac{1}{x_1}\frac{\partial}{\partial x_i}\sum_{k_1,\ldots,k_n\geq0}\frac{m_{2k_1+2k_i,I_n\setminus\{k_i\}}}{x_1^{2k_1+2k_i+1}x_2^{k_2+1}\cdots x_{i-1}^{k_{i-1}+1}x_{i+1}^{k_{i+1}+1}\cdots x_n^{k_n+1}}\left(\frac{x_1}{x_i}\right)^{2k_i}
\\&=-\frac{1}{x_1}\frac{\partial}{\partial x_i}\sum_{k_2,\ldots,k_n\geq0}\sum_{k_1\geq k_i}\frac{m_{2k_1,I_n\setminus\{k_i\}}}{x_1^{2k_1+1}x_2^{k_2+1}\cdots x_{i-1}^{k_{i-1}+1}x_{i+1}^{k_{i+1}+1}\cdots x_n^{k_n+1}}\left(\frac{x_1}{x_i}\right)^{2k_i}
\\&=-\frac{1}{x_1}\frac{\partial}{\partial x_i}\sum_{k_1,\ldots,k_{i-1},k_{i+1},\ldots,k_n\geq0}\sum_{k_i=0}^{k_1}\frac{m_{2k_1,I_n\setminus\{k_i\}}}{x_1^{2k_1+1}x_2^{k_2+1}\cdots x_{i-1}^{k_{i-1}+1}x_{i+1}^{k_{i+1}+1}\cdots x_n^{k_n+1}}\left(\frac{x_1}{x_i}\right)^{2k_i}
\\&=-\frac{1}{x_1}\frac{\partial}{\partial x_i}\sum_{k_1,\ldots,k_{i-1},k_{i+1},\ldots,k_n\geq0}\frac{m_{2k_1,I_n\setminus\{k_i\}}}{x_1^{2k_1+1}x_2^{k_2+1}\cdots x_{i-1}^{k_{i-1}+1}x_{i+1}^{k_{i+1}+1}\cdots x_n^{k_n+1}}\frac{1-(x_1/x_i)^{2k_1+2}}{1-x_1^2/x_i^2}
\\&=\frac{1}{x_1}\frac{\partial}{\partial x_i}\frac{x_i^2}{x_1^2-x_i^2}\Bigg(\sum_{k_1,\ldots,k_{i-1},k_{i+1},\ldots,k_n\geq0}\frac{m_{2k_1,I_n\setminus\{k_i\}}}{x_1^{2k_1+1}x_2^{k_2+1}\cdots x_{i-1}^{k_{i-1}+1}x_{i+1}^{k_{i+1}+1}\cdots x_n^{k_n+1}}
\\&\hspace{10em}-\sum_{k_1,\ldots,k_{i-1},k_{i+1},\ldots,k_n\geq0}\frac{x_1}{x_i}\frac{m_{2k_1,I_n\setminus\{k_i\}}}{x_i^{2k_1+1}x_2^{k_2+1}\cdots x_{i-1}^{k_{i-1}+1}x_{i+1}^{k_{i+1}+1}\cdots x_n^{k_n+1}}\Bigg).
\end{align*}
We may now replace $2k_1$ by $k_1$ in the summands without any problem since $m_{k_1,I_n\setminus\{k_i\}}=0$ whenever $k_1$ is odd. Rewriting the sums in terms of the unconnected correlators according to equation \eqref{eq4.2.3} and performing some basic algebra then yields the right-hand side of equation \eqref{eqC.1.6}, as required.

We do not display the proofs of equations \eqref{eqC.1.3}, \eqref{eqC.1.5}, \eqref{eqC.1.7}--\eqref{eqC.1.9} since they can be proven using extensions of the ideas given above.

\setcounter{equation}{0}
\section{Loop Equations on the $\tilde{m}_{k_1,\ldots,k_n}^{(\mathcal{H}_1)}$} \label{appendixC.2}
Let $m_{k_1,\ldots,k_n}$ and $U_n(x_1,\ldots,x_n)$ now be as defined in equations \eqref{eq4.3.3} and \eqref{eq4.3.4}, respectively. Taking $i,j,h$ to be pairwise distinct, the averages $\mathcal{A}_{i',j',h'}$ of equation \eqref{eq4.3.19} are given by
\begin{multline} \label{eqC.2.1}
\mathcal{A}_{1,1,1}=\sum_{p_1+\cdots+p_4=k_1-1}m_{p_1,\ldots,p_4,I_n}+\frac{k_1(k_1-1)}{2}\sum_{p_1+p_2=k_1-1}m_{p_1,p_2,I_n}
\\+2\sum_{p_1+p_2+p_3=k_1-1}p_1m_{p_1+p_2,p_3,I_n},
\end{multline}
\begin{multline}
\mathcal{A}_{1,1,i}=\mathcal{A}_{1,i,1}=\mathcal{A}_{i,1,1}
\\=\frac{k_ik_1(k_1-1)}{2}m_{k_1+k_i-1,I_n\setminus\{k_i\}}+k_i\sum_{p_1+p_2+p_3=k_1-1}m_{p_1+k_i,p_2,p_3,I_n\setminus\{k_i\}},
\end{multline}
\begin{align}
\mathcal{A}_{1,i,i}&=\mathcal{A}_{i,i,1}=k_i\sum_{p_1+p_2=k_1-1}\sum_{q_1+q_2=k_i-1}m_{p_1+q_1+1,p_2,q_2,I_n\setminus\{k_i\}},
\\\mathcal{A}_{i,1,i}&=k_1k_i^2m_{k_1+k_i-1,I_n\setminus\{k_i\}},
\\\mathcal{A}_{i,i,i}&=\frac{k_i^2(k_i-1)}{2}m_{k_1+k_i-1,I_n\setminus\{k_i\}}+k_i\sum_{p_1+p_2+p_3=k_i-1}m_{k_1+p_1,p_2,p_3,I_n\setminus\{k_i\}},
\\\mathcal{A}_{1,i,j}&=\mathcal{A}_{i,j,1}=k_ik_j\sum_{p_1+p_2=k_1-1}m_{p_1+k_i+k_j,p_2,I_n\setminus\{k_i,k_j\}},
\\\mathcal{A}_{i,1,j}&=k_ik_j\sum_{p_1+p_2=k_1-1}m_{p_1+k_i,p_2+k_j,I_n\setminus\{k_i,k_j\}},
\\\mathcal{A}_{i,i,j}&=\mathcal{A}_{j,i,i}=k_ik_j\sum_{p_1+p_2=k_i-1}m_{k_1+p_1+k_j,p_2,I_n\setminus\{k_i,k_j\}},
\\\mathcal{A}_{i,j,i}&=k_ik_j\sum_{p_1+p_2=k_i-1}m_{k_1+p_1,p_2+k_j,I_n\setminus\{k_i,k_j\}},
\\\mathcal{A}_{i,j,h}&=k_ik_jk_hm_{k_1+k_i+k_j+k_h-1,I_n\setminus\{k_i,k_j,k_h\}}. \label{eqC.2.10}
\end{align}

We do not provide proofs for equations \eqref{eqC.2.1}--\eqref{eqC.2.10}, as they can be computed using Lemma \ref{L4.5} via the same ideas as in the proof of Lemma \ref{L4.3}.

Now, substituting equation \eqref{eq4.3.19} into equation \eqref{eq4.3.16} while observing that some of the above expressions are unchanged upon reordering indices shows that
\begin{align}
0&=-N^3m_{k_1+1,I_n}+k_1\sum_{p_1+p_2=k_1-1}m_{p_1,p_2,I_n}+2\sum_{2\leq i<j\leq n}k_ik_jm_{k_1,k_i+k_j-1,I_n\setminus\{k_i,k_j\}} \nonumber
\\&\quad+\sum_{i=2}^nk_i\left(2k_1m_{k_1+k_i-1,I_n\setminus\{k_i\}}+\sum_{p_1+p_2=k_i-1}m_{k_1,p_1,p_2,I_n\setminus\{k_i\}}\right)+\mathcal{A}_{1,1,1} \nonumber
\\&\quad+\sum_{i=2}^n\left(3\mathcal{A}_{1,1,i}+2\mathcal{A}_{1,i,i}+\mathcal{A}_{i,1,i}+\mathcal{A}_{i,i,i}\right)+2\sum_{2\leq i<j\leq n}\left(2\mathcal{A}_{1,i,j}+\mathcal{A}_{i,1,j}\right) \nonumber
\\&\quad+\sum_{\substack{2\leq i,j\leq n,\\i\neq j}}\left(2\mathcal{A}_{i,i,j}+\mathcal{A}_{i,j,i}\right)+6\sum_{2\leq i<j<h\leq n}\mathcal{A}_{i,j,h}. \label{eqC.2.11}
\end{align}
Inserting the expressions \eqref{eqC.2.1}--\eqref{eqC.2.10} into this equation then produces the loop equation on the mixed moments,
\begin{align}
0&=-N^3m_{k_1+1,I_n}+k_1\sum_{p_1+p_2=k_1-1}m_{p_1,p_2,I_n}+2\sum_{2\leq i<j\leq n}k_ik_jm_{k_1,k_i+k_j-1,I_n\setminus\{k_i,k_j\}} \nonumber
\\&\quad+\sum_{i=2}^nk_i\left(2k_1m_{k_1+k_i-1,I_n\setminus\{k_i\}}+\sum_{p_1+p_2=k_i-1}m_{k_1,p_1,p_2,I_n\setminus\{k_i\}}\right)+\sum_{p_1+\cdots+p_4=k_1-1}m_{p_1,\ldots,p_4,I_n} \nonumber
\\&\quad+\frac{k_1(k_1-1)}{2}\sum_{p_1+p_2=k_1-1}m_{p_1,p_2,I_n}+2\sum_{p_1+p_2+p_3=k_1-1}p_1m_{p_1+p_2,p_3,I_n} \nonumber
\\&\quad+3\sum_{i=2}^n\left(\frac{k_ik_1(k_1-1)}{2}m_{k_1+k_i-1,I_n\setminus\{k_i\}}+k_i\sum_{p_1+p_2+p_3=k_1-1}m_{p_1+k_i,p_2,p_3,I_n\setminus\{k_i\}}\right) \nonumber
\\&\quad+\sum_{i=2}^n\left(2k_i\sum_{p_1+p_2=k_1-1}\sum_{q_1+q_2=k_i-1}m_{p_1+q_1+1,p_2,q_2,I_n\setminus\{k_i\}}+k_1k_i^2m_{k_1+k_i-1,I_n\setminus\{k_i\}}\right) \nonumber
\\&\quad+\sum_{i=2}^n\left(\frac{k_i^2(k_i-1)}{2}m_{k_1+k_i-1,I_n\setminus\{k_i\}}+k_i\sum_{p_1+p_2+p_3=k_i-1}m_{k_1+p_1,p_2,p_3,I_n\setminus\{k_i\}}\right) \nonumber
\\&\quad+2\sum_{2\leq i<j\leq n}k_ik_j\left(2\sum_{p_1+p_2=k_1-1}m_{p_1+k_i+k_j,p_2,I_n\setminus\{k_i,k_j\}}+\sum_{p_1+p_2=k_1-1}m_{p_1+k_i,p_2+k_j,I_n\setminus\{k_i,k_j\}}\right) \nonumber
\\&\quad+\sum_{\substack{2\leq i,j\leq n,\\i\neq j}}k_ik_j\left(2\sum_{p_1+p_2=k_i-1}m_{k_1+p_1+k_j,p_2,I_n\setminus\{k_i,k_j\}}+\sum_{p_1+p_2=k_i-1}m_{k_1+p_1,p_2+k_j,I_n\setminus\{k_i,k_j\}}\right) \nonumber
\\&\quad+6\sum_{2\leq i<j<h\leq n}k_ik_jk_hm_{k_1+k_i+k_j+k_h-1,I_n\setminus\{k_i,k_j,k_h\}}. \label{eqC.2.12}
\end{align}

Multiplying both sides of this equation by $x_1^{-k_1-1}\cdots x_n^{-k_n-1}$ and then taking the sum over $k_1,\ldots,k_n\geq0$ yields the loop equation on the unconnected correlators $U_n(x_1,\ldots,x_n)$ given in Proposition~\ref{prop4.5}. We apply this prescription term by term using extensions of the ideas underlying the proofs presented in Appendix \ref{appendixC.1} above (recall again that $J_n=(x_2,\ldots,x_n)$):
\begin{equation} \label{eqC.2.13}
\sum_{k_1,\ldots,k_n\geq0}\frac{m_{k_1+1,I_n}}{x_1^{k_1+1}\cdots x_n^{k_n+1}}=x_1U_n(x_1,J_n)-NU_{n-1}(J_n),
\end{equation}
\begin{equation} \label{eqC.2.14}
\sum_{k_1,\ldots,k_n\geq0}k_1\sum_{p_1+p_2=k_1-1}\frac{m_{p_1,p_2,I_n}}{x_1^{k_1+1}\cdots x_n^{k_n+1}}=-\frac{\partial}{\partial x_1}x_1U_{n+1}(x_1,x_1,J_n),
\end{equation}
\begin{multline} \label{eqC.2.15}
\sum_{k_1,\ldots,k_n\geq0}k_ik_j\frac{m_{k_1,k_i+k_j-1,I_n\setminus\{k_i,k_j\}}}{x_1^{k_1+1}\cdots x_n^{k_n+1}}
\\=\frac{\partial^2}{\partial x_i\partial x_j}\left\{\frac{x_iU_{n-1}(x_1,J_n\setminus\{x_i\})-x_jU_{n-1}(x_1,J_n\setminus\{x_j\})}{x_i-x_j}\right\},
\end{multline}
\begin{equation} \label{eqC.2.16}
\sum_{k_1,\ldots,k_n\geq0}k_1k_i\frac{m_{k_1+k_i-1,I_n\setminus\{k_i\}}}{x_1^{k_1+1}\cdots x_n^{k_n+1}}=\frac{\partial^2}{\partial x_1\partial x_i}\left\{\frac{x_1U_{n-1}(J_n)-x_iU_{n-1}(x_1,J_n\setminus\{x_i\})}{x_1-x_i}\right\},
\end{equation}
\begin{equation} \label{eqC.2.17}
\sum_{k_1,\ldots,k_n\geq0}k_i\sum_{p_1+p_2=k_i-1}\frac{m_{k_1,p_1,p_2,I_n\setminus\{k_i\}}}{x_1^{k_1+1}\cdots x_n^{k_n+1}}=-\frac{\partial}{\partial x_i}x_iU_{n+1}(x_1,x_i,J_n),
\end{equation}
\begin{multline} \label{eqC.2.18}
\sum_{k_1,\ldots,k_n\geq0}\frac{\mathcal{A}_{1,1,1}}{x_1^{k_1+1}\cdots x_n^{k_n+1}}=x_1^2U_{n+3}(x_1,x_1,x_1,x_1,J_n)
\\\hspace{4em}+\left(\frac{1}{2}x_1^2\frac{\partial^2}{\partial x_1^2}+2x_1\frac{\partial}{\partial x_1}+1\right)U_{n+1}(x_1,x_1,J_n)
\\+\lim_{\xi\to x_1}\left(x_1^2\frac{\partial^2}{\partial x_1^2}+2x_1\frac{\partial}{\partial x_1}\right)U_{n+1}(\xi,x_1,J_n),
\end{multline}
\begin{multline} \label{eqC.2.19}
\sum_{k_1,\ldots,k_n\geq0}\frac{\mathcal{A}_{1,1,i}}{x_1^{k_1+1}\cdots x_n^{k_n+1}}=\frac{\partial}{\partial x_i}\frac{x_1x_i}{x_1-x_i}\big\{U_{n+1}(x_1,x_1,x_1,J_n\setminus\{x_i\})-U_{n+1}(x_1,x_1,J_n)\big\}
\\+\frac{1}{2}\frac{\partial^3}{\partial x_1^2\partial x_i}\left\{\frac{x_1x_iU_{n-1}(x_1,J_n\setminus\{x_i\})-x_1^2U_{n-1}(J_n)}{x_1-x_i}\right\},
\end{multline}
\begin{equation} \label{eqC.2.20}
\sum_{k_1,\ldots,k_n\geq0}\frac{\mathcal{A}_{1,i,i}}{x_1^{k_1+1}\cdots x_n^{k_n+1}}=\frac{\partial}{\partial x_i}\left\{\frac{x_1x_iU_{n+1}(x_1,x_1,J_n)-x_i^2U_{n+1}(x_1,x_i,J_n)}{x_1-x_i}\right\},
\end{equation}
\begin{multline} \label{eqC.2.21}
\sum_{k_1,\ldots,k_n\geq0}\frac{\mathcal{A}_{i,1,i}}{x_1^{k_1+1}\cdots x_n^{k_n+1}}
\\=\left(x_i\frac{\partial}{\partial x_i}+1\right)\frac{\partial^2}{\partial x_1\partial x_i}\left\{\frac{x_iU_{n-1}(x_1,J_n\setminus\{x_i\})-x_1U_{n-1}(J_n)}{x_1-x_i}\right\},
\end{multline}
\begin{multline} \label{eqC.2.22}
\sum_{k_1,\ldots,k_n\geq0}\frac{\mathcal{A}_{i,i,i}}{x_1^{k_1+1}\cdots x_n^{k_n+1}}=\frac{1}{2x_1}\left(x_i\frac{\partial}{\partial x_i}+1\right)\frac{\partial^2}{\partial x_i^2}\left\{\frac{x_i^2U_{n-1}(x_1,J_n\setminus\{x_i\})-x_1x_iU_{n-1}(J_n)}{x_1-x_i}\right\}
\\+\frac{\partial}{\partial x_i}\frac{x_i^2}{x_1-x_i}\big\{U_{n+1}(x_1,x_i,J_n)-U_{n+1}(x_i,x_i,J_n)\big\},
\end{multline}
\begin{multline} \label{eqC.2.23}
\sum_{k_1,\ldots,k_n\geq0}\frac{\mathcal{A}_{1,i,j}}{x_1^{k_1+1}\cdots x_n^{k_n+1}}=\frac{\partial^2}{\partial x_i\partial x_j}x_ix_j\left\{\frac{U_{n-1}(x_1,J_n\setminus\{x_i\})}{(x_1-x_j)(x_i-x_j)}-\frac{U_{n-1}(x_1,J_n\setminus\{x_j\})}{(x_1-x_i)(x_i-x_j)}\right.
\\+\left.\frac{U_{n-1}(x_1,x_1,J_n\setminus\{x_i,x_j\})}{(x_1-x_i)(x_1-x_j)}\right\},
\end{multline}
\begin{multline} \label{eqC.2.24}
\sum_{k_1,\ldots,k_n\geq0}\frac{\mathcal{A}_{i,1,j}}{x_1^{k_1+1}\cdots x_n^{k_n+1}}=\frac{\partial^2}{\partial x_i\partial x_j}\frac{x_ix_j}{(x_1-x_i)(x_1-x_j)}
\\\hspace{9em}\times\big\{U_{n-1}(x_1,x_1,J_n\setminus\{x_i,x_j\})-U_{n-1}(x_1,J_n\setminus\{x_j\})
\\+U_{n-1}(J_n)-U_{n-1}(x_1,J_n\setminus\{x_i\})\big\},
\end{multline}
\begin{multline} \label{eqC.2.25}
\sum_{k_1,\ldots,k_n\geq0}\frac{\mathcal{A}_{i,i,j}}{x_1^{k_1+1}\cdots x_n^{k_n+1}}=\frac{\partial^2}{\partial x_i\partial x_j}x_ix_j\left\{\frac{U_{n-1}(J_n)}{(x_1-x_j)(x_i-x_j)}-\frac{U_{n-1}(x_i,J_n\setminus\{x_j\})}{(x_1-x_i)(x_i-x_j)}\right.
\\+\left.\frac{U_{n-1}(x_1,J_n\setminus\{x_j\})}{(x_1-x_i)(x_1-x_j)}\right\},
\end{multline}
\begin{multline} \label{eqC.2.26}
\sum_{k_1,\ldots,k_n\geq0}\frac{\mathcal{A}_{i,j,i}}{x_1^{k_1+1}\cdots x_n^{k_n+1}}=\frac{\partial^2}{\partial x_i\partial x_j}\frac{x_ix_j}{(x_1-x_i)(x_i-x_j)}
\\\hspace{6em}\times\big\{U_{n-1}(x_1,J_n\setminus\{x_j\})-U_{n-1}(x_i,J_n\setminus\{x_j\})
\\+U_{n-1}(J_n)-U_{n-1}(x_1,J_n\setminus\{x_i\})\big\},
\end{multline}
\begin{multline} \label{eqC.2.27}
\sum_{k_1,\ldots,k_n\geq0}\frac{\mathcal{A}_{i,j,h}}{x_1^{k_1+1}\cdots x_n^{k_n+1}}
\\\hspace{-4em}=\frac{1}{x_1}\frac{\partial^3}{\partial x_i\partial x_j\partial x_h}\left\{\frac{x_ix_jx_hU_{n-3}(x_1,J_n\setminus\{x_i,x_j,x_h\})}{(x_1-x_i)(x_1-x_j)(x_1-x_h)}-\frac{x_1x_jx_hU_{n-3}(J_n\setminus\{x_j,x_h\})}{(x_1-x_i)(x_i-x_j)(x_i-x_h)}\right.
\\+\left.\frac{x_1x_ix_hU_{n-3}(J_n\setminus\{x_i,x_h\})}{(x_1-x_j)(x_i-x_j)(x_j-x_h)}-\frac{x_1x_ix_jU_{n-3}(J_n\setminus\{x_i,x_j\})}{(x_1-x_h)(x_i-x_h)(x_j-x_h)}\right\}.
\end{multline}
Equation \eqref{eqC.2.13} is simply equation \eqref{eqC.1.4}, while equation \eqref{eqC.2.14} can be obtained by applying the operator $-\frac{\partial}{\partial x_1}x_1$ to both sides of equation \eqref{eqC.1.3}. Likewise, equation \eqref{eqC.2.17} can be obtained by first interchanging $x_1\leftrightarrow x_i$ in both sides of equation \eqref{eqC.1.3} before applying $-\frac{\partial}{\partial x_i}x_i$ to both sides of said equation. Equation \eqref{eqC.2.16} is proven in a similar way to equation \eqref{eqC.1.6}, with steps relating to the factor $[k_i+1]_{\mathrm{mod}\,2}$ being skipped. All other computations \eqref{eqC.2.15}--\eqref{eqC.2.27}, bar a caveat on equation \eqref{eqC.2.18}, can be proven using the methods demonstrated in Appendix \ref{appendixC.1} above and the reasonings just given. We now conclude this appendix with a partial proof of equation \eqref{eqC.2.18}:

The last term on the right-hand side of equation \eqref{eqC.2.1} stands out in that it cannot be treated in the same manner as the other expressions considered in this appendix. Multiplying this term by $x_1^{-k_1-1}$, summing over $k_1\geq0$, and simplifying appropriately shows that 
\begin{align*}
\sum_{k_1\geq0}2\sum_{p_1+p_2+p_3=k_1-1}p_1\frac{m_{p_1+p_2,p_3,I_n}}{x_1^{k_1+1}}&=2\sum_{k_1\geq0}\sum_{p_1=0}^{k_1-1}\sum_{p_2=0}^{k_1-p_1-1}p_1\frac{m_{p_1+p_2,k_1-p_1-p_2-1,I_n}}{x_1^{k_1+1}}
\\&=2\sum_{p_1=0}^{\infty}\sum_{k_1=p_1+1}^{\infty}\sum_{p_2=0}^{k_1-p_1-1}p_1\frac{m_{p_1+p_2,k_1-p_1-p_2-1,I_n}}{x_1^{k_1+1}}
\\&=2\sum_{p_1=0}^{\infty}\sum_{k_1=0}^{\infty}\sum_{p_2=0}^{k_1}p_1\frac{m_{p_1+p_2,k_1-p_2,I_n}}{x_1^{k_1+p_1+2}}
\\&=2\sum_{p_1,p_2=0}^{\infty}\sum_{k_1=p_2}^{\infty}p_1\frac{m_{p_1+p_2,k_1-p_2,I_n}}{x_1^{k_1+p_1+2}}
\\&=2\sum_{p_1,p_2,k_1=0}^{\infty}p_1\frac{m_{p_1+p_2,k_1,I_n}}{x_1^{k_1+p_1+p_2+2}}
\\&=2\sum_{q,k_1=0}^{\infty}\sum_{p_1+p_2=q}p_1\frac{m_{q,k_1,I_n}}{x_1^{k_1+1}x_1^{q+1}}
\\&=\sum_{q,k_1=0}^{\infty}q(q+1)\frac{m_{q,k_1,I_n}}{x_1^{k_1+1}x_1^{q+1}}.
\end{align*}
Further multiplying this result by $x_2^{-k_2-1}\cdots x_n^{-k_n-1}$ and summing over $k_2,\ldots,k_n\geq0$ then shows that
\begin{multline*}
\sum_{k_1,\ldots,k_n\geq0}2\sum_{p_1+p_2+p_3=k_1-1}p_1\frac{m_{p_1+p_2,p_3,I_n}}{x_1^{k_1+1}\cdots x_n^{k_n+1}}=\sum_{q,k_1,\ldots,k_n\geq0}q(q+1)\frac{m_{q,k_1,I_n}}{x_1^{k_1+1}x_1^{q+1}x_2^{k_2+1}\cdots x_n^{k_n+1}}
\\=\lim_{\xi\to x_1}\left(x_1^2\frac{\partial^2}{\partial x_1^2}+2x_1\frac{\partial}{\partial x_1}\right)\sum_{k_1,\ldots,k_n\geq0}\frac{m_{q,k_1,I_n}}{\xi^{k_1+1}x_1^{q+1}x_2^{k_2+1}\cdots x_n^{k_n+1}}.
\end{multline*}
The final expression here is manifestly the third line of equation \eqref{eqC.2.18}.

\chapter{Evaluation of Some Cumulants at Leading Order}\label{appendixD}
\section{Evaluation of $c_{k_1}^{(\mathcal{J}_1),0}$, $c_{k_1}^{(\mathcal{J}_1),1}$, and $c_{k_1,k_2}^{(\mathcal{J}_1),0}$} \label{appendixD.1}
As discussed in \S\ref{s4.2.3}, setting $x(z_i)=1/(z_i^3+z_i)$ \eqref{eq4.2.50} in the residue formula \eqref{eq4.2.57}
\begin{equation} \label{eqD.1.1}
c_{k_1,\ldots,k_n}^{(\mathcal{J}_1),l}=(-1)^n\underset{z_1,\ldots,z_n=0}{\mathrm{Res}}W_n^{(\mathcal{J}_1),l}(x(z_1),\ldots,x(z_n))\prod_{i=1}^nx(z_i)^{k_i}x'(z_i)\,\mathrm{d}z_i
\end{equation}
and substituting in $W_1^{(\mathcal{J}_1),0}(x(z_1))=z_1$ along with the specifications \eqref{eq4.2.53} and \eqref{eq4.2.52} of, respectively, $W_1^{(\mathcal{J}_1),1}(x(z_1))$ and $W_2^{(\mathcal{J}_1),0}(x(z_1),x(z_2))$ enables the computation of the cumulant genus expansion coefficients $c_{k_1}^{(\mathcal{J}_1),0}$, $c_{k_1}^{(\mathcal{J}_1),1}$, and $c_{k_1,k_2}^{(\mathcal{J}_1),0}$ defined implicitly through equation \eqref{eq4.2.54}.

Letting $\tilde{\mathcal{J}}_1$ be drawn from the global scaled $(N,N)$ antisymmetrised Laguerre ensemble defined in Propositions \ref{prop3.11} and \ref{prop3.13}, we have from equations \eqref{eq1.1.27} and \eqref{eq1.1.28} that
\begin{align}
c_{k_1}^{(\mathcal{J}_1),0}&=\lim_{N\to\infty}\frac{1}{N}\mean{\Tr\,\tilde{\mathcal{J}}_1^{k_1}}, \label{eqD.1.2}
\\ c_{k_1}^{(\mathcal{J}_1),1}&=\lim_{N\to\infty}\left(\mean{\Tr\,\tilde{\mathcal{J}}_1^{k_1}}-Nc_{k_1}^{(\mathcal{J}_1),0}\right); \label{eqD.1.3}
\end{align}
the aforementioned propositions also give combinatorial interpretations of these cumulant expansion coefficients as signed enumerations of particular ribbon graphs and topological hypermaps (see also Figure \ref{fig4.3}). Due to the antisymmetric nature of $\tilde{\mathcal{J}}_1$, the cumulant expansion coefficients $c_{k_1}^{(\mathcal{J}_1),0}$ and $c_{k_1}^{(\mathcal{J}_1),1}$ vanish for odd values of $k_1$, while for $k_1$ even, the former relates to the $m=2$ Fuss--Catalan numbers \eqref{eq1.3.17} through the relation \eqref{eq4.2.47}
\begin{equation} \label{eqD.1.4}
c_{k_1}^{(\mathcal{J}_1),0}=\tilde{M}_{k_1,0}^{(\mathcal{J}_1)}=\mathrm{i}^{k_1}m_{k_1/2}^{(FC_2)}=\frac{\mathrm{i}^{k_1}}{k_1+1}\binom{3k_1/2}{k_1/2}.
\end{equation}
We now present the values of $c_{k_1}^{(\mathcal{J}_1),0}$ and $c_{k_1}^{(\mathcal{J}_1),1}$ computed through the residue formula \eqref{eqD.1.1} with $0\leq k_1\leq 18$ even --- our calculations are consistent with equation \eqref{eqD.1.4} above.

\begin{center}
\small
\begin{tabular}{c|c|c|c|c|c|c|c|c|c|c} 
$k_1$&$0$&$2$&$4$&$6$&$8$&$10$&$12$&$14$&$16$&$18$ \\ \hline
$c_{k_1}^{(\mathcal{J}_1),0}$&$1$&$-1$&$3$&$-12$&$55$&$-273$&$1\,428$&$-7\,752$&$43\,263$&$-246\,675$ \\ \hline
$c_{k_1}^{(\mathcal{J}_1),1}$&$0$&$1$&$-5$&$28$&$-165$&$1\,001$&$-6\,188$&$38\,760$&$-245\,157$&$1\,562\,275$
\end{tabular}
\end{center}
Comparing the second row of the above table to equation \eqref{eqD.1.4} suggests that for $k_1\in2\mathbb{N}$,
\begin{multline*}
c_{k_1}^{(\mathcal{J}_1),1}=-\mathrm{i}^{k_1}\binom{3k_1/2-1}{k_1/2-1}+\chi_{k_1=0}=\frac{\chi_{k_1=0}}{3}-\frac{(k_1+1)}{3}c_{k_1}^{(\mathcal{J}_1),0}
\\\iff W_1^{(\mathcal{J}_1),1}(x(z_1))=\frac{x(z_1)}{3x'(z_1)}\frac{\partial}{\partial z_1}W_1^{(\mathcal{J}_1),0}(x(z_1))+\frac{1}{3x(z_1)}=\frac{x(z_1)}{3x'(z_1)}+\frac{1}{3x(z_1)},
\end{multline*}
which is precisely in keeping with equation \eqref{eq4.2.53}.

For $n=2$, the analogue of equations \eqref{eqD.1.2} and \eqref{eqD.1.3} is
\begin{equation} \label{eqD.1.5}
c_{k_1,k_2}^{(\mathcal{J}_1),0}=\lim_{N\to\infty}\left(\mean{\Tr\,\tilde{\mathcal{J}}_1^{k_1}\Tr\,\tilde{\mathcal{J}}_1^{k_2}}-\mean{\Tr\,\tilde{\mathcal{J}}_1^{k_1}}\mean{\Tr\,\tilde{\mathcal{J}}_1^{k_2}}\right);
\end{equation}
see Propositions \ref{prop3.11} and \ref{prop3.13} in addition to Figure \ref{fig4.4} for combinatorial interpretations of this cumulant expansion coefficient. Note that the right-hand side of equation \eqref{eqD.1.5} is identically zero whenever either of $k_1,k_2$ are odd, since $\tilde{\mathcal{J}}_1$ is antisymmetric, and also whenever either of $k_1,k_2$ equal zero, since $\Tr\,\tilde{\mathcal{J}}_1^0=N$ factors out of the first covariance therein. Thus, we present the values of $c_{k_1,k_2}^{(\mathcal{J}_1),0}$ computed through equation \eqref{eqD.1.1} with $2\leq k_1,k_2\leq 14$ even --- the asterisks represent redundant data, as $c_{k_2,k_1}^{(\mathcal{J}_1),0}=c_{k_1,k_2}^{(\mathcal{J}_1),0}$.

\begin{center}
\small
\begin{tabular}{c c|c|c|c|c|c|c|c}
&$k_1$&$2$&$4$&$6$&$8$&$10$&$12$&$14$
\\$k_2$&$c_{k_1,k_2}^{(\mathcal{J}_1),0}$&&&&&&&
\\ \hline $2$&&$12$&$-80$&$504$&$-3\,168$&$20\,020$&$-127\,296$&$813\,960$
\\ \hline $4$&&$*$&$600$&$-4\,032$&$26\,400$&$-171\,600$&$1\,113\,840$&$-7\,235\,200$
\\ \hline $6$&&$*$&$*$&$28\,224$&$-190\,080$&$1\,261\,260$&$-8\,316\,672$&$54\,698\,112$
\\ \hline $8$&&$*$&$*$&$*$&$1\,306\,800$&$-8\,808\,800$&$58\,810\,752$&$-390\,700\,800$
\\ \hline $10$&&$*$&$*$&$*$&$*$&$60\,120\,060$&$-405\,437\,760$&$2\,715\,913\,200$
\\ \hline $12$&&$*$&$*$&$*$&$*$&$*$&$2\,756\,976\,768$&$-18\,597\,358\,080$
\\ \hline $14$&&$*$&$*$&$*$&$*$&$*$&$*$&$126\,196\,358\,400$
\end{tabular}
\end{center}
Comparing this table to Table 1 of \citep{DF20} suggests that $c_{k_1,k_2}^{(\mathcal{J}_1),0}$ relates to the analogous quantity $c_{k_1,k_2}^{(c\mathcal{W}_2),0}$ of the $(N,N,N)$ complex Wishart product ensemble (Definition \ref{def1.12}) according to
\begin{equation}
c_{2k_1,2k_2}^{(\mathcal{J}_1),0}=4\,(-1)^{k_1+k_2}\,c_{k_1,k_2}^{(c\mathcal{W}_2),0},\qquad k_1,k_2\in\mathbb{N},
\end{equation}
which is in agreement with Proposition \ref{prop1.8}.

\setcounter{equation}{0}
\section{Evaluation of $c_{k_1}^{(\mathcal{H}_1),0}$, $c_{k_1}^{(\mathcal{H}_1),2}$, and $c_{k_1,k_2}^{(\mathcal{H}_1),0}$} \label{appendixD.2}
In parallel to Appendix \ref{appendixD.1}, let us recall from \S\ref{s4.3.3} that setting $x(z_i)=(z_i^2+1)^2/z_i$ \eqref{eq4.3.34} in the residue formula \eqref{eq4.3.38}
\begin{equation} \label{eqD.2.1}
c_{k_1,\ldots,k_n}^{(\mathcal{H}_1),l}=(-1)^n\underset{z_1,\ldots,z_n=0}{\mathrm{Res}}W_n^{(\mathcal{H}_1),l}(x(z_1),\ldots,x(z_n))\prod_{i=1}^nx(z_i)^{k_i}x'(z_i)\,\mathrm{d}z_i
\end{equation}
and then substituting in $W_1^{(\mathcal{H}_1),0}(x(z_1))=y(z_1)$ \eqref{eq4.3.33} along with the expressions for $W_1^{(\mathcal{H}_1),2}(x(z_1))$ and $W_2^{(\mathcal{H}_1),0}(x(z_1),x(z_2))$ given respectively in equations \eqref{eq4.3.36} and \eqref{eq4.3.35} gives us a means to compute the coefficients $c_{k_1}^{(\mathcal{H}_1),0}$, $c_{k_1}^{(\mathcal{H}_1),2}$, and $c_{k_1,k_2}^{(\mathcal{H}_1),0}$ of the genus expansions \eqref{eq4.3.32} of the associated mixed cumulants $\tilde{c}_{k_1,\ldots,k_n}^{(\mathcal{H}_1)}$.

With $\tilde{\mathcal{H}}_1$ drawn from the global scaled $(N,N)$ Hermitised Laguerre ensemble defined in Propositions \ref{prop3.10} and \ref{prop3.12}, the relevant analogues of equations \eqref{eqD.1.2}, \eqref{eqD.1.3}, \eqref{eqD.1.5} are
\begin{align}
c_{k_1}^{(\mathcal{H}_1),0}&=\lim_{N\to\infty}\frac{1}{N}\mean{\Tr\,\tilde{\mathcal{H}}_1^{k_1}}, \label{eqD.2.2}
\\ c_{k_1}^{(\mathcal{H}_1),2}&=\lim_{N\to\infty}\left(N\mean{\Tr\,\tilde{\mathcal{H}}_1^{k_1}}-N^2c_{k_1}^{(\mathcal{H}_1),0}\right), \label{eqD.2.3}
\\ c_{k_1,k_2}^{(\mathcal{H}_1),0}&=\lim_{N\to\infty}\left(\mean{\Tr\,\tilde{\mathcal{H}}_1^{k_1}\Tr\,\tilde{\mathcal{H}}_1^{k_2}}-\mean{\Tr\,\tilde{\mathcal{H}}_1^{k_1}}\mean{\Tr\,\tilde{\mathcal{H}}_1^{k_2}}\right). \label{eqD.2.4}
\end{align}
Here, we have used the fact (recall the discussion following the proof of Proposition \ref{prop3.10}) that the $l=1$ coefficient $c_{k_1}^{(\mathcal{H}_1),1}$ within the expansion \eqref{eq4.3.32} is zero. Like in the case of the antisymmetrised Laguerre ensemble, the cumulant expansion coefficients \eqref{eqD.2.2}--\eqref{eqD.2.4} have combinatorial interpretations as enumerations of certain ribbon graphs and topological hypermaps due to Propositions~\ref{prop3.10} and \ref{prop3.12} (see also Figures~\ref{fig3.22} and~\ref{fig4.5}).

Recall from Lemma \ref{L3.5} that $c_{k_1}^{(\mathcal{H}_1)}$, consequently $\tilde{c}_{k_1}^{(\mathcal{H}_1)}$ and, in turn, $c_{k_1}^{(\mathcal{H}_1),0}$ and $c_{k_1}^{(\mathcal{H}_1),2}$, vanish whenever $k_1$ is odd. Hence, we display the values of these expansion coefficients obtained through equation \eqref{eqD.2.1} for $0\leq k_1\leq18$ even.

\begin{center}
\small
\begin{tabular}{c|c|c|c|c|c|c|c|c|c|c} 
$k_1$&$0$&$2$&$4$&$6$&$8$&$10$&$12$&$14$&$16$&$18$ \\ \hline
$c_{k_1}^{(\mathcal{H}_1),0}$&$1$&$1$&$4$&$22$&$140$&$969$&$7\,084$&$53\,820$&$420\,732$&$3\,362\,260$ \\ \hline
$c_{k_1}^{(\mathcal{H}_1),2}$&$0$&$1$&$34$&$645$&$9\,828$&$133\,620$&$1\,694\,154$&$20\,490\,470$&$239\,545\,800$&$2\,729\,482\,668$
\end{tabular}
\end{center}

In keeping with the discussion below equation \eqref{eq4.3.27}, it is expected from the relation
\begin{equation}
W_1^{(\mathcal{H}_1),0}(x_1)=x_1W_1^{(FC_3)}(x_1^2)
\end{equation}
between the resolvents \eqref{eq1.1.18} of the limiting eigenvalue density $\rho^{(\mathcal{H}_1),0}(\lambda)$ of the global scaled $(N,N)$ Hermitised Laguerre ensemble (recall Definition \ref{def1.13}) and the $m=3$ Fuss--Catalan distribution \eqref{eq1.3.16}, which is a consequence of equation \eqref{eq1.3.22}, that for even integers $k_1$, $c_{k_1}^{(\mathcal{H}_1),0}$ relates to the $m=3$ Fuss--Catalan numbers \eqref{eq1.3.17} according to
\begin{equation}
c_{k_1}^{(\mathcal{H}_1),0}=m_{k_1/2}^{(FC_3)}=\frac{1}{3k_1/2+1}\binom{2k_1}{k_1/2}.
\end{equation}
It can readily be checked that this is in line with the values of $c_{k_1}^{(\mathcal{H}_1),0}$ given in the first row of the above table.

For the same reason as given below equation \eqref{eqD.1.5}, $c_{k_1,k_2}^{(\mathcal{H}_1),0}$ vanishes when either of $k_1,k_2$ are zero. Thus, we give evaluations of this cumulant expansion coefficient for $1\leq k_1,k_2\leq 9$ --- the asterisks have the same meaning as in the table for $c_{k_1,k_2}^{(\mathcal{J}_1),0}$ displayed earlier.

\begin{center}
\small
\begin{tabular}{c c|c|c|c|c|c|c|c|c|c}
&$k_1$&$1$&$2$&$3$&$4$&$5$&$6$&$7$&$8$&$9$
\\$k_2$&$c_{k_1,k_2}^{(\mathcal{H}_1),0}$&&&&&&&&&
\\ \hline $1$&&$2$&$0$&$15$&$0$&$120$&$0$&$1\,001$&$0$&$8\,568$
\\ \hline $2$&&$*$&$12$&$0$&$112$&$0$&$990$&$0$&$8\,736$&$0$
\\ \hline $3$&&$*$&$*$&$150$&$0$&$1\,350$&$0$&$12\,012$&$0$&$107\,100$
\\ \hline $4$&&$*$&$*$&$*$&$1\,176$&$0$&$11\,088$&$0$&$101\,920$&$0$
\\ \hline $5$&&$*$&$*$&$*$&$*$&$12\,960$&$0$&$120\,120$&$0$&$1\,101\,600$
\\ \hline $6$&&$*$&$*$&$*$&$*$&$*$&$108\,900$&$0$&$1\,029\,600$&$0$
\\ \hline $7$&&$*$&$*$&$*$&$*$&$*$&$*$&$1\,145\,144$&$0$&$10\,720\,710$
\\ \hline $8$&&$*$&$*$&$*$&$*$&$*$&$*$&$*$&$9\,937\,200$&$0$
\\ \hline $9$&&$*$&$*$&$*$&$*$&$*$&$*$&$*$&$*$&$101\,959\,200$
\end{tabular}
\end{center}

Observe that $c_{k_1,k_2}^{(\mathcal{H}_1),0}=0$ whenever $k_1+k_2$ is odd, which agrees precisely with what is known from Lemma \ref{L3.5}.
\end{appendices}
\end{document}